%% file: these.tex
\documentclass{these}
\usepackage[utf8]{inputenc} %spécification de l'encodage du fichier .tex, à
\usepackage[T1]{fontenc}
\usepackage{graphicx}
\usepackage{float}
\usepackage{vmargin}
\usepackage{amsmath}
\usepackage{amssymb}
\usepackage{amsthm}
\usepackage{amsfonts}
\usepackage{indentfirst}
\usepackage{upgreek}
\usepackage[font=small,labelfont=bf,up]{caption}
\usepackage{pstricks}
\usepackage[loose,bf]{subfigure}
\usepackage{multirow}
\usepackage{array}
\usepackage{ae}
\usepackage{bbm}
\usepackage{bm}
\usepackage{cite}     % pour les citations multiples

\newcommand{\icomp}{i}
\newcommand{\sint}{\sin\theta}
\newcommand{\p}[1]{\left({#1}\right)}
\newcommand{\pq}[1]{\left[{#1}\right]}
\newcommand{\ut}{u_\theta}
\newcommand{\up}{u_\phi}
\newcommand{\ulm}{u_{lm}}
\newcommand{\Ylm}{Y_l^m}
\newcommand{\average}[1]{\left<{#1}\right>}
\newcommand{\Hl}{\widetilde H_l}
\newcommand{\bs}{\bar\sigma}
\newcommand{\rr}{\mathbf{r}}
\newcommand{\derpart}[2]{\frac{\partial #1}{\partial #2}}

\newcommand{\diagH}[1]{
\begin{tabular}{c}
 \includegraphics[angle=0,scale=.5]{#1}
\end{tabular}
}

\newcommand{\diaggd}[1]{
\begin{tabular}{c}
 \includegraphics[angle=0,scale=.18]{#1}
\end{tabular}
}

\newcommand{\diagpq}[1]{
\begin{tabular}{c}
 \includegraphics[angle=0,scale=.115]{#1}
\end{tabular}
}

\newcommand{\diag}[1]{
\begin{tabular}{c}
 \includegraphics[angle=0,scale=.13]{#1}
\end{tabular}
}

\title{Forces et fluctuations en\\ \vspace{0.2cm}
  membranes planes, sphériques et\\ \vspace{0.4cm}
  tubulaires}
\author{Camilla BARBETTA}
\date{07/09/2010}
\labo{Matière et Systèmes Complexes}
\school{Matière Condensée et Interfaces}
\speciality{Physique théorique}
\university{l'UNIVERSITÉ PARIS.DIDEROT (Paris7)}{\Large{UNIVERSITÉ PARIS.DIDEROT (Paris7)}}
\advisor[M]{Jean-Baptiste FOURNIER} %Directeur de thèse
\coadvisor[M]{} %Co-directeur de thèse, si nécessaire
\jury{  %définition des membres du jury
\jurymember[Mme]{Patricia BASSEREAU}{Président du jury}\\
\jurymember[M]{Andrea PARMEGGIANI}{Rapporteur}\\
\jurymember[M]{Dominique MOUHANNA}{Rapporteur}\\
\jurymember[M]{Laurent LIMOZIN}{Examinateur}\\
\jurymember[M]{Alberto IMPARATO}{Examinateur}\\
\juryadvisor\\    % inclut le directeur de thèse dans le jury
}

\begin{document}
\frontmatter
\maketitle

\input{resume.tex}
\input{preamble.tex}
\input{remerciements.tex}
\tableofcontents

\mainmatter
\input{introduction.tex}
\part{Planar membranes}
\input{chap1.tex}
\input{chap11.tex}

\part{Vesicles}
\label{Partie_2}
\input{chap2.tex}

\part{Nanotubes of membrane}
\input{chap3.tex}
\input{chap4.tex}

\part{Preliminar works on a 2-D simulation}
\input{chap5.tex}

\input{conclusion.tex}

%\nocite{*}

\appendix
\input{appendix1.tex}

\input{appendix2.tex}
\input{appendix21.tex}

\input{appendix3.tex}
\input{appendix4.tex}
\input{appendix5.tex}

\bibliography{biblio}

\backmatter
%\makeback
\end{document}

%% file: resume.tex
\resume{
Les membranes lipidiques constituent des mat\'eriaux tr\`es particuliers: d'une part, elles sont tr\`es peu r\'esistantes aux \'etirements microscopiques; d'autre part elles sont extr\^emement flexibles, pr\'esentant des d\'eformations m\^eme \`a des petites \'echelles.
En cons\'equence, une portion de membrane poss\`ede un exc\`es d'aire relatif \`a l'aire optiquement visible, qu'on appelle l'aire projet\'ee.
D'un point de vue m\'ecanique, on peut alors distinguer trois tensions associ\'ees aux membranes lipidiques: la tension m\'ecanique effective $\tau$, associ\'ee \`a l'augmentation de l'aire projet\'ee et au lissage des fluctuations; la tension $\sigma$, associ\'ee \`a l'aire microscopique de la membrane et donc non-mesurable, mais couramment utilis\'ee dans les pr\'edictions th\'eoriques; et son \'equivalent macroscopique mesur\'e \`a travers du spectre des fluctuations, $r$.
Jusqu'au moment, pour interpr\'eter les donn\'ees exp\'erimentales, on suppose l'\'egalit\'e entre ces quantit\'es.
Dans cette th\`ese, nous avons \'etudi\'e, en utilisant le tenseur des contraintes projet\'e, si et sous quelles conditions il est justifi\'e d'assumer $\tau = \sigma$.
Nous avons \'etudi\'e trois g\'eom\'etries (planaire, sph\'erique et cylindrique) et obtenu la relation $\tau \approx \sigma - \sigma_0$, o\`u $\sigma_0$ est une constante qui d\'epend seulement du plus grand vecteur d'onde de la membrane et de la temp\'erature.
En cons\'equence, nous concluons que n\'egliger la diff\'erence entre $\tau$ et $\sigma$ est justifiable seulement pour des membranes sous grande tension: pour des tensions faibles, il faut consid\'erer des corrections.
Nous avons \'etudi\'e les implications de ce r\'esultat \`a l'interpr\'etation des exp\'eriences d'extraction de nanotubes de membrane.
En ce que concerne $r$, nous questionnons une d\'emonstration pr\'ec\'edente de son \'egalit\'e avec $\tau$. 
Finalement, la fluctuation des forces pour les membranes planes et pour des nanotubes de membranes a \'et\'e quantifi\'ee pour la premi\`ere fois. 
}{
Lipid membranes constitute very particular materials: on the one hand, they break very easily under microscopical stretching; on the other hand, they are extremely flexible, presenting deformations even at small scales.
Consequently, a piece of membrane has an area excess relative to its optically resolvable area, also called projected area.
From a mechanical point of view, we can thus identify three tensions associated to lipid membranes: the mechanical effective tension $\tau$, associated to an increase in the projected area and to the flattening of the fluctuations; the tension $\sigma$, associated to the microscopical area of the membrane and thus non measurable, but commonly used in theoretical predictions; and its macroscopical counterpart measured through the fluctuation spectrum, $r$.
Up to now, the equality between these quantities was taken for granted when analyzing experimental data.
In this dissertation, we have studied, using the projected stress tensor, whether and under which conditions it is justified to assume $\tau = \sigma$.
We studied three geometries (planar, spherical and cylindrical) and obtained the relation $\tau \approx \sigma - \sigma_0$, where $\sigma_0$ is a constant depending only on the membrane's high frequency cutoff and on the temperature.
Accordingly, we conclude that neglecting the difference between $\tau$ and $\sigma$ is justifiable only to membranes under large tensions: in the case of small tensions, corrections must be taken into account.
We have studied the implications of this result to the interpretation of experiments involving membrane nanotubes.
Regarding $r$, we have questioned a former demonstration concerning its equality with $\tau$.
Finally, the force fluctuation for planar membranes and membrane nanotubes was quantified for the first time. 
}

%% file: preamble.tex
\chapter*{Pr\'eambule}

Les membranes biologiques sont constitu\'ees principalement par des mol\'ecules lipidiques amphiphiles, i. e., des mol\'ecules qui poss\`edent un groupe hydrophile et un groupe hydrophobe.
Le caract\`ere amphiphile de ces mol\'ecules donne des propri\'et\'es tr\`es particuli\`eres aux membranes lipidiques: d'une part, leur coh\'esion est assur\'ee principalement par des interactions hydrophobes-hydrophiles entre les lipides et l'eau et d'autre par, la membrane pr\'esente une rigidit\'e de courbure.
Le fait que les membranes s'associent surtout par r\'epulsion \`a l'eau et non par des liaisons chimiques a pour cons\'equence la grande mobilit\'e de ces mol\'ecules \`a l'int\'erieur de la membrane (ce qui explique l'expression \textit{liquide bi-dimensionnel} fr\'equemment utilis\'ee pour d\'esigner de membranes lipidiques) et une grande fragilit\'e \`a des \'etirements au niveau mol\'eculaire.
Une autre cons\'equence est leur grande flexibilit\'e: les membranes lipidiques sont facilement d\'eformables, m\^eme \`a des \'echelles plus petites que celles accessibles exp\'erimentalement.
Une portion de membrane poss\`ede alors un exc\`es d'aire relatif \`a l'aire optiquement discernable, qu'on appelle l'aire projet\'ee.

Imaginons maintenant une exp\'erience: une portion de membrane est attach\'ee \`a un cadre.
Une force lat\'erale $\tau$ est appliqu\'ee sur le cadre de fa\c{c}on \`a augmenter l'aire projet\'ee de la membrane.
La force $\tau$ correspond \`a la force n\'ecessaire pour lisser les fluctuations de la membrane.
Elle est donc d'origine purement entropique et nous l'appelons alors tension m\'ecanique effective.
Exp\'erimentalement, elle correspond \`a la tension appliqu\'ee \`a travers des micropipettes sur des v\'esicules, par exemple.
Malheureusement, les pr\'edictions th\'eoriques concernent normalement la tension $\sigma$ associ\'ee \`a l'aire microscopique de la membrane, qui n'est pas mesurable.
En effet, \`a travers l'analyse du spectre de fluctuation des membranes, il est uniquement possible de mesurer l'\'equivalent macroscopique de $\sigma$, la tension $r$.
En cons\'equence, lors de l'interpr\'etation des r\'esultats exp\'erimentaux, l'\'egalit\'e entre ces tensions est couramment admise.
Dans cette dissertation, j'ai alors examin\'e, sous la direction de Jean-Baptiste Fournier, si et sous quelles conditions cette hypoth\`ese est valable pour des g\'eom\'etries diverses.

Dans la premi\`ere partie de cette th\`ese, nous \'etudions le cas d'une pi\`ece plane de membrane (r\'esultats pr\'esent\'es dans le chapitre \ref{chapitre/planar_membrane} et publi\'es en~\cite{Fournier_08_eu}).
Dans la litt\'erature scientifique, nous trouvons quelques relations entre $\tau$, $\sigma$ et $r$, sans qu'il aie un consensus~\cite{Cai_94},~\cite{Imparato_06}.
Toutes ces d\'erivations, cependant, partent de l'\'energie libre, ce qui peut-\^etre tr\`es subtile, comme on discutera dans la section~\ref{subsection_1_free_energy}.
Nous avons alors utilis\'e comme base de nos calculs le tenseur des contraintes projet\'e, un outil d\'evelopp\'e r\'ecemment~\cite{Fournier_07} et introduit en~\ref{section_projected_stress}, \`a partir duquel on a pu obtenir directement la tension $\tau$ en fonction de $\sigma$:

\begin{equation}
\tau \simeq \sigma - \sigma_0 \, ,
\label{relat}
\end{equation}

\noindent o\`u $\sigma_0$ est une constante qui ne d\'epend que de la temp\'erature et du plus grand vecteur d'onde de la membrane.
Ainsi, il est justifiable de consid\'erer $\tau \approx \sigma$ seulement pour des fortes tensions.
Dans ce chapitre, nous questionnons aussi une ancienne d\'emonstration de l'\'egalit\'e entre $r$ et $\tau$~\cite{Cai_94}: nous attendons donc en g\'en\'eral $\tau \neq \sigma \neq r$.

\`a partir du tenseur des contraintes projet\'e, il est aussi possible d'examiner la corr\'elation des contraintes sur la membrane.
Ces r\'esultats in\'edits sont pr\'esent\'es dans chapitre~\ref{Fluct_plan} et indiquent que ces corr\'elations d\'ecroissent tr\`es rapidement, ind\'ependemment de la tension de la membrane.
Les calculs d\'evelopp\'es dans ce chapitre ont \'et\'e fondamentaux pour introduire et ma\^itriser une repr\'esentation diagrammatique des moyennes propos\'ee par nous et inspir\'ee des diagrammes de Feynmann.
Ces outils sont repris dans le chapitre~\ref{Fluct_TUBE} et simplifient grandement les \'evaluations.

Exp\'erimentalement, une pi\`ece planaire de membrane est difficile \`a manipuler.
Plus populairement, des v\'esicules de grande taille qui peuvent \^etre manipul\'ees avec de micropipettes sont utilis\'ees en laboratoire.
Nous avons alors \'etudi\'e le cas des v\'esicules quasi-sph\'eriques ferm\'ees (dont le volume interne est fixe) et perc\'ees (dont le volume n'est pas contraint) dans la partie~\ref{Partie_2} de cette th\`ese (r\'esultats obtenus en collaboration avec Alberto Imparato et publi\'es sur~\cite{Barbetta_10}).
Apr\`es avoir d\'eriv\'e le tenseur des contraintes pour cette g\'eom\'etrie, nous concluons que la diff\'erence entre $\tau$ et $\sigma$ est bien approxim\'ee aussi dans le cas des v\'esicules sph\'eriques (perc\'ees ou non) par la relation montr\'e dans l'eq.(\ref{relat}).
Une cons\'equence int\'eressante est la possibilit\'e d'avoir une v\'esicule dont la pression interne est plus petite que la pression externe, ce qui est impossible pour le cas d'une goutte de liquide.

Dans la troisi\`eme partie de la th\`ese, nous examinons les cons\'equences de nos r\'esultats pour les exp\'eriences d'extraction de nanotube de membrane (r\'esultats publi\'es en \cite{Barbetta_09}).
Dans ces exp\'eriences, une bille en verre, par exemple, est attach\'ee \`a la membrane.
Avec une pince optique, une force est appliqu\'ee \`a la bille et un nanotube de membrane est extrait.
Jusqu'au moment, ces exp\'eriences ont \'et\'e interpr\'et\'ees en supposant la validit\'e de deux hypoth\`eses: la diff\'erence entre entre $\tau$ et $\sigma$ est n\'egligeable et les fluctuations thermiques des tubes peuvent \^etre \'egalement n\'eglig\'ees.
Dans ces conditions, une relation tr\`es simple relie la force appliqu\'ee par la pince optique \`a la tension de la membrane.
Il a \'et\'e cependant montr\'e que ces fluctuations sont tr\`es importantes d\^u \`a la pr\'esence de modes tr\`es peu \'energ\'etiques~\cite{Fournier_07a} (modes de Goldstone).
Nous avons d\'edi\'e le chapitre~\ref{TUBE} \`a l'\'etude les effets des fluctuations thermiques sur la relation entre la force et la tension de la membrane.
Curieusement, nous concluons que par co\"incidence, ces effets sont compens\'es par l'hypoth\`ese $\tau \approx \sigma$ dans le r\'egime de forte tension, justifiant \`a posteriori le traitement habituel des donn\'ees exp\'erimentales.

Avec le montage exp\'erimental utilis\'e pour extraire des tubes de membrane, nous pouvons non seulement mesurer la force n\'ecessaire pour l'extraire, mais aussi la d\'eviation quadratique moyenne de cette force dans la direction de l'axe du tube.
Cette quantit\'e pourrait fournir des informations suppl\'ementaires sur les caract\'eristiques m\'ecaniques des membranes.
Utilisant les outils diagrammatiques introduits pr\'ec\'edemment, nous \'evaluons alors la d\'eviation quadratique moyenne de la force n\'ecessaire pour extraire un tube dans le chapitre suivant.
Nous pr\'edisons une d\'ependance tr\`es faible de cette quantit\'e en fonction de la tension et de la rigidit\'e de courbure.
En cons\'equence, la fluctuation de la force est peu utile pour la caract\'erisation m\'ecanique des membranes.
Par contre, nous discutons dans ce chapitre une possible utilisation dans la caract\'erisation de l'activit\'e des pompes actives ins\'er\'ees dans les membranes.

Finalement, dans la derni\`ere partie, nous pr\'esentons les premiers r\'esultats concernant une simulation num\'erique propos\'ee par nous dans l'objectif de v\'erifier nos pr\'evisions.
Nous consid\'erons une pi\`ece planaire de membrane attach\'ee \`a un cadre circulaire.
La membrane a \'et\'e mod\'elis\'ee de fa\c{c}on tr\`es simplifi\'ee par des particules effectives reli\'ees par un r\'eseau triangulaire dont les liaisons se r\'earrangeaient pour assurer la liquidit\'e de la membrane.
L'objectif de cette simulation est de reproduire les conditions d'une vraie exp\'erience: la tension $\tau$ \'etait contr\^ol\'ee par la tension appliqu\'ee au cadre circulaire et le spectre de fluctuation pouvait \^etre mesur\'e.
En outre, dans ce cas, la tension interne $\sigma$ pouvait \^etre li\'e \`a l'extension des liens entre les particules effectives et donc estim\'ee.
D\^u \`a des contraintes de temps, nous pr\'esentons dans ce chapitre seulement quelques r\'esultats pr\'eliminaires.

%% file: remerciements.tex
\chapter*{Remerciements}

Initialement, je voudrais remercier Jean-Baptiste Fournier pour m'avoir permis de faire cette th\`ese sur un sujet que je trouve tr\`es int\'erressant.
J'ai beaucoup appr\'eci\'e le fait que mon travail soit th\'eorique, mais jamais tr\`es \'eloign\'e des applications/v\'erifications exp\'erimentales.
Merci pour m'avoir pouss\'e \`a m'am\'eliorer, surtout dans la clart\'e des mes textes - j'ai beaucoup r\^al\'e, mais j'ai beaucoup appris!

Je voudrais aussi remercier aux deux collaborateurs avec qui j'ai travaill\'e: Alberto Imparato et Luca Peliti.
Vous m'avez fait sentir suffisamment \`a l'aise pour dire \`a voix haute les id\'ees qui me traversaient la t\^ete.

Dans la suite, je remercie \`a tous mes incroyables coll\`egues de bureau et d'enseignement pour la bonne ambiance et les bons souvenirs:

\begin{itemize}
  \item Yann, pour sa patience toute orientale;
    \item Franck, pour son attitude ``no stress''; 
  \item Kristina, ma copine de bureau et amie. Je t'admire beaucoup!;
  \item Benoit, pour tous les d\'ejeuners et rigolades lors des go\^uters.
  \item Damien, un fran\c{c}ais tr\`es br\'esilien, pour m'avoir motiv\'e \`a participer \`a des activit\'es tr\`es fran\c{c}aises - les manifestations.
  Merci aussi pour ton implication dans l'enseignement et dans notre petit projet p\'edagogique;
\item G\'erald, pour les id\'ees assez pol\'emiques qui engendraient des grandes discussions;
\item Milad, pour sa compagnie dans notre trajet de bus;
\item Anne-Florence, pour sa bonne humeur, son talent culinaire et pour avoir arros\'e mes plantes;
\item Alexandre, pour les d\'ejeuners et caf\'es en langue maternelle.
\end{itemize}

\noindent Je remercie aussi tous les autres coll\`egues que je n'ai pas eu la chance de conna\^itre mieux, mais qui ont tous \'et\'e tr\`es sympas avec moi.

J'embrasse aussi tr\`es fort le groupe fid\`ele d'amis que j'ai fait pendant mon master: Sophie Aimon, Fabien et C\'eline Paillusson, Thomas Gu\'erin et Fanny, Laura Messio et S\'ebastien Boyaval.
Gr\^ace \`a vous je me suis sentie finalement bien int\'egr\'ee ici.

Pensant au pass\'e, un jour de ma vie a chang\'e d\'efinitivement mon futur: le jour o\`u j'ai pass\'e les examens d'admission \`a Polytechnique.
Merci aux examinateurs pour m'avoir fait confiance.
Je suis tr\`es reconnaissante \`a la petite communaut\'e br\'esilienne qui a \'et\'e une deuxi\`eme famille pour moi pendant mes deux premier ann\'ees en France et qui reste tr\`es soud\'ee.
Un grand merci en particulier \`a Ricardo Malaquias.

Je remercie ma lointaine famille pour avoir toujours donn\'ee priorit\'e \`a  mon \'education et pour avoir toujours respect\'e mes choix.  
Finalement, je suis infiniment reconnaissante \`a la personne la plus dou\'ee et g\'en\'ereuse du monde, dont j'ai la chance de partager le quotidien depuis presque une dizaine d'ann\'ees: Vitor Sessak.

%% file: introduction.tex
\chapter{Introduction}
\label{introd}

\vspace{-0.5cm}

In this section we will present membranes, first from a biological and
historical point of view (section~\ref{bio_hist}) and secondly from a modern
physical perspective (section~\ref{model_membranes}). 
In section~\ref{mechanical}, we define the quantities that are the focus of this work: the mechanical tension $\tau$, the Lagrange-multiplier $\sigma$ and its measurable counterpart $r$.
The main theoretical models for membranes, as well as their validity, are presented in section~\ref{model_model}.
There we derive the first
fundamental results for planar membranes in contact with a lipid reservoir.
Section~\ref{exp} summarizes the most current experimental techniques used to
characterize the mechanics of membranes.
Finally, the stress tensor for planar membranes is introduced in section~\ref{stress_plan}. 

\section{Biological membranes}
\label{bio_hist}

\begin{figure}[H]
\begin{center}
\includegraphics[width=0.9\columnwidth]{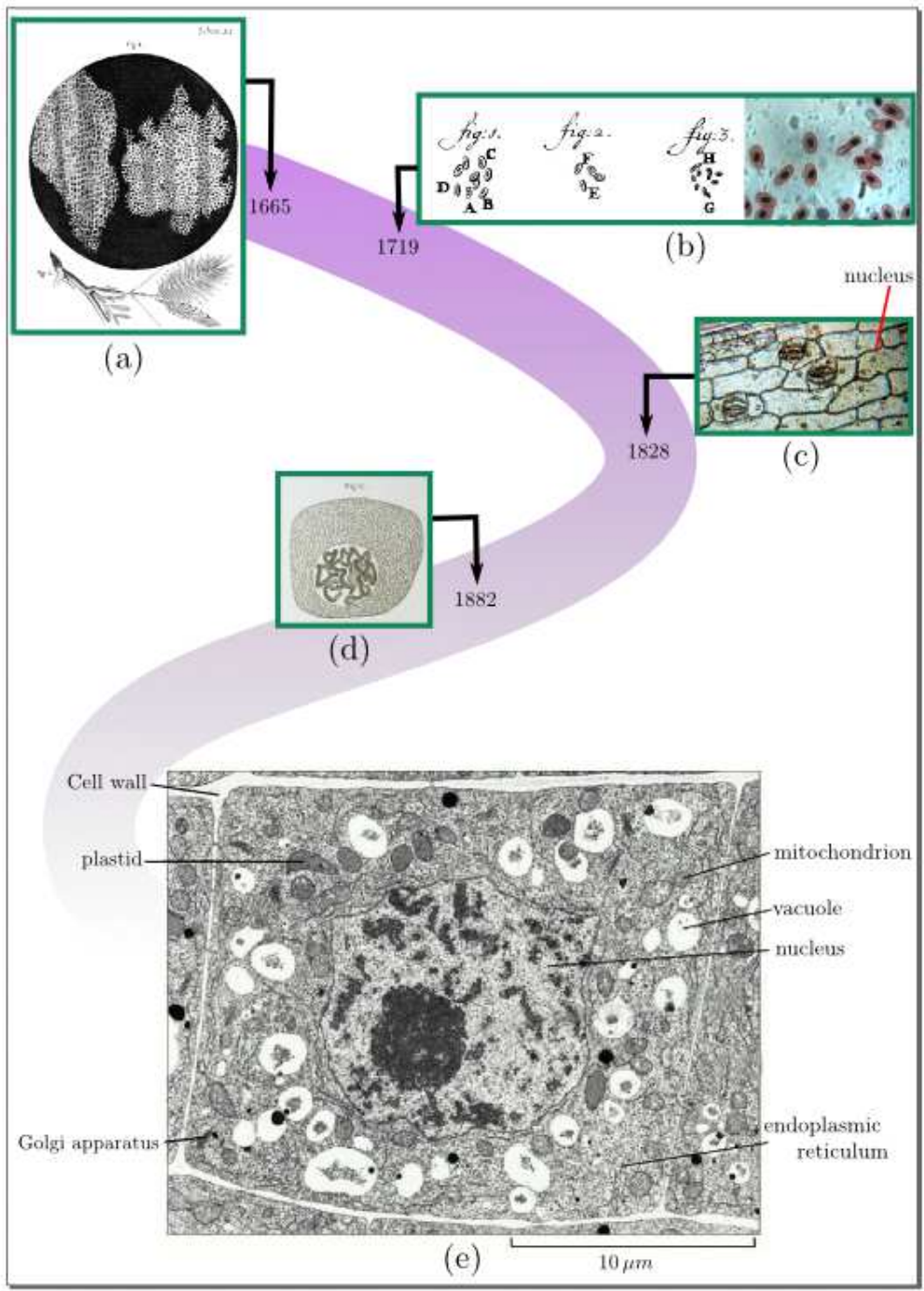}
\caption{The cell through history.
Inset (a) shows the first representation
of the cell, or more precisely, of the cell walls, by Hooke in 1665. Inset
(b) left shows the first drawing of the nucleus accurately made by Leeuwenhoek in 1719,
while he studied the salmon red blood cells (which are nucleated in
fishes).
The right photo shows a contemporary optical microscope image of the
same cells for comparison.
The next inset shows 
a modern photograph of plant tissue taken with roughly the same technology as in
1828, five years before Brown, best known for his observations of the Brownian
motion, named the nucleus.
Inset (d) depicts the chromosomes inside the nucleus (Flemming).
Finally, sub-figure (e) shows a nowadays electron
micrography of a plant cell~\cite{Alberts}.
We can see the complicated internal
structure and identify some organelles.  
}
\label{history}
\end{center}      
\end{figure}

During the last four hundred years, the image of the cell has become more and more complex~\cite{Baker_52} (see Fig.~\ref{history}).
As experimental techniques evolved,  
many questions were answered -- and many others were raised.
In particular, we have learned a lot about the cell's boundary. We will start
thus by a brief non-exhaustive historical review (further details can be found
on~\cite{Baker_52},~\cite{Robertson_81},~\cite{Edidin_03},~\cite{Heimburg_07}).

Up to the $19$th century, 
living beings were believed to have a sponge-like microscopical structure. 
There would be two continuous substances: a membranous meshwork, as one can
see in inset~\ref{history}(a), and a fluid filling the communicating
cells. 
The meshwork was considered the true essential constituent, while the fluid
had a mere nourishing function~\cite{Baker_52}.
At that time, the term membrane named already the cell's boundary, although it corresponded more to what we nowadays call the
cell wall, the rigid cellulose structure that encapsulates
plant cells.
This image changed in 1807, when Link showed that cells were in fact separated.
He observed that colored fluids did not diffuse through the surrounding cells,
as one would expect with the former theory. He concluded thus that the essential component of life was the unitary cell itself. 

Due to the 
low numerical aperture of the microscopical objectives available at that time (see inset~\ref{history}(c) for a typical image), animal cells were also believed to have a {\it membrane}, i. e., a cell wall.
Cells from both animals and plants had then the same features: a nucleus, an aqueous plasma and a cell wall.
This apparent universality was an important support to the cellular
theory proposed by Schwann and Schleiden, which stated that every living being
was constituted by cells.

\begin{figure}[H]
\begin{center}
\includegraphics[width=0.9\columnwidth]{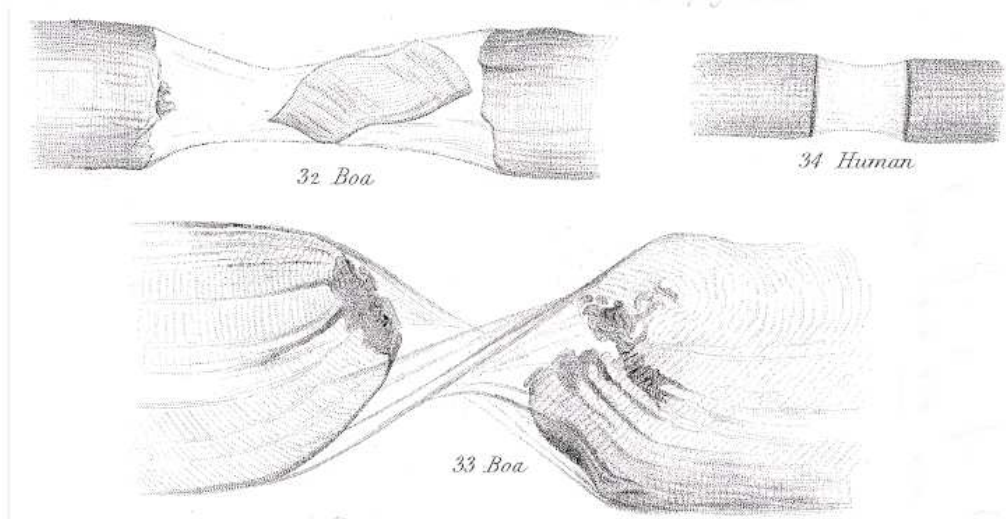}
\caption{Drawings from Bowman's work representing boa and human striated
  muscle cells~\cite{Bowman_1840}.
  The disrupted fibers are enclosed by a
  {\it sheath} called sarcolemma: the cell membrane.}
\label{bowman}
\end{center}      
\end{figure}

Some years later, the histologist and physiologist
William Bowman~\cite{Bowman_bio}, best known for his work in nephrology (the
Bowman's capsule is named after him), represented for the first time an actual cell membrane.
In his 1840 work, he studied the striated
muscle cells~\cite{Bowman_1840}.
He noticed that by stretching those cells, he could disrupt 
their internal fibers leaving a transparent {\it sheath} called
sarcolemma intact (see Fig.~\ref{bowman}).
This {\it sheath}, in that time 
thought a cell wall, is actually what we currently call the cell membrane.    

Subsequent experiments on the effects of osmotic pressure on plant cells
showed that the cell could pull away from the cell wall. 
This phenomenon is called plasmolysis. 
It is caused by the selective permittivity of the cell
membrane, which is permeable to water, but not to ions, sugars and other water
soluble molecules. 
Although specialist of that time could have interpreted it as an indirect
evidence of a membrane encapsulating the cell, they blamed osmotic effects on
vacuoles and  
proposed the naked-cell theory: the cell was defined as {\it a small naked lump of protoplasm with
  nucleus}, in Schultze's words (1860).
It could eventually be encapsulated by a
non-essential {\it skin} (de Bary, 1861)~\cite{Baker_52}.
%In order toThe protoplasm would be non hydrosoluble in order to assure the stability of the cell. 

\begin{figure}[H]
\begin{center}
\includegraphics[width=0.4\columnwidth]{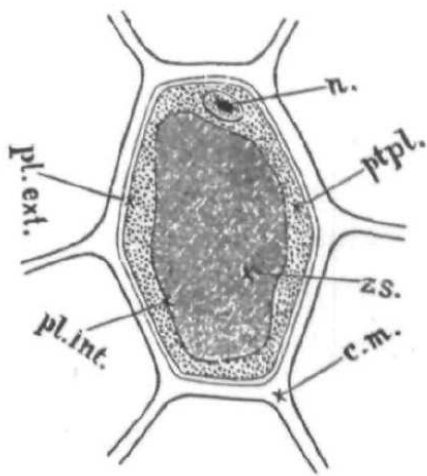}
\caption{Schematic representation of a plant-cell made by Overton in his 1895
  work~\cite{Overton}. The cell wall (indicated by {\it c.m.}) and the cell membrane (indicated by {\it pl.ext.}) are represented as
  different entities. Note that Overton decided arbitrarily the thickness of
  the membrane.}
\label{overton}
\end{center}      
\end{figure}

The first one to realize that there was effectively a {\it skin} of different
nature around the protoplasm was 
Ernest Overton in 1895~\cite{Overton}.
He noticed that
plant cells under a $8 \%$ sugar solution suffered plasmolysis.
This indicated that sugar molecules could not easily penetrate the cell even though the protoplasm was composed by water. 
He repeated the experiment using successive solutions of alcohols, ether, acetone and phenol with the same osmotic pressure as the sugar solution.
He remarked that in some of these cases there was no plasmolysis, depending on
the substance solubility in water.
He concluded that water insoluble substances
penetrated easily the external part of the protoplasm ({\it Grenzschicht}),
whereas water soluble molecules, as sugar, did not.
He inferred thus that the boundary region was distinct from the rest of the
protoplasm (see Fig.~\ref{overton}) and that it was impregnated by a substance of the same nature of
fatty oils.

Indeed, today we know that biological membranes are mostly composed by amphiphilic lipids, having a
hydrophobic tail and a hydrophilic head. 
This feature make them self-assembly
in aqueous media in large bilayers, even though there are no chemical bonds
between them. 
This explains the relative impermeability of membranes to water soluble substances.
The most abundant kind of lipid present on membranes are 
phospholipids. 
These molecules have one or two long hydrocarbon chains, which may contain only simple
bonds (saturated) or double bonds (unsaturated). 
As we shall see in section~\ref{model_membranes}, the stability of the bilayer
is assured by the fact that these molecules have an effective
shape close to a cylinder, whose dimensions are
typically $0.5 \,\mathrm{nm}$ for the radius and $1.0 - 1.5\, \mathrm{nm}$ for
the length~\cite{Mouritsen}. 
In addition to phospholipids, membranes may also contain cholesterol, an amphiphilic lipid that unlike phospholipids has a ring-like tail.

%Experiments followed showing that membranes
%were also relatively impermeable to ions (see ref.~\cite{Robertson_81} for a complete bibliography).
The first indirect estimation of the membrane thickness was made by Hugo Fricke in 1925.
He measured the capacitance of blood cells and, supposing that they were composed by lipids, deduced a thickness of
$3.3 \, \mathrm{nm}$~\cite{Fricke}.
This is a remarkable result, since posterior direct measures give a thickness
between $5 \, \mathrm{nm}$ and $10 \, \mathrm{nm}$~\cite{Robertson_59}.
Meanwhile, Gorter and Grendel gave
a fundamental step towards the comprehension of how lipids arrange themselves
within the membrane~\cite{Gorter_25}.
They extracted the lipids from a known
number of red blood cells.
Using a Langmuir trough, they measured the
area covered by lipids and it corresponded to twice the estimated surface area of the
erythrocytes.
They deduced that the membrane was constituted by a bilayer of lipids whose
polar heads pointed outward.

\begin{figure}[H]
\begin{center}
\includegraphics[width=0.8\columnwidth]{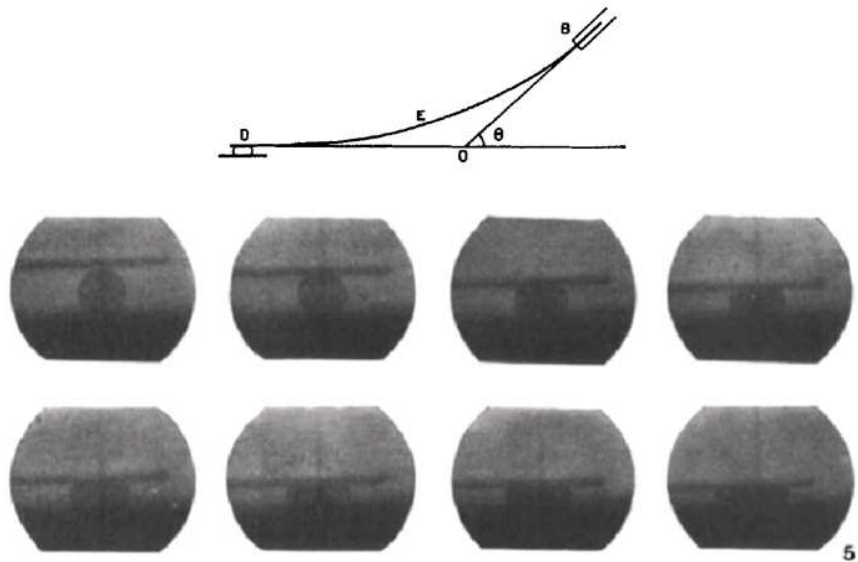}
\caption{Original drawing and photos from Cole's experiment on urchin eggs: the egg (D) is compressed by a gold fiber (E).}
\label{cole}
\end{center}      
\end{figure}

In 1932, a puzzling experiment suggested that there was more than
lipids on a membrane.
Kenneth Cole studied urchin eggs by 
compressing them between a plate and a gold fiber with a known force, much in the same way as in modern experiments~\cite{Balland_06} (see Fig.~\ref{cole}).
He deduced the tension of the egg's membrane by studying its degree of
flattening.
His measures
yielded a tension of $0.08 \, \mathrm{dyn}/\mathrm{cm}$, which is a hundred times smaller than the tension of oily films~\cite{Cole}.
This was a surprising result, if one believed the membrane to be constituted only by lipids. 

\begin{figure}[H]
\begin{center}
\includegraphics[width=0.37\columnwidth]{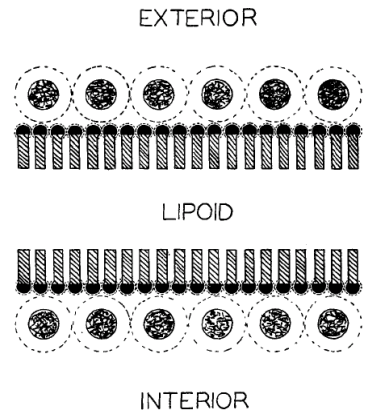}
\caption{Original drawing from Danielli-Davson's paper representing the membrane structure:
  a bulk of lipids bordered by ordered lipid monolayers covered with proteins.}
\label{davson}
\end{center}      
\end{figure}

Two years later, Danielli and Harvey solved the paradox.
They centrifugated
smashed mackerel eggs in order to separate lipids from the aqueous
phase.
First, they measured the surface tension of the oily phase and obtained
$\approx 9 \, \mathrm{dyn}/\mathrm{cm}$.
Then they added the aqueous phase
and observed a tension lowering.
By studying the time evolution and the influence of
temperature on their mixture, they deduced that proteins were responsible for the tension
lowering~\cite{Danielli_34}.
Danielli and Davson proposed in the following
year the first model of the membrane structure: every plasma membrane would 
have a core of lipids bordered by two monolayers of lipids whose polar head
pointed outward and the whole would be coated by a layer of
proteins~\cite{Danielli_35} (see Fig.~\ref{davson}).
It was an important step, since today we know that proteins are responsible for almost every membrane function but enclosing, such as active transport of molecules, binding to cytoskeleton and reception of chemical
signals from the outer environment.

\begin{figure}[H]
\begin{center}
\includegraphics[width=0.7\columnwidth]{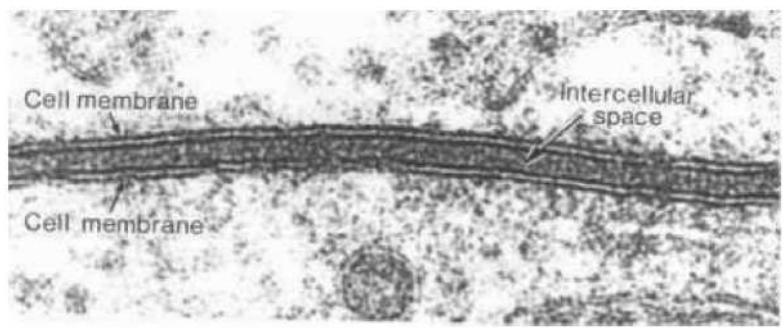}
\caption{Electron micrography of the cell membrane~\cite{Fawcett}. The
  hydrophilic heads of lipids are colored, while the
  hydrophobic tails are not. This results on the characteristic three
  lines pattern seen above.}
\label{fawcett}
\end{center}      
\end{figure}

In the 1950s the new technology of electron micrography allowed to make the
first direct images of the cell membrane (see Fig.~\ref{fawcett}).
The use of permanganate fixation was also important, since it stained only the hydrophilic head of lipids.
Robertson observed a three line pattern of about $7.5 \, \mathrm{nm}$ corresponding to a simple bilayer. 
This excluded the possibility of a bulk of lipids in membranes.
He observed the three lines pattern not only on the cell boundary, but also
encapsulating organelles from different animals and bacteria~\cite{Robertson_59}.
Moreover,
using innovative staining processes, he showed the
asymmetry of some membranes' coating, the external surface containing also 
carbohydrates.  
These new features were incorporated in the unit membrane model: every
biological membrane shared the same architecture - a lipid bilayer, asymmetrically coated by proteins and carbohydrates.

\begin{figure}[H]
\begin{center}
\subfigure[Micrography of a fractured human erythrocyte membrane. The left surface
  corresponds to the external fracture (EF) and the right corresponds to the
  protoplasmic fracture (PF). The tiny particles on both surfaces measure between $5$ and $10\,
\mathrm{nm}$ and correspond to proteins.
Note that they are more numerous on the PF face due to the presence of peripheral proteins attaching the membrane to the cytoskeleton.]{
  \includegraphics[width=0.55\columnwidth]{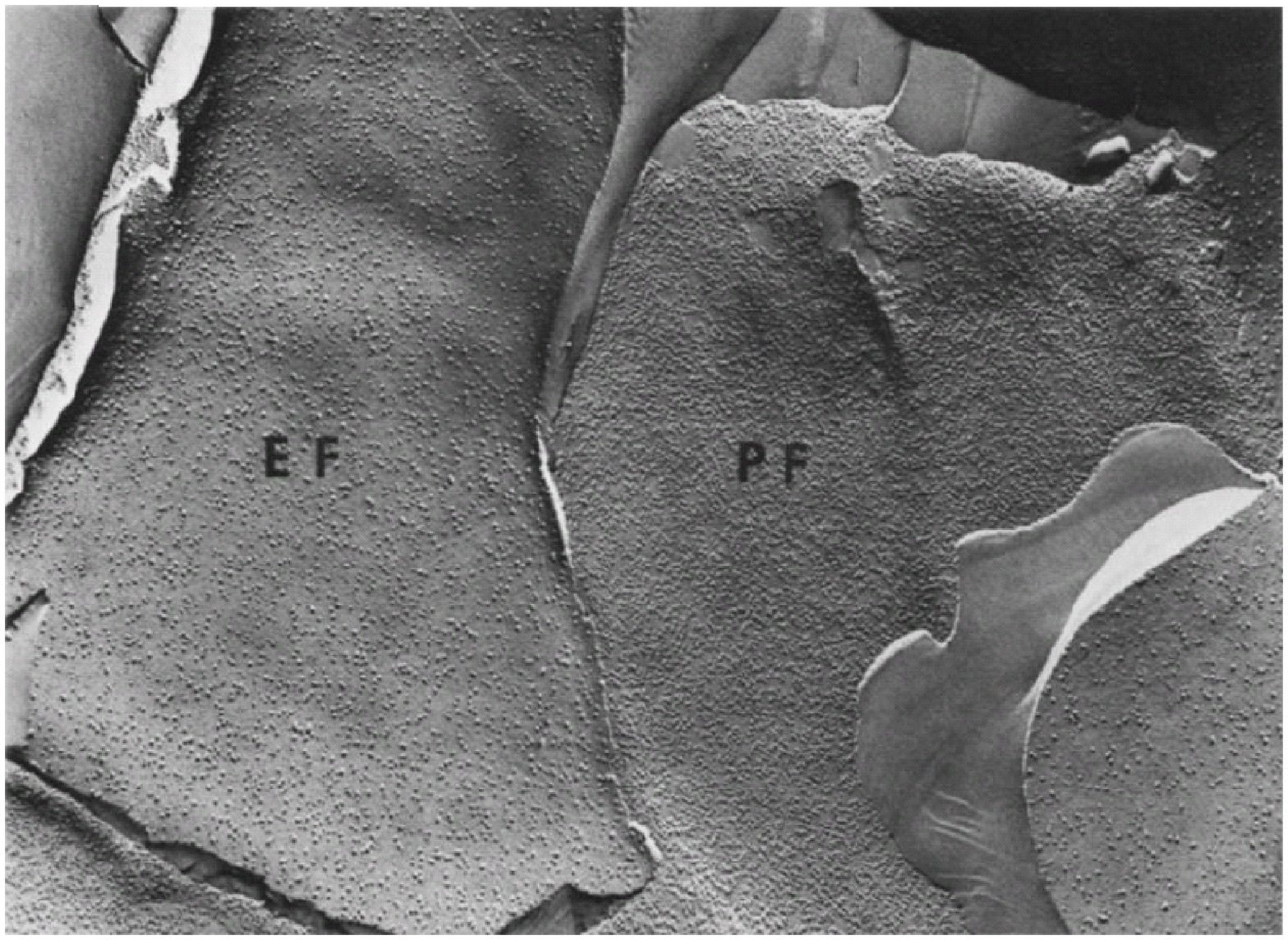}
  \label{da_silva}
}
\subfigure[Original picture of the fluorescence
labeled antigens experiment which showed that proteins diffuse on the cell membrane\cite{Frye_70}.
Antigens were labeled in red (lower half) and green (upper half).
After some minutes, the colors were uniformly distributed over the cell.]{
  \includegraphics[width=0.35\columnwidth]{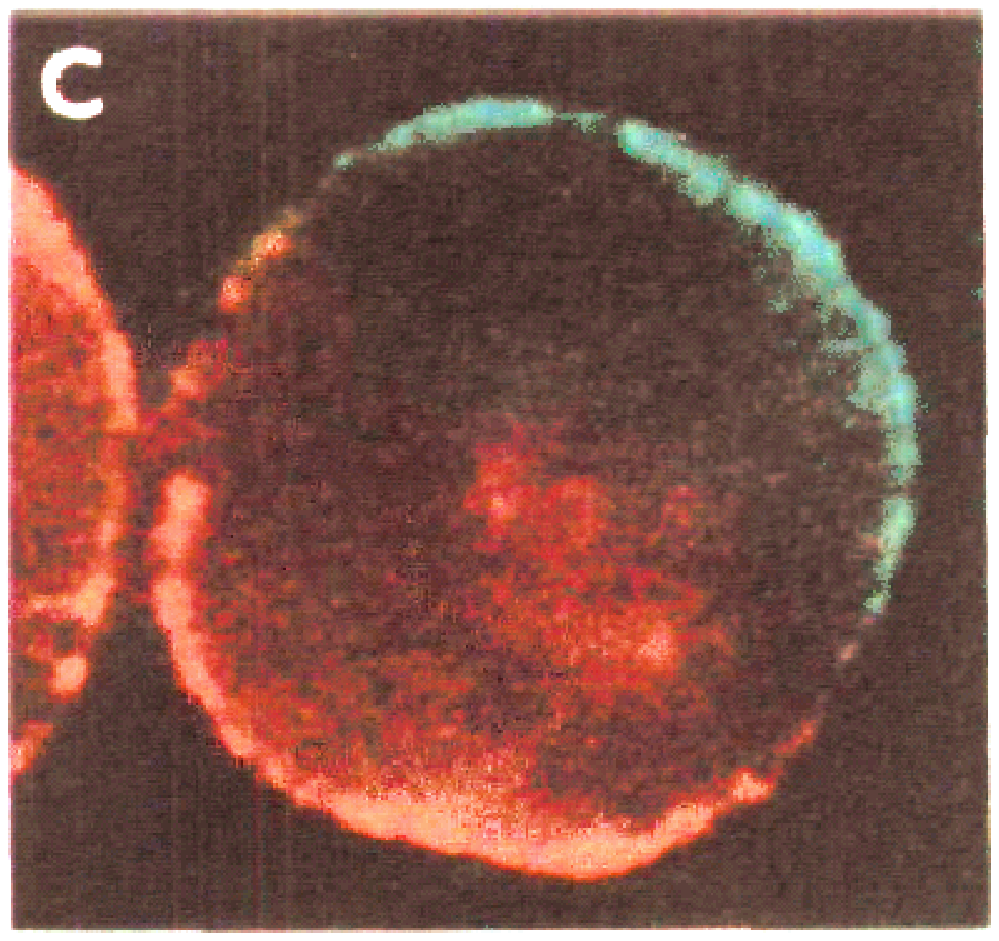}
  \label{frye}
}
\caption{Evidences in favor of the Mosaic Fluid model.}
\end{center}      
\end{figure}

Some years later, the advent of the freeze
fracture etch (FFE) electron microscopy brought new advances.
This technique consists on freezing cells by immersion on nitrogen. 
The block of cells is then fractured.
By deposing carbon and platinum vapor, a replica of the fractured surface is constituted.
At last, the replica can be examined by a transmission electron microscope.
In the late 1960s, da Silva and Branton showed that on biological membranes, these fractures tended
to pull apart the lipid bilayer~\cite{Branton_66},~\cite{Silva_70}. They obtained
the micrography shown in Fig.~\ref{da_silva}, which suggested that proteins
were actually embedded in the lipid bilayer. Besides, another work using
fluorescent labeling showed that proteins diffuse, implying that membranes were in
fact fluid~\cite{Frye_70} (see Fig.~\ref{frye}). 

\begin{figure}[H]
\begin{center}
  \vspace{1cm}
\includegraphics[width=0.8\columnwidth]{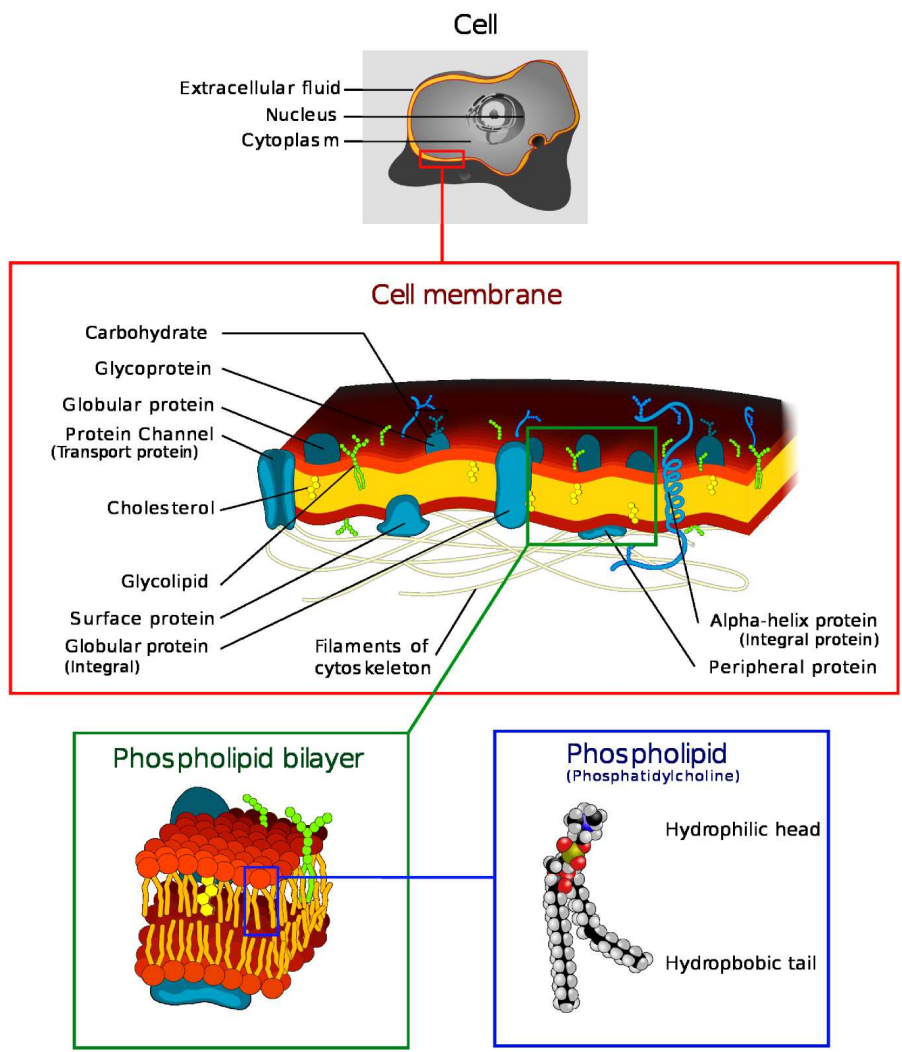}
\caption{Schematic view of a biological membrane as proposed by the fluid
  mosaic model. 
The green inset shows the bilayer
  structure in detail and the blue inset represents of a phospholipid
  (phosphatidylcholine), also known as lecithin, present in egg yolk. 
Remark the unsaturated
  bond in one of its tails. }
\label{memb_schema}
\end{center}      
\end{figure}

Finally, bearing in mind these experiments, Singer and Nicholson proposed the
mosaic fluid model of membranes (see Fig.~\ref{memb_schema} for a sum-up), which is the basis to the modern vision of
biological membranes. 
They made the distinction between peripheral proteins,
i. e., those loosely attached to the membrane like those that bind the membrane
to the cytoskeleton, and the integral proteins, which are embedded in the lipid
bilayer. 
They postulated that
lipids and proteins diffuse freely inside the membrane's surface, as a
two-dimensional liquid.
Consequently, membranes should have no long-range order. 
They noted that the membrane leaflets were probably asymmetrical with respect
to lipid and protein composition due to the energy barrier of moving the polar
head from the aqueous interface into the bilayer interior.
Later experiments confirmed that asymmetry was indeed present on biological membranes~\cite{Rothmann_77}.

The model was elegant, but experiments rapidly showed that it was
oversimplified. 
Biological membranes are not so homogeneous as the fluid mosaic model
implies. 
Already in the $80$'s, indirect measures showed that polarized epithelial cells present
membrane domains, i. e., the cell membrane that faces a cavity has
different composition from the other faces~\cite{Simons_88}. 
Moreover, biological membranes
present also smaller domains, ranging from dozens to hundreds of
nanometers~\cite{Edidin_97}. 
A simple statistical reasoning suggests that heterogeneities should indeed be
expected: a random lipid and protein distribution means that the pairwise interactions between
lipid-lipid, lipid-protein and protein-protein should be within thermal
energies, which is rather improbable, given their wide variety~\cite{Engelman_05}.

\begin{figure}[H]
\begin{center}
  \vspace{1cm}
\subfigure[Photo of a vesicle composed by a $1:1$ mixture of DLPC and
  DPPC, two kind of phospholipids. The green and red patches represent respectively the fluid and
  the gel phase.]{
\includegraphics[width=0.4\columnwidth]{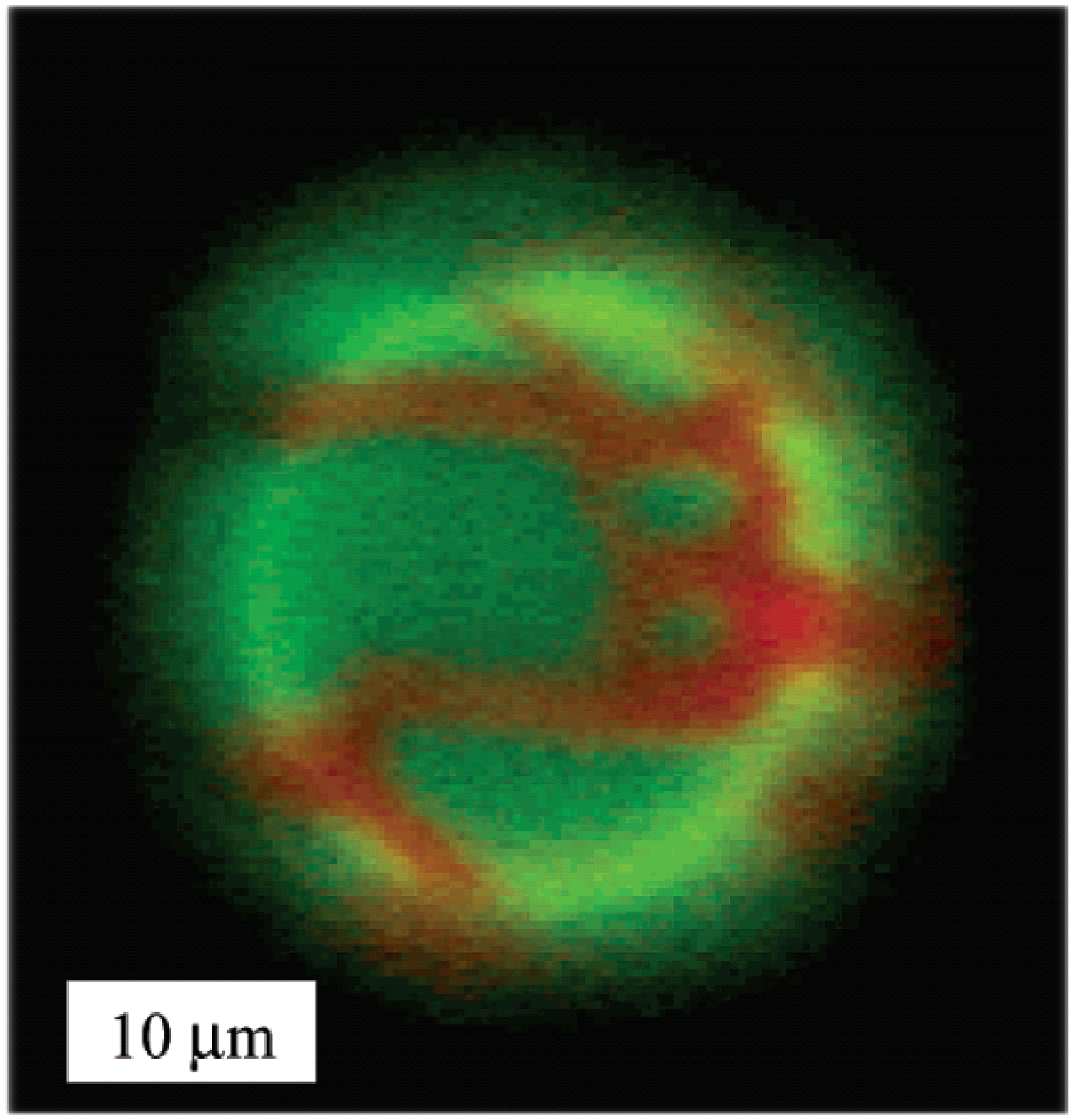}
\label{segreg}
}
\subfigure[Time evolution of a vesicle composed by a $1:1$ DOPC and DPPC
(phospholipids) 
mixture added of $35\%$ cholesterol. We see also a phase separation, but this
time both phases are liquid. Remark the domains widen with time.]{
\includegraphics[width=0.55\columnwidth]{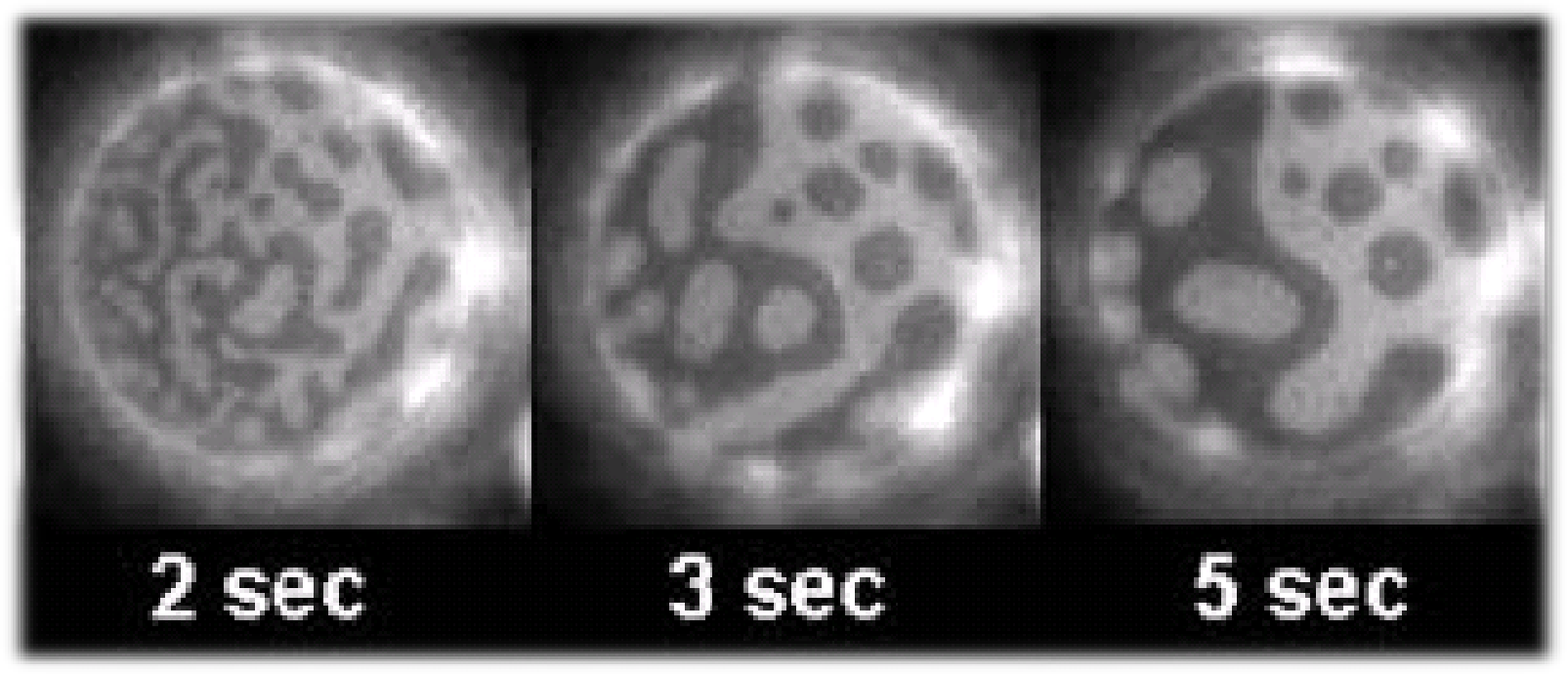}
\label{fluid_fluid}
}
\caption{Phase separation on membranes.}
\end{center}
\end{figure}

At this point, model membranes, i.e., lipid bilayers reconstituted in laboratory to mimic biological
membranes (see section~\ref{model_membranes}), confirmed and gave
new insight to the question of lipid phase separation.
A mixture of phospholipids in the gel/liquid phases segregates,
as shown in Fig.~\ref{segreg}~\cite{Donofrio_03}.
At the interface between
these phases, a line tension builds up and they tend to
separate in order to minimize energy. 
More interestingly regarding cells, it is possible to have phase separation between two liquid
phases: the liquid-disordered and the liquid-ordered~\cite{Mouritsen}.
Experimentally, this may be achieved in a mixture of phospholipid and
cholesterol~\cite{Almeida_92},~\cite{Veatch_03} (see Fig.~\ref{fluid_fluid}). 

\vspace{1cm}

\begin{figure}[H]
\begin{center}
\subfigure[typical trajectories of gold particles attached to
  certain proteins on a cell surface (followed for $30\, \mathrm{s}$). Trajectory A
  corresponds to the stationary mode, B, E and F to simple diffusion, C to directed
diffusion and D to restricted diffusion~\cite{Kusumi_93}.]{
\includegraphics[width=0.45\columnwidth]{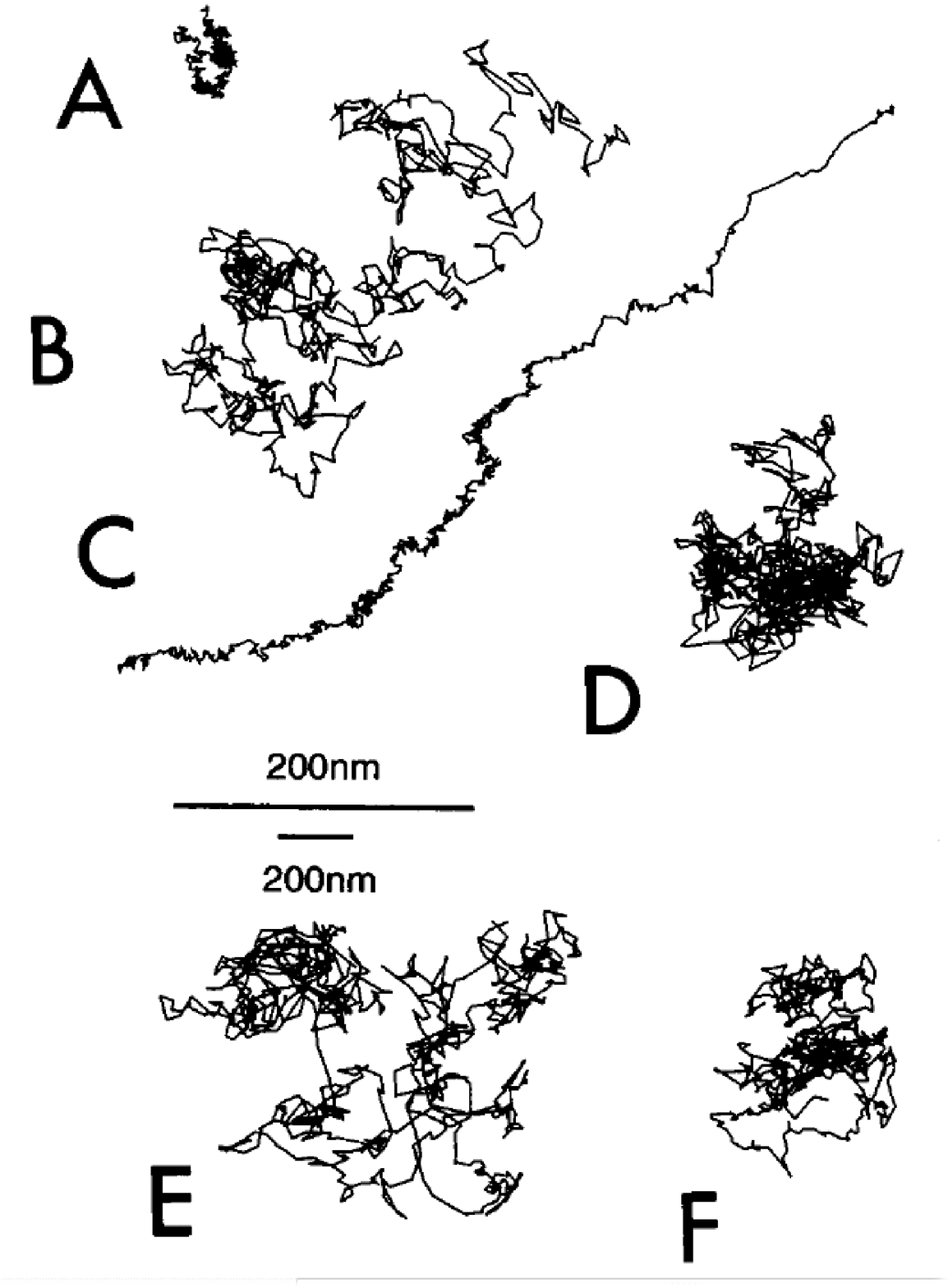}
%\label{}
}
\subfigure[Here, single particle tracking is used to study a model lipid
monolayer divided in two phases: the liquid-ordered (dark gray) and liquid-disordered (light gray). The arrow in (a) indicates the polystyrene bead that
was tracked. Fig.(b) shows the bead's random walk and (c) shows a detail of
this walk. Remark that the bead remains on the liquid-disordered phase~\cite{Forstner_03}.]{
\includegraphics[width=0.37\columnwidth]{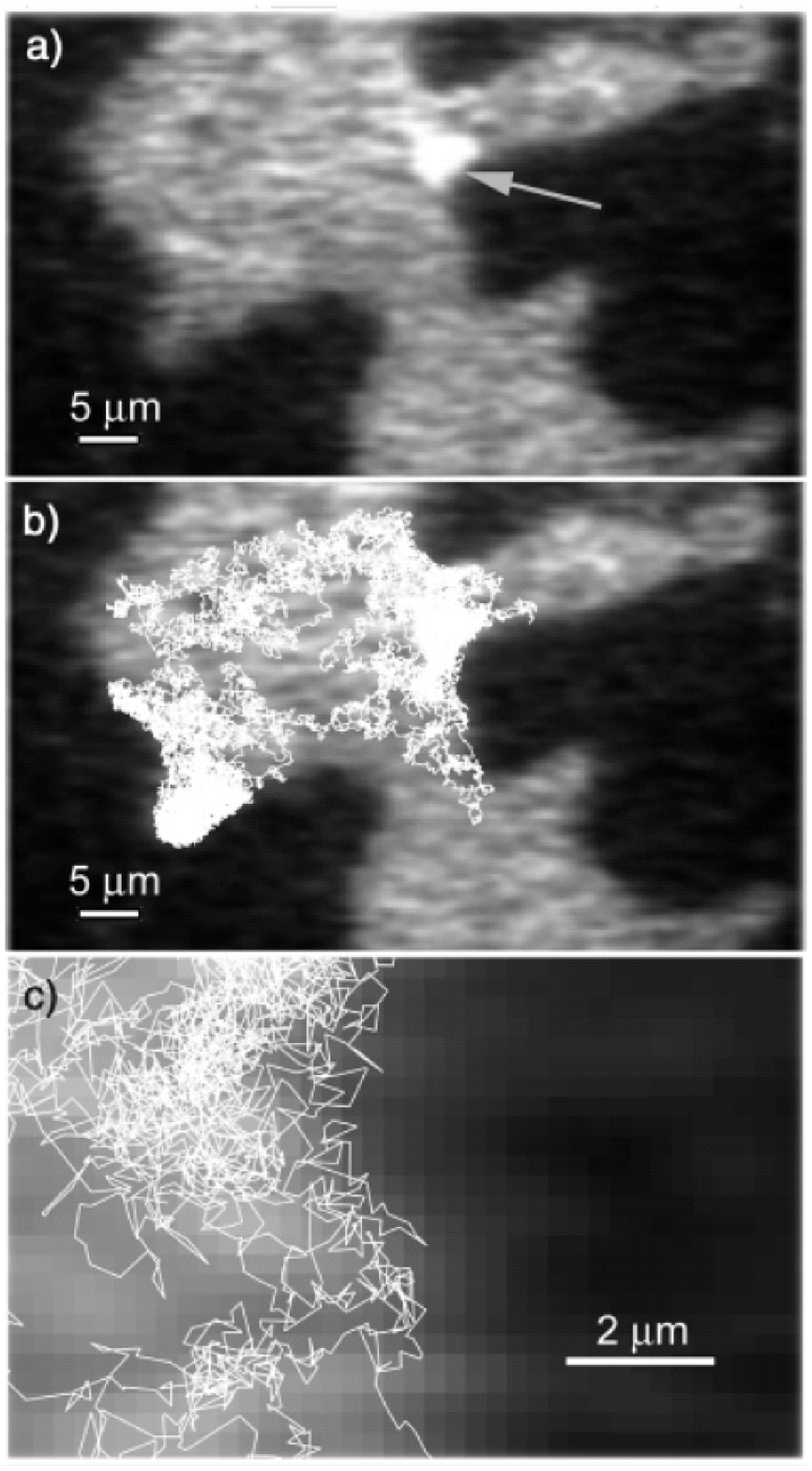}
%\label{fluid_fluid}
}
\caption{Single particle tracking.}
\label{tracking}
\end{center}      
\end{figure}

Besides the segregation of lipids, it was also shown that some proteins cluster in
model membranes.
In these cases, the aggregation depends on the length of the lipids constituting
the bilayer where proteins are embedded~\cite{Kusumi_82}.
In addition, modern techniques, such as single particle tracking, show that
lipids and proteins in living cells can have anomalous movements, such as directed
or confined motion and anomalous diffusion, possibly due to the
cytoskeleton or to restrictions imposed
by lipid domains~\cite{Saxton_97} (see Fig.~\ref{tracking}).

\begin{figure}[H]
\begin{center}
\includegraphics[width=0.65\columnwidth]{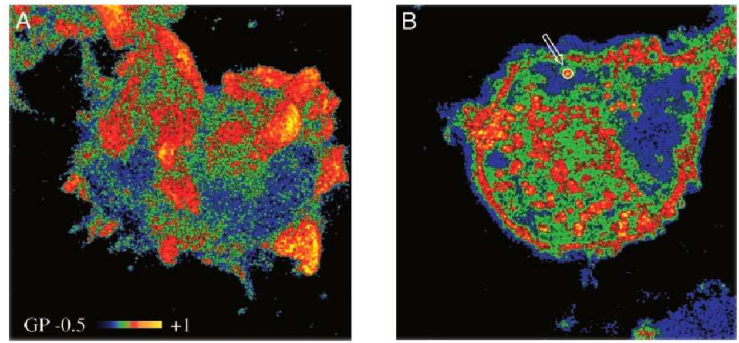}
\caption{GP images of living macrophages (mouse $RAW264.7$ and human $THP-1$,
  respectively): red stands for the liquid-ordered phase and blue to the
  liquid-disordered. The circled area indicated in B shows the
  pixilation of the image ($167$ pixels inside the circle). Note in A that liquid-ordered phase tend to be
  observed on the tip of filopodia~\cite{Gaus_03}.}
\label{cell_patches}
\end{center}      
\end{figure}

\begin{figure}[H]
\begin{center}
\includegraphics[width=0.4\columnwidth]{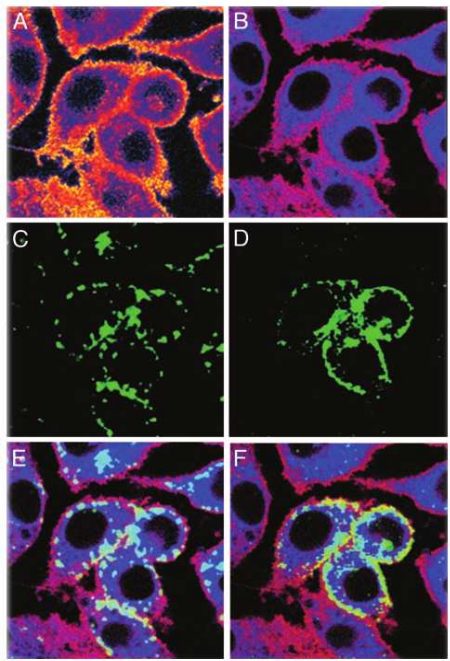}
\caption{Images of $RAW264.7$ living cells. Figure A shows the GP image and B
  shows the corresponding dual-colored image. Figs.C and D show the
  fluorescence images for transferrin receptor and caveolin-1,
  respectively. Figs.E and F correspond to the superposition of B with C and
  D, respectively: light blue patches indicate colocalization with liquid-disordered
phases and yellow patches indicates colocalization with liquid-ordered phases\cite{Gaus_03}.}
\label{rafts}
\end{center}      
\end{figure}

Finally, modern fluorescence techniques allow the direct visualization of
domains in living cells~\cite{Gaus_03}. In Fig.~\ref{cell_patches}, Fig.~\ref{rafts}.A and Fig.~\ref{rafts}.B, one can see patches whose
overall structure is different.
The discriminating feature is the GP
(generalized polarization): red for liquid-ordered phase, richer in
cholesterol, and blue the
liquid-disordered phase.
The images show clearly that there is coexistence of
liquid phases in cells. Moreover,
it is possible to detect certain types of proteins.
In Fig.~\ref{rafts}.C and in
Fig.~\ref{rafts}.D, transferrin receptor and caveolin-1 are respectively shown
by fluorescence.
Fig.~\ref{rafts}.E and Fig.~\ref{rafts}.F show the merge of C and D with B, respectively.
%The light blue patches indicate a colocalization of these proteins with
%liquid-disordered domains, whereas yellow indicates the colocalization with
%liquid-ordered domains. 
We can see that the transferrin receptor is found
mostly on the liquid-disordered phase, while caveolin-1 is found mostly on the
liquid-ordered phase.
Indeed, for a long time liquid structures called rafts composed mostly by glycosphingolipids,
cholesterol and some proteins were suspected to exist~\cite{Mouritsen}. 
These rafts would help protein sorting and be involved in cell signaling. These images present however
only some evidence of their existence and the subject is still
debated~\cite{Edidin_03b},~\cite{Pike_09}.

\begin{figure}[H]
\begin{center}
  \vspace{1cm}
\includegraphics[width=0.45\columnwidth]{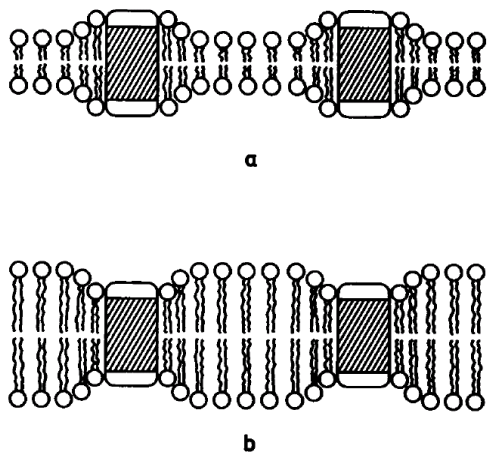}
\caption{Drawing representing the {\it Mattress model}: hydrophobic mismatch
  deforms lipids. Proteins tend thus to aggregate.}
\label{mattress}
\end{center}      
\end{figure}

In order to account for some of these results, refinements of the fluid mosaic
model were proposed. 
For instance, to explain protein clustering
Mouritsen proposed the {\it Mattress Model}~\cite{Mouritsen_84} (see
Fig.~\ref{mattress}). 
The model comes from the
observation that the bilayer thickness may be smaller or larger than the
length of the hydrophobic part of embedded protein. 
This mismatch would expose hydrophobic parts of the
protein or of the lipids, which would in consequence deform.
The deformation would give rise to a line tension which would tend to cluster proteins and
aggregate some kinds of lipids.
Another model was proposed by Erich Sackmann  
to explain the confinement of proteins
observed during single particle tracking. 
It stresses the interactions between the membrane and the cytoskeleton. 
To the moment however,
there is no model that accommodates all recent results about biological membranes.   

\section{Model membranes and mechanical probing}
\label{model_membranes}

\begin{figure}[H]
\begin{center}
  \vspace{1cm}
\subfigure[Fluorescence micrography of a self-tightened knot (center of
  image) made of lipid nanotubes extracted from membrane vesicles. The vesicle
  on the left is about $10 \, \mathrm{ \mu m}$~\cite{Lobovkina_04}.]{
\includegraphics[width=0.40\columnwidth]{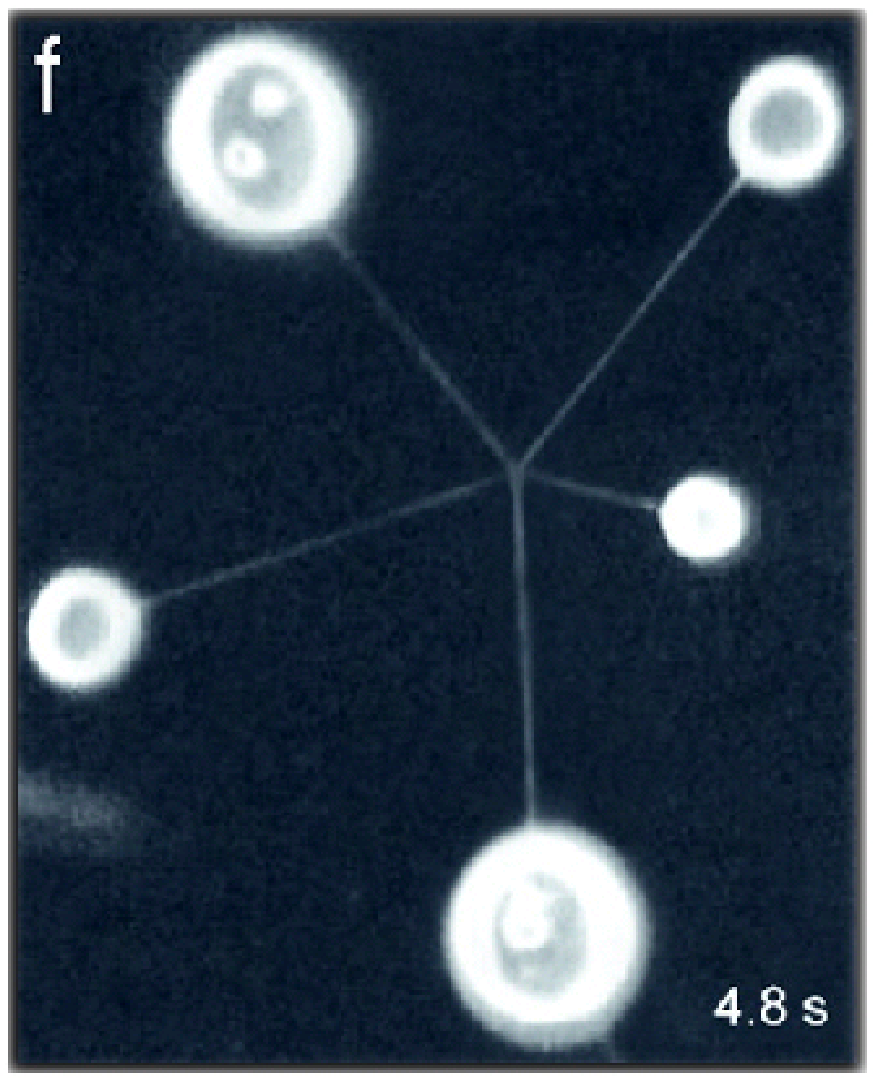}
\label{Lobovkina}
}
\subfigure[Artistic representation of a bio-electronic device composed by a nanowire $30 \, \mathrm{nm}$ wide (gray) covered by a lipid
membrane (blue/orange). In this membrane, proteins that control ion passage
were incorporated (pink) (image by Scott Dougherty,~\cite{Misra_09}).]{
\includegraphics[width=0.45\columnwidth]{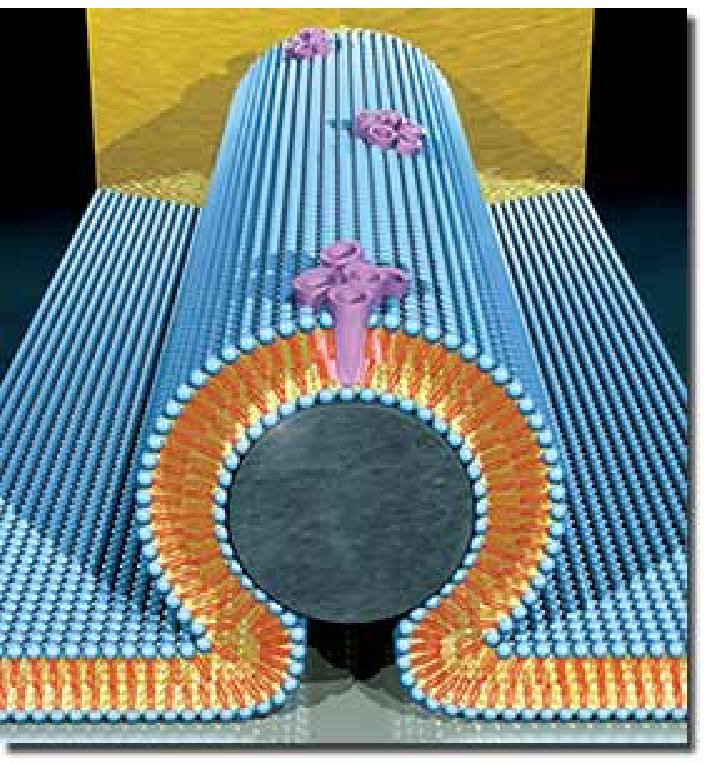}
%\label{fluid_fluid}
}
\caption{Technological applications of membranes.}
\label{techno}
\end{center}      
\end{figure}

As we have seen in the last section, biological membranes are very complex.
Simpler membranes reconstituted in laboratory called model membranes
are thus doubly interesting.
First, they give insight to the comprehension of phenomena in living cells.
Model membranes are advantageous because they have both chemical composition and environment controlled, allowing reproducible experiments.
Moreover, these membranes are generally in thermodynamic equilibrium, which is
impossible to achieve in living cells.
By consequence, these experiments may be described using 
conventional statistical mechanics tools. 
Secondly, model membranes have technological interest in their own: they are
used to improve drug
delivery, to build micro-chambers for chemical reactions~\cite{Orwar_03},~\cite{Lobovkina_04} and
even to build bio-electronic devices~\cite{Misra_09} (see Fig.~\ref{techno}).

\begin{figure}[H]
\begin{center}
  \vspace{1.5cm}
\includegraphics[width=0.8\columnwidth]{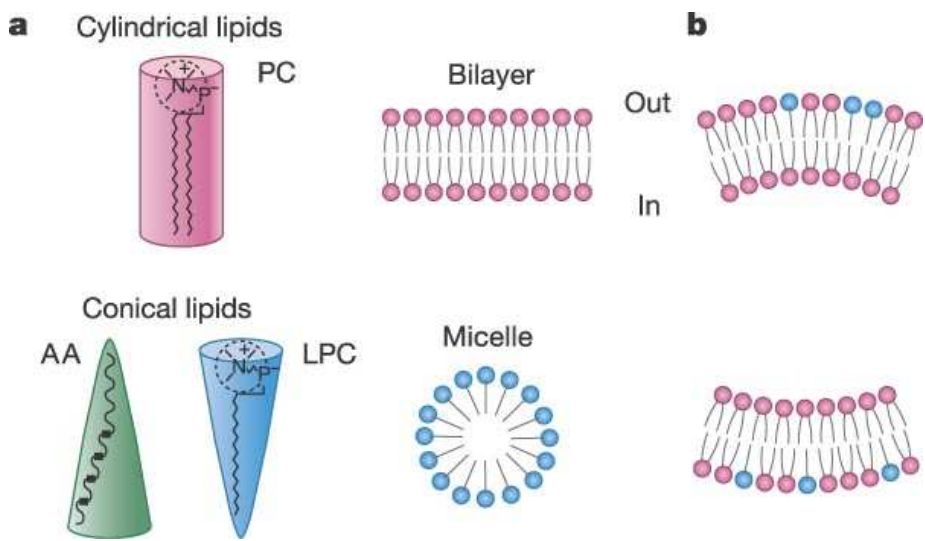}
\caption{Inset (a) shows the effective shape of some phospholipids: in pink,
  phosphatidylcholine (PC), in blue lysophosphatidylcholine (LPC), with only
  one hydrophobic tail, and in green arachidonic acid (AA), with an
  unsaturated tail. The upper surface corresponds to the hydrophilic head. In
  the center, we can see some self-assembled structures. Inset (b) shows an asymmetrical bilayer whose composition leads to a natural bending tendency.}
\label{packing}
\end{center}      
\end{figure}

Model membranes are prepared by dissolving phospholipids in an aqueous solution.
In order to minimize the exposition of their hydrophobic tails to water, they
self-assemble in a large variety of forms, from small micelles to vesicles and
bilayers, depending on temperature, on concentration and on the effective shape of
phospholipids, which is a measure of their average cross section of as a function of how profoundly buried they are on a membrane.
In Fig.~\ref{packing}(a), we can see examples of effective shapes. 
Note that to form a bilayer, lipids must have roughly a cylindrical shape.
In Fig.~\ref{packing}(b), we can see an asymmetrical bilayer, which naturally tends to bend.
We remark that the leaflets' asymmetry is stable over time, since spontaneous passage of
lipids from one monolayer to the other, known as flip-flop, is very slow in
pure lipid bilayers (of the order of several hours~\cite{Kornberg_71}).
Indeed, there is a high energetic barrier for the hydrophilic head to traverse the hydrophobic core of the membrane. 

\vspace{0.5cm}

An essential point is the very weak water solubility of phospholipids.
This implies that once a structure such as a vesicle is formed, the number of
phospholipids it contains is constant.
Besides, phospholipids do not resist to stretching, as we will see in
section~\ref{mechanical} and thus the total area of these structures is also constant.  

\begin{figure}[H]
\begin{center}
\subfigure[Drawing of the experimental set used in black film
  experiments.]{
\includegraphics[width=0.5\columnwidth]{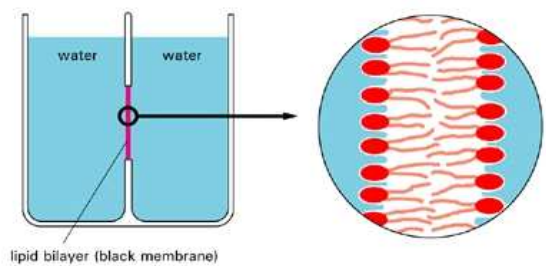}
\label{black_film}
}
\subfigure[Photo of the Teflon foil used to separate the two chambers shown above. The hole is about $1 \, \mathrm{mm}$
large. The lipid membrane is deposed over this aperture~\cite{Hole_www}.]{
\includegraphics[width=0.5\columnwidth]{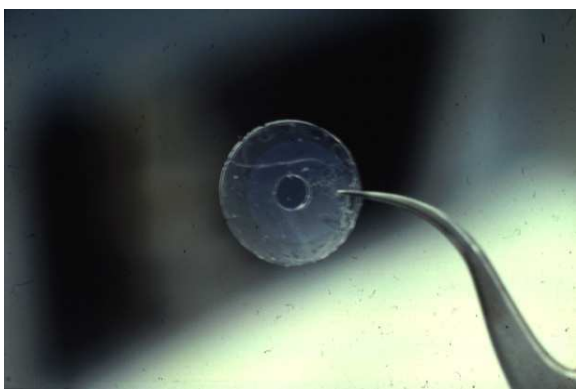}
\label{hole}
}
\caption{Black lipid films.} 
\end{center}      
\end{figure}

In this work, we are interested in the mechanical properties of liquid
membranes.
To this aim, three structures are usually studied: planar membranes, vesicles
and membrane tubes.
Planar membranes are also called black film membranes (BLM). 
They are used since the $60$'s and their name come from the destructive
interference that a light beam suffers due to the thinness of the lipid membrane.
The experimental set is constituted by two aqueous chambers separated by a plate (see Fig.~\ref{black_film}).
This plate, usually made of hydrophobic materials to assure the adherence
of lipid molecules, has a hole ranging from micrometers to several millimeters
(see Fig.~\ref{hole}). 
A bilayer can be deposed over this aperture through a variety of techniques~\cite{Castellana_06}.
BLMs are widely used to characterize the electrical properties of
membrane spanning proteins, since one can control the composition of both
aqueous solutions.
It has also been used in single particle tracking~\cite{Sonnleitner_99}. 

\vspace{1.2cm}

Sadly, the technique presents many disadvantages for mechanical probing.
First, one cannot control the tension of the frame: it depends on the film
deposition.
If the tension is too small, the membrane fluctuates a lot and is unstable. 
So, usually the film is relatively tense.
If however it is too tense, a minimum osmotic difference between the two cavities leads
to the rupture of the membrane~\cite{Winterhalter_00}.
Another problem is the film deposition technique, which may involve solvents
that contaminate the membrane~\cite{Aimon_per}.

The most popular objects used for membrane mechanical probing are uni-lamellar vesicles, which are self-assembled
{\it bags} of a single bilayer containing fluid.
They are obtained from several techniques and 
range in size from a few tens of nanometers to tens 
of micrometers. 
In the last case, they are also called GUV (giant uni-lamellar
vesicles) and they are of special interest, since they have roughly the same
size of cells, they are easy to manipulate and they are directly visible with
light microscopy techniques~\cite{Dimova_06}.
Moreover, they are stable and they can be deflated by changing the 
osmotic pressure.
 Each vesicle has a constant surface, since phospholipids are weakly soluble in
water and their volume is also constant, as long as the osmotic pressure is kept constant.

\begin{figure}[H]
\begin{center}
\subfigure{
\includegraphics[scale=.63,angle=0]{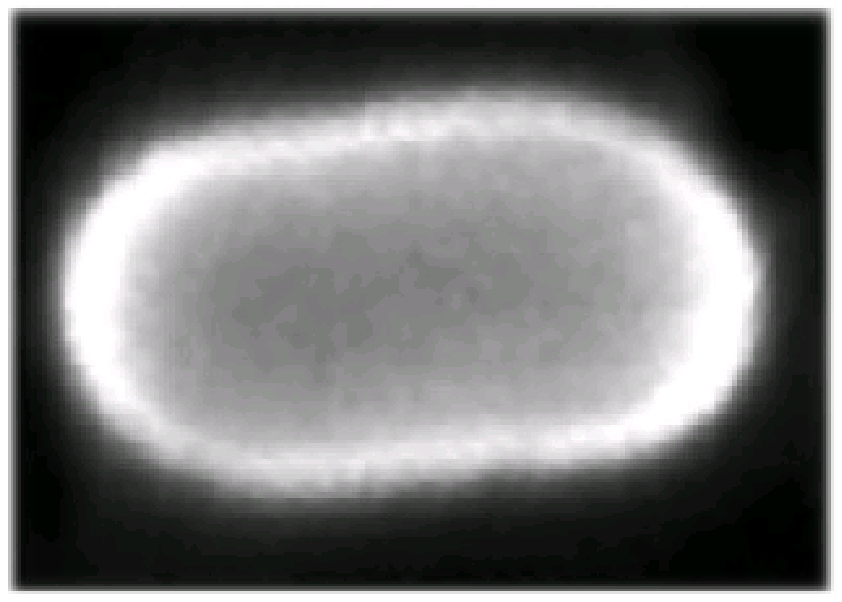}
%\label{}
}
\subfigure{
\includegraphics[scale=.5,angle=0]{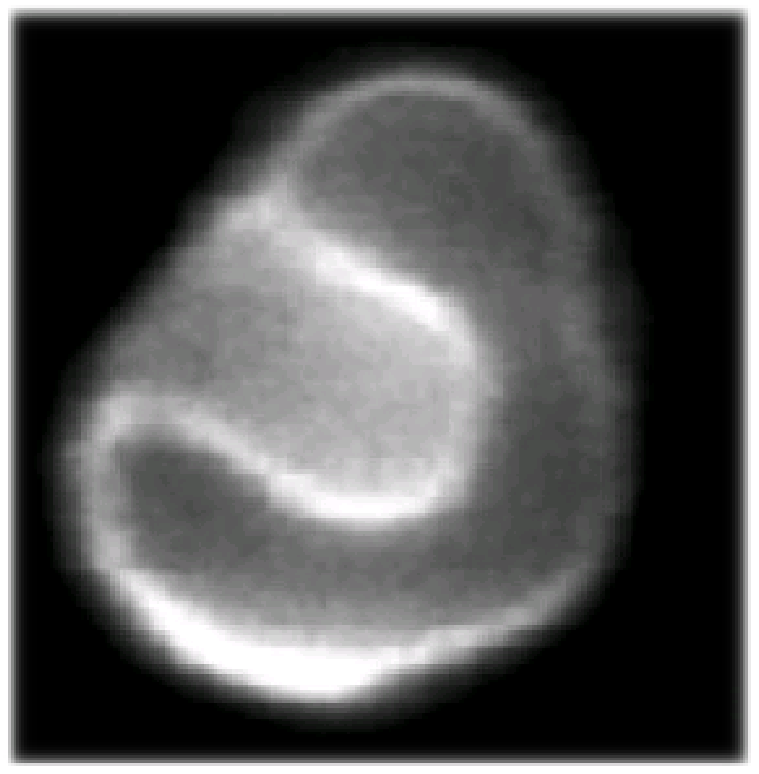}
%\label{fluid_fluid}
}
\subfigure{
\includegraphics[scale=.63,angle=0]{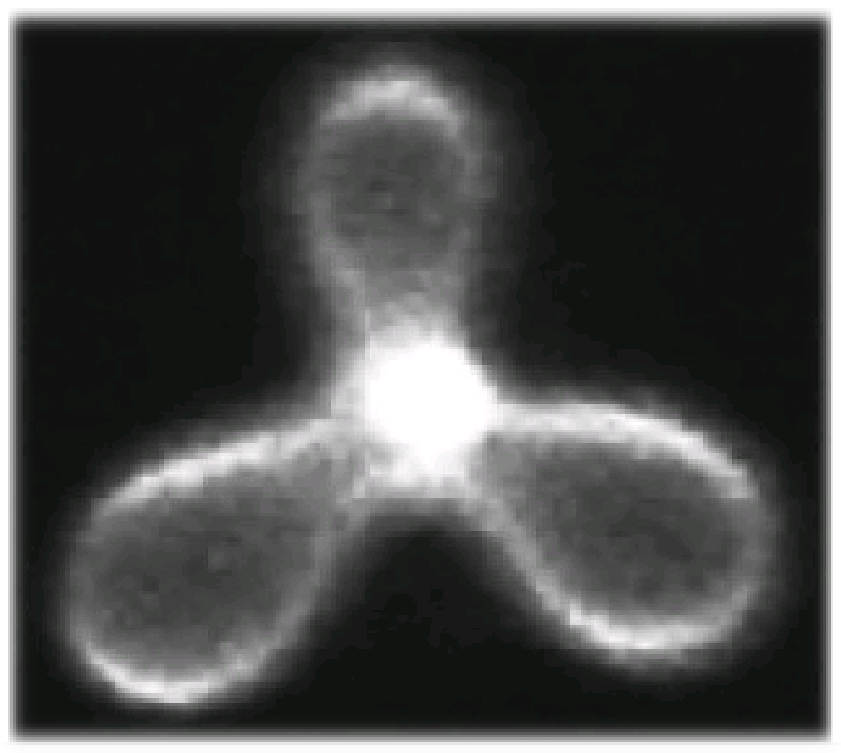}
%\label{fluid_fluid}
}
\caption{Prolate, stomatocyte and starfish vesicles made of a ternary mixture
  of phospholipids and cholesterol~\cite{Tresset_09}. The high
  frequency deformations are not optically resolvable.}
\label{shapes}
\end{center}      
\end{figure}

They appear in a variety of fluctuating shapes (see
Fig.~\ref{shapes}), whose average form depends on
the enclosed volume, total area and the asymmetry between
the leaflets that form the bilayer~\cite{Mouritsen},~\cite{Dobereiner_97}.
Note however that these images are coarse-grained, as the wave length of light, about half a micron, 
is much bigger than the membrane thickness. 
So, only low-frequency fluctuations
are visible. 
    
In the last ten years, another structure used to characterize a membrane are
membrane nanotubes, such as those seen of Fig.~\ref{Lobovkina}. 
These tubes are
formed when a point force is applied to a lipid bilayer. 
Their radii range
from a few to hundreds of nanometers. 
We will discuss them in detail in
section~\ref{tube_exp}.

\subsection{How do you characterize mechanically a membrane?}
\label{mechanical}

\begin{figure}[H]
\begin{center}
\subfigure{
\includegraphics[scale=.2,angle=0]{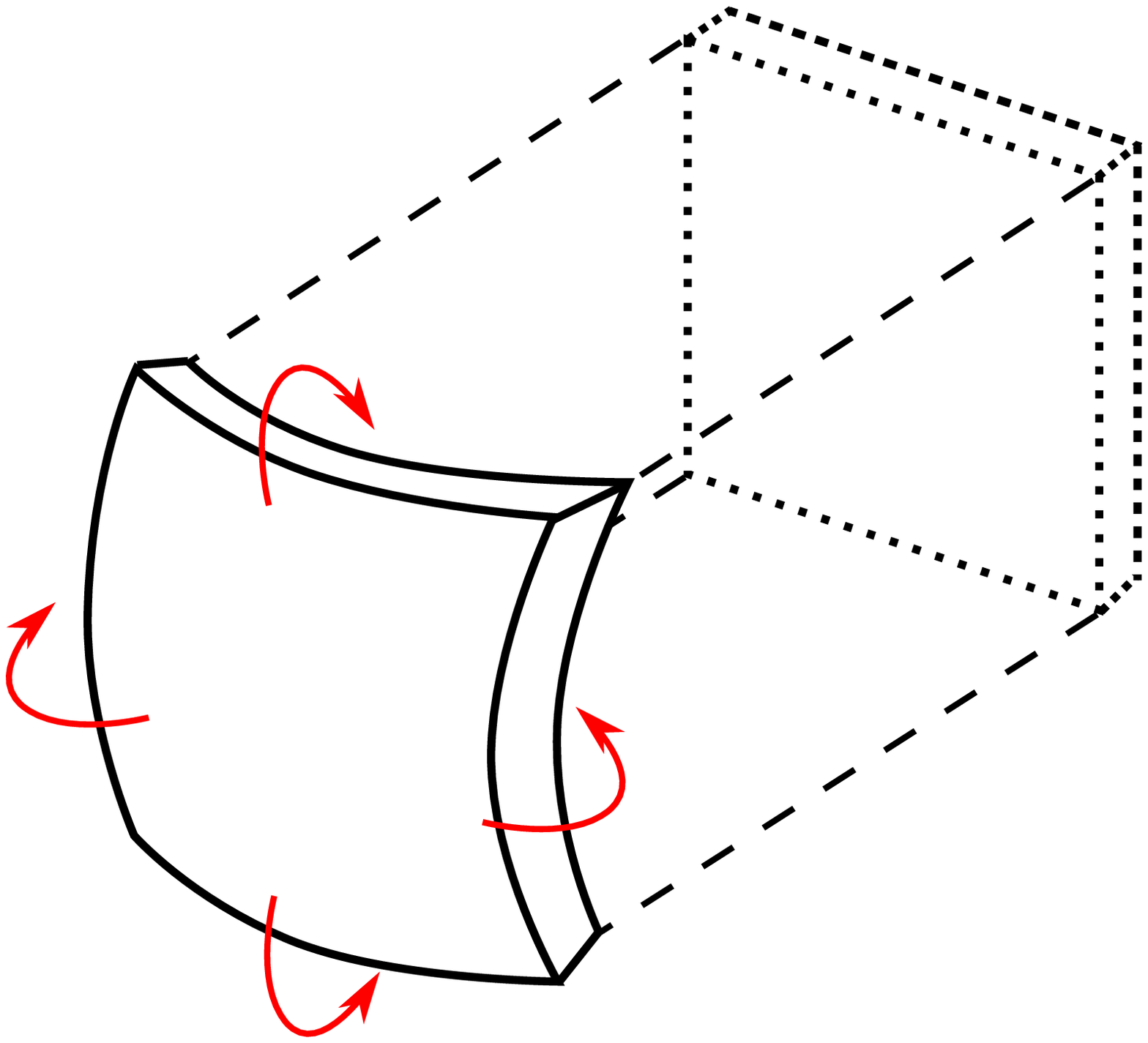}
\label{bending}
}
\subfigure{
\includegraphics[scale=.2,angle=0]{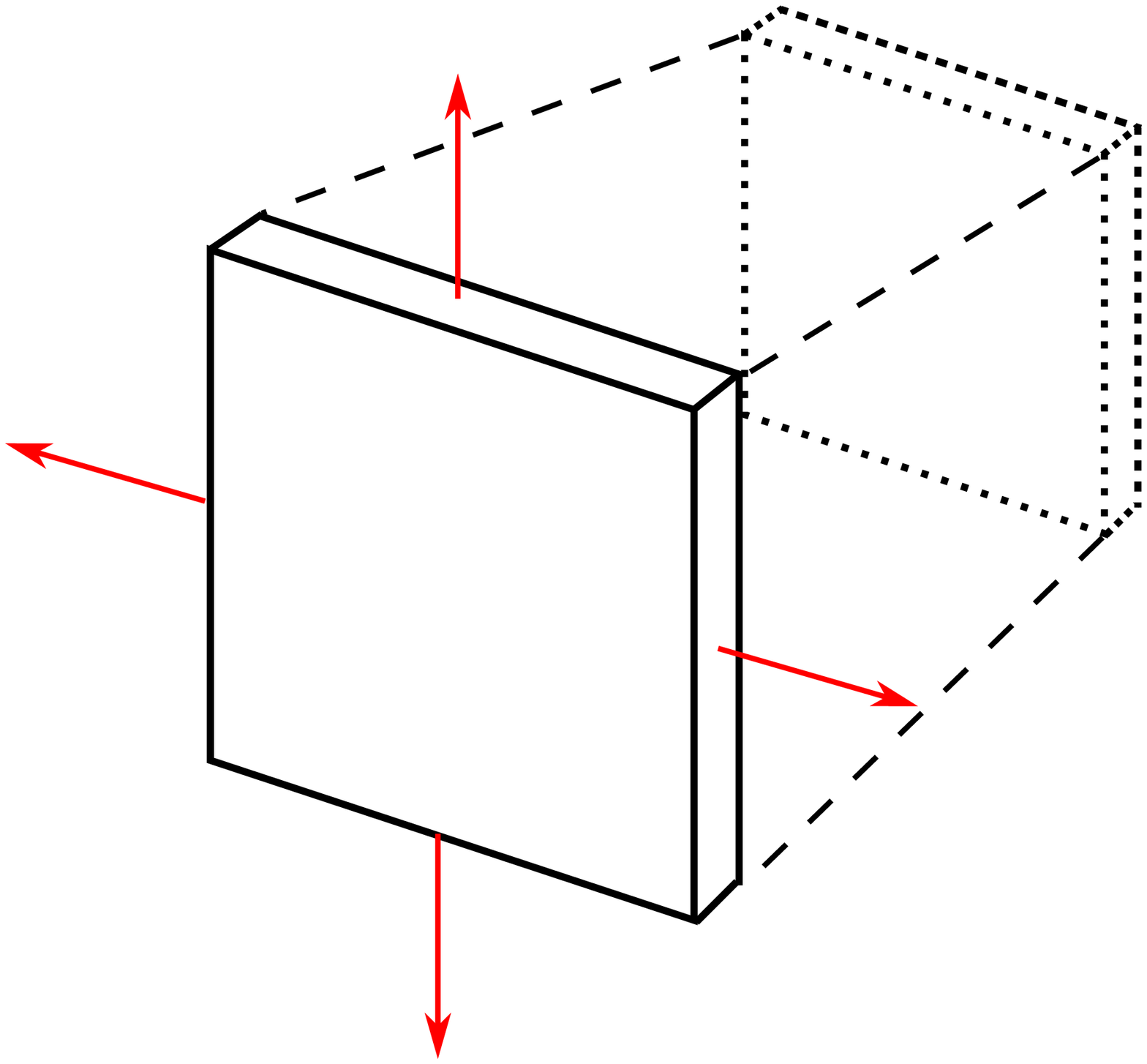}
\label{stretching}
}
\subfigure{
\includegraphics[scale=.2,angle=0]{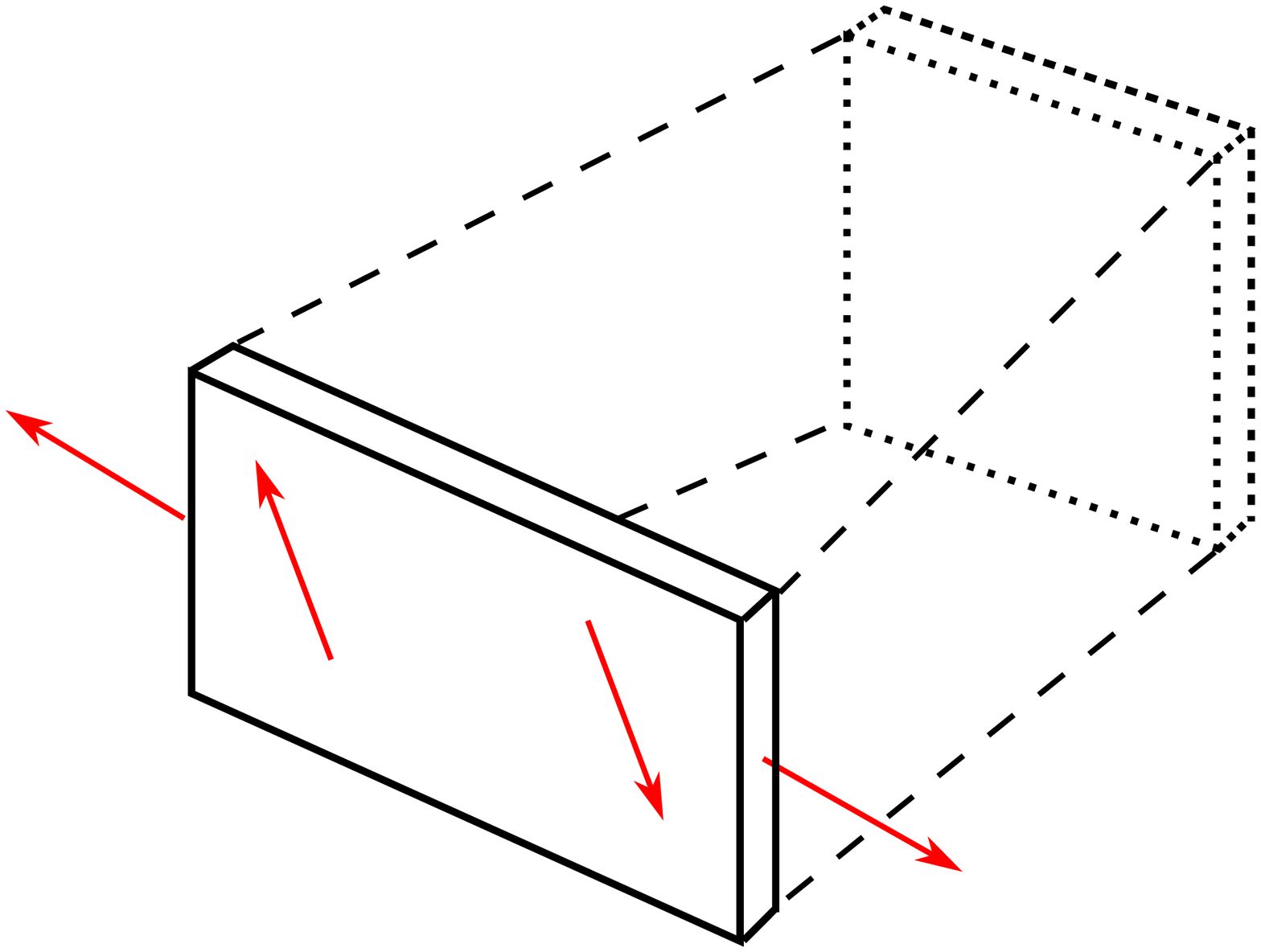}
\label{shearing}
}
\vspace{-1cm}
\caption{Three mesoscopic elastic deformations used to characterize a thin interface: bending, stretching and shearing.}
\end{center}      
\end{figure}

The first way to characterize the mechanical behavior of a material is by studying how it behaves under a reversible deformation, i. e., by studying its elastic deformations.
On a mesoscopic scale, i. e., on length scales bigger than the material
thickness, but smaller than the persistence length, which we will define in
the following, one can imagine three of these deformations: bending, stretching and shearing.
For thin interfaces, such as lipid membranes, 
bending means changing the curvature of a piece of material keeping its area
constant (see Fig.~\ref{bending}), stretching means increasing the average area
per molecule that composes the material by applying a tangential stress (Fig.~\ref{stretching}) and shearing
means changing the shape without changing its area (Fig.~\ref{shearing}). 
As lipid membranes are composed by two leaflets, one should also consider the
friction between these layers.
In the following, we will deal only with static measures, so friction will not be important.  

In liquid interfaces, such as membranes in the liquid state, molecules are free to move. Consequently, there is no resistance to shearing and
we will not study this kind of deformation.
The resistance to stretching is measured by the compression modulus $K$.
It is defined by the amount of energy $E_K$ per unit area needed to increase a piece of surface $A_0$ of $\Delta A$:

\begin{equation}
E_K = \frac{K}{2} \left(\frac{\Delta A}{A_0}\right)^2\, .
\end{equation}

\begin{figure}[H]
\begin{center}
\includegraphics[scale=.4,angle=0]{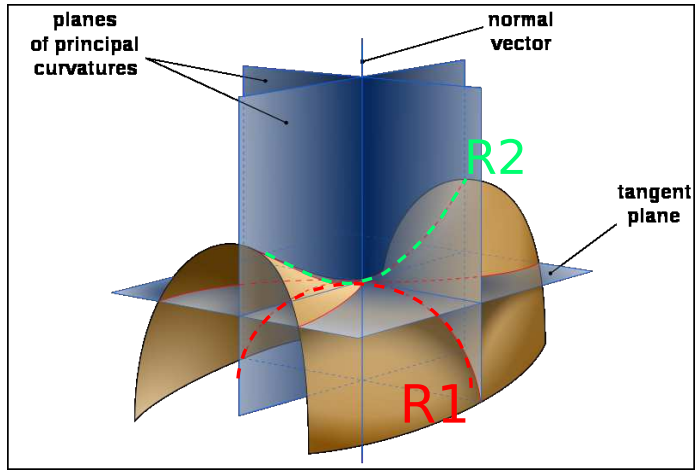}
\caption{Drawing representing the principal curvature radii of a surface.}
\label{radii}
\end{center}      
\end{figure}

Similarly, the capacity of bending is measured by the bending rigidity modulus
$\kappa$ and the Gaussian curvature modulus $\kappa_G$ defined by

\begin{equation}
E_\mathrm{curv} = E_\kappa + E_{\kappa_G} = 2 \kappa \left(H - H_0\right)^2 + \frac{\kappa_G}{R_1 R_2}\, ,
\end{equation}

\noindent where $E_\mathrm{curv}$ is the energy per unit area needed to bend, $R_1$
and $R_2$ are the two principal curvature radii seen on Fig.~\ref{radii}, $H$ is
the mean curvature, given by 

\begin{equation}
H = \frac{1}{2} \left( \frac{1}{R_1} + \frac{1}{R_2}\right) \,,
\end{equation}

\noindent and $H_0$ is the spontaneous mean curvature.
Due to the liquidity, the spontaneous mean radius $R_0$ is isotropic and $H_0
= 1/R_0$.

Equivalently, the bending rigidity of a material is reflected by its persistence length
$\xi$, defined as the length beyond which the correlation in the
direction of the tangent to the surface is lost.
For a free membrane, it relates to the bending modulus through

\begin{equation}
\xi \approx a \, \exp\left(\frac{c \, \kappa}{k_\mathrm{B} T} \right) \, ,
\end{equation}

\noindent where $a$ is a molecular length of the same order of the lipid's length, $c$ is a constant, $k_\mathrm{B}$ is the Boltzmann constant and $T$ is the temperature.

As a consequence of the Gauss-Bonnet theorem, the total contribution of the Gaussian curvature for {\bf closed} surfaces is constant:

\begin{equation}
E^\mathrm{tot}_{\kappa_G} = \kappa_G \oint_S \frac{1}{R_1R_2} \, dA = 4\pi\kappa_G (1 - g)
\, ,
\end{equation}

\noindent where the integral runs over a {\bf closed} surface $S$ and $g$ is the genus number, which describes the topology of the
surface. 
As we will not consider topology changes in this work, we will not consider
this contribution to the energy of closed membranes henceforth.

Now, let's see the figures for a typical phospholipid membrane.
We will describe in section~\ref{exp} how these quantities are measured.
First, membranes are extremely flexible, with $\kappa \approx 20 \, k_\mathrm{B}T$~\cite{Rawicz_00}, which is about a quarter of million smaller than the bending rigidity of a sheet of polystyrene of the same thickness.
This implies that they fluctuate a lot even in small scales, as it has indeed
been measured through NMR and X-rays techniques on stacks of bilayers~\cite{Salditt_03}.
Secondly, they have a high compressibility modulus $K \approx 240 \,  \mathrm{mN/m}$~\cite{Rawicz_00}, which means that the stretching due to thermal fluctuation is $\sim 10^{-8} \, \%$ and thus negligible.
Measures indicate that membranes rupture under a charge of only $\tau_\mathrm{rup}
\approx 10 \,
\mathrm{mN/m}$~\cite{Rawicz_00}, which means that it can only stretch
about $4 \%$ before breaking apart.  
Indeed, as phospholipids are bonded to each other only by
entropic forces and not by chemical bonds,
it is relatively easy to break cohesion.
Therefore, under a stress smaller than the rupture charge, it is a good
approximation to consider the total area of the membrane constant, as we have
discussed in last section.
Throughout this work, unless explicitly said, we shall place ourselves under this condition and thus only the bending energy will be considered.

The great flexibility of membranes has an important experimental consequence: one cannot measure the true surface of a bilayer.
Indeed, up to the moment, we are not able to control exactly the number of phospholipids within a membrane.
Moreover, they fluctuate on a nanometric scale, not resolvable with optical
microscopy techniques.
Thus, it is useful to define two macroscopic quantities: the excess area and
the effective or entropic mechanical tension.

\begin{figure}[H]
\begin{center}
\includegraphics[scale=.4,angle=0]{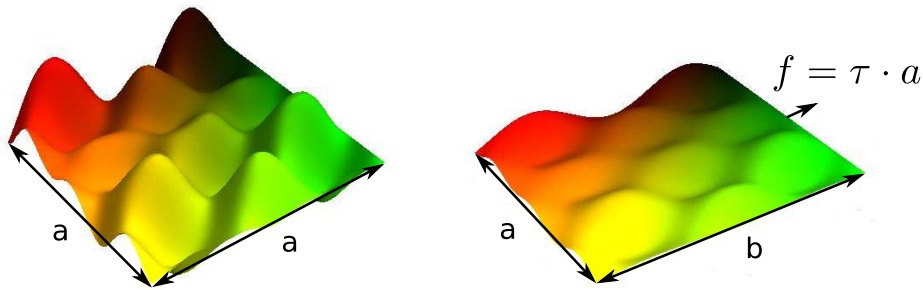}
\caption{On the left we see an illustration of the undulating surface of a membrane. The membrane
  area $A$ corresponds to the surface area of the colored membrane, while the
  macroscopically resolvable projected area $A_p$ is $a^2$. 
On the right, we see an illustration of the diminution of the excess area due to a mechanical
tension applied on the same membrane patch: the area $A$ remains the same, but the
projected area is now $a \cdot b$.}
\label{superf}
\end{center}      
\end{figure}

The excess area $\alpha$ is simply a measure of the average membrane crumpling which is not optically resolvable. It is defined by

\begin{equation}
\alpha = \frac{\langle A \rangle - A_p}{A_p}\, ,  
\end{equation}

\noindent where $A$ is the microscopic membrane area and $A_p$ is the
optically resolvable area, which we will also call in the following the projected area (see Fig.~\ref{superf}).
Experimentally, we have access only to variations on the excess area.

The entropic mechanical tension $\tau$ is a measurable macroscopic averaged
tension associated to the diminution of the excess area, i.e., to the flattening of fluctuations.
It is a pure entropic force, arising from the diminution of accessible
configurations.
It is defined as 

\begin{equation}
\tau = \left(\frac{\partial \mathcal{F}}{\partial A_p}\right)_{N_p} \, ,
\label{eq_intro_tau}
\end{equation}

\noindent where $\mathcal{F} = - k_\mathrm{B} T \ln(\mathcal{Z})$ is the free-energy of
the membrane, $\mathcal{Z}$ is the partition function and the symbol $N_p$
indicates that the derivative is taken under the condition that the total number of
lipids is constant.
In the following, as we shall consider tensions a lot weaker than the rupture
tension, it is justified to consider that the total mechanical tension
corresponds simply to $\tau$. 
%To sum up, a membrane is mechanically charactized by its
%thickness, the asymmetric on its leaflets, its bending rigidity and its excess area.   

\subsection{How different are membranes from liquid interfaces?}
\label{diff_liq}

\begin{figure}[H]
  \begin{center} 
\includegraphics[scale=.09,angle=0]{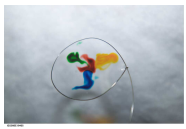}
\caption{Photo taken by International Space Station Science Officer Don Pettit of a pure water film held by
  a metal loop under micro-gravity. Food coloring was added only for visualizing~\cite{Nasa_03}. We just do not observe this phenomenon in everyday life due to gravity.  }
\label{nasa}
\end{center}      
\end{figure}

The aim of this section is to highlight the differences between membranes and
other macroscopic materials.
First, it is easy to understand why membranes behave differently from
solid membranes, such as rubber membranes, since molecules of the later are not free to move.
The difference is much subtler with liquid interfaces.
Indeed, both present two-dimensional disorder, high deformability, form
thin films (see Fig.~\ref{nasa}) and in both cases the term surface tension is currently used in the
literature.
We shall see, however, that this expression has a different meaning in each context
and that liquid interfaces are fundamentally different from membranes. 

The surface tension $\gamma$ on liquids is a constant of the material, depending only
on the molecular composition and on temperature.
At microscopical level, molecules of liquids, such as water, are strongly
chemically bonded to each other and these bonds are energetically favorable.
Molecules on the surface have less neighbors and are thus energetically
costly: the surface tension tends to make the interface as small as possible,
which results in a certain interfacial stiffness.

In terms of free-energy $\mathcal{F}$, the surface tension is conjugated to the total area
of the interface:

\begin{equation}
\gamma = \left(\frac{\partial \mathcal{F}}{\partial A}\right)_V \, ,
\end{equation}

\noindent where $V$ indicates that the derivative
is taken at constant volume.

In lipid membranes, however, if there is no stretching, there is no contribution to the free-energy
coming from the interface area.
So, $\gamma \rightarrow 0$: the lipids will form a bilayer without a bulk of
lipids.
A noteworthy confusion in literature arises from misleadingly calling the
mechanical tension $\tau$ also surface tension.
The tension $\tau$ is also associated to a surface, but to the surface of the
projected area.
Moreover, it is not a material constant since it has an entropic origin.
In the following, we will avoid the use of the ambiguous expression surface tension.
 
\begin{figure}[H]
\begin{center}
\subfigure[Metallic liquid drop over a solid substrate. Note the sharp edges at the contact between the drop and the solid~\cite{Chatain_04}.]{
\includegraphics[scale=.4,angle=0]{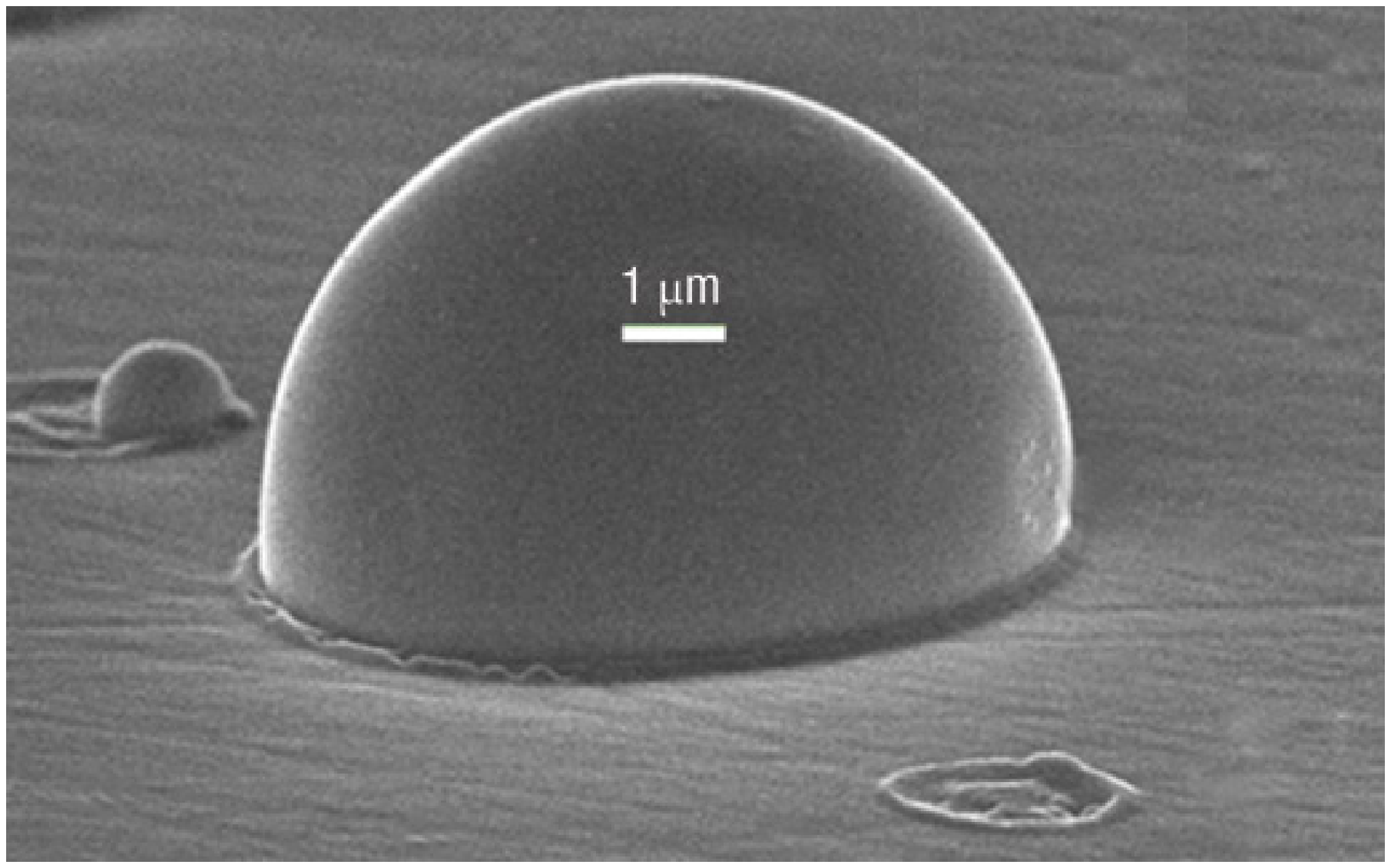}
%\label{}
}
\subfigure[An optical micrography of two vesicles adhering to a pure glass substrate, which reflects the vesicles~\cite{Gruhn_07}.
The rounded shape of the vesicle near to the glass is due to the bending rigidity.]{
\includegraphics[scale=.4,angle=0]{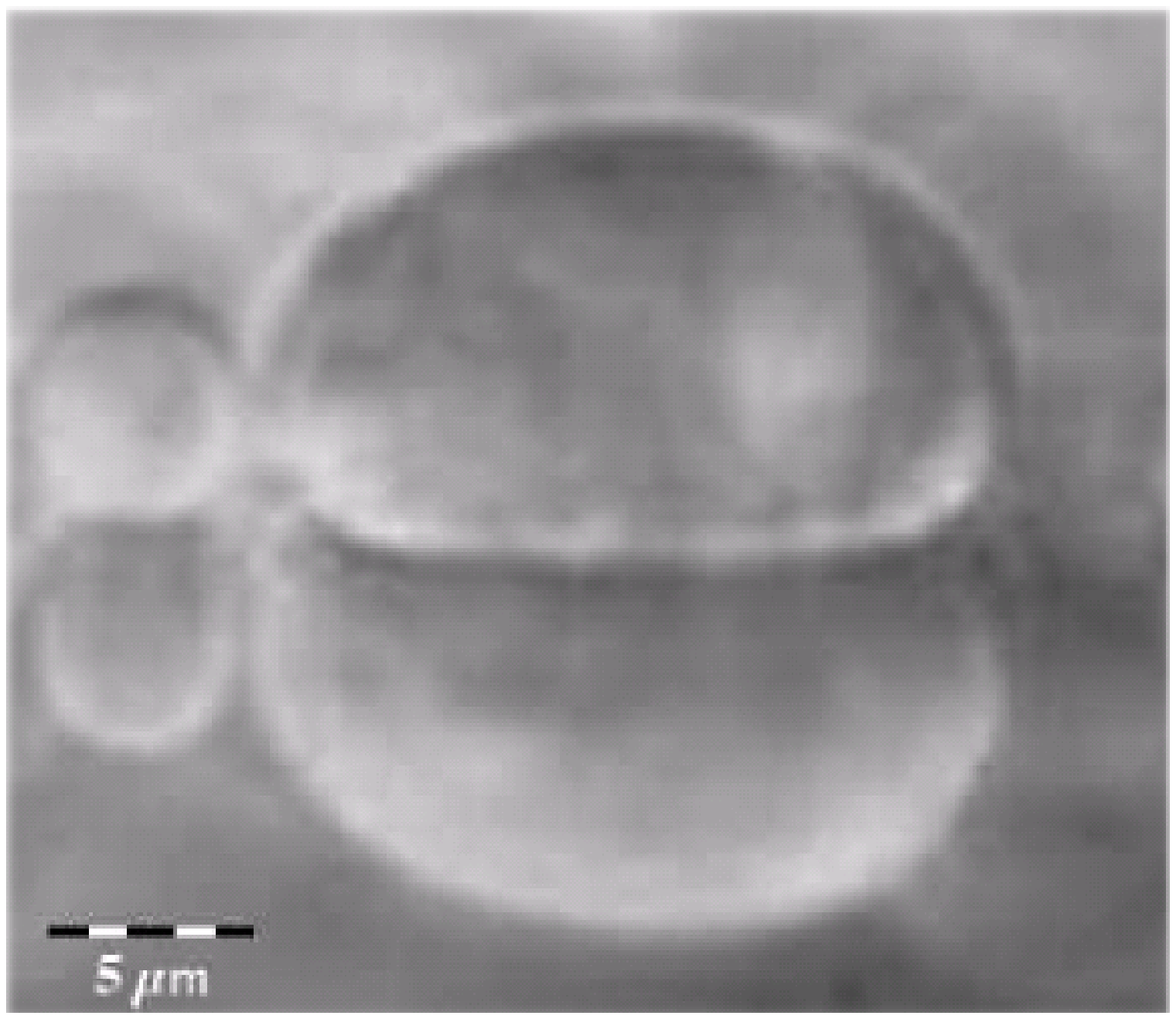}
%\label{fluid_fluid}
}
\caption{Effect of the bending rigidity.}
\label{gota_adhes}
\end{center}      
\end{figure}

Finally, another obvious difference is the bending rigidity. 
Lipids in a membrane are arranged in a particular ordered way due to the amphiphilic
nature of lipids, which leads to a bending rigidity.
In liquids, it is not the case.
We can see a consequence in Fig.~\ref{gota_adhes}: while liquid drops present a sharp contact with a substrate, membrane vesicles have a rounded contact region.
Besides, one cannot expect to extract
tubes by applying point forces to liquid films, as one does in membranes (see
section~\ref{tube_exp}). 

\section{A model for model membranes}
\label{model_model}

Here we present the three main theoretical models for liquid lipid
bilayers.
They describe membranes in a length scale much larger than the membrane
thickness, so that it can be seen as a mathematical surface.
They differ mainly in the description of the membrane's two-leaflet structure~\cite{Seifert_95}. 
In the following three descriptions, the microscopic area $A$ is kept constant.
In the case of vesicles, there is an additional constraint on the volume
enclosed by the surface.

\subsection{Spontaneous curvature (SC) model}

This model was introduced by Helfrich is 1973~\cite{Helfrich_73} and is the very simplest one.
The membrane is seen as an infinitely thin surface and its internal bilayer structure is described by a spontaneous
mean curvature $H_0$. The Hamiltonian of a bilayer is simply given by the
bending energy $E_\mathrm{curv}$ 

\begin{equation}
\mathcal{H}_\mathrm{SC} = \int_S \left[2 \kappa (H - H_0)^2 + \frac{\kappa_G}{R_1 R_2}\right] \, dA\, ,
\label{curvature}
\end{equation}

\noindent where the integral runs over the membrane surface $S$.

\subsection{Bilayer couple (BC)  model}

\begin{figure}[H]
\begin{center}
\includegraphics[scale=.45,angle=0]{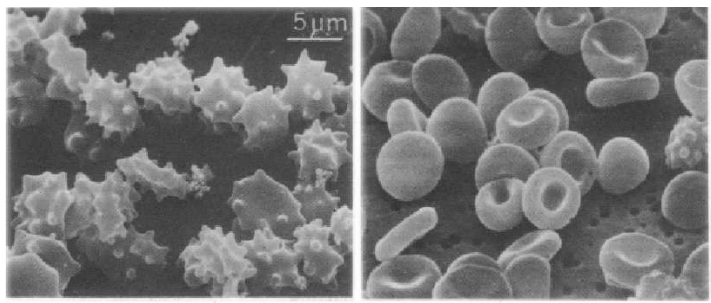}
\caption{Original electron micrographs of erythrocytes treated with $0.2 \,
  \mathrm{mM}$ of metho-chlorpromazine (at left) and $6 \, \mathrm{\mu M}$ of
  chlorpromazine (at right)~\cite{Sheetz_74}. In the first case, red blood cells become spiked,
  while in the later they adopt a cup-like shape.}
\label{RBC}
\end{center}      
\end{figure}

In this model, the two leaflets of a bilayer may respond differently to an
external perturbation, such as chemical substances, while remaining coupled~\cite{Svetina_89}.
It was first introduced in 1974 to explain qualitatively experiments on red
blood cells~\cite{Sheetz_74},~\cite{Evans_74}.  
It had been observed that erythrocytes treated with amphiphilic drugs change of
shape, becoming more cup-like or instead, spiked (see Fig.~\ref{RBC}).
The authors proposed that spike-inducing drugs tended to bind to the external
leaflet of the bilayer, while the cup-inducers binded mainly to the cytoplasmic
leaflet.
Each monolayer would thus have a different area, which would force a
curvature. 
As flip-flop transitions are very slow, this area difference would be constant
over time.
Another evidence in favor of this model comes from vesicles: if the SC model
were correct, vesicles composed by a single kind of phospholipid and 
similarly prepared should behave the same way, since the natural bending tendency comes exclusively from the chemical asymmetry of the monolayers.
Experiments show the contrary: vesicles prepared the same way have different
preferred curvatures, possibly due to the fact that the two monolayers had
different areas when they closed to form a vesicle~\cite{Miao_94}.

The model proposes that the preferred curvature of a membrane depends on two contributions: the
spontaneous curvature of each monolayer that adds up to a local spontaneous
curvature of the membrane $H_0$ and the area difference between the two
monolayers, which gives a non-local contribution~\cite{Dobereiner_00}.
The energy in this model is still given by equation (\ref{curvature}), but
there is an additional constraint in the area-difference between the neutral
surfaces of the outer and inner leaflet, the neutral surface being the imaginary surface within a
bent leaflet where there is no compression or extension. 
This means keeping a hard constraint on

\begin{equation}
\Delta A = A_\mathrm{out} - A_\mathrm{in} = 2 D M = 2 D \int_S H \, dA \, ,
\end{equation}

\noindent where $D$ is the distance between neutral surfaces and $M$ is the
total mean curvature. 

\subsection{The area-difference elasticity (ADE) model}

\begin{figure}[H]
\begin{center}
\includegraphics[scale=.4,angle=0]{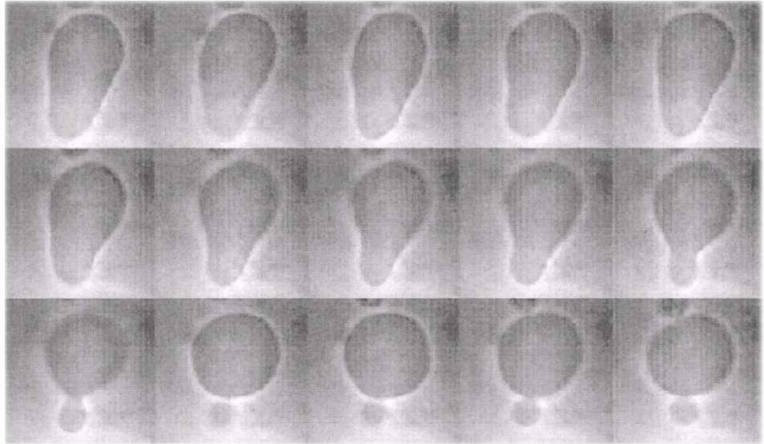}
\caption{Time sequence showing the budding transition from a pear-shaped
  vesicle. In this case, the pear-like vesicle is instable, but it is not always the
  case~\cite{Dobereiner_97}.}
\label{budding}
\end{center}      
\end{figure}

The ADE model is the generalization of the two preceding models.
It was introduced in the $90$'s to explain the budding transition of some
vesicles, i. e., when a vesicle adopts the shape of a parent vesicle
attached through a neck to a smaller vesicle (see Fig.~\ref{budding}).
The BC model predicts that this transition should be continuous, while in some
experiments discontinuous transitions were observed~\cite{Miao_94}. 

The ADE model accounts for the fact that small relative compressions or extensions of the
bilayer have an energetic cost comparable to the bending energy.   
Instead of a hard constraint on the area-difference between leaflets, the
area-difference is regulated through an additional quadratic term in the energy, leading
to

\begin{equation}
\mathcal{H}_\mathrm{ADE} = \mathcal{H}_\mathrm{SC} + \frac{\bar{\kappa}}{2} \frac{\pi}{A D^2} \left(\Delta A - \Delta
A_0\right)^2\, , 
\end{equation}  

\noindent where $\Delta A_0$ is the optimal area difference, also defined
through

\begin{equation}
\Delta A_0 = \frac{N_\mathrm{out} - N_\mathrm{in}}{\phi_0} \, ,
\end{equation}

\noindent where $N_\mathrm{out/in}$ is the number of lipids on the
outer/inner monolayer and $\phi_0$ is the equilibrium density of lipid molecules.
In the limiting case where $\bar{\kappa} \rightarrow 0$, we recover the SC
model, whereas in the limit $\bar{\kappa} \rightarrow \infty$, we recover the
BC model.

\subsection{Validity of models}

\begin{figure}[H]
\begin{center}
\includegraphics[scale=.75,angle=0]{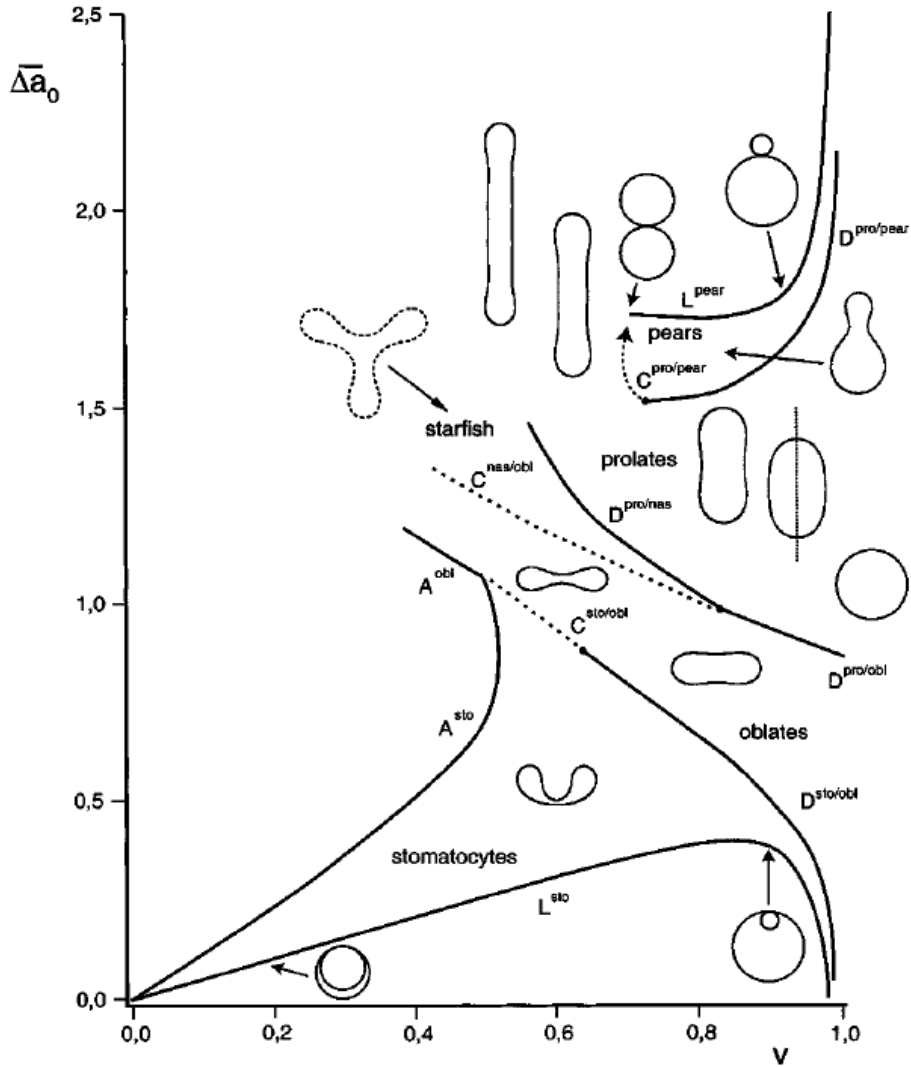}
\caption{Phase diagram of the stationary shapes of vesicles as predicted from
  the ADE-model. 
The quantity $v$ is the volume-to-area ratio and $\overline{\Delta a_0}$ is the
effective area difference (dimensionless)~\cite{Dobereiner_00}.}
\label{eq_formes}
\end{center}      
\end{figure}

The question of which model describes the best a bilayer was not easy to
answer.
The main difficulty when studying vesicles comes from the fact that the three models have the same
equilibrium shapes~\cite{Miao_94},~\cite{Seifert_91} (one can see a map of these shapes for the ADE model on Fig.~\ref{eq_formes}~\cite{Dobereiner_00}).
To complicate things, in certain cases, such as quasi-spherical vesicles, even the thermal
fluctuations of the three models is the same~\cite{Seifert_95}.

The three models predict however different stabilities for equilibrium shapes.
For instance, the SC model predicts that pear-like vesicles should be always
unstable, the BC model predicts these shapes to be always stable and the ADE
model predicts stability for large values of $\bar{\kappa}$.
Another difference is the nature of shape transitions induced by changes in
temperature or osmotic pressure, which are in general continuous
in BC and ADE and discontinuous SC.

A careful work by D\"obereiner et al.~\cite{Dobereiner_97} showed that
experimental data was indeed compatible with the theoretical phase diagram for
ADE-model shown in Fig.~\ref{eq_formes}. 
This result was corroborated by good experimental and theoretical
compatibility of the analysis of the stability and of trajectories on
this phase diagrams.
Nowadays, the ADE-model is accepted as the best description for closed
bilayers.

There are however some situations where using SC is justifiable. 
First, as we have said, the models are equivalent on the study of thermal
fluctuations of quasi-spherical vesicles. 
It is justifiable and simpler to use the SC model in this case.  
Besides, there is no area-difference between monolayers when these are both in
contact with the same lipid reservoir and thus SC is suitable also in this case.

\subsection{From canonical ensemble to macrocanonical}
\label{sig_intro}

The Hamiltonian presented in the previous sections have an additional
constraint in the number of lipids per membrane, or equivalently, in the total
surface.
In statistical mechanics, the ensemble of these constrained configurations is known as the canonical ensemble.
It is a standard procedure to pass to the macrocanonical ensemble and let the
area fluctuate.
Physically, it means that the system we are interested in is in contact with a
large reservoir of lipids, which may be a justified supposition in some cases.
In this ensemble, one can control the average area by adding a term 

\begin{equation}
\mathcal{H}_A = \sigma A 
\end{equation}

\noindent to the Hamiltonian. The constant $\sigma$ is a Lagrange-multiplier
analogous to a chemical potential used to impose a certain value to the average area a posteriori, once

\begin{equation}
\langle A \rangle = \left(\frac{\partial \mathcal{F}}{\partial \sigma}\right) \, ,
\end{equation} 

\noindent where $\mathcal{F}$ is the free-energy. The Lagrange-multiplier
$\sigma$ has the dimension of a
tension, so sometimes it is called the surface tension, term we will avoid
here.
It is important to note that it is not generally measurable in experiments.
It has however a measurable macroscopic counterpart $r$, that we will define in
the next section.

\subsection{Fluctuation spectrum for small fluctuations}
\label{grand_can}

We suppose here that we study a symmetrical planar bilayer with squared projected area
$A_p \equiv L^2$, in contact with a lipid reservoir and
well described by the SC model.
The energy is given by

\begin{equation}
\mathcal{H} = \int_S \left(2\kappa H^2 + \frac{\kappa_G}{R_1 R_2} + \sigma \right) dA\, .
\label{Helfrich}
\end{equation}

\begin{figure}[H]
\begin{center}
\includegraphics[scale=.35,angle=0]{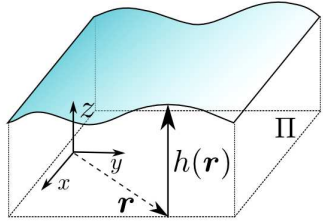}
\caption{Weakly fluctuating planar membrane described in the Monge gauge.}
\label{monge}
\end{center}      
\end{figure}

\noindent This energy will be used many times throughout this work. We will call it
simply the Helfrich Hamiltonian.
Consider now that the membrane is reasonably tense, so that fluctuations
are small. 
Its position can be described by its height $h(\bm{r})$ with respect to a plan $\Pi$ parallel to the
average plan, where $\bm{r} = (x,y)$ (see Fig.~\ref{monge}).
This is known as the Monge gauge. 

Under these assumptions, eq.(\ref{Helfrich}) becomes

\begin{eqnarray}
\mathcal{H} &=& \mathcal{H}_0 + \int_{A_p} \left[\frac{\kappa}{2}\left(\nabla^2 h\right)^2 + 
  \frac{\sigma}{2} \left(\nabla h\right)^2  + \kappa_G \, \det(h_{ij}) + \mathcal{O}(h^4)\right] dx dy \, , \nonumber\\
&=& \mathcal{H}_0 + \frac{1}{2} \int_{A_p} h(\bm{r}) \, \mathcal{L}\,  h(\bm{r}) \, dx dy + \mathcal{O}(h^4)
\label{Hel_free}
\end{eqnarray}

\noindent where $\mathcal{H}_0 = \sigma A_p$ is a constant, $h_{ij} \equiv \partial ^2 h/\partial i \partial j$, where $i, j \in \{x,y\}$, $\det(h_{ij}) = h_{xx}\, h_{yy} - h_{xy}^2$ and $\mathcal{L} \equiv \kappa \Delta^2 - \sigma \Delta$ is the operator associated to the quadratic terms of the energy. 

In order to evaluate averages involving $h(\bm{r})$, in field theory one usually adds a term proportional to an imaginary external field $m(\bm{r})$ to the Hamiltonian, obtaining the Hamiltonian

\begin{equation}
  \mathcal{H}' = \mathcal{H} - \int_{A_p} h(\bm{r})\, m(\bm{r}) \, dx\, dy \, .
\end{equation}

\noindent The corresponding free-energy is 

\begin{equation}
  \mathcal{F} = -k_\mathrm{B} T \, \ln(\mathcal{Z}) \, ,
\end{equation}

\noindent with the partition function given by the functional integral

\begin{equation}
  \mathcal{Z} = \int \mathcal{D}[h] \, e^{- \beta \, \mathcal{H}'[h]} \, ,
  \label{eq_intro_Z}
\end{equation}

\noindent where $\beta = 1/(k_\mathrm{B} T)$. It follows 

\begin{equation}
  \langle h(\bm{r}) \rangle \equiv \left.\frac{\int \mathcal{D}[h] \, h(\bm{r}) \,
    e^{-\beta \mathcal{H}[h]} \, e^{-\beta \int_{A_p} h \, m\, 
      dA}}{\mathcal{Z}}\right|_{m=0} =  -\left. \frac{\delta \mathcal{F}}{\delta m(\bm{r})}\right|_{m=0}
  \label{eq_intro_ave}
\end{equation}

\noindent and

\begin{equation}
  \langle h(\bm{r})\, h(\bm{r}') \rangle - \langle h(\bm{r}) \rangle \, \langle h(\bm{r}') \rangle = -\left. \frac{1}{\beta} \frac{\delta^2 \mathcal{F}}{\delta m(\bm{r})\delta m(\bm{r}')}\right|_{m=0} \, ,
   \label{eq_intro_corr}
\end{equation}

\noindent where $\delta \mathcal{F}/\delta m(\bm{r})$ represents
the functional derivative of the free-energy with respect to the field
$m$ at the point $\bm{r}$.

One has now to choose an appropriate measure $\mathcal{D}[h]$, which is in general a complex task~\cite{Cai_94}, \cite{David_91}.
Up to first order on the temperature and up to second order on $h$, it is
justified to consider simply a discretization of the projected plan on $N^2$
squares of area $\bar{a}^2$ and let 

\begin{equation}
  \mathcal{D}_\mathrm{naive}[h] = \prod_{p_x,p_y} \frac{dh^{p_x,p_y}}{\lambda} \, ,
\end{equation}

\noindent where $h^{p_x,p_y}$ is the height at the point $\bm{r}_{p_x,p_y} =
p_x \, \bar{a} \, \bm{e}_x + p_y \, \bar{a} \, \bm{e}_y $,  $\lambda$ is a
length introduced to render the measure dimensionless and both $p_x$ and $p_y$
$\in
\{1,\cdots, N\}$ ~\cite{Cai_94}.
This measure, which we will call {\it naive} as in ref.\cite{Cai_94}, yields no supplemental term to the Hamiltonian.

Evaluating the Gaussian integrals on eq.(\ref{eq_intro_Z}), one obtains

\begin{equation}
  \mathcal{Z} = C \,  e^{\frac{\beta}{2} \int_{A_p} m(\bm{r}) \, \mathcal{H}^{-1}(\bm{r}, \bm{r}') \, m(\bm{r}') \, dx dy}\, ,
\end{equation}

\noindent with $C$ a constant and $\mathcal{H}^{-1}(\bm{r}, \bm{r}') \equiv \Gamma(\bm{r}' - \bm{r})$ solution of

\begin{equation}
  \mathcal{L} \, \Gamma(\bm{r}) = \delta(\bm{r}) \, ,
  \label{eq_intro_gamma}
\end{equation}

\noindent where $\delta(\bm{r})$ is Dirac's delta function. 
Using the Fourier transform

\begin{equation}
\Gamma(\bm{r}) = \frac{1}{L} \sum_{\bm{q}} \Gamma_{n,m} \, e^{\icomp \, \bm{q}\cdot\bm{r}} \, , 
\end{equation}

\noindent with $\bm{q} = 2\pi/L \, (n,m)$, $n,m \in \mathbb{N}$ and

\begin{equation}
\sum_{\bm{q}} \equiv \sum_{|n| \, \leq \, N_\mathrm{max}} \, \sum_{|m| \, \leq \, N_\mathrm{max}} \, ,
\end{equation}

\noindent where $N_\mathrm{max} = L/(2 \bar{a})$ corresponds to smallest possible wave-length, eq.(\ref{eq_intro_gamma}) yields

\begin{equation}
  \Gamma_{n,m} = \frac{1}{\sigma q^2 + \kappa q^4}\, .
\end{equation}

From eq.(\ref{eq_intro_ave}) and eq.(\ref{eq_intro_corr}), one obtains
respectively $\langle h(\bm{r}) \rangle = 0$ and the correlation function

\begin{equation}
\langle h(\bm{r}) h(\bm{r}') \rangle = \frac{k_\mathrm{B} T}{A_p} \sum_{\bm{q}} \frac{e^{i \, \bm{q} \cdot (\bm{r} - \bm{r'})}}{\sigma q^2 + \kappa q^4} \,.
\label{eq_intro_correl}
\end{equation}

\noindent Note that the Gaussian curvature contribution vanishes.
Applying the Fourier transform as defined above to $h(\bm{r})$ and $h(\bm{r}')$ in eq.(\ref{eq_intro_correl}), one obtains 

%% \begin{equation}
%% h(\bm{r}) = \frac{1}{L} \sum_{\bm{q}} h_{n,m} \, e^{\icomp \, \bm{q}\cdot
%%    \bm{r}} \, , 
%% \end{equation}

%% \noindent with $\bm{q} = 2\pi/L \, (n,m)$, $n,m \in \mathbb{N}$ and

%% \begin{equation}
%% \sum_{\bm{q}} \equiv \sum_{|n| \, \leq \, N_\mathrm{max}} \, \sum_{|m| \, \leq \, N_\mathrm{max}} \, ,
%% \end{equation}

%% \noindent 

\begin{equation}
\langle |h_{n,m}|^2 \rangle = \frac{ k_\mathrm{B} T}{\sigma q^2 + \kappa q^4}
\, ,
\label{spect_base}
\end{equation}

%\noindent to eq.(\ref{Hel_free}), one obtains up to order two in $h$
 
%% \begin{equation}
%% \mathcal{H} = \mathcal{H}_0 + \frac{1}{2}\sum_{\bm{q}} \left(\sigma q^2 + \kappa q^4\right) |h_{n,m}|^2 \,
%% ,
%% \end{equation}

\noindent where the wave vectors range from $q_\mathrm{min} = 2 \pi/L$
to an upper cut-off $\Lambda = 2\pi N_\mathrm{max}/L \approx 1/\bar{a}$.
Throughout this work, we shall consider that numerically $\bar{a} \equiv a$, where $a$ is of the order of the membrane thickness.
Remark that since $h(\bm{r})$ is real, we have $h_{-n,-m} = h_{n,m}^*$, where the symbol $^*$ stands for the complex conjugate, yielding $|h_{-n,-m}| = |h_{n,m}|$.

Similar calculations can be carried out for non-planar membranes, yielding a
correlation function similar eq.(\ref{eq_intro_correl}).
Consequently, by measuring the fluctuations of a membrane, one could deduce its bending rigidity and the tension $\sigma$.
The problem is that experimentally we have access only to a coarse-grained vision of the
membrane. 
One has thus to consider that the values deduced from the fluctuation
spectrum are in fact renormalized values, which we will call $\kappa_\mathrm{eff}$ for
the bending rigidity and $r$ for the effective tension.   
Renormalization calculations~\cite{Peliti_85} indicate that

\begin{equation}
  \kappa_\mathrm{eff} = \kappa - \frac{3 k_\mathrm{B}T}{4\pi}
  \ln\left(\frac{L}{a}\right)\, ,
\label{kappa_effect}
\end{equation}

\noindent where $L$ is the size of the membrane and $a$ is a microscopical cut-off.
Experimentally, the dependence of $\kappa_\mathrm{eff}$ as a function of $L$ is very difficult to measure, since the dependence is logarithmic.
Numerical simulations however confirmed the logarithmic dependence on $L$~\cite{Gompper_96}.
For a rough numerical estimate, if we consider a vesicle with radius $R
\approx 10 \, \mathrm{\mu m}$ and we consider the cut-off of the order of the membrane
thickness $a \approx 5 \, \mathrm{nm}$, we obtain $\kappa_\mathrm{eff} \approx \kappa - 2 k_\mathrm{B} T$.
The correction is thus one order of magnitude smaller than typical values of $\kappa$.
Henceforth, we will assume $\kappa_\mathrm{eff} \equiv \kappa$.
The distinction between $r$ and $\sigma$ will however be kept and discussed
throughout this paper.

\section{Experiments}
\label{exp}

Here we present some current experimental apparatus and techniques used to measure the
relevant mechanical parameters $\kappa$, $K$, $\tau$, $r$ and $\alpha$.
We will not mention neither measures of $\bar{\kappa}$, since we will focus on
symmetrical membranes, neither measures of $\kappa_G$, since we will work with
closed membranes of fixed topology (in the case of non-closed membranes, we
will show it is not relevant in section~\ref{stress_plan}).
In section~\ref{sum_exp}, we sum up these experiments.

\subsection{Micropipette experiments}
\label{subsection_0_micro}

\begin{figure}[H]
\begin{center}
\subfigure{
\includegraphics[scale=.35,angle=0]{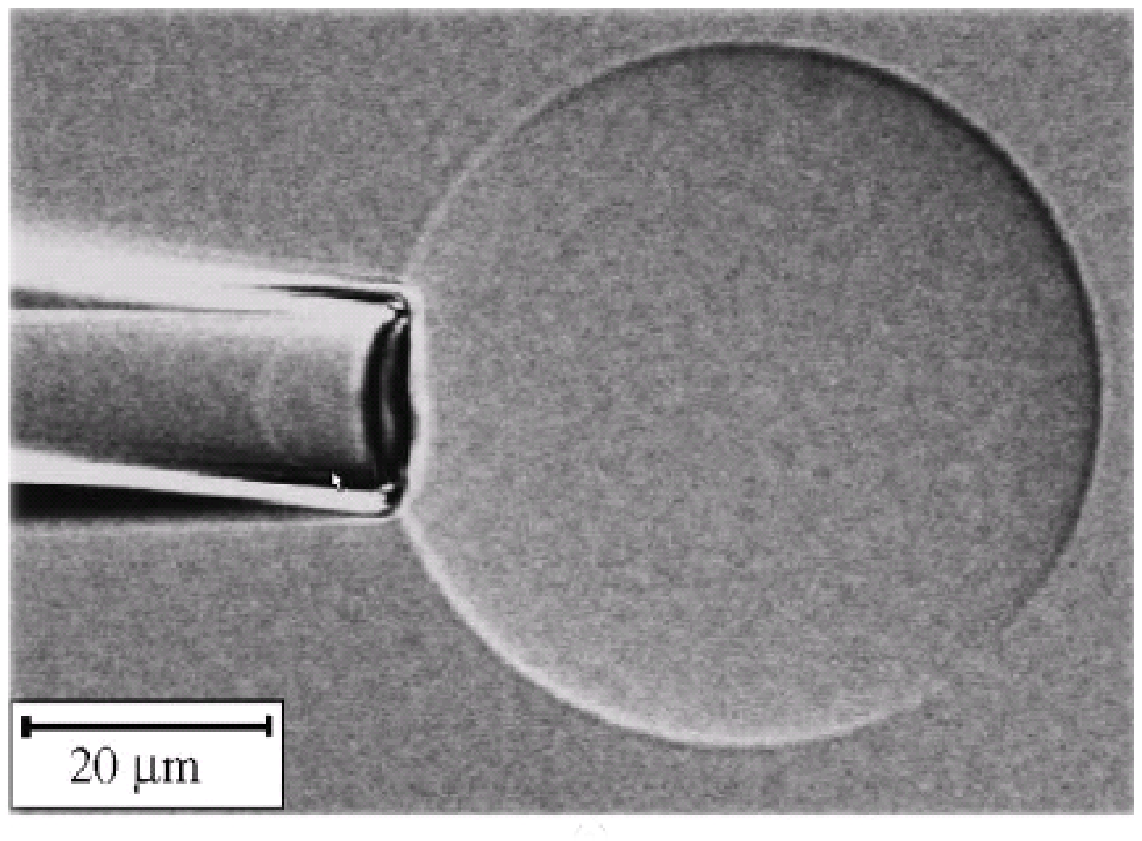}
%\label{}
}
\subfigure{
\includegraphics[scale=.35,angle=0]{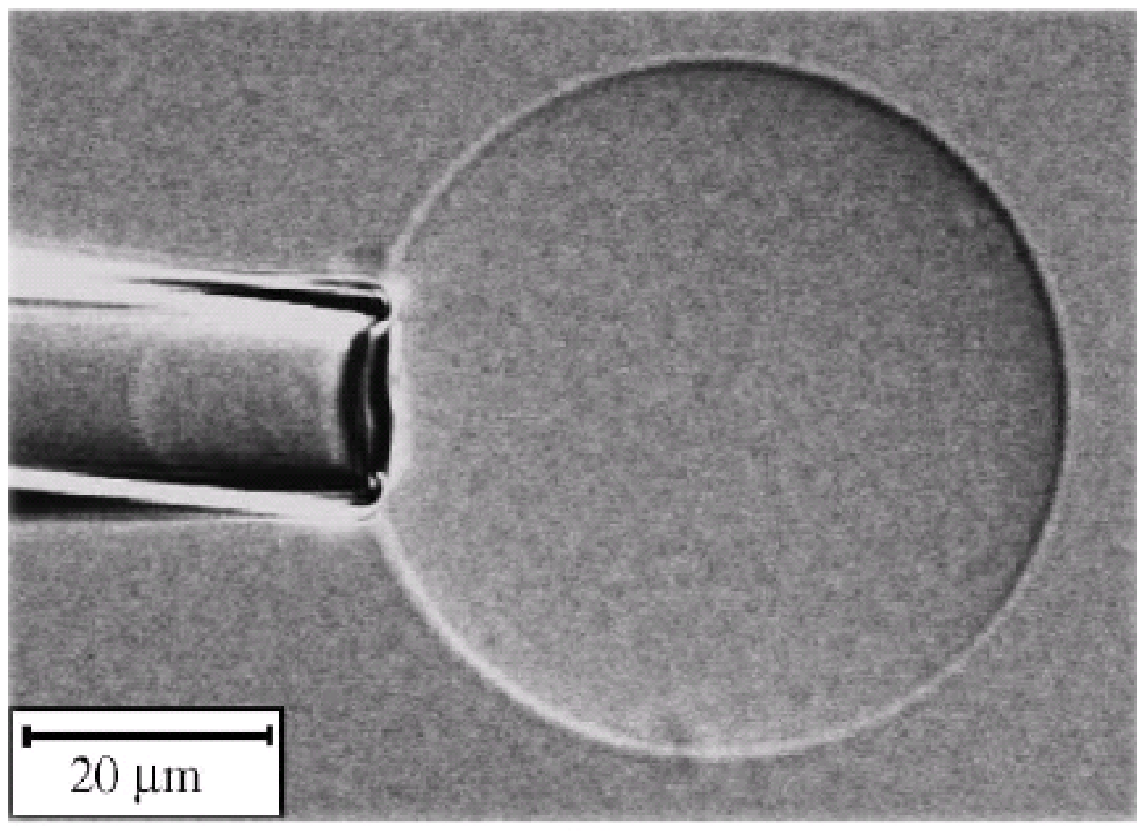}
%\label{fluid_fluid}
}
\subfigure{
\includegraphics[scale=.45,angle=0]{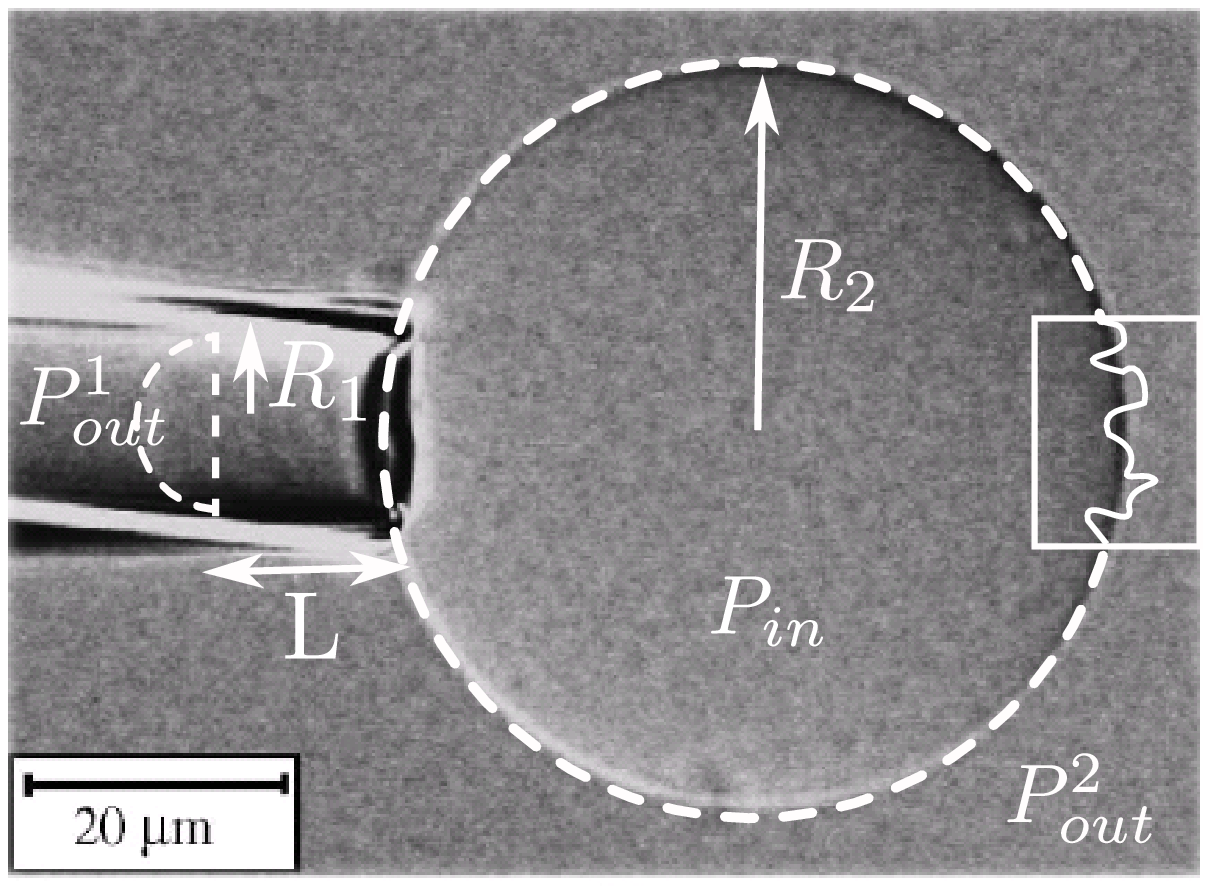}
%\label{fluid_fluid}
}
\caption{The upper figures show typical micrographs of vesicles under suction
  for increasing values of $\Delta P$ (extracted from~\cite{Monteiro_04}). The lower drawing indicates the
  measurable quantities. The inset on the right represents a zoom of the
  microscopic fluctuations that are averaged out in these micrographs.}
\label{micropipet}
\end{center}      
\end{figure}

In these experiments, a micropipette of some micrometers of diameter is held in contact with a vesicle.
One increases the membrane tension by decreasing the pressure
$P_\mathrm{out}^1$ on the pipette. 
A portion of the membrane of length $L$ is then
sucked inside the pipette and the optical resolvable surface $A_p$ increases (see
Fig.~\ref{micropipet}).

Usually, successive measures with increasing pressure are taken.
The first configuration, when the vesicle is just grabbed by the pipette and $L$
is small, is the reference configuration.
The projected area of this configuration is $A_p^{i}$, $R_1$ is the radius of
the micropipette and $R_2$ is the radius of the vesicle in the reference configuration. 
Under the condition of constant volume and $R_2 \ll R_1$, the percent
difference on the projected area of a posterior measure whose projected area is
$A_p^f$ is given by

\begin{equation}
\frac{A_p^f - A_p^i}{A_p^i} = \frac{1}{2} \left[\left(\frac{R_1}{R_2}\right)^2 - \left(\frac{R_1}{R_2}\right)^3\right] \frac{\Delta
L}{R_1} \gamma\, ,
\end{equation}

\noindent where $\Delta L$ is the length variation of the cylinder sucked inside the pipette relative to the reference measure and $\gamma$ is a corrective factor which
arises when $L$ is non zero in the reference
configuration~\cite{Henriksen_04}.

Meanwhile, the average applied tension can be related to the difference of
pressure $\Delta P = P_\mathrm{out}^2 - P_\mathrm{out}^1$ through the Young--Laplace equation~\cite{Evans_90}.
For a very thin interface under tension $\tau$ and whose principal curvature radii are $R'$ and $R''$, it states

\begin{equation}
  \Delta P = \tau\left(\frac{1}{R'} + \frac{1}{R''}\right) \, .
  \label{Young_Laplace}
  \end{equation}

\noindent For the system shown in Fig.~\ref{micropipet}, one obtains thus the relation

\begin{equation}
  P_\mathrm{in} - P_\mathrm{out}^2 = \frac{2 \tau}{R_2}
  \label{intro_lapla2}
\end{equation}

\noindent for the largest part of the vesicle and

\begin{equation}
  P_\mathrm{in} - P_\mathrm{out}^1 = \frac{2 \tau}{R_1}
  \label{intro_lapla1}
\end{equation}

\noindent for the spherical cap inside the micropipette.
Finally, by subtracting eq.(\ref{intro_lapla2}) from eq.(\ref{intro_lapla1}), one obtains

\begin{equation}
\Delta P = P_\mathrm{out}^2 - P_\mathrm{out}^1 =  2 \tau \left(\frac{1}{R_1} - \frac{1}{R_2}\right) \, .
\label{Laplace}
\end{equation}

\noindent For very small pipettes or for vesicles under very weak tension, this relation
must be corrected~\cite{Fournier_08}.
Through this technique, one can apply a wide range of tensions on membranes,
from very small ones ($\sim 10^{-9}\,  \mathrm{N/m}$) up to 
rupture tensions ($\sim 10^{-2} \, \mathrm{N/m}$)~\cite{Rawicz_00},~\cite{Olbrich_00}.

Theoretically, calculations in the macrocanonical ensemble predict
two regimes: one at low tension, where it comes from the flattening of
fluctuations and thus~\cite{Dimova_06},~\cite{Evans_90} (see section~\ref{subsection_1_discuss} for a detailed derivation)

\begin{equation}
\frac{A_p^f - A_p^i}{A_p^i} = \frac{k_\mathrm{B}T}{8 \pi \kappa} \ln \left(\frac{\sigma_f}{\sigma_i}\right)\, ,
\label{eq_0_alpha}
\end{equation}

\noindent where $\sigma_{i/f}$ is respectively the Lagrange-multiplier for the initial/final configuration; and one at high tension, where it arises mainly through
stretching and thus

\begin{equation}
\frac{A_p^f - A_p^i}{A_p^i} = \frac{\sigma_f}{K} \,.
\label{eq_0_K}
\end{equation}

\begin{figure}[H]
\begin{center}
\includegraphics[scale=.47,angle=0]{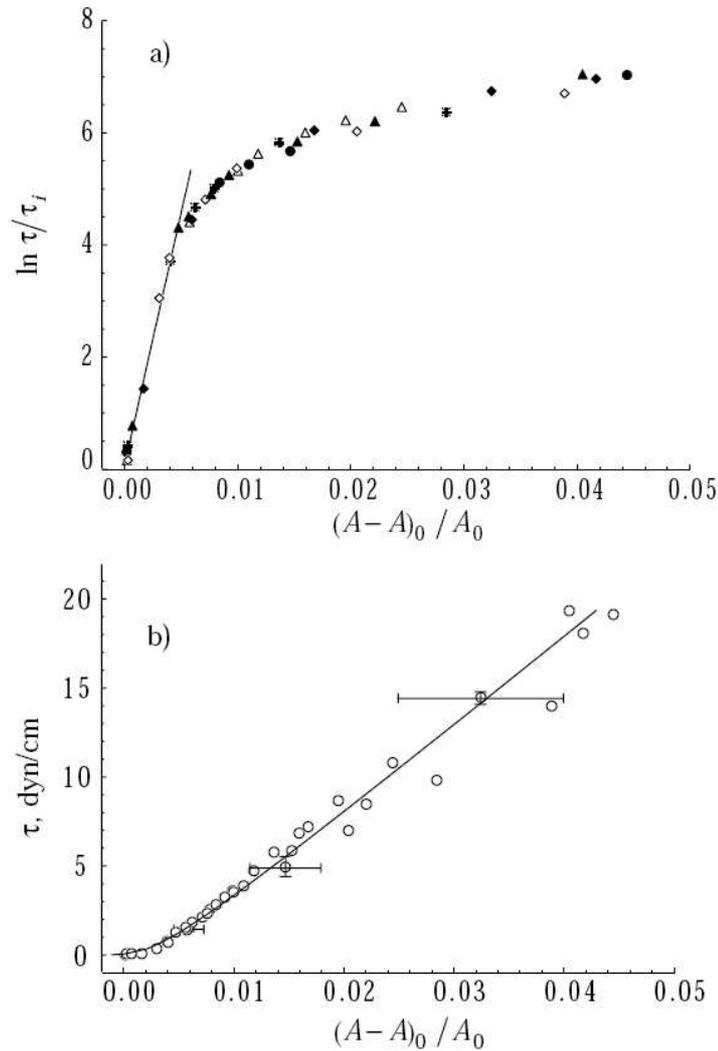}
\caption{Graphics showing a typical data analysis of micropipette
  experiments~\cite{Dimova_02}. Fig.(a) and Fig.(b) show the same data with different scales for $\tau$.
  In Fig.(a), the tensions were displayed in log-scale in order to highlight the logarithmic behavior in the region of low tension.
  Through the fit shown in Fig.(a) and using eq.(\ref{eq_0_alpha}), one measures the bending rigidity ($\kappa = 42 \pm 5 \, k_\mathrm{B}T$ in these measures).
  Fig.(b) shows the same data in the linear scale.
  The area compressibility $K$ is obtained through a
  fit in the region of high tensions using eq.(\ref{eq_0_K}) ($K = 450 \pm 85 \, \mathrm{mN/m}$ here).
  Remark that it was assumed that $\sigma \approx \tau$.}
\label{dimova}
\end{center}      
\end{figure}

Even though these previsions involve the non-measurable Lagrange-multiplier
$\sigma$, in experiments it is currently assumed that $\sigma \approx \tau$
~\cite{Henriksen_04},~\cite{Evans_90},~\cite{Dimova_02}.
As a consequence, by plotting $\tau$ as a function of the variation of the projected, one can measure $\kappa$ and $K$ (see an example in Fig.~\ref{dimova}).

\subsection{Contour analysis experiments}
\label{fluct}

\begin{figure}[H]
\begin{center}
\includegraphics[scale=.4,angle=0]{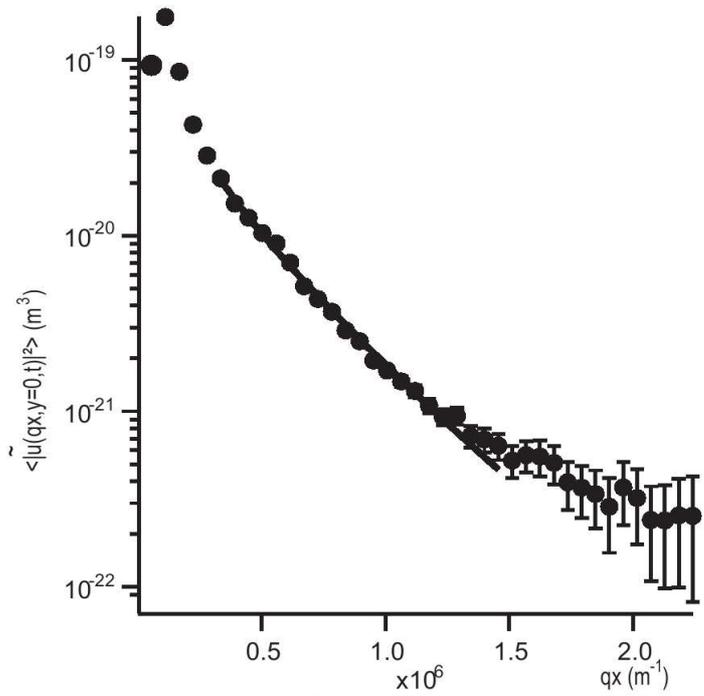}
\caption{Intensity of a fluctuating mode as a function of the wave vector for
  a quasi-spherical vesicle. The solid line represents the fit of the data with
  the equivalent of eq.(\ref{spect_base}) for spherical geometry, yielding $\kappa = 9.44 \times 10^{-20} \, \mathrm{J}$ and $r
= 1.74 \times 10^{-7} \, \mathrm{N/m}$~\cite{Pecreaux_04}.}
\label{spect}
\end{center}      
\end{figure}

The aim of these experiments is to determine $r$ and $\kappa$ by studying
the mean squared amplitude of fluctuating modes, as seen on
section~\ref{grand_can}.
Some experiments were made in planar geometry, using BLM. 
As explained before, these membranes tend to be too tense. 
Indeed, a 1999 experiment by Hirn et al. found $r = (0.42\pm 0.03) \,
\mathrm{mN/m}$~\cite{Hirn_99}.
The fluctuations are then dominated by the tension and one cannot observe the
effects of the bending rigidity.
For this reason vesicles are usually preferred in spectrum fluctuation measures.
A typical spectrum with the fitting can be seen in Fig.~\ref{spect}.

\vspace{1cm}

\subsection{Adhesion of vesicles}
\label{subsection_adhesion}

\vspace{1cm}

The adhesion of membranes is very important for tissue formation.
At the cellular level, the adhesion between the membrane and the cytoskeleton helps to regulate the formation of vesicles and lamellipodia~\cite{Sheetz_01}.
It also plays an important role in the exocytosis and endocytosis~\cite{Seifert_90}.
In the context of the physics of liquids, the adhesion of liquid drops to solid substrates is traditionally used to study tensions.
Inspired by these experiments, one finds in the literature a wide variety of papers on the adhesion of vesicles among themselves~\cite{Evans_85},~\cite{Bailey_90} and on the adhesion of vesicles with a solid substrate~\cite{Evans_80},~\cite{Raedler_95},~\cite{Bruinsma_00},~\cite{Puech_04}.

\vspace{1cm}

Here, we will focus on experiments dealing with the interaction of vesicles of radius $R_\mathrm{ves}$ with a flat solid substrate.
The vesicle, usually filled with an aqueous solution fluid denser than the suspension medium, is attracted towards the bottom surface.
At equilibrium, it adopts a deformed shape as shown in Fig.~\ref{gota_adhes}, with a flat region near the substrate.
This region, with a radius $R_a$, is considered adhered to the substrate.
Frequently, the bottom surface is composed of glass, since the technique of RICM (reflection interference contrast microscopy) is very popular to obtain images of the adhesion region. 
The adhesion is ruled by the interplay of attractive and repulsive interactions, such as:

\begin{itemize}
  \item the short-range van der Waals attractive potential between the membrane and the substrate.
    It can be corrected to take into account the screening due to the presence of ions in the suspension medium;
  \item the attractive gravitational potential;
  \item the repulsive effective interaction coming from the reduction of the entropy of the membrane.
    Indeed, the substrate imposes a spatial restriction that limits the membrane fluctuations;
  \item the short-range steric repulsion coming from the lipids;
  \item in some cases, the substrate can be coated by polymers~\cite{Sengupta_10}.
    One must thus consider a supplemental steric repulsion coming from the polymer coating of the substrate;
  \item it is also possible to cover the substrate with a piece of membrane containing proteins that attach to specific proteins embedded in the vesicle's membrane~\cite{Bruinsma_00},~\cite{Albersdorfer_97}.
    In this case, there are supplemental attractive interactions.
\end{itemize}

\begin{figure}[H]
\begin{center}
\subfigure[An adhesion potential including the gravity contribution $V_\mathrm{grav}$, a Van der Waals contribution $V_\mathrm{VdW}$ and an entropic repulsive term $V_\mathrm{steric}$~\cite{Raedler_95}.
In this case, we find only one shallow minimum.]{
\includegraphics[width=0.55\columnwidth]{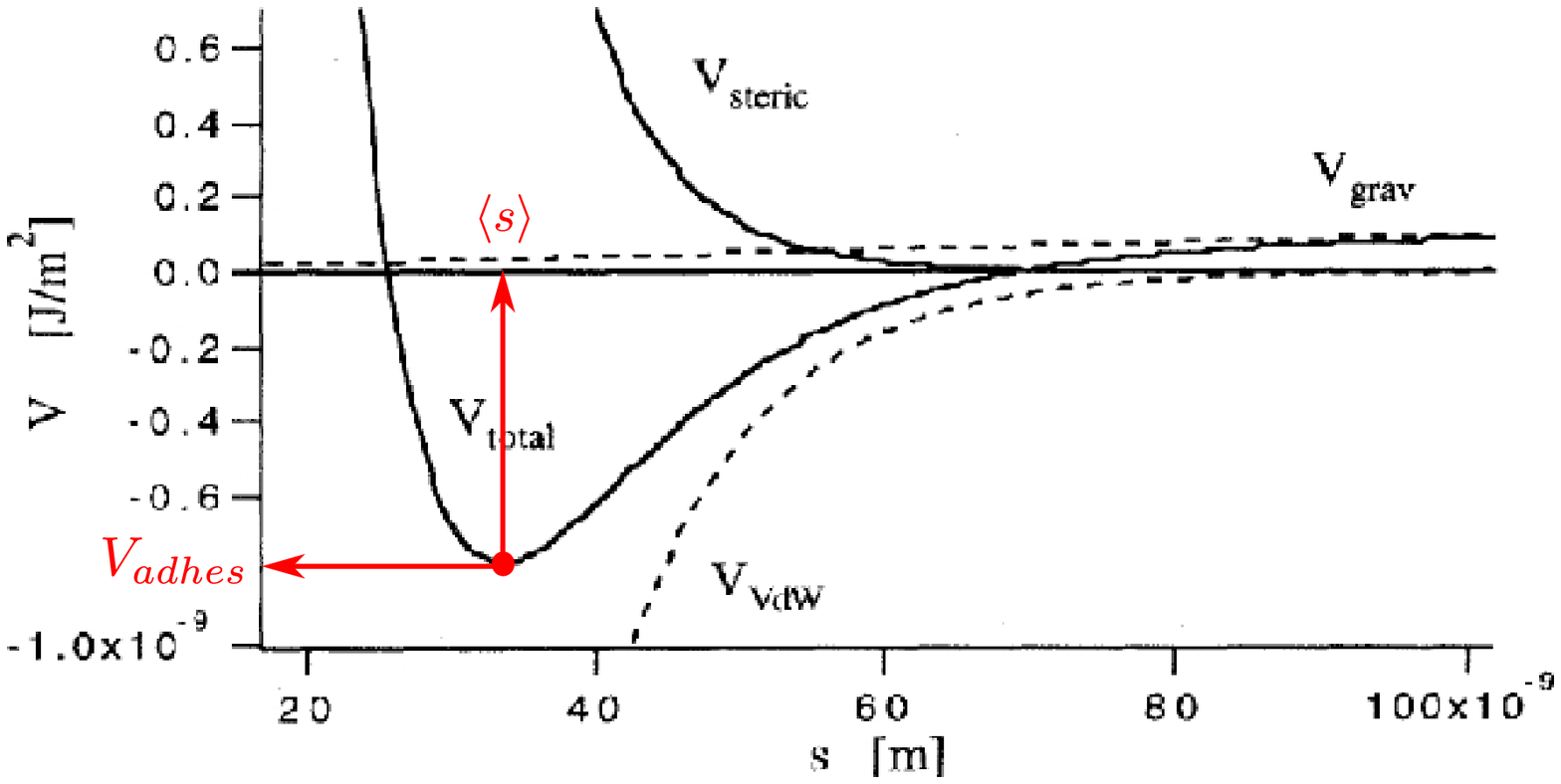}
\label{adhesion}
}
  \subfigure[Predicted adhesion potential for an experiment involving three different polymer coatings of the substrate~\cite{Sengupta_10}.
    The red
  circles indicate the deep minima and the star indicates the shallow
  minimum.]{
  \includegraphics[width=0.55\columnwidth]{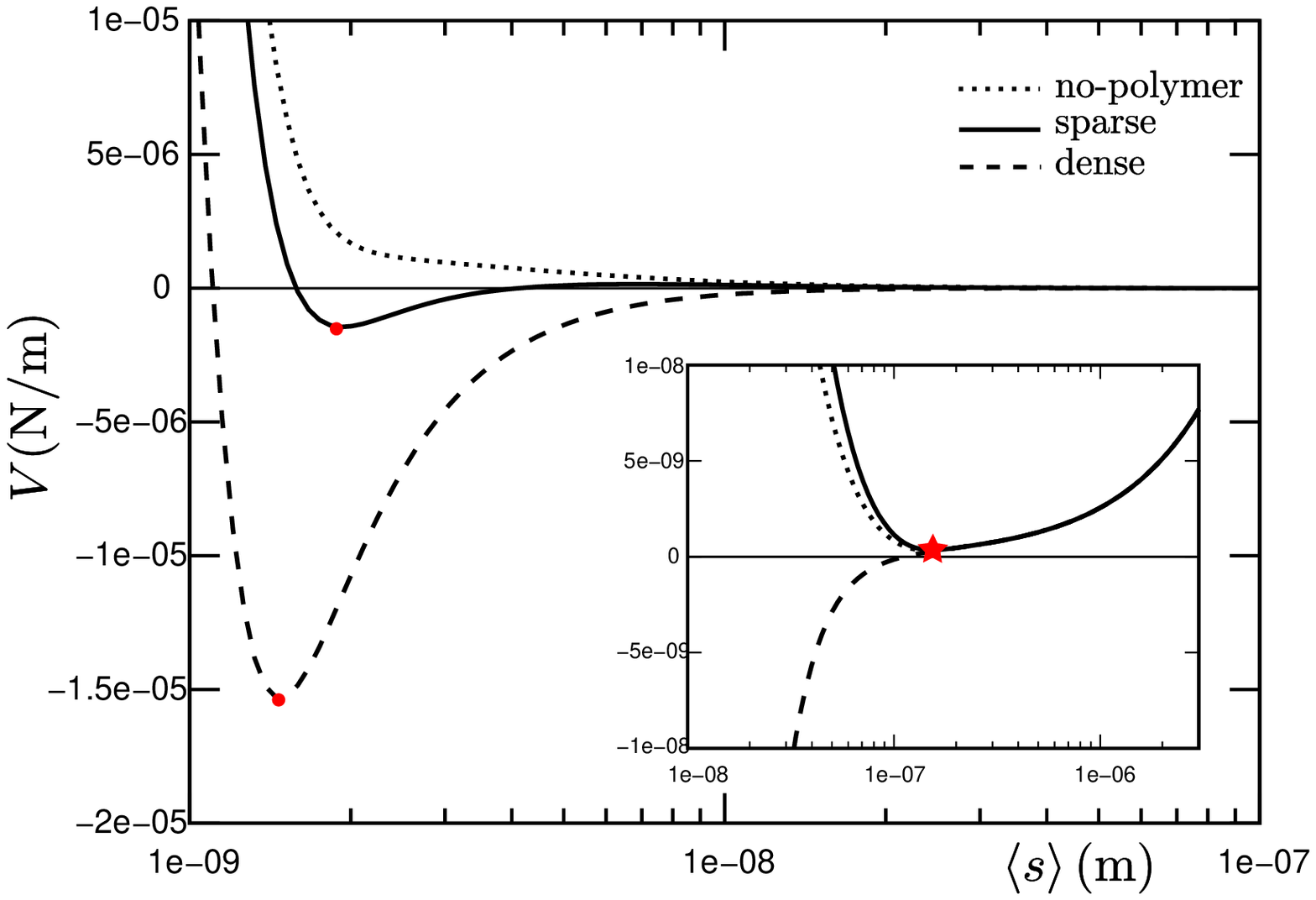}
  \label{limozin}
}
\end{center}      
  \caption{Examples of adhesion potentials.} 
  \label{fig_adhes_pot}
\end{figure}

Two examples of resulting potentials can be seen in Fig.~\ref{fig_adhes_pot}.

Depending on the resulting potential, one can find two types of adhesion:

\begin{enumerate}
  \item Weak adhesion: in this case, the adhering patch fluctuates strongly at a distance $s(x)$ well above the surface, as shown in Fig.~\ref{radler}.
    It corresponds to a shallow minimum of the adhesion potential (see Fig.~\ref{fig_adhes_pot}).
    When it is a local minimum, it is also said that the vesicle is in a pre-nucleation state.

    In this case, one can measure the fluctuation spectrum of the adhering region.
Considering the rest of the vesicle as a lipid reservoir and approximating the
energy of adhesion as quadratic near $\langle s \rangle$, which is justified given Fig.~\ref{fig_adhes_pot}, the Hamiltonian
up to order two is given by~\cite{Raedler_95}

\begin{equation}
\mathcal{H} = \int_{S_\mathrm{adhe}} \left[\frac{\kappa}{2}\left(\nabla^2 h\right)^2 +
  \frac{\sigma}{2}\left(\nabla h\right)^2 + \frac{V''}{2} h^2\right] d^2\bm{x}
\, ,
\label{hamilt_adhes}
\end{equation}

\noindent where $h(\bm{x}) = s(\bm{x}) - \langle s \rangle$, $V''$ is the
coefficient of the harmonic approximation of the adhesion energy and $S_\mathrm{adhe}$ is the projected surface of the adhering portion of the vesicle.  
Following the same reasoning presented in section~\ref{grand_can}, the mean
square amplitude of each mode is given by

\begin{equation}
\langle |h(\bm{q})|^2 \rangle = \frac{k_\mathrm{B} T}{V'' + r q^2 +
  \kappa q^4} \, .
\label{spect_fluct_adhes}
\end{equation}

\noindent Note that we have substituted $\sigma$ by its macroscopical counterpart $r$, which is experimentally measurable (see further details in the end of section~\ref{grand_can}). By a measuring the fluctuation spectrum, it is possible thus to determine $r$, $\kappa$ and $V''$.
    
\item Strong adhesion: the vesicle is very near the substrate (less than $10 \, \mathrm{nm}$ of distance).
  The membrane fluctuations are barely detectable.
  The adhesion energy is higher, corresponding to a deep minimum of the adhesion potential.
\end{enumerate}

\begin{figure}[H]
\begin{center}
\includegraphics[scale=.4,angle=0]{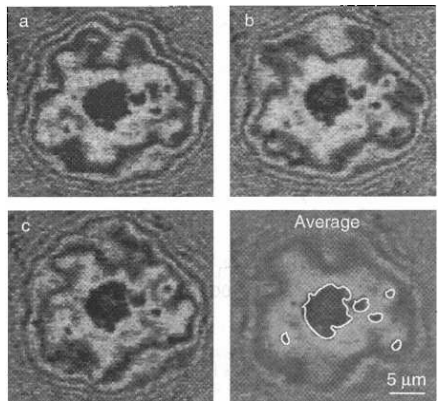}
\caption{RICM micrograph of a vesicle adhering to a solid substrate.
The adhering region is surrounded by the fringes.
The gray shading is inversely proportional to the vesicle-substrate distance: the dark gray patch is very near to the substrate, while the light gray patch is well above the solid.
Pictures (a), (b) and (c) were taken at $\sim 0.1 \mathrm{s}$ of interval.
Note the strong fluctuations of the light gray region.
The last picture shows the average over $64$ snapshots.
From these images, we can conclude that the gray region adheres only weakly to the substrate, while the dark patch is strongly adhered~\cite{Albersdorfer_97}.} 
\label{albers}
\end{center}      
\end{figure}

In Fig.~\ref{albers}, we show some typical RICM images from an adhering vesicle that has weakly and strongly adhering patches.
From these images, the height profile of the adhesion region $s(x)$ can be reconstructed.
One can subsequently study the mechanics of the membrane.

\subsubsection{Adhesion mechanics}

\begin{figure}[H]
\begin{center}
\includegraphics[scale=.5,angle=0]{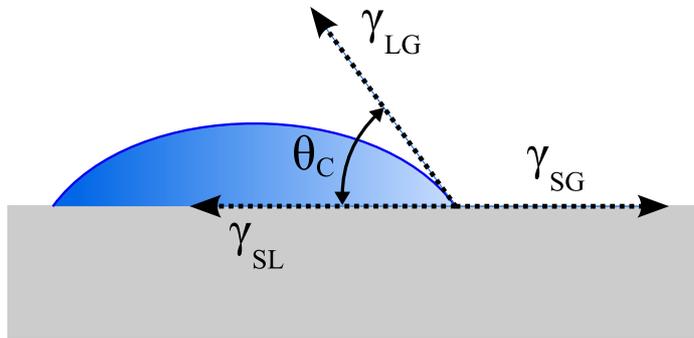}
\caption{Illustration of the Young's equation: $\gamma_{SL}$ is the solid-liquid tension, $\gamma_{SG}$ is the solid-gas tension and $\gamma \equiv \gamma_{LG}$ is the liquid-gas tension.
  The contact angle $\theta_c$ is defined by the mechanical equilibrium.} 
\label{Contact_angle}
\end{center}      
\end{figure}

Let's first recall some results concerning liquid drops.
As discussed in section~\ref{diff_liq}, the contact angle $\theta_c$ that liquid drops do with solids substrate is very sharp.
It is defined by the solid-liquid tension $\gamma_{SL}$, the solid-gas tension $\gamma_{SG}$ and the liquid-gas solid $\gamma \equiv \gamma_{LG}$.
The mechanical equilibrium, illustrated in Fig.~\ref{Contact_angle}, gives the Young relation

\begin{equation}
  \gamma_{SG} = \gamma_{SL} + \gamma \cos\theta_c \, .
\end{equation}

\noindent The energy variation per unit of contact area between the liquid and the solid, also known as the adhesion energy per unit area, is given by the Young-Dupree relation:

\begin{equation}
  W_A = \gamma(1 - \cos\theta_c) \, .
\end{equation}

Adhering vesicles were first theoretically studied in details by Seifert et Lipowsky in 1990~\cite{Seifert_90}.
They considered a free-energy containing a contribution from curvature, a term $-W_A \, (\pi R_a^2)$ corresponding to the adhesion energy plus the area and volume constraints.
By minimizing the free-energy, they derived the equilibrium shapes, that shared two features:

\begin{itemize}
  \item a contact angle $\theta_c = \pi$ due to the bending rigidity;
  \item a curvature at the contact given by

    \begin{equation}
      R_c = \left(\frac{2 \, W_A}{\kappa}\right)^{1/2} \, .
      \label{Rc}
    \end{equation}
\end{itemize}

\begin{figure}[H]
\begin{center}
\subfigure[Definition of the effective contact angle.
The red circle at left indicates the curvature radius $R_c$ at the contact point.
Note that $R_c < R_\mathrm{ves}$~\cite{Seifert_90}.]{
\includegraphics[scale=.3,angle=0]{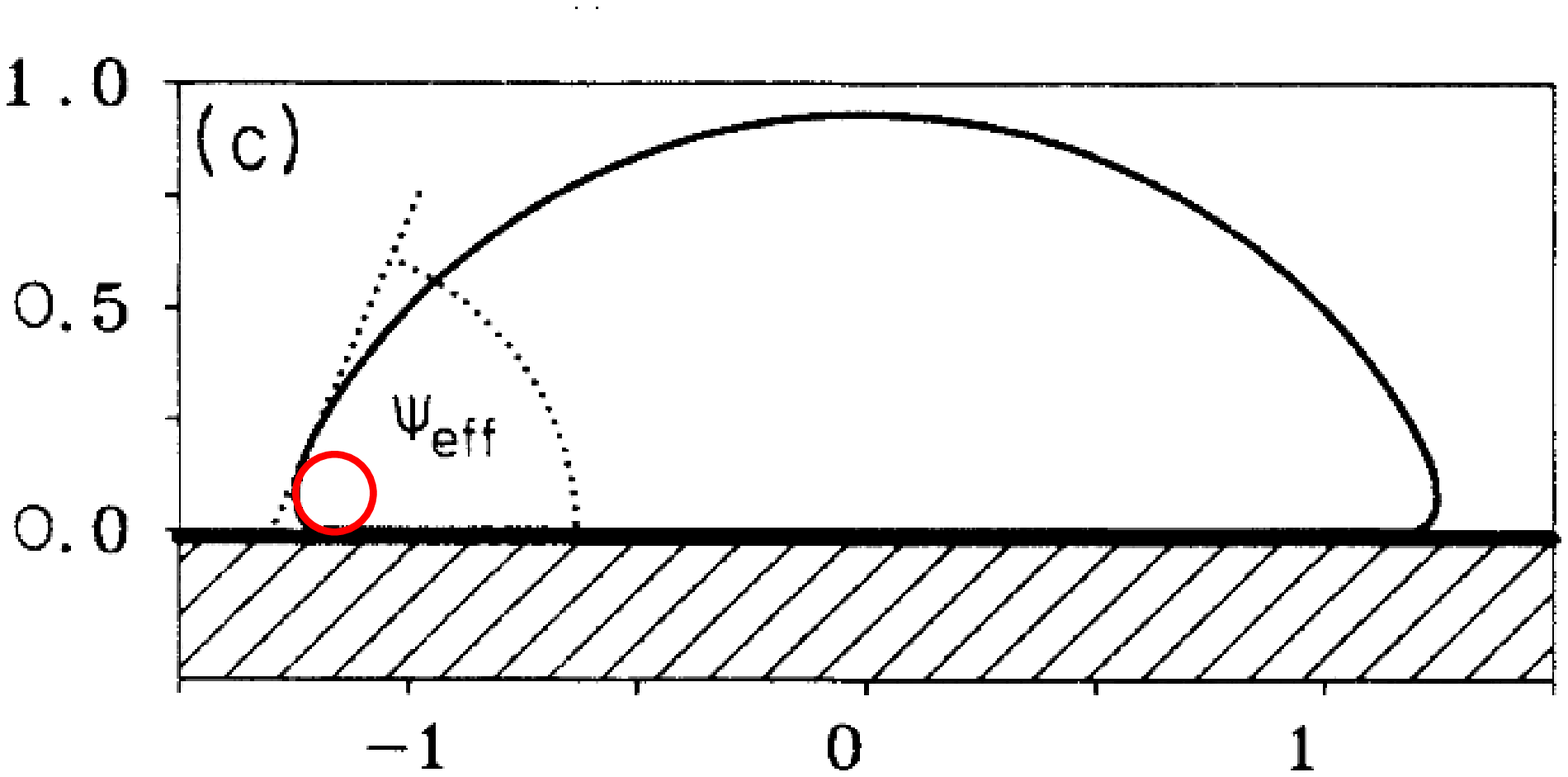}
\label{def_angle}
}
\subfigure[Drawing of a vesicle weakly adhering to a substrate with the
  relevant measurable parameters~\cite{Raedler_95}.]{
\includegraphics[scale=.3,angle=0]{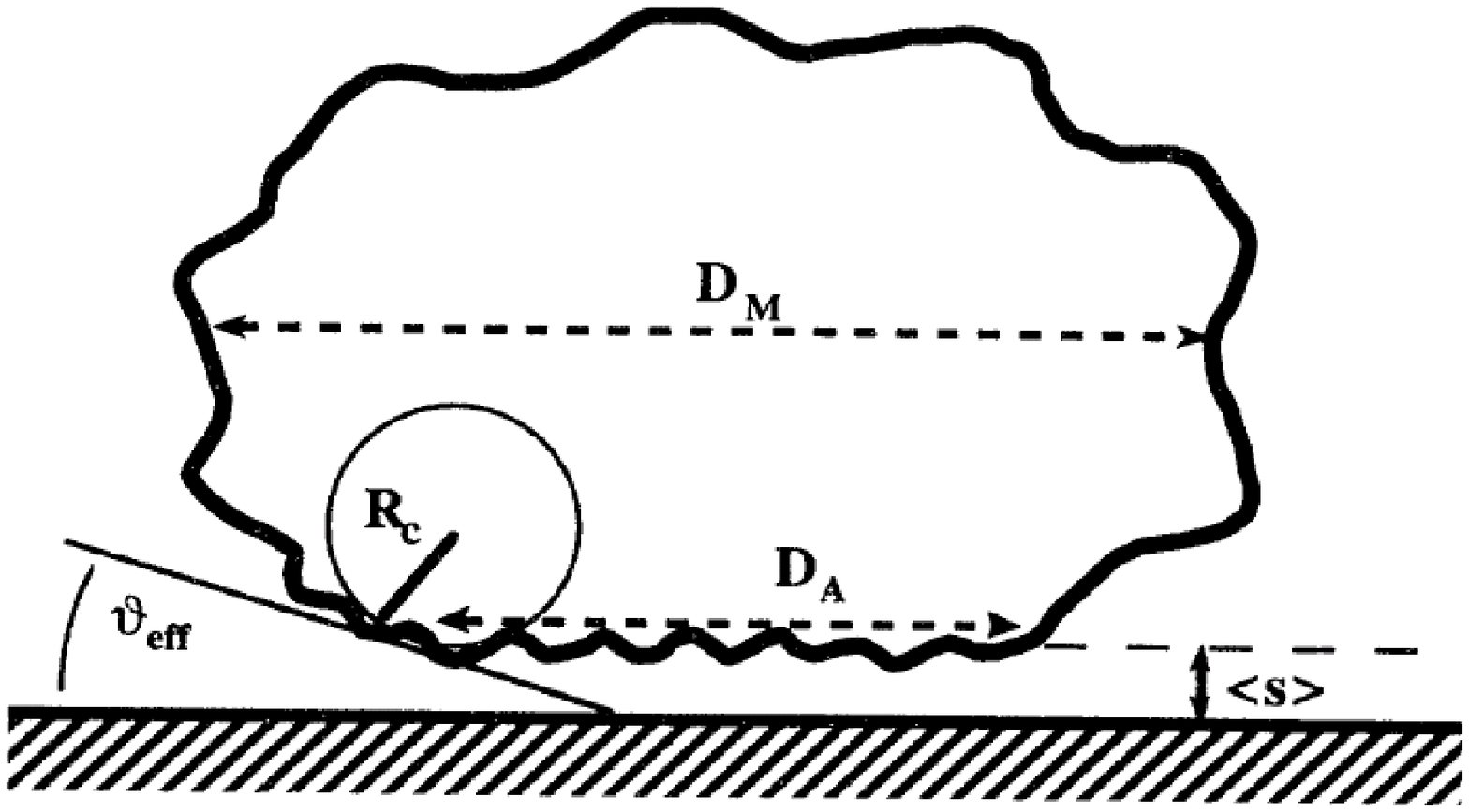}
\label{radler}
}
\caption{Effective contact angle.} 
\end{center}      
\end{figure}

\noindent They argued that, in general, one could not expect to use the Young-Dupree relation to link $W_A$ and the lateral tension $\tau$ of the membrane due to the effects of the bending rigidity.
In the limit, however, of small bending rigidity, the vesicle becomes a spherical cap for an internal pressure bigger than the outer, with a rounded contact region of length $R_c < R_\mathrm{ves}$.
One can thus define an effective contact angle $\theta_\mathrm{eff}$ (see Fig.~\ref{def_angle}) that obeys an analogous of the Young-Dupree equation  

\begin{equation}
W_A = \tau\left[ 1 - \cos(\theta_\mathrm{eff})\right]\, .
\label{WA}
\end{equation}

In principle, for a vesicle under these conditions, one could measure $R_c$ and $\theta_\mathrm{eff}$ and thus deduce $W_A$ and $\tau$.
In practice, however, as we will see later, measuring these quantities can be very tricky and imprecise.

An alternative was thus proposed by Bruinsma~\cite{Bruinsma_95}.
He studied the equilibrium of the forces due to the bending rigidity and tension near the rim of the contact region and obtained

\begin{equation}
  h(x) = \left\{ \begin{array}{ccc} \theta_\mathrm{eff} \, x - \theta_\mathrm{eff} \, \lambda \left[ 1- e^{-\left(\frac{x}{\lambda}\right)}\right]\,\,\,& $for$ &\, x > 0\, ,\\
    \\
    0\,\,\, &$for$ &\, x<0\, ,\end{array} \right.
\end{equation}

\noindent where $h(x)$ is the height of the membrane, $x = 0$ at the contact point and

\begin{equation}
  \lambda = \sqrt{\frac{\kappa}{\tau}} \, .
  \label{eq_lambda}
\end{equation}

\begin{figure}[H]
\begin{center}
\includegraphics[scale=.5,angle=0]{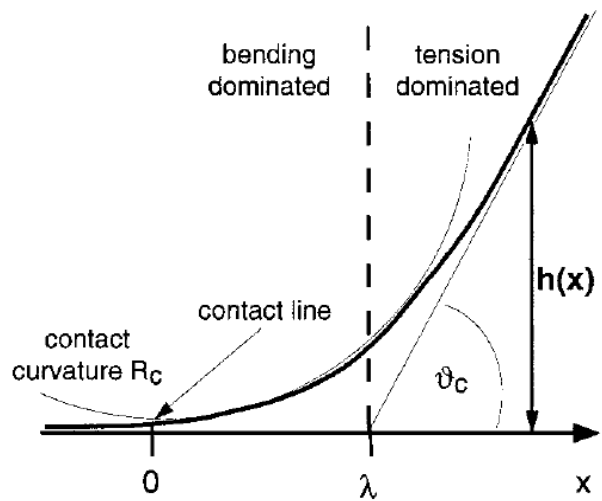}
\caption{Shape of the membrane near the contact region.
For $x < 0$, the membrane adheres to the substrate.
The contact angle $\theta_\mathrm{eff}$, here shown as $\theta_c$, is determined by the linear fit.
The characteristic length $\lambda$ is obtained from the intersection of the linear fit with the $x$-axis~\cite{Simson_98}.}
\label{Albers1}
\end{center}      
\end{figure}

\noindent The characteristic length $\lambda$ separates two regions: for $x < \lambda$, the bending rigidity dominates and the membrane is thus curve; for $x > \lambda$, the tension dominates and the membrane approaches a straight line (see Fig.~\ref{Albers1}).

Experimentally, from the height profile of the membrane, one can obtain $\theta_\mathrm{eff}$ and $\lambda$.
Using eq.(\ref{eq_lambda}) one is able thus to deduce $\tau$ and subsequently $W_A$ through the Young-Dupree's relation~\cite{Albersdorfer_97}.
This method, however, presents a serious limitation: if the tension is very large, the length $\lambda$ is undetectable~\cite{Puech_04}. 
In the next chapter, we will discuss some experiments using these techniques.

\subsection{Nanotube extraction}
\label{tube_exp}

\begin{figure}[H]
\begin{center}
\includegraphics[scale=.3,angle=0]{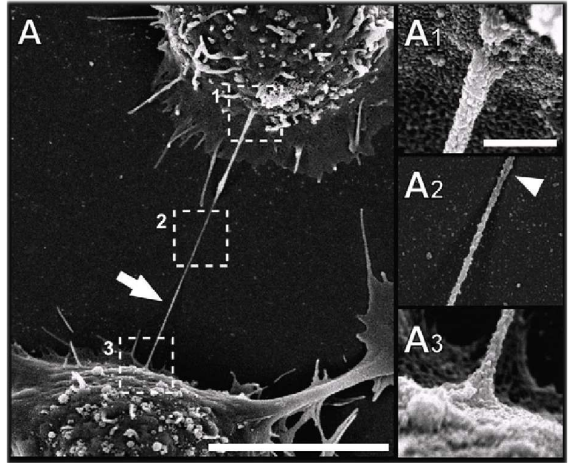}
\caption{Scanning electron microscopic image of a tunneling tube between two
  cultured PC12 cells. The boxed areas are sections of $80 \, \mathrm{nm}$
  enlarged in A1, A2 and A3, respectively~\cite{Gerdes_07}.}
\label{cell_com}
\end{center}      
\end{figure}

Membrane nanotubes, also called tethers, are cylindrical structures whose radius range
from a few up to hundreds of nanometers, while their length can
reach tens of micrometers. 
In living cells, they are formed by localized forces generated by molecular
motors or by polymerizing cytoskeleton filaments, such as microtubules.
These tethers are suspected to play a major hole in the intracellular
transport of vesicles~\cite{Iglic_03} and in the communication between
cells, since they form also between different cells and proteins were shown to
pass through these tunneling nanotubes (TNT)~\cite{Onfelt_04} (see
Fig.~\ref{cell_com}).
Recently, it has been found some evidence that TNT may even be crucial in the
HIV virus spreading~\cite{Eugenin_09}.

\begin{figure}[H]
\begin{center}
\subfigure[Typical sequence of nanotube extraction from a
  GUV. The image was obtained through differential interference contrast
  microscopy to enhance contrast, since the tube is not optically
  resolvable~\cite{Koster_05}.]{
\includegraphics[scale=.4,angle=0]{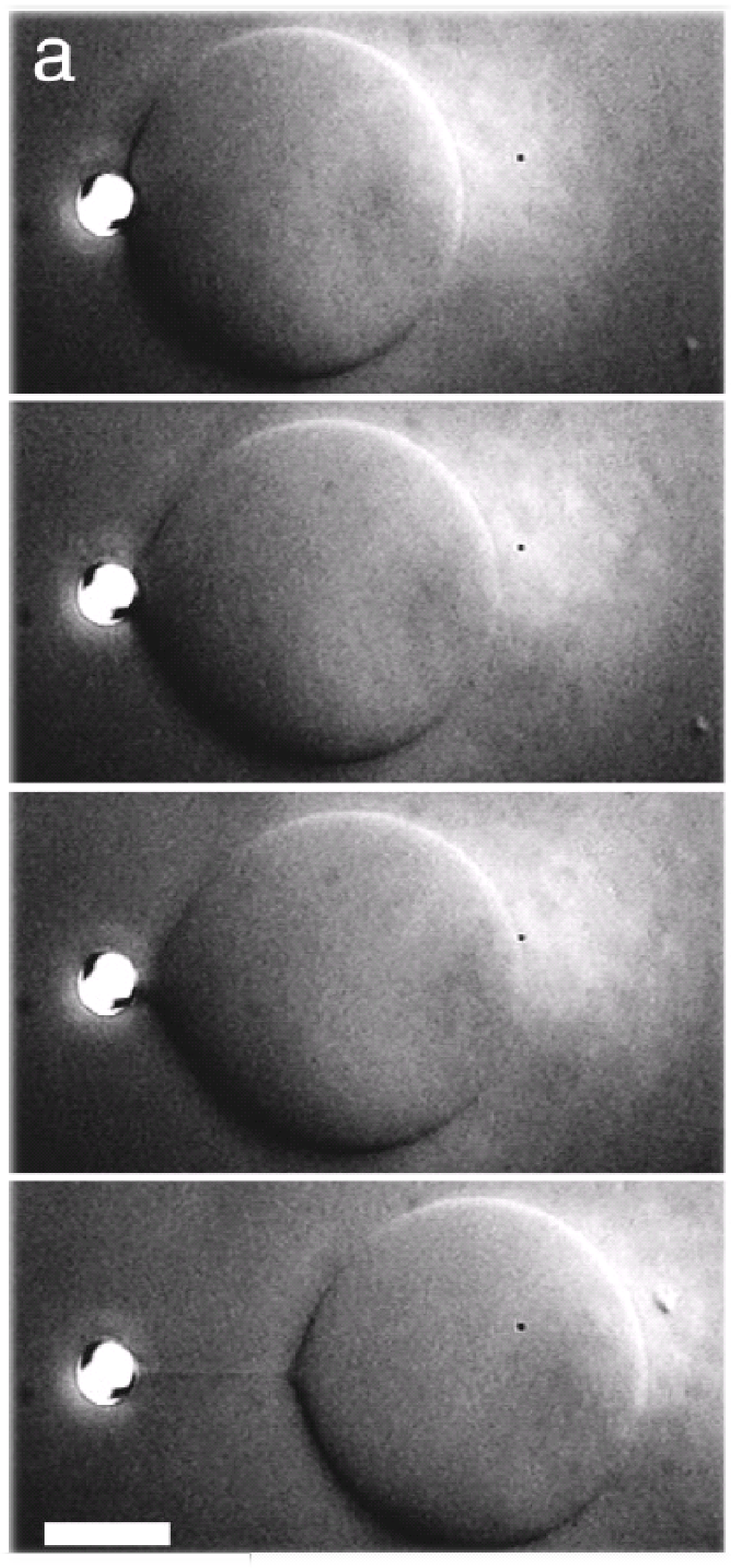}
\label{koster}
}
\subfigure[Some methods used to
  extract tubes from GUV: (a) vesicles under hydrodynamic flow~\cite{Rossier_03}, (b) vesicles held by micropipette and attached to
  a mobile glass or magnetic bead~\cite{Heinrich_96},
  ~\cite{Bo_89} and (c) nanotube extraction with molecular
  motors~\cite{Leduc_04}.
]{
%  blahblahblah
\includegraphics[scale=.5,angle=0]{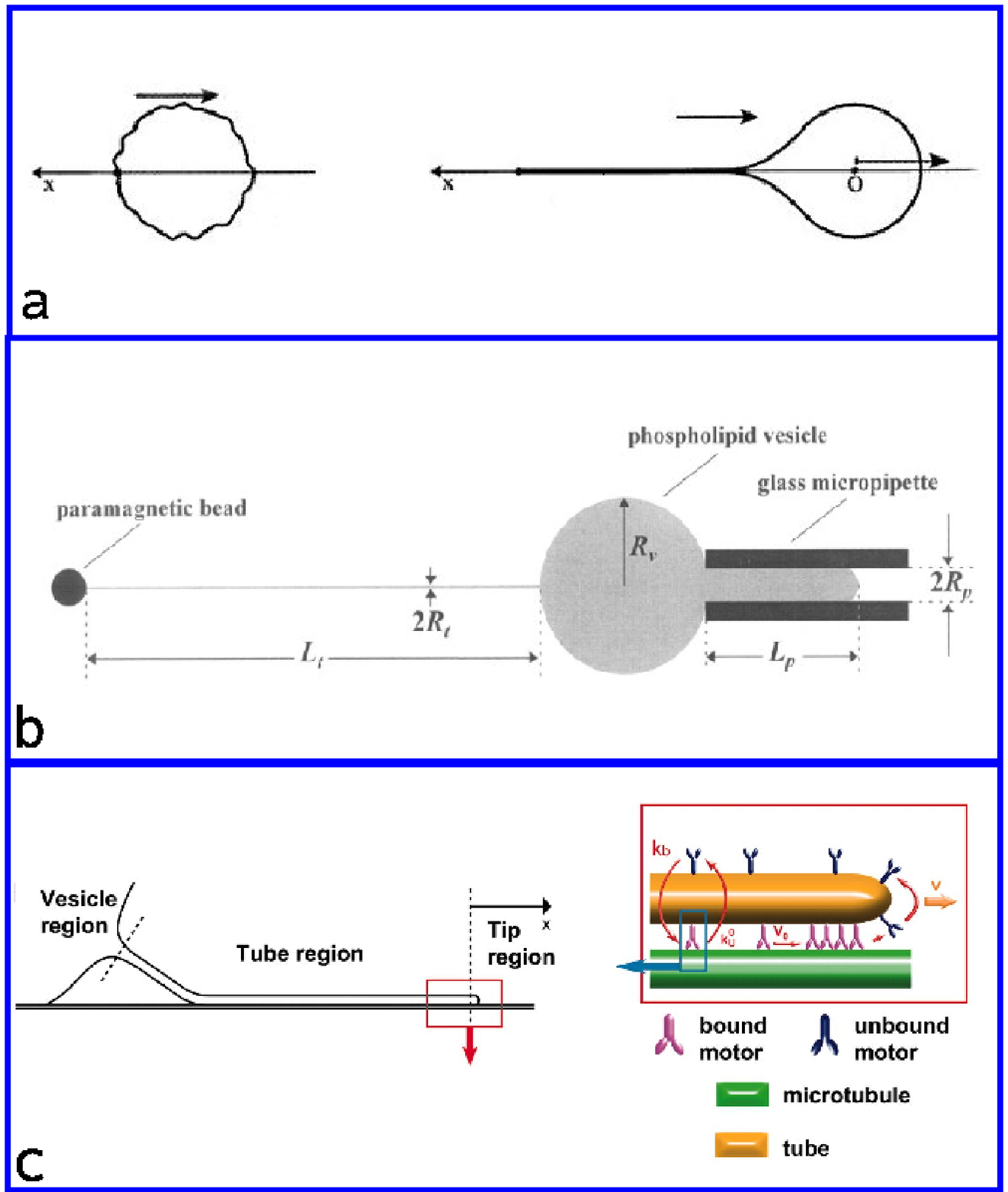}
\label{methods}
}
\subfigure[Experimental setting used to extract nanotubes from BLM~\cite{Bashirov_07}.]{
\includegraphics[scale=.5,angle=0]{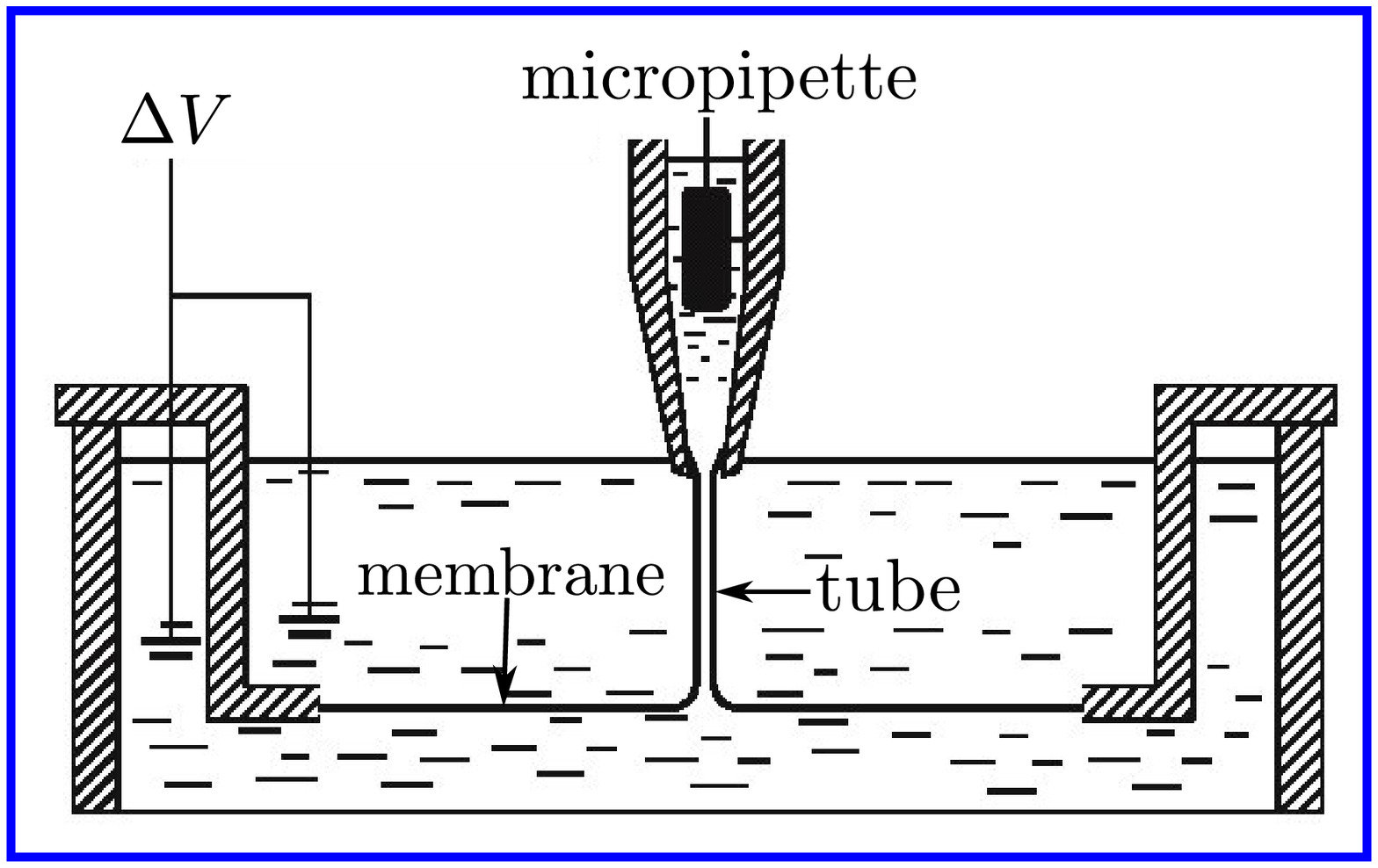}
\label{methods_1}
}
\caption{Nanotube extraction techniques.}
\label{nano_extract}
\end{center}      
\end{figure}

In vitro tube extraction was used to evaluate the adhesion energy between
the cell cytoskeleton and the cell membrane. 
In these experiments, it was also shown that these tethers do not contain
cytoskeleton~\cite{Sheetz_01},~\cite{Waugh_95}.
Here we will restrain ourselves to experiments of tube extraction from model
membranes (see Fig.~\ref{koster}).
Figs.~\ref{methods}-\ref{methods_1} sum up the main techniques used in laboratory to extract tubes.

In typical experiments with GUVs, one cannot optically resolve the tube, even though its length is readily measurable (see Fig.~\ref{koster}).
The force needed to extract the tube $f$ can be directly measured by the force applied over the glass bead (for an optical tweezer) or over the magnetic bead.
Moreover, if the vesicle is held by a micropipette, as in the experimental apparatus shown in Fig.~\ref{methods}, subfigure b, the tension $\tau$ is measured through the
applied pressure from eq.(\ref{Laplace})~\cite{Heinrich_96}.
Another technique consists in extracting nanotubes with controlled length from BLMs.
With this configuration, it is possible to apply a difference of electrical potential between the interior and exterior of the tube.
One can thus deduce the radius of the tube and the tension $\tau$ of the membrane~\cite{Bashirov_07}. 

From the theoretical point of view, as these tubes are so thin, it is reasonable to consider the adjacent GUV or
BLM as a lipid reservoir.
For a symmetrical membrane, the tube free energy is thus given by
eq.(\ref{Helfrich}) plus a contribution coming from the force $f$ that holds the
tube.
For a cylindrical tube of radius $R$ and length $L$, one has

\begin{equation}
\mathcal{H} = \left(\frac{\kappa}{2R^2} + \sigma\right)2\pi R L - f L\, .
\label{hel_tube}
\end{equation}

The equilibrium radius $R_0$ and force $f_0$ are given by the minimization of
eq.(\ref{hel_tube}) with respect to $R$ and $L$ respectively, yielding

\begin{equation}
R_0 = \sqrt{\frac{\kappa}{2 \sigma}} 
\label{radius}
\end{equation}

\noindent and 

\begin{equation}
f_0 = 2\pi \sqrt{2 \sigma \kappa} = 2 \times 2\pi R_0 \, \sigma \, .
\label{force_tube}
\end{equation}

\noindent Eq.(\ref{radius}) shows clearly that the radius is determined by a
competition between the tension, which tends to create a very thin tube, and
the bending rigidity, which opposes to high curvatures~\cite{Derenyi_02}.
Interestingly, the result given in eq.(\ref{force_tube}) highlights the difference between a membrane and a liquid interface: if we had a tube constituted by a film of liquid, we should expect $f = 2 \pi R_0 \gamma$.
The factor $2$ in $f_0$ comes from the curvature energy present only in membranes.
One must keep in mind that these results hold only if thermal
fluctuations are neglected and the tube is a perfect cylinder.
We will discuss this point in chapter~\ref{TUBE}.

\begin{figure}[H]
\begin{center}
\includegraphics[scale=.35,angle=0]{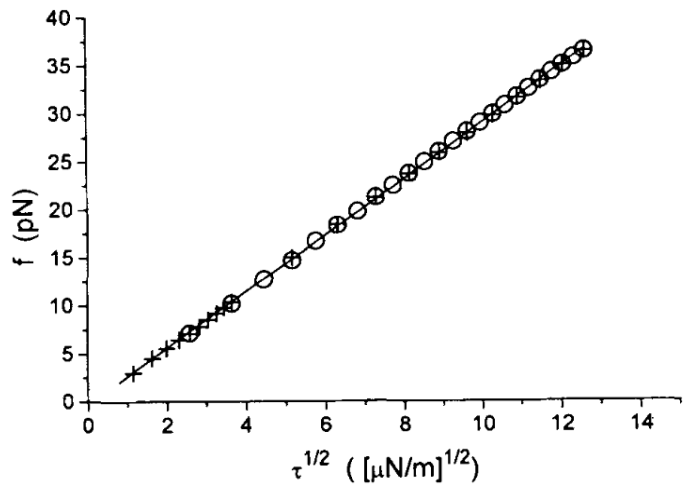}
\caption{Data from the extraction of two successive tubes from the same
  vesicle of diameter $18.8\, \mathrm{\mu m}$. The vertical axis shows the force needed
  to extract the tube while the horizontal axis shows the square root of the
  tension $\tau$ measured through the difference of pressure in the
  micropipette. The data show a good linear fit, indicating that in this
  experiment, it was apparently justified to neglect thermal undulations and suppose $\sigma \approx \tau$. The bending rigidity obtained through the fit is approximately $25 \, k_\mathrm{B}T$~\cite{Heinrich_96}. By using eq.(\ref{radius}), we can estimate that the tube radius are rather large, ranging from a thousand to two hundred nanometers in this particular experiment.}
\label{heinrich}
\end{center}      
\end{figure}

From eq.(\ref{force_tube}), we see that one could obtain $\kappa$ by measuring $f_0$ and $\sigma$.
In GUV experiments, it is usually assumed that the measured force to extract the tube $f$ is well approximated by $f_0$, which means neglecting thermal undulations.
Besides, the tension $\tau$ measured through the micropipette aspiration is assumed equivalent to $\sigma$.
One obtains thus the curve shown in Fig.~\ref{heinrich} and deduces $\kappa$ by a linear fit.
The other way to obtain $\kappa$, using eq.(\ref{radius}), is explored in the BLM experiments by once more assuming that $\tau \approx \sigma$.
A recent experiment by Bashkirov~\cite{Bashirov_07} found very thin tubes of
about $10 \, \mathrm{nm}$ thick, indicating a very tense membrane, and deduced
values to $\kappa$ compatible with previous results.

\subsection{Sum-up}
\label{sum_exp}

In table~\ref{table_1}, we see a sum-up of the techniques presented in this
section and used to
mechanically probe membranes. 
In the last column, we can see the usual assumption made in order to deduce
results from the third column.
It generally involves the three tensions $\tau$, $\sigma$ and $r$.
In this work, we will try to examine in detail these assumptions. 
In special, we will quantify the difference between $\tau$ and $\sigma$ for the
three mainly studied geometries and discuss under which conditions these
suppositions are justifiable. 
In chapter~\ref{TUBE}, we will also examine the role of thermal fluctuations
on the force needed to extract a tube.

\vspace{1cm}

\begin{table}[H]
\label{table_1}
  \begin{center}
\begin{tabular}{|m{2.5cm}|c|c|m{2.5cm}|}
\hline
\bf{{\red Technique}} & \bf{{\red Direct measure}} & \bf{{\red Used to deduce}} & \bf{{\red Usual assumption}} \\ 
\hline
micropipette & $\Delta P$, $A_p$ & $\tau$, $\Delta\alpha$, $\kappa$, K & $\sigma \approx \tau$ \\
\hline
contour analysis & $\langle|h(\bm{q})|^2\rangle$ & r, $\kappa$ & \\
\hline
adhesion & $\theta_\mathrm{eff}$, $\langle|h(\bm{q})|^2\rangle$, $\langle s\rangle$ &
r, $\kappa$, $W_\mathrm{adhes}$ & $\tau \approx r$\\
\hline
tube extraction (GUV) & $\Delta P$, $L$, $f$ & $\tau$, $\kappa$ & $\tau \approx
\sigma$ and $f \approx f_0$\\
\hline
tube extraction (BLM) & $\Delta V$, $L$ & $R_0$, $\kappa$ & $\tau \approx
\sigma$ and $R \approx R_0$\\
\hline
\end{tabular}
\caption{Sum-up of experimental techniques and measured quantities. In the
  second column, the quantities directly measurable are listed, while in the
  third column we list the quantities deduced from a fit or from the use of
  theoretical equations. In the last column, we present the main assumptions
  made to obtain the results from the previous column.}
\end{center}
\end{table}

\section{Stress tensor for a planar membrane}
\label{stress_plan}

Here we introduce the stress tensor for planar membranes.
In particular, we will derive the projected stress tensor.
This tool is very useful, since it allows not only the direct calculation of the
average mechanical tension $\tau$, but also the evaluation
of the fluctuation of this tension due to thermal fluctuations, which has
never been done.
The derivation presented here will inspire our derivation of the
projected stress tensor for other geometries in the following chapters.
Note that even though one has no reason not to consider the Gaussian curvature on the
energy for open membranes, we will show that the stress tensor does not depend on it. 

\subsection[{Stress tensor on the local frame $\tilde{\Sigma}$}]{Stress tensor on the local frame $\tilde{\bm{\Sigma}}$}
\label{stress_local_chap}

\begin{figure}[H]
\begin{center}
\subfigure[Local tangent frame $(X,Y,Z)$. The imaginary infinitesimal cut of
length $d\ell$ in green, whose
normal $\bm{\nu}$ is contained in the $(X,Y)$ plane, separates the regions $1$
and $2$. Region $1$ exerts a three-dimensional force $d\bm{\phi}_\mathrm{1\rightarrow
  2}$ over $2$.]{
\includegraphics[scale=.5,angle=0]{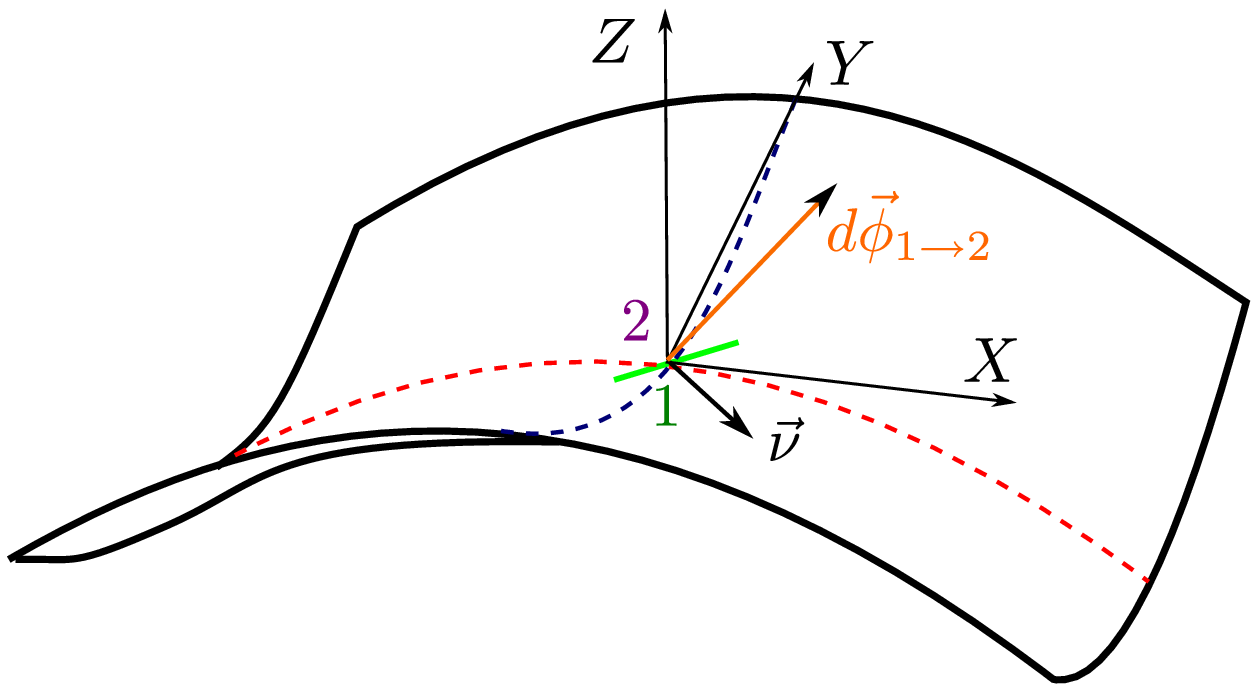}
\label{stress_local}
}
\subfigure[Components of the stress tensor on the local frame]{
\includegraphics[scale=.4,angle=0]{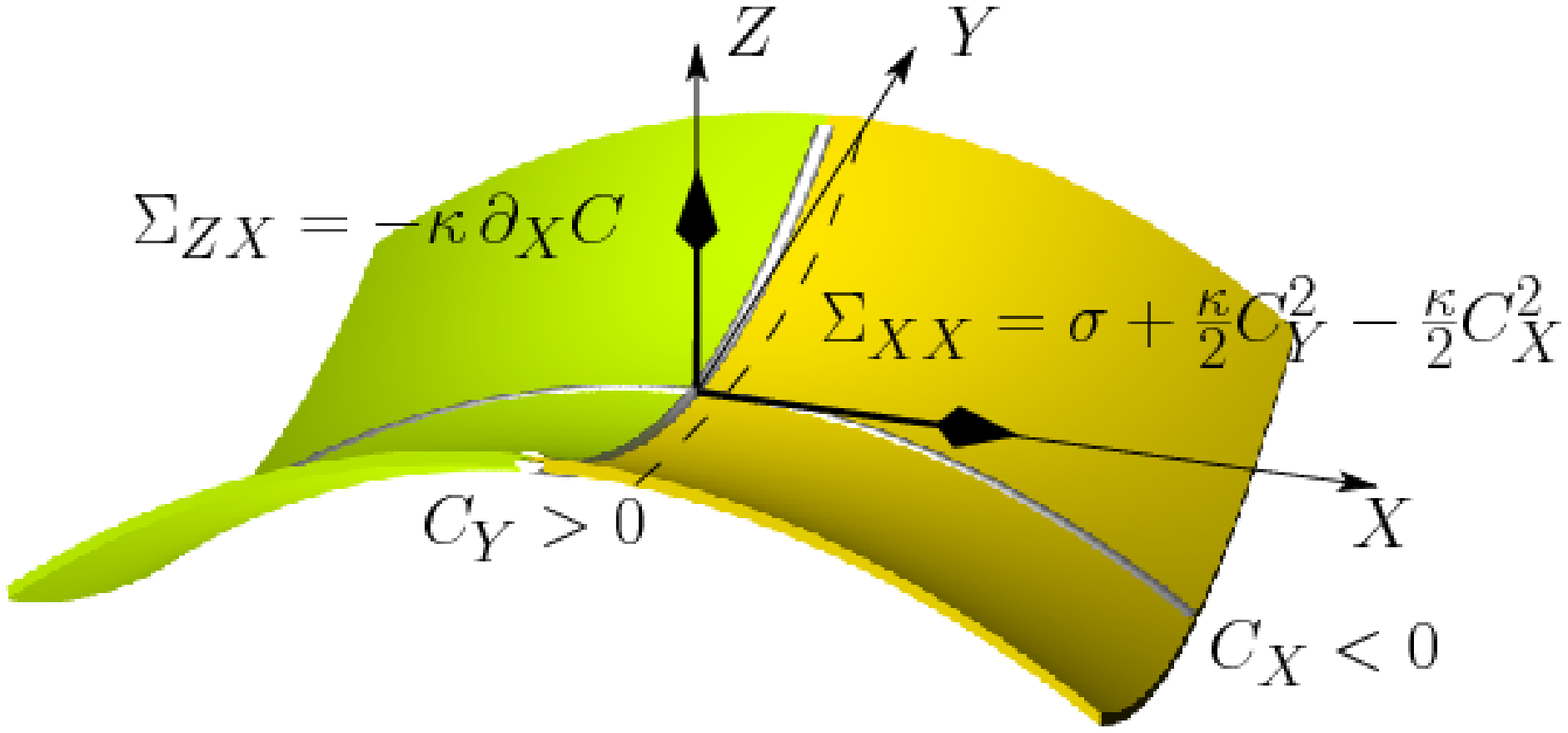}
}
\caption{Stress tensor on the local frame.}
\end{center}      
\end{figure}

Consider a local frame on a membrane $(X,Y,Z)$, whose the first two axes are parallel to the
principal curvature directions and the third one is parallel to the normal of the membrane.
Consider now an imaginary infinitesimal cut of length $d\ell'$ and normal
$\bm{\nu} = \nu_X \, \bm{e}_X + \nu_Y \, \bm{e}_Y$ that
separates the membrane on regions $1$ and $2$ (see Fig.~\ref{stress_local}).
The region $1$ exerts a force $d\bm{\phi}_\mathrm{1\rightarrow 2}$ over the region
$2$ given by

\begin{equation}
d\bm{\phi}_{1 \rightarrow 2} = \tilde{\bm{\Sigma}} \cdot \bm{\nu} \, d\ell'\,.
\end{equation}

\noindent This relation defines the local stress tensor
$\tilde{\bm{\Sigma}}$, a tensor with $3 \times 2 = 6$ components, since the
vector $\bm{\nu}$ has only two components.
For the Helfrich Hamiltonian, one has~\cite{Capovilla_02},~\cite{Fournier_07}
 
\begin{eqnarray}
\tilde{\bm{\Sigma}} &=& \left(\sigma + \frac{\kappa}{2}\,  C_Y^2 - \frac{\kappa}{2}
 \,  C_X^2\right) \, \bm{e}_X \otimes \bm{e}_X + \left(\sigma + \frac{\kappa}{2}
 \, C_X^2 - \frac{\kappa}{2}\, C_Y^2\right) \, \bm{e}_Y \otimes \bm{e}_Y \nonumber\\
&-& \kappa \left(\partial_XC\right)\, \bm{e}_Z\otimes \bm{e}_X - \kappa
\left(\partial_Y C\right) \, \bm{e}_Z \otimes \bm{e}_Y\, ,
\label{Sigma_local_frame}
\end{eqnarray}
 
\noindent where $C_X$ and $C_Y$ are the principal curvatures parallel to
$\bm{e}_X$ and $\bm{e}_Y$ respectively, $C = C_X + C_Y$ and 
$\partial_i$ stands for the derivative with respect to $i$.  

\subsection[Projected stress tensor $\Sigma$]{Projected stress tensor $\bm{\Sigma}$}
\label{section_projected_stress}

\begin{figure}[H]
\begin{center}
\includegraphics[scale=.4,angle=0]{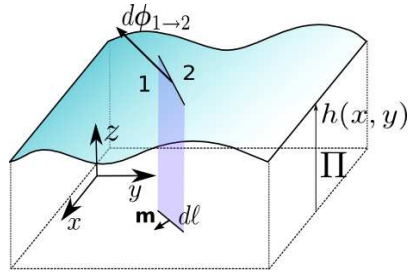}
\caption{The projected stress tensor relates the three-dimensional force
  $d\bm{\phi}_\mathrm{1\rightarrow2}$ to the projection of an imaginary cut on a
  fixed plane.}
\label{S1}
\end{center}      
\end{figure}

 Due to thermal fluctuations, both the tangent frame and $d \ell '$ are not
 constant.
It is thus convenient to introduce the {\it projected} stress tensor $\bm{\Sigma}$, which
relates the force through an imaginary infinitesimal projected cut to the force
$d\bm{\phi}_\mathrm{1\rightarrow 2}$ through

\begin{equation}
d\bm{\phi}_{1\rightarrow2} = \bm{\Sigma} \cdot \bm{m}\,  d\ell\, ,
\end{equation}

\noindent where $d\ell$ is the length of the cut's projection on a reference
fixed plane $\Pi$, $(x,y,z)$ is a orthonormal basis and $\bm{m} = m_x \, \bm{e}_x + m_y \, \bm{e}_y$ is the normal
to the cut's projection on the plane pointing towards region $1$ (see Fig.~\ref{S1}).  
As before, $\bm{\Sigma}$ is a $6$-component tensor.
The advantage of this definition is that one can evaluate the average of the
force exerted through two regions by simply evaluating $\langle \bm{\Sigma}
\rangle$.
It gives thus a straightforward tool to evaluate $\tau$. 

\begin{figure}[H]
\begin{center}
\includegraphics[scale=.4,angle=0]{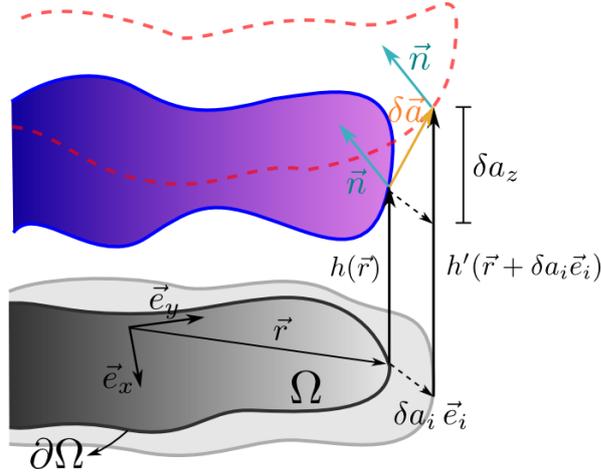}
\caption{The upper surface, shaded in purple,  represents the membrane, while  the respective projected surface
  $\Omega$ is represented by the lower gray shaded surface.
  The red dashed curve represents the new position of the membrane
  after deformation.}
\label{B1}
\end{center}      
\end{figure}

We derive $\bm{\Sigma}$ by studying the work needed 
to produce a deformation~\cite{Fournier_07}.
An alternative derivation is given in appendix~\ref{annexe1}.
First, we consider a piece of membrane weakly departing from a plane described
in the Monge gauge by its height $h(\bm{r}) = h(x,y)$, so that
we can neglect derivatives of order higher than two on $h$.
The general energy is thus given, up to order two, by

\begin{equation}
\mathcal{H} = \int_{\Omega} f(\{h_i\},\{h_{ij}\}) \, dxdy\, ,
\end{equation} 

\noindent where $\Omega$ is the domain of the projected plan over which the membrane is defined, $\partial
\Omega(x,y)$ being the curve that delimits $\Omega$ (see Fig.~\ref{B1}).
In this section, latin indices $\in \{x,y\}$, $h_i \equiv \partial h/\partial i$, $h_{ij} = \partial^2 h/(\partial i \partial j)$ and summation over repeated indices
will be implicit.  

Imagine now that we impose an arbitrary small displacement $\delta \bm{a} = \delta
a_x \, \bm{e}_x + \delta a_y \, \bm{e}_y + \delta a_z \, \bm{e}_z$ to every
point of the membrane, so that the
$h(\bm{r}) \rightarrow h'(\bm{r}) = h(\bm{r}) + \delta h(\bm{r})$. 
Besides, imagine that this displacement keeps the normal along the boundaries
of the membrane constant, so that torques perform no work (see Fig.~\ref{B1}).

On one hand, at equilibrium, the energy variation reads

\begin{equation}
\delta \mathcal{H}  = \int_{\partial\Omega} m_i\left[f\, \delta a_i + \left(\frac{\partial
    f}{\partial h_i} - \partial_i\frac{\partial f}{\partial
    h_{ij}}\right)\delta h + \frac{\partial f}{\partial h_{ij}}\delta
h_j\right] ds \, ,
\label{var_energy}
\end{equation}

\noindent where $ds$ is a length of an infinitesimal element of the curve $\partial\Omega(x,y)$. On the other hand, we have in terms of the stress tensor

\begin{equation}
\delta \mathcal{H} = \int_{\partial\Omega} \delta\bm{a} \cdot \bm{\Sigma} \cdot \bm{m}\, ds\, .
\label{var_stress}
\end{equation}

\noindent One can then obtain $\bm{\Sigma}$ by comparing eq.(\ref{var_energy})
and eq.(\ref{var_stress}). In order to do so, one must express $\delta h$ and
$\delta h_j$ over the boundary in terms of $\delta \bm{a}$.
As shown in Fig.~\ref{B1}, we have $h'(\bm{r} + \delta a_i(\bm{r})\,
\bm{e}_i) = h(\bm{r}) + \delta a_z(\bm{r})$. Up to first order on $\delta h$,
it is easy to deduce $\delta h = \delta a_z - \delta a_j \, h_j$. 
Finally, one has to impose that the normal $\bm{n}$ at the boundary is kept
constant, which yields $h'_k(\bm{r} + \delta a_i(\bm{r})\, \bm{e}_i) =
h_k(\bm{r})$.
Again, up to first order, one has $\partial_k\delta h = - \delta a_j\,  h_{jk}$. 
These results put together lead to 

\begin{eqnarray}
\label{Sigma_interm_ij}
\Sigma_{ij} &=& f \, \delta_{ij} - \left[\frac{\partial f}{\partial h_j} -
  \partial_k\left(\frac{\partial f}{\partial h_{jk}}\right)\right] h_i -
\frac{\partial f}{\partial h_{jk}} h_{ik} \, ,\\
\nonumber\\
\Sigma_{zj} &=& \frac{\partial f}{\partial h_j} - \partial_k\left(\frac{\partial
    f}{\partial h_{jk}}\right) \, .
\label{Sigma_interm_zj}
\end{eqnarray}

In the case of the Helfrich Hamiltonian, one has

\begin{equation}
f = \sigma + \frac{\sigma}{2}(\nabla h)^2 + \frac{\kappa}{2}(\nabla^2 h)^2  +
\bar{f}\, ,
\end{equation}

\noindent where $\bar{f} = \kappa_G \, det(h_{ij})$ is the contribution from the Gaussian curvature.
Eqs.(\ref{Sigma_interm_ij})-(\ref{Sigma_interm_zj}) become thus

\begin{eqnarray}
\label{sigma_xx}
\Sigma_{xx} &=& \sigma + \frac{\sigma}{2}\left(h_y^2 - h_x^2\right) +
  \frac{\kappa}{2}\left(h_{yy}^2 - h_{xx}^2\right) + \kappa \, h_x
  \partial_x\nabla^2 h\, ,\\
\nonumber\\
\Sigma_{xy} &=& -\sigma \, h_xh_y - \kappa \, h_{xy}\nabla^2h + \kappa \, h_x
\partial_y\nabla^2h\, ,\\
\nonumber\\
\Sigma_{zx} &=& \sigma \, h_x - \kappa \, \partial_x\nabla^2 h\, .
\label{sigma_zx}
\end{eqnarray}

\noindent The other components of the tensor can be obtained by permutation of $x$ and $y$.
Remark that these expressions are valid up to order two in $h$ and that the Gaussian curvature gives no contribution to the stress tensor.

%% file: chap1.tex
\chapter{Planar membrane}
\label{chapitre/planar_membrane}

As we have seen in the last chapter, membranes are very particular systems from the mechanical point of view: they are liquid, but rigid; they disrupt very easily under stretching and they fluctuate a lot in even in sub-optical levels.

Accordingly, the term {\it surface tension} has always been rather confusing.
First, it refers to the energy needed to bring a bunch of phospholipids in contact with the aqueous media, which we denote $\gamma$.
As these molecules are amphiphilic, there is almost no energetic cost for creating an interface.
Consequently, one can find in the literature statements like ``a membrane has
vanishing surface tension'' \cite{Seifert_95}.
Secondly, the expression stands for the tension $\tau$ that one can mechanically apply to a membrane, for instance, by aspiring it with a micropipette or by extracting a nanotube.
As discussed in section~\ref{mechanical}, unless in extreme situations, this tension has entropic origin, coming from the flattening of thermal fluctuations.
Thus, it is also called effective mechanical tension.
At last, surface tension denotes also the multiplier Lagrange $\sigma$ one adds to the Hamiltonian in order to fix the total membrane's area, as we have done in section~\ref{grand_can}.
In this case, the tension is more like a chemical potential associated to the total membrane area.
The tension $\sigma$ is not experimentally measurable, but its large-scale counterpart $r$, renormalized by fluctuations, is measurable through the $q^2$ dependence of the spectrum fluctuation.

From an experimental point of view, it is fundamental to determine the relation between $r$, $\sigma$ and $\tau$.
In particular, it is very important to determine under which conditions these
quantities can be assumed identical. 
Indeed, experimentally, one measures usually $r$ or $\tau$, whereas the theoretical predictions involve frequently $\sigma$, which is non measurable.
One takes currently for granted the equality between $\tau$,
$\sigma$ and $r$ to interpret data, as one can see in the sum-up presented in
table~\ref{table_1}, even though there is no support to this premise.

Many theoretical articles were written in order to clarify this question~\cite{Cai_94}, \cite{David_91}, \cite{Farago_04}, \cite{Imparato_06}.
In most cases, the authors tried to derive $r$ and $\tau$ from the free-energy~$\mathcal{F}$.
This route is however very tricky, since one needs to consider terms up to $\mathcal{O}(h^4)$ in order to evaluate $r$.
In this case, the {\it naive} measure presented in section~\ref{grand_can} must be subtly corrected~\cite{Cai_94}.
Besides, the definition of the effective mechanical tension $\tau$ from the
free-energy is not so clear and slightly different alternatives for the definition presented in eq.(\ref{eq_intro_tau}) were proposed.

In this chapter, we try to address the question of the relation between $\tau$ and $\sigma$ for
symmetrical planar membranes in contact with a lipid reservoir.
We evaluate $\tau$ using the projected stress tensor introduced in the end of the last chapter in section~\ref{section_projected_stress}.
This calculation is much more straightforward, since one avoids problems
related to the choice of the measure.
Besides, the definition of $\tau$ in terms of the projected stress tensor is
unique: $\tau$ is simply given by the average of the latter.
In this section, we show that in general, we can assume $\tau = \sigma - \sigma_0$, where $\sigma_0$ is a constant non negligible for small tensions.

In section~\ref{subsection_1_free_energy}, we compare our result to the ones derived by Cai et al.~\cite{Cai_94} and by
Imparato~\cite{Imparato_06} by differentiating the free-energy with respect to the projected area $A_p$.
In his derivation, Imparato used the definition presented in eq.(\ref{eq_intro_tau}), while Cai et al. used slightly different definition, obtaining consequently a different result.
In this section, we show that our result coincides with the one from Imparato,
which gives support to the definition presented in the last chapter

\begin{equation}
\tau = \left(\frac{\partial \mathcal{F}}{\partial A_p}\right)_{N_p} \, ,
\end{equation}

\noindent where the derivative is taken with the total number of lipids
constant.
Besides, we question the previous demonstration by Cai et al. that $\tau =
r$, since their definition of $\tau$ seems less suitable.
We propose then that in general, we should have three different values
for $\tau$, $\sigma$ and $r$.
In order to check this prediction, we present a simple numerical experiment in section~\ref{subsection_1_simu_1D}.

In section~\ref{subsection_1_discuss} and ~\ref{section_1_evidences} we discuss some consequences to
experiments, namely to those involving micropipettes, introduced in section~\ref{subsection_0_micro}.
As $\tau$ is indeed different from $\sigma$, we propose corrections to the eq.(\ref{eq_0_alpha}) presented in the last chapter.
We conclude in section~\ref{section_1_evidences} with the description of the
first recent numerical and experimental evidences that $\tau \neq \sigma$.
All results presented in this chapter were obtained under the direction of Jean-Baptiste Fournier and published in~\cite{Fournier_08_eu}.

\section[Evaluation of ${\tau}$ from the stress tensor]{Evaluation of ${\bm \tau}$ from the stress tensor}
\label{section_1_tau}

\begin{figure}[H]
\begin{center}
\includegraphics[scale=.4,angle=0]{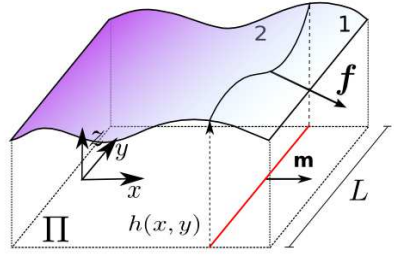}
\caption{Weakly fluctuating planar membrane described in the Monge's gauge. The
  force $\bm{f}$ is exchanged through the cut of projected
  length $L$ (red). Note that we have chosen an orthonormal basis in order to
  have $\bm{m} = \bm{e}_x$.}
\label{fig_1_cut}
\end{center}      
\end{figure}

Consider a planar membrane whose projected area on a plan $\Pi$ parallel to
the average plan of the membrane is $A_p$ and well described by the Helfrich Hamiltonian given in eq.(\ref{Helfrich}).
This membrane is not stretched and departs very weakly from a plane.
Therefore, we use the Monge's gauge and develop $\mathcal{H}$ up to order
two, obtaining the Hamiltonian given in eq.(\ref{Hel_free}), the average
$\langle |h(\bm{q})|^2\rangle$ given in eq.(\ref{spect_base}) and 
the projected stress tensor is given in eqs.(\ref{sigma_xx})-(\ref{sigma_zx}).
Consider now a cut of length $L$ on $\Pi$ parallel to $\bm{e}_y$, so that the normal to the projected cut is simply $\bm{m} = \bm{e}_x$, as shown in Fig.~\ref{fig_1_cut}.  

The force exchanged through the cut is 

\begin{equation}
  \bm{f} = \int_{-L/2}^{L/2} \left(\Sigma_{xx} \, \bm{e}_x + \Sigma_{yx} \,
    \bm{e}_y + \Sigma_{zx} \, \bm{e}_z \right) \, dy \, .
\end{equation}

\noindent The thermal average of $\bm{f}$, denoted by the brackets $\langle \rangle$, is given by 

\begin{equation}
\langle\bm{f}\rangle = L\left(\langle \Sigma_{xx} \rangle \, \bm{e}_x +
  \langle \Sigma_{yx} \rangle \, \bm{e}_y + \langle \Sigma_{zx} \rangle \, \bm{e}_z \right) \, .
\end{equation}

\noindent One must evaluate the average of each term of eqs.(\ref{sigma_xx})-(\ref{sigma_zx}). 
Hence, it is interesting to use the Fourier transform introduced in section~\ref{grand_can}

\begin{equation}
  h(\bm{r}) = \frac{1}{\sqrt{A_p}}\sum_{\bm{q}} h_{n,m} \, e^{\icomp \, \bm{q} \cdot \bm{r}} \, ,
\end{equation}

\noindent with $\bm{q} = 2\pi/\sqrt{A_p} \, (n,m)$, $n,m \in \mathbb{N}$ and

\begin{equation}
  \sum_{\bm{q}} \equiv \sum_{|n| \, \leq \, N_\mathrm{max}} \, \sum_{|m| \, \leq \, N_\mathrm{max}} \, .
\end{equation}

\noindent Note that as $h(\bm{r})$ is real, one has $h_{-n,-m} = h_{n,m}^*$, where the symbol $^*$ indicates the complex conjugate. 
The mode $n=0$ and $m=0$ corresponds to a simple translation and
gives no contribution to the energy. It will be therefore omitted throughout this section.

Using this definition of the Fourier transform, the Hamiltonian for a weakly fluctuating membrane in contact with a lipid reservoir introduced in section~\ref{grand_can} becomes

\begin{equation}
\mathcal{H} = \sigma \, A_p + \frac{1}{2}\sum_{\bm{q}} \left(\sigma q^2 + \kappa 
  q^4\right)|h_{n,m}|^2 \, ,
\label{eq_1_hamilton}
\end{equation}

\noindent where $q$ varies between $q_\mathrm{min} = 2\pi/\sqrt{A_p}$ up to $\Lambda  = 2\pi N_\mathrm{max}/\sqrt{A_p} \approx 1/a$, where $a$ is a microscopical cut-off comparable to the membrane thickness.
The correlation function is given by 

\begin{eqnarray}
  G(\bm{r} - \bm{r}') &\equiv& \langle h(\bm{r})h(\bm{r}')\rangle \, \nonumber\\
  &=& \frac{1}{A_p} \sum_{\bm{q}}\sum_{\bm{k}} \langle h_{n,m}h_{n',m'} \rangle \, e^{\icomp \, \bm{q}\cdot \bm{r}}e^{\icomp \, \bm{k} \cdot \bm{r'}} \, \nonumber \\
  &=& \frac{k_\mathrm{B} T}{A_p} \sum_{\bm{q}} \frac{e^{\icomp \, \bm{q}\cdot
      (\bm{r} - \bm{r'})}}{\sigma\,  q^2 + \kappa \, q^4} \, , 
\label{eq_1_correl}
\end{eqnarray}

\noindent where we have used the result displayed on eq.(\ref{spect_base}) to obtain the last passage.

It is a straightforward calculation to evaluate averages using the correlation function.
As an example, we do a step-by-step evaluation the average of $\Sigma_{xx}$:

\begin{eqnarray}
  \langle \Sigma_{xx} \rangle &=& \sigma + \kappa \, \langle h_xh_{xxx} + h_xh_{xyy}\rangle \, \nonumber\\
  &=& \sigma + \kappa \, \partial_x\partial_{x'}\left(\partial_{x'}^2 + \partial_{y'}^2\right)G\left(\bm{r} - \bm{r'}\right)\vert_{\bm{r} = \bm{r}'}\nonumber\\
  &=& \sigma + \kappa \, \frac{k_\mathrm{B}T}{A_p} \, \sum_{\bm{q}} \frac{(\icomp \, q_x)(-\icomp \, q_x)\left(-q_x^2 - q_y^2 \right)}{\sigma \, q^2 + \kappa \, q^4}  \nonumber\\
  &=& \sigma - \frac{k_\mathrm{B} T}{A_p} \sum_{\bm{q}} \frac{\kappa\, q_x^2}{\sigma + \kappa \, q^2} \nonumber \\
  &=& \sigma - \frac{k_\mathrm{B} T}{2 A_p} \sum_{\bm{q}} \frac{\kappa\, q^2}{\sigma + \kappa \, q^2} \, ,
\end{eqnarray}

\noindent where we used in the first and in the last passage the fact that by symmetry $\langle h_x^2 \rangle = \langle h_y^2 \rangle$ and $\langle h_{xx}^2 \rangle = \langle h_{yy}^2 \rangle$.
By the same reasoning, one can demonstrate that $\langle \Sigma_{yx}\rangle = 0$ and $\langle \Sigma_{zx} \rangle = 0$, as expected given the symmetry of the system.
We have thus the effective tension $\tau$, which relates to the average of the
force $\bm{f}$ through

\begin{equation}
  \tau \equiv \frac{\langle \bm{f} \rangle \cdot \bm{e}_x}{L}  = \sigma - \frac{k_\mathrm{B}T}{2 A_p} \sum_{\bm{q}} \frac{\kappa \, q^2}{\sigma + \kappa \, q^2} \, .
  \label{eq_1_diff_complet}
\end{equation}

\begin{figure}[H]
\begin{center}
\includegraphics[scale=.55,angle=0]{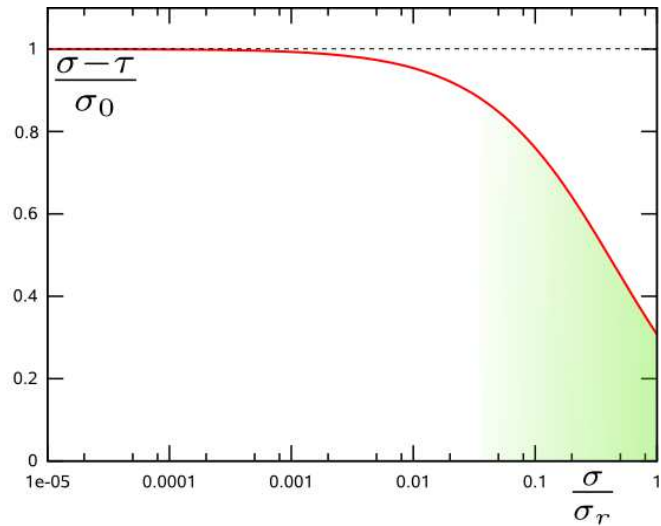}
\caption{The difference $\sigma - \tau$ normalized by $\sigma_0$ as a function
  $\sigma/\sigma_r$. For tensions smaller than $10^{-2} \, \sigma_r$, we see
  that $\tau \simeq \sigma - \sigma_0$. The green shaded area corresponds to the region where we expect our theory to need corrections due to the stretching of the membrane.}
\label{fig_1_diff_per}
\end{center}      
\end{figure}

In the thermodynamic limit, $A_p$ is very large and the sum over $\bm{q}$ becomes an integral whose calculation leads to

\begin{equation}
  \tau - \sigma = - \frac{k_\mathrm{B} T \, \Lambda^2}{8 \pi} \left[1 - \frac{\sigma}{\sigma_r}\ln\left(1 + \frac{\sigma_r}{\sigma}\right)\right] \, ,
  \label{eq_1_diff}
\end{equation}

\noindent where $\sigma_r = \kappa \Lambda^2$.
Numerically, for typical values $a \approx 5 \, \mathrm{nm}$ and $\kappa
\approx 10^{-19} \, \mathrm{J}$, one obtains $\sigma_r \approx 5 \times
10^{-3} \, \mathrm{N/m}$, which is  of the same order of magnitude of the rupture tension of membranes~\cite{Rawicz_00}.
Fig.~\ref{fig_1_diff_per} shows the difference $\sigma - \tau$ normalized by

\begin{equation}
  \sigma_0 = \frac{k_\mathrm{B} T \, \Lambda^2}{8 \pi} = \frac{\sigma_r}{8\pi \beta\kappa} \, ,
  \label{eq_1_sigma0}
\end{equation}

\noindent as a function of $\sigma/\sigma_r$.

\begin{figure}[H]
\begin{center}
\includegraphics[scale=.5,angle=0]{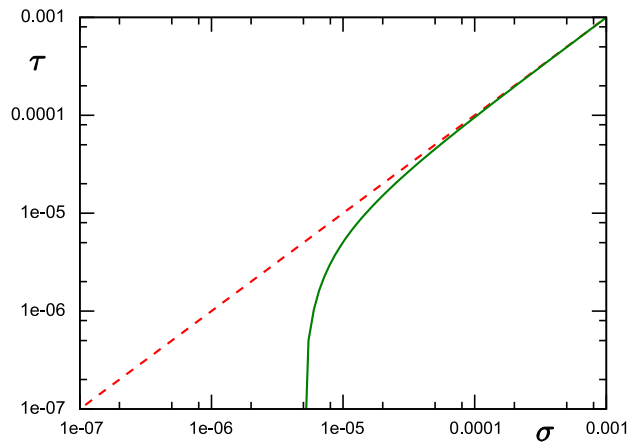}
\caption{The red dashed straight line shows the standard approximation $\tau \approx
  \sigma$, while the green curve shows the behavior of $\tau$ predicted by
  our theory. For small tensions, the correction is relevant. Curves for $a =
  5\, \mathrm{nm}$, $\kappa = 10^{-19} \, \mathrm{J}$ and $k_\mathrm{B}T = 4 \times 10^{-21} \, \mathrm{J}$.}
\label{fig_1_diff}
\end{center}      
\end{figure}

As we can see, it is a good approximation to set $\tau \simeq \sigma -
\sigma_0$ for $\sigma < 10^{-2} \, \sigma_r$.
Beyond this limit, the tension is relatively high and we expect corrections
coming from the stretching of the membrane.
For the previous values of $a$ and $\kappa$ and taking $k_\mathrm{B} T \approx 4\times 10^{-21} \, \mathrm{J}$, we obtain $\sigma_0 \approx 5 \times 10^{-6} \, \mathrm{N/m}$.
As tensions as small as $\tau \approx 10^{-8} \, \mathrm{N/m}$ are measured in
micropipette experiments, this correction may be non negligible (see Fig.~\ref{fig_1_diff}).
We will discuss the consequences of this prediction to experiments in section~\ref{subsection_1_discuss}.

\section{Derivation from the free-energy}
\label{subsection_1_free_energy}

Here we re-derive the result given in eq.(\ref{eq_1_diff}) by
differentiating the free-energy.
To begin, we evaluate the free-energy as in
references~\cite{Cai_94} and~\cite{Imparato_06}.
By definition, we have 

\begin{equation}
\mathcal{F} = -k_\mathrm{B} T \ln(\mathcal{Z}) \, ,
\end{equation}

\noindent where $\mathcal{Z}$ is the partition function given by

\begin{equation}
\mathcal{Z} = \int \mathcal{D}[h] \, e^{-\beta \mathcal{H}[h]} \, .
\label{eq_1_partition}
\end{equation}

As discussed in section~\ref{grand_can}, one needs a measure to evaluate this integral.
We will consider here the {\it naive} measure, which is justified up to first order on the temperature and up to second order on $h$ (further discussion on the
subtleties of the measure can be found in \cite{Cai_94} and \cite{Nelson_93}).
We discretize the projected plane $\Pi$ in
$N^2$ squares of area $\bar{a}^2$, so that the height of each one of these squares is denoted $h^{p_x,p_y}\equiv h(p_x \, \bar{a} \, \bm{e}_x + p_y \, \bar{a} \, \bm{e}_y)$.
The {\it naive} measure reads

\begin{equation}
  \mathcal{D}_\mathrm{naive}[h] =  \prod_{p_x = 1}^{N}\prod_{p_y=1}^{N}
  \frac{dh^{p_x,p_y}}{\lambda} \, 
\label{eq_1_direct}
\end{equation}

\noindent The factor $\lambda$ is a vertical quantum introduced to keep
$\mathcal{Z}$ dimensionless.

\subsection{{\it Naive} measure in the Fourier space}

In section~\ref{grand_can}, we have used this measure to evaluate averages, but we have not derived explicitly $\mathcal{F}$.
Here, to do so, we prefer to work in the Fourier space.
We need thus to derive the equivalent of the measure above in this space.
As in the last section, let's consider the Fourier transform

\begin{equation}
h^{p_x,p_y} = \frac{1}{\sqrt{A_p}} \sum_{\bm{q}} h_{n,m} e^{\frac{2
  \pi i n \, p_x }{N}} \, e^{\frac{2 \pi i m \, p_y}{N}}
\label{eq_1_denovo}
\end{equation}

\noindent Remark that only half of the total number of modes are
independent, since $h_{-n,-m} = h_{n,m}^*$.
In terms of these independent modes, the measure is thus 

\begin{equation}
  \mathcal{D}_\mathrm{naive}[h] =  \left( \prod_{|n| \, \leq \, \frac{N}{2}, \, n \, \neq\,  0} \, \prod_{m = 0}^\frac{N}{2}
  \int \frac{dh_{n,m}^{\mathcal{R}} }{\lambda}
  \, \frac{dh_{n,m}^{\mathcal{I}}}{\lambda }\right) \times J\,
  \label{eq_1_measure}
\end{equation}

\noindent where the superscripts $\mathcal{R}$ and $\mathcal{I}$ stand,
respectively, for the real and imaginary part of $h_{n,m}$ and $J$ is the
Jacobian of the transformation.
To simplify notations, in the following we will simply denote

\begin{eqnarray}
  \prod_n \prod_{m\geq 0} &\equiv& \prod_{|n| \, \leq \, \frac{N}{2},\, n \, \neq \, 0} \, \prod_{m = 0}^\frac{N}{2} \, ,\\
  \sum_n \sum_{m\geq 0} &\equiv& \sum_{|n| \, \leq \, \frac{N}{2}, \, n\, \neq\, 0} \, \sum_{m = 0}^\frac{N}{2} \, .
\end{eqnarray}

To determine $J$, we will evaluate the partition function for a simple
Gaussian Hamiltonian.
With the measure given in eq.(\ref{eq_1_direct}), we have

\begin{equation}
\mathcal{Z} = \int \prod_{p_x = 1}^{N}\prod_{p_y=1}^{N}
  \frac{dh^{p_x,p_y}}{\lambda} \, e^{-\alpha \sum_{p_x=1}^N \sum_{p_y = 1}^N \,
    {(h^{p_x,p_y}})^2} =
  \left(\sqrt{\frac{\pi}{\alpha}}\right)^{N^2}
\label{eq_1_jac1}
\end{equation}

\noindent Using the definition presented in eq.(\ref{eq_1_denovo}), the
quadratic Hamiltonian becomes

\begin{eqnarray}
\alpha \sum_{p_x = 1}^N \sum_{p_y = 1}^N {(h^{p_x, p_y})}^2 &=&
\frac{\alpha}{L^2} \sum_{\bm{q}}\sum_{\bm{q}'} h_{n,m} \, h_{n', m'} \times
N^2 \, \delta_{n', -n} \, \delta_{m', -m} \nonumber\\
&=& \frac{\alpha}{\bar{a}^2} \sum_{\bm{q}} \left[(h_{n,m}^\mathcal{R})^2 + (h_{n,m}^\mathcal{I})^2\right]
\end{eqnarray}

\noindent and the partition function is 

\begin{equation}
\mathcal{Z} = J \times \int  \prod_{n} \, \prod_{m \geq 0}
  \int \frac{dh_{n,m}^{\mathcal{R}} }{\lambda}
  \, \frac{dh_{n,m}^{\mathcal{I}}}{\lambda }\,e^{-\frac{\alpha}{\bar{a}^2}
    \sum_{\bm{q}} \left[(h_{n,m}^\mathcal{R})^2 +
      (h_{n,m}^\mathcal{I})^2\right]} = \left(\sqrt{\frac{\pi \,
          \bar{a}^2}{\alpha}}\right)^{N^2} \times J
\label{eq_1_jac2}
\end{equation}

\noindent The partition function should be the same in both
eq.(\ref{eq_1_jac1}) and eq.(\ref{eq_1_jac2}), which implies
$J = 1/(\bar{a})^{N^2}$.
Summing up, in the Fourier space, for a weakly fluctuating membrane, the {\it naive} measure is equivalent to 
%Eq.(\ref{eq_1_partition}) becomes

%\begin{equation}
%\mathcal{Z} \simeq \prod_{i = 0}^{i = N-1} \int \frac{d h^i}{\lambda} \, e^{-\beta
 % \mathcal{H}(\{h^i\})} \, ,
%\end{equation}

\begin{equation}
  \mathcal{D}_\mathrm{naive}[h] = \prod_{n}
  \, \prod_{m \geq 0} \frac{dh_{n,m}^{\mathcal{R}}}{\lambda
    \bar{a}} \frac{dh_{n,m}^{\mathcal{I}}}{\lambda \bar{a}}\, .
  \label{eq_1_measure1}
\end{equation}

\subsection[Evaluation of $\mathcal{F}$ and discussion]{Evaluation of   $\bm{\mathcal{F}}$ and discussion}

Using the Hamiltonian given in eq.(\ref{eq_1_hamilton}), the partition function given in eq.(\ref{eq_1_partition}) with the measure (\ref{eq_1_measure1}) becomes

\begin{eqnarray}
\mathcal{Z} &\simeq& \prod_{n} \, \prod_{m \geq 0}
  \int \frac{dh_{n,m}^{\mathcal{R}}}{\lambda \bar{a}}\frac{dh_{n,m}^{\mathcal{I}}}{\lambda \bar{a}}
  \, e^{-\beta \mathcal{H}(\{h_{n,m}\})} \, , \nonumber \\
&\simeq& e^{-\beta \sigma A_p} \prod_{n} \, \prod_{m \geq 0}
  \left(\int \frac{dh_{n,m}^{\mathcal{R}}}{\lambda \bar{a}} \,  e^{-\frac{\beta}{2} \sum_n \sum_{m\geq0} \left(\sigma q^2 + \kappa q^4\right)\, {h_{n,m}^{\mathcal{R}}}^2} \right) \nonumber \\
&\times& \left(\frac{dh_{n,m}^{\mathcal{I}}}{\lambda \bar{a}} \,  e^{-\frac{\beta}{2} \sum_n \sum_{m\geq0} \left(\sigma q^2 + \kappa q^4\right)\,  {h_{n,m}^{\mathcal{I}}}^2}\right) \, .
\end{eqnarray}

\noindent Finally, carrying out the Gaussian integrals, we have

\begin{equation}
\mathcal{Z} \simeq e^{-\beta \sigma A_p} \prod_{n} \, \prod_{m \geq 0}
\frac{2\pi}{\beta \bar{a}^2
  \lambda^2\left(\sigma q^2 + \kappa q^4\right)}\, .
\end{equation}

\noindent Accordingly, the free-energy is given, to lowest order in $T$, by

\begin{equation}
  \mathcal{F} = \sigma \, A_p + k_\mathrm{B} T \sum_{n} \, \sum_{m\geq0}
  \ln\left[\left(\sigma \, q^2 + \kappa \, q^4 \right) \frac{\bar{a}^2 \lambda^2}{2\pi k_\mathrm{B} T}\right]\, ,
  \label{eq_1_free_energy}
  \end{equation}

\noindent where we remind that $\lambda$ is a quantum discretizing the
membrane vertical displacements.
Equivalently, highlighting the dependence of $\mathcal{F}$ on $A_p$, one obtains

\begin{equation}
  \mathcal{F} = \sigma \, A_p + k_\mathrm{B} T \sum_{n} \, \sum_{m\geq0}
  \ln\left[\left(\sigma \, \tilde{q}^2 + \frac{\kappa \, \tilde{q}^4}{A_p} \right) \frac{\lambda^2}{2\pi N k_\mathrm{B} T}\right]\, ,
  \label{eq_1_free_energy_Ap}
   \end{equation}

\noindent where we have used $\bar{a}^2 = A_p/N = 4\pi/\Lambda^2$, $N$ being the total number of modes or degrees of freedom, and $\tilde{\bm{q}} = 2\pi
(n,m)$.

With the definition presented in section~\ref{mechanical}

\begin{equation}
  \tau = \left(\frac{\partial \mathcal{F}}{\partial A_p}\right)_{N_p} \, ,
  \label{eq_1_def}
\end{equation}

\noindent one obtains from eq.(\ref{eq_1_free_energy_Ap})

\begin{eqnarray}
  \left(\frac{\partial \mathcal{F}}{\partial A_p}\right)_{N_p} &=&
  \sigma - \frac{k_\mathrm{B}T}{2 A_p} \sum_{\bm{q}} \frac{ \kappa \,
  q^2}{\sigma  + \kappa q^2} \,  ,
\label{eq_1_Alb}
\end{eqnarray}

\noindent where the derivation was taken keeping the number of modes constant.
Indeed, once the cutoff $\Lambda$ is fixed, having a total number of particles
fixed is equivalent to  having a total number of modes fixed.
This result coincides with our previous derivation (eq.(\ref{eq_1_diff})) and gives some evidence of the correctness of the derivation presented in~\cite{Imparato_06}. 
Instead, in their work, Cai et al. used a slightly different definition for the effective tension, assuming 

\begin{equation}
  \tau_\mathrm{Cai} = \frac{\partial \mathcal{F_\mathrm{lim}}}{\partial A_p} \, ,
\end{equation}

\noindent where $\mathcal{F}_\mathrm{lim}$ is the free-energy in the limit of very large membranes.
In this case, the sum in eq.(\ref{eq_1_free_energy}) becomes an integral and one obtains

\begin{equation}
  \mathcal{F}_\mathrm{lim} = \sigma \, A_p + \frac{k_\mathrm{B} T \, A_p}{2} \int \frac{d^2 q}{(2\pi)^2} \ln\left[\left(\sigma  q^2 + \kappa  q^4 \right) \frac{\bar{a}^2 \lambda^2}{2\pi k_\mathrm{B} T}\right]\, .
\end{equation}

\noindent It follows

\begin{equation}
  \frac{\partial \mathcal{F}_\mathrm{lim}}{\partial A_p} = \sigma + \frac{k_\mathrm{B} T }{2} \int \frac{d^2 q}{(2\pi)^2} \ln\left[\left(\sigma  q^2 + \kappa  q^4 \right) \frac{\bar{a}^2 \lambda^2}{2\pi k_\mathrm{B} T}\right]\, .
  \label{eq_1_Cai}
\end{equation}

%%   \frac{\partial
%%     \mathcal{F}}{\partial A_p} + \frac{\partial
%%     \mathcal{F}}{\partial q^2}\frac{\partial q^2}{\partial A_p} + \frac{\partial
%%     \mathcal{F}}{\partial \bar{a}^2}\frac{\partial \bar{a}^2}{\partial A_p}\, ,\nonumber\\
%% &=& \sigma - \frac{k_\mathrm{B} T}{2 A_p}
%% \sum_{\bm{q}} \frac{\sigma q^2 + 2 \kappa q^4}{\sigma q^2 + \kappa q^4}
%% + \frac{N\, k_\mathrm{B} T}{2 A_p}\, ,\nonumber \\

\noindent As this result disagrees with the one obtained in eq.(\ref{eq_1_diff}), we conclude that the definition eq.(\ref{eq_1_def}) is more appropriate: one  
must first differentiate with respect to $A_p$ keeping the number of modes
constant and only after take the thermodynamic limit, if needed.
Note that with the projected stress tensor these subtleties are avoided, once
one deals only with the straightforward evaluation of averages.

\subsection{What about r?}
\label{subsubsection_1_r}

In their work, Cai et al. showed also that one should have $r = \tau$.
Here we will present in detail their reasoning and argue that their conclusion
follow from the fact that their definition of $\tau$ is slightly different
from ours (compare eq.(\ref{eq_1_Cai}) with the thermodynamical limit of eq.(\ref{eq_1_Alb})).
Thus, in general, one should have $r \neq \tau$.

First of all, as in section~\ref{grand_can}, they introduced a conjugated
field $m(\bm{r})$ to the Hamiltonian in order to fix a general average shape $\langle h(\bm{r}) \rangle = \bar{h}(\bm{r}) \equiv \bar{h}$, obtaining

\begin{equation}
  \mathcal{H}' = \mathcal{H} - \int_S h(\bm{r}) \, m(\bm{r}) \, dA \, ,
\end{equation}

\noindent where $\mathcal{H}$ is the physical Hamiltonian given in eq.(\ref{Helfrich}).
The corresponding partition function is

\begin{equation}
  \mathcal{Z} = \int \mathcal{D}[h] \, e^{-\beta \mathcal{H}[h]}\, e^{\beta \int_S h \, m\,  dA}
\end{equation}

\noindent and the effective action, i. e., the Legendre transform of the free-energy, is given by

\begin{equation}
\mathcal{F}_\mathrm{eff} = -k_\mathrm{B} T \ln \mathcal{Z} + \int_S \bar{h} \, m \, dA \, .
\end{equation}

%\noindent Note that when $m = 0$, one recovers the partition function $\mathcal{Z}$ and the free-energy $\mathcal{F}$ introduced in the beginning of this section.
\noindent The average height of the membrane is given by

\begin{eqnarray}
  \langle h(\bm{r}) \rangle_m \equiv \frac{\int \mathcal{D}[h] \, h(\bm{r}) \, e^{-\beta \mathcal{H}[h]} \, e^{\int_S h\, m \, dA}}{\mathcal{Z}} = \frac{k_\mathrm{B} T}{\mathcal{Z}} \frac{\delta \mathcal{Z}}{\delta m(\bm{r})}  = \bar{h}(\bm{r})\, ,
  \label{eq_1_hbar}
\end{eqnarray}

\noindent where $\delta \mathcal{Z}/\delta m(\bm{r})$ stands for the functional derivative of the effective partition function with respect to the field $m$ at the point $\bm{r}$.
For $m=0$, we have a simple Gaussian integral and thus $\bar{h}(\bm{r})
|_{m=0} = 0$, which corresponds to the case of a planar membrane.
Differentiating the free-energy with respect to $\bar{h}(\bm{r})$ and using eq.(\ref{eq_1_hbar}), one obtains

\begin{equation}
  \frac{\delta \mathcal{F}_\mathrm{eff}}{\delta \bar{h}(\bm{r})} = m(\bm{r}) - \frac{k_\mathrm{B} T}{\mathcal{Z}} \int \frac{\delta m(\bm{r}')}{\delta \bar{h}(\bm{r})} \, \frac{\delta \mathcal{Z}}{\delta m(\bm{r}')} \, dA + \int \frac{\delta m(\bm{r}')}{\delta \bar{h}(\bm{r})} \, \bar{h}(\bm{r}') \, dA = m(\bm{r})\, .
  \label{eq_1_j}
  \end{equation}

\noindent For the case $m=0$, the correlation function is given by

\begin{eqnarray}
  \langle h(\bm{r}) \, h(\bm{r}') \rangle_{m = 0} &=& \frac{(k_\mathrm{B} T)^2}{\mathcal{Z}} \, \left. \frac{\delta^2 \mathcal{Z}}{\delta m(\bm{r}) \delta m(\bm{r}')}  \right|_0 \, ,\nonumber \\
  &=& \frac{k_\mathrm{B} T}{\mathcal{Z}} \, \left.\frac{\delta \left[\mathcal{Z}\,  \bar{h}(\bm{r}')\right]}{\delta m(\bm{r})}\right |_0 \, , \nonumber\\
  &=& k_\mathrm{B} T \left[\left. \frac{\delta m(\bm{r})}{\delta \bar{h}(\bm{r}')} \right |_0 \right]^{-1} \, , \nonumber \\
  &=& k_\mathrm{B} T \left[\left. \frac{\delta
        \mathcal{F}_\mathrm{eff}}{\delta \bar{h}(\bm{r}) \, \delta
        \bar{h}(\bm{r}')} \right |_0 \right]^{-1} \, ,
\label{eq_1_r}
  \end{eqnarray}

\noindent where we have used eq.(\ref{eq_1_hbar}) and eq.(\ref{eq_1_j}) in the
second and third passage, respectively.

Meanwhile, as we have seen above, Cai et al. defined the tension as

\begin{equation}
  \tau_\mathrm{Cai} = \frac{\partial \mathcal{F}_\mathrm{lim}}{\partial A_p} =
  \frac{\partial \mathcal{F}_{\mathrm{eff,lim}} \, [\bar{h} = 0]}{\partial
    A_p} \, ,
%= \frac{\partial \mathcal{F}_{\mathrm{eff,lim}}^0 }{\partial A_p} \, ,
\end{equation}

\noindent where the $\mathcal{F}_\mathrm{eff,lim}$ is the effective action for
the limit of large membranes. % and the superscript $0$ stands for $m=0$.
Suppose now that the average shape $\bar{h}(\bm{r})$ is tilted.
The free-energy should remain the same, since the physical area of the membrane has not changed.
The dependence of the free-energy on $\bar{h}$ to lowest order should thus be\cite{Nelson_93}

\begin{equation}
\mathcal{F}_\mathrm{eff} = \tau_\mathrm{Cai} \int_S \left[\left(1 + \frac{1}{2} (\vec{\nabla} \bar{h})^2 + \cdots\right) + \cdots\right] \, dA \, ,
\end{equation}

\noindent where the first ellipsis involves terms $\mathcal{O}(\bar{h}^4)$ and the second involves high order derivatives on $\bar{h}$.
Note that the dependence should remain the same if one takes the thermodynamical limit.
One has thus

\begin{equation}
\left. \frac{\delta \mathcal{F}_\mathrm{eff}}{\delta \bar{h}(\bm{r}) \, \delta
    \bar{h}(\bm{r}')} \right |_{m = 0}   = - \tau_\mathrm{Cai} \,
\Delta_{\bm{r}}\, 
\delta(\bm{r} - \bm{r}') + \cdots.
\end{equation}

\noindent where $\delta (x)$ is the Dirac delta function and $\Delta_{\bm{r}}$
is the Laplacian calculated at the point $\bm{r}$.
By definition, the inverse of an operator $M(\bm{r})$ is given by

\begin{equation}
\int M(\bm{r} - \bm{r}') \, M^{-1}(\bm{r}' - \bm{r}'') \, d\bm{r}' = \delta(\bm{r} -
\bm{r}'')\, ,
\end{equation}
 
\noindent which yields 

\begin{equation}
  \left[ \left.\frac{\delta \mathcal{F}_\mathrm{eff}}{\delta \bar{h}(\bm{r})
        \, \delta \bar{h}(\bm{r}')} \right |_{m=0} \right]^{-1}   = \frac{1}{A_p}
  \sum_{\bm{q}} \frac{e^{i \, \bm{q} \cdot(\bm{r} -
      \bm{r}')}}{\tau_\mathrm{Cai} \, q^2 + \mathcal{O}(q^4)} \, .
\label{eq_1_inv}
\end{equation}

\noindent Let's look again at eq.(\ref{eq_1_r}): the term on the left is the
correlation function for a planar membrane, given in general by 

\begin{equation}
\langle h(\bm{r}) \, h(\bm{r}') \rangle = \frac{k_B T}{A_p} \sum_{\bm{q}}
\frac{e^{i\, \bm{q} \cdot (\bm{r} - \bm{r}')}}{r \, q^2 + \mathcal{O}(q^4)} \, .
\end{equation}

\noindent Note that here the coefficient of the quadratic term $r$ is in
general different from $\sigma$.
Indeed, to obtain the correlation function given in
section~\ref{grand_can}, we have used the {\it naive} measure, while the
discussion presented here remains general and valid for any measure. 

Eq.(\ref{eq_1_r}) combined with eq.(\ref{eq_1_inv}) imply thus that
one should have always $r = \tau_\mathrm{Cai}$.
Indeed, after a careful study taking into account measure subtleties, Cai et al. succeed to prove this assertion.
We do not question their proof, but rather their definition of $\tau$, which seems less appropriated, since it does not yield the same results as with the stress tensor.
With our definition of $\tau$, we have $\tau \neq \tau_\mathrm{Cai}$ and thus in general we should expect $r \neq \tau \neq \sigma$.
We will show that it is indeed the case in a simple numerical experiment in section~\ref{subsection_1_simu_1D}.

\section{1-D Numerical experiment}
\label{subsection_1_simu_1D}

Here we present a simple numerical experiment proposed to check
the results of the two last sections.
We have chosen for simplicity to simulate the $1$-d equivalent of a membrane,
i. e., a $1$-d filament fluctuating in a
$2$-d space.
%The former results for
%a filament are re-derived in section \ref{subsection_1_theo_1D}.
Despite the plainness of our numerical system, described in section~\ref{subsection_1_simu_describe}, we have access to the three tensions $r$,
$\sigma$ and $\tau$.
In section~\ref{subsection_1_results} we present and discuss the
compatibility of the numerical data with the theoretical predictions for a
filament, derived in section~\ref{subsection_1_theo_1D}.

\subsection[The tension $\tau$ for a $1$-d filament]{The tension \bm{$\tau$} for a \bm{$1$}-d filament}
\label{subsection_1_theo_1D}

Let's call $\bm{e}_\parallel$ the average direction of the filament and $\bm{e}_\perp$ the
perpendicular direction.
The filament's length $L$ is fixed by adjusting the conjugated variable $\sigma$ and the projected length on $\bm{e}_\parallel$ is denoted $L_p$.
In the Monge's gauge, its shape is described by the height $h(x) \, \bm{e}_\perp$, where $x$ is the ordinate in the direction $\bm{e}_\parallel$.
For a weakly fluctuating filament, the energy is given by the $1$-d counterpart of eq.(\ref{Helfrich})

\begin{equation}
\mathcal{H}_\mathrm{1D} = \sigma \, L_p + \int_{L_p} \left[\frac{\kappa}{2}\, h_{xx}^2 +
  \frac{\sigma}{2} \, h_x^2 \right] \, dx \, .
\label{eq_1_H1D}
\end{equation}

\noindent Accordingly, with the Fourier transform

\begin{equation}
h(x) = \frac{1}{\sqrt{L_p}} \sum_{q} h_n \, e^{\icomp \, q x} \, ,
\end{equation}

\noindent where

\begin{equation}
  \sum_q \equiv \sum_{|n| \, \in \, [1,N_\mathrm{max}]} \, ,
\end{equation}

\noindent and $q = 2\pi n/L_p$, $n \in \mathbb{N}^*$, one has

\begin{equation}
\mathcal{H}_\mathrm{1D} = \mathcal{H}_0 + \frac{1}{2} \sum_q \left(\sigma q^2 +
  \kappa q^4 \right)|h(q)|^2 \, ,
\end{equation} 

\noindent where $2 \pi N_\mathrm{max}/L_p = \Lambda \approx 1/a$, where $a$ is a microscopical cut-off.
It follows that $\langle |h(q)|^2\rangle$ is given by the equivalent of eq.(\ref{spect_base}).

In appendix~\ref{annexe2}, we derive the projected stress tensor for a  $1$-d
filament.
There, we show that it has just two components, one tangent to the filament
direction $\Sigma_\parallel^\mathrm{1D}$, developed up to order two on $h$,
and other perpendicular to it $\Sigma_\perp^\mathrm{1D}$, developed up to
first order in $h$,
yielding

\begin{eqnarray}
  \label{eq_1_Sigma1D}
\Sigma^\mathrm{1D}_\parallel &=& \sigma - \frac{\sigma}{2} h_x^2 - \frac{\kappa}{2} h_{xx}^2 +
\kappa \, h_{xxx}h_x \, ,\\
\nonumber\\
\Sigma^\mathrm{1D}_\perp &=& \sigma \, h_x - \kappa \, h_{xxx} \, .
\end{eqnarray}

\noindent Note that these equations are equivalent to $\Sigma_{xx}$ and
$\Sigma_{zx}$ (eqs.(\ref{sigma_xx}) and (\ref{sigma_zx}), respectively) for $h_y = 0$ and $h_{yy} = 0$. 
In order to evaluate $\tau_\mathrm{1D} \equiv \langle \Sigma^\mathrm{1D}_{\parallel} \rangle$, we
introduce the correlation function for $\mathcal{H}_\mathrm{1D}$:

\begin{equation}
G(x - x') = \frac{k_BT}{L_p} \sum_q \frac{e^{\icomp \, q(x - x')}}{\sigma q^2
  + \kappa q^4}\, .
\end{equation}

\noindent We have thus

\begin{eqnarray}
  \label{eq_1_Tau1D}
\tau_\mathrm{1D} &=& \sigma - \frac{\sigma}{2}\langle h_x^2\rangle
-\frac{\kappa}{2}\langle h_{xx}^2 \rangle + \kappa \langle h_x h_{xxx}\rangle \, ,\\
&=& \sigma - \frac{k_\mathrm{B}T}{2 L_p} \sum_q \frac{\sigma + 3 \kappa
  q^2}{\sigma + \kappa q^2} \, ,\nonumber\\
\label{eq_1_passage}
&=& \sigma - \frac{k_\mathrm{B}T}{2} \int \frac{dq}{2\pi}\frac{\sigma + 3
  \kappa q^2}{\sigma + \kappa q^2} \, , \\
&=& \sigma - \frac{3 \, k_\mathrm{B}T \, \Lambda}{2\pi} \left[1 -
  \frac{2}{3\Lambda\xi} \arctan \left(\Lambda \xi\right)\right] \, , 
\label{eq_1_tau_1D}
\end{eqnarray}

\noindent where we have taken the thermodynamic limit in
eq.(\ref{eq_1_passage}) and $\xi = \sqrt{\kappa/\sigma}$. 
For $\xi \ll a$, i.e., for non-extreme tensions, we have simply

\begin{equation}
\tau_{1D} \approx \sigma - \frac{3 \, k_\mathrm{B}T\, \Lambda}{2 \pi} \, .
\label{eq_1_diff_1D}
\end{equation}

\noindent As for a  two-dimensional membrane, the effective tension is smaller
than the tension $\sigma$ and the difference is well approximated by a constant.

Numerically, to evaluate $\tau_\mathrm{1D} = \langle \Sigma_\parallel^\mathrm{1D}
\rangle$, one should evaluate each one of the three averages of
eq.(\ref{eq_1_Tau1D}) ($\langle h_x^2\rangle$, $\langle h_{xx}^2 \rangle$ and
$\langle h_x \, h_{xxx} \rangle$) at the point $x = L_p$.
If however one imposes the filament to remain horizontal at its end, i. e. $h_x|_{L_p} = 0$, eq.(\ref{eq_1_Tau1D}) becomes simply

\begin{equation}
\tau_\mathrm{1D} = \sigma - \frac{\kappa}{2} \, \langle C^2_{L_p} \rangle\, \equiv \langle \sigma_t \rangle ,
\label{eq_1_tau_curv}
\end{equation}

\noindent where $C_{L_P} = h_{xx}|_{L_p}$ is the curvature at the filament's end and where $\sigma_t$ stands for the tangential tension.
Eq.(\ref{eq_1_tau_curv}) is much simpler to check numerically, since one has just to evaluate one average.
We shall thus impose in our simulation $h_x|_{L_p} = 0$ and verify
independently both equations
(\ref{eq_1_tau_1D}) and (\ref{eq_1_tau_curv}).

\subsection{Numerical system and dynamics}
\label{subsection_1_simu_describe}

We considered a discretized version of a $1$-d filament constituted of a chain of $N$ rod-like segments of natural length
$a$, each rod representing a coarse-graining of several lipids.
We assumed in an approximation that all segments had the same length $a(1 +
\epsilon)$ and thus the total length of the system was $L = N\, a(1 +
\epsilon)$.
We wanted a filament with $L = N\, a$.
Imposing $\epsilon = 0$ would not allow us however to measure $\sigma$, which
is a fundamental point of this simulation.
We have thus considered that the chain was connected to a lipid reservoir, so that
$\epsilon$ was free to vary.
In order to fix $\langle \epsilon \rangle = 0$, the conjugated variable $\sigma$ had to be properly adjusted, as in eq.(\ref{eq_1_H1D}).

\begin{figure}[H]
\begin{center}
\includegraphics[scale=.3,angle=0]{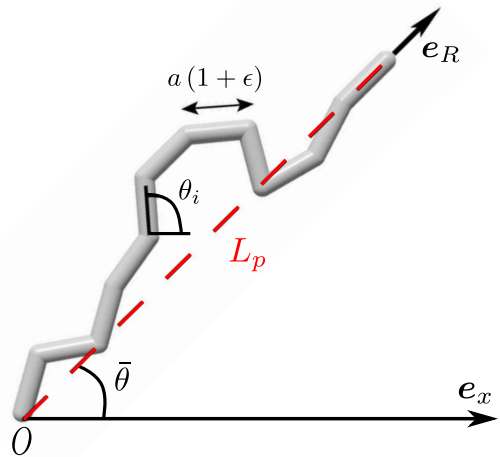}
\caption{Parameters of the numerical experiment. The angle $\bar{\theta}$ is
  the average angle that the chain does with the horizontal axis (see eq.(\ref{eq_1_thetabar})) and
  $\bm{e}_R$ is the end-to-end direction. The projected length, indicated in red
  and denoted $L_p$, is the length of the chain projected on $\bm{e}_R$. Note
  that we impose the last segment to be parallel to $\bm{e}_R$.}
\label{fig_1_simu_1D}
\end{center}      
\end{figure}

Each configuration $\Omega_i$ of the chain was described by the set
$\{\theta_1,...,\theta_{N-1},\epsilon\}$, where $\theta_i$ stands for the angle
that the segment $i$ makes with the horizontal axis (see
Fig.~\ref{fig_1_simu_1D}). 
%In order to verify eq.(\ref{eq_1_tau_curv}),
The last segment was imposed always parallel to the vector $\bm{R}$, defined by

\begin{equation}
\bm{R} = \frac{L}{N}\sum_{i=1}^{i = N-1} \bm{u}_i \, ,
\label{eq_1_R}
\end{equation}

\noindent where $\bm{u}_i$ is the unitary vector in the direction of the $i$-th
segment so that $\tau_\mathrm{1D} = \langle \Sigma_\parallel^\mathrm{1D} \rangle$ can be easily checked through eq.(\ref{eq_1_tau_curv}). 
Associated to the vector $\bm{R}$, we define the average direction $\bm{e}_R
\equiv \bm{R}/R \equiv \bm{e}_\parallel$, where $R \equiv |\bm{R}|$.
Thus we impose $\bm{u}_N = \bm{e}_R$ and

\begin{equation}
\theta_N = \bar{\theta} = \frac{1}{N}\sum_{i = 1}^{i = N} \theta_i \, .
\label{eq_1_thetabar}
\end{equation}

\noindent The projected end-to-end length is given by $L_p \equiv |
\bm{R}| \equiv (L/N) \sum_{i=1}^{N} \cos(\theta_i - \bar{\theta})$ (see Fig.~\ref{fig_1_simu_1D}).

\begin{figure}[H]
\begin{center}
\subfigure[Move $A$: changing the angle of one segment with respect to the
horizontal. The last segment must be adjusted so that its direction is
parallel to the vector $\bm{e}_R$.] {
\includegraphics[scale=.3,angle=0]{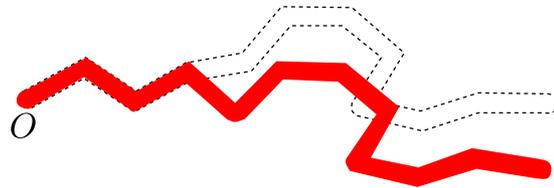}
\label{fig_1_kind_A}
}  
\subfigure[Move $E$: changing the extension $\epsilon$ of segments.]{ 
  \includegraphics[scale=.3,angle=0]{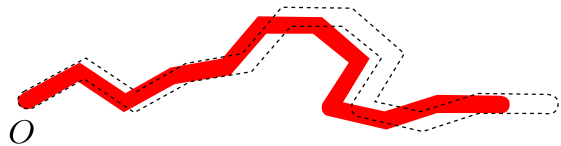}
  \label{fig_1_kind_E1}
}
\caption{Two kinds of movements considered for the chain.}
\end{center}      
\end{figure}

During the simulation, we considered two kinds of moves: 
\begin{enumerate}
\item Move $A$: changing one segments
angle $\theta_i$, which corresponds to the effects of thermal fluctuations on
the chain's shape (see Fig.~\ref{fig_1_kind_A}). 
In this case, the direction $\bm{e}_R$ is changed and 
consequently, the last segment must have its direction corrected;
\item Move $E$: changing the extension of segments through $\epsilon$, which
represents the exchange of lipids with the reservoir (see
Fig.~\ref{fig_1_kind_E1}).
\end{enumerate}

In addition, an external force $\bm{f} = \tau \cdot \bm{e}_R$ always
parallel to the last segment is exerted over the chain.
The chain's free-energy is given by a bending contribution, a contribution
coming from the Lagrange multiplier $\sigma$ plus a contribution from the
external force

\begin{equation}
\mathcal{H}_\mathrm{discret} = \sigma L + \sum_{i = 1}^{N-1}
\frac{1}{2}\, 
\kappa \,a\, (1 + \epsilon)\,  C_i^2 - \tau \, L_p \, ,
\label{eq_1_Hchain}
\end{equation}

\noindent where 

\begin{equation}
C_i = \frac{\theta_{i+1} - \theta_i}{a\, (1+\epsilon)}\, 
\end{equation}

\noindent is an approximation of the curvature between two successive
segments. 
Note that the problem is invariant under rotation around the origin $O$, so
that at any moment we can describe the configuration on Monge's gauge.

The imposed parameters of the simulation were $N$, $\tau$ in units of $\beta \, a$  and
$\kappa$ in units of $\beta$.
For each $\tau$, $\sigma$ was adjusted in order to fix $\langle \epsilon \rangle = 0$, as discussed above.
We detail how it was done in the following.

In the end of section~\ref{subsection_1_theo_1D}, we have argued that in general, one should expect $r \neq \tau \neq \sigma$, where $r$ is the coefficient in $q^2$ of the spectrum fluctuation

\begin{equation}
  \langle |h_n|^2 \rangle = \frac{k_\mathrm{B} T}{r q^2 + \kappa q^4} \, .
  \end{equation}

\begin{figure}[H]
\begin{center}
\includegraphics[scale=.5,angle=0]{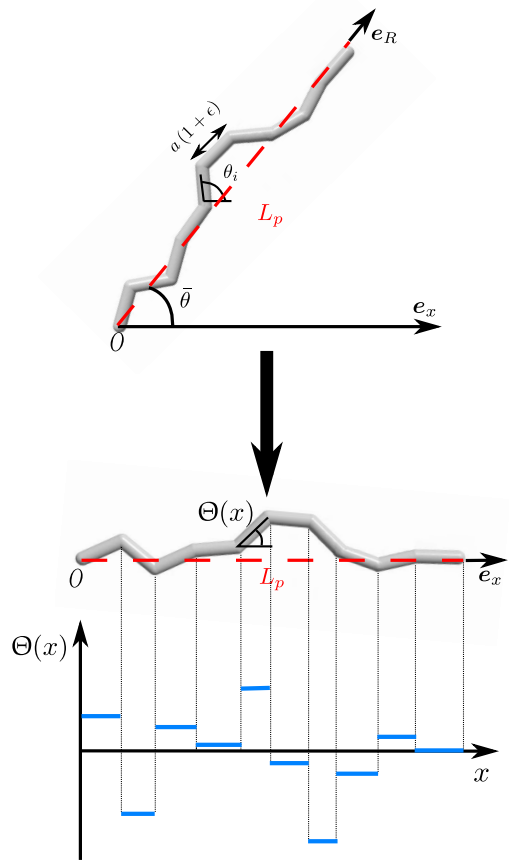}
\caption{Construction of function $\Uptheta(x)$. First, the initial
  configuration is rotated of $-\, \bar{\theta}$. The tension $r$ was
  deduced from the average of the Fourier transform of this curve over a large
  sample of configurations.} 
\label{fig_1_Theta}
\end{center}      
\end{figure}

\noindent In order to measure $\langle |h_n|^2 \rangle$ (and thus $r$), we have first performed a rotation so that we were in the same
situation as in section~\ref{subsection_1_theo_1D}.
We defined $\Uptheta(x) = \theta_i - \bar{\theta}$, where $x$ is the ordinate in the axis $\bm{e}_x$.
The function $\Uptheta(x)$ is a series of steps of length $\cos(\theta_i -
\bar{\theta})$, 
as shown in Fig.~\ref{fig_1_Theta}.
Note that this function is well-defined only when there are no overhangs ($|\theta_i - \bar{\theta}| <
\pi/2$ for all segments), unlike the configuration seen on Fig.~\ref{fig_1_simu_1D}.

\begin{figure}[H]
\begin{center}
\includegraphics[scale=.5,angle=0]{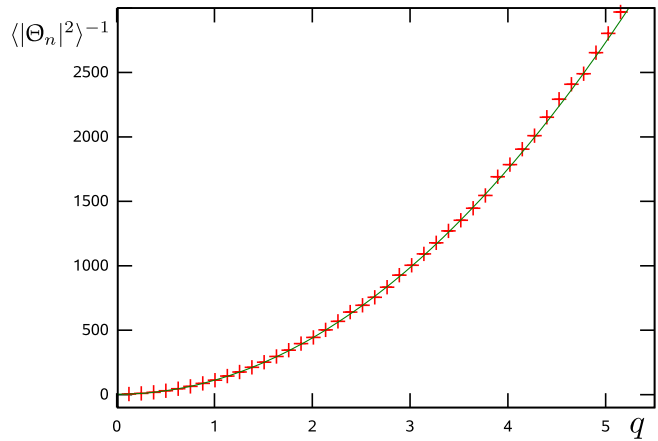}
\caption{Example the fit (solid line) of $\langle |\Uptheta_n|^2 \rangle^{-1}$
  (crosses) with eq.(\ref{eq_1_fit}) for $N = 50$, $\kappa \beta = 125$ and $\tau\beta a = 2$ and $\sigma\beta a
  = 2.45535$. The fit yields $r \beta a = 2.83 \pm 0.03$ and $\kappa \beta= 109.28
  \pm 0.09$. Fit with $\chi^2_\mathrm{red} = 0.82$, indicating a good fit.}
\label{fig_1_fit}
\end{center}      
\end{figure}

With the Fourier transform

\begin{equation}
\Uptheta_n = \int_0^{L_p} \Uptheta(x) \, e^{-\icomp \, \frac{2\pi n x}{L_p}} \, dx
\, ,
\end{equation}

\noindent we expect 

\begin{equation}
\langle |\Uptheta_n|^2\rangle^{-1} = \beta\left[r + \kappa q^2 +
  \mathcal{O}(q^4)\right]\, ,
\label{eq_1_fit}
\end{equation}

\noindent with $q = 2\pi n/\langle L_p \rangle$.
The strategy to obtain $r$ was to average $|\Uptheta_n|^2$ over a large set of
configurations and fit the data with eq.(\ref{eq_1_fit}).
Fig.~\ref{fig_1_fit} shows a representative fit. 

A sum-up of the variables can be seen in table~\ref{table_1_simu_tension}.

\vspace{1cm}

\begin{table}[H]
    \begin{center}
\begin{tabular}{|c|c|c|}
\hline
\bf{{\red Tension}} & \bm{{\red Status}} & \bm{{\red How}} \\ 
\hline
& & \\
$\tau$ & imposed & -- \\
& &  \\
\hline
& &  \\
$\sigma$ & adjusted & $\langle L \rangle = N\, a$\\
& & \\
\hline
& & \\
$r$ & measured & $\langle |\Uptheta_n|^2 \rangle$\\
& & \\
\hline
\end{tabular}
\caption{Sum-up of how we have dealt with tensions in the numerical experiment.}
\label{table_1_simu_tension}
\end{center}
\end{table}

\subsubsection{Numerical dynamics}

We used a Monte Carlo method to generate a sufficiently large sample of chain's
configuration so that we could evaluate averages with a good precision~\cite{Krauth}.
The configurations were generated through a Markov chain algorithm: from a
certain $\Omega_i$, a random move of kind $A$ or $E$ was
proposed, generating a new state $\Omega_{i+1}$. 
In order to respect the detailed balance, the configuration $\Omega_{i+1}$ was
accepted with the probability $P(\Omega_i \rightarrow \Omega_{i+1})$ given by the Metropolis algorithm

\begin{equation}
P(\Omega_i \rightarrow \Omega_{i+1}) =
\min\left[1,\frac{p(\Omega_{i+1})}{p(\Omega_i)}\right] \, .
\end{equation}

In the case of thermodynamic equilibrium, the probability of each
configuration is given by the Boltzmann distribution

\begin{equation}
p(\Omega_i) = \frac{e^{-\beta \, \mathcal{H}_\mathrm{discret}(\Omega_i)}}{\mathcal{Z}} \, ,
\end{equation}

\noindent where $\mathcal{Z}$ is the partition function.
The probability of transition is then simply given by

\begin{equation}
P(\Omega_i \rightarrow \Omega_{i+1}) = \min\left[1, e^{-\beta\,
    \Delta\mathcal{H}}\right]\, ,
\end{equation}

\noindent where $\Delta \mathcal{H} = \mathcal{H}_\mathrm{discret}(\Omega_{i+1}) - \mathcal{H}_\mathrm{discret}(\Omega_i)$.
In our case, we have

\begin{enumerate}
\item move $A$: an angle $\theta_i$ of the set $\{\theta_1, ...,
  \theta_{N-1}\}$ is randomly chosen. We propose a new angle $\theta_i' =
  \theta_i + \Delta\theta$, where $\Delta\theta = \delta_\theta \times
  \,\mathrm{rand}(-1,1)$, with $\mathrm{rand}(a,b)$ a random number with uniform
  distribution of probability between $a$ and $b$. 
  The new $\bar{\theta}'$ is
  evaluated and consequently $\theta_N' = \bar{\theta}'$.
  The variation of free-energy is thus

\begin{eqnarray}
\Delta \mathcal{H}_A &=& \frac{\kappa}{2 \, a(1 + \epsilon)}
\left[\left(\theta_{i+1} - \theta_i'\right)^2 - \left(\theta_{i+1} -
    \theta_i\right)^2 + \left(\theta_i' - \theta_{i-1}\right)^2\right.\nonumber\\  
&-& \left.\left(\theta_i - \theta_{i-1}\right)^2+\left(\theta_N'- \theta_{N-1}\right)^2 - \left(\theta_N -
    \theta_{N-1}\right)^2\right]\nonumber\\
&-& \tau \, \frac{L}{N} \left[\sum_{j=0}^{j = N}\cos\left(\theta_j -
    \bar{\theta}'\right) - \sum_{j=0}^{j = N}\cos\left(\theta_j
    - \bar{\theta}\right)\right] \, .
\end{eqnarray}

If $i = 0$, the third and the fourth terms should not be taken into account.
The value of $\delta_\theta$ is chosen in order to have $\approx 50\%$ of
acceptance of this kind of move.

\item move $E$: a new extension $\epsilon' = \epsilon + \Delta\epsilon$, with
  $\Delta \epsilon = \delta_\epsilon \times\, \mathrm{rand}(-1,1)$ is proposed.
The free-energy variation reads

\begin{eqnarray}
\Delta \mathcal{H}_E &=& \sigma \, a\, \Delta \epsilon - \tau \, a\, \Delta
\epsilon \sum_{j =0}^{j = N}\cos\left(\theta_j -
  \bar{\theta}\right)\nonumber\\
&+& \frac{\kappa \, \Delta\epsilon}{2\, a(1+\epsilon)(1 +
  \epsilon')}\sum_{j=0}^{j=N-1}\left(\theta_{j+1} - \theta_j\right)^2\, .
\end{eqnarray}

Again, $\delta_\epsilon$ was chosen in order to have $\approx 50\%$ of
acceptance for moves of kind $E$.

\end{enumerate} 

To each move of kind $E$, $N-1$ moves of kind $A$ were tried in order to
assure that in average every degree of freedom is equally modified.
In the following, we will call this sequence a Monte Carlo step.

\subsubsection{Equilibration criterion}

In order to obtain meaningful averages, we had to be sure that our numerical experiment reached equilibrium.
Usually, it is enough to examine the number of Monte Carlo steps needed to decorrelate the longest modes on the Fourier space, which are the slowest to relax,  and then choose a number of steps much larger for the simulation\cite{Krauth}.
As our system is really simple (in the sense that the energy do not have several local minima) and that we have not chosen too long chains, 
we have chosen two criterion that together are stronger than the relaxation of the longest modes.
First, for the equilibration of angles, we have required the average of $\bm{R}$, given in eq.(\ref{eq_1_R}), to be $\sim 0$.
One could imagine the case of a rotating fixed configuration, which would also yield $\langle \bm{R} \rangle \sim 0$. 
To exclude this improbable situation, we have visually checked a set of configurations.
Secondly, to study the equilibration of the extension $\epsilon$, we have examined the evolution of $\langle \epsilon \rangle$ over time: when it reached a plateau, we considered the system at equilibrium.

In our experiments, we have taken $N=50$ and a larger $\beta \kappa = 125$ to
assure that the chain departs weakly from a straight line.
For typical values ranging from $\tau = -0.2\, \beta a$ up to $\tau = 5\,
\beta a$, the equilibration was attained after $5 \times 10^{5}$ steps.
Currently, we have made $8 \times 10^{6}$ steps to be sure that the sampled
configurations had an equilibrium distribution. 
%so that the we had a plateau for $\langle \epsilon \rangle$ (with $\sqrt{\langle
 % \epsilon^2 \rangle - \langle \epsilon \rangle^2} \ll 0.001$ stable over time) and $\langle \bm{R} \rangle$ was very small. 
At the end of each step, we have calculated $\Uptheta(x)$ (when there were no overhangs).

\subsubsection{Adjusting \bm{$\sigma$}}

\begin{figure}[H]
\begin{center}
\includegraphics[scale=.4,angle=0]{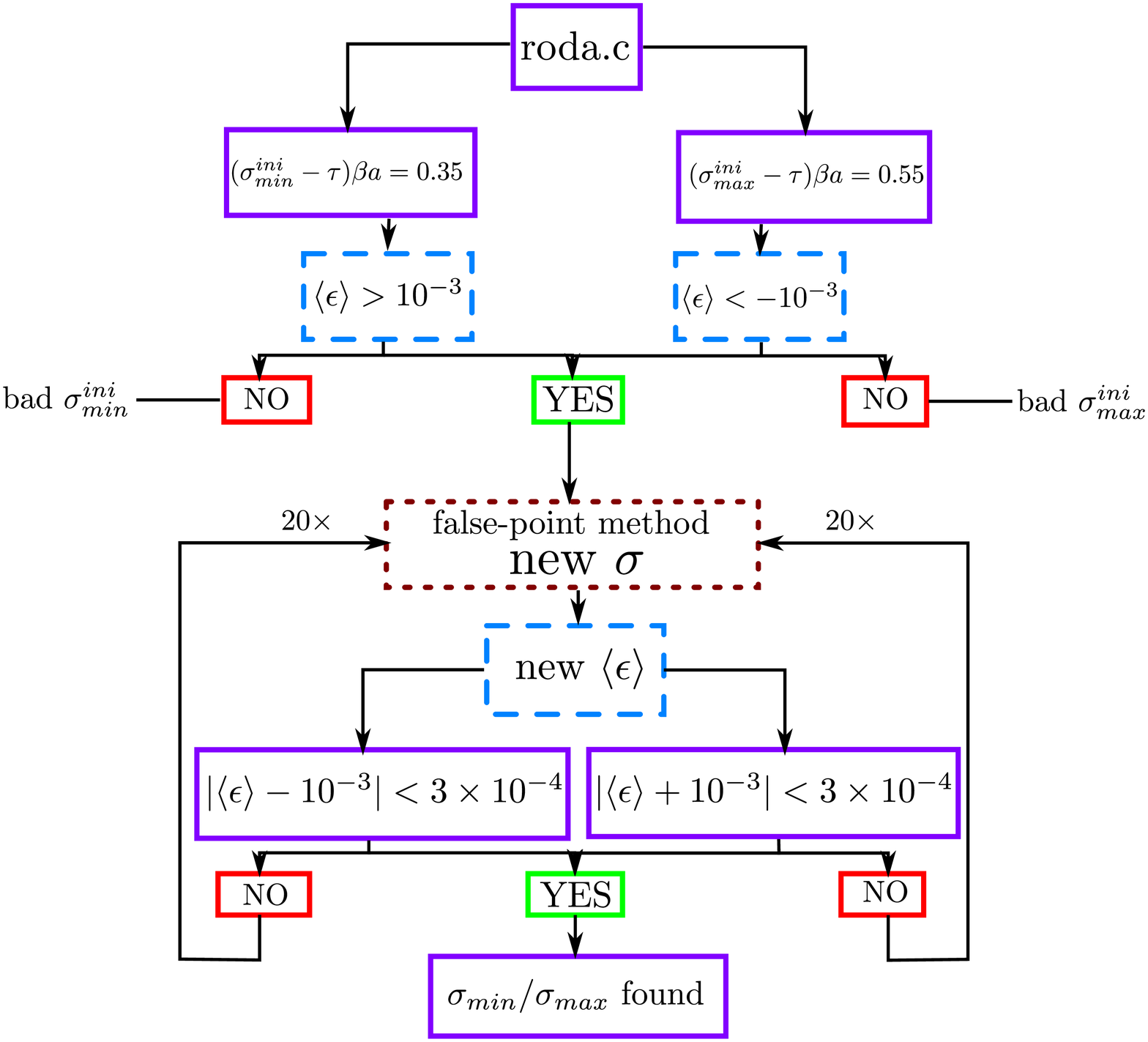}
\caption{Fluxogram of the adjustment procedure. The processes taking place in
  the main function roda.c are inside violet boxes (solid lines). Each blue box (dashed lines) indicates 
  a run of Monte Carlo of $8 \times 10^{6}$ steps. The brown box (dotted lines) indicates
  that the
  false-point algorithm was used to suggest a new $\sigma$. It is called at most $20$ times for each boundary
  (usually, less than $5$ times were enough).} 
\label{fig_1_flux}
\end{center}      
\end{figure}

As we applied $\tau$ to the membrane, we had to adjust $\sigma$ in order to
have $\langle \epsilon \rangle \sim 0$.
To estimate also the uncertainty of $\sigma$, for each
pair $\kappa, \tau$ we determined $\sigma_\mathrm{min}$ corresponding to $\epsilon_\mathrm{max}= \langle
\epsilon \rangle = 10^{-3}$ and $\sigma_\mathrm{max}$ corresponding to
$\epsilon_\mathrm{min} = \langle
\epsilon \rangle = - 10^{-3}$.
To do so, we inspired ourselves on eq.(\ref{eq_1_diff_1D}) 

\begin{equation}
\left(\sigma -
\tau_\mathrm{1D} \right)\beta a = \frac{3}{2\pi} \simeq 0.5\, ,
\end{equation}
 
\noindent and started with two guesses $(\sigma_\mathrm{min}
- \tau)
\beta a = 0.35$ and $(\sigma_\mathrm{max} - \tau)\beta a = 0.55$.
The approach of the boundaries $\epsilon_\mathrm{max}$ and $\epsilon_\mathrm{min}$ was made
through a false-point algorithm in $20$ iterations at maximum~\cite{Numerical_Recipes}. 
The stop criterion for $\sigma_\mathrm{min}$ was $|\langle \epsilon \rangle - \epsilon_\mathrm{min}| < 3\times 10^{-4}$ and equivalently $|\langle \epsilon \rangle - \epsilon_\mathrm{max}| < 3\times 10^{-4}$ for $\sigma_\mathrm{max}$.
A fluxogram of the adjustment procedure is presented in Fig.~\ref{fig_1_flux}.
For the typical values presented in the last section, the difference $\sigma_\mathrm{max} - \sigma_\mathrm{min}$ was systematically of $\approx 0.001\,
\beta a$.

\subsubsection{Buckling transition}

\begin{figure}[H]
\begin{center}
\subfigure[Buckled configuration for $\tau = -0.3\, \beta a$.]{ 
  \includegraphics[scale=.4,angle=0]{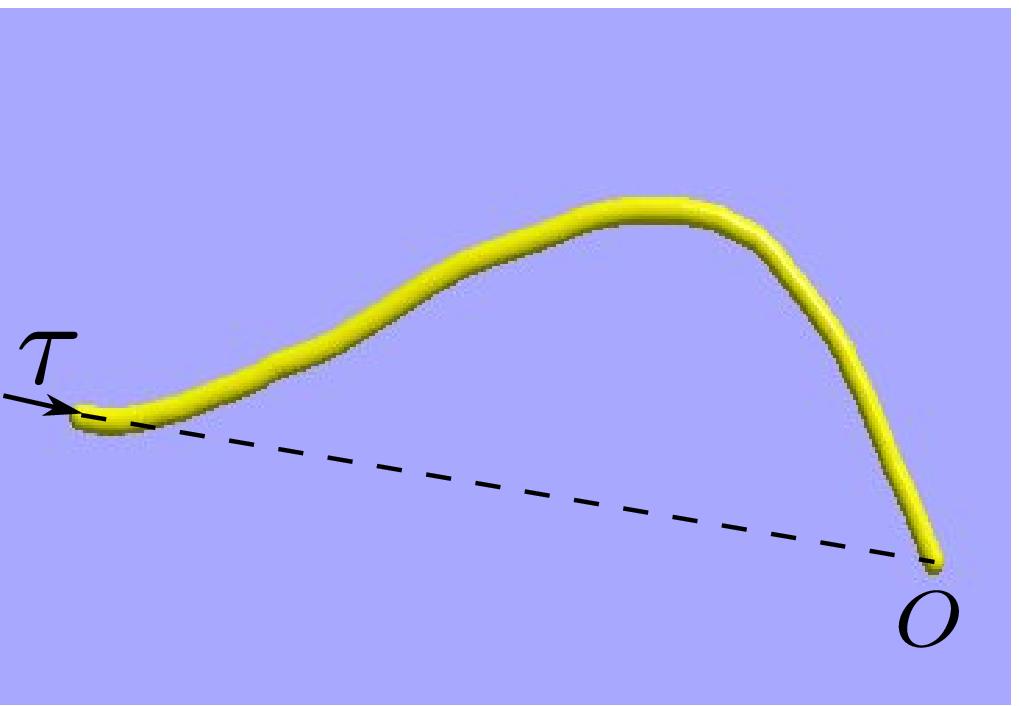}
}
\subfigure[Non-buckled configuration for $\tau = 0.5\, \beta a$.] {
\includegraphics[scale=.4,angle=0]{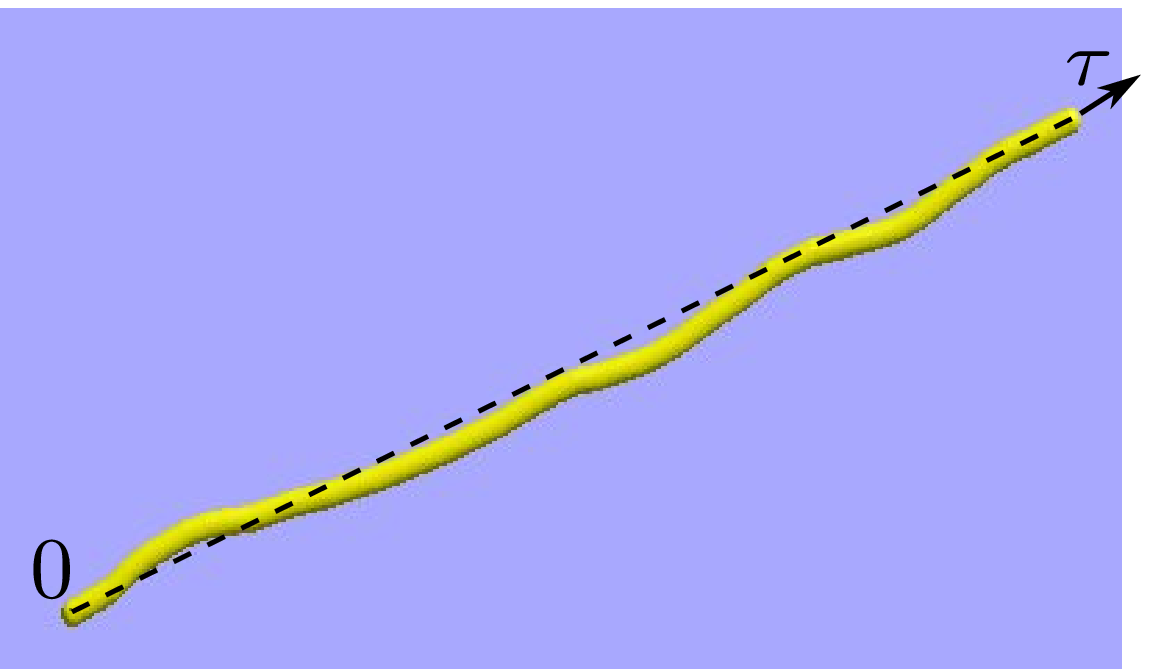}
}  
\caption{Buckling transition.}
\label{fig_1_buckled}
\end{center}      
\end{figure}

In order to verify the correctness of our simulation, we have also applied negative tensions to the filament.
Indeed, for compressive tensions bigger than a certain limit, we expect our filament to fluctuate around a curved line, instead than around a straight line, as shown in Fig.~\ref{fig_1_buckled}.
This transition is known as the buckling transition.

\begin{figure}[H]
\begin{center}
\includegraphics[scale=.45,angle=0]{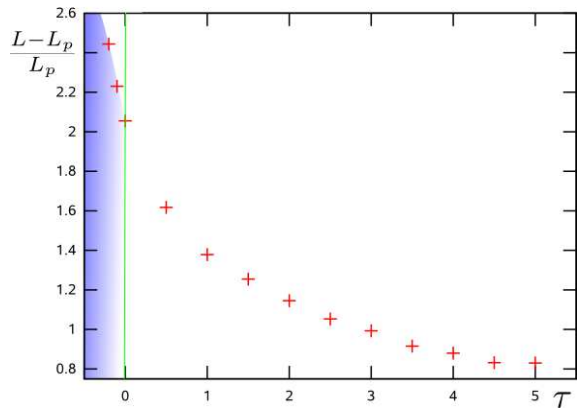}
\caption{Percent length excess as a function of $\tau$. The shaded region on
the left represents the approach of the buckling transition.}
\label{fig_1_alpha_1D}
\end{center}      
\end{figure}

This transition is also characterized by an increase and a discontinuity 
on the length excess, the
equivalent of $\alpha$, defined as 

\begin{equation}
\alpha_\mathrm{1D} = \frac{L - \langle L_p \rangle}{\langle L_p \rangle} \, ,
\end{equation}

\noindent as the compressive tension increases.
In Fig.~\ref{fig_1_alpha_1D} we can see that we have effectively a relatively
abrupt increase of $\alpha_\mathrm{1D}$ as the tension approaches $\tau = -0.3\beta a$.
We have thus some evidence of a buckling transition for negative tensions.
We have not however made a systematic study of the transition, since it was not the aim our numerical experiment.
Moreover, we have found some technical problems for $\tau \leq -0.3 \beta a$,
since the projected length was highly fluctuating due to the alternating
presence of buckled and straight configurations near the transition.
The averages varied a lot and thus, with the parameters presented in the
last section, the false-point method failed to converge in $20$ iterations.
Therefore, we have just considered $\tau \geqslant -0.2 \beta a$, situation in which we are sure that we had small fluctuations around a straight line.  

\subsection{Results}
\label{subsection_1_results}

\begin{figure}[H]
\begin{center}
\includegraphics[scale=.5,angle=0]{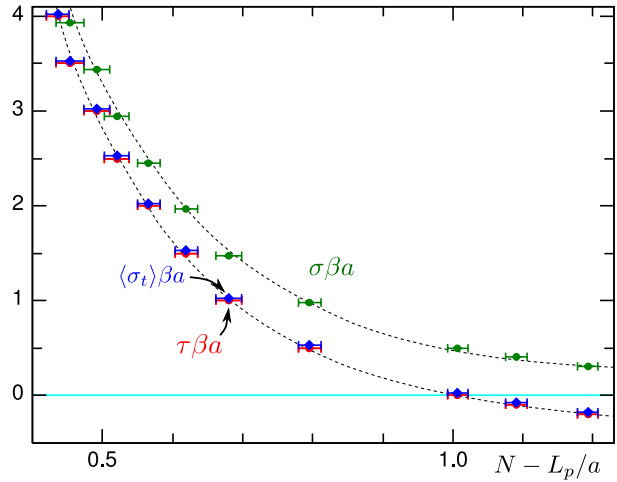}
\caption{Tensions $\tau$ (diamonds) and $\sigma$ (circles on top) as a function of the shortening of
the projected length. Note that as predicted, the average of the tangential
  stress tensor at the extremity $\langle \sigma_t \rangle$ (circles superposed with the diamonds) coincides with
  $\tau$. The dashed lines are just guides to the eye.} 
\label{fig_1_diff_prl}
\end{center}      
\end{figure}

Once $\sigma_\mathrm{min}$ and $\sigma_\mathrm{max}$ were found, we have performed
the numerical experiment three times with each value. 
We have taken  the average of these three runs
for the spectrum and for the average curvature of the last segment.  
For each averaged spectrum, we have made a fit using gnuplot to obtain
$r$ (see Fig.~\ref{fig_1_fit} for an example).
We obtained $\kappa \beta \approx 110$ in all fits, which we remind is a bit
different from the microscopical $\kappa \beta = 125$.
This result is coherent with what one should expect given
eq.(\ref{kappa_effect}), which takes into account corrections due to the renormalization.
The final results are summed up in table~\ref{table_1_resfin}.
In this table, $\sigma$ and $r$ are the averages of $\sigma_\mathrm{max}$ and
$\sigma_\mathrm{min}$ and the corresponding $r$.
From table~\ref{table_1_resfin}, it is evident that $\tau \neq \sigma$ (see also Fig.~\ref{fig_1_diff_prl}).
In the same graphic, we have also plotted $\langle \sigma_t \rangle = \sigma -
\kappa\langle C_{L_p}^2 \rangle/2$. 
As predicted, we have indeed $\langle \sigma_t \rangle = \tau$.

\begin{table}[H]
 \begin{center}
\begin{tabular}{|c|c|c|c|c|}
\hline
\bm{{\red $\tau$}} & \bm{{\red $\sigma$}} & \bm{{\red $r$}} & \bm{{\red
    $\kappa/2 \langle C_{L_p}^2\rangle$}}&\bm{{\red $\langle L_p \rangle$}}\\ 
\hline
{-0.2}& 0.31&0.01&0.48&48.81\\
{-0.1}& 0.40&0.17&0.47&48.91\\
{0}& 0.50&0.32&0.47&48.99\\
{0.5}& 0.98&0.97&0.46&49.20\\
{1}& 1.47&1.63&0.45&49.32\\
{1.5}& 1.96&2.24&0.44&49.38\\
{2}& 2.46&2.81&0.43&49.43\\
{2.5}& 2.95&3.45&0.42&49.48\\
{3}& 3.44&4.11&0.42&49.51\\
{3.5}& 3.94&4.53&0.41&49.55\\
{4}& 4.43&5.19&0.41&49.56\\
{4.5}& 4.93&5.71&0.41&49.59\\
{5}& 5.42&6.16&0.4&49.59\\
\hline
\end{tabular}
\caption{Sum-up of the results obtained from our numerical experiment for $N =
  50$ and $\beta \kappa = 125$. The
  values of the four first columns are in units of $\beta a$, while the last
  column is in units of $a$. The
   errors on the second and third column are of $\approx 0.001$ and
$\approx 0.02$ respectively. The fourth column is the average of the curvature
energy of the last segment.}
\label{table_1_resfin}
\end{center}
\end{table}

\begin{figure}[H]
\begin{center}
\includegraphics[scale=.5,angle=0]{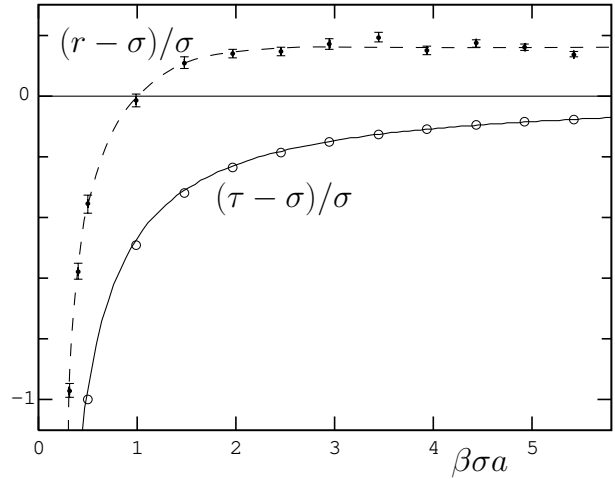}
\caption{Lower data: comparison between $\tau$ and $\sigma$ (error bars given
  by the symbol size); the solid line is a fit using eq.(\ref{eq_1_tau_1D}).
Upper data: comparison between $r$ and $\sigma$; the dashed is only a guide
for the eye.}
\label{fig_1_fit_prl}
\end{center}      
\end{figure}

Quantitatively, the fit with the theoretical equation for the difference
$\sigma - \tau$ given in eq.(\ref{eq_1_tau_1D}) can be seen in
Fig.~\ref{fig_1_fit_prl} (solid line).
The agreement is excellent for $\Lambda = 1.1 \, a^{-1}$ (one parameter fit),
which is a very reasonable value for the cut-off.
In this graphics, we have also plotted the percent difference between
$\sigma$ and $r$. 
The behavior is non-trivial: the sign of $\sigma - r$ changes at low tensions
and we have $\sigma \neq r$ even at high tensions.

To conclude, with this simple numerical experiment, we could accede to
the tension $\sigma$ needed to fix the length of the filament, which one
cannot usually measure in true experiments, simultaneously to the tension $\tau$ and $r$.
Our data corroborates the prediction that $\tau \neq \sigma$ and verify eq.(\ref{eq_1_tau_curv}).
In addition, the difference between $\tau$ and $\sigma$ was well fitted by eq.(\ref{eq_1_Tau1D}), giving some support to our theory.
Regarding $r$, as discussed in section~\ref{subsubsection_1_r}, we expected in general $\tau \neq \sigma \neq r$, which our data seems to confirm.
The way in which $r$ depends on $\sigma$ seems to be non-trivial (see Fig.~\ref{fig_1_fit_prl}) and further studies have to be done in order to understand it.

\section{Some experimental implications}
\label{subsection_1_discuss}

In this section, we discuss the implications of eq.(\ref{eq_1_diff}) to
micropipette experiments.
Indeed, in these experiments, it is usually assumed that $\sigma \approx
\tau$, which, as we have seen, is not justified in the limit of low tensions.
We will assume that we are dealing with very large GUV and that the difference
of pressure between the inside and the outside of the vesicle is very small, so that
the membrane is locally equivalent to a flat membrane.
A more detailed derivation for quasi-spherical vesicles of any size taking
into account the
pressure difference will be done in
chapter~\ref{chapitre_vesicle}.

In the limit of small fluctuations, the excess area is given by

\begin{eqnarray}
\alpha = \frac{\langle A\rangle - A_p}{A_p} &\simeq&  \frac{1}{2} \, \langle (\nabla
h)^2 \rangle\, ,\nonumber \\
&\simeq& \frac{k_\mathrm{B} T}{2 A_p} \sum_{\bm{q}} \frac{1}{\sigma + \kappa
  q^2} \, ,
\label{eq_1_alpha_quase}
\end{eqnarray}

\noindent where we have used the correlation function given in eq.(\ref{eq_1_correl}).
In the thermodynamic limit, we have

\begin{eqnarray}
\label{eq_1_alpha11}
\alpha &\simeq& \frac{k_\mathrm{B}T}{8\pi\kappa} \ln\left(\frac{\sigma +
    \sigma_r}{\sigma + \kappa \frac{(2\pi)^2}{A_p}}\right) \, ,\\
&\simeq&\frac{k_\mathrm{B}
  T}{8\pi\kappa}\ln\left(\frac{\sigma_r}{\sigma}\right) \, ,
\end{eqnarray}

\noindent where the last approximation is valid in the limit $\kappa/A_p \ll
\sigma \lesssim 10^{-2} \, \sigma_r$.
\noindent Using the fact that in this limit $\sigma \simeq \tau + \sigma_0$, we have finally

\begin{equation}
\alpha \simeq \frac{k_\mathrm{B}
  T}{8\pi\kappa}\ln\left(\frac{\sigma_r}{\sigma_0 + \tau}\right)\, .
\label{eq_1_alpha}
\end{equation}

In micropipette experiments, one measures the percent difference of projected area
between the initial configuration $A_p^i$ and the final configuration $A_p^f$:

\begin{eqnarray}
\frac{A_p^f - A_p^i}{A_p^i} &=& \left(\frac{\langle A \rangle -
    A_p^i}{A_p^i}\right) - \left(\frac{\langle A \rangle -
    A_p^f}{A_p^i}\right)\, ,\nonumber \\
&\simeq& \alpha_i - \alpha_f\, \nonumber\\
\label{eq_1_alpha_ancient}
&\simeq&
\frac{k_\mathrm{B}T}{8\pi\kappa}\ln\left(\frac{\sigma_f}{\sigma_i}\right)\,,
\\
&\simeq&\frac{k_\mathrm{B}T}{8\pi\kappa}\ln\left(\frac{\sigma_0 +
    \tau_f}{\sigma_0 + \tau_i}\right)\,.
\label{eq_1_alpha_new}
\end{eqnarray}

\noindent Eq.(\ref{eq_1_alpha_ancient}) corresponds to the result presented in
section~\ref{subsection_0_micro} and usually used to deduce $\kappa$ by
considering $\tau \approx \sigma$. 
In eq.(\ref{eq_1_alpha_new}), we see the explicit relation as a function of
$\tau$.

\begin{figure}[H]
\begin{center}
\includegraphics[scale=.65,angle=0]{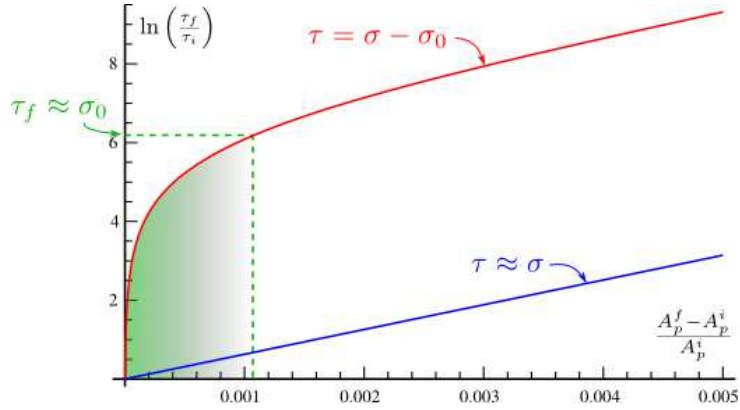}
\caption{Theoretical relation linking the tension to the difference of the
  projected area: in red (upper curve), we see the curve where the distinction between $\tau$
and $\sigma$ is taken in account (eq.(\ref{eq_1_alpha_new})), while in blue (lower straight line) we
see the usual law, supposing that $\tau = \sigma$
(eq.(\ref{eq_1_alpha_ancient})). Numerical values: $k_\mathrm{B}T = 4 \times
10^{-21} \, \mathrm{J}$, $\kappa = 10^{-19} \, \mathrm{J}$, $a = 5 \,
\mathrm{nm}$. We considered a vesicle under initial tension $\tau^i = 10^{-8}
\, \mathrm{N/m}$. The tension is increased up to $\tau = 10^{-4} \,
\mathrm{N/m}$. Note that for $\tau < \sigma_0$, we have no more a linear behavior.}
\label{fig_1_alpha}
\end{center}      
\end{figure}

In Fig.~\ref{fig_1_alpha}, we imagine a typical micropipette experiment with a
vesicle initially under very small tension $\tau^i = 10^{-8} \, \mathrm{N/m}$
and $\kappa = 25 \, k_\mathrm{B}T$. 
We increase the tension up to $10^{-4} \, \mathrm{N/m}$ by aspiring the
vesicle.
The curves in Fig.~\ref{fig_1_alpha} represent the expected relation between the logarithm
of $\tau^f/\tau^i$ and the percent of increase in the projected area for
$\tau = \sigma$ and for $\tau = \sigma - \sigma_0$. 

First of all, note that the percent increase of the projected area is very
small (less than $0.5 \%$), which corresponds to the validity range of our
results (no stretching).
For $\tau > \sigma_0$, we have a linear dependence of the logarithm of
$\tau$ on the percent projected area with roughly the same slope whether
one takes into account the difference between $\tau$ and $\sigma$
(eq.(\ref{eq_1_alpha_new})) or not (eq.(\ref{eq_1_alpha_ancient})).
Thus, it is justified to deduce $\kappa$ by fitting a straight line to data on
this region, as it is usually experimentally done (see section~\ref{chapitre_vesicle}, Fig.~\ref{dimova}.(a)).
We predict however a different behavior for small tensions ($\tau <
\sigma_0$): as we can see in Fig.~\ref{fig_1_alpha}, in the shaded area, we do
not have a linear relation between the logarithm of $\tau$ and the area
excess.
Sadly, we cannot identify this behavior in the data of Fig.~\ref{dimova}, but our prediction can be tested by further experiments using vesicles under
small tension.

\subsection{Natural excess area}
\label{subsection_1_nat_excess_area}

Another related consequence concerns the natural excess area, i. e., the measure of the fluctuations of a membrane
under no external force ($\tau = 0$).
Using eq.(\ref{eq_1_alpha}), we have

\begin{equation}
\alpha_\mathrm{eq} \simeq \frac{\ln(8\pi\beta\kappa)}{8\pi\beta\kappa} \, ,
\label{eq_1_alpha_eq}
\end{equation}

\noindent which yields $\alpha_\mathrm{eq} \simeq 0.03, 0.01, 0.005$ for $\beta
\kappa = 5,25,50$, respectively.

Traditionally, however, one makes $\sigma = 0$ in eq.(\ref{eq_1_alpha_quase}),
which leads to 

\begin{equation}
\alpha_\mathrm{eq}^\mathrm{trad} \simeq \frac{1}{4 \pi
  \beta\kappa}\ln\left(\frac{\Lambda\sqrt{A_p}}{2\pi}\right) \, .
\label{eq_1_alpha_eq_trad}
\end{equation}

\noindent The main difference between these equations is the dependence in terms
of the projected area $A_p$, since for the last equation one expects an
explicit logarithmic dependence.
Eq.(\ref{eq_1_alpha_eq}) presents also a hidden dependence on $A_p$ through
eq.(\ref{kappa_effect}) due to renormalization, although it should be a far weaker dependence.
This result is well suited for numerical verifications.

\section[First evidences that $\tau \neq \sigma$]{First evidences that \bm{$ \tau \neq \sigma$}}
\label{section_1_evidences}

Here we present the first strong numerical and experimental evidences of the
correctness of our results.
In the first part, we present the results of recent numerical experiments far more complex
than the one presented in section~\ref{subsection_1_simu_1D}.
In the second part, we discuss experiments on the adhesion of vesicles to
solid substrate.
We begin by mentioning a previous puzzling result by R\"adler et
al.~\cite{Raedler_95}, already introduced in section~\ref{subsection_adhesion}.
We report the attempts to understand this result made by
Seifert~\cite{Seifert_95b}.
Finally, we describe a recent experiment that seems to corroborate our previsions~\cite{Sengupta_10}.
%The reader familiar with this subject may skip this first section.

\subsection{Numerical experiments}

\begin{figure}[H]
\begin{center}
\includegraphics[width=.25\columnwidth]{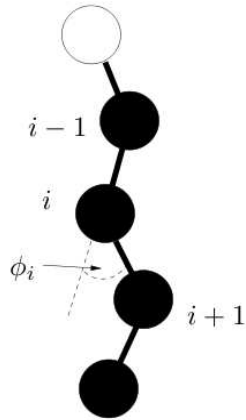}
\caption{Coarse-grained amphiphilic molecule used in the numerical experiment proposed in ref.~\cite{Imparato_06}. The black beads represent the molecule's hydrophobic tail, while the hydrophilic head is represented by the white bead.}
\label{fig_1_amp}
\end{center}      
\end{figure}

In the same ref.~\cite{Imparato_06}, discussed in
section~\ref{subsection_1_free_energy}, a $2$-d numerical experiment was
proposed to check the author's predictions.
The numerical system consisted on coarse-grained amphiphilic lipids represented by chains of
beads (see Fig.~\ref{fig_1_amp}).
The black beads represent the hydrophobic tail, while the white one
stands for the hydrophilic head.
In addition, single beads stood for water molecules.

The total energy was composed by four terms:

\begin{enumerate}
\item the hydrophobic interaction between the beads of the tail and
  water/hydrophilic head;
\item an attractive interaction between two molecules, given by a
  Lennard-Jones potential;
\item a harmonic potential between beads along a single molecule;
\item a three-body bending potential that models the effects of hydrocarbon
  chain stiffness.
\end{enumerate} 

Both molecules of lipids and water were free to move inside a fixed cuboidal box with periodic boundary conditions.
At each realization of the simulation, the size of the box could be changed, implying a change in the membranes tension $\tau$.
The dynamics alternated sequences of Monte Carlo steps with sequences of molecular dynamics steps.

\begin{figure}[H]
\begin{center}
  \vspace{-2cm}
\includegraphics[width=.7\columnwidth]{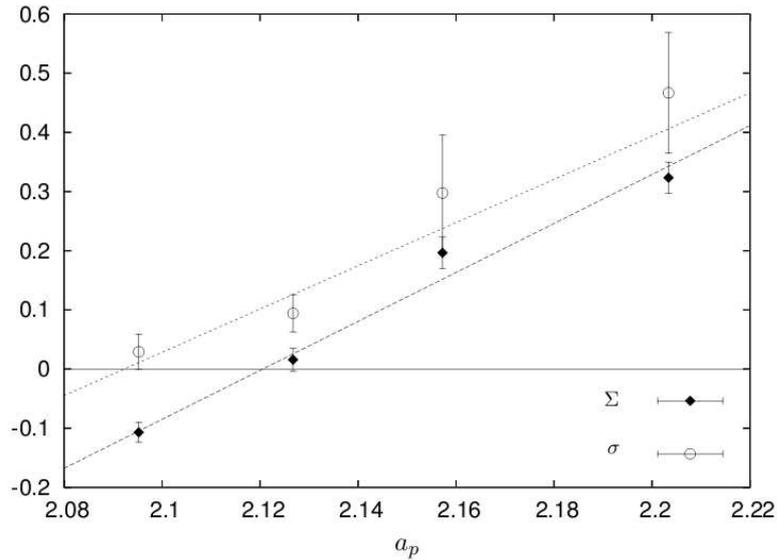}
\caption{Plot of $r$ (indicated by a circle) and $\tau$ (black circle) as a function of the projected area per molecule $a_p$ obtained in ref.~\cite{Imparato_06}. As $a_p$ increases, the fluctuations are flattened and the tension $\tau$ increases. The tension is displayed in units of $\epsilon/\ell^2$ and $a_p$ in units of $\ell^2$, with $\epsilon = 1/3 \times 10^{-20} \, \mathrm{J}$ and $\ell = 1/3 \mathrm{nm}$. }
\label{fig_1_data}
\end{center}      
\end{figure}

In order to measure $\tau$, the forces exchanged through imaginary cuts perpendicular to the membrane plane were averaged for different box sizes.
As in real experiments, the tension $r$ was measured through the fluctuation
spectrum.
Similarly with our simple $1$-d simulation, the buckling transition was observed for high compressive tensions. 
For the non-buckled regime, the results for a simulation involving $1152$ amphiphilic molecules and $7200$ water molecules can be seen in Fig.~\ref{fig_1_data}.
In agreement with the discussion of section~\ref{subsection_1_free_energy}, the author obtained indeed $\tau \neq r$.
Moreover, negative tensions are observed for non-buckled membranes, as in our case.

\begin{figure}[H]
\begin{center}
\includegraphics[width=.5\columnwidth]{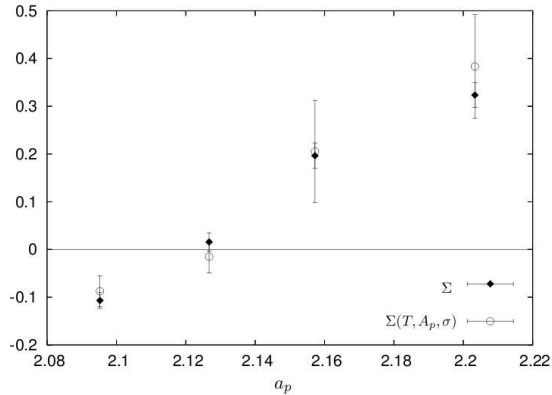}
\caption{The black squares indicate the directly measured tension $\tau$ in ref.~\cite{Imparato_06}. The circles represent the fitted values using eq.(\ref{eq_1_diff_complet}) and assuming $r \approx \sigma$. The units are the same as in the last figure.}
\label{fig_1_Imp2}
\end{center}      
\end{figure}

In this work, the relation given in eq.(\ref{eq_1_diff_complet}) was obtained by differentiating the free-energy with respect to the projected area.
Assuming that $r \approx \sigma$, as usually done in laboratory experiments, the author fitted eq.(\ref{eq_1_diff_complet}) to $\tau$ by adjusting one parameter related to the upper wave-length cutoff.
As we an see in Fig.~\ref{fig_1_Imp2}, the agreement is very good, supporting the predicted relation between $\tau$ and $\sigma$ given in eq.(\ref{eq_1_diff_complet}).

\begin{figure}[H]
\begin{center}
\includegraphics[width=0.7\columnwidth]{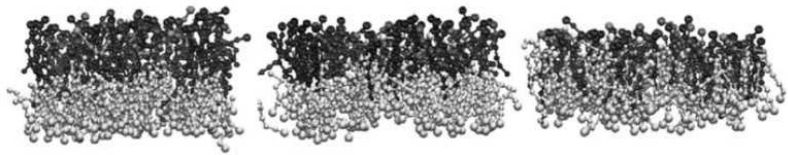}
\caption{Snapshots of bilayer configurations in the liquid phase~\cite{Neder_10}. The dark gray molecules point upward from tail to head while the light gray point downward. In the first snapshot at left, the membrane is tensionless. In the following two snapshots, the tension is increased ($0.01 \, \mathrm{J/m^2}$ and $0.02 \, \mathrm{J/m^2}$, respectively). Remark the interdigitations in the configuration of highest tension due to the stretching of the membrane.}
\label{fig_1_neder}
\end{center}      
\end{figure}

Very recently, a similar simulation was performed by Neder et al.~\cite{Neder_10}.
They have also used coarse-grained amphiphilic molecules similar to the one shown in Fig.~\ref{fig_1_amp} and the energy contributions were roughly the same as in~\cite{Imparato_06}, added of a term $- \tau \, A_p$.
Thus, the main difference in this simulation is the fact that $\tau$ is imposed (and not measured) and the box size was free to change.
In other words, the simulation was performed at $\tau$ and $N_p$, the number of lipids, fixed.
The advantage of this method is the possibility of controlling directly $\tau$, while in the method used in~\cite{Imparato_06}, the tension was imposed by the size of the box.
The configurations were generated through a Monte Carlo algorithm, since only static measures were done.
Different phases of the membrane were observed, depending on the temperature of the system.
In particular, we can see some snapshots for the liquid phase in Fig.~\ref{fig_1_neder}.

As before, $r$ was measured through the fluctuation spectrum for the tensions above mentioned (results presented in table $III$ of~\cite{Neder_10}).
For the tensionless state, they obtained $r = (0.11 \pm 0.19)\times 10^{-4}\, \mathrm{J/m^2}$.
This result seems to agree with our prediction that $r$ should be bigger than $\tau$, even though one should be cautious given the large error-bars.
For the systems under higher tension, however, the trend was inverted.
This fact does not contradict our predictions, since stretching was not taken into account in our theory.
Indeed, by measuring an overlap parameter, as well as the nematic order parameter for the liquid phase, the authors confirmed that stretching takes place for $\tau \gtrsim 0.01 \, \mathrm{J/m^2}$.
Further simulations in the regime of low tension should be useful to compare with our predictions.

\subsection{Adhesion experiments: a puzzling result}

\begin{figure}[H]
\begin{center}
\subfigure[High tension (the bar corresponds to $10 \, \mathrm{\mu m}$).]{
\includegraphics[scale=.35,angle=0]{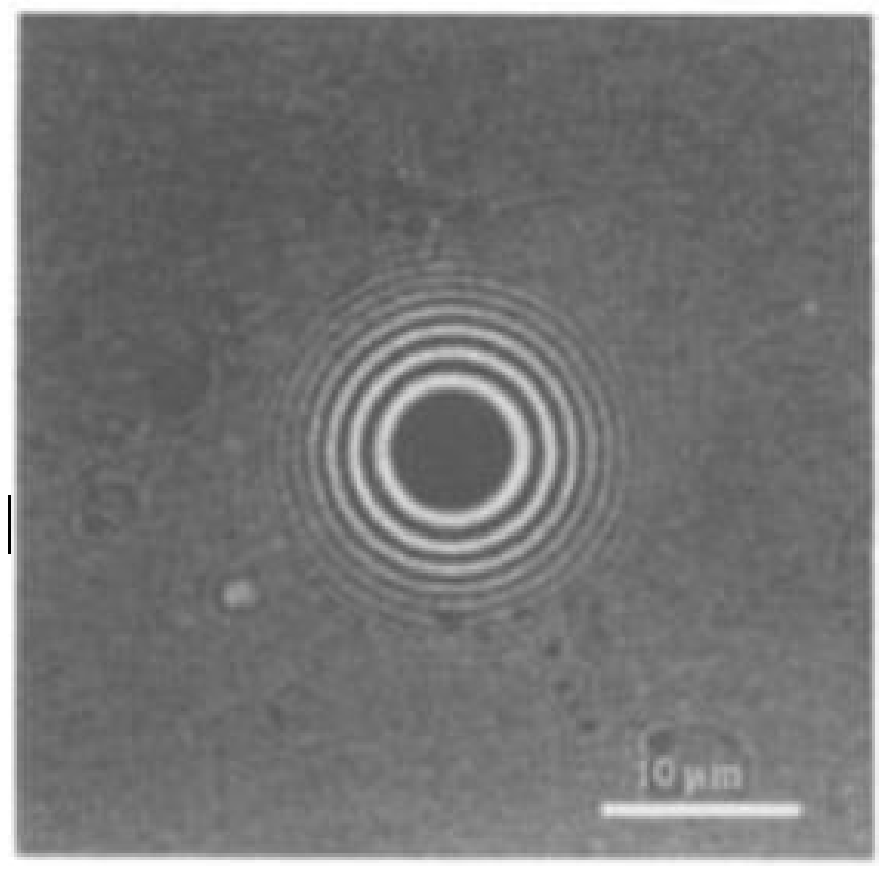}
}
\subfigure[Small tension]{
\includegraphics[scale=.5,angle=0]{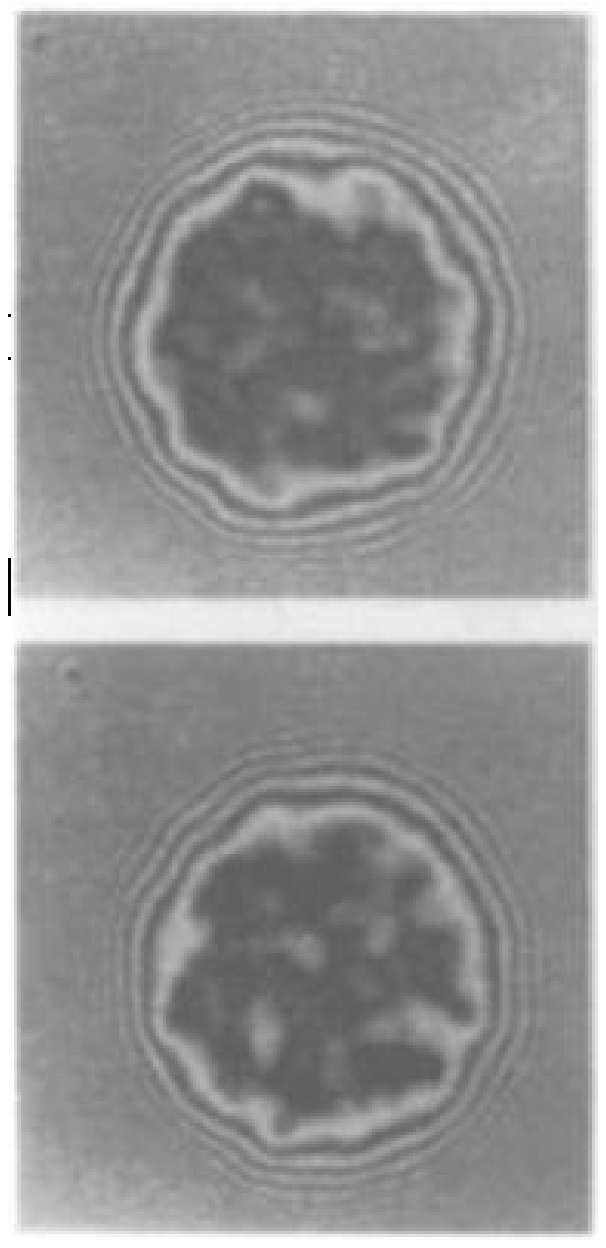}
}
\caption{RICM micrographs of adhering vesicles under different tensions~\cite{Raedler_95}. The height profile is obtained by measuring the intensity of light through a line that passes by the center of the contact region. The fringes at the edge of the round area indicate the end of the contact region.}
\label{fig_1_adhes_exp}
\end{center}      
\end{figure}

Here we will comment on some experiments involving the adhesion of vesicles to solid flat substrates, discussed in section~\ref{subsection_adhesion}.
In 1995, R\"adler et al.~\cite{Raedler_95} studied the adhesion of GUVs to solid substrates.
They constituted GUVs of stearoyl-oleoyl-phosphatidylcholine (SOPC) in a $100\, \,\mathrm{mM}$ sucrose solution, so that the vesicles were denser than the buffer solution and sank to
the bottom of the chamber, where a glass cover slip coated with a thin film of $MgF^2$ and bovine serum albumin had been deposed.
The vesicle then floated above the glass slip with a height $s(\bm{r})$, as shown in Fig.~\ref{radler}, in a weakly adhered state.
Using reflection interference contrast microscopy (RICM) and phase contrast microscopy (see Fig.~\ref{fig_1_adhes_exp}), the group could measure the radius of the vesicle $R_\mathrm{ves}$, the radius of the contact region $R_a$, and reconstruct the height profile of the adhered patch.

Under the supposition that energy of the contact region was well described by eq.(\ref{hamilt_adhes}) and defining $h(\bm{r}) = s(\bm{r}) - \langle s \rangle$, they could infer:

\begin{enumerate}
\item the fluctuation spectrum $\langle |h(\bm{q})|^2 \rangle$, which once fitted with eq.(\ref{spect_fluct_adhes}) allowed to obtain $r$ and $V''$ (see Fig.~\ref{fig_1_spectrum}). The bending rigidity for SOPC, obtained in previous experiments, was assumed to be $35 \, k_\mathrm{B}T$;

  \begin{figure}[H]
\begin{center}
\includegraphics[scale=.3,angle=0]{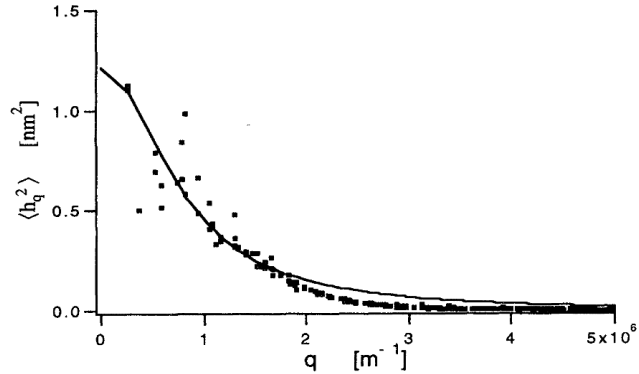}
\caption{Fluctuation spectrum of the contact region of the adhering vesicle~\cite{Raedler_95}. The solid line corresponds to the fit of eq.(\ref{spect_fluct_adhes}) from which $V''$ and $r$ are deduced (it was assumed that $\kappa = 35 \, k_\mathrm{B}T$).}
\label{fig_1_spectrum}
\end{center}      
\end{figure}
  
\item the correlation function $\langle h(x)h(0) \rangle$, which can be approximated by an exponential asymptote
  
  \begin{equation}
    G(\bm{r}, \bm{r}') = \langle h(\bm{r})h(\bm{r}') \rangle \sim \xi_\mathrm{\perp}^2 \, e^{-\frac{\bm{r} - \bm{r}'}{\xi_\mathrm{\parallel}}} \, ,
  \end{equation}
  
  \noindent where $\xi_\parallel$ is the distance beyond which two pieces of membrane are uncorrelated and $\xi_\perp$ is a measure of the membrane roughness.
  From the experimental data, the authors deduced $\xi_\parallel$ through a fit and $\xi_\perp$ from the value of $G(0)$ (see Fig.~\ref{fig_1_correlation});

    \begin{figure}[H]
\begin{center}
\includegraphics[scale=.35,angle=0]{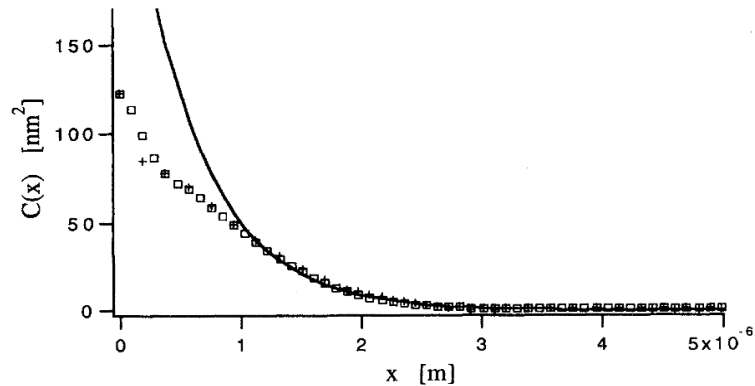}
\caption{Correlation $G(x)$ ~\cite{Raedler_95}. The solid curve shows the fit from which $\xi_\parallel$ is deduced, while $\xi_\perp$ is deduced from $G(0)$.}
\label{fig_1_correlation}
\end{center}      
\end{figure}
    
\item as we have discussed in section~\ref{subsection_adhesion}, under some conditions, the vesicle behaves as a spherical cap and we can define an effective contact angle that respects an analogous to the Young-Dupree relation.
  After a reconstruction of the average height profile of the adhesion patch from the RICM images, R\"adler et al. tried to obtain the effective contact angle $\theta_\mathrm{eff}$ by a linear fit near the edge of the contact region (see Fig.~\ref{fig_1_contact_angle}).
  They have also tried to fit a circle in the contact region in order to obtain the curvature radius $R_c$, which relates to the adhesion energy per unit area $W_A$ through eq.(\ref{Rc}) in the case $R_c < R_\mathrm{ves}$.

      \begin{figure}[H]
\begin{center}
\includegraphics[scale=.35,angle=0]{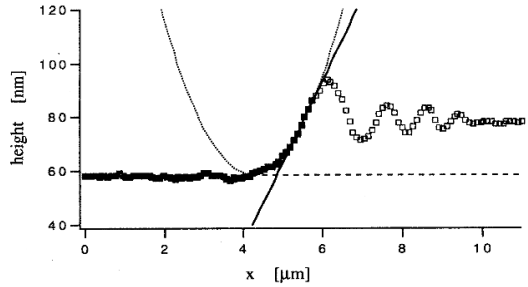}
\caption{Average contour of a vesicle near the contact region obtained from the RICM images.
  The dark points correspond to the regime $s < 100 \, \mathrm{nm}$, while the open data correspond to higher order fringes and are not further considered.
  The solid line shows a linear fit from which $\theta_\mathrm{eff}$ is derived and the dotted line shows the fit of a circle to the contact region aiming to obtain $R_c$.
  Note that the scales are distorted and that $\theta_\mathrm{eff} \approx 1^o$ - the membrane is extremely flat.}
\label{fig_1_contact_angle}
\end{center}      
\end{figure}

      As the vesicle was extremely flat and rounded near the contact point, one could not obtain $\theta_\mathrm{eff}$ precisely from the height profile.
      Indeed, measuring the contact angle in larger scales would lead to larger values for $\theta_\mathrm{eff}$.
      Concerning $R_c$, the fit was made difficult by the thermal fluctuations that remain even after averaging.
      Sadly, the values of $R_c$ obtained were comparable to the vesicle's radius $R_\mathrm{ves}$.
      Accordingly, eq.(\ref{Rc}) could not be used to obtain the value of the adhesion energy $W_A$.

\item the average height $\langle s \rangle$.
\end{enumerate}

To sum up, R\"adler et al. were able to obtain $r$ from the fluctuation spectrum and to made a rough estimate of the effective contact angle.
They could not however measure directly $\tau$ nor the adhesion energy $W_A$.
So, they made a theoretical estimate of the adhesion energy to check the self-consistency of their results.

\subsubsection{Theoretical estimate of the adhesion energy per unit area \bm{$W_A^\mathrm{theo}$}}

Here we explain just in general lines how the value of the adhesion energy per unit area was theoretically estimated.
The details are given in appendix~\ref{annexe21}.
R\"adler et al. considered that the contact region of the vesicle was submitted to three potentials: two attractive, coming from the van der Waals interaction and gravity, and one repulsive with steric origin.
They considered the screened van der Waals potential, since some part of the $MgF^2$ coating of the glass
cover slip is expect to be present in small concentration in the buffer solution.   

To obtain the repulsive contribution coming from the reduction of the configurations due to the substrate, they had to determine whether the adhesion was dominated by the bending rigidity or by the tension, which was done by studying $\xi_\perp$ (see appendix~\ref{annexe21} for details).
They concluded that the behavior of the membrane was dominated by tension.
Furthermore, they could also conclude that it was reasonable to assume $\sigma \approx r$ in this experiment. 

A plot of $V_\mathrm{total} = V_\mathrm{vdW} + V_\mathrm{steric} + V_\mathrm{grav}$ for typical experimental values ($\sigma = 1.7 \times 10^{-5} \, \mathrm{N/m}$, $\kappa = 35 \, k_\mathrm{B} T$, $b = 0.085$, $A_H = 2.6 \times 10^{-21} \, \mathrm{J}$, $D_M = 20 \, \mathrm{\mu m}$, $D_A = 10 \, \mathrm{\mu m}$, $\Delta \rho = 7 \, \mathrm{kg/m^3}$) can be seen in Fig.~\ref{adhesion}.
The potential presents a minimum whose depth can be considered as a first estimate of the adhesion energy per area $W_A^\mathrm{theo} \approx 10^{-9} \, \mathrm{N/m}$.

\subsubsection{Coherence test: estimate of the adhesion energy through Young-Dupree relation and discussion}

The second strategy of the authors was to estimate the energy of adhesion through the Young-Dupree relation

\begin{equation}
  W_A^\mathrm{Young} = \tau\left(1 - \cos \theta_\mathrm{eff}\right)\, ,
  \label{eq_1_young}
\end{equation}
  
\noindent where $\theta_\mathrm{eff}$ is the effective contact angle obtained through the fit shown in Fig.~\ref{fig_1_contact_angle}.
Assuming that $\tau \approx r$, their results are summarized in table~\ref{table_1_results}.
Despite the imprecision in the measures of the effective contact angle, there seems to be an incoherence between the theoretical estimate and the value of the adhesion energy obtained from the Young-Dupree relation, which was initially blamed on the simplified theoretical framework that did not account for the constraints on area and volume. 

\vspace{1cm}

\begin{table}[H]
  \begin{center}
%\begin{tabular}{|c|c|c|c|c|c|c|}
%\hline
%& & & \\
%\bm{{\red $D_M$}} & \bm{{\red $D_A$}}&\bm{{\red $\theta_{eff}$}} & \bm{{\red $r$}} & \bm{{\red $W_A^{Young}$}} & \bm{{\red $W_A^{theo}$}} & \bm{{\red $\tau^{theo}$}}\\
%\tiny{$(\mu m)$} & \tiny{$(\mu m)$} &  \tiny{(deg)} & \tiny{$(10^{-6} \, \mathrm{N/m})$} & \tiny{$(10^{-9} \, \mathrm{N/m})$}& \tiny{$(10^{-9} \, \mathrm{N/m})$} & \tiny{$10^{-6} \, \mathrm{N/m})$}\\ 
%\hline
%52&17&1.4 & 87.3 & 26.1 & &3.4\\
%58&10&2.1 & 51.1 & 34.3 &&1.5\\
%55&18&2.1 & 4.2 & 2.8 &&1.5\\
%88&45&0.8 & 8.1 & 0.8& $\sim 1$&10.3\\
%62&27&1.1 & 14.5 & 2.7 &&5.4\\
%53&15&0.7 & 17.3 & 1.3&&13.4\\
%83&64&2.1 & 27.3 & 18.3&&1.5\\
%91&31&0.9 & 15.9 & 2&&8.1\\
%\hline
\begin{tabular}{|c|c|c|c|c|c|}
\hline
%& & & \\
\bm{{\red $2 \, R_\mathrm{ves}$}} & \bm{{\red $2 \, R_a$}}&\bm{{\red $\theta_\mathrm{eff}$}} & \bm{{\red $r$}} & \bm{{\red $W_A^\mathrm{Young}$}} & \bm{{\red $W_A^\mathrm{theo}$}} \\
\tiny{$(\mu m)$} & \tiny{$(\mu m)$} &  \tiny{(deg)} & \tiny{$(10^{-6} \, \mathrm{N/m})$} & \tiny{$(10^{-9} \, \mathrm{N/m})$}& \tiny{$(10^{-9} \, \mathrm{N/m})$} \\ 
\hline
52&17&1.4 & 87.3 & 26.1 & \\
58&10&2.1 & 51.1 & 34.3 &\\
55&18&2.1 & 4.2 & 2.8 &\\
88&45&0.8 & 8.1 & 0.8& $\sim 1$\\
62&27&1.1 & 14.5 & 2.7 &\\
53&15&0.7 & 17.3 & 1.3&\\
83&64&2.1 & 27.3 & 18.3&\\
91&31&0.9 & 15.9 & 2&\\
\hline
\end{tabular}
\caption{The first four columns show the measured parameters from eight
  different vesicles. The fifth column is the adhesion energy evaluated
  through eq.(\ref{eq_1_young}) assuming $\tau \approx r$. The sixth columns
  shows the theoretical estimate (the minimum of $V_\mathrm{tot}$). }
    \label{table_1_results}
%The last column is the of $\tau$ obtained by supposing that the values of the preceding column are right and inverting eq.(\ref{eq_1_young}). Note that $\tau$ thus estimated is generally far smaller than $r$.}
\end{center}
\end{table}

\subsubsection{A refined theory does not solve the problem...}

%% \begin{equation}
%%   V_\mathrm{steric}^\mathrm{Seifert} = \frac{ 6 \, b^2 k_\mathrm{B} T}{\kappa \langle s \rangle^2}\frac{y^2}{\sinh^2(y)}\, ,
%%   \label{eq_1_Vsteric_Seifert}
%% \end{equation}

%% \noindent where $b$ is the same dimensionless numerical constant as before and

%% \begin{equation}
%%   y \equiv \left(\frac{\sigma}{b \, k_\mathrm{B}T}\right)^{\frac{1}{2}} \, \frac{\langle s\rangle}{2} \, .
%% \end{equation}

Shortly after, Udo Seifert proposed a refined theory that considered also the constraints on area and volume~\cite{Seifert_95b}.
His calculations yielded a different repulsive potential $V_\mathrm{steric}^\mathrm{Seifert}$.
Considering only $V_\mathrm{steric}^\mathrm{Seifert}$ and $V_\mathrm{vdW}$, he concluded that the vesicle should present tension-induced adhesion, i. e, the potential should present a local quadratic minimum like in the experiment of R\"adler when

\begin{equation}
  a \equiv \frac{2 \kappa (1 - \cos\varphi_\mathrm{eff})}{3\,  b \, k_\mathrm{B} T} < \frac{1}{3} \, ,
\end{equation}

\noindent where $\varphi_\mathrm{eff} \approx R_a/R_\mathrm{ves}$ and $b$ is a constant.
Taking $b = 1/2\pi$ and $\kappa = 35 \, k_\mathrm{B}T$, this condition is satisfied for $a \simeq 150[1 - \cos(\varphi_\mathrm{eff})]$ and thus $\varphi_\mathrm{eff}^\mathrm{max} \simeq 0.07 \, \mathrm{rad}$.
In R\"adler experiment, $\varphi_\mathrm{eff} \gtrsim 1/4$ (see table~\ref{table_1_results}) and thus the refined theory could still not explain the data.

Finally, Seifert examined the possibility that gravity could reconcile his theory with R\"adler's experiment. 
He compared the total contribution to the potential energy coming from the bending rigidity and from gravity:

\begin{eqnarray}
  \frac{V_\mathrm{grav}^\mathrm{tot}}{V_\kappa^\mathrm{tot}} &=& \frac{g \, \Delta \rho \, V_D \, h_\mathrm{CM}}{\int \frac{1}{2}\frac{\kappa}{R^2} \, dA}\, ,\nonumber\\
  \nonumber \\
  &\simeq& \frac{g \, \Delta\rho \, R_\mathrm{ves}^4}{\kappa} \, ,
  \end{eqnarray}

\noindent where $V_D$ is the vesicle's volume, $g$ is the gravitational acceleration, $\Delta \rho$ is the difference of density between the liquid contained in the vesicle and the suspension medium and $h_{CM}$ is the height of the vesicle's center of mass.
As a rough estimate, he assumed the vesicle a sphere and $h_\mathrm{CM} \approx R_\mathrm{ves}$.
For typical experimental values, the ratio is of approximately one hundred: gravity is thus very important to determine the shape of a vesicle.
Neglecting the adhesion energy, which is justified in the case of weak adhesion, and neglecting the bending energy, the contact angle of the vesicle should be zero and the tension should simply be given by the balance between gravity and the mechanical tension:

\begin{equation}
  \tau \approx \frac{2}{3} \, g \, \Delta \rho \, R_\mathrm{ves}^2 \left(\frac{R_\mathrm{ves}}{R_a}\right)^2 \, ,
  \label{eq_Sei}
\end{equation}

\noindent where we have approximated $h_\mathrm{CM} \approx R_\mathrm{ves}$.
Numerically, for the experimental data of R\"adler et al., one obtains $\tau \simeq 3 \times 10^{-7} \, \mathrm{N/m}$, which
is still far smaller than the values of $r$ measured (see table~\ref{table_1_results}).
R\"adler's data remained unexplainable.

\subsubsection{The solution and a posterior confirmation}
\label{subsection_1_Limozin}

In 1995, the fact that $r \approx \sigma$ was very significantly different from $\tau$ was totally unexpected. 
Let's now re-examine the experimental data of that time under our theoretical framework.
Our theory predicts that $\tau$ should indeed be different from $\sigma$.
As a first approximation, let's assume that we are in the conditions where $\tau = \sigma - \sigma_0$.
If we look at the values of table~\ref{table_1_results}, we have $r \approx
\sigma \simeq 10^{-5} \, \mathrm{N/m}$. 
If we suppose that the adhesion energy per unit area is indeed $\sim 10^{-9}
\, \mathrm{N/m}$ and we invert eq.(\ref{eq_1_young}), we obtain $\tau \simeq
10^{-6} \, \mathrm{N/m}$, yielding
$\sigma_0 \simeq 10^{-5} \, \mathrm{N/m}$.
Recalling the definition of $\sigma_0$ presented on eq.(\ref{eq_1_sigma0}), this result implies that the microscopic cut-off is $a \approx 4 \, \mathrm{nm}$, which is very reasonable.
Therefore, our theory could explain the results of R\"adler et al.

\begin{figure}[H]
\begin{center}
\includegraphics[scale=.7,angle=0]{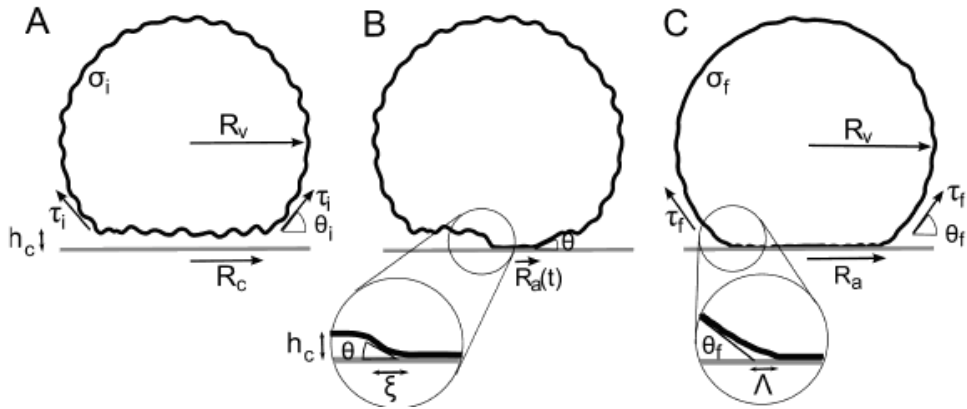}
\caption{Three states studied during the experiment proposed by Sengupta et al.~\cite{Sengupta_10}: at left, the vesicle is in a pre-nucleation state.
  It fluctuates at a height $h_c$ given by a local minimum of the adhesion potential.
  In the middle, we see the nucleation: a part of the adhesion patch adheres strongly to the substrate, which corresponds a transition to the deep minimum of the adhesion potential.
  Finally, at right, we see the final strongly adhered state.
  The effective contact angle, here indicated by $\theta_f$, and the characteristic length $\lambda \equiv \Lambda$ introduced in section~\ref{subsection_adhesion} are also represented.}
\label{fig_states}
\end{center}      
\end{figure}

Recently, Sengupta and Limozin made a careful study on the adhesion of vesicles\cite{Sengupta_10}.
They examined the adhesion of stiffer GUVs composed by phosphatydilcholine and cholesterol filled with a $200 \, \mathrm{mM}$ sucrose solution on a substrate coated with polymers in three different concentrations: without polymer (no-polymer coating), with $c_\mathrm{pol} = 0.75 \, \mathrm{ \mu m^{-2}}$ (sparse polymer coating) and with $c_\mathrm{pol} = 1 \, \mathrm{\mu m^{-2}}$ (dense polymer coating).
They observed systematically the pre-nucleation state (weak adhesion), the nucleation, i. e., the formations of the first patch of strongly adhered membrane, the growth of these patches and the mechanics in the final state of strong adhesion (see Fig.~\ref{fig_states}). 

\begin{figure}[H]
\begin{center}
\includegraphics[scale=.7,angle=0]{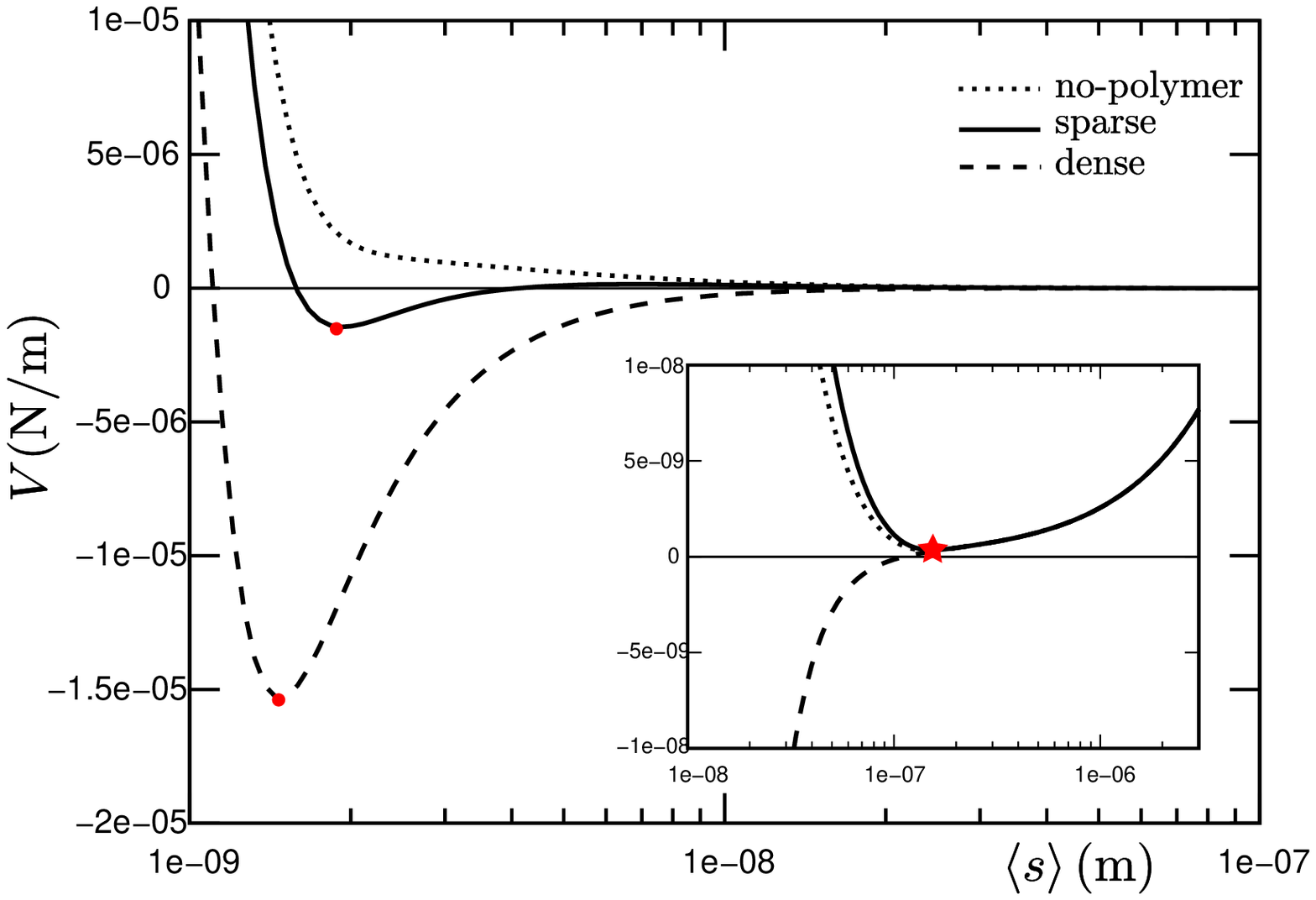}
\caption{Predicted adhesion potential for three different coatings. The red
  circles indicate the deep minima and the star indicates the shallow
  minimum. The curves were obtained for $R_\mathrm{ves} = 10 \, \mathrm{\mu m}$,
  $\ell_\mathrm{ster} = 0.2 \, \mathrm{nm}$, $d_\mathrm{lip} = 0.9 \,
  \mathrm{nm}$, $\Delta \rho = 12.8 \, \mathrm{kg/m^3}$, $A_H = 2 \times
  10^{-21} \, \mathrm{J}$, $\kappa = 100 \, k_\mathrm{B} T$, $a = 4 \, \mathrm{nm}$, $b = 0.1$ and $R_f = 87 \, \mathrm{nm}$. }
\label{fig_1_limozin}
\end{center}      
\end{figure}

From a theoretical point of view, the predicted adhesion potential for the three coatings is shown in Fig.~\ref{limozin} and reproduced in Fig.~\ref{fig_1_limozin}.
For no-polymer coating, there is just a deep minimum and thus strong adhesion, while for sparse polymer coating, there is also a shallow minimum at $\approx 100 \, \mathrm{nm}$ corresponding to weak adhesion.
For the dense polymer coating, only the shallow minimum remains and only weak adhesion is predicted.
The nucleation represents thus the passage of the shallow minimum to the deeper one.

Experimentally, as in former adhesion studies, RICM images using two different wave-lengths were used to obtain an intensity map of the adhered region.
This time, however, a major improvement was introduced in the reconstruction of the height profile from these images: the case of profiles with high and variable curvature was addressed for the first time.
Indeed, up to now, only the deviations caused by pure tilts and by profiles of constant curvature were accounted for.
With this new method, the membrane profile was described by a succession of small curved segments and the reconstruction was made fringe by fringe (see a description of the method in~\cite{Sengupta_10}).
The advantage of this method is that it allows a more reliable profile reconstruction even for steeper profiles, allowing thus to obtain the contact angle more precisely. 

The results concerning the membrane mechanics can be summarized as follows:

\begin{enumerate}
  \item Pre-nucleation state: in this state, the membranes presents strong undulations in the adhesion region.
    The vertical roughness $\xi_\perp \ \simeq 15 \, \mathrm{nm}$ was measured, from which $\sigma \sim 10^{-5} \, \mathrm{N/m}$ could be deduced (the relation between these quantities is given in appendix~\ref{annexe21}).
%Once more, it was verified that the experiment was dominated by tension
%($\sigma > \sigma^*$), so that $\sigma$ could be deduced from the first
%equation of the last line of table~\ref{table_1_xi}, yielding $\sigma \sim 10^{-5} \, \mathrm{N/m}$.
    As Seifert had shown in his work, gravity is dominating in the case of weak adhesion, so that the tension $\tau$ can be deduced from eq.(\ref{eq_Sei}), leading to $\tau = 10^{-7} - 10^{-6} \, \mathrm{N/m}$.
    Sengupta and Limozin verified that this discrepancy is compatible with $\tau = \sigma - \sigma_0$ for $a \sim 5 \, \mathrm{nm}$.

    \item Saturation of growth of the strongly adhered patch: the vesicle gains energy by increasing the contact area.
      In the process, its excess are decreases up to the equilibrium represented in C of Fig.~\ref{fig_states}.
    As in micropipette experiments, the excess area before and after strong adhesion should verify eq.(\ref{eq_1_alpha_new}).
    This relation was verified for all vesicles studied in this work within a factor between five and ten for $\sigma_0 \approx 10^{-6} \, \mathrm{N/m}$.

    \item Final state of strong adhesion: instead of measuring the effective contact angle $\theta_\mathrm{eff}$ and the curvature radius $R_c$ as in R\"adler's work, the authors used the second method proposed in section~\ref{subsection_adhesion} to obtain the adhesion energy $W_A$ and the membrane tension $\tau$, by measuring the contact angle $\theta_\mathrm{eff}$ and the length $\lambda$ (see Fig.~\ref{fig_states}).
      This time, the results obtained were more reliable due to the new reconstruction method and to the fact that the effective angle is more easily defined in the case of strong adhesion.
      From $\lambda$, the tension $\tau$ could be directly derived (eq.(\ref{eq_lambda})).
      Using the Young-Dupree relation and the measured values of $\theta_\mathrm{eff}$, the adhesion energy $W_A$ for each polymer coating could be obtained.
      The values obtained for $W_A$ for the different coatings were compatible with the theoretical values, corresponding to the deep minima of the adhesion potential (see Fig.~\ref{fig_1_limozin}).
      Sadly, in this case one cannot measure neither $r$ nor $\sigma$ by measuring the fluctuations of the membrane, since the membrane is too near to the substrate.
\end{enumerate}

The results described in the points $1$ and $2$ are the first strong evidences in agreement with our predictions.

\section{In a nutshell}

In this chapter, we have discussed the difference between the mechanical
tension $\tau$ one applies through micropipettes, for instance, and the
tension $\sigma$ usually added to the Hamiltonian in theoretical calculations.
Quantitatively, for large membranes, we have found 

\begin{eqnarray}
\tau &=& \sigma - \sigma_0 \left[ 1 - \frac{\sigma}{\kappa \Lambda^2} \ln
  \left(1 + \frac{\kappa \Lambda^2}{\sigma}\right)\right] \, ,\nonumber\\
&\simeq& \sigma - \sigma_0\, ,
\end{eqnarray}

\noindent where the last approximation is valid for small tension ($\sigma <
10^{-2} \, \sigma_r$, where $\sigma_r = \kappa \Lambda^2$ is of the order of the rupture tension).
The constant $\sigma_0$ depends only on the temperature and on the upper wave-vector cutoff $\Lambda$ through

\begin{equation}
\sigma_0  = \frac{k_\mathrm{B} T \, \Lambda^2}{8\pi} \, .
\end{equation}

\noindent The cutoff $\Lambda$ is related to a microscopical length of the same order of the membrane thickness.
Numerically, at room temperature and assuming $\Lambda = 1/(5 \, \mathrm{nm})$, we obtain $ \approx 5 \times 10^{-6} \, \mathrm{N/m}$, which is a not so small.
Indeed, we predict non-negligible corrections for experiments involving small tensions.
We have also questioned a former demonstration asserting that the coefficient of the $q^2$ term of the fluctuation spectrum, measured in contour analysis experiments, was equal to the mechanical tension.
We have presented some results supporting our predictions: a simple numerical experiment and a recent experiment on the adhesion of GUVs.

%% file: chap11.tex
\chapter{Fluctuation of forces in planar membranes}
\label{Fluct_plan}

In the last chapter, we have examined the average force exerted through a cut of projected length $L$ on a fluctuating planar membrane

\begin{equation}
  \langle \bm{f} \rangle = \tau \, L \, \bm{e}_x \, ,
\end{equation}

\noindent where $\bm{e}_x$ is the direction perpendicular to the cut.
Using the projected stress tensor, we have obtained $\tau$ as a function of the tension $\sigma$, introduced in the Hamiltonian in order to fix the average area of a membrane.

In this chapter, we would like to study the mean square deviation of this force, defined as

\begin{equation}
(\Delta \bm{f})^2 = \langle \bm{f}^2 \rangle - \langle \bm{f} \rangle^2 \, .
\end{equation}

\noindent In the following, we will call $\Delta \bm{f}$ simply the fluctuation of the force.
The results exposed here were obtained in the company of Jean-Baptiste Fournier and remain unpublished. 
Our motivation is three-fold:

\begin{enumerate}
  \item first, as experimentally one can measure the average of forces, it should also be possible to measure its fluctuations.
    Experimentally, for planar membranes or GUVs, this may be technically difficult, since one does not control $\tau$ directly (figures are different for experiments involving membrane nanotubes, as we shall see in the following chapters).
    Numerically, however, it should be reasonably simple to obtain $\Delta \bm{f}$ using systems similar to the one proposed in ref.~\cite{Imparato_06}, introduced in section~\ref{section_1_evidences}.
    
\item secondly, there is no theoretical prediction on the matter.
Up to now, as we have seen in last chapter, calculations on $\tau$ involved differentiations of the free-energy, which is very tricky.
The projected stress tensor simplifies calculations, allowing one to obtain more directly the mean square deviation of forces;

\item at last, this chapter is an intermediate step towards the calculation of the fluctuation of the force needed to hold a membrane nanotube, which will be done in chapter~\ref{Fluct_TUBE}.
  Indeed, this geometry is far more interesting, presenting highly fluctuating Goldstone modes~\cite{Fournier_07a}.
  Moreover, from an experimental point of view, in order to extract or hold a tube, one applies directly a point force using optical tweezer or using magnetic beads, as we have seen in section~\ref{tube_exp}.
  These techniques are very precise and one should thus be able to measure $\Delta \bm{f}$, even if it is of the order of some $\mathrm{pN}$.
\end{enumerate}
 
In the first section, we will define precisely $\Delta \bm{f}$ and remind some results obtained in the last chapter.
After, in section~\ref{section_11_rules} we shall introduce some diagrammatic tools, which are very useful since they make calculations visual.
Using diagrams, one can easily identify terms whose contribution is zero and group
rapidly other terms. 
It will prove specially useful in the calculation of the fluctuation of the
force.
To gain familiarity with these diagrams, we recover the result given in eq.(\ref{eq_1_diff_complet}) in section~\ref{section_11_ave}. 
In section~\ref{section_11_corr}, the most technical one, we shall evaluate the correlation of each term of the stress tensor.
These results are finally used in section~\ref{section_11_fluct} to obtain $\Delta \bm{f}$.

\section{Definitions and former results}

\begin{figure}[H]
\begin{center}
\includegraphics[scale=.4,angle=0]{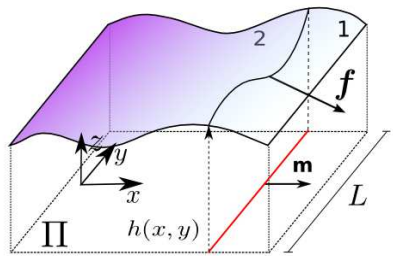}
\caption{Weakly fluctuating planar membrane described in the Monge's gauge. The
  force $\bm{f}$ is exchanged through the cut of projected
  length $L$ (red). Note that we have chosen an orthonormal basis in order to
  have $\bm{m} = \bm{e}_x$.}
\label{fig_11_cut}
\end{center}      
\end{figure}

Let us consider the same weakly fluctuating planar membrane described in
chapter~\ref{chapitre/planar_membrane}, whose projected area on a plane $\Pi$ parallel to the average plane of the membrane is $A_p$ (see Fig.~\ref{fig_1_cut}, which we reproduce in Fig.~\ref{fig_11_cut}).

The Hamiltonian is given by eq.(\ref{eq_1_hamilton}) and the we remind that the corresponding correlation function reads

\begin{equation}
  G(\bm{r} - \bm{r}') \equiv \langle h(\bm{r})h(\bm{r}')\rangle  = \frac{k_\mathrm{B} T}{A_p} \sum_{\bm{q}} \frac{e^{\icomp \, \bm{q}\cdot
      (\bm{r} - \bm{r'})}}{\sigma\,  q^2 + \kappa \, q^4} \, ,
\label{eq_11_correl}
\end{equation}

\noindent where $\bm{q} = 2 \pi/\sqrt{A_p} (m,n)$ and

\begin{equation}
  \sum_{\bm{q}} \equiv \sum_{|n| \leq N_\mathrm{max}} \sum_{|m| \leq N_\mathrm{max}}
\end{equation}

\noindent with $N_\mathrm{max} = \sqrt{A_p}/(2\pi a)$, corresponding to a maximum wave-vector $q_\mathrm{max} = 1/a$, with $a$ a microscopical length of the order of the membrane thickness.

As shown in Fig.~\ref{fig_11_cut}, we consider a cut of projected length $L$ parallel to $\bm{e}_y$. 
The average force exchanged through the cut is 

\begin{equation}
\langle\bm{f}\rangle = L\left(\langle \Sigma_{xx} \rangle \, \bm{e}_x +
  \langle \Sigma_{yx} \rangle \, \bm{e}_y + \langle \Sigma_{zx} \rangle \,
  \bm{e}_z \right) \, ,
\label{eq_11_fmoy}
\end{equation}

\noindent where $\Sigma_{ij}$ are the terms of the projected stress tensor for planar
membranes introduced in section~\ref{section_projected_stress}.
In chapter~\ref{chapitre/planar_membrane}, we have obtained

\begin{equation}
  \langle \bm{f} \rangle   = \left(\sigma - \frac{k_\mathrm{B}T}{2 A_p}
    \sum_{\bm{q}} \frac{\kappa \, q^2}{\sigma + \kappa \, q^2}\right) L \,
  \bm{e}_x \, .
\label{eq_11_f}
\end{equation}

\noindent In section~\ref{section_11_ave}, we will recover this result using diagrammatic tools.

The squared fluctuation of the force is given by

\begin{eqnarray}
(\Delta \bm{f})^2&=&\langle \bm{f}^2 \rangle-\langle \bm{f}
\rangle^2 \, ,\nonumber\\
&=& \left(\Delta f_x\right)^2 + \left(\Delta f_y \right)^2 + \left(\Delta
  f_z\right)^2 \, ,
\end{eqnarray}

\noindent where 

\begin{eqnarray}
  \label{eq_11_def_corryx}
(\Delta f_x)^2 &=& \iint^{L/2}_{-L/2} \left[\,\left\langle\Sigma_{xx}(x,y)\Sigma_{xx}(x,y')\right\rangle -
      \langle \Sigma_{xx}\rangle^2
      \, \right] \, dy dy' \, ,\\
  (\Delta f_y)^2 &=&\iint^{L/2}_{-L/2}
\left\langle\Sigma_{yx}(x,y)\Sigma_{yx}(x,y') \right\rangle \, dydy'\, ,\\
(\Delta f_z)^2&=&
\iint^{L/2}_{-L/2} \left\langle \Sigma_{zx}(x,y)\Sigma_{zx}(x,y')
\right\rangle \, dydy' \, ,
\label{eq_11_def_corrzx}
\end{eqnarray}

\noindent are the squared fluctuation of the forces perpendicular to
the cut, parallel to the cut and 
normal to the average
membrane's plane, respectively.
Note that we have omitted $\langle \Sigma_{yx} \rangle$ and $\langle \Sigma_{zx} \rangle$ in eq.(\ref{eq_11_def_corryx}) and eq.(\ref{eq_11_def_corrzx}), respectively, as these averages vanish (see section~\ref{section_1_tau}).
The evaluation of the force fluctuation is made two steps: first, we will evaluate the correlations

\begin{eqnarray}
  C_{xx}(y - y') &=& \langle \Sigma_{xx}(x,y) \Sigma_{xx}(x,y') \rangle - \langle \Sigma_{xx} \rangle^2 \,, \\
  C_{yx}(y - y') &=& \langle \Sigma_{yx}(x,y) \Sigma_{yx}(x,y') \rangle\, ,\\
  C_{zx}(y - y') &=& \langle \Sigma_{zx}(x,y) \Sigma_{zx}(x,y') \rangle \, ,
\end{eqnarray}

\noindent in section~\ref{section_11_corr}.
After, in section~\ref{section_11_fluct}, we will integrate these correlations twice over the cut's length.
First of all, let's introduce some diagrammatic tools.

\section{Diagrammatic tools}
\label{section_11_rules}

In physics, the word field is used to denote any physical quantity that varies in space.
Accordingly, the height of the membrane $h(\bm{r})$ is a field.
Inspired from the Feynman diagrams used in statistical field theory, we associate graphical representations to the fields in order to make calculations visual, allowing quicker simplifications. 
Each field is represented by a simple straight line with a point appended to it.
This point, called a vertex, represents the point $\bm{r}$ in which the field is evaluated. 
When two or more fields are evaluated at the same point, we represent them connected by the same vertex.
Besides, we represent the differentiation
with respect to $x$ or $y$ by a slash or a dot over the lines.
We present a basic diagrammatic vocabulary in table~\ref{table_11_vocab}.

\begin{table}[H]
  \begin{center}
\begin{tabular}{|c|c|}
\hline
\red{Usually} & \red{Diagrammatically} \\
\hline
&\\
$h(\bm{r})$ & \diaggd{./figures_Chap11/simple_0.epsi} \\
&\\
\hline
&\\
$h_x(\bm{r})$ & \diaggd{./figures_Chap11/simple_1.epsi} \\
&\\
\hline
&\\
$h_y(\bm{r})$ & \diaggd{./figures_Chap11/simple_4.epsi} \\
&\\
\hline
&\\
$h(\bm{r})\, h(\bm{r})$ & \diaggd{./figures_Chap11/champ_0.epsi} \\
&\\
\hline
&\\
$h(\bm{r})\, h(\bm{r}')$ & $\diaggd{./figures_Chap11/simple_0.epsi} \diaggd{./figures_Chap11/simple_0.epsi}$ \\
&\\
\hline
\end{tabular}
\caption{Basic {\it translation} rules from the usual notation into diagrams.}
\label{table_11_vocab}
\end{center}
\end{table}

The thermal averages of fields are
performed using Wick's theorem, which states that the average of an even number of fields is given by the sum of all possible complete contractions.
By a complete contraction, we mean linking the free ends of a set of fields, two by two, in a way that no single field remains.
The continuous line formed after the contraction between two fields represents the correlation function $G(\bm{r})$ (in this context also called propagator), suitably differentiated.
If the number of fields is uneven, the theorem states that the average vanishes.

Let's see an example of the simplest case, involving only two fields:

\begin{equation}
  \left\langle
  \begin{array}{c}
\diaggd{./figures_Chap11/simple_1.epsi}\\
{}^{\bm{r}}
\end{array}
    \begin{array}{c}
\diaggd{./figures_Chap11/simple_4.epsi}\\
{}^{\bm{r}'}
\end{array}
    \right \rangle =
\diaggd{./figures_Chap11/reto_3.epsi} =
\partial_x |_{\bm{r}} \, \partial_y |_{\bm{r}'} \, [G(\bm{r}' - \bm{r})] \, ,
\label{eq_11_dig}
\end{equation}

\noindent where $\partial_x |_{\bm{r}}$ stands for the derivation with respect to $x$ at the point $\bm{r}$.
The arrow indicates that the propagator {\it leaves} at the vertex $\bm{r}$ and {\it enters} at the vertex $\bm{r}'$.
It's direction is arbitrary: by inverting it, we would obtain $\partial_x |_{\bm{r}} \, \partial_y |_{\bm{r}'} \, [G(\bm{r} - \bm{r}')]$, which yields the same result as in eq.(\ref{eq_11_dig}), since $G(\bm{r} - \bm{r}')$ is a function of $|\bm{r} - \bm{r}'|$.
For the propagator given in eq.(\ref{eq_11_correl}), we obtain

\begin{equation}
  \left\langle
  \begin{array}{c}
\diaggd{./figures_Chap11/simple_1.epsi}\\
{}^{\bm{r}}
\end{array}
    \begin{array}{c}
\diaggd{./figures_Chap11/simple_4.epsi}\\
{}^{\bm{r}'}
\end{array}
    \right \rangle =
\frac{k_\mathrm{B} T}{A_p} \sum_{\bm{q}} \frac{(-i \, q_x)(i\, q_y)\, e^{i \, \bm{q} \cdot(\bm{r}' - \bm{r})}}{\sigma q^2 + \kappa q^4} \, .
\end{equation}

\noindent In other words, every slash (resp.\  dot) contributes to the sum a factor $i \, q_x$ (resp.\
$i \, q_y$) if the propagator enters the vertex to which it is attached and $-i \, q_x$ (resp.\ $-i\, q_y$) otherwise.

From this result, it is easy to show that whenever we have correlation function of the same kind of the one given in eq.(\ref{eq_11_correl}), we can group slashes and dots using the following rule: in
any propagator branch, one can shift a slash or a dot from one vertex side to the
other if one multiplies the diagram's coefficient by $-1$; once all
derivatives are on the same side, the side matters no more.
All the derivatives can be taken at the same point, contributing $(i\,q_x)$ for a slash or $(i\,q_y)$ for a dot, and we represent them in the center of the propagator:

\begin{equation}
\diaggd{./figures_Chap11/reto_3.epsi} =
(-1) \times \diaggd{./figures_Chap11/reto_4.epsi} = (-1) \times \partial_x \partial_y  \, [G(\bm{r}' - \bm{r})] \, .
\end{equation} 

As a second example, let's see a typical case of the average of two fields evaluated at the same point.
We have

\begin{eqnarray}
\left\langle \diaggd{./figures_Chap11/champ_4.epsi} \right\rangle &=&
\diaggd{./figures_Chap11/wick_3.epsi} =
(-1) \times
\diaggd{./figures_Chap11/wick_7.epsi} \, ,\nonumber\\
&=& - \partial_{yy} G(\bm{r}) \, ,\nonumber\\
&=& - \frac{k_\mathrm{B} T}{A_p} \sum_{\bm{q}} \frac{(i\, q_y)^2\, e^{i \, \bm{q} \cdot\bm{r}}}{\sigma q^2 + \kappa q^4} \, .
\end{eqnarray}

Finally, in section~\ref{section_11_corr}, we will deal with averages involving four fields.
A representative example follows, where we have numbered each field to highlight all the possible complete contractions:

\begin{eqnarray}
  \left\langle
  \begin{array}{c}
\diaggd{./figures_Chap11/champ_7_n1.epsi}\\
{}^{\bm{r}}
\end{array}
    \begin{array}{c}
\diaggd{./figures_Chap11/champ_7_n2.epsi}\\
{}^{\bm{r}'}
\end{array}
    \right \rangle
    &=&
     \begin{array}{c}
\diaggd{./figures_Chap11/wick_n1.epsi}\\
{}^{\bm{r}}
\end{array}
    \begin{array}{c}
\diaggd{./figures_Chap11/wick_n2.epsi}\\
{}^{\bm{r}'}
\end{array}
+
\diaggd{./figures_Chap11/peixe_n11.epsi}
+
\diaggd{./figures_Chap11/peixe_n21.epsi} \, ,\nonumber \\
&=&    
\diaggd{./figures_Chap11/wick_n1_1.epsi}
\diaggd{./figures_Chap11/wick_n1_1.epsi}
+ 
\diaggd{./figures_Chap11/peixe_n1_1.epsi}
+
\diaggd{./figures_Chap11/peixe_n2_1.epsi}\, ,\nonumber\\
\label{eq_11_avediag}
\end{eqnarray}

\noindent which one can readily read by noting the equivalence

\begin{equation}
\diaggd{./figures_Chap11/peixe_n2_1.epsi} = \diaggd{./figures_Chap11/reto_4.epsi} \times \diaggd{./figures_Chap11/reto_4.epsi} \, .
\label{eq_11_peixe}
\end{equation}

\section[Getting familiar: evaluating $\langle f \rangle$ with diagrammatic tools]{Getting familiar: evaluating $\langle \bm{f} \rangle$ with diagrammatic tools}
\label{section_11_ave}

From eq.(\ref{eq_11_fmoy}), we see that to evaluate $\langle \bm{f} \rangle$, one needs to evaluate the average of some component of the projected stress tensor.
These components, introduced in section~\ref{section_projected_stress}, can be written in terms of diagrams as 

\begin{eqnarray}
  \label{eq_11_Sigmaxxdiag1}
\Sigma_{xx} = \sigma &+& 
\frac{\sigma}{2}\diaggd{./figures_Chap11/champ_4.epsi} -
\frac{\sigma}{2}\diaggd{./figures_Chap11/champ_3.epsi} +
\frac{\kappa}{2}\diaggd{./figures_Chap11/champ_6.epsi}
\nonumber\\
&-&
\frac{\kappa}{2}\diaggd{./figures_Chap11/champ_5.epsi} +
\kappa\diaggd{./figures_Chap11/champ_1.epsi} +
\kappa \diaggd{./figures_Chap11/champ_2.epsi} \, ,\\
\Sigma_{yx} =  
&-& \, \sigma\diaggd{./figures_Chap11/champ_7.epsi} -
\kappa\diaggd{./figures_Chap11/champ_8.epsi} -
\kappa\diaggd{./figures_Chap11/champ_9.epsi}
\nonumber\\
&+&
\kappa\diaggd{./figures_Chap11/champ_10.epsi} +
\kappa \diaggd{./figures_Chap11/champ_11.epsi} \, ,\\
\Sigma_{zx} =  
&&\sigma\diaggd{./figures_Chap11/simple_1.epsi} -
\kappa\diaggd{./figures_Chap11/simple_2.epsi} -
\kappa\diaggd{./figures_Chap11/simple_3.epsi} \, .
\label{eq_11_Sigmaxxdiag}
\end{eqnarray} 

The average of eq.(\ref{eq_11_Sigmaxxdiag}) is the simplest one to evaluate: since each term has only an uneven number of fields, Wick's theorem imply directly a vanishing average.
We shall evaluate in details the average of $\Sigma_{xx}$ as an example.
We have

\begin{eqnarray}
\langle \Sigma_{xx} \rangle = \sigma &+& 
\frac{\sigma}{2}
\diaggd{./figures_Chap11/wick_3.epsi}
-\frac{\sigma}{2}
\diaggd{./figures_Chap11/wick_4.epsi}
+ \frac{\kappa}{2}
\diaggd{./figures_Chap11/wick_5.epsi}
\nonumber\\
&-&
\frac{\kappa}{2}
\diaggd{./figures_Chap11/wick_6.epsi} 
+ \kappa
\diaggd{./figures_Chap11/wick_1.epsi} 
+\kappa
\diaggd{./figures_Chap11/wick_2.epsi} \, ,
\end{eqnarray}
 
\noindent where the vertex indicates the point $\bm{r}$ in which the average is calculated.
Note that the average does not depend on it, given the isotropy of the system.
Grouping the differentiations, we obtain

\begin{eqnarray}
\langle \Sigma_{xx} \rangle = \sigma &-& 
\frac{\sigma}{2}
\begin{array}{c}
\diaggd{./figures_Chap11/wick_7.epsi}
\end{array}
+\frac{\sigma}{2}
\begin{array}{c}
\diaggd{./figures_Chap11/wick_8.epsi}
\end{array} 
+ \frac{\kappa}{2}
\begin{array}{c}
\diaggd{./figures_Chap11/wick_9.epsi}
\end{array} 
\nonumber\\
&-&
\frac{\kappa}{2}
\begin{array}{c}
\diaggd{./figures_Chap11/wick_10.epsi} 
\end{array} 
- \kappa
\begin{array}{c}
\diaggd{./figures_Chap11/wick_10.epsi} 
\end{array} 
-\kappa
\begin{array}{c}
\diaggd{./figures_Chap11/wick_11.epsi} \, .
\end{array} 
\end{eqnarray}

\noindent Now, in the particular case of these diagrams, with only one vertex, we have

\begin{eqnarray}
  \begin{array}{c}
\diaggd{./figures_Chap11/wick_7.epsi}
\end{array}
= \frac{k_\mathrm{B} T}{A_p} \sum_{\bm{q}} \frac{(i\, q_y)^2}{\sigma q^2 + \kappa q^4} =  \frac{k_\mathrm{B} T}{A_p} \sum_{\bm{q}} \frac{(i\, q_x)^2}{\sigma q^2 + \kappa q^4} =
\begin{array}{c}
\diaggd{./figures_Chap11/wick_8.epsi}
\end{array} \, , \\
  \begin{array}{c}
\diaggd{./figures_Chap11/wick_9.epsi}
\end{array}
= \frac{k_\mathrm{B} T}{A_p} \sum_{\bm{q}} \frac{(i\, q_y)^4}{\sigma q^2 + \kappa q^4} =  \frac{k_\mathrm{B} T}{A_p} \sum_{\bm{q}} \frac{(i\, q_x)^4}{\sigma q^2 + \kappa q^4} =
\begin{array}{c}
\diaggd{./figures_Chap11/wick_10.epsi}
\end{array} \, .
\end{eqnarray}

\noindent In fact, as the correlation function is calculated at $\bm{r} = \bm{r}'$, it follows that more generally
we can exchange globally slashes and dots in a diagram.
Sadly, this nice property does not hold for the evaluation of the kind of diagram shown in eq.(\ref{eq_11_peixe}), particularly important to evaluate the force fluctuation in the next section.
For the present case, it follows

\begin{eqnarray}
\langle \Sigma_{xx} \rangle = \sigma &-& 
\frac{\sigma}{2}
\begin{array}{c}
\diaggd{./figures_Chap11/wick_8.epsi}
\end{array}
+\frac{\sigma}{2}
\begin{array}{c}
\diaggd{./figures_Chap11/wick_8.epsi}
\end{array} 
+ \frac{\kappa}{2}
\begin{array}{c}
\diaggd{./figures_Chap11/wick_10.epsi}
\end{array} 
\nonumber\\
&-&
\frac{\kappa}{2}
\begin{array}{c}
\diaggd{./figures_Chap11/wick_10.epsi} 
\end{array} 
- \kappa
\begin{array}{c}
\diaggd{./figures_Chap11/wick_10.epsi} 
\end{array} 
-\kappa
\begin{array}{c}
\diaggd{./figures_Chap11/wick_11.epsi} \, , 
\end{array} \nonumber\\
= \sigma &-& 
\kappa
\begin{array}{c}
\diaggd{./figures_Chap11/wick_10.epsi}
\end{array}
-\kappa
\begin{array}{c}
\diaggd{./figures_Chap11/wick_11.epsi} \, ,
\end{array} 
\end{eqnarray}

\noindent which reads

\begin{eqnarray}
\langle \Sigma_{xx} \rangle &=& \sigma - \frac{\kappa \, k_\mathrm{B} T}{A_p}
\sum_{\bm{q}} 
\frac{\left[(i \, q_x)^4 + (i \, q_x)^2(i\,
    q_y)^2\right]\, e^{i \bm{q}(\bm{r} - \bm{r}')}}{\sigma q^2
  + \kappa q^4} \left\lvert_{\bm{r} = \bm{r}'} \phantom{\frac{1}{1}}\right. \,
,\nonumber\\
&=& \sigma - \frac{k_\mathrm{B} T}{2 \, A_p} \sum_{\bm{q}} \frac{\kappa \, q^2}{\sigma + \kappa q^2} \, . 
\end{eqnarray}

\noindent As expected, we have recovered $\tau$ given in eq.(\ref{eq_1_diff_complet}).

Taking into account the rules introduced in the last section, the average of $\Sigma_{yx}$ is very simple to evaluate.
Grouping the derivatives, we have 

\begin{eqnarray}
\langle \Sigma_{yx} \rangle &=& \sigma 
\diaggd{./figures_Chap11/wick_n1_1.epsi}
-\kappa
\diaggd{./figures_Chap11/wick_12.epsi}
-\kappa
\diaggd{./figures_Chap11/wick_13.epsi}
-\kappa
\diaggd{./figures_Chap11/wick_12.epsi} 
- \kappa
\diaggd{./figures_Chap11/wick_13.epsi} \, ,\nonumber\\
&=& \sigma 
\diaggd{./figures_Chap11/wick_n1_1.epsi}
-2 \kappa
\diaggd{./figures_Chap11/wick_12.epsi}
-2 \kappa
\diaggd{./figures_Chap11/wick_13.epsi}\, .
\end{eqnarray}

\noindent Let's evaluate the first diagram:

\begin{equation}
  \diaggd{./figures_Chap11/wick_n1_1.epsi} = \frac{k_B T}{A_p} \sum_{\bm{q}} \frac{(i \, q_x)(i\, q_y)}{\sigma q^2 + \kappa q^4} \, .
\end{equation}

\noindent Recalling that $q_x = 2 \pi m/\sqrt{A_p}$ and that $q_y = 2\pi n/\sqrt{A_p}$ and that the sum over $\bm{q}$ stands for two sums, one on $m$, and other on $n$, both running from $-N_\mathrm{max}$ up to $N_\mathrm{max}$, one can readily show that the contribution of this diagram vanishes.
More generally, for this kind of diagram, a uneven number of slashes or dots imply a vanishing contribution.
So, we conclude that $\langle \Sigma_{yx} \rangle = 0$ and we re-obtain the result given in eq.(\ref{eq_11_f}).

\section{Evaluation of the projected stress tensor correlation}
\label{section_11_corr}

In this section, we will evaluate the correlation of the stress tensor at two general points over the projected cut $\bm{r} = x \, \bm{e}_x + y \, \bm{e}_y$ and $\bm{r}' = x \, \bm{e}_x + y'\, \bm{e}_y$.
As we discussed in section~\ref{section_11_rules}, these calculations will involve mostly diagrams of the general family

\begin{equation}
\diaggd{./figures_Chap11/peixe_gen.epsi}\, .
\end{equation}
  
\noindent We begin thus by reminding two properties of these diagrams:

\begin{enumerate}
  \item they may be separated in two components of the form

    \begin{eqnarray}
      \label{eq_11_Gnm1}
      &&\diaggd{./figures_Chap11/reto_2.epsi} \equiv G_{n,m} \, ,\\
      &&= \frac{k_\mathrm{B} T}{A_p} \sum_{\bm{q}}\frac{(i)^{n+m} \, q_x^n \, q_y^m \, e^{i\, (y-y')q_y}}{\sigma q^2 + \kappa q^4} \, ,\nonumber\\
      \nonumber\\
      &&= k_\mathrm{B} T \, (i)^{n+m} \int \frac{d^2 q}{(2\pi)^2} \frac{q_x^n \, q_y^m \, e^{i\, (y-y')q_y}}{\sigma q^2 + \kappa q^4} \, ,\nonumber\\
      \nonumber\\
      &&= (i)^{n+m} \frac{k_\mathrm{B} T}{(2\pi)^2} \left[\int_0^\Lambda \frac{q^{n+m-1}}{\sigma + \kappa q^2} \left(\int_0^{2\pi} \cos^n\theta \, \sin^n\theta \, e^{i \, (y-y')\, \sin\theta} \, d\theta\right) \, dq\right] \, ,\nonumber\\
    \label{eq_11_Gnm}
    \end{eqnarray}

    \noindent where the two last passages are good approximations for very large membranes.
    We remind that $\Lambda$ is the upper wave-vector cutoff given by $1/a$, where $a$ is a microscopical length of the order of the membrane thickness.
    
  \item the contribution of propagators with an uneven number of {\bf slashes} -- and only slashes -- vanishes.
    Indeed, one can easily proof this by remarking that the sum over $\bm{q}$ is symmetrical.
    Note that this property would not hold if we didn't have $x = x'$.
    %This however does not change the generality of our calculation, since the membrane is isotropic and one can always chose an appropriate basis in order to satisfy this condition.
\end{enumerate}

\subsection[Evaluation of $C_{xx}$]{Evaluation of $\bm{C_{xx}}$}

Let's begin by calculating 

\begin{equation} 
C_{xx}(y - y') = \langle\Sigma_{xx}(x,y) \Sigma_{xx}(x,y')\rangle -\langle\Sigma_{xx}\rangle^2\, .
\end{equation}

\noindent Here we can finally understand how diagrams can simplify this calculation, since normally, from eq.(\ref{eq_11_Sigmaxxdiag1}), one should have to evaluate about $7 \times 7$ terms like those shown in eq.(\ref{eq_11_avediag}), implying $\approx 150$ terms in total.
Diagrammatically, however, many terms vanish and other can be simplified, yielding in the end 

\begin{eqnarray}
C_{xx}(y-y') &=&
\frac{\sigma^2}{2}\diagpq{./figures_Chap11/peixe_1.epsi}
 +
\frac{\sigma^2}{2}\diagpq{./figures_Chap11/peixe_2.epsi}
-
\sigma\kappa\diagpq{./figures_Chap11/peixe_3.epsi}
\nonumber\\
&+&
\sigma \kappa\diagpq{./figures_Chap11/peixe_4.epsi}
 -
 2\sigma \kappa\diagpq{./figures_Chap11/peixe_8.epsi}
 -
 2\sigma \kappa\diagpq{./figures_Chap11/peixe_9.epsi}
 \nonumber\\
 &+&
\frac{\kappa^2}{2}\diagpq{./figures_Chap11/peixe_7.epsi}
 +
\frac{3 \, \kappa^2}{2}\diagpq{./figures_Chap11/peixe_5.epsi}
 +
 2 \kappa^2 \diagpq{./figures_Chap11/peixe_6.epsi}
 \nonumber\\
 &+&
 \kappa^2 \diagpq{./figures_Chap11/peixe_11.epsi}
 +
 2\kappa^2\diagpq{./figures_Chap11/peixe_10.epsi}
 +
 \kappa^2 \diagpq{./figures_Chap11/peixe_12.epsi} \, .\nonumber\\
 \end{eqnarray}

\noindent In terms of the $G_{n, m}$ defined in eq.(\ref{eq_11_Gnm1}), we obtain

\begin{eqnarray}
  C_{xx}(y - y') &=& \frac{\sigma^2}{2} \left(G_{2,0}^2 + G_{0,2}^2\right) - \sigma \kappa \left(G_{2,1}^2 -  G_{0,3}^2\right) - 2\sigma \kappa \, G_{2,0}\left(G_{4,0} + G_{2,2}\right) \nonumber\\
  &+&\frac{\kappa^2}{2} \left(G_{0,4}^2 + 3 \, G_{4,0}^2 + 4 \, G_{4,0}\, G_{2,2}\right)
  +\kappa^2 \, G_{2,0}\left(G_{6,0} + 2 \, G_{4,2} + G_{2,4}\right) \, .\nonumber\\
  \label{eq_11_Cxx_G}
\end{eqnarray}

\noindent Considering very large membranes and performing the angular integral in eq.(\ref{eq_11_Gnm}) for each $G_{n,m}$, eq.(\ref{eq_11_Cxx_G}) becomes

\begin{eqnarray}
  C_{xx}(y - y') &=& \sigma^2 \left[\frac{1}{2} \, B_{21}^2(Y) - \frac{1}{Y}\,  B_{10}(Y)\, B_{21}(Y) + \frac{1}{Y^2} \, B_{10}^2(Y) \right]\nonumber\\
&+& \sigma \kappa \, \left[B_{12}^2(Y) - \frac{2}{Y} \, B_{12}(Y)\, B_{21}(Y) + \frac{2}{Y^2} \, B_{10}(Y)\,B_{12}(Y) \right] \nonumber\\
&+& \kappa^2 \left[\frac{1}{2}\,  B_{03}^2(Y) - \frac{2}{Y} \, B_{03}(Y) \, B_{12}(Y) + \frac{2}{Y^2} \, B_{12}^2(Y) \right.\nonumber\\
&+&\left. \frac{1}{Y^2}\, B_{10}(Y)\, B_{14}(Y) + \frac{3}{Y^2} \, B_{03}(Y) \, B_{21}(Y)\right]\, ,
  \label{eq_11_Cxx}
\end{eqnarray}

\noindent where $Y = |y - y'|$ and

\begin{equation}
  B_{ij}(y) = {k_\mathrm{B} T} \int_0^\Lambda \frac{dq}{2\pi} \frac{q^j \, J_i(q\, y)}{\sigma + \kappa q^2}  \, ,
  \label{eq_11_Bij}
  \end{equation}

\noindent with $J_i$ standing for the first kind Bessel function of order $i$.
Note that, as expected given the isotropy of the system, $C_{xx}$ depends only on the distance between the points.
At this point, it is useful to rewrite eq.(\ref{eq_11_Bij}):

\begin{equation}
  B_{ij}(y) = \frac{k_\mathrm{B} T}{\kappa} \, \Lambda^{j-1} \, \int_0^1 \frac{dq}{2\pi} \frac{q^j \, J_i(q \, \Lambda \, y)}{r + q^2} = \frac{k_\mathrm{B} T}{\kappa} \, \Lambda^{j-1} \, \tilde{B}_{ij}(\Lambda y) \, ,
  \label{eq_11_Bijtilde}
\end{equation}

\noindent where $\tilde{B}_{ij}$ is dimensionless.
To simplify notations, we have omitted the dependence of $\tilde{B}_{ij}$ on $r$, given by 

\begin{equation}
  r = \frac{\sigma}{\kappa \Lambda^2} = \frac{\sigma}{\sigma_r} \, ,
\end{equation}

\noindent with $\sigma_r$ of the order of the membrane rupture tension.
Eq.(\ref{eq_11_Cxx}) can be rewritten as

\begin{eqnarray}
  C_{xx}(y - y') &=& 64 \pi^2 \sigma_0^2 \left\{r^2\left[\frac{1}{2} \, \tilde{B}_{21}^2(\Lambda Y) - \frac{\tilde{B}_{10}(\Lambda Y)\, \tilde{B}_{21}(\Lambda Y)}{\Lambda Y} + \frac{\tilde{B}_{10}^2(\Lambda Y)}{(\Lambda Y)^2} \right]\right.\nonumber\\
&+& r\, \left[\tilde{B}_{12}^2(\Lambda Y) - 2 \, \frac{\tilde{B}_{12}(Y)\, \tilde{B}_{21}(\Lambda Y)}{\Lambda Y} + 2\, \frac{\tilde{B}_{10}(\Lambda Y)\,\tilde{B}_{12}(\Lambda Y)}{(\Lambda Y)^2} \right] \nonumber\\
&+& \left[\frac{1}{2}\,  \tilde{B}_{03}^2(\Lambda Y) - 2\, \frac{\tilde{B}_{03}(\Lambda Y) \, \tilde{B}_{12}(\Lambda Y)}{\Lambda Y} + 2\,  \frac{\tilde{B}_{12}^2(\Lambda Y)}{(\Lambda Y)^2} \right.\nonumber\\
&+& \left. \left. \frac{\tilde{B}_{10}(\Lambda Y)\, \tilde{B}_{14}(\Lambda Y)}{(\Lambda Y)^2} + 3 \, \frac{\tilde{B}_{03}(\Lambda Y) \, \tilde{B}_{21}(\Lambda Y)}{(\Lambda Y)^2}\right]\right\} \, ,
  \label{eq_11_Cxx1}
\end{eqnarray}

\noindent where

\begin{equation}
  \sigma_0 = \frac{\kappa \Lambda^2}{8 \pi \beta \kappa}
\end{equation}

\noindent was already introduced in the last chapter.
Note that the terms inside the brackets are dimensionless and that $C_{xx}$ depends actually only on $\sigma_0$, on $r$ and on $\Lambda Y \equiv |y - y'|/a$.
%The three first parameters are fixed for a given membrane and only the last one is a free variable.

\begin{figure}[H]
\begin{center}
\includegraphics[width=0.5\columnwidth]{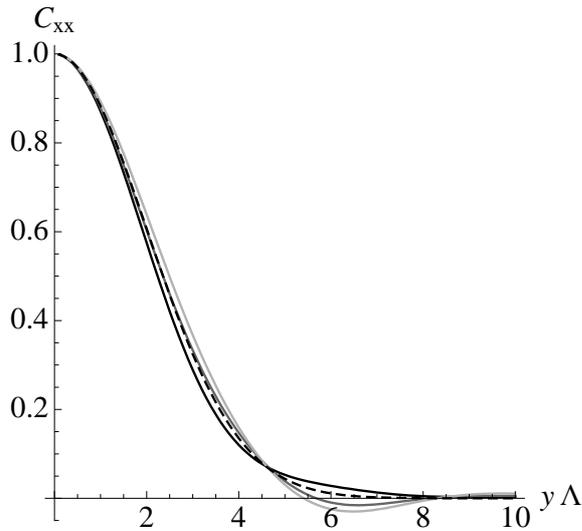}
\caption{Plot of the correlation of $\Sigma_{xx}$ as a function of the distance in units of the microscopical length $a=\Lambda^{-1}$ (normalized by the auto-correlation  given in eq.(\ref{eq_11_Cxx0})).
  The black line stands for $r = 1$, the dark gray corresponds to $r = 10^{-2}$ and the light gray corresponds to $r = 10^{-4}$.
  For $\kappa \approx 25 \, k_\mathrm{B} T$, it corresponds to $\sigma = 4 \times 10^{-3}\, \mathrm{N/m}$, $\sigma = 4 \times 10^{-5} \, \mathrm{N/m}$ and $\sigma = 4 \times 10^{-7} \, \mathrm{N/m}$, respectively.
  The dashed line shows that in any case, the curves are well approximated by the Gaussian $e^{-(\Lambda Y)^2/8}$.}
\label{fig_11_Cxx}
\end{center}      
\end{figure}

In Fig.(\ref{fig_11_Cxx}) we can see $C_{xx}$ as a function of $Y$ in units of $a$ for different tensions.
These curves were normalized by 

\begin{equation}
  C_{xx}(0) = \left(\frac{\sigma_0^2}{2}\right) \left\{3 + \left[ 1 - 4 r + 3r^2 \ln\left(1 + \frac{1}{r}\right)\right]\ln\left(1 + \frac{1}{r}\right)\right\} \, ,
  \label{eq_11_Cxx0}
\end{equation}

\noindent which we have obtained analytically from eq.(\ref{eq_11_Cxx1}).

\noindent First of all, we notice that the curves do almost not depend on $r$ -- and consequently on the tension.
Accordingly, the decrease of the correlation is dominated by the bending rigidity, which
is not evident, since the $\langle \Sigma_{xx} \rangle$ depends strongly on the tension. 
Secondly, $C_{xx}$ decreases relatively fast: about five times the microscopical length $a$ for any tension.
At last, in the following we will need to integrate $C_{xx}$.
It will be thus useful to remark that for any tension, it is very well approximated by

\begin{equation}
  C_{xx}(y - y') \simeq C_{xx}(0) \, e^{-\frac{\Lambda^2\, (y - y')^2}{8}} \, .
  \label{eq_11_Cxxapprox}
\end{equation}

\subsection[Evaluation of $C_{yx}$]{Evaluation of $\bm{C_{yx}}$}

Following the same route as in the last section,

\begin{equation}
  C_{yx}(y - y') \equiv \langle \Sigma_{yx}(x, y)\, \Sigma_{yx}(x,y') \rangle
  \end{equation}

\noindent can be written in terms of diagrams as

\begin{eqnarray}
C_{yx}(y - y') &=&
\sigma^2\diagpq{./figures_Chap11/peixe_13.epsi}
-
2 \sigma \kappa \diagpq{./figures_Chap11/peixe_21.epsi}
-
2 \sigma \kappa \diagpq{./figures_Chap11/peixe_22.epsi}
\nonumber\\
&+&
2 \sigma \kappa \diagpq{./figures_Chap11/peixe_3.epsi}
+
2 \sigma \kappa \diagpq{./figures_Chap11/peixe_23.epsi}
-
2 \kappa^2 \diagpq{./figures_Chap11/peixe_17.epsi}
\nonumber\\
&-&
2 \kappa^2 \diagpq{./figures_Chap11/peixe_18.epsi}
-
2 \kappa^2 \diagpq{./figures_Chap11/peixe_24.epsi}
-
2 \kappa^2 \diagpq{./figures_Chap11/peixe_25.epsi}
\nonumber\\
&+&
\kappa^2 \diagpq{./figures_Chap11/peixe_6.epsi}
+
2 \kappa^2 \diagpq{./figures_Chap11/peixe_15.epsi}
+
\kappa^2 \diagpq{./figures_Chap11/peixe_16.epsi}
\nonumber\\
&+&
\kappa^2 \diagpq{./figures_Chap11/peixe_26.epsi}
+
2 \kappa^2 \diagpq{./figures_Chap11/peixe_27.epsi}
+
\kappa^2\diagpq{./figures_Chap11/peixe_28.epsi}
\, ,\nonumber\\
\phantom{}
\end{eqnarray}

\noindent which reads

\begin{eqnarray}
  C_{yx}(y - y') &=& \sigma^2 \, G_{0,2}\, G_{2,0} + 2 \sigma \kappa \left[G_{2,1}\left(G_{2,1} + G_{0,3}\right) - G_{0,2}\left(G_{4,0} + G_{2,2}\right)\right] \nonumber\\
  &-&2\,  \kappa^2 \left(G_{2,1} + G_{0,3}\right)\left(G_{4,1} + G_{2,3}\right) + \kappa^2 \, G_{2,2} \left(G_{4,0} + 2 \, G_{2,2} + G_{0,4} \right) \nonumber\\
  &+& \kappa^2 \,  G_{0,2} \, \left(G_{6,0} + 2\, G_{4,2} + G_{2,4} \right)\, . \nonumber\\
\end{eqnarray}

In the thermodynamical limit, we obtain

\begin{eqnarray}
  C_{yx}(y - y') &=& 64 \pi^2 \sigma_0^2 \left\{ r^2\, \left[\frac{\tilde{B}_{10}^2(\Lambda Y)}{(\Lambda Y)^2}  - \frac{\tilde{B}_{10}(\Lambda Y) \, \tilde{B}_{21}(\Lambda Y)}{\Lambda Y}\right]+ 2 r \, \frac{\tilde{B}_{10}(\Lambda Y) \, \tilde{B}_{12}(\Lambda Y)}{(\Lambda Y)^2} \right. \nonumber\\
 &+& \left[2 \, \frac{\tilde{B}_{12}(\Lambda Y) \, \tilde{B}_{23}(\Lambda Y)}{\Lambda Y} - \frac{\tilde{B}_{03}(\Lambda Y) \, \tilde{B}_{32}(\Lambda Y)}{\Lambda Y} - \frac{\tilde{B}_{21}(\Lambda Y) \, \tilde{B}_{14}(\Lambda Y)}{\Lambda Y}\right. \nonumber\\
    &+& \left. \left. \frac{\tilde{B}_{03}(\Lambda Y) \, \tilde{B}_{21}(\Lambda Y)}{(\Lambda Y)^2} + \frac{\tilde{B}_{10}(\Lambda Y) \, \tilde{B}_{14}(\Lambda Y)}{(\Lambda Y)^2}\right] \right\}\, ,
\label{eq_11_Cyx}
\end{eqnarray}

\noindent where $Y = |y - y'|$ and $\tilde{B}_{ij}$ is defined in eq.(\ref{eq_11_Bijtilde}).

\begin{figure}[H]
  \centerline{\includegraphics[width=0.5\columnwidth]
{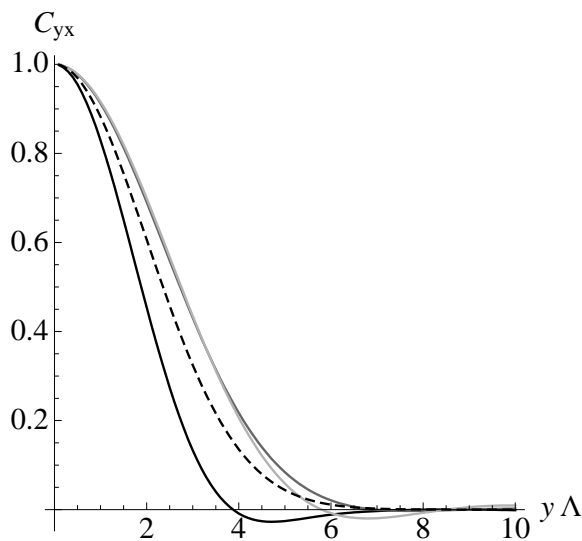}}
\caption{Plot of the normalized correlation of the transverse component of the stress
tensor for different values of $r$ as a function of the distance in units of $\Lambda^{-1}$.
The black line corresponds to $r=1$, dark gray corresponds to
$r = 10^{-2}$ and light gray
corresponds to $r =10^{-4}$.
For $\kappa \approx 25 \, k_\mathrm{B} T$, it corresponds to $\sigma = 4 \times 10^{-3}\, \mathrm{N/m}$, $\sigma = 4\times 10^{-5} \, \mathrm{N/m}$ and $\sigma = 4 \times 10^{-7} \, \mathrm{N/m}$, respectively.
The dashed line
shows a rough Gaussian approximation.}
\label{fig_11_Cyx}
\end{figure}

At last, we show in Fig.~\ref{fig_11_Cyx} the behavior of $C_{yx}$ normalized by the auto-correlation

\begin{equation}
  C_{yx}(0) = \left(\frac{\sigma_0^2}{2}\right) \left\{1 + \left[ 1 + r^2 \ln\left(1 + \frac{1}{r}\right)\right]\ln\left(1 + \frac{1}{r}\right)\right\} \, ,
  \label{eq_11_Cyx0}
\end{equation}

\noindent obtained analytically from eq.(\ref{eq_11_Cyx}).

The correlation $C_{yx}$ is very similar to $C_{xx}$, sharing with it three features:

\begin{enumerate}
\item as before, $C_{yx}$ normalized by the auto-correlation depends only weakly on the tension, specially in the regime of low tensions. The shape of the correlation is dominated by the bending rigidity;
\item $C_{yx}$ relaxes over approximately five times the microscopical length $a$;
\item the same approximation
  \begin{equation}
    C_{yx}(y - y') \simeq C_{yx}(0) \, e^{-\frac{\Lambda^2 (y - y')^2}{8}}
    \label{eq_11_Cyxapprox}
  \end{equation}
  \noindent holds, although it is less good.
\end{enumerate}

\subsection[Evaluation of $C_{zx}$]{Evaluation of $\bm{C_{zx}}$}

The correlation of the normal component

\begin{equation}
  C_{zx}(y - y') = \langle \Sigma_{zx}(x,y) \, \Sigma_{zx}(x,y') \rangle 
  \end{equation}

\noindent is the simplest one to evaluate.
Diagrammatically, we have

\begin{eqnarray}
C_{zx}(y - y') = 
&-&
\sigma^2 \diag{./figures_Chap11/ligne_1.epsi}
+2 \sigma \kappa
\diag{./figures_Chap11/ligne_2.epsi}
+2 \sigma \kappa
\diag{./figures_Chap11/ligne_3.epsi}\nonumber\\
&-& \kappa^2
\diag{./figures_Chap11/ligne_4.epsi}
- 
2 \kappa^2 \diag{./figures_Chap11/ligne_5.epsi}
 - \kappa^2
\phantom{1}\diag{./figures_Chap11/ligne_6.epsi} \, ,\nonumber \\
&=& - \sigma^2 \, G_{2,0} + 2 \sigma \kappa \left(G_{4,0} + G_{2,2}\right) - \kappa^2 \left(G_{6,0} + 2 G_{4,2} + G_{2,4}\right) \, . \nonumber\\
\end{eqnarray}

This time, in the thermodynamical limit, one can integrate $C_{zx}$ not only over the angular coordinate, but also over $q$, obtaining

\begin{equation}
C_{zx}(y - y') = 32\pi\beta\kappa \, \sigma_0^2 \left\{\frac{r \left[1 - J_0(\Lambda Y)\right] + J_2(\Lambda Y)}{(\Lambda Y)^2}\right\}\,, 
\label{Czx_B}
\end{equation}

\noindent and accordingly

\begin{equation}
  C_{zx}(0) = 8\pi\beta\kappa \, \left(\frac{\sigma_0^2}{2}\right) \left(1 + 2\, r\right) \, .
  \label{eq_11_Czx0}
\end{equation}

\begin{figure}[H]
\centerline{\includegraphics[width=0.5\columnwidth]
{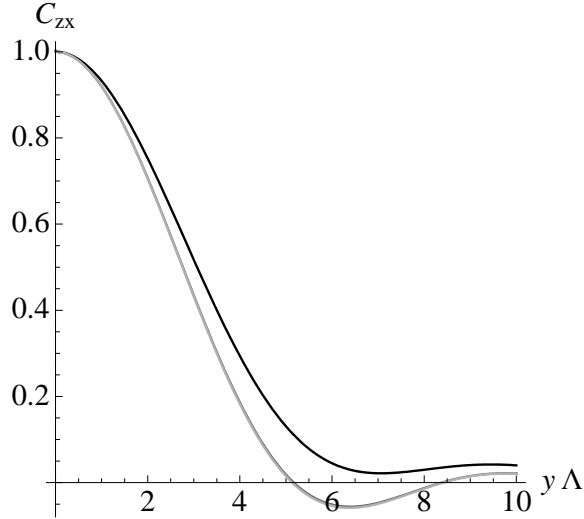}}
\caption{Plot of the normalized correlation of the normal component of the stress
tensor for different $r$ as a function of $\Lambda
y$.
As before, the black line corresponds to $r=1$, dark gray corresponds to
   $r=10^{-2}$ and light gray
   corresponds to $r=10^{-4}$ (superposed).}  
\label{fig_11_Czx}
\end{figure}

As we can see in Fig.~\ref{fig_11_Czx}, $C_{zx}$ normalized by $C_{zx}(0)$ has roughly the same be features of the former correlations: it does almost not depend on the tension and it becomes negligible for distances bigger than $\approx 5 \, a$.
This time, however, as $C_{zx}$ is very simple, directly given by an analytical function.

\subsection{Summing-up}
\label{subsection_11_disc}

Here we sum-up some important results obtained in this section.
First, the three correlations normalized by it's value at $y = y'$ share the following features: 

\begin{enumerate}
  \item the normalized correlation depends only weakly on the tension;
  \item they present roughly a Gaussian behavior. Moreover, $C_{xx}$ and $C_{yx}$ are well approximated by

    \begin{equation}
      \frac{C_{xx}(y - y')}{C_{xx}(0)} = \frac{C_{yx}(y - y')}{C_{yx}(0)} = e^{-\frac{\Lambda^2 (y - y')^2}{8}} \, ,
      \label{eq_11_norm}
    \end{equation}

    \noindent where $\Lambda^{-1} = a$ is the smallest wave-length cut-off;
   \item the correlation is negligible for distances larger than $5 \, a$, which is really small, considering $a \approx 5 \, \mathrm{nm}$.
\end{enumerate}

\begin{figure}[H]
\centerline{\includegraphics[width=0.7\columnwidth]
{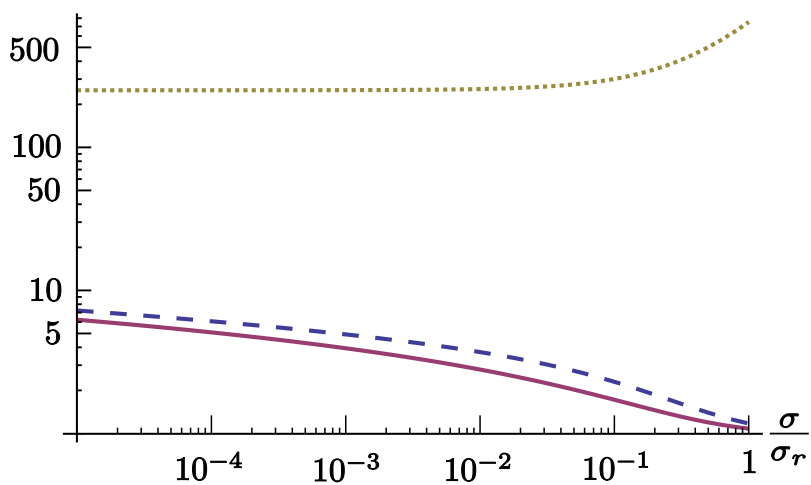}}
\caption{Plot of $C_{xx}(0)$ (blue dashed line), $C_{yx}(0)$ (red solid line) and $C_{zx}(0)$ (yellow dotted line) in units of $\sigma_0$ as a function of $r = \sigma/\sigma_r$. For the plot of $C_{zx}(0)$, we have chosen $8 \pi \beta \kappa = 500$ as a typical value.}
\label{fig_11_C0}
\end{figure}

Finally, as the dependence on the tension happens mainly through the correlation at $y = y'$, it is interesting to plot $C_{xx}(0)$, $C_{yx}(0)$ and $C_{zx}(0)$, given in eq.(\ref{eq_11_Cxx0}), eq.(\ref{eq_11_Cyx0}) and eq.(\ref{eq_11_Czx0}), respectively, as a function of the tension (see Fig.~\ref{fig_11_C0}).
Two important features of these curves will be reflected in the force fluctuation:

\begin{enumerate}
\item first, in the three cases, the dependence on the tension is not accentuated, implying that one could actually simply neglect from start every diagram proportional to $\sigma^2$ and $\sigma \kappa$ in the last sections;
\item secondly, the correlation of the component of the stress tensor normal to the membrane $C_{zx}$ is far bigger than the two other contributions, which are comparable among them. 
\end{enumerate}

\section{Fluctuation of the force}
\label{section_11_fluct}

To obtain square of the force fluctuation in each direction, defined in eqs.(\ref{eq_11_def_corryx})--(\ref{eq_11_def_corrzx}), we must integrate the correlation function twice over the cut's length:

\begin{eqnarray}
  \label{eq_11_Dfx}
(\Delta f_x)^2 &=& \iint^{L/2}_{-L/2} C_{xx}(y - y') \, dy dy' \, ,\\
(\Delta f_y)^2 &=&\iint^{L/2}_{-L/2} C_{yx}(y - y') \, dydy'\, ,\\
(\Delta f_z)^2&=&
\iint^{L/2}_{-L/2} C_{zx}(y - y')\, dydy' \, .
\label{eq_11_Dfz}
\end{eqnarray}
 
\begin{figure}[H]
\centerline{\includegraphics[width=0.85\columnwidth]
{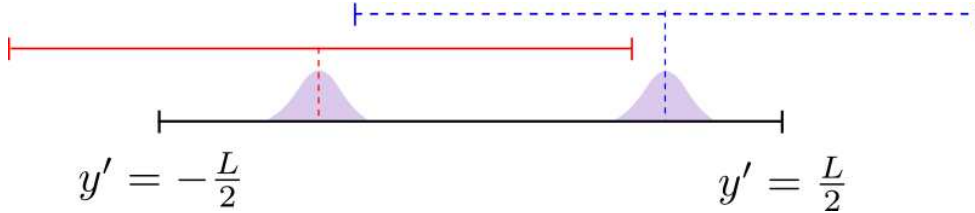}}
\caption{To obtain the square of the force fluctuation in each direction, one has to integrate the correlation function, here represented in purple, from $-L/2$ to $L/2$ 
  at each point of the cut (domain shown by the blue dashed horizontal line and by the solid red line) and then sum all the contributions over the black line.
As the correlation function decreases quickly compared to the length of the cut, remark that the final result would change only minimally if the colored lines where far lengthier.
}
\label{fig_11_aide}
\end{figure}

\noindent In the last section, we have seen that the correlations decrease very quickly, with a characteristic length of about $\ell = 5 \, a \approx 25 \, \mathrm{nm}$.
Recalling that $L$ is the length of the projected cut, it is reasonable thus to assume $L \gg \ell$.
In Fig.~\ref{fig_11_aide}, we can see a graphical representation of the integrals of eqs.(\ref{eq_11_Dfx})--(\ref{eq_11_Dfz}) for the case where $L \gg \ell$.

\noindent In Fig.~\ref{fig_11_aide}, we see that for $L \gg \ell$, eqs.(\ref{eq_11_Dfx})--(\ref{eq_11_Dfz}) can be well approximated by

\begin{eqnarray}
  (\Delta f_x)^2 &\simeq& L\, \int_{-\infty}^{\infty} C_{xx}(y) \, dy \, , \\
  (\Delta f_y)^2 &\simeq& L\, \int_{-\infty}^{\infty} C_{yx}(y) \, dy \, , \\
  (\Delta f_z)^2 &\simeq& L\, \int_{-\infty}^{\infty} C_{zx}(y) \, dy \, .
\end{eqnarray}

\noindent For the two first cases, it is not possible to obtain an analytical equation from the exact expression of the correlation (eq.(\ref{eq_11_Cxx}) and eq.(\ref{eq_11_Cyx})).
We will thus use the Gaussian approximation given in eq.(\ref{eq_11_norm}), yielding

\begin{eqnarray}
  (\Delta f_x)^2 &\simeq& \sqrt{8\pi}\,  \frac{L}{\Lambda}\, C_{xx}(0) \nonumber  \\
  &\simeq& \sqrt{8\pi}\,  \frac{L}{\Lambda}\, \left(\frac{\sigma_0^2}{2}\right) \left\{3 + \left[ 1 - 4 r + 3r^2 \ln\left(1 + \frac{1}{r}\right)\right]\ln\left(1 + \frac{1}{r}\right)\right\} \, ,\nonumber \\
  \\
  (\Delta f_y)^2 &\simeq& \sqrt{8\pi}\,  \frac{L}{\Lambda}\, C_{yx}(0) \nonumber \\
  &\simeq& \sqrt{8\pi}\,  \frac{L}{\Lambda}\, \left(\frac{\sigma_0^2}{2}\right) \left\{1 + \left[ 1 + r^2 \ln\left(1 + \frac{1}{r}\right)\right]\ln\left(1 + \frac{1}{r}\right)\right\} \, , \nonumber \\
  \\
  (\Delta f_z)^2 &\simeq& \frac{128 \pi \beta \kappa}{3} \, \frac{L}{\Lambda} \left(\frac{\sigma_0^2}{2}\right) \,  (1 + 3r) \, .
\end{eqnarray}

\begin{figure}[H]
\centerline{\includegraphics[width=0.7\columnwidth]
{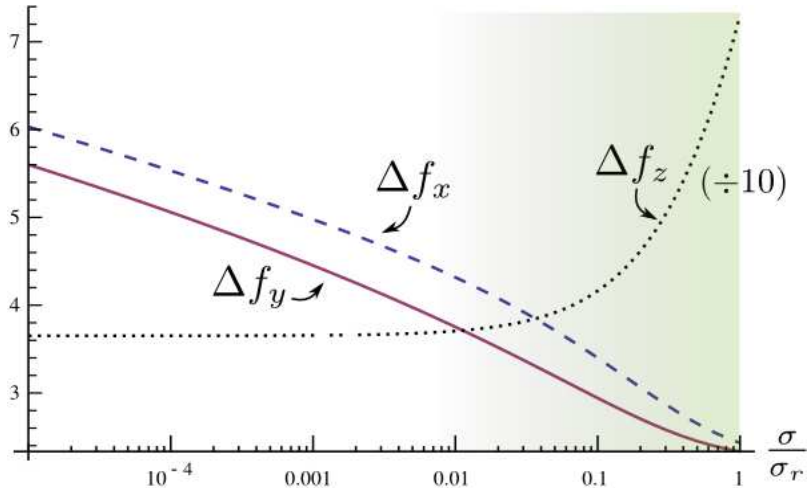}}
\caption{Plot of $\Delta f_x$ (blue dashed line), $\Delta f_y$ (red solid line) and $\Delta f_z$ (black dotted line) in units of $\sigma_0 \, \sqrt{L/\Lambda}$ as a function of $r = \sigma/\sigma_r$. In the plot of $\Delta f_z$, we have chosen $8 \pi \beta \kappa = 500$ as a typical value. Note that the $\Delta f_z$ is much intenser -- the yellow curve corresponds to the actual values divided by ten.
The shaded green area indicates the region where the effects of stretching, which we have neglected, should be important.}
\label{fig_11_fluct}
\end{figure}

\noindent Not surprisingly, the dependence of the force fluctuation per unit length in terms of the tension shown in Fig.~\ref{fig_11_fluct} is very similar to the trend shown in Fig.~\ref{fig_11_C0}. 

Essentially, with $\Lambda^{-1} = a$, we can say that

\begin{equation}
  \Delta f_x \approx \Delta f_y \propto \sqrt{\frac{L}{a}} \, \frac{k_\mathrm{B} T}{a}  \,
\end{equation}

\noindent and

\begin{equation}
  \Delta f_z \propto \sqrt{\frac{L}{a}} \, \sqrt{\frac{k_\mathrm{B} T \, \kappa}{a^2}} \,.
\end{equation}

\noindent In both cases, the dependence on $L/a$ can be understood by remembering that the correlations of the projected stress tensor decreases over a characteristic length of approximately $5\, a$ for any component.
For a cut of length $L$, we have thus roughly $L/a$ uncorrelated patches of membrane, which with the Central Limit Theorem explains the factor $\sqrt{L/a}$.
Numerically, each patch contributes approximately $k_\mathrm{B} T/a \sim (4\times 10^{-21})/(5\times 10^{-9}) \, \mathrm{N} \sim  1 \, \mathrm{pN}$ for the transverse and parallel components of the force fluctuation and $\sqrt{k_\mathrm{B} T \, \kappa/a^2} \sim \sqrt{(4\times 10^{-21} \times 10^{-19})/(5\times 10^{-9})^2} \, \mathrm{N} \sim 4 \, \mathrm{pN}$ for the normal component of the fluctuation. 

A remarkable point is the fact that both the transverse $\Delta f_x$ and the parallel $\Delta f_y$ components of the fluctuation depend only on the temperature and on the microscopical cutoff $a$, regardless of the rigidity or tension of the membrane.

\section{In a nutshell}

In this chapter, we have evaluated for the first time the fluctuation of the force exchanged through a cut of projected length $L$ in a planar membrane.
To do so, we have introduced some diagrammatic tools useful in the following chapters.
The calculation was done in two steps: first, we have evaluated the correlation of some elements of the projected stress tensor and after we have integrated them over the cut.
These correlations present some interesting features: their shape do almost not depend on the tension and they decrease very quickly, becoming negligible for distances larger than $5 \, a \approx 25 \, \mathrm{nm}$, with $a$ of the order of the membrane thickness. 
%Consequently, the force fluctuation is also almost independent on the tension.
For the fluctuation of the force component transverse to the cut, $\Delta f_x$,  and parallel to it, $\Delta f_y$, we have obtained the same scaling behavior

\begin{equation}
  \Delta f_x \approx \Delta f_y \propto \sqrt{\frac{L}{a}} \, \frac{k_\mathrm{B} T}{a}  \, ,
\end{equation}

\noindent whereas for the component perpendicular to the membrane, $\Delta f_z$, we have obtained

\begin{equation}
  \Delta f_z \propto \sqrt{\frac{L}{a}} \, \frac{k_\mathrm{B} T}{a} \, \sqrt{\frac{k_\mathrm{B} T}{\kappa}}  \,. 
\end{equation}

\noindent These equations hold up to a numerical factor of the order of the unity that depends very weakly on the tension.
Interestingly, the scaling law for $\Delta f_x$ and $\Delta f_y$ depends neither on the bending rigidity.

%% file: chap2.tex
\chapter{Quasi-spherical vesicles}
\label{chapitre_vesicle}

In chapter~\ref{introd}, we have seen that vesicles are widely used in experiments, since they are easy to assemble and to manipulate.
Vesicles are used both in micropipette and adhesion experiments, in which one
increases the mechanical tension $\tau$ by flattening the membrane's
fluctuations and in contour analysis experiments, in which one measures $r$,
the large-scale counterpart of the tension $\sigma$, through the fluctuation spectrum.
In the chapter~\ref{chapitre/planar_membrane}, we have derived $\tau$ as a
function of $\sigma$ for planar membranes, obtaining

\begin{equation}
\tau = \sigma - \sigma_0 \left[ 1 - \frac{\sigma}{\sigma_r} \ln \left(1 +
    \frac{\sigma_r}{\sigma}\right)\right] \, 
\label{eq_2_tauplan}
\end{equation}

\noindent in the limit of large membranes.
In this equation, $\sigma_r = \kappa \Lambda^2$ is a tension of the order of
the rupture tension, $\Lambda = 1/a$, where $a$ is a microscopical cut-off of
the order of the membrane thickness and $\sigma_0 = \sigma_r/(8 \pi \beta \kappa)$.
This relation reduces simply to $\tau \simeq \sigma - \sigma_0$ for membranes under small tensions ($\sigma < 10^{-2} \, \sigma_r$).
We do not know, however, if eq.(\ref{eq_2_tauplan}) still holds for vesicles since they have a different geometry and they present a supplementary volume constraint. 

In this chapter, we shall thus calculate $\tau$ from the projected stress tensor for the case of quasi-spherical vesicles.
We shall examine both the usual case of a closed vesicle whose volume is constrained and the case of poked vesicles.
We call poked vesicles those vesicles that are free to exchange liquid with
the outer media.
Experimentally, it can be achieved by embedding special proteins in
the  membrane or by making holes in it with a micropipette.
They can however keep a pressure difference with the outer media if the
inner/outer fluid contains molecules bigger than the holes, so they can not
transit across the membrane.

In particular, we will address the following interesting questions, the first
three having experimental implications while the last question deals with a
more theoretical issue:

\begin{enumerate}
  \item What is the difference between $\tau$ for a quasi-spherical vesicle
    (closed or poked) and $\tau$ for a planar membrane? Is there a
    characteristic radius over which they coincide, in which case one can
    simply consider the relation given in eq.(\ref{eq_2_tauplan})?
  \item How does the volume constraint affect the expression for $\tau$?
  \item Can $\tau$ be negative, in which case the inner pressure of the vesicle would be smaller than the outer?
  \item Can $\tau$ be obtained by differentiating the free-energy with respect to the projected area? If so, what does projected area mean in the case of a vesicle?
\end{enumerate}

Usually, as discussed in chapter~\ref{introd}, one should use the
ADE-model Hamiltonian.
We will however use the simpler Helfrich Hamiltonian, introduced in
section~\ref{section_2_param}.
There, this choice will be justified.
Following the same reasoning as in section~\ref{section_projected_stress}, we derive the projected stress tensor for a quasi-spherical geometry in section~\ref{section_2_tensor}.
In section~\ref{section_2_correlations} we present some averages and correlations for the case of closed vesicles, which are used in section~\ref{section_2_tau_closed} to evaluate $\tau_{\mathrm{closed}}$.
The results of the last two sections are easily transposed to the case of
poked vesicles in section~\ref{section_2_poked}, where we obtain
$\tau_{\mathrm{poked}}$.

Finally, we discuss the first three questions in
section~\ref{section_2_discussion}.
We show that it is justified to use the relation given in eq.(\ref{eq_2_tauplan}) for quasi-spherical vesicles, closed or poked, with a radius bigger than $1 \,
\mathrm{\mu m}$.
Besides, the volume constraint seems to be unimportant for the dependence of
$\tau$ on $\sigma$.
Experimentally, however, $\sigma$ is not measurable.
With vesicles, the true control parameter is the area excess, which depends
considerably more on the volume constraint.
We expect thus some difference between closed and poked vesicles, specially in
the case of small vesicles. 
Lastly, we show that negative values of $\tau$ are expected well before
the transition to oblate shapes in both cases, implying that vesicles may
support an internal pressure smaller than the outer.

At last, in section~\ref{section_2_differentiation} we shall address the more
theoretical question of recovering $\tau$ for closed and poked vesicles by
differentiating the free-energy.
Differently for the case of planar membranes, the sense of the term projected
area for a vesicle is not clear: it can refer to the area of a sphere with
the average radius or the area of a sphere of volume $V$, for instance.
In this section, we shall see that indeed the term is not well-defined,
since it corresponds to different area depending on whether the vesicle is
closed or not.

All calculations and discussions presented here were done with the collaboration of Jean-Baptiste Fournier and Alberto Imparato.
The main results can be found in ref.~\cite{Barbetta_10}.

\section{Parametrization and effective Hamiltonian}
\label{section_2_param}

\begin{figure}[H]
  \begin{center}
    \includegraphics[scale=0.7,angle=0]{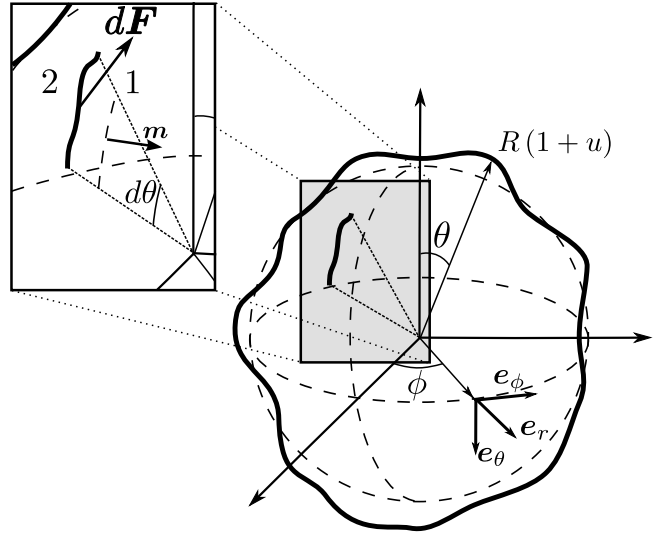}
    \caption{Parametrization in spherical coordinates of a vesicle (bold line) fluctuating around a reference sphere (dashed line). The inset shows the force exchanged through a cut that separates region $1$ and region $2$.}
  \label{fig_2_param}
  \end{center}      
\end{figure}

We consider a quasi-spherical vesicle whose area $A$ and volume $V$ are fixed (for closed vesicles).
Its shape is parametrized by

\begin{equation}
  \bm{r} = R \left[1 + u\left(\theta, \phi\right)\right]\, \bm{e}_r\, ,
\end{equation}

\noindent where $u \ll 1$ (see Fig.~\ref{fig_2_param}). 
For closed vesicles,
as in Seifert's work \cite{Seifert_95}, we choose the sphere of volume $V$
as the reference sphere, so that $R = \left(\frac{3}{4} V/\pi\right)^{1/3}$. 
Indeed, in experiments, one can control $V$ by lowering the ion concentration
of the outer media of the vesicle, so it inflates at its maximum.
Equivalently, using a micropipette, one can apply a large pressure difference
through the membrane.
In both cases, the excess area is negligible and thus from the optically
resolvable shape, one deduces $V$.
In the case of poked vesicles, as there are no volume constraints, one cannot control $V$.
Instead, one can measure the average radius of the vesicle and deduce the
average volume of the vesicle.
In this case, we choose thus simply the average vesicle's shape as the reference sphere, so that $\langle u(\theta, \phi) \rangle = 0$.

The area constraint reads

\begin{equation}
 A = \int_S dA \, ,
\end{equation}

\noindent with 

\begin{equation}
  dA = \left| \partial_{\theta}\bm{r} \times \partial_{\phi}\bm{r}\right| \, d\theta d\phi \,,
\label{eq_2_diff_area}
\end{equation}

\noindent where $\partial_\theta \bm{r} \equiv \partial \bm{r}/\partial \theta$, $\partial_\phi \bm{r} \equiv \partial \bm{r} /\partial \phi$, yielding, in terms of $u$

\begin{equation}
A=R^2 \int_0^{\pi} d\theta \int_0^{2\pi} d\phi \,  [1 + u(\theta,\phi)]\sqrt{\left[1 + u(\theta,\phi)\right]^2\sin^2 \theta + u_\phi^2 + u_\theta^2 \sin^2\theta}\, .
\label{eq_2_Area_0}
\end{equation}

\noindent Here and throughout this section, $u_i \equiv \partial u/\partial i$, $u_{ij} \equiv \partial^2 u/\partial i \partial j$, where $i, j \in \{\theta, \phi\}$. Latin indices will denote either $\theta$ or $\phi$, not $r$.
The volume constraint, important for closed vesicles, reads 

\begin{equation}
V = \frac{1}{3}R^3 \int_0^{\pi} d\theta \int_0^{2\pi} d\phi \,  [1 +
u(\theta,\phi)]^3 \sin\theta \, . 
\label{eq_2_Volume_0}
\end{equation}

As we have seen in section \ref{model_model}, the energy of a vesicle is best described by the area-difference elasticity ($ADE$) model.
Seifert \cite{Seifert_95} has however shown that in the quasi-spherical limit, the $ADE$ Hamiltonian was equivalent to the minimal Helfrich model, i. e., the spontaneous curvature ($SC$) model with vanishing spontaneous curvature.
Hence, we adopt the latter, which corresponds to an effective Hamiltonian

\begin{equation}
  \mathcal{H} = \int_S 2 \kappa H^2 \, dA \, ,
\end{equation}

\noindent supplemented by the area and (if necessary) volume constraints given in eqs.(\ref{eq_2_Area_0})--(\ref{eq_2_Volume_0}).
While the volume constraint is quite easy to implement, it is difficult to handle the surface constraint exactly \cite{Seifert_95}.
We shall therefore use the traditional approach, namely introducing a Lagrange multiplier $\sigma$ playing the role of a tension in order to take into account the area constraint.
Again, as discussed by Seifert \cite{Seifert_95}, this approach gives correct results in the small excess area limit, in which we shall place ourselves in the following.
The effective Hamiltonian thus reads

\begin{equation}
\mathcal{H} = \int_S \left(2 \kappa H^2 + \sigma\right) \, dA \,,
\end{equation}

\noindent with the additional constraint given by eq.(\ref{eq_2_Volume_0}) for closed vesicles.

From differential geometry, $dA$ is given by eq.(\ref{eq_2_diff_area}) and

\begin{equation}
  H = \frac{2bB - cA - aC}{AC - B^2} \, ,
  \label{eq_2_H}
\end{equation}

\noindent with $A = (\partial_\theta \bm{r})^2$, $B = (\partial_\theta \bm{r})\cdot (\partial_\phi \bm{r})$, $C = (\partial_\phi \bm{r})^2$, $a = \bm{n} \cdot \partial_\theta^2 \bm{r}$, $b = \bm{n} \cdot \partial_\theta\partial_\phi \bm{r}$, $c = \bm{n} \cdot \partial_\phi^2 \bm{r}$, $\bm{n} = (\partial_\theta \bm{r} \times \partial_\phi \bm{r})/|\partial_\theta \bm{r} \times \partial_\phi \bm{r}|$ being the normal to the surface.
Up to the second order on $u(\theta, \phi)$, we have thus

\begin{equation}
  \mathcal{H} = \int_S h\left(u, \{u_i\}, \{u_{ij}\}\right) \, d\theta d\phi \, ,
\end{equation}

\noindent with \cite{Helfrich_86} \cite{Milner_87} 

\begin{eqnarray}
h&=&(2 \kappa +R^2 \sigma)\sint  \nonumber \\
&+& 2 \sint \left[R^2 \sigma u - \kappa \p{u_{\phi\phi} \csc^2 \theta
    + u_\theta \cot \theta  +u_{\theta\theta}}\right] \nonumber\\
&+& \frac{1}{2}\sint \left[ 2 R^2 \sigma u^2 +(2 \kappa +R^2 \sigma) (\ut^2
  +\up^2 \csc^2 \theta )   \right. \nonumber\\
&+& \kappa \p{\ut \cot\theta +u_{\phi\phi} \csc^2 \theta}^2  +\kappa \,
  u_{\theta\theta} \, (u_{\theta\theta}+4 u) \nonumber\\ 
&+& \left. 2\kappa \, (u_{\theta\theta} +2 u)\p{\ut \cot\theta
    + u_{\phi\phi} \csc^2 \theta} \right] + \mathcal{O}(u^3)\, .
\label{eq_2_energie_2}
\end{eqnarray}

\section{Derivation of the stress tensor for a quasi-spherical geometry}
\label{section_2_tensor}

Let us consider an infinitesimal cut at constant
longitude ($\phi$ constant) separating a region $1$ from a region $2$ (see Fig.~\ref{fig_2_param}).
The normal to the projection of this cut onto the reference sphere
is $\bm{m} = \bm{e}_\phi$.
Analogously to the case of planar membranes presented in section~\ref{section_projected_stress}, the projected stress tensor $\bm{\Sigma}$ in spherical geometry relates by definition linearly the
force $d\bm{F}$ that region $1$ exerts on region $2$
to the angular length $ds = d\theta$ of the projection of the cut onto the reference sphere:

\begin{eqnarray}
d\bm{F} &=& \bm{\Sigma} \cdot \bm{m} \, ds \,, \nonumber\\ 
&=& \left(\Sigma_{\theta\phi} \, \bm{e}_\theta +
\Sigma_{\phi\phi} \, \bm{e}_\phi + \Sigma_{r\phi} 
\, \bm{e}_r \right) \, d\theta \, .
\end{eqnarray}

Likewise, for a cut at constant latitude ($\theta$ constant),
with $\bm{m} = \bm{e}_\theta$ and $ds = d\phi$, we have

\begin{eqnarray}
d\bm{F} &=& \bm{\Sigma} \cdot \bm{m} \, d\phi \,, \nonumber\\ 
&=& \left(\Sigma_{\theta\theta}  \, \bm{e}_\theta +
\Sigma_{\phi\theta} \, \bm{e}_\phi + \Sigma_{r\theta} \, \bm{e}_r \right) \, d\phi \, .
\end{eqnarray}

\noindent For an oblique cut, $d \bm{F}$ is obtained by decomposing $\bm{m}$ along $\bm{e}_\theta$ and $\bm{e}_\phi$.

\begin{figure}[H]
  \begin{center}
    \includegraphics[scale=0.35,angle=0]{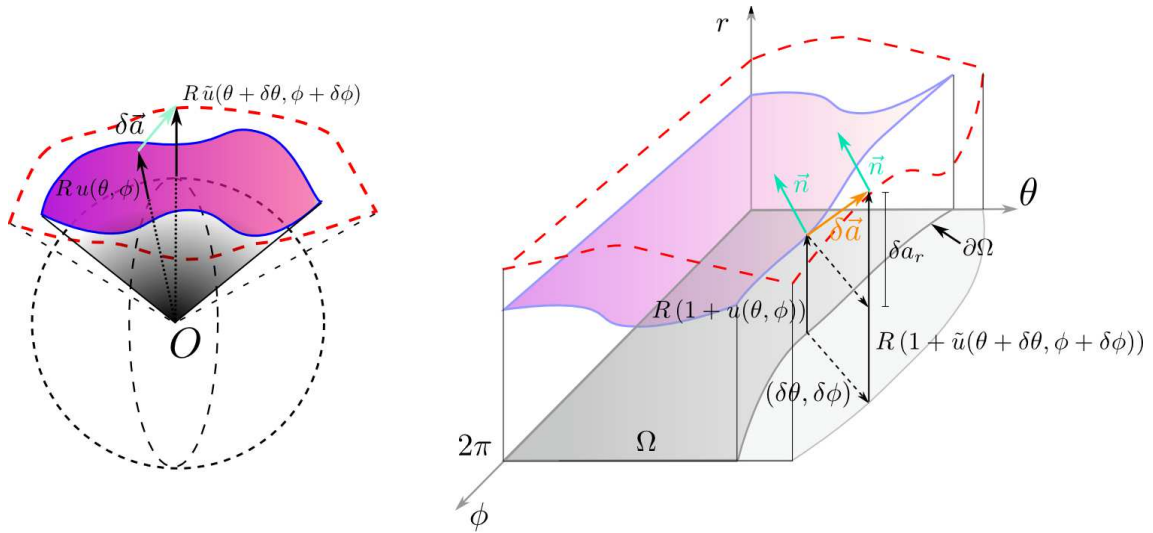}
    \caption{On the left we see a patch of quasi-spherical membrane before (shaded shape) and after (dashed red shape) the displacement $\delta \bm{a}$. At right, we show the same displacement in the $(\theta,\phi, r)$ space. From this drawing, it is easier to see the relation between $\delta \bm{a}$, $\delta u$, $\delta \theta$ and $\delta \phi$.}
  \label{fig_2_sphere}
  \end{center}      
\end{figure}

The derivation of the projected stress tensor in spherical geometry follows the same route as for planar geometry (section~\ref{section_projected_stress}).
We consider a patch of membrane delimited by a closed curve,
corresponding to a domain $\Omega$ on the reference sphere enclosed by the curve
$\partial \Omega$.
The membrane within the patch is assumed to be deformed, at equilibrium, by means of a distribution of surface and boundary forces (and a distribution of boundary torques).
To each point of this patch, we impose an
arbitrary displacement $\delta \bm{a} = \delta a_r \, \bm{e}_r + \delta
a_\theta \, \bm{e}_\theta + \delta a_\phi \, \bm{e}_\phi$ that keeps the
orientation of the membrane's normal $\bm{n}$ constant along the boundary, so that the torques produce no work (see Fig.~\ref{fig_2_sphere}).

On the one hand, the boundary energy variation, after integration by parts, reads

\begin{equation}
\delta \mathcal{H} = \int_{\partial \Omega} m_i \left[ h \, \delta i + \left(\frac{\partial h}{\partial u_i} -
    \partial_i\frac{\partial h}{\partial u_{ij}}\right)\delta u +
    \frac{\partial h}{\partial u_{ij}} \, \delta u_j\right]\, ds \,,
\label{eq_2_delta_h_1}
\end{equation} 

\noindent where $ds$ is the arc-length in the $(\theta,\phi)$ space, $h$ is given by eq.(\ref{eq_2_energie_2}), $\bm{m}$ is the normal to $\partial \Omega$, and $\delta i \in \{\delta \theta,\delta \phi\}$ corresponds
to the variation of $\partial \Omega$.
On the other hand, the
work of the force exerted through the boundary reads

\begin{equation}
\delta \mathcal{H} = \int_{\partial\Omega} \delta \bm{a} \cdot \bm{\Sigma} \cdot \, \bm{m}\,ds\, .
\label{eq_2_delta_h_2}
\end{equation} 

By comparing eqs.(\ref{eq_2_delta_h_1}) and (\ref{eq_2_delta_h_2}), one can obtain $\bm{\Sigma}$.
Accordingly, one has to derive $\delta \theta$, $\delta \phi$, $\delta u$, $\delta u_\theta$
and $\delta u_\phi$ in terms of $\delta \bm{a}$.
The first three can be
obtained by identifying the new membrane's shape $\tilde{u} = u + \delta
u$ with the translation of the old one: $R[1 + \tilde{u}(\theta + \delta \theta, \phi + \delta
\phi)] \, \bm{e}_r(\theta + \delta \theta, \phi + \delta \phi) = R[1 +
u(\theta,\phi)]\, \bm{e}_r + \delta \bm{a}$ (see Fig.~\ref{fig_2_sphere}).
This leads to 

\begin{eqnarray}
\label{eq_2_delta_a_r}
\delta a_r &=& R\, (u_\theta \, \delta \theta + u_\phi \, \delta\phi + \delta
u)\, ,\\
\delta a_\theta &=& R\, (1 + u) \, \delta \theta \, ,\\
\delta a_\phi &=& R\, (1 + u) \sin\theta \, \delta \phi\, .
\label{eq_2_delta_a_phi}
\end{eqnarray}

Let $\delta \bm{n} = \tilde{\bm{n}}(\theta + \delta
\theta, \phi + \delta \phi) - \bm{n}(\theta,\phi)$ be the variation of the
normal, where $\tilde{\bm{n}}$ is the normal to the shape defined by  
$\tilde{u}(\theta,\phi)$ and $\bm{n} = \bm{t}_\theta \times \bm{t}_\phi/|\bm{t}_\theta
\times \bm{t}_\phi|$, with $\bm{t}_i = \partial_i \bm{r}$.
The variation of the normal vanishes (implying no work of the torques) if $\delta \bm{n} \cdot \bm{t}_\theta = 0$ and $\delta\bm{n} \cdot \bm{t}_\phi = 0$ over the
border, yielding

\begin{eqnarray}
\delta u_\theta &=& \frac{1}{1+u}\Big\{\delta \theta \left[(1+u)^2+ 2 u_\theta^2
    -(1+u)\, u_{\theta\theta}\right] \phantom{\frac{\delta}{\delta}} \nonumber \\
&+& \frac{\delta \phi}{\sin\theta}\left[(1+u)\, u_\phi\cos\theta
    -(1+u) \, u_{\theta\phi}\sin\theta 
  \right.  \nonumber \\
&+& \left. 2 u_\phi u_\theta \sint
  \right] + \delta u \,  u_\theta\Big\}\, ,\nonumber\\
\phantom{1}
\label{eq_2_eq4}\\
\delta u_\phi &=& \frac  1{1+u} \Big\{ \delta\theta \left[(1+u)\, u_\phi \cot\theta
    + 2 u_\theta u_\phi \right.\nonumber \\ 
 &-&\left. (1+u)\, u_{\theta\phi}\right] + \delta\phi \left[(1+u)^2 \sin^2\theta+
   2 u_\phi^2 \right. \nonumber \\
&-&\left. (1+u)(u_\theta \cos\theta
     \sin\theta  +u_{\phi\phi}) \right] + \delta u \, u_\phi \Big\} \, .
\label{eq_2_eq5}
\end{eqnarray} 
 
These equations, combined with eqs.(\ref{eq_2_delta_a_r})-(\ref{eq_2_delta_a_phi}),
allow us to write $\delta u_\theta$ and $\delta u_\phi$
in terms of $\delta \bm{a}$. Up to order $u^2$, we obtain

\begin{eqnarray}
\delta\theta&=&\frac{\delta a_\theta}{R} (1-u+u^2) \, ,\\
\delta\phi&=&\frac{\delta a_\phi}{R\sint } (1-u+u^2) \, ,\\
\delta u&=&\frac{1}{R} \left[\delta a_r-(1-u)\, (\delta a_\phi u_\phi \csc\theta
  +\delta a_\theta  u_\theta)\right] \, ,\\
\delta u_\theta&=&\frac{1}{R} \Big\{\delta a_r \, (1-u)\, u_\theta+\delta
    a_\theta\left[1+u_\theta^2-(1-u)\, u_{\theta\theta}\right] \nonumber\\
&+& \delta a_\phi\csc\theta \big[u_\phi [(1-u)\cot \theta
    +u_\theta]-(1-u)\, u_{\theta\phi}\big]\Big\} \,,  \nonumber \\
\phantom{1}\\
\delta u_\phi&=&\frac{1}{R} \Big\{\delta a_\theta\big[u_\phi \,  [(1-u)\,\cot \theta + u_\theta]-(1-u)\, u_{\theta\phi}\big]\nonumber\\
&+& \delta a_r \, (1-u)\, u_\phi + \delta a_\phi\csc\theta\big[\sin^2 \theta +u_\phi^2 \nonumber \\
&-& (1 - u)\, (u_{\phi\phi}+ u_\theta \cos\theta \sin\theta
    )\big]\Big\}\, . 
\end{eqnarray}

These expressions are to be inserted into eq.(\ref{eq_2_delta_h_1}).
Note that it is necessary to expand $h$ up to $\mathcal{O}(u^3)$ in order to obtain 
$\partial h/\partial u_i$ and $\partial h/\partial u_{ij}$ consistently at $\mathcal{O}(u^2)$ in eq.(\ref{eq_2_delta_h_1}). This means adding

\begin{eqnarray}
h_{3} &=& - \kappa \sin\theta \, \big\{ 4 \csc^4 \theta \, u_\phi \, u_\theta\, 
  u_{\theta \phi} + 2 u_\theta^2 \, u_{\theta\theta} \nonumber \\
&-& 2 u_\phi^2 \, \csc^2\theta\, \big(u_{\phi\phi}\, \csc^2 \theta - u_\theta \, \cot\theta\big) \nonumber \\
&+& u\left[u_{\theta\theta}^2 + u_\theta^2\, \left(2 + \cot^2\theta\right) +
  2\, u_\phi^2\, \csc^2\theta  + u_{\phi\phi}^2 \, \csc^4\theta\right.\nonumber\\
&+& \left. 2 \, u_\theta \, u_{\theta\theta}\, \cot\theta + 2 \, u_\theta \, u_{\phi\phi}\, \csc^2\theta \cot\theta + 2 \, u_{\theta\theta}\, u_{\phi\phi}\csc^2\theta\right]\nonumber\\
&+& 2 u^2 \left(u_{\theta\theta} + u_\theta \cot\theta + u_{\phi\phi}\csc^2\theta\right)\big\} 
\end{eqnarray} 

\noindent to eq.(\ref{eq_2_energie_2}) before calculating the derivatives. Finally, comparing eq.(\ref{eq_2_delta_h_1}) and eq.(\ref{eq_2_delta_h_2}), we obtain

\begin{eqnarray}
\Sigma_{\theta\theta}&=& \frac{1}{2 R} \Big\{ R^2 \sigma \sint \left(2 + 2 u +
  \up^2 \csc^2 \theta - \ut^2\right) \nonumber \\
  &&+\kappa \pq{u_{\phi\phi}^2  \, \csc^3 \theta - u_{\theta\theta}^2\, \sint + 2 \ut\, \left(u_{\theta \phi\phi}\, \csc\theta+ u_{\theta\theta\theta} \, \sint \right)} \nonumber \\
&&+  \kappa \, \csc \theta \pq{2 \up^2 - \ut^2\, \cos^2\theta - 2
    u_{\phi\phi} \, (1+\ut \, \cot \theta) + 4 u \, \left(u_{\phi\phi} - u_{\theta\theta} \, \sin^2\theta\right)} \nonumber\\
&&+ 2 \kappa\big[  u_{\theta \theta}\, (\sint + \ut \cos \theta)
    + (2u - 1)\, \ut \, \cos\theta\big] \Big\} \, , \nonumber
\label{eq_2_stt}\\
\end{eqnarray}

\begin{eqnarray}
\Sigma_{\phi\theta} &=& - R \sigma u_\theta u_\phi + \frac{\kappa}{R} \left(2
  u_{\theta \phi} - u_\phi\cot\theta\right) + \frac{\kappa}{R} \Big(4
  u\, u_\phi\, \cot\theta - u_\phi \, u_{\phi\phi} \, \cot\theta \csc^2\theta \nonumber \\
  &+& u_\theta u_\phi - 4 u \, u_{\theta \phi} - u_{\phi\phi}\, u_{\theta\phi}\, \csc^2\theta
- u_\theta \, u_{\theta\phi}\, \cot\theta + u_\phi \, u_{\theta\phi\phi}\, \csc^2\theta \nonumber \\
&+& 2 u_\phi \, u_{\theta\theta}\, \cot\theta  -
  u_{\theta\phi} \, u_{\theta\theta} + u_\phi \, u_{\theta\theta\theta}\Big) \, , \label{eq_2_spt}
\end{eqnarray}

\begin{eqnarray}
\Sigma_{r\theta}&=& R \sigma \sin\theta \, u_{\theta} -
\frac{\kappa}{R}\Big[\left(1 - 2u\right)\, u_\theta \, \left(2\sin\theta - \csc\theta\right) +
  u_{\theta\theta} \, \cos\theta - 2 u_{\phi\phi}\, \cot\theta \csc\theta \nonumber\\
&+& u_{\theta\phi\phi} \, \csc\theta + u_{\theta\theta\theta} \, \sin\theta \Big]
- \frac{\kappa}{R} \Big[ 2 u_\phi^2\, \cot\theta \csc\theta -
  u_\theta^2\, \cos\theta + 4 u\, u_{\phi\phi}\, \cot\theta \csc\theta \nonumber\\
&-& 2 u\, u_{\theta\theta} \cos\theta 
- 2 u \, u_{\theta\phi\phi}  \csc\theta - 2 u \, u_{\theta\theta\theta} \sin\theta  + u_{\phi\phi} \, u_\theta \csc\theta - 3 u_\theta \, u_{\theta\theta} \sin\theta\Big] \, , \nonumber\\
\label{eq_2_srt} \phantom{}\\
\nonumber\\
\Sigma_{\phi\phi}&=& \frac{1}{2 R} \Big\{R^2 \sigma \big(2 + 2 u -\up^2 \csc^2
  \theta + \ut^2\big) \nonumber \\
  &+&\kappa \big[-u_{\phi\phi}^2  \csc^4 \theta +
    u_{\theta\theta}^2 + 2 \up\csc^2\theta\big(u_{\phi\phi\phi}\csc^2\theta + u_{\theta \theta\phi}\big)\big]\nonumber \\
&+& \kappa \csc^2 \theta \pq{2 \up^2+ \ut^2\, \left(3 \sin^2\theta - 1\right)+ 2 u_{\phi\phi}\left(1-\ut\cot\theta -2 u\right)}\nonumber\\
  &+& 2 \kappa (1- 2u)\ut\cot\theta  -2 \kappa  (1 - 2u)u_{\theta\theta} 
+2 \kappa \up \, u_{\theta\phi} \csc^2 \theta \cot\theta \Big\} \,, \nonumber\\
\label{eq_2_spp}  \\
\nonumber\\
\Sigma_{\theta\phi} &=& - R \sigma \csc\theta \, u_\theta u_\phi + 
\frac{\kappa \csc\theta}{R} \Big[ \left(u_\phi \cot\theta  - u_{\theta\phi}\right)\, \left(-2 + 4 u +
      u_{\theta\theta} + u_{\phi\phi} \csc^2\theta \right) \nonumber \\
      &+& u_\theta u_\phi \left(1 +\csc^2\theta\right)
+ u_\theta \left( u_{\phi\phi\phi} \csc^2\theta 
  + u_{\theta\theta\phi}\right)\Big] \, ,\nonumber \\
\\
\nonumber\\
\Sigma_{r\phi} &=& \frac{\csc\theta}{R} \Big[ u_\phi \left(R^2 \sigma - 2
    \kappa + 4 \kappa u + 3 u_{\phi\phi}\csc^2 \theta + 3 \ut \cot\theta 
    - u_{\theta\theta}\right) \nonumber\\
    &-& \kappa (1 - 2u)\left(u_{\theta\phi}\cot\theta  +
  u_{\theta\theta\phi} + u_{\phi\phi\phi}\csc^2\theta \right)\Big] \, .\nonumber
\label{eq_2_srp}\\
\end{eqnarray}

These expressions are valid up to $\mathcal{O}(u^2)$.
A verification of these results is presented in appendix~\ref{annexe4}.

\section{Closed vesicles}

In this section we shall derive the effective tension for closed vesicles, $\tau_{\mathrm{closed}}$,  using the stress tensor and the free-energy.
As these results are readily transposable to the case of poked vesicles, we shall present here a more detailed account of our derivations. 

\subsection{Thermal averages and correlations for closed vesicles}
\label{section_2_correlations}

In order to calculate the effective tension, we will see in section~\ref{section_2_tau_closed} that we need to evaluate the thermal average of $\Sigma_{\theta\theta}$.
To this aim, we do the standard decomposition of $u(\theta,\phi)$ in spherical harmonics ~\cite{Seifert_95} \cite{Helfrich_86}\cite{Milner_87} 

\begin{equation}
u(\theta,\phi) = \frac{u_{0,0}}{\sqrt{4\pi}} + \sum_\omega u_{l,m} Y_l^m(\theta,\phi)\, ,
\end{equation}

\noindent with $u_{l, -m} = (-1)^m u_{l,m}^*$ and 

\begin{equation}
\sum_\omega = \sum_{l=2}^{L}\sum_{m = -l}^l\,, 
\end{equation}

\noindent where $L$ is a high wave-vector cutoff (see discussion on the following).
Note that the
modes $l=1$, which correspond to simple translations, are discarded.

In terms of $u_{l,m}$ and up to order $u^2$, eq.(\ref{eq_2_Area_0}) and eq.(\ref{eq_2_Volume_0}) take the form, respectively,

\begin{equation}
  A = R^2 \left\{4 \pi \left(1 + \frac{u_{0,0}}{\sqrt{4\pi}} \right)^2 + \sum_\omega \left[ 1+ \frac{l(l+1)}{2}\right] |u_{l,m}|^2 \right\} \,
  \label{eq_2_Area}
\end{equation}

\noindent and

\begin{equation}
V = R^3\left[\frac{4\pi}{3}\left( 1 + \frac{u_{0,0}}{\sqrt{4 \pi}}\right)^3 +
  \sum_\omega |u_{l,m}|^2 \right]\, .
\label{eq_2_Volume}
\end{equation}

\noindent The volume constraint $V = \frac{4}{3} \pi R^3$ (recall the definition of $R$ for closed vesicles) implies therefore \cite{Seifert_95}

\begin{equation}
u_{0,0} = -\frac{1}{\sqrt{4 \pi}}\sum_\omega |u_{l,m}|^2\, .
\label{eq_2_u00}
\end{equation}

\noindent With the help of the relation

\begin{equation}
\cot\theta \,  \frac{\partial Y_l^m}{\partial \theta} + \csc^2\theta \, 
\frac{\partial^2 Y_l^m}{\partial \phi^2} = - \frac{\partial^2 Y_l^m}{\partial
  \theta^2} - l(l+1) Y_l^m\, ,
\end{equation}

\noindent using eq.(\ref{eq_2_u00}) and integrating over $\theta$ and $\phi$, the Hamiltonian for closed vesicles in terms of $u_{l,m}$ is given by~\cite{Seifert_95}

\begin{equation}
\mathcal{H}_\mathrm{closed} = 4 \pi R^2 \sigma + \frac{1}{2} \sum_\omega \tilde{H}_l \, |u_{l,m}|^2 + \mathcal{O}(u^3) \, ,
\label{eq_2_energie_closed}
\end{equation}

\noindent where 

\begin{equation}
  \tilde{H}_l = \kappa \, \left(l-1\right)\left(l+2\right)\left(l^2 + l + \bar{\sigma}\right) \, .
\end{equation}

\noindent Here

\begin{equation}
  \bar{\sigma} = \frac{\sigma}{\kappa/R^2}
\label{eq_2_def_sigmabar}
\end{equation}

\noindent is the reduced tension.
Note that we have discarded in $\mathcal{H}_\mathrm{closed}$ a constant energy term, $8\pi\kappa$.

We emphasize that negative values of $\bar{\sigma}$ are allowed \cite{Seifert_95}.
Indeed, the minimum of the Hamiltonian $\mathcal{H}_\mathrm{closed}$ given in eq.(\ref{eq_2_energie_closed}) corresponds for $\bar{\sigma} > -6$ to a perfectly spherical vesicle ($u_{l,m} = 0$, $\forall \, l \geq 2$).
The mean-field transition to an oblate shape occurs thus at $\bar{\sigma} = -6$ (non harmonic terms being then needed to stabilize the system).

Standard statistical mechanics yields $\langle u_{l,m} \rangle = 0$, $\forall \, l\neq 0$ and

\begin{equation}
\langle u_{l,m} u_{l',m'} \rangle = (-1)^m \frac{k_\mathrm{B} T}{\tilde{H}_l}\,
\delta_{l,l'}\, \delta_{m,-m'}\, ,
\label{eq_2_correlation}
\end{equation}

\noindent where $k_\mathrm{B} T$ is the temperature in energy units.

We may now calculate the fluctuation amplitudes.
Using eq.(\ref{eq_2_correlation}) and the Addition Theorem for spherical harmonics:

\begin{equation}
\sum_{m=-l}^l Y_l^m(\theta, \phi) {Y_l^m}^*(\theta, \phi) = \frac{2l +
  1}{4\pi}\, ,
\end{equation}

\noindent we obtain

\begin{eqnarray}
\label{u}
\langle u \rangle &=& \frac{\langle u_{0,0} \rangle}{\sqrt{4\pi}} = -
\frac{1}{4\pi} \sum_\omega \langle |u_{l,m}|^2 \rangle \nonumber\\
&=& -  \frac{k_{\mathrm{B}} T}{4 \pi} \sum_{l = 2}^L \frac{2l + 1}{\tilde{H}_l}\, \\
\label{ucarre}\nonumber\\
\langle u^2 \rangle &=& \sum_\omega \sum_{\omega'} Y_l^m(\theta, \phi)
Y_{l'}^{m'}(\theta, \phi) \langle u_{l,m} u_{l',m'} \rangle \nonumber \\ 
&=& \sum_{\omega} \frac{k_{\mathrm{B}}T}{\tilde{H}_l} Y_l^m(\theta,\phi)
{Y_l^m}^*(\theta,\phi) \nonumber \\
&=& \frac{k_{\mathrm{B}} T}{4 \pi} \sum_{l = 2}^L \frac{2l + 1}{\tilde{H}_l} =
- \langle u \rangle\,
, \\
\nonumber\\
\langle u_\phi^2 \rangle &=& \sin^2 \theta \, \frac{k_{\mathrm{B}} T}{4 \pi}
  \sum_{l=2}^L \frac{l(l+1)(2l+1)}{2 \tilde{H}_l}\, ,\\
  \nonumber\\
\langle u_\theta^2 \rangle &=& \frac{k_{\mathrm{B}} T}{4 \pi}
  \sum_{l=2}^L \frac{l(l+1)(2l+1)}{2 \tilde{H}_l}\, , 
\end{eqnarray}

The correlations of the other derivatives of $u$ are given in appendix~\ref{annexe5}.
Note that $\langle u \rangle$ is negative and that $\langle u\rangle = -\langle u^2 \rangle$, which shows how the temperature-dependent fluctuations affect the mean shape.

\subsubsection{Cutoff}
\label{subsection_2_cutoff}

The large wavenumber cutoff $L$ should be related to the smallest wave vector allowed, $\Lambda \approx a^{-1}$, where $a$ is a length comparable to the membrane thickness (i.e., $\pi/\Lambda$ of the order of a few times $a$).
With spherical harmonics, however, this is not easy to implement.
The requirement that we should recover the planar limit for large values of $R$ will guide us.

For a square patch of fluctuating flat membrane with reference area $A_p$ and periodic boundary conditions, the wave vectors are quantified according to $\bm{q} = 2
\pi /\sqrt{A_p}\,  (n_x,n_y)$, where $n_x$ and $n_y$ are integers and $|\bm{q}| < \Lambda$.
The number of modes is then approximately $\pi \Lambda^2/(2\pi/\sqrt{A_p})^2$ and the number of modes per unit area is

\begin{equation}
\frac{N_\mathrm{modes}}{A_p} \approx \frac{\Lambda^2}{4 \pi}\,.
\label{eq_2_Nmodes}
\end{equation}

\noindent For a vesicle, we have

\begin{equation}
  \frac{N_\mathrm{modes}}{A_p} = \frac{1}{4 \pi R^2} \sum_{l=2}^L\left(2l+1\right) = \frac{(L-1)(L-3)}{4 \pi R^2} \, .
\end{equation}

\noindent Asking that the number of degrees of freedom per unit area (per lipid, in some sense) be the same in both cases, we require these two quantities to be equal.
Hence we get

\begin{equation}
(L+3)(L-1) = \Lambda^2 R^2\,,
\label{eq_2_L}
\end{equation}

\noindent which gives $L = \lfloor \sqrt{4 + R^2 \Lambda^2} - 1\rfloor$ ($\lfloor x \rfloor$ is the integer part of $x$).
In the limit $R \gg \Lambda^{-1}$, this gives simply $L \simeq \Lambda R$.

\subsubsection{Validity of the Gaussian approximation}
\label{subsection_2_validity}

Since our calculations are limited to $\mathcal{O}(u^2)$, we should check, in
principle, that higher order terms are negligible.
In practice this not feasible.
To check the smallness of $u$ (which is especially critical in the case
$\sigma \leq 0$) we propose a necessary, but not sufficient condition, requiring:

\begin{equation}
\langle u^2 \rangle = \frac{k_{\mathrm{B}}T}{4 \pi}\sum_{l=2}^L
\frac{2l +1}{\tilde{H}_l} = \frac{k_{\mathrm{B}}T}{4 \pi \kappa}\sum_{l=2}^L
\frac{2l +1}{(l+2)(l-1)(l^2 + l +\bar{\sigma})} \leq U_\mathrm{max}^2\, .
\label{eq_2_cond_sig}
\end{equation}

\noindent In the following, we shall take 

 \begin{equation}
U_\mathrm{max} = 5 \, \% \,.
\end{equation}

\noindent Note that the presence of the factor $(l^2 + l + \bar{\sigma})$ in
the denominator of eq.(\ref{eq_2_cond_sig}), together with the condition $l
\geq 2$, implies $\bar{\sigma} \in [-6, \infty[$, as already discussed.

Solving condition (\ref{eq_2_cond_sig}) for the typical values $\Lambda^{-1}
\simeq 5 \, \mathrm{nm}$, $\kappa = 25 \, k_\mathrm{B}T$, and taking $U_\mathrm{max}
= 0.05$, we find

\begin{equation}
\bar{\sigma} \geq \bar{\sigma}_\mathrm{min} \approx -4 \, ,
\label{eq_2_sigmabar_min}
\end{equation}

\noindent almost independently of $R$.
%Indeed, the sum over $l$ is roughly dominated by the modes $l=2$ and $l=3$. 
%Keeping only these two terms, eq.(\ref{eq_2_cond_sig}) reduces to a simple
%quadratic equation on $\bar{\sigma}$ without dependance on $R$.
It follows that for closed vesicles, negative tensions $\sigma$ are in fact within the validity
range of our Gaussian approximation.
However, a negative $\sigma$ does not imply, in principle, negative effective tensions $\tau$.

In experiments, the actual control parameter is the excess area

\begin{equation}
\alpha = \frac{\langle A \rangle - A_p}{A_p}\, .
\end{equation}

\noindent The average of eq.(\ref{eq_2_Area}) for closed vesicles up to order two yields

\begin{eqnarray}
  \langle A \rangle &=& R^2 \left\{4 \pi + 8\pi \, \frac{\langle u_{0,0} \rangle}{\sqrt{4 \pi}} + \sum_\omega \left[1 + \frac{l\left(l+1\right)}{2}\right] \langle |u_{l,m}|^2\rangle \right\} \, , \\
 \label{eq_2_area_1} \phantom{} \nonumber\\
  &=& 4\pi R^2 + \frac{k_\mathrm{B} T \, R^2}{2} \sum_\omega \frac{(l+2)(l-1)}{\tilde{H}_l} \, .
\end{eqnarray}

\noindent Consequently, taking $A_p = 4 \pi R^2$ (the area of the vesicle with volume $V$), one obtains

\begin{equation}
\alpha_\mathrm{closed} = \frac{k_{\mathrm{B}}T}{8\pi \kappa}
\sum_{l=2}^L \frac{2l+1}{l^2 + l + \bar{\sigma}}\, .
\label{eq_2_def_alpha}
\end{equation}

\begin{figure}[H]
  \begin{center}
    \includegraphics[scale=0.75,angle=0]{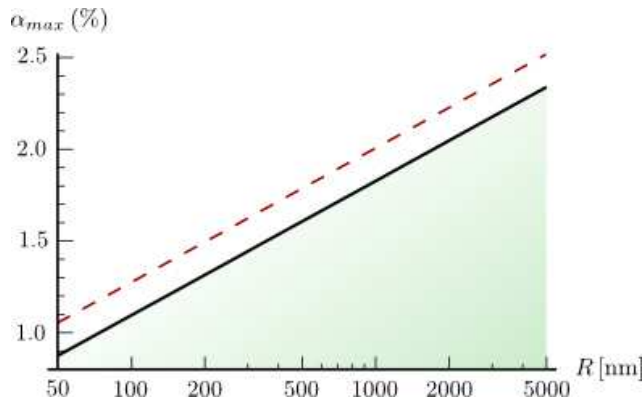}
    \caption{The solid line represents the maximum excess area corresponding
      to $\sqrt{\langle u^2 \rangle} = 0.05$ for closed vesicles taking $A_p$
      the area of the sphere of volume $V$, which we take as the validity criterion of our Gaussian
      approximation. The dashed line shows the maximum excess area for closed
      vesicles taking $A_p$ the area associated to the average radius.
      In abscissa is the vesicle's radius. Here, $\Lambda^{-1}
      \simeq 5 \, \mathrm{nm}$ and $\kappa = 25 \, k_\mathrm{B}T$.}
  \label{fig_2_alpha}
  \end{center}      
\end{figure}

\noindent Our validity condition (\ref{eq_2_sigmabar_min}) implies $\alpha_\mathrm{closed}
\leq \alpha_\mathrm{max}$ ($\alpha_\mathrm{max}$ corresponding to $\alpha$ for
$\bar{\sigma} = -\, 4$), with $\alpha_\mathrm{max}$ shown in Fig.~\ref{fig_2_alpha}. 
One can see that $\alpha_\mathrm{max} \approx
c_1 + c_2 \, \ln R$, where $c_1$ and $c_2$ are constants. 
Indeed, the sum in eq.(\ref{eq_2_def_alpha}) is dominated by the
modes $l=2$ and $l=3$, the rest of the sum being well
approximated for $\bar{\sigma} = \mathcal{O}(1)$ by an integral proportional
to $\ln (R)$.
Note that if one takes $A_p = 4 \pi R^2 ( 1 + \langle u \rangle)^2$ (the area
associated to the average radius) one obtains $\alpha_\mathrm{max}$ just slightly
bigger (see Fig.~\ref{fig_2_alpha}). 

\subsection[Evaluation of $\tau_{\mathrm{closed}}$ from the projected stress tensor]{Evaluation of $\bm{\tau_{\mathrm{closed}}}$ from the projected stress tensor}
\label{section_2_tau_closed}

\begin{figure}[H]
\begin{center}
\subfigure[The vesicle fluctuates around the average
      shaded spherical shell. The dashed line represents the reference
      sphere. Note that $\langle u \rangle$ is negative.]{
\includegraphics[scale=.5,angle=0]{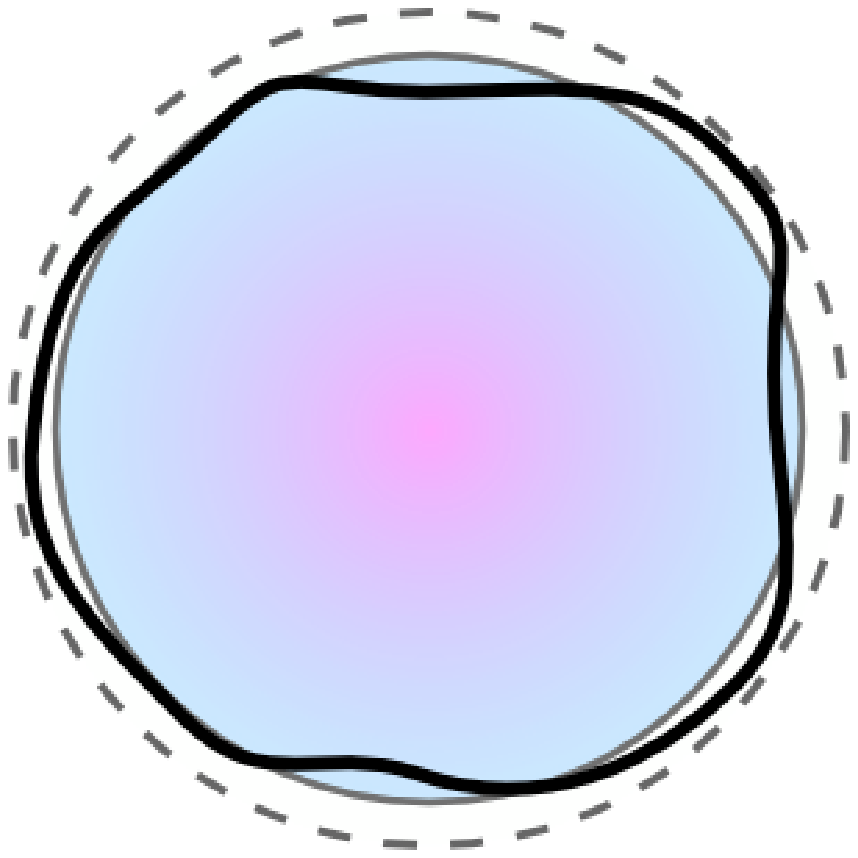}
\label{fig_2_ref_sphere}
}
\subfigure[Force exchanged through an infinitesimal cut at $\theta$ constant. The length of the projected cut is $R \sin \theta d \phi$, while the true length of the membrane is indicated in red.]{ 
  \includegraphics[scale=.4,angle=0]{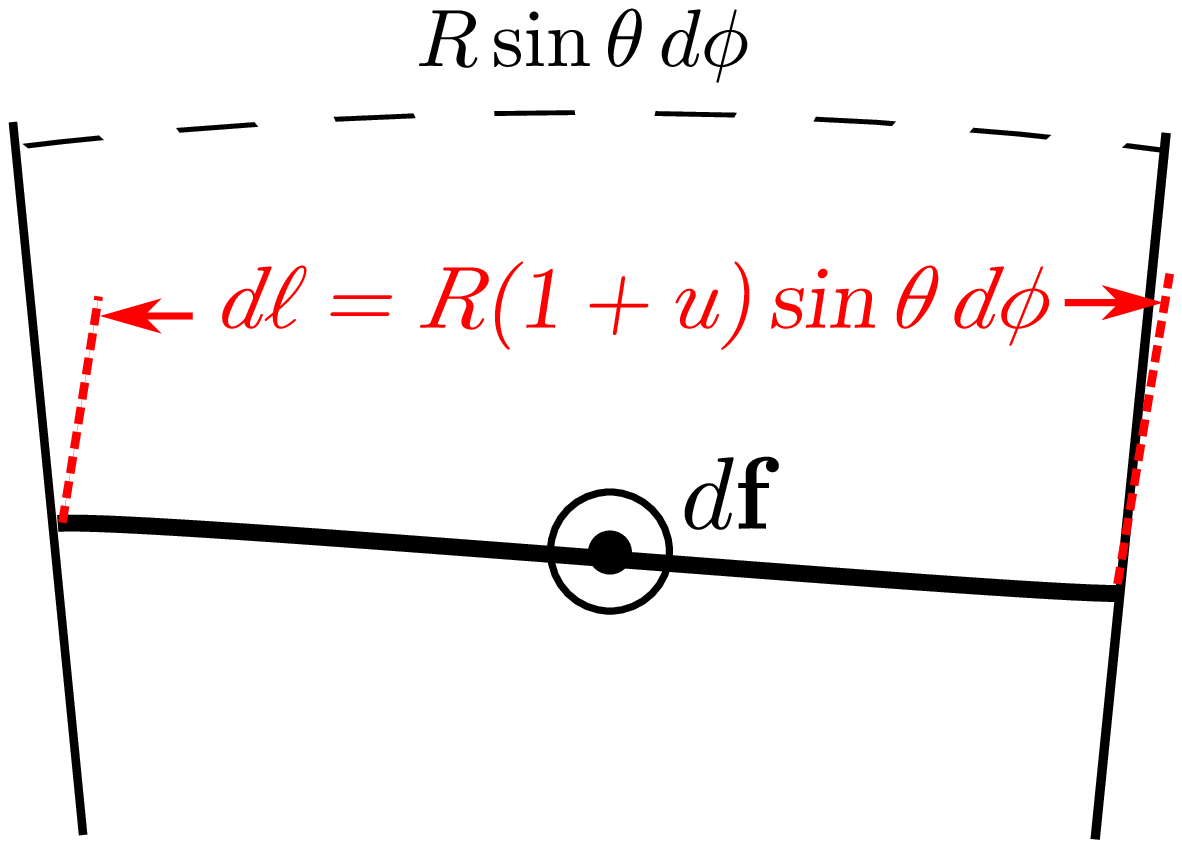}
  \label{fig_1_kind_E}
}
\caption{The vesicle is represented by a solid thick line, while the reference sphere is represented in with a dashed line.}
\end{center}      
\end{figure}

Imagine replacing the fluctuating vesicle by a 
shell coinciding with its average shape (see Fig.~\ref{fig_2_ref_sphere}).

The effective tension $\tau$ is the average force per unit length that is
exchanged tangentially to the shell's surface.
Because of the spherical symmetry, $\tau$ depends neither on the point
$(\theta, \phi)$ nor in the direction in which it is calculated.
Let us thus consider an infinitesimal cut with $\theta$ constant of extension $d \phi$.
The component along $\bm{e}_\theta$ of the force exchanged through the cut is
on average $d \bm{f} = \langle \Sigma_{\theta\theta} \, d \phi \rangle$.
The length of the cut is on average $\langle R(1+u)\sin\theta \, d \phi\rangle$.
Hence,

\begin{equation}
\tau = \frac{\langle \Sigma_{\theta\theta} \rangle}{R\sin\theta(1 + \langle u
\rangle)}\, .
\label{eq_2_tau_def}
\end{equation}

\noindent Since $\Sigma_{\theta\theta} = \sigma R \sin\theta +
\mathcal{O}(u)$, we obtain equivalently 

\begin{equation}
\tau = \frac{1}{R\sin\theta} \langle \Sigma_{\theta\theta} \rangle - \sigma
\langle u \rangle + \mathcal{O}(u^3) \, .
\end{equation}

\noindent Using eq.(\ref{eq_2_stt}), the results presented in section~\ref{section_2_correlations} and in appendix~\ref{annexe5}, 
we obtain for closed vesicles

\begin{eqnarray}
\tau_\mathrm{closed} &=& \sigma - \frac{k_\mathrm{B} T \kappa}{8\pi
  R^2} \sum_{l=2}^L \frac{(l+2)(l+1)l(l-1)(2l+1)}{\tilde{H}_l} \, ,
\label{eq_2_resgen}\\
&=& \sigma - \frac{k_\mathrm{B} T}{8\pi
  R^2} \sum_{l=2}^L \frac{l(l+1)(2l+1)}{l^2 + l + \bar{\sigma}} \, .
\label{eq_2_tau}
\end{eqnarray}  

\noindent Note that the terms on $\langle u \rangle$ vanish and, as expected, $\tau_\mathrm{closed}$ is independent of the point $(\theta,
\phi)$ in which it is calculated.
It is interesting to examine $\tau_\mathrm{closed}$ in the limit of large vesicles.
In this
case, the sum on eq.(\ref{eq_2_tau}) may be substituted by an integral:

\begin{eqnarray}
\tau_\mathrm{closed} - \sigma &\approx& -\frac{k_{\mathrm{B}}T}{8 \pi R^2} \int_2^{R \Lambda}
\frac{(l+1)l(2l+1)}{l^2 + l + \bar{\sigma}}  \, , \nonumber \\
\nonumber\\
&\approx& -\frac{k_\mathrm{B}T \, \Lambda^2}{8 \pi} \left[ 1 -
  \frac{\sigma}{\kappa \Lambda^2} \log\left(1 + \frac{\kappa \Lambda^2}{\sigma}\right)\right]
+ \mathcal{O}\left(\frac{\Lambda}{R}\right)\, .
\label{eq_2_tau_0}
\end{eqnarray} 

\noindent The dominant term in eq.(\ref{eq_2_tau_0}) correctly matches the
difference $\tau - \sigma$ for flat membranes given in chapter~\ref{chapitre/planar_membrane}, eq.(\ref{eq_1_diff}).

We have also calculated the normal and orthogonal components of the tension.
Both vanish: $\langle \Sigma_{r\theta} \rangle = 0$ and $\langle
\Sigma_{\phi\theta} \rangle = 0$.
While the latter result is obvious on symmetry grounds, the former one is
interesting, implying that the shell mentioned above can indeed be considered
as a purely tense surface.
This would probably not hold for a vesicle with non-spherical average shape.
As a consequence, the Laplace law can be used without curvature corrections
for a fluctuating quasi-spherical vesicle, provided that one uses $\tau$
instead of $\sigma$.
Indeed, this could be expected from renormalization arguments, since the
Laplace law is exact (despite the curvature energy) for a perfectly spherical
membrane \cite{Fournier_08}.

\section{Poked vesicles}
\label{section_2_poked}

The route to obtain $\tau_{\mathrm{poked}}$ is the same as with a closed vesicle, with some minor changes.
We remind that the reference sphere in poked vesicles is the sphere where
$\langle u \rangle = 0$, since there is no constraint on volume.
Accordingly, instead of eq.(\ref{eq_2_u00}), we have simply $u_{0,0} = 0$.
The Hamiltonian in the Gaussian approximation becomes

\begin{equation}
\mathcal{H} = 4 \pi R^2 \sigma + \frac{1}{2} \sum_\omega \tilde{H}_l' \, |u_{l,m}|^2 + \mathcal{O}(u^3) \, ,
\label{eq_2_energie_poked}
\end{equation}

\noindent where $\tilde H'_l=\tilde H_l+4\kappa\bar\sigma$ \cite{Henriksen_04}.
The correlations given in section~\ref{section_2_correlations} and in appendix~\ref{annexe5} remain correct, provided one replaces $\tilde{H}_l$ by $\tilde{H}_l + 4 \kappa \bar{\sigma}$.
Note that $\langle u^2 \rangle \neq \langle u\rangle \equiv 0$ here and that, differently from the case of closed vesicles, one must have $\bar\sigma \in [-3, \infty[$ in order to assure that correlations are positive.
The discussion about the cutoff of section~\ref{subsection_2_cutoff} remains valid for poked vesicles and the validity condition for the Gaussian approximation given in eq.(\ref{eq_2_cond_sig}) becomes

\begin{equation}
\langle u^2 \rangle = \frac{k_{\mathrm{B}}T}{4 \pi \kappa}\sum_{l=2}^L
\frac{2l +1}{(l+2)(l-1)(l^2 + l +\bar{\sigma}) + 4 \bar{\sigma}} \leq U_{max}^2\, .
\label{eq_2_cond_sig_poked}
\end{equation}

\noindent With the same $U_\mathrm{max} = 5\%$ as before, we have $\bar\sigma_\mathrm{min} \approx -2$ for poked vesicles.
The average area is given by eq.(\ref{eq_2_area_1}) with $\langle u_{0,0} \rangle =0$, yielding 

\begin{equation}
  \langle A \rangle 
  = 4\pi R^2 + \frac{k_\mathrm{B}T\, R^2}{2} \sum_\omega \frac{l^2 + l +2}{\tilde{H}_l + 4\kappa \bar{\sigma}} \, .
\end{equation}

\noindent Consequently, 

\begin{equation}
\alpha_\mathrm{poked} = \frac{k_{\mathrm{B}}T}{8\pi \kappa}
\sum_{l=2}^L \frac{(l^2 + l + 2)(2l+1)}{(l-1)(l+2)(l^2 + l + \bar{\sigma}) + 4 \bar{\sigma}}\, .
\label{eq_2_def_alpha_poked}
\end{equation}

\begin{figure}[H]
  \begin{center}
    \includegraphics[scale=0.75,angle=0]{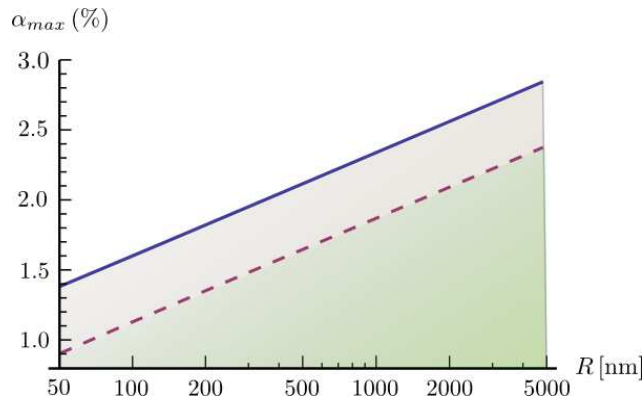}
    \caption{The blue line stands for the maximum excess area corresponding to
      $\sqrt{\langle u^2 \rangle} = 0.05$ for poked vesicles, while the red dashed
      line stands closed vesicles.
      In abscissa is the vesicle's radius. Here, $\Lambda^{-1}
      \simeq 5 \, \mathrm{nm}$ and $\kappa = 25 \, k_\mathrm{B}T$.}
  \label{fig_2_alpha_poked}
  \end{center}      
\end{figure}

\noindent Fig.~\ref{fig_2_alpha_poked} shows $\alpha_\mathrm{max}$, i. e., $\alpha_\mathrm{poked}$ with $\bar{\sigma} = -2$, as a function of the vesicle's radius.
The excess area is somewhat larger than in the case of closed vesicles, but the general behavior is the same.

Eq.(\ref{eq_2_resgen}), which gives $\tau_{\mathrm{closed}}$ by the stress
tensor method, is valid whatever the form of $\tilde H_l$, since $\tau$ bears
no term on $u$. 
Hence, we need just to replace $\tilde H_l$ by $\tilde H_l+4\kappa\bar\sigma$, which yields:

\begin{equation}
\tau_{\mathrm{poked}}=\sigma- \frac{k_B T}{8 \pi R^2}\sum_{l=2}^L
\frac{l\p{l+1}\p{2l+1}}{l^2+l+\bs+\frac{4\bs}{\p{l-1}\p{l+2}}}\,.
\label{eq_2_tau2}
\end{equation}

\noindent In the limit of large vesicles, we recover again the result for flat membranes.

\section[Discussion on $\tau$ for closed and poked vesicles]{Discussion on \bm{$\tau$} for closed and poked vesicles}
\label{section_2_discussion}

\begin{figure}[H]
  \begin{center}
    \includegraphics[scale=1,angle=0]{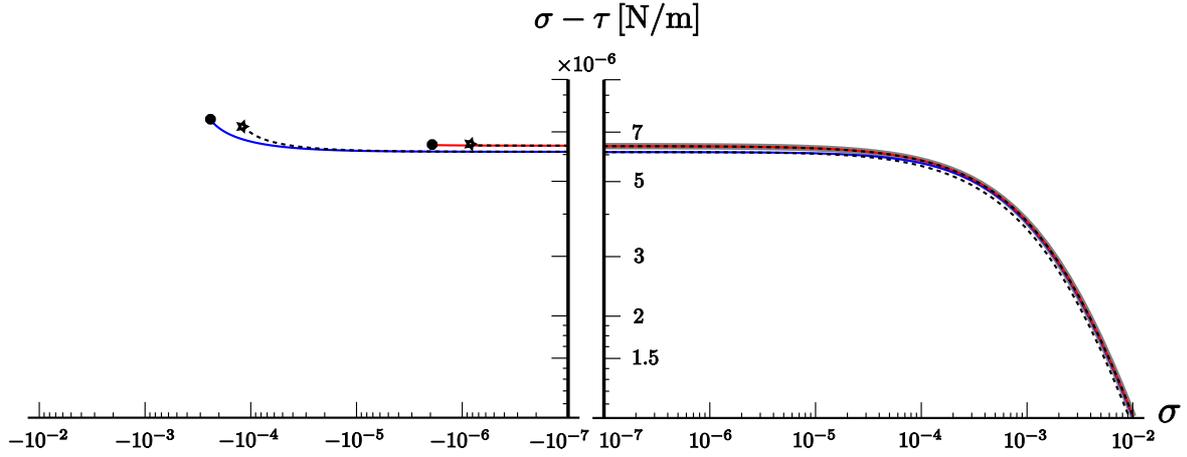}
    \caption{Difference between the Lagrange multiplier $\sigma$ and the effective mechanical tension $\tau$ as a function of $\sigma$  for $\kappa = 10^{-19} \, \mathrm{J}$, $k_\mathrm{B} T = 4 \times
      10^{-21} \, \mathrm{J}$ and $\Lambda^{-1} = 5 \, \mathrm{nm}$. The
      colored solid lines correspond to closed vesicles of $R = 50 \, \mathrm{nm}$
      (blue) and $R = 0.5 \, \mu m$ (red), whereas the corresponding colored dashed lines represent the results for a poked vesicle.
      The end-points indicate the limit beyond which our Gaussian
      approximation is no longer valid according to section
      \ref{subsection_2_validity} (circles) or according to the discussion in section \ref{section_2_poked} (stars).
      The thick gray line corresponds to a flat membrane (eq.(\ref{eq_2_tau_0})).
      }
  \label{fig_2_tau_sigma}
  \end{center}      
\end{figure}

We show in Fig.~\ref{fig_2_tau_sigma} the behavior of $\sigma - \tau$ as a function of the
Lagrange multiplier $\sigma$ for closed and poked vesicles, as well as the limiting case of planar membranes.
First of all, although eqs.(\ref{eq_2_tau2}) and (\ref{eq_2_tau}) differ mathematically, it turns out that their difference as a function of $\sigma$ is numerically irrelevant (see Fig.~\ref{fig_2_tau_sigma}).
Indeed the extra term $4\bar\sigma/[(l-1)(l+2)]$ in the denominator of eq.(\ref{eq_2_tau2}) is only important for small $l$'s,  while the sum is dominated by large $l$'s.

\begin{figure}[H]
  \begin{center}
    \includegraphics[scale=0.65,angle=0]{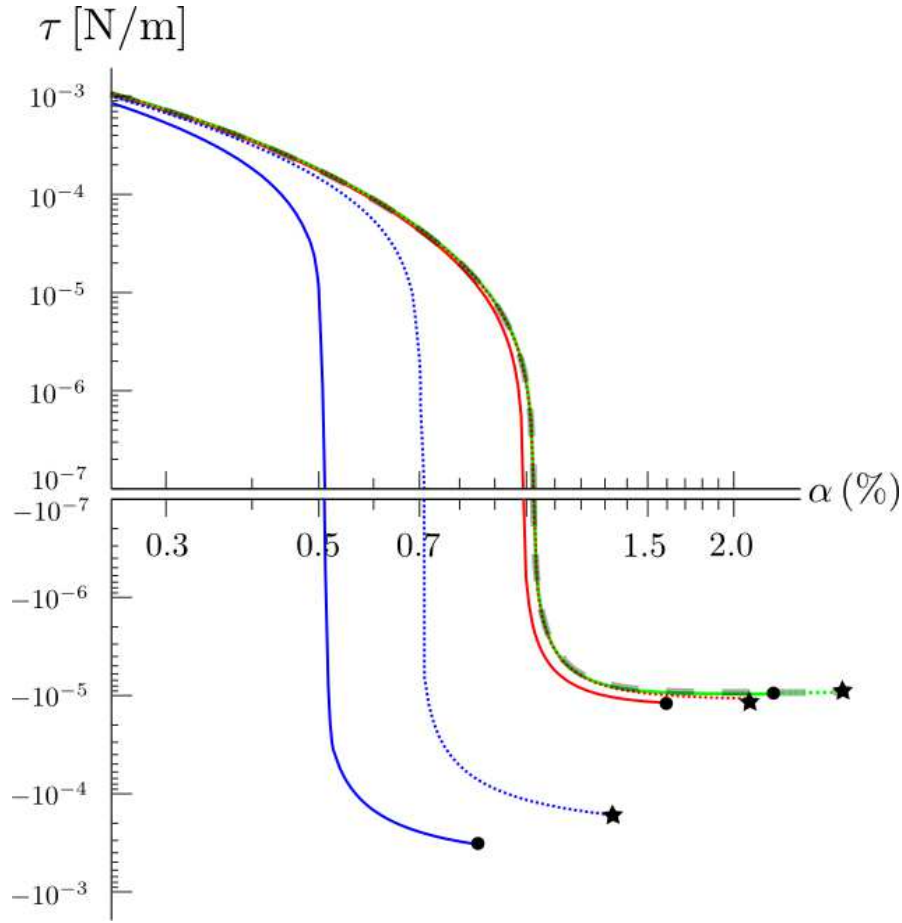}
    \caption{Effective mechanical tension $\tau$ as a function of the excess
      area $\alpha$ for $\kappa = 10^{-19} \, \mathrm{J}$, $k_\mathrm{B} T = 4 \times
      10^{-21} \, \mathrm{J}$ and $\Lambda^{-1} = 5 \, \mathrm{nm}$. The
      colored solid lines correspond to closed vesicles of $R = 50 \, \mathrm{nm}$
      (blue leftmost curves), $R = 0.5 \, \mu m$ (red central curves) and $R = 5 \, \mu m$ (green rightmost curves), while the corresponding dotted lines represent $\tau_\mathrm{poked}$.
      The end-points/stars indicate the limit beyond which our Gaussian
      approximation is no longer valid.
      The gray dashed line corresponds to the flat membrane limit given in eq.(\ref{eq_2_tau_alpha_flat}). } 
  \label{fig_2_tau_alpha}
  \end{center}      
\end{figure}

This representation is however not very useful, since $\sigma$ is not a control parameter.
The most physical representation is shown in Fig.~\ref{fig_2_tau_alpha}, where we see the behavior of $\tau$ as a function of the
excess area $\alpha$.
We show also the limiting case $R \rightarrow \infty$.
In this case, the relation between $\alpha$ and $\sigma$ is analytical and
given in eq.(\ref{eq_1_alpha11}).
In the limit of large membranes, we obtain

\begin{equation}
\sigma = \frac{\kappa \Lambda^2}{e^{8 \pi \beta \kappa \alpha} - 1} \, .
\end{equation}

\noindent Applying this result to eq.(\ref{eq_2_tauplan}), we obtain an
analytical expression for $\tau$ as
a function of the area excess, given by

\begin{equation}
\tau_{\mathrm{flat}}(\alpha) = \frac{\kappa \Lambda^2 (1 + \alpha)}{e^{8 \pi \beta
    \kappa \alpha} - 1} - \frac{k_\mathrm{B} T \Lambda^2}{8\pi} \, .
\label{eq_2_tau_alpha_flat}
\end{equation}

\noindent For vesicles, $\alpha$ given in eq.(\ref{eq_2_def_alpha}) (or in eq.(\ref{eq_2_def_alpha_poked})) is numerically inverted in order to obtain $\sigma(\alpha)$.
The result is then applied to eq.(\ref{eq_2_tau}) (respectively, eq.(\ref{eq_2_tau2})), yielding the curves of Fig~\ref{fig_2_tau_alpha}.

There are several salient points:

\begin{enumerate}
\item Even though $\tau_\mathrm{closed}$ and $\tau_\mathrm{poked}$ are almost indistinguishable as a function of $\sigma$, $\alpha_\mathrm{closed}$ and $\alpha_\mathrm{poked}$ present different dependences in terms of $\sigma$, as shown in Fig.~\ref{fig_2_alpha_sigma}, especially for small values of $R$.
  This indicates that the volume constraint affects mainly the excess area and
  explains the differences shown in Fig.~\ref{fig_2_tau_alpha}.

\begin{figure}[H]
  \begin{center}
    \includegraphics[scale=0.95,angle=0]{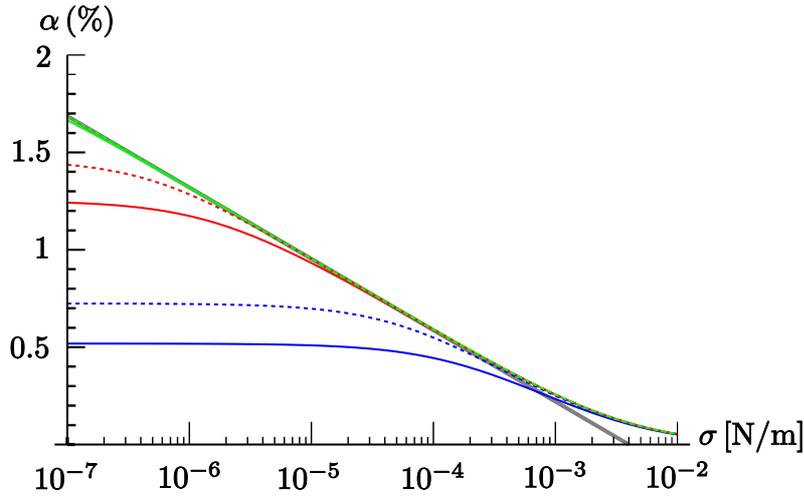}
    \caption{Solid/dashed lines stand for the excess area for closed/poked vesicles for $R = 50 \, \mathrm{nm}$ (blue lower curves), $R = 0.5 \, \mu m$ (red central curves) and $R = 5 \, \mu m$ (green upper curves). The limit of large planar membranes is shown in gray. Curves for $\kappa = 10^{-19} \, \mathrm{J}$, $k_\mathrm{B} T = 4 \times 10^{-21} \, \mathrm{J}$ and $\Lambda^{-1} = 5 \, \mathrm{nm}$. }
  \label{fig_2_alpha_sigma}
  \end{center}      
\end{figure}
  
\item The results for $\tau$ deviate from the flat limit ($R \rightarrow
  \infty$) essentially for $R \leq 1 \mu m$ for both closed and poked vesicles (see Fig.~\ref{fig_2_tau_alpha}).
  Consequently, for GUVs, one is allowed to use simply the relation given in
  eq.(\ref{eq_2_tauplan}).
Moreover, for small tensions, it is justified to assume $\tau \simeq \sigma - \sigma_0$, justifying the assumptions made on \cite{Sengupta_10}  and presented in section~\ref{subsection_1_Limozin}.

\item Negative and quite large values of $\tau$ are indeed accessible within the validity range of our Gaussian analysis in both cases (see Fig.~\ref{fig_2_tau_min}).

  \begin{figure}[H]
  \begin{center}
    \includegraphics[scale=0.95,angle=0]{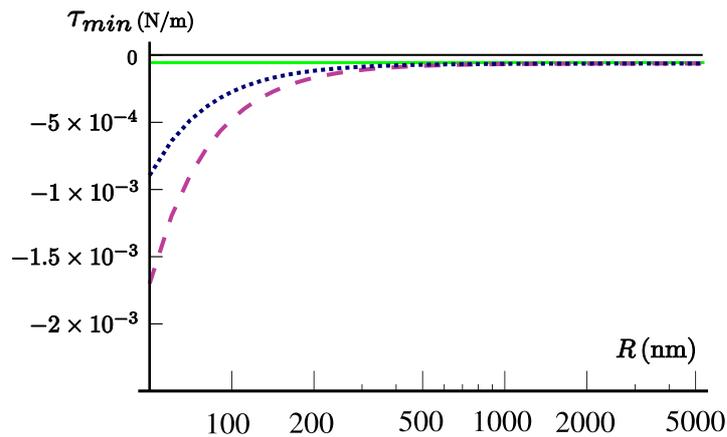}
    \caption{Largest negative tension within our validity condition (for $U_{max} = 5\%$) as a function of the vesicle's radius for $\kappa = 10^{-19} \, \mathrm{J}$, $k_\mathrm{B} T = 4 \times 10^{-21} \, \mathrm{J}$ and $\Lambda^{-1} = 5 \, \mathrm{nm}$.
      The violet dashed line stands for closed vesicles, while the blue dotted one stands for poked vesicles.
    The green solid line represents the smallest value of $\tau$ for the limit of large planar membranes.}
  \label{fig_2_tau_min}
  \end{center}      
\end{figure}

%Our analysis shows that $\tau$ may indeed become negative within the validity range of our Gaussian analysis.
%For instance, small vesicles with $R=50\,\mathrm{nm}$ can reach negatives tensions of $-10^{-4}\,\mathrm{N/m}$ with an excess area still less than $1\%$ (see fig.~\ref{tau_sigma}).

  From Fig.~\ref{fig_2_tau_min}, we note that the biggest negative tension $\tau_\mathrm{min}$ that GUVs (with $R\ge1\mathrm{\mu m}$) may sustain coincide with the biggest negative tension that large planar membranes sustain:

  \begin{equation}
    \tau_\mathrm{min} = - \frac{k_\mathrm{B} T \, \Lambda^2}{8 \pi} \, .
  \end{equation}
  
\noindent Depending on the uncertainty on the value of the cutoff, $\tau_\mathrm{min}$ may be of order $-10^{-6}\,\mathrm{N/m}$ or $-10^{-5}\,\mathrm{N/m}$.

Let's recall the Young--Laplace equation

\begin{equation}
  \Delta P = P_\mathrm{inner} - P_\mathrm{outer} = \tau \left(\frac{1}{R_1} + \frac{1}{R_2} \right)\, ,
  \label{eq_2_Young_Laplace}
\end{equation}

\noindent where $P_\mathrm{inner/outer}$ is, respectively, the inner and the
outer pressure of the vesicle, $R_1$ and $R_2$ are the two principal radii ($R_1 = R_2$ in the spherical case).
As our analysis shows that $\tau$ may indeed become negative, this would imply that vesicles could sustain an inner pressure \textit{lower} than the outer pressure.
For liquid drops, this situation is impossible, since the surface tension is a true material constant always positive.

The possibility to sustain negative tensions, or negative pressure differences, might be experimentally investigated:
i) by controlling the outer osmotic pressure, in the case of small vesicles,
or ii) by poking a giant vesicle with a micropipette to which it would adhere and gently decreasing its inner pressure.

\item $\tau$ has a plateau at large values of $\alpha$ for both closed and poked vesicles, which probably
  corresponds to the actual transition to oblate shapes: when $\tau$ reaches a critical value $\tau_c<0$, the excess area rises dramatically. 
For small closed vesicles we find roughly $\tau_c R^2/\kappa\approx-5$ while for giant closed vesicles it is given by $\tau_c\simeq-k_\mathrm{B}T\Lambda^2/(8\pi)$, i. e., below the
mean-field threshold (see discussion after eq.(\ref{eq_2_def_sigmabar})).
The high symmetry phase (spherical vesicle) is thus stabilized by its entropic
fluctuations, as one might have expected.

Experimentally, this transition might be tested by controlling the pressure outside the vesicle.
Indeed, applying the Young--Laplace pressure formula given in eq.(\ref{eq_2_Young_Laplace}) for a vesicle of radius $R$ and at $\tau_c$, we find that the critical pressure difference yielding the shape transition is
$\Delta P_c\approx-\sup[10\kappa/R^3,k_\mathrm{B}T\Lambda^2/(4\pi R)]$.
Numerically, for a closed spherical vesicle with $\kappa = 25 \, k_\mathrm{B} T$, $T = 300 \, K$ and $\Lambda = (1/5 \, \mathrm{nm})$, we find $\Delta P_c = - \, 8 \times 10^3 \, \mathrm{Pa}$ and $\Delta P_c = - 25 \, \mathrm{Pa}$, for vesicles with radius $50 \, \mathrm{nm}$ and $5\, \mathrm{\mu m}$, respectively.

\item There exists a well defined excess area $\alpha_0$ corresponding to a vanishing lateral tension $\tau = 0$ (see Fig.~\ref{fig_2_alpha_0}).
  This corresponds to the case where the pressure difference between the inner and the outer media vanishes.
%  Since, as we shall show, the curves in fig.\ref{fig_2_tau_alpha} are the same
%without the volume constraint, $\alpha_0$ corresponds to the spontaneous
%excess area taken up by a poked vesicle (vanishing Laplace pressure).
Its value is very much radius dependent for $R \leq 1 \, \mu m$, but one
recovers for $R \geq 2 \, \mu m$ the flat membrane limit given in eq.(\ref{eq_1_alpha_eq}).

\begin{figure}[H]
  \begin{center}
    \includegraphics[scale=0.75,angle=0]{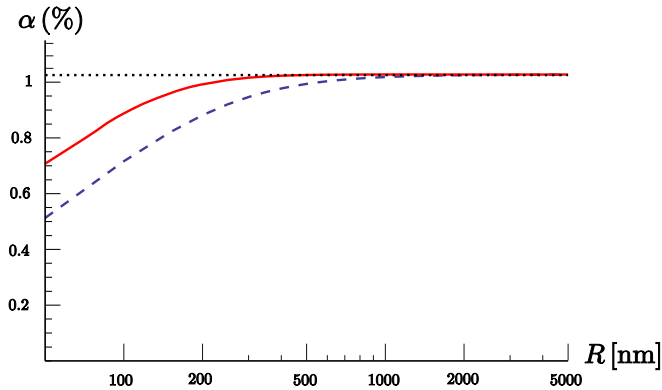}
    \caption{Spontaneous excess area $\alpha_0$, corresponding to $\tau = 0$, as a
      function of the vesicle radius for closed (blue dashed line) and poked (red solid line) vesicles. The dotted horizontal line gives the flat membrane
      limit. The material parameters are the same as in
      Fig.~\ref{fig_2_tau_alpha}.}
  \label{fig_2_alpha_0}
  \end{center}      
\end{figure}

\end{enumerate}

\section[Derivation of $\tau$ using the free-energy]{Derivation of \bm{$\tau$} using the free-energy}
\label{section_2_differentiation}

For a flat membrane, one may also obtain $\tau$ by differentiating
the free-energy with respect to the projected
area $A_p$, as we have shown in section~\ref{subsection_1_free_energy}~\cite{Henriksen_04},\cite{Cai_94}, \cite{Imparato_06}, but there are two pitfalls.
One must: i) take the thermodynamic limit $A_p\to\infty$ only \textit{after} the differentiation, and ii) introduce a variation of the cutoff in order that the total number of modes remains constant during the differentiation, as discussed in~\cite{Imparato_06}.

Let us investigate the free-energy method in the case of quasi-spherical vesicles. The free-energy, $\mathcal{F}$, is given by

\begin{equation}
\mathcal{F}=-\frac1\beta \ln\!\int\!\mathcal{D}[\bm{r}]\, e^{-\beta \mathcal{H}},
\label{eq_2_fe1}
\end{equation}

\noindent the integral running over all the configurations of the vesicle.
At the Gaussian level and for closed vesicles, $\mathcal{H}$ is given in terms of spherical harmonics by eq.(\ref{eq_2_energie_closed}), and since 
$\bm{r}=R\,\bm{e}_r+R\,u(\theta,\phi)\,\bm{e}_r$, we may write (in agreement with ref.~\cite{Seifert_95}):

\begin{eqnarray}
\mathcal{D}[\bm{r}]
=\prod_{l=2}^L
\p{\prod_{m=0}^l\,R\,du_{l,m}^\mathcal{R}}
\p{\prod_{m=1}^l\,R\,du_{l,m}^\mathcal{I}},
\label{eq_2_jacob}
\end{eqnarray}

\noindent where the superscripts $\mathcal{R}$ and $\mathcal{I}$ signify real part and imaginary part, respectively.
This measure corresponds to the so-called normal gauge, which is known to be correct for small fluctuations~\cite{Seifert_95}. 
We note that the radius $R$ of the reference sphere appears explicitly and that for each value of $l$, only half of the allowed values of $m$ have to be considered, as $\rr$ is real.
Performing the Gaussian integrals, one obtains

\begin{equation}
\mathcal{F}= 4 \pi R^2 \sigma +k_B T \sum_{l=2}^L\frac{2 l +1}{2}\ln\p{\frac{\beta \Hl}{R^2}} \, .
\label{eq_2_eq_free}
\end{equation}

In order to obtain $\tau_\mathrm{closed}$, we must differentiate
$\mathcal{F}$ with respect to the vesicle's ``projected area" $A_p$.
Which one, however? The area $A_V = 4\pi R^2$ of the reference sphere (i.e., the sphere having the same volume as the vesicle's), or the area  of the vesicle's average shape, defined as $A_m=4\pi\langle R(1+u)\rangle^2$?
It will turn out that the former choice is the correct one.
In a sense, this is natural because it corresponds to our parametrization.
However, it is not that obvious, because the definition of $\tau_\mathrm{closed}$ in eq.(\ref{eq_2_tau_def}) involves the area of the \textit{average} vesicle's shape.

Let us thus pick $A_p= A_V \equiv 4\pi R^2$. It is worth noticing
that $\Hl$ depends on $A_p$ only through $\bar{\sigma}=\sigma R^2/\kappa$, yielding

\begin{equation}
\derpart{\Hl}{R^2}=\p{l-1}\p{l+2}\sigma\,.
\end{equation}

\noindent With this choice:

\begin{equation}
\tau_\mathrm{closed}=\derpart{\mathcal{F}}{A_p}=\frac1{4\pi}\derpart{\mathcal{F}}{R^2}\,,
\end{equation}

\noindent we obtain

\begin{equation}
\tau_\mathrm{closed}
=\sigma- \frac{k_B T}{8 \pi R^2}\sum_{l=2}^L
\frac{l\p{l+1}\p{2l+1}}{l^2+l+\bs}\,,
\end{equation}

\noindent which is identical to the result obtained from the stress tensor approach, eq.(\ref{eq_2_tau}).
How about the pitfalls mentioned above?
First, we didn't take the thermodynamic limit before differentiating.
Actually, this would not be problematic, since the quantification of the modes does not involve the size of the system, like it is the case for planar membranes.
Second, we have kept $L$ (hence the number of modes) constant during the differentiation, in agreement with the fact that $L = \lfloor \sqrt{4 + R^2 \Lambda^2} - 1\rfloor$ is constant for a mathematically infinitesimal change of $R$.

We may obtain a more intrinsic expression for $\tau_\mathrm{closed}$. With $\average{A}=A_p\, (1+\alpha)$, and $N_\mathrm{modes} = \sum_{l=2}^L (2l+1)$, we
may rewrite $\tau_\mathrm{closed}$ as

\begin{equation}
\tau_\mathrm{closed} \, A_p=\average A \sigma -\frac{k_B T}{2 }N_\mathrm{modes}\,.
\end{equation} 

\noindent The quickest way to obtain this result is to keep separate, when differentiating with respect to $R^2$, the two terms coming from $\ln(\tilde H_l)$ and $\ln(1/R^2)$ in eq.(\ref{eq_2_eq_free}).
The interpretation of this equation is not straightforward, because $\frac12 k_\mathrm{B}T$ is the internal energy per mode (not the free-energy per mode).
Note that the same form for $\tau$ is also valid in the planar case, as shown in~\cite{Imparato_06}.

In addition, let's see what happens if we take $A_p = A_m \equiv 4 \pi R^2(1 + \langle u \rangle)^2$:

\begin{equation}
  \tau_{A_m} = \frac{\partial \mathcal{F}}{\partial A_p} = \frac{\partial \mathcal{F}}{\partial A_V} \left(\frac{\partial A_m}{\partial A_V}\right)^{-1}\, 
\end{equation}

\noindent where we remind $A_V = 4\pi R^2$. Using $\langle u \rangle = - \langle u^2 \rangle$, one obtains

\begin{equation}
  \left(\frac{\partial A_m}{\partial A_V}\right)^{-1} = \frac{1}{1 - \langle u^2 \rangle} = 1 + \langle u^2 \rangle + \mathcal{O}(u^3)\, .
\end{equation}

\noindent Clearly, differentiating with respect to $A_m$ yields supplemental terms of order $u^2$
The result is thus wrong, in the sense that it differs from the result obtained by the stress tensor method.

Let's now re-derive eq.(\ref{eq_2_tau2}) by deriving the free-energy.
In the case of poked vesicles, it is given by the same expression as eq.(\ref{eq_2_eq_free}) with $\tilde H_l$ replaced by $\tilde H'_l$: 

\begin{equation}
\mathcal{F'} = 4 \pi R^2 \sigma + k_\mathrm{B} T
\sum_{l=2}^L \frac{2l+1}{2} \ln\left(\frac{\beta \tilde H'_l}{R^2}\right)\, .
\label{eq_2_eq_free2}
\end{equation}

\noindent It turns out, again, that $\tau_\mathrm{poked}=\partial \mathcal{F}'/\partial (4\pi R^2)$ exactly. 
This result is satisfying, but at the same time it shows how slippery the free-energy approach can be: differentiating with respect to the area of the average vesicle is correct in the case of poked vesicles but not in the case of closed vesicles. The stress tensor method is thus a much safer.  

Our expression for $\tau_\mathrm{poked}$ differs from that obtained in ref.\cite{Henriksen_04}, where the authors considered also a quasi-spherical membrane without volume constraint.
In particular, the mechanical tension obtained in that reference cannot take negative values.
We believe that the discrepancy between the two results comes from the omission in ref.\cite{Henriksen_04} of the factors $R$ within the measure.
Indeed the factor $1/R^2$ in the logarithm of our eq.(\ref{eq_2_eq_free2}) is absent in the corresponding expression~(A.9) of ref.\cite{Henriksen_04}.

\section{In a nutshell}

In this chapter, we have compared the mechanical tension $\tau$ 
one applies by aspiring a vesicle with a micropipette, for instance, with the
tension $\sigma$ theoretically introduced in the Hamiltonian to fix the membrane's area in
the case of quasi-spherical vesicles.
We have studied both the case of usual closed vesicles and the case of poked vesicles, free to exchange liquid with the outer media.
We conclude that in both cases, for GUVs, the relation between $\tau$ and
$\sigma$ is very well approximated by the relation obtained in the case of
planar membranes, given in eq.(\ref{eq_2_tauplan}).
Accordingly, for GUVs under small tensions, we can assume simply
$\tau \simeq \sigma - \sigma_0$, as in the case of planar membranes.
Moreover, in both cases, we predict the possibility of an internal pressure
smaller than the outer, situation impossible in the case of liquid drops. 
Regarding comparatively the behavior of closed and poked vesicles, we expect
the excess area of both to differ for small vesicles.
At last, we have shown that the concept of projected area for vesicles is not
clear.
Thus, we conclude that it is much safer to derive $\tau$ by averaging the
projected stress tensor.

%% file: chap3.tex
\chapter{Force needed to extract a fluctuating nanotube}
\label{TUBE}

In this and in the following chapter, we shall study the membrane nanotubes presented in section~\ref{tube_exp}.
The main results of both chapters were obtained with Jean-Baptiste Fournier and were published in~\cite{Barbetta_09}.
As we have seen, these tubes are very thin, with a radius ranging from dozens up to hundreds of nanometers, while their length may achieve micrometers.
They are very current in living cells and seem to play an important role in cell transport and communication~\cite{Onfelt_04}.

\begin{figure}[H]
\begin{center}
\includegraphics[scale=.7,angle=0]{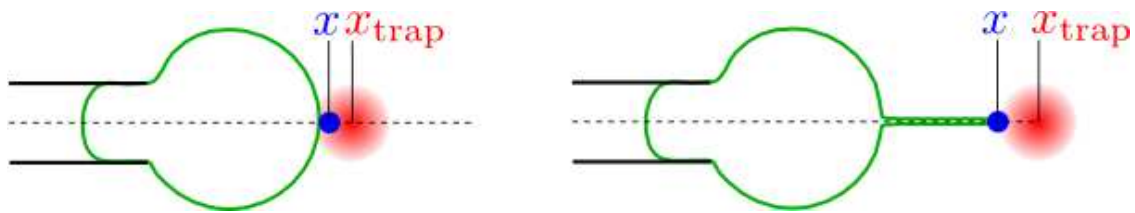}
\caption{Extraction of nanotubes using optical tweezers. The laser beam, represented in red, traps the glass bead attached to the membrane (in blue). The laser is then displaced and a tube is pulled. By controlling the position of the center of the laser beam $x_\mathrm{trap}$ and of the bead $x$, one deduces the applied force in the direction of the extracted tube.}
\label{fig_3_trap}
\end{center}      
\end{figure}

In laboratory, nanotubes can be extracted by applying very localized forces to membranes.
In Fig.~\ref{nano_extract} of chapter~\ref{introd}, we have presented a brief sum-up of the more popular methods used to extract nanotubes.
Here we interest ourselves in the force needed to extract (and hold) these tubes, which can be precisely measured in experiments using optical tweezer. 
The experimental procedure in this case consists in attaching a small glass
bead to a vesicle held by a micropipette~\cite{Heinrich_96},~\cite{Bo_89}.
A laser is pointed to the glass bead, which is thus attracted to the center of
the beam with a force that depends linearly on the distance between the bead
and the center of the beam.
In experiments, one displaces the position of the center of the beam, denoted
$x_{\mathrm{trap}}$, while measuring the position $x$ of the bead, as shown in Fig.~\ref{fig_3_trap}.
One then deduces the applied force through $f = -k_\mathrm{trap}(x -
x_\mathrm{trap})$, where $k_\mathrm{trap}$ is a constant that characterizes
the stiffness of the optical trap.

\noindent Remark that usually, as in the case of Fig.~\ref{fig_3_trap}, one measures only the force in the
axis of the tube, which is by symmetry the only component with non-vanishing average.
Note also that nanotubes are not stable: if one stops applying the point force, the membrane will evolve to a less curve configuration and the tube will be re-absorbed in the vesicle. 

Former theoretical works studied both the formation mechanism of nanotubes~\cite{Derenyi_02}, \cite{Powers_02} and their (dynamical) stability~\cite{Bozic_01}, \cite{Ou-Zhong_89}, ~\cite{Nelson_95}. 
As nanotubes are very thin compared to the GUVs from which they are usually pulled, it is usually assumed that the vesicle acts as a lipid reservoir to the tube.
In this case, as discussed in refs.~\cite{Derenyi_02} and ~\cite{Powers_02}, one can neglect the pressure difference across the tube.
The effective energy $\mathcal{H}'$ is thus simply given by the Helfrich Hamiltonian (eq.(\ref{Helfrich})) plus the work of the force that keeps the tube.
For a symmetrical membrane, i. e., a membrane whose spontaneous curvature vanishes, the energy for a perfectly cylindrical tube with radius $R$ and length $L$ is~\cite{Derenyi_02} 

\begin{equation}
\mathcal{H}' = \left(\frac{\kappa}{2R^2} + \sigma\right)2\pi R L - f L\, .
\end{equation}

\noindent where $\kappa$ is the bending rigidity and $\sigma$ is the Lagrange multiplier associated to the microscopical area of the membrane, which we remind is not directly measurable.
The energy coming from the Gaussian curvature is omitted, since we do not consider topological changes.
Minimizing this energy with respect to $R$ and $L$, one obtains, respectively

\begin{equation}
R_0 = \sqrt{\frac{\kappa}{2 \sigma}} 
\end{equation}

\noindent and

\begin{equation}
  f_0 = 2\pi \sqrt{2 \kappa \sigma} \, .
\end{equation}

\noindent These values correspond to the mean-field values of the radius and of the force needed to hold a tube, in the sense that thermal fluctuations relatives to the cylindrical shape were totally neglected. 

\begin{figure}[H]
\begin{center}
\includegraphics[scale=.5,angle=0]{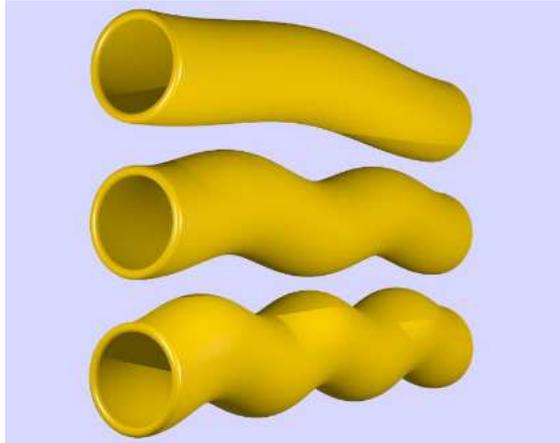}
\caption{The first soft, Goldstone modes~\cite{Fournier_07a}.}
\label{fig_3_Goldstone}
\end{center}      
\end{figure}

At first glance, one may think that neglecting the effects of thermal fluctuations is largely justified, since it is a reasonable assumption for planar membranes: as the correlation length is $\propto \sigma^{-1/2}$, fluctuations are quickly suppressed as the tension increases~\cite{Powers_02}. 
Recently, however, it has been shown that the tubular geometry implied a substantially different behavior:
tubes should present very strong shape fluctuations due to a one-dimensional set of extremely soft modes (Goldstone modes, see Fig.~\ref{fig_3_Goldstone})~\cite{Fournier_07a}.

\noindent Accordingly, it is natural to ask how the average force along the tube's axis $f$, taking into account its fluctuations, differs from the mean-field value $f_0$.
It is our aim this chapter to settle this question.
To do so,  we follow roughly the same steps as in the last chapter, starting by introducing the parametrization and the energy in \ref{section_3_param}.
Afterwards, we shall derive the projected stress tensor for quasi-cylindrical geometry in section~\ref{section_3_stress}.
As this calculation is totally new, we propose some verifications in the same section.
In section~\ref{section_3_average}, we average the stress tensor and evaluate $f$.  
At last, in section~\ref{section_3_discussion}, we compare $f$ with $f_0$ and discuss in which cases one is allowed to assume $f \simeq f_0$.
There we discuss also experimental consequences and re-interpret the curve shown in Fig.~\ref{heinrich}.

\section{Parametrization and Hamiltonian}
\label{section_3_param}

We shall restrict our attention to deformed tubes weakly departing from the cylinder corresponding the mean-field approximation whose radius, as we have shown above, is given by

\begin{equation}
  R_0 = \sqrt{\frac{\kappa}{2\sigma}} \, .
  \label{eq_3_R0}
\end{equation}

\begin{figure}[H]
\begin{center}
\includegraphics[width=0.35\columnwidth]{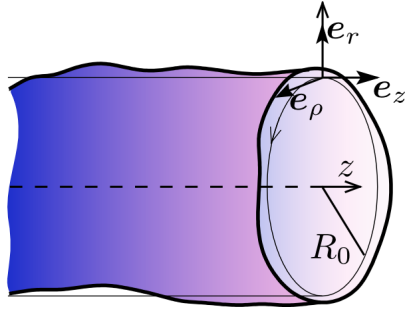}
\caption{Parametrization of the fluctuating tube (thick line). The thin solid line represents the reference cylinder with radius $R_0$.}
\label{fig_3_param}
\end{center}      
\end{figure}

Let's consider a cylindrical coordinate system $(O; r, \theta, z)$ aligned with the tube (see Fig.~\ref{fig_3_param}).

\noindent The shape of the fluctuating tube is parametrized by

\begin{equation}
  \bm{r}(\rho,z) = R_0 \left[ 1 + u(\rho,z)\right] \bm{e}_r(\rho) + z \, \bm{e}_z \, ,
  \label{eq_3_param}
\end{equation}

\noindent with $u \ll 1$ and $z \in [0,L]$, where $L$ is the total length of the tube.
Note that instead of $\theta$, we have used $\rho = R_0\, \theta \in [0, 2\pi R_0]$, in order to have $u$ as a function of two variables with the same dimension.

We shall consider the situation of a relatively short tubule extracted from a giant vesicle of radius $R_\mathrm{ves} \gg R_0$ (or from a vesicle connected to a lipid reservoir), so that  each monolayer of the tubule is actually exchanging material with a very
large reservoir and the standard Helfrich model (see section~\ref{model_model}) is sufficient for the calculation of equilibrium and statistical
properties~\cite{Derenyi_02},~\cite{Powers_02}.
As we do not consider topology changes, the Hamiltonian is simply given by

\begin{equation}
  \mathcal{H} = \int_S \left ( \sigma + \frac{\kappa}{2} H^2 \right) \, dA\, ,
  \end{equation}

\noindent where $S$ is the tube's surface.
Note that taking into account the area-difference elasticity, as is done in the $ADE$ model, is essential in the situation where very long tubules are extracted from small
vesicles~\cite{Waugh_92}, or when studying the formation of small
tethered vesicles under the action of an axial load~\cite{Heinrich_99}.
Remark also that one should usually consider the pressure difference across the membrane by adding a term $- \Delta P \, V$ to the Hamiltonian, where $V$ is the volume of the tube and $\Delta P = P_\mathrm{in} - P_\mathrm{out}$, with $P_\mathrm{in}$ (resp. $P_\mathrm{out}$) is the pressure inside (resp. outside) the vesicle from which the tube is extracted.
From the Young-Laplace equation (eq.(\ref{Laplace})), $\Delta P$ relates to the vesicle's radius and tension through $\Delta P = 2 \tau/R_\mathrm{ves}$.
Let's compare the contribution of this term with the contribution coming the term proportional to $\sigma$ for a tube of radius $R \ll R_\mathrm{ves}$ and length $L$:

\begin{equation}
  \frac{\Delta P \, V}{\sigma \, A} = \frac{\Delta P \, \pi R^2 \, L}{\sigma \, 2\pi R \, L} = \frac{\tau}{\sigma} \times \frac{R}{R_\mathrm{ves}} \ll 1 \, ,
\end{equation}

\noindent since we are extracting tubes far smaller than the vesicle and since $\tau < \sigma$.
It is justified thus to neglect the pressure difference across the tubule~\cite{Derenyi_02},~\cite{Powers_02}.

Differential geometry yields the general $dA$ given in eq.(\ref{eq_2_diff_area}) and $H$ given in eq.(\ref{eq_2_H}).
For the case of quasi-cylindrical geometry, we obtain up to order two on $u$

\begin{equation}
  \mathcal{H} \simeq \int_\Omega  h(u,\{u_i\}, \{u_{ij}\}) \, d\rho \, dz ,
  \label{eq_3_H}
  \end{equation}

\noindent where $i,j \in \{\rho,z\}$, $u_i \equiv \partial_i u = \partial u/\partial i$, $\Omega$ corresponds to the domain of the reference cylinder of radius $R_0$ over which the membrane stands, and~\cite{Fournier_07a}

\begin{eqnarray}
h&=&\sigma\left[
2+u^2-2 \, R_0^2\, (u_{\rho\rho}+u_{zz})+R_0^4\, (u_{\rho\rho}+u_{zz})^2
\right.\nonumber\\
\nonumber\\
&+&\left.2\, R_0^2\, u_\rho^2+4\, R_0^2\, u\, u_{\rho\rho}\right]+\mathcal{O}(u^3).
\label{eq_3_h2}
\end{eqnarray}

\section{Derivation of the stress tensor for a cylindrical\\
  geometry}
\label{section_3_stress}

\begin{figure}[H]
\begin{center}
\includegraphics[scale=.05,angle=0]{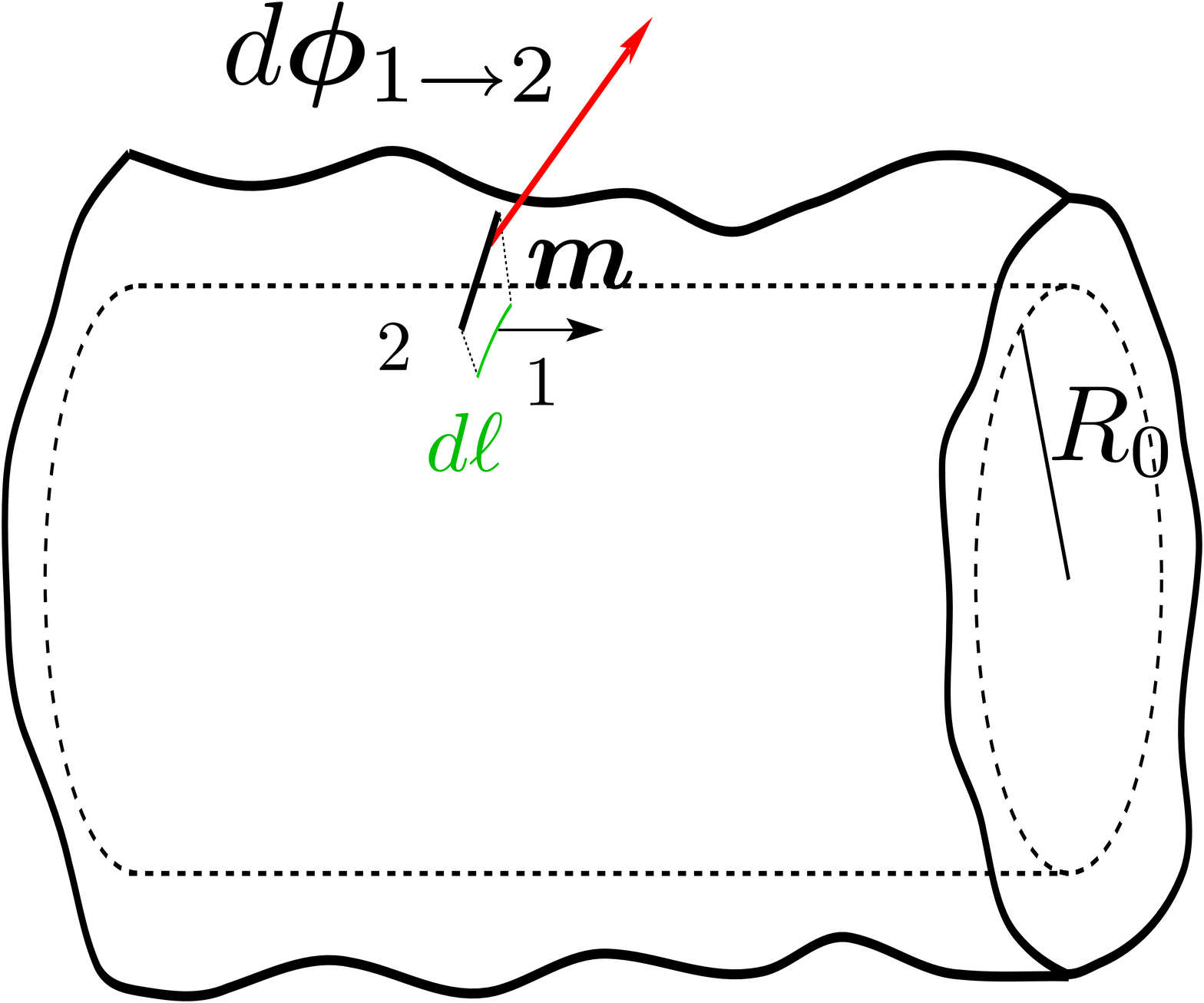}
\caption{The upper vector in red represents the three-dimensional force exchanged through an infinitesimal cut on the membrane. In green, we see the projection of the cut onto the reference cylinder. Note that the vector $\bm{m}$, normal to the projected cut, is contained on the reference cylinder (dashed lines).}
\label{fig_3_force_schema}
\end{center}      
\end{figure}

In analogy to the planar and quasi-spherical cases, the projected stress tensor relates linearly the force that the region $1$ exerts over region $2$ to the length of the projected cut $d \ell$ through

\begin{equation}
  d \bm{\phi}_\mathrm{1 \rightarrow 2} = \bm{\Sigma} \cdot \bm{m} \, d\ell \, ,
\end{equation}

\noindent where $\bm{m} = m_\rho \, \bm{e}_\rho + m_z \, \bm{e}_z$ is the normal to the cut on the reference cylinder (see Fig~\ref{fig_3_force_schema}).

\begin{figure}[H]
\begin{center}
\subfigure[Coordinate space used in the parametrization of the tube's shape. The shaded area represents $\Omega$, the thick solid line represents $\partial \Omega$ and the dashed line represents $\partial \Omega$ after an infinitesimal displacement $\delta \bm{a}$ of the membrane. The line in the center represents the projected cut.]{
\includegraphics[scale=.5,angle=0]{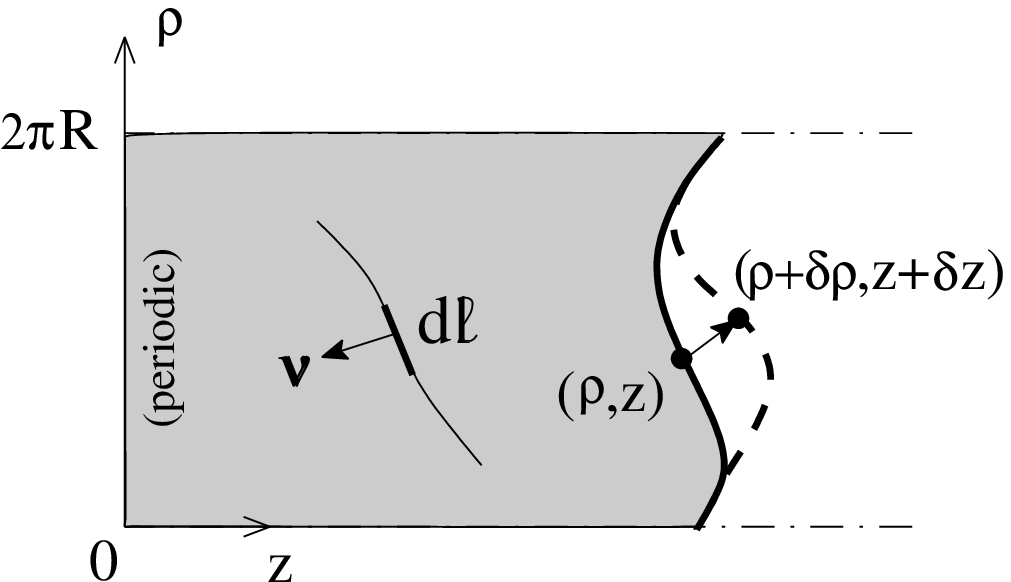}
\label{fig_3_stressfig}
}
\subfigure[Fluctuating tube before (shaded) and after (dashed red) a general displacement. The reference cylinder is shown by the dashed black line.]{
\includegraphics[scale=.75,angle=0]{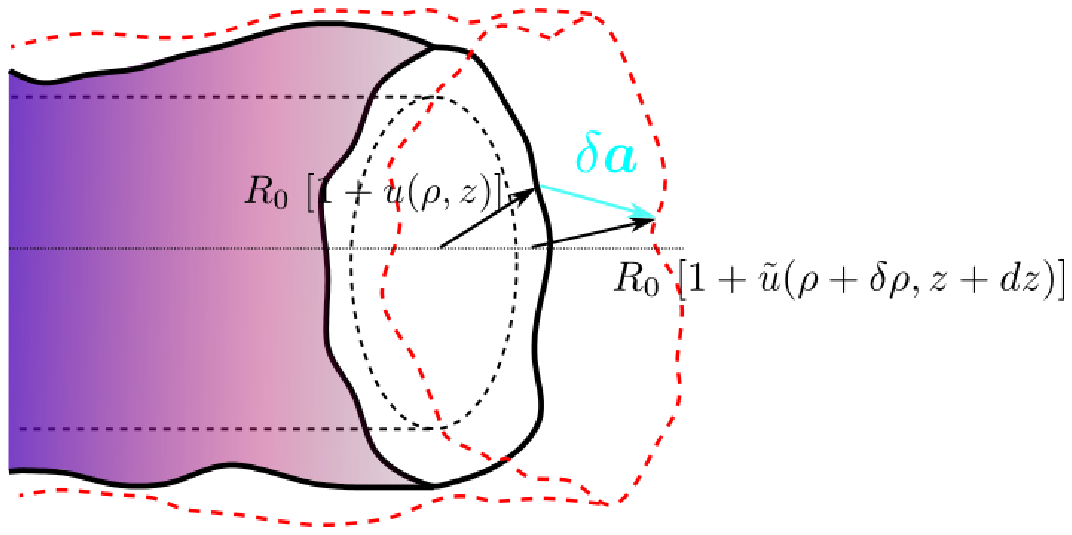}
\label{fig_3_before_after}
}
\caption{Deriving the projected stress tensor.}
\label{fig_3_deform}
\end{center}      
\end{figure}

In order to derive $\bm{\Sigma}$, we consider at each point of the
membrane an arbitrary infinitesimal displacement
$\delta\bm{a}=\delta a_\rho\,\bm{e}_\rho+\delta a_z\,\bm{e}_z+\delta a_r\,\bm{e}_r$
corresponding to a variation $(\delta\rho,\delta z)$ on the projected
cylinder (see Fig.~\ref{fig_3_deform}).
Accordingly, the membrane's shape becomes $\tilde{u}(\rho,z)=u(\rho,z)+\delta
u(\rho,z)$.

As one can see in Fig.~\ref{fig_3_before_after}, the new edge's position satisfies $R_0[1+\tilde{u}(\rho+\delta\rho,z+\delta z)]\,\bm{e}_r(\rho+\delta\rho) +(z+\delta z)\,\bm{e}_z =R_0[1+u(\rho,z)]\,\bm{e}_r(\rho)+z\,\bm{e}_z+\delta\bm{a}$, which implies

\begin{eqnarray}
\label{eq_3_first}
\delta a_\rho&=&\delta\rho\left(1+u\right),\\
\delta a_z&=&\delta z\,,\\
\delta a_r&=&R_0\left(\delta u+u_z\delta z+u_\rho\delta\rho\right).
\label{eq_3_firstend}
\end{eqnarray}

We now impose that $\delta\bm{a}$ is done at fixed orientation of the
membrane's normal: in this way, only the boundary forces work,
not the torques.
The normal to the membrane is
$\bm{n}=\bm{t}_\rho\times\bm{t}_z/|\bm{t}_\rho\times\bm{t}_z|$, with
$\bm{t}_\rho=\partial_\rho\bm{r}=R_0 \, u_\rho\,\bm{e}_r+(1+u)\,\bm{e}_\rho$ and
$\bm{t}_z=\partial_z\bm{r}=R_0\,u_z\, \bm{e}_r+\bm{e}_z$.
This gives 
$\bm{n}=\left[1-R_0^2\left(u_\rho^2+u_z^2\right)/2\right]\,\bm{e}_r(\rho)
+R_0 \left(u-1\right) \, u_\rho\,\bm{e}_\rho(\rho) -R_0 \, u_z\,\bm{e}_z+\mathcal{O}(u^3)$.
The normal variation is given by $\delta\bm{n}=\tilde{\bm{n}} (\rho+\delta\rho,z+\delta z)-\bm{n}(\rho,z)$, $\tilde{\bm{n}}$ being the analog of $\bm{n}$ for $\tilde{u}$ instead of $u$.
In order to impose $\delta\bm{n}=0$, we require that $\delta\bm{n}\cdot\bm{t}_\rho=0$ and
$\delta\bm{n}\cdot\bm{t}_z=0$, yielding, up to order $u^2$

\begin{eqnarray}
\label{eq_3_second}
\delta u_z&=&
\left(
u_z \, u_\rho-u_{\rho z}
\right)\delta\rho-u_{zz}\, \delta z\,, \nonumber\\
\\
\delta u_\rho&=&
\left[
R_0^{-2}\left(1+u\right)+2 \, u_\rho^2-u_{\rho\rho}
\right]\delta\rho\nonumber\\
&+&\left( u_z\, u_\rho-u_{\rho z} \right)\delta z
+u_\rho\, \delta u\,.
\label{eq_3_secondend}
\end{eqnarray}

Using eqs.(\ref{eq_3_first})--(\ref{eq_3_secondend}), we may now express the
variations $\{\delta\rho,\delta z,\delta u,\delta u_\rho,\delta
u_z\}$ at the boundary in terms of the components of $\delta\bm{a}$.
We obtain, to order $u^2$

\begin{eqnarray}
\delta\rho&=&\left(1-u+u^2\right)\delta a_\rho\,,\\
\nonumber\\
\delta z&=&\delta a_z\,,\\
\nonumber\\
\delta u&=&(u-1)u_\rho\,\delta a_\rho-u_z\,\delta a_z+R_0^{-1}\,\delta a_r\,,\\
\nonumber\\
\delta u_\rho&=&
\left[
R_0^{-2}+u_\rho^2+\left(u-1\right)u_{\rho\rho}
\right]
\,\delta a_\rho\nonumber\\
&-&u_{\rho z}\,\delta a_z+R_0^{-1}\left(1-u\right)u_\rho\,\delta a_r\,,\\
\nonumber\\
\delta u_z&=&
\left[
u_\rho \, u_z+\left(u-1\right)u_{\rho z}
\right]
\,\delta a_\rho-u_{zz}\,\delta a_z\,.
\end{eqnarray}

To obtain $\bm{\Sigma}$, we study the variation of the energy after the displacement $\delta \bm{a}$.
On the one hand, in terms of $h$, one has

\begin{equation}
\delta \mathcal{H}_\mathrm{bulk} = \int_\Omega \left[\frac{\partial h}{\partial u}
  - \partial_i \left(\frac{\partial h}{\partial u_i}\right)
  + \partial_i \partial_j \left(\frac{\partial h}{\partial u_{ij}}\right)
  \right] \, d\rho \, dz \, 
\label{eq_3_deltaH_bulk}
\end{equation}

\noindent for the bulk of the membrane and

\begin{equation}
\delta \mathcal{H}_\mathrm{boundary}  =  \int_{\partial\Omega}\! m_i\left[
h \, \delta i+
\left(\frac{\partial h}{\partial u_i}-\partial_j\frac{\partial
h}{\partial u_{ij}}\right) \delta u
+\frac{\partial h}{\partial u_{ij}} \, \delta u_j
\right] d\ell \, ,
\end{equation}

\noindent for the boundary energy variation.
On the other hand, the work of forces at the boundary is given by

\begin{equation}
\delta \mathcal{H}_\mathrm{boundary} = \!\!\int_{\partial\Omega}\!
\delta\bm{a}\cdot\bm{\Sigma}\cdot \bm{m} \, d \ell \,.
\end{equation}

\noindent By comparing the last two equations, we obtain $\Sigma_{zi}$, $\Sigma_{ri}$ and
$\Sigma_{\rho i}$ ($i$ being either $\rho$ or $z$):

\begin{eqnarray}
\Sigma_{zi}&=&h \, \delta_{zi}
-\left(\frac{\partial h}{\partial u_i}-\partial_j\frac{\partial
h}{\partial
u_{ij}}\right)u_z-\frac{\partial h}{\partial u_{i\rho}}\, u_{\rho z}
-\frac{\partial h}{\partial u_{iz}}\, u_{zz}+\mathcal{O}(u^3)\, ,
\label{eq_3_sigmazi}
\\
\phantom{}\nonumber\\
\nonumber\\
\Sigma_{ri}&=&
\frac{1}{R_0}\left(\frac{\partial h}{\partial
u_i}-\partial_j\frac{\partial
h}{\partial u_{ij}}\right)
+\frac{1}{R_0}\left(1-u\right)\frac{\partial h}{\partial u_{i\rho}}\, u_\rho
+\mathcal{O}(u^3)\, ,\\
\phantom{}\nonumber\\
\nonumber \\
\Sigma_{\rho i}&=&\left(1-u+u^2\right)h\,\delta_{\rho i}
+\left(\frac{\partial h}{\partial u_i}-\partial_j\frac{\partial
h}{\partial
u_{ij}}\right)\left(u-1\right)u_\rho \nonumber\\
\nonumber\\
&+&\frac{\partial h}{\partial
u_{iz}}\left[
u_\rho \, u_z+\left(u-1\right)u_{\rho z}
\right] + \frac{\partial h}{\partial u_{i\rho}}\left[
\frac{1}{R_0^2}+u_\rho^2+\left(u-1\right)u_{\rho\rho}
\right]+\mathcal{O}(u^3)\,. \nonumber
\label{eq_3_sigmaroi}\\
\end{eqnarray}

Note that due to the presence of terms such as $R_0^{-2}\,\partial h/\partial
u_{i\rho}$ (see, e.g., the expression of $\Sigma_{\rho i}$), it is
in general necessary to have $h$ at $\mathcal{O}(u^3)$ in order to get the
stress components at $\mathcal{O}(u^2)$.
This means adding

\begin{eqnarray}
h_3 &=& -u^3 - R_0^2 \, \left(6 \, u \, u_{\rho\rho} +  5 \, u_\rho^2 - u_z^2 \right)\, u \nonumber\\
&+& R_0^4 \left(u_{zz}^2 - 3 \, u_{\rho\rho}^2 - 2 \, u_{zz}\, u_{\rho\rho}\right) + 4 \, R_0^4 \, u_\rho \, u_z \, u_{\rho z} \nonumber\\
&+& R_0^4 \left(u_{\rho\rho} - u_{zz}\right)u_\rho^2 + R_0^4\left(u_{\rho\rho} + 3 u_{zz}\right)u_z^2
\end{eqnarray}

\noindent to $h$ given in eq.(\ref{eq_3_h2}) before evaluating the derivatives in eqs.(\ref{eq_3_sigmazi})--(\ref{eq_3_sigmaroi}).
For $\Sigma_{zi}$, however, one
may check the $\mathcal{O}(u^2)$ terms in $h$ are sufficient.
Explicitly, we obtain up to order two on $u$

\begin{eqnarray}
\Sigma_{zz}&=&\sigma\left\{2+u^2+2R_0^2\left[u_\rho^2+\left(2u-1\right)u_{\rho\rho}\right]\right. \nonumber \\
&+&R_0^4 \left. \left[u_{\rho\rho}^2-u_{zz}^2+2u_z\left(u_{zzz}+u_{\rho\rho
z}\right)\right]\right\} \, ,
\label{eq_3_stress1}\\
\nonumber\\
\Sigma_{z\rho} &=& 2 \, R_0^2\left(1 - 2 \, u\right)u_{\rho z} + 2 \, R_0^4 \left[u_z\left(u_{\rho z z} + u_{\rho\rho\rho}\right) - u_{\rho z}\left(u_{zz} + u_{\rho\rho}\right)\right] \, , \\
\nonumber \\
\Sigma_{rz} &=& 2 \, R_0 \, u \, u_z - 2 \, R_0^3\left[\left(1+u\right)u_{zzz} +
  \left(1 - u\right)u_{\rho\rho z} + u_z \, u_{zz} - u_\rho \, u_{z\rho} \right] \, , \\
\nonumber \\
\Sigma_{r\rho} &=& - 2 \, R_0 \left(1 -4 \, u\right)u_\rho \nonumber \\
&-& 2 \, R_0^3 \, \left[\left(1 - 3\, u\right)u_{\rho\rho\rho} + \left(1 - u\right)u_{\rho z z } - 4 u_\rho \, u_{\rho\rho} + u_\rho \, u_{zz} + u_z \, u_{\rho z}\right] \, ,\\
\nonumber \\
\Sigma_{\rho z} &=& 2 \, R_0^2 \left[\left( 1- u\right)u_{\rho z} + u_{\rho} \, u_z \right] \nonumber \\
&-& 2 \, R_0^4 \left[\left(u_{zz} + u_{\rho\rho}\right)u_{\rho z} - \left(u_{\rho\rho z} + u_{zzz} \right)u_\rho\right] \,,\\
\nonumber\\
\Sigma_{\rho \rho} &=& 2 \, u - 3\, u^2 + R_0^2 \left[u_z^2 + u_\rho^2 + 2 \left( 1- 3\, u\right)u_{\rho\rho}\right] \nonumber \\
&+& R_0^4 \left[u_{zz}^2 - u_{\rho\rho}^2 + 2 \left(u_{\rho z z } + u_{\rho\rho\rho}\right)u_\rho\right]\, .
\label{eq_3_sigmarhorho}
\end{eqnarray}

Note that we may easily recover from $\Sigma_{zz}$ the
mean-field force needed to hold a tubule.
Indeed, $u=0$ yields
$\Sigma_{zz}=2\sigma$~\cite{Fournier_07}, i.e., $f_0=2\pi
R\times2\sigma=2\pi\sqrt{2\kappa\sigma}$.
As we have discussed in section~\ref{tube_exp}, this result is very interesting: if the mechanical tension were due only to the tension $\sigma$, we should expect $f_0 = \sigma \times 2\pi R_0$.
In reality, the curvature yields a supplementary term $1/2 \, (\kappa/R_0^2)$ to the mechanical tension $\tau$, which explains the factor two in our result.
In the next two sections we shall propose some tests to verify the correctness of these equations.

\subsection{Verification: stress tensor in the tangent frame}

\begin{figure}[H]
\begin{center}
\includegraphics[width=.5\columnwidth,angle=0]{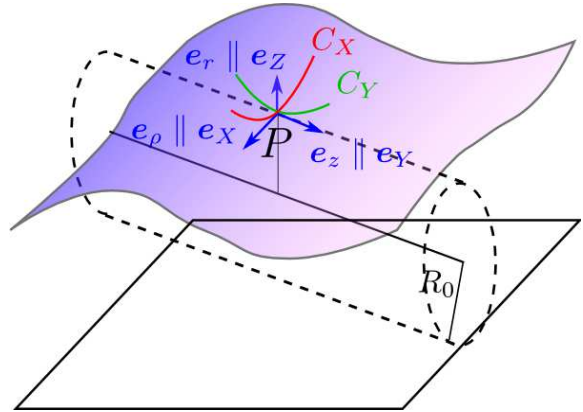}
\caption{The shaded surface represents a piece of membrane. The principal curvatures, as well as the principal directions $\bm{e}_X$ and $\bm{e}_Y$ in $P$, are shown in red and green.
The dashed cylinder represents the reference cylinder.}
\label{fig_3_tangent}
\end{center}      
\end{figure}

Here we propose a first check by showing that from eqs.(\ref{eq_3_stress1})--(\ref{eq_3_sigmarhorho}), one re-obtains the stress tensor in the local frame given in eq.(\ref{Sigma_local_frame}).
We consider a general membrane, not necessarily tubular.
At a general point $P$ of the membrane, there are two principal curvatures $C_X$ and $C_Y$ whose principal directions $\bm{e}_X$ and $\bm{e}_Y$ are orthogonal.
We place our reference cylinder tangent to the membrane at $P$, with its axis direction 
$\bm{e}_z$ parallel to $\bm{e}_Y$ and $\bm{e}_\rho$ parallel to $\bm{e}_X$ , as shown in Fig.~\ref{fig_3_tangent}.

\begin{figure}[H]
\begin{center}
\includegraphics[width=.45\columnwidth,angle=0]{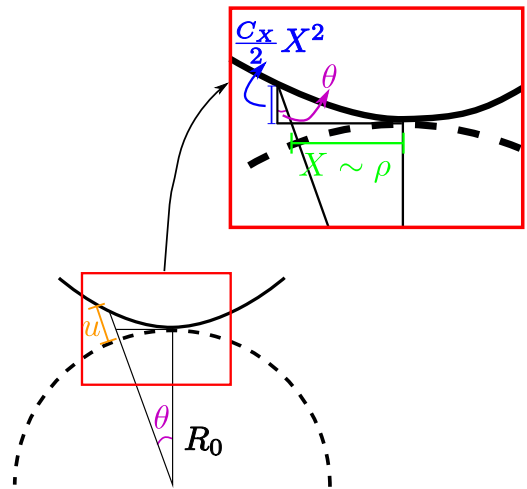}
\caption{Schematic representation highlighting the geometrical quantities needed to determine $u(\rho,z)$ in the plan perpendicular to $\bm{e}_z$.}
\label{fig_3_tangent_geom}
\end{center}      
\end{figure}

By geometry (see Fig.~\ref{fig_3_tangent_geom}), one determines the shape of the membrane near to $P$ in the cylindrical coordinate system, yielding

\begin{equation}
  \bm{r} = R_0 \, \left[1 + \frac{1}{2}\left(C_X + \frac{1}{R_0} \right) \frac{\rho^2}{R_0} + \frac{1}{2} \, C_Y \frac{z^2}{R^2} \right] \bm{e}_r \, .
  \label{eq_3_tangent}
  \end{equation}

\noindent Comparing eq.(\ref{eq_3_param}) and eq.(\ref{eq_3_tangent}), we identify

\begin{equation}
  u(\rho,z) = \frac{1}{2}\left(C_X + \frac{1}{R_0} \right) \frac{\rho^2}{R_0} + \frac{1}{2} \, C_Y \frac{z^2}{R^2} \, .
  \label{eq_3_u}
  \end{equation}

Applying eq.(\ref{eq_3_u}) to eqs.(\ref{eq_3_stress1})--(\ref{eq_3_sigmarhorho}), one obtains

\begin{eqnarray}
  \Sigma_{zz} &=& \sigma - \frac{\kappa}{2} \, C_Y^2 + \frac{\kappa}{2} \, C_X^2 \, ,\\
  \Sigma_{\rho\rho} &=& \sigma + \frac{\kappa}{2} \, C_Y^2 - \frac{\kappa}{2} \, C_X^2 \, ,\\
  \Sigma_{rz} &=& -\kappa \, \frac{\partial C}{\partial z} \, ,\\
  \Sigma_{r\rho} &=& -\kappa \, \frac{\partial C}{\partial \rho} \, ,\\
  \Sigma_{z\rho} &=& 0 \, ,\\
  \Sigma_{\rho z} &=& 0 \, ,
  \end{eqnarray}

\noindent where $C = C_X + C_Y$ is the total curvature.
Noting that we have the equivalences $X \equiv \rho$, $Y \equiv z$ and $Z \equiv r$ near $P$, these equations are identical to eq.(\ref{Sigma_local_frame}).

\subsection{Verification: force between two rings constraining the tube}

\begin{figure}[H]
\begin{center}
\includegraphics[width=.65\columnwidth,angle=0]{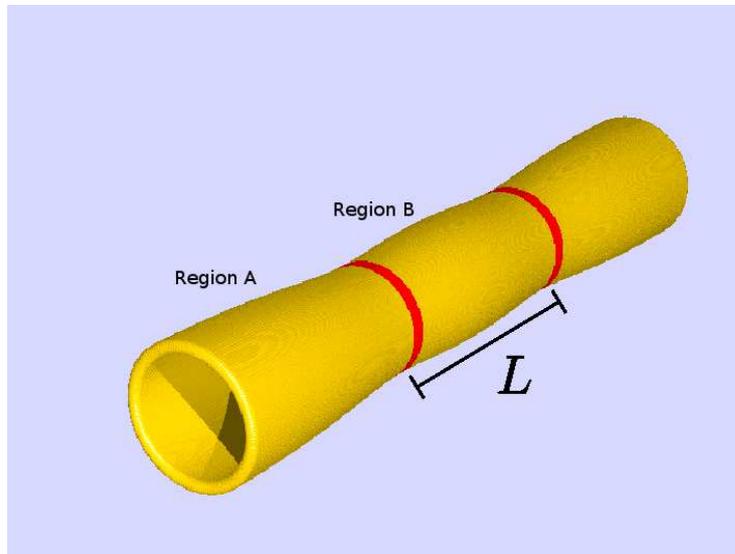}
\caption{Tube constrained by two rings separated by a distance $L$. We have also considered the case of ``undulating rings''. The region $A$ goes from very far to the first ring up to it, while the region $B$ stands for the region between the rings.}
\label{fig_3_anneaux}
\end{center}      
\end{figure}

In order to control the validity of the formula giving $\Sigma_{zz}$, which will be the only component used in the next sections, let us
calculate the force acting between two ``undulating rings"
separated by a distance $L$ (see Fig.~\ref{fig_3_anneaux}) by deriving the free-energy and compare to the force obtained using the projected stress tensor.

The rings are described by the boundary conditions

\begin{equation}
u(\rho,\pm L/2)=U_0\, \cos\left(\frac{n\rho}{R}\right)\,,
\end{equation}

\noindent and $\partial_zu(\rho,\pm L/2)=0$, for $n>1$.
By symmetry, we assume
$u(\rho,z)=U(z) \, \cos(n\rho/R)$.
Thus the distortion
energy~(\ref{eq_3_H})--(\ref{eq_3_h2})
between the rings takes the form:

\begin{equation}
\mathcal{H}=\pi R_0 \, \sigma\int_{-\frac{L}{2}}^{\frac{L}{2}}
\left[\left(n^2-1\right)^2U^2
-2 \, n^2 R_0^2 \, U \, U''+R_0^4 \, U''^2 \right] \, dz \, .
  \label{eq_3_Hbetween}
\end{equation}

The equilibrium shape is given by the Euler-Lagrange equation given in eq.(\ref{eq_3_deltaH_bulk}), yielding

\begin{equation}
  (n^2-1)^2U(z)-2n^2R^2U''(z)+R^4U''''(z)=0 \, .
  \label{eq_3_diff}
\end{equation}

\noindent The solution satisfying the boundary conditions is 

\begin{equation}
U(z)= U_0  \, \frac{\left[n_+\sinh(n_+\ell)\cosh(n_-z/R_0) -n_-\sinh(n_-\ell)\cosh(n_+z/R_0)\right]}{A(\ell)} \, ,
\label{eq_3_Usol}
\end{equation}

\noindent where $\ell=L/(2R_0)$, $A(\ell) = n_+\sinh(n_+\ell)\cosh(n_-\ell)-n_-\sinh(n_-\ell)\cosh(n_+\ell)$ and $n_\pm=(n^2\pm\sqrt{2n^2-1})^{1/2}$.

To evaluate the balance of the forces acting over first ring, which is symmetrical from the force acting over the second one, we have to consider the force exerted by region $A$, $\bm{f}_A$, and the force exerted by region $B$, $\bm{f}_B$.
Each region is in equilibrium, implying that the integral of $\Sigma_{zz}$ over $\rho$ is constant in each one of them.
We may thus consider an arbitrary projected path with $\bm{m} = \bm{e}_z$ in each region to evaluate forces.
For the region $A$, we will consider a path very far from the ring, so that $u \rightarrow 0$.
We have

\begin{eqnarray}
  \bm{f}_A &=& \int_0^{2 \pi R_0} \bm{\Sigma} \cdot \bm{e}_z \, d\rho \, , \nonumber\\
  &=& \int_0^{2 \pi R_0} \Sigma_{\rho z} \, d\rho \, \bm{e}_\rho  + \int_0^{2 \pi R_0} \Sigma_{r z} \, d\rho \, \bm{e}_r + \int_0^{2 \pi R_0} \Sigma_{z z} \, d\rho \, \bm{e}_z\, , \nonumber\\
  &=& \int_0^{2\pi R_0} 2\sigma \, d\rho \, \bm{e}_z \, .
\end{eqnarray}

\noindent In the last passage, we have used the fact that $\Sigma_{zz} \to 2 \sigma$, $\Sigma_{\rho z} \to 0$ and $\Sigma_{rz} \to 0$  as $u \to 0$.
For the region $B$, we consider a path at $z = 0$.
We obtain

\begin{eqnarray}
  \bm{f}_B &=& \int_0^{2 \pi R_0} \bm{\Sigma}|_{z=0} \cdot \bm{e}_z \, d\rho \, , \nonumber\\
  &=& \int_0^{2 \pi R_0} \Sigma_{\rho z}|_{z=0} \, d\rho \, \bm{e}_\rho  + \int_0^{2 \pi R_0} \Sigma_{r z}|_{z=0} \, d\rho \, \bm{e}_r + \int_0^{2 \pi R_0} \Sigma_{z z}|_{z=0} \, d\rho \, \bm{e}_z\, , \nonumber\\
  &=& \int_0^{2\pi R_0} \Sigma_{zz}|_{z=0} \, d\rho \, \bm{e}_z \, .
\end{eqnarray}

\noindent The forces in the other directions, $\bm{e}_\rho$ and $\bm{e}_r$, vanish after integration, as expected given the symmetry of the system.
The resulting force is then

\begin{equation}
f(L) \, \bm{e}_z = \bm{f}_A - \bm{f}_B = \int_0^{2\pi R_0}\left(2\sigma-\Sigma_{zz}|_{z=0}\right) \, d\rho \, \bm{e}_z\, .
\end{equation}

Using eq.(\ref{eq_3_stress1}) to calculate $\Sigma_{zz}$, we obtain

\begin{equation}
f(L)=-B\left(n^2-1\right)^2
 \, \frac{\cosh\left(2n_+\ell\right)-\cosh\left(2n_-\ell\right)}
{A(\ell)^2}\,,
\end{equation}

\noindent where $B=\pi R\sigma U_0^2\sqrt{2n^2-1}$.
Intuitively, one should expect the rings to collapse in order to minimize the tube's deformation.
Indeed, the resulting force between the rings is always attractive.

In order to check this result, we propose ourselves to re-derive $f(L)$ using a different method that does not involve the stress tensor.
First, we will evaluate the energy stored between the rings, given in eq.(\ref{eq_3_Hbetween}).
As the tube is in equilibrium, it's shape obeys eq.(\ref{eq_3_diff}).
Applying this equation to the first term of eq.(\ref{eq_3_Hbetween}), we obtain

\begin{equation}
  \mathcal{H} = \pi R_0^5 \, \sigma \int_{-\frac{L}{2}}^{\frac{L}{2}} \left(- U \, U'''' + U''^2\right) \, dz \, .
\end{equation}

\noindent Integrating by parts the term on $U\, U''''$ and reminding that $U(z)$ is an even function, that $U(\pm L/2) = U_0$ and that $U'(\pm L/2) = 0$, we obtain  

\begin{eqnarray}
\mathcal{H}&=&2\pi  \, R_0^5 \, \sigma \, U_0 \, U'''(-L/2) \, , \nonumber\\
\nonumber\\
&=&4BR_0\,\frac{n_+n_-\sinh(n_+\ell)\sinh(n_-\ell)}{A(\ell)}\,, 
\end{eqnarray}

\noindent where we have used the solution given in eq.(\ref{eq_3_Usol}) to obtain the last passage.
From the stored energy, the resulting force between the rings is given by

\begin{equation}
f(L)=-\frac{d\mathcal{H}(L)}{dL} \, .
\end{equation}

\noindent After a careful calculation, we recover the result obtained from the stress tensor (eq.(\ref{eq_3_stress1})), testifying of the correctness of the component $\Sigma_{zz}$ of the stress tensor.

\section{Evaluation of the average force}
\label{section_3_average}

In order to hold a nanotube, one must apply a force exactly equivalent to the force that the rest of the fluctuating tubule exerts.
Thus, considering a section of the tube with $\bm{m} = \bm{e}_z$, the average force needed to hold a fluctuating tube is

\begin{eqnarray}
f \, \bm{e}_z &=&  \left\langle\int_0^{2 \pi R_0} \bm{\Sigma} \cdot \bm{e}_z \, d \rho \right\rangle \, ,\nonumber \\
&=&\int_0^{2\pi R_0} \left(\langle \Sigma_{rz} \rangle \, \bm{e}_r + \langle \Sigma_{\rho z} \rangle \, \bm{e}_\rho + \langle \Sigma_{zz} \rangle \, \bm{e}_z \right)\, d\rho \, , \nonumber\\
&=& \int_0^{2\pi R_0} \langle\Sigma_{zz}\rangle \, d\rho \, \bm{e}_z 
= (f_0+f_\mathrm{fl}) \, \bm{e}_z \, ,
\end{eqnarray}

\noindent where we remind that $f_0=2\pi\sqrt{2\kappa\sigma}=2\pi\kappa/R_0$ is the mean-field force and $f_\mathrm{fl}$ is the correction due to
fluctuations.
Note that in average, there is no force perpendicular to the tube's axis, as expected by symmetry reasons.

\subsection{Correlation function}

Let us consider a tubule of length $L$ with periodic boundary conditions for simplicity.
The fluctuations of the tube's
shape may be decomposed in Fourier modes:

\begin{equation}
u(\rho,z)=\sqrt{\frac{R_0}{2\pi L}}\,{\sum_{m,\bar{q}}}u_{m,\bar q}\,
e^{i R_0^{-1}(m\rho+\bar{q}z)},
\end{equation}

\noindent where $m=0,\pm1,\ldots,\pm M$ and $\bar{q}=2\pi nR_0L$, with
$n=0,\pm1,\ldots,\pm N$.
As the modes with $m=\pm1$ and $\bar{q} = 0$ correspond to pure 
translation, they will be omitted in the following.
The cutoffs $M$ and $\bar q_\mathrm{max}$ (or $N$) are related to the high
wave-vector cutoff $\Lambda$ through $M=\Lambda R_0$ and $\bar{q}_\mathrm{max}=2\pi NR_0/L=\Lambda R_0$.
As in the last chapters, we assume that $\pi/\Lambda$ is
somewhat larger than the membrane thickness $a\approx5$~nm and we take
$\Lambda\approx1/a$.
Note that there is an uncertainty on $\Lambda$ of a
factor of order unity.

In terms of the Fourier modes, the Hamiltonian given in eq.(\ref{eq_3_H})
 becomes~\cite{Fournier_07a}, \cite{Santangelo_02} 

 \begin{equation}
\mathcal{H}\simeq\frac{\kappa}{2}\sum_{m,\bar {q}}\left[\left(m^2-1\right)^2+\bar q^2\left(\bar
q^2+2m^2\right)\right]|u_{m,\bar q}|^2 \, .
\end{equation}

\noindent Using the equipartition of energy, we have

\begin{equation}
\langle u_{m, \bar{q}} \, u_{n, \bar{k}} \rangle = \frac{k_\mathrm{B} T}{\kappa} \, \frac{1}{(m^2 - 1)^2 + \bar{q}^2(\bar{q}^2 + 2 m^2)} \, \delta_{m,n} \delta_{\bar{q},\bar{k}} \, ,
  \label{eq_3_corterm}
\end{equation}
 
\noindent where $\delta$ stands for the delta of Kronecker.
Hence, with $u\equiv u(\rho,z)$ and
$u'\equiv u(\rho',z')$, the correlation function of the
tubule thermal fluctuations is given by

\begin{eqnarray}
  G(\rho - \rho', z - z') &\equiv& \langle u\,u'\rangle \, , \nonumber\\
&=&\frac{k_\mathrm{B}T \, R_0}{2\pi\kappa L}
\sum_{m, \bar{q}}
\frac{
e^{iR_0^{-1}\left[m\left(\rho-\rho'\right)+\bar
q\left(z-z'\right)\right]}
}{
 \left(m^2-1\right)^2+\bar q^2\left(\bar q^2+2m^2\right)
 } \, , \\
&=&\frac{k_\mathrm{B}T}{(2\pi)^2\kappa}
\sum_m\int 
\frac{
e^{iR_0^{-1}\left[m\left(\rho-\rho'\right)+\bar
q\left(z-z'\right)\right]}
}{
 \left(m^2-1\right)^2+\bar q^2\left(\bar q^2+2m^2\right)
 } \, d\bar{q} \,  .
\label{eq_3_corel}
\end{eqnarray}

\noindent Here, as in the last sections, $k_\mathrm{B}T$ is the temperature and the brackets indicate the thermal average.
In the last passage, we have transformed the sum over $n$ into an integral,
which is legitimate for tubes longer than a few times $R_0$, so that both the sum and the integral run from $-\Lambda R_0$ to
$\Lambda R_0$.
Using this correlation, one can easily derive other correlations involving derivatives with respect to $\rho$ or $z$.
Let's see an example in detail: first, let's evaluate an average without using the correlation function

\begin{eqnarray}
  &&\langle u_z(\rho,z) \, u_{\rho\rho z}(\rho',z') \rangle = \nonumber \\
  &=& \frac{R_0}{2 \pi L} \left\langle \partial_z\left(\sum_{m,\bar{q}} u_{m,\bar{q}}\,  e^{i \, R_0^{-1}(m\rho + \bar{q}z)}\right) \times \partial_\rho^2\partial_z\left(\sum_{n,\bar{k}} u_{n,\bar{k}} \, e^{i \, R_0^{-1}(n\rho' + \bar{k}z')}\right) \right\rangle \, ,\nonumber\\
  &=& \frac{R_0}{2 \pi L} \left\langle \left[\sum_{m,\bar{q}} \left(\frac{i \, \bar{q}}{R_0}\right) u_{m,\bar{q}}\,  e^{i \, R_0^{-1}(m\rho + \bar{q}z)}\right] \times \left[\sum_{n,\bar{k}} \left(\frac{i\, n}{R_0}\right)^2 \left(\frac{i\, \bar{k}}{R_0}\right) u_{n,\bar{k}} \, e^{i \, R_0^{-1}(n\rho' + \bar{k}z')}\right] \right\rangle \, ,\nonumber\\
  &=&\frac{R_0}{2\pi L} \sum_{m,\bar{q}} \sum_{n, \bar{k}} \left(\frac{n^2 \bar{q} \bar{k}}{R_0^4}\right) \langle u_{m,\bar{q}} \, u_{n,\bar{k}}\rangle e^{i \, R_0^{-1}(m\rho + \bar{q}z) }\, e^{i \, R_0^{-1}(n\rho' + \bar{k}z') } \, , \nonumber\\
  &=& - \frac{k_\mathrm{B} T \, R_0}{2\pi \kappa L} \sum_{m,\bar{q}} \left(\frac{m^2 \bar{q}^2}{R_0^4}\right) \frac{ e^{i \, R_0^{-1}[m(\rho-\rho') + \bar{q}(z-z')] }}{(m^2 -1)^2 + \bar{q}^2(\bar{q}^2 + 2 m^2)}  \, ,
\label{eq_3_example}
\end{eqnarray}

\noindent where we have used eq.(\ref{eq_3_corterm}) to obtain eq.(\ref{eq_3_example}).
One can easily verify that this result can be simply obtained from eq.(\ref{eq_3_corel}) through

\begin{equation}
  \langle u_z(\rho,z) \, u_{\rho\rho z}(\rho',z') \rangle = \partial_z|_{(\rho,z)} \left(\partial_\rho^2\partial_z\right)|_{(\rho', z')} \left[G(\rho - \rho', z- z')\right] \, ,
\end{equation}

\noindent where $|_{(\rho,z)}$ indicates that the derivative is taken at the point $(\rho,z)$.
This method can be generalized to the calculation of similar averages.

\subsection{Average force}
\label{subsection_ave_force}

Calculating the average of each term of eq.(\ref{eq_3_stress1}), we obtain

\begin{equation}
  \langle \Sigma_{zz} \rangle = 2 \sigma + \frac{k_\mathrm{B} T}{4 \pi R_0 L} \sum_{m, \bar{q}} \frac{(m^2 - 1)^2 - \bar{q}^2(3 \bar{q}^2 + 2 m^2)}{(m^2 -1 )^2 + \bar{q}^2(\bar{q}^2 + 2m^2)} \, .
  \label{eq_3_Sigmamoy}
\end{equation}

\noindent The average force is thus

\begin{eqnarray}
  f &=& 4\pi\sigma R_0 + \frac{k_\mathrm{B} T}{2 L} \sum_{m, \bar{q}} \frac{(m^2 - 1)^2 - \bar{q}^2(3 \bar{q}^2 + 2 m^2)}{(m^2 -1 )^2 + \bar{q}^2(\bar{q}^2 + 2m^2)} \,, \nonumber\\
  &=& f_0 + \frac{k_\mathrm{B} T}{2 L} \sum_{m, \bar{q}} \frac{(m^2 - 1)^2 - \bar{q}^2(3 \bar{q}^2 + 2 m^2)}{(m^2 -1 )^2 + \bar{q}^2(\bar{q}^2 + 2m^2)} \,,
\end{eqnarray}

\noindent with, in tubes whose length is bigger than $R_0$, 

\begin{equation}
f_\mathrm{fl}=\frac{k_\mathrm{B}T}{4\pi R_0}\sum_{m=-M}^M
\int_{-\Lambda R_0}^{\Lambda R_0}
\frac{\left(m^2-1\right)^2-\bar q^2\left(3\bar q^2+2m^2\right)}{
\left(m^2-1\right)^2+\bar q^2\left(\bar q^2+2m^2\right)}\,d \bar{q} \, .
\label{eq_3_flnum}
\end{equation}

\begin{figure}[H]
\begin{center}
\includegraphics[width=.5\columnwidth,angle=0]{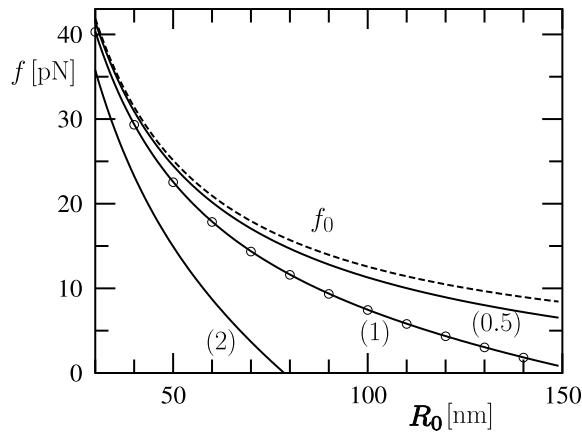}
\caption{Average force $f$ as a function of the tubule radius $R_0$.
  The dashed line corresponds to the mean-field force $f_0$.
  The solid lines correspond to the renormalized force, as given by
  eq.(\ref{eq_3_res}), with, from top to bottom $\Lambda a=0.5$, $1$,
  $2$, as indicated.
  The circles are obtained using the exact numerical
  sum (\ref{eq_3_flnum}), showing the quality of the approximation (\ref{eq_3_res}).
  The parameters used are $a \simeq 5 \, \mathrm{nm}$,
  $k_\mathrm{B}T \simeq 4 \times 10^{-21} \, \mathrm{J}$ and $\kappa \simeq 50\,k_\mathrm{B}T$.
}
\label{fig_3_force}
\end{center}      
\end{figure}

\noindent For $|m|\gg1$ the integral yields $\simeq\!-6M+8m\arctan(M/m)$.
A crude approximation may then be obtained by replacing the sum over $m$ by
$\int_{-M}^{M}\,[-6M+8m\arctan(M/m)] \, dm =-4M^2$.
It turns out that this
approximation is excellent in the regimes of interest (see
Fig.~\ref{fig_3_force}).
It follows 

\begin{equation}
f_\mathrm{fl}\simeq-\frac{k_\mathrm{B}T}{\pi}\Lambda^2R_0 \, ,
\label{eq_3_flapp}
\end{equation}

\noindent and consequently

\begin{equation}
f\simeq\frac{2\pi\kappa}{R_0}\left[1-
\frac{k_\mathrm{B}T}{2\pi^2\kappa}\,R_0^2 \, \Lambda^2\right] \, .
\label{eq_3_res}
\end{equation}

Equivalently, using the definition of $R_0$ given in eq.(\ref{eq_3_R0}), we obtain in terms of $\sigma$

\begin{equation}
f\simeq 2\pi\sqrt{2\kappa\sigma}\left[1-
\frac{k_\mathrm{B}T}{4\pi^2}\,\frac{\Lambda^2}{\sigma}\right] \, .
\label{eq_3_res1}
\end{equation}

\noindent Hence, we find that the actual force $f$ is significantly smaller than the
mean-field approximation $f_0$, the correction being more important when
$R_0$ is large (Fig.~\ref{fig_3_force}).
Note, however, the strong influence of the uncertainty on $\Lambda$.  

\subsection{Discussion on the validity of our results}

Let us comment on the validity of our results.
First, we should recall that eq.(\ref{eq_3_res}) actually corresponds to the first
term in a power series expansion of the form
$f=f_0[1+(k_\mathrm{B}T/\kappa)f_1(\Lambda
R_0)+(k_\mathrm{B}T/\kappa)^2f_2(\Lambda R_0)+\ldots]$, the higher-order
terms arising from terms beyond $\mathcal{O}(u^2)$ within the
expressions of $h$ and $\Sigma_{zz}$.
The fact that $k_\mathrm{B}T/\kappa\ll1$ for biological membranes is
good for the convergence of the series, but $R_0$ should not
become too large.
Obviously, $f$ must be positive, implying the upper bound condition

\begin{equation}
R_0 < a\,\sqrt{\frac{2\pi^2\kappa}{k_\mathrm{B}T}}
\approx 150\, \mathrm{nm}\,,
\end{equation}

\noindent with $a\equiv\Lambda^{-1}\approx5\, \mathrm{nm}$ and
$\kappa \simeq 50\,k_\mathrm{B}T$.
This condition, essentially due to the existence of an upper wave-vector cutoff, is normally verified
(see, e.g., Ref.~\cite{Waugh_92}).

At the same time, we must require
$\langle u^2\rangle\ll1$ for the harmonic approximation to be valid.
As shown in ref.~\cite{Fournier_07}, eq.~(\ref{eq_3_corel}) is well approximated
by

\begin{equation}
\langle u(0,0)\,u(\rho,z)\rangle\simeq
\frac{k_\mathrm{B}T}{4\pi\kappa}\left(
\frac{L}{6R_0}-\frac{|z|}{R_0}+\frac{z^2}{L\,R_0}
\right)\cos\left(\frac{\rho}{R_0}\right)\,.
\end{equation}

\noindent Requiring, e.g., $\sqrt{\langle u^2\rangle}<0.2$ corresponds to the
condition $L/R_0<\pi\kappa/(k_\mathrm{B}T)$, i.e., $L/R_0<200$ for
$\kappa\simeq50\,k_\mathrm{B}T$.
When $R_0\le50 \, \mathrm{nm}$ this corresponds
to $L<10 \, \mathrm{\mu m}$.
These ranges, together with the requirement that
the vesicle from which the tubule is extracted should be very large,
define the conditions of validity of our analysis.
To conclude, let us comment on the influence of the boundary conditions.
Due to the force conservation principle, $f$ cannot depend on the
position at which it is measured.
Therefore, the boundary conditions are not important to the average force and it is justified to choose periodic boundary conditions, as we have done here.

\section{Discussion on experiments}
\label{section_3_discussion}

In this chapter, we have analyzed the influence of the thermal fluctuations
on the force exerted by a nanotube which is pulled from a membrane
with bending rigidity $\kappa$ and internal tension $\sigma$.
Two other parameters play a role: the thermal energy $k_\mathrm{B}T$ and the upper
wave-vector cutoff $\Lambda\approx1/a$ (up to a prefactor of order
unity), where $a$ is the membrane thickness.
While $\kappa$,
$\Lambda$ and $k_\mathrm{B}T$ are rather fixed, $\sigma$, the in-plane stress,
may span several decades as it depends on the way the membrane is
tangentially stressed.
As we have seen previously, the problem is that $\sigma$ itself is not exactly a control parameter.
Instead, one usually controls the effective mechanical tension $\tau$.

Let's examine a typical experiment involving nanotubes, as presented in section~\ref{tube_exp}.
To a giant vesicle, held by a micropipette, one attaches a glass or magnetic bead.
This bead is subsequently displaced, forming a tube, while the vesicle is held at the same position.
By measuring the difference of pressure between the interior of the micropipette and the aqueous solution, the tension $\tau$ can be obtained, using eq.(\ref{Laplace}).
Let's suppose one is interested in studying the force $f$ needed to extract a tube as a function of the membrane's tension. 
As we have discussed in section~\ref{tube_exp}, two assumptions are usually made:

\begin{enumerate}
  \item firstly, one considers $\sigma \approx \tau$;
  \item secondly, one neglects the thermal fluctuations of the tube, implying that the force to extract a tube is simply $f_0$.
\end{enumerate}

\noindent Thus, under these assumptions, the force needed to extract a tube, which we will call $f_0'$, is given by

\begin{equation}
  f_0' = 2 \pi \sqrt{2 \pi \tau} \, .
  \label{eq_3_f0prime}
\end{equation}

\begin{figure}[H]
\centerline{\includegraphics[width=.65\columnwidth]{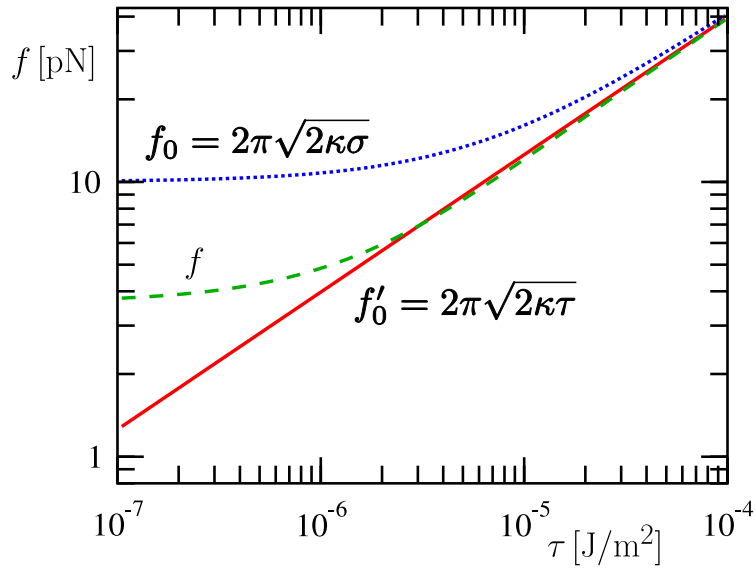}}
\caption{
Average force $f$ as a function of
the effectively applied mechanical tension $\tau$.
The red solid line shows the usual relation given in eq.(\ref{eq_3_f0prime}), under the assumption that $\tau \approx \sigma$ and neglecting thermal fluctuations.
The upper blue dotted line shown the expected relation if one considers $\tau = \sigma - \sigma_0$, but neglects the thermal fluctuations (eq.(\ref{eq_3_f0})).
Finally, the green dashed curve represents the expected relation taking into account both the difference between $\tau$ and $\sigma$ and the thermal fluctuations (eq.(\ref{eq_3_ffin})).
The
parameters used are $\Lambda a=1$ with $a\simeq5\, \mathrm{nm}$,
$k_BT\simeq4\times10^{-21}\, \mathrm{J}$ and $\kappa\simeq50\,k_BT$.  
}
\label{fig_3_fentau}
\end{figure}

In Fig.~\ref{fig_3_fentau}, we have plotted this relation, which is simply linear in log-scale (line in red).
In experiments, as one can see for instance in the Fig.~\ref{heinrich} in section~\ref{tube_exp}, this linear behavior seems to be indeed verified and consequently, up to now, these two assumptions were held as justified~\cite{Heinrich_96},\cite{Koster_05},\cite{Cuvelier_05}.

In the chapter~\ref{chapitre/planar_membrane}, however, we have seen that $\tau$ was considerably different from $\sigma$, since 
$\tau$ has additional contributions arising from the curvature
strains excited by the thermal undulation.
For a planar membrane, we have in general

\begin{equation}
\tau-\sigma\simeq - \sigma_0 = -\frac{k_\mathrm{B}T\Lambda^2}{8\pi}\,,
\label{eq_3_tau}
\end{equation}

\noindent relation still valid for large vesicles (see chapter~\ref{chapitre_vesicle}).
Taking into account this difference, but still neglecting thermal fluctuations,
the force needed to extract a tube should be

\begin{equation}
  f_0 = 2 \pi \sqrt{2 \kappa \sigma} = 2\pi \sqrt{2 \kappa (\tau + \sigma_0)} \, .
  \label{eq_3_f0}
\end{equation}

\noindent This curve is shown in blue in Fig.~\ref{fig_3_fentau}, which seems to be completely incompatible with the linear trend of experimental data.

Finally, we have seen in the previous section that the contribution of the thermal fluctuations to the force may be important.
Taking into account the difference between $\tau$ and $\sigma$ as well as the thermal fluctuations, we obtain from eq.(\ref{eq_3_res1})

\begin{equation}
  f \simeq 2\pi \sqrt{2 \kappa(\tau + \sigma_0)} \left[1 - \frac{2}{\pi} \, \frac{\sigma_0}{\tau + \sigma_0} \right] \, .
  \label{eq_3_ffin}
\end{equation}

\noindent This curve is plotted in green in Fig.~\ref{fig_3_fentau}.

We observe that thermal fluctuations are indeed important: the average force $f$ differs significantly from
the mean-field approximation $f_0=2\pi\sqrt{2\kappa\sigma}$.
The relative error $(f_0-f)/f$ is of order $5\%$ at
$\tau=10^{-4}\, \mathrm{J/m}^2$, of $30\%$ at
$\tau=10^{-5}\, \mathrm{J/m}^2$ and it reaches $100\%$ at
$\tau=10^{-6}\, \mathrm{J/m}^2$ (see Fig.~\ref{fig_3_fentau}).
Interestingly, the relative error $(f_0'-f)/f$ is much smaller than $(f_0-f)/f$ (see Fig.~\ref{fig_3_fentau}).
Indeed, it is less than $1\%$ for
$\tau>10^{-5}\, \mathrm{J/m}^2$, and it becomes larger than $20\%$ only for
$\tau<10^{-6}\, \mathrm  {J/m}^2$.
Hence $f'_0$ appears to be indeed a good approximation of the average force: for $\tau > 10^{-6} \, \mathrm{J/m}^2$, one should expect a linear behavior.
This happens however by a happy coincidence, since one makes two non justified assumptions.

Let us discuss what could be done experimentally in order to
test these predictions.
The difference between $f$ and
$f_0'$ will be difficult to evidence, because one
should detect a difference of the order of a few $\mathrm{pN}$ while
measuring precisely the tension in the range $\tau<10^{-6}\, \mathrm{J/m}^2$.
It should be easier to detect the difference between $f$ and
$f_0=2\pi\sqrt{2\kappa\sigma}$, since it is already significant at
$\tau\simeq10^{-5}\, \mathrm{J/m}^2$.
This could be done if the tension $r$ were
measured simultaneously from the thermal fluctuation spectrum of the
vesicle from which the tubule is drawn and then assuming that $r \approx \sigma$.
It would be interesting to
measure $R_0$ as a function of $\tau$ directly, in order to check the
difference between $R_0$ and the usually assumed relation $\sqrt{\kappa/(2\tau)}$ (we expect $\sqrt{\kappa/[2(\tau + \sigma_0)]}$).
This would require a
specific experiment, since $R_0$ is normally below optical resolution.

\section{In a nutshell}

The mean-field force needed to extract a membrane nanotube in terms of the membrane rigidity $\kappa$ and it's microscopical tension $\sigma$ is well known and given by

\begin{equation}
  f_0 = 2\pi \sqrt{2 \kappa \sigma} \, .
\end{equation}

\noindent Assuming that thermal fluctuations are negligible and that the mechanical tension $\tau$ coming from the flattening of the membrane's fluctuation is a good approximation for $\sigma$, this relation seemed to be successfully experimentally verified.
Recently, however, it was shown that these nanotubes, due to their geometry, present very soft modes and should thus have strong fluctuations, implying 
that the actual force $f$ needed to extract a tube should be somewhat different from $f_0$.
To evaluate this difference, we have derived the stress tensor for quasi-cylindrical geometry and averaged it appropriately, yielding

\begin{equation}
  f \simeq f_0 \left[1 - \frac{2}{\pi} \, \frac{\sigma_0}{\sigma}\right] \, ,
\end{equation}

\noindent where

\begin{equation}
  \sigma_0 = \frac{\kappa \Lambda^2}{8 \pi \kappa \beta} \, .
\end{equation}

\noindent Numerically, the difference between $f$ and $f_0$ is non-negligible.
The fact that it has not been previously noticed comes from a happy coincidence: the assumption that $\sigma \approx \tau$ seems to make up for the neglected thermal fluctuations.

%% file: chap4.tex
\chapter{Fluctuation of the force needed to extract a membrane nanotube}
\label{Fluct_TUBE}

\begin{figure}[H]
\begin{center}
\includegraphics[scale=.3,angle=0]{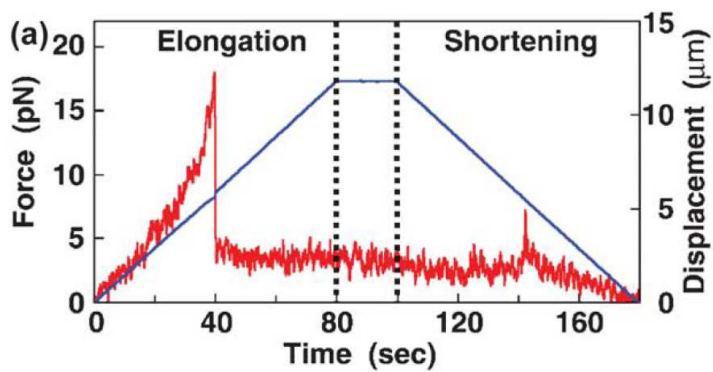}
\caption{Extraction and retraction of a nanotube using a glass bead and an optical
  trap~\cite{Inaba_05}. 
The blue well-defined line represents the displacement of the glass bead as a function of
time.
The bead was moved in order to elongate the vesicle and create a tube (from
$0$ up to $80 \, \mathrm{s}$), then kept roughly at the same point for $20 \, \mathrm{s}$ and finally
moved in the opposite direction. 
The red fluctuating line shows the force applied to the bead.
Note that there is a force barrier to form a tube (region before $40 \, \mathrm{s}$), but afterwards, the force does
almost not depend on the length of the tube.}
\label{fig_4_inaba}
\end{center}      
\end{figure}

As discussed in section~\ref{tube_exp}, nanotubes are
extracted from vesicles by applying point forces.
In the last chapter, we have described a popular method for pulling nanotubes, which consists in attaching a glass bead to the membrane and displacing the bead with a laser.
The advantage of this method is that one deduces with precision the applied
force by measuring the position of the center of the beam and the position of
the bead along the tube's axis (see Fig.~\ref{fig_3_trap}).
In general, only the force in the direction of the tube's axis $f_z$ is
measured, since by symmetry the averages of the transversal components of the force vanish.
A typical time-sequence of the bead's position along the tube's axis and force $f_z$ can be seen in Fig.~\ref{fig_4_inaba}.

In the last chapter and here, we are interested in the situation
where the tube's length is kept constant (results published in~\cite{Barbetta_09}).
In Fig.~\ref{fig_4_inaba}, this corresponds to the interval between the dashed
lines, where we can see that the force is roughly a plateau.
We have studied the average value of this plateau in the last chapter, obtaining

\begin{equation}
\langle f_z \rangle \simeq 2\pi\sqrt{2\kappa\sigma}\left[1-
\frac{k_\mathrm{B}T}{4\pi^2}\,\frac{\Lambda^2}{\sigma}\right] \, ,
\end{equation}

\noindent where $\kappa$ is the bending rigidity, $\Lambda$ is a wave-vector
cutoff and $\sigma$ is the tension associated to the microscopical area of the membrane.

Interestingly, in Fig.~\ref{fig_4_inaba}, we see that the measures of $f_z$ present a considerable dispersion,
mainly coming from the fluctuations of the bead's position (we cannot see this
fluctuation on the blue curve due to the length scale).
In effect, the bead is subjected to many sources of thermal fluctuations,
such as the fluctuating forces that the solvent applies to the bead, producing a Brownian movement, and the thermal fluctuations of the membrane to which the bead is attached. 

As membrane nanotubes present very soft
Goldstone modes~\cite{Fournier_07a}, the membrane fluctuation is possibly
responsible for an important part of the dispersion, in which case measures of force fluctuation could be used to characterize membranes.
Indeed, in section~\ref{tube_exp}, we have seen that from measures of the average force
$\langle f_z \rangle$, one can deduce the bending rigidity of a membrane.
Likewise, the fluctuation of the force in the direction of the tube's axis, easily accessible with the same experimental setting, could provide supplemental informations.
Accordingly, our aim in this chapter is to study the contribution of the membrane fluctuations to the fluctuation of the force in the direction of the tube's axis, defined through

\begin{eqnarray}
  (\Delta f_z)^2 &=& \langle f_z^2 \rangle - \langle f_z \rangle^2 \, . 
\end{eqnarray}

As in chapter~\ref{TUBE}, we will consider a tube small enough so that we can
consider the vesicle from which it is extracted as a lipid reservoir and we
can neglect volume constraints.
In section~\ref{section_4_param} we remind some important results deduced in the last chapter.
To evaluate $\Delta f_z$, we will use the diagrammatic tools introduced in
chapter~\ref{Fluct_plan}.
We will thus sum-up some properties of these diagrams and write the stress
tensor using them in section~\ref{section_4_diag}.
In section~\ref{section_4_fluct_calcul}, we evaluate $\Delta f_z$.
Finally, we discuss our results in section~\ref{section_4_discussion}.

\section{Some important definitions and results}
\label{section_4_param}

As in chapter~\ref{TUBE}, we shall restrict our attention to deformed tubes
weakly departing from the mean-field cylinder, whose radius is given by

\begin{equation}
  R_0 = \sqrt{\frac{\kappa}{2\sigma}} \, .
\end{equation}

\noindent We will keep the same coordinate system presented in Fig.~\ref{fig_3_param} and the tube's shape will be parametrized by eq.(\ref{eq_3_param}).  

As before, we consider a tube relatively short compared to the vesicle from which it is extracted, so that the vesicle can be treated as a lipid reservoir.
In the case of short tubes, one can also neglect the pressure difference across
the tube's membrane (see discussion in section~\ref{section_3_param}).
The energy is simply given by the Helfrich Hamiltonian, given in eq.(\ref{eq_3_H}) and eq.(\ref{eq_3_h2}).
The corresponding correlation function is given by

\begin{eqnarray}
  G(\rho - \rho', z- z') &=& \langle u(\rho, z) \, u(\rho', z')\rangle \, ,\nonumber\\
 &=& \frac{k_\mathrm{B}T \, R_0}{2\pi \kappa L} \sum_m \sum_{\bar{q}} \frac{e^{\frac{i}{R_0}\left[m(\rho - \rho') + \bar{q}(z -z')\right]}}{(m^2 -1)^2 + \bar{q}^2(\bar{q}^2 + 2 m^2)} \, ,
\end{eqnarray}

\noindent where $m \in \{-M, \cdots, M\}$ and $\bar{q} = 2\pi n R_0/L$, with $n \in \{-N, \cdots, N\}$.
We remind that the upper bounds $N$ and $M$ are given by $M = \Lambda R_0$ and $\bar{q}_\mathrm{max} = 2 \pi N R_0/L = \Lambda R_0$, where $\Lambda = 1/a$ is the high wave-vector cutoff and $a$ is of the order of the membrane thickness.
%For tubes whose length is a few times bigger than $R_0$, one can transform the sum over $\bar{q}$ in an integral, yielding

%\begin{equation}
% G(\rho - \rho', z- z') =  \frac{k_\mathrm{B}T }{(2\pi)^2 \kappa} \sum_m \int_{-\Lambda R_0}^{\Lambda R_0} \frac{e^{\frac{i}{R_0}\left[m(\rho - \rho') + \bar{q}(z -z')\right]}}{(m^2 -1)^2 + \bar{q}^2(\bar{q}^2 + 2 m^2)} \, .
%\end{equation}

The force needed to extract a tube is given by

\begin{equation}
  \bm{f} = \left(\int_0^{2\pi R_0} \Sigma_{rz} \, d\rho\right) \bm{e}_r + \left(\int_0^{2 \pi R_0} \Sigma_{\rho z} \, d\rho \right) \bm{e}_\rho + \left(\int_0^{2 \pi R_0} \Sigma_{zz} \, d\rho \right) \bm{e}_z \, , 
\end{equation}

\noindent where $\Sigma_{rz}$, $\Sigma_{\rho z}$ and

\begin{eqnarray}
  \Sigma_{zz}&=&\sigma\left\{2+u^2+2R_0^2\left[u_\rho^2+\left(2u-1\right)u_{\rho\rho}\right]\right. \nonumber \\
&+&R_0^4 \left. \left[u_{\rho\rho}^2-u_{zz}^2+2u_z\left(u_{zzz}+u_{\rho\rho
z}\right)\right]\right\}
  \label{eq_4_Sigmazz}
\end{eqnarray}

\noindent are the components of the projected stress tensor for quasi-cylindrical geometry derived in section~\ref{section_3_stress}.
As in the last chapter, the subscript $\rho$ (resp. $z$) indicates the derivative with respect to $\rho$ (resp. $z$).
Here we will evaluate the fluctuation of the force in the direction of the tube's axis:

\begin{eqnarray}
  (\Delta f_z)^2 &=& \langle f_z^2 \rangle - \langle f_z \rangle^2 \, , \nonumber\\
  &=& \int_0^{2 \pi R_0} \int_0^{2\pi R_0} \langle \Sigma_{zz}(\rho,z) \Sigma_{zz}(\rho', z) \rangle - \langle \Sigma_{zz} \rangle^2  \, d\rho d\rho' \, .
  \label{eq_4_Deltafz}
\end{eqnarray}

\noindent As in chapter~\ref{Fluct_plan}, the first step is to evaluate the correlation function of the stress tensor over the same section of tube:

\begin{equation}
  C(\rho, \rho') = \langle \Sigma_{zz}(\rho,z) \Sigma_{zz}(\rho', z) \rangle - \langle \Sigma_{zz} \rangle^2 \, .
\end{equation}

\noindent To do so, we will use another time the diagrammatic tools introduced in chapter~\ref{Fluct_plan}.
We recall their properties in the next section.

\section{Diagrammatic tools}
\label{section_4_diag}

Throughout this chapter, we will use notations similar to those introduced in section~\ref{section_11_rules}.
Each field $u(\rho,z)$ is represented by a straight line.
The derivatives with respect to $\rho$ are represented by a dot over the field, while a derivative with respect to $z$ is represented by a slash.
An adapted diagrammatic vocabulary is presented in table~\ref{table_4_vocab}. 

\begin{table}[H]
  \begin{center}
\begin{tabular}{|c|c|}
\hline
\red{Usually} & \red{Diagrammatically} \\
\hline
&\\
$u(\rho,z)$ & \diaggd{./figures_Chap4/simple_0.epsi} \\
&\\
\hline
&\\
$u_z(\rho,z)$ & \diaggd{./figures_Chap4/simple_1.epsi} \\
&\\
\hline
&\\
$u_\rho(\rho,z)$ & \diaggd{./figures_Chap4/simple_4.epsi} \\
&\\
\hline
&\\
$u(\rho,z)\, u(\rho,z)$ & \diaggd{./figures_Chap4/champ_0.epsi} \\
&\\
\hline
&\\
$u(\rho, z)\, u(\rho',z')$ & $\diaggd{./figures_Chap4/simple_0.epsi} \diaggd{./figures_Chap4/simple_0.epsi}$ \\
&\\
\hline
\end{tabular}
\caption{Basic {\it translation} rules from the usual notation into diagrams.}
\label{table_4_vocab}
\end{center}
\end{table}

Averages are performed using Wick's theorem, i. e., by adding all complete contractions of fields.
Each contraction yields a propagator and, as in the case of section~\ref{section_11_rules}, one can pass a derivative from one branch of the propagator to the other by multiplying the diagram's coefficient by $-1$.
For instance, with $\bm{r} = (\rho, z)$ and $\bm{r}' = (\rho',z')$, we have

\begin{eqnarray}
  \left\langle
  \begin{array}{c}
\diaggd{./figures_Chap4/simple_1.epsi}\\
{}^{\bm{r}}
\end{array}
    \begin{array}{c}
\diaggd{./figures_Chap4/simple_4.epsi}\\
{}^{\bm{r'}}
\end{array}
    \right \rangle &=&
\diaggd{./figures_Chap4/reto_3.epsi} \, ,\nonumber \\
&=&
(-1) \times \diaggd{./figures_Chap4/reto_4.epsi} \, ,\nonumber\\
\nonumber \\
&=& (-1) \times \partial_z \partial_\rho  \, [G(\bm{r}' - \bm{r})] \, ,\nonumber \\
\nonumber\\
 &=& \frac{k_\mathrm{B}T \, R_0}{2\pi \kappa L} \sum_m \sum_{\bar{q}} \frac{(i \, m/R_0)(i \, \bar{q}/R_0) \, e^{\frac{i}{R_0}\left[m(\rho - \rho') + \bar{q}(z -z')\right]}}{(m^2 -1)^2 + \bar{q}^2(\bar{q}^2 + 2 m^2)} \, .
\end{eqnarray}

\noindent This time, once the derivatives are grouped, every slash contributes with a factor $i \, \bar{q}/R_0$ and every dot contributes with a factor $i \,  m/R_0$.

In the following section, we will evaluate the propagators between points over the same section of tube, i. e., with $z = z'$.
In this case, as the sum over $\bar{q}$ is symmetrical, an uneven number of slashes over a propagator implies a vanishing contribution.
Here follows a typical example of the terms that we will need to evaluate:

\begin{eqnarray}
  \left\langle
  \begin{array}{c}
\diaggd{./figures_Chap4/champ_7_n1.epsi}\\
{}^{\bm{r}}
\end{array}
    \begin{array}{c}
\diaggd{./figures_Chap4/champ_7_n2.epsi}\\
{}^{\bm{r}'}
\end{array}
    \right \rangle
    &=&
     \begin{array}{c}
\diaggd{./figures_Chap4/wick_n1.epsi}\\
{}^{\bm{r}}
\end{array}
    \begin{array}{c}
\diaggd{./figures_Chap4/wick_n2.epsi}\\
{}^{\bm{r}'}
\end{array}
+
\diaggd{./figures_Chap4/peixe_n11.epsi}
+
\diaggd{./figures_Chap4/peixe_n21.epsi} \, ,\nonumber \\
&=&    
\diaggd{./figures_Chap4/wick_n1_1.epsi}
\diaggd{./figures_Chap4/wick_n1_1.epsi}
+ 
\diaggd{./figures_Chap4/peixe_n1_1.epsi}
+
\diaggd{./figures_Chap4/peixe_n2_1.epsi}\, ,\nonumber\\
\label{eq_4_avediag}
\end{eqnarray}

\noindent which one can readily read by noting the equivalence

\begin{equation}
\diaggd{./figures_Chap4/peixe_n2_1.epsi} = \diaggd{./figures_Chap4/reto_4.epsi} \times \diaggd{./figures_Chap4/reto_4.epsi} \, .
\label{eq_4_peixe}
\end{equation}

\noindent As the number of slashes over these propagators is uneven, the contribution of this diagram vanishes.
The first term of eq.(\ref{eq_4_avediag}) is also composed by diagrams with an uneven number of slashes whose contribution vanishes.
At the end, one obtains simply

\begin{eqnarray}
  \left\langle
  \begin{array}{c}
\diaggd{./figures_Chap4/champ_7_n1.epsi}\\
{}^{\bm{r}}
\end{array}
    \begin{array}{c}
\diaggd{./figures_Chap4/champ_7_n2.epsi}\\
{}^{\bm{r}'}
\end{array}
    \right \rangle
    &=&
\diaggd{./figures_Chap4/peixe_n1_1.epsi} \, .
\end{eqnarray}

\noindent In the next section, we will re-derive $\langle \Sigma_{zz} \rangle$
in terms of diagrams in order to gain familiarity with these tools.

\subsection[Getting familiar: re-deriving $\langle \Sigma_{zz} \rangle$]{Getting familiar: re-deriving $\bm{\langle \Sigma_{zz} \rangle}$}

The component $\Sigma_{zz}$, given in eq.(\ref{eq_4_Sigmazz}), can be written in terms of diagrams as

\begin{eqnarray}
  \sigma^{-1} \, \langle \Sigma_{zz} \rangle &=& 2 + 
\diaggd{./figures_Chap4/champ_0.epsi}
+ 2 R_0^2 \diaggd{./figures_Chap4/champ_4.epsi}
+ 4 R_0^2 \diaggd{./figures_Chap4/champ_12.epsi}
- 2 R_0^2 \diaggd{./figures_Chap4/simple_5.epsi} \nonumber\\
&+& R_0^4 \diaggd{./figures_Chap4/champ_6.epsi}
- R_0^4 \diaggd{./figures_Chap4/champ_5.epsi}
+ 2 R_0^4 \diaggd{./figures_Chap4/champ_1.epsi}
+  2 R_0^4 \diaggd{./figures_Chap4/champ_2.epsi} \, .
\label{eq_4_Sigmadiag}
\end{eqnarray}

\noindent Using Wick's theorem to evaluate the average, we obtain

\begin{eqnarray}
    \sigma^{-1} \, \langle \Sigma_{zz} \rangle &=& 2 + 
\diaggd{./figures_Chap4/wick_0.epsi}
+ 2 R_0^2 \diaggd{./figures_Chap4/wick_3.epsi}
+ 4 R_0^2 \diaggd{./figures_Chap4/wick_14.epsi}\nonumber\\
&+& R_0^4 \diaggd{./figures_Chap4/wick_5.epsi}
- R_0^4 \diaggd{./figures_Chap4/wick_6.epsi}
+ 2 R_0^4 \diaggd{./figures_Chap4/wick_1.epsi}
+  2 R_0^4 \diaggd{./figures_Chap4/wick_15.epsi} \, .
\end{eqnarray}

\noindent Grouping the derivatives, we have

\begin{eqnarray}
    \sigma^{-1} \, \langle \Sigma_{zz} \rangle &=& 2 + 
\diaggd{./figures_Chap4/wick_0_0.epsi}
- 2 R_0^2 \diaggd{./figures_Chap4/wick_7.epsi}
+ 4 R_0^2 \diaggd{./figures_Chap4/wick_7.epsi}\nonumber\\
&+& R_0^4 \diaggd{./figures_Chap4/wick_9.epsi}
- R_0^4 \diaggd{./figures_Chap4/wick_10.epsi}
- 2 R_0^4 \diaggd{./figures_Chap4/wick_10.epsi}
-  2 R_0^4 \diaggd{./figures_Chap4/wick_11.epsi} \, , \nonumber\\
\nonumber \\
&=& 2 + 
\diaggd{./figures_Chap4/wick_0_0.epsi}
+ 2 R_0^2 \diaggd{./figures_Chap4/wick_7.epsi}
+ R_0^4 \diaggd{./figures_Chap4/wick_9.epsi}
- 3 R_0^4 \diaggd{./figures_Chap4/wick_10.epsi}
-  2 R_0^4 \diaggd{./figures_Chap4/wick_11.epsi} \, . \nonumber\\
\label{eq_4_aveSigma}
\end{eqnarray}

\noindent Let's read eq.(\ref{eq_4_aveSigma}):

\begin{eqnarray}
  \langle \Sigma_{zz} \rangle &=& 2 \sigma  + \sigma \frac{k_\mathrm{B} T \, R_0}{2 \pi \kappa L} \sum_m \sum_{\bar{q}} \frac{1 + 2(i \, m)^2 + (i \, m)^4 - 3 (i \, \bar{q})^4 -2 (i \, m)^2(i \, \bar{q})^2}{(m^2 - 1)^2 + \bar{q}^2(\bar{q}^2 + 2 m^2)} \, , \nonumber\\
  &=& 2 \sigma  + \frac{k_\mathrm{B} T }{4 \pi R_0 L} \sum_m \sum_{\bar{q}} \frac{(1-m^2)^2 - \bar{q}^2(3 \bar{q}^2 + 2 m^2)}{(m^2 - 1)^2 + \bar{q}^2(\bar{q}^2 + 2 m^2)} \, ,
\end{eqnarray}

\noindent which coincides, as it should, with eq.(\ref{eq_3_Sigmamoy}).

\section{Evaluation of the fluctuation of the force}
\label{section_4_fluct_calcul}

Aiming to obtain $\Delta f_z$, we start this section by 
evaluating the correlation function of the component $\Sigma_{zz}$ of the stress tensor.
In section~\ref{section_4_Df}, we integrate twice this correlation and derive
the force fluctuation.
There, we discuss also some approximations in order to obtain a simple analytical expression.
Finally, we conclude by a short discussion on the validity of our final result in section~\ref{section_4_validity}.

\subsection[Correlation of ${\Sigma_{zz}}$]{Correlation of $\bm{\Sigma_{zz}}$}
\label{subsection_4_corr}

Here we will evaluate the correlation function

\begin{equation}
  C(\rho, \rho') = \langle \Sigma_{zz}(\rho,z) \Sigma_{zz}(\rho', z) \rangle - \langle \Sigma_{zz} \rangle^2 \, .
\end{equation}

\noindent From eq.(\ref{eq_4_Sigmadiag}), using Wick's Theorem and the rules presented in section~\ref{section_4_diag}, we obtain

\begin{eqnarray}
\sigma^{-2} \, C(\rho, \rho') &=&
2 \diagH{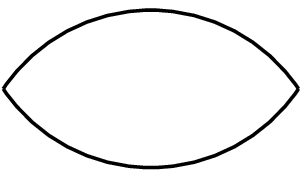}
+ 8 R_0^2 \diagH{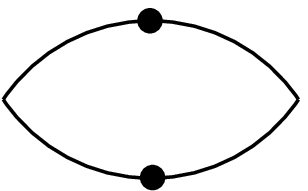}
+ 16 R_0^2 \diagH{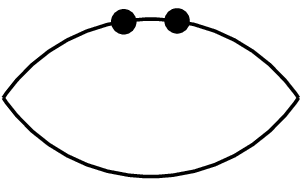}
+ 28 R_0^4 \diagH{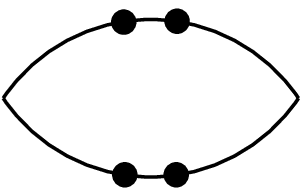} \nonumber\\
&+& 32 R_0^4 \diagH{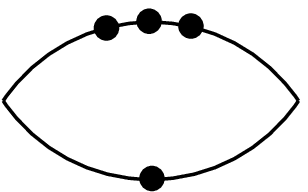}
+ 16 R_0^4 \diagH{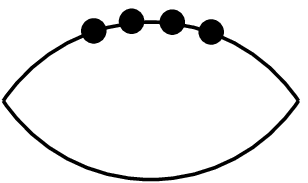}
- 4 R_0^4 \diagH{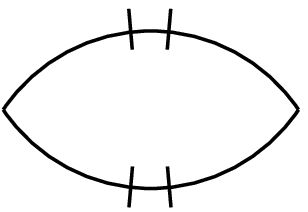}
+ 8 R_0^6 \diagH{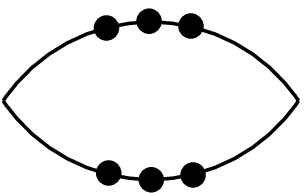} \nonumber\\
&-&8 R_0^6 \diagH{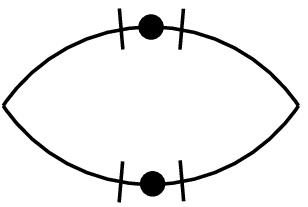}
+ 16 R_0^6 \diagH{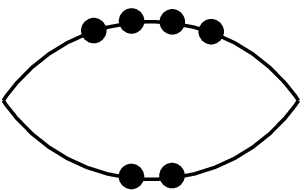}
- 16 R_0^6 \diagH{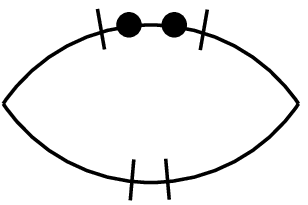}
+2 R_0^8 \diagH{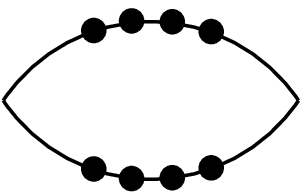}\nonumber\\
&+& 6 R_0^8 \diagH{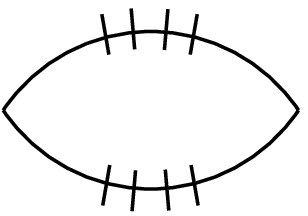} 
+8 R_0^8 \diagH{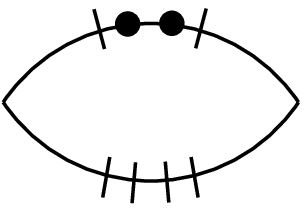}
+4 R_0^8 \diagH{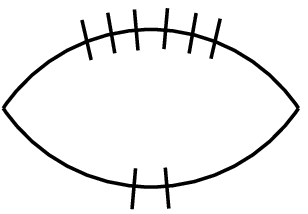}
+ 4 R_0^8 \diagH{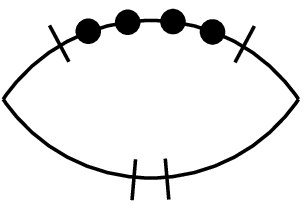} \nonumber\\
&+& 8 R_0^8 \diagH{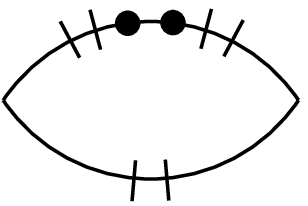}
+4 R_0^4 \diagH{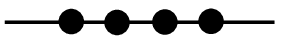} \, .
\label{eq_4_fluctdiag}
\end{eqnarray}

\noindent Note that all the terms involving diagrams of the kind $\diaggd{./figures_Chap4/wick_0_0.epsi}$ vanish.
Explicitly, we obtain

\begin{eqnarray}
  C(\rho, \rho') &=& \left(\frac{k_B T}{4\pi R_0 L} \right)^2 \sum_{m,\bar{q}} \sum_{n,\bar{k}} f_{n,m,\bar{q},\bar{k}} \, e^{i R_0^{-1} (m+n)(\rho - \rho')} \nonumber \\
  &+& 2 \, \frac{\kappa}{R_0^2} \left(\frac{k_\mathrm{B} T}{4\pi R_0 L}\right) \sum_{m,
    \bar{q}} \frac{m^4 \, e^{iR_0^{-1} m(\rho - \rho')}}{(m^2 -1)^2 +
    \bar{q}^2(\bar{q}^2 + 2 m^2)} \, ,
\label{eq_4_C}
\end{eqnarray}

\noindent where $f_{n,m,\bar{q}, \bar{q}}$ is a complicated coefficient
depending on $m$, $n$, $\bar{q}$ and $\bar{k}$:

\begin{equation}
f_{n,m,\bar{q},\bar{k}} = \frac{g_{n,m,\bar{q}, \bar{k}}}{\left[(m^2 -1)^2 +
    \bar{q}^2(\bar{q}^2 + 2 m^2)\right]\left[(n^2 -1)^2 + \bar{k}^2(\bar{k}^2
    + 2 n^2)\right]} \, ,
\end{equation}

\noindent with

\begin{eqnarray}
g_{n,m,\bar{q},\bar{k}} &=& 2 - 16\,  m^2 + 16 \, m^4 - 8 \, m\, n + 32 \, m^3
\, n + 28\, m^2 \, n^2 - 16 \, m^4 \, n^2 \nonumber\\
&-& 8 \, m^3 \, n^3 + 2 \, m^4\, n^4 - 4 \, \bar{q}^2 \, \bar{k}^2 + 16 \, m^2
\, \bar{q}^2 \, \bar{k}^2 + 8 \, m^2 \, \bar{q}^2 \, \bar{k}^4 \nonumber\\
&+& 4 \, m^4\, \bar{q}^2 \bar{k}^2 + 8 \, m\, n \, \bar{q}^2 \, \bar{k}^2 + 6
\, \bar{q}^4 \, \bar{k}^4 + 8 \, m^2 \, \bar{q}^4 \, \bar{k}^2 + 4 \,
\bar{q}^6 \, \bar{k}^2 \, .
\end{eqnarray}

\begin{figure}[H]
\begin{center}
  \vspace{2cm}
\includegraphics[width=0.6\columnwidth]{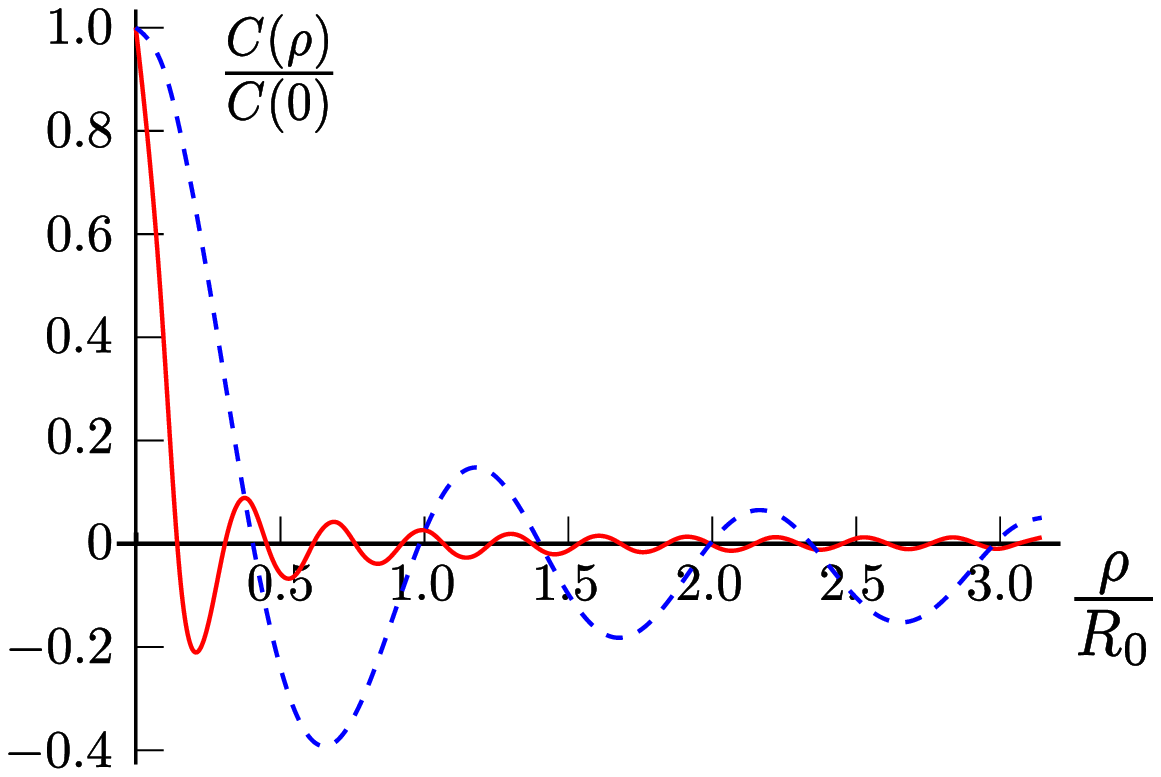}
\caption{Correlation function of the component $\Sigma_{zz}$ of the stress tensor normalized by it's value at $\rho = 0$ as a function of $\rho/R_0$.
  In both curves, we considered tubes of length $500 \, \mathrm{nm}$ and bending rigidity $\kappa = 50 \, k_\mathrm{B}T$.
  We have chosen $\Lambda^{-1} = 5 \,  \mathrm{nm}$ as the value of the microscopical cut-off.
  The blue dashed curve stands for a thin tube with $R_0 = 30 \, \mathrm{nm}$, while the red solid curve stands for a tube with $R_0 = 100 \, \mathrm{nm}$.}
\label{fig_4_correl}
\end{center}      
\end{figure}

\noindent The second term of eq.(\ref{eq_4_C}), with an unique sum over the wavenumbers, is the contribution given by the last diagram of eq.(\ref{eq_4_fluctdiag}).
As expected, the correlation depends only on $\rho - \rho'$.
In Fig.~\ref{fig_4_correl}, we show the behavior of the correlation normalized by it's value at $\rho = 0$ for two different tubes with the same length, bending rigidity and wave-length cutoff.

First of all, we note that even though the stress tensor correlation decreases with the distance, we have no more the fast decay found in the case of planar membranes (see section~\ref{section_11_corr}).
In both cases, the function $C(\rho)$ presents oscillations that remain non negligible throughout the whole section of the
tube, indicating that the stress tensor is correlated all over the length of a
tube's cross section.
This is a signature of the fact that the fluctuations in the shape of membrane
tubes are themselves correlated over a whole cross section, whatever the tube's radius \cite{Fournier_07a}.
Moreover, we observe that the oscillations in Fig.~\ref{fig_4_correl} take
place  over a roughly constant wave-length $\lambda$.
For the tube with $R_0 = 30 \, \mathrm{nm}$, we have $6$ oscillations
distributed over the perimeter, which gives a wave-length $\lambda \approx 31 \, \mathrm{nm} \sim 6 \, \Lambda^{-1}$, with $\Lambda^{-1} \sim 5 \, \mathrm{nm}$.
Interestingly, we find the same value for the larger tube.
This characteristic wave-length corresponds to the length beyond which the correlation of the stress tensor in planar membranes becomes negligible (see section~\ref{section_11_corr}).
It is thus probably an universal quantity, valid for any value of $R_0$.

\begin{figure}[H]
\begin{center}
\includegraphics[width=0.6\columnwidth]{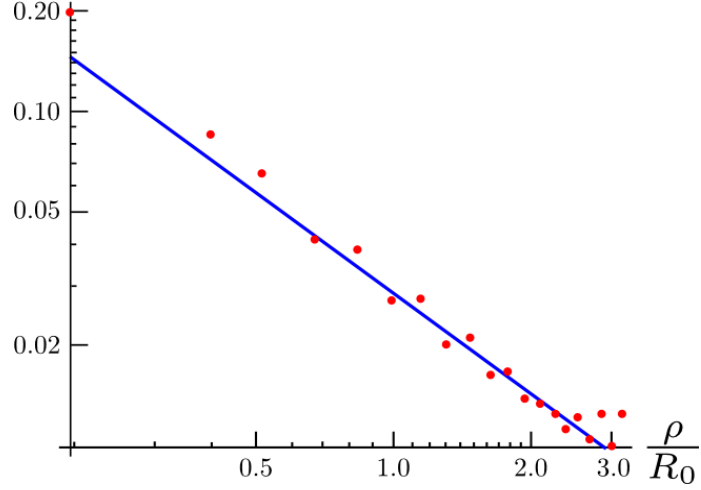}
\caption{The red dots represent the absolute value of the extrema of the red solid curve in Fig.~\ref{fig_4_correl} as a function of $\rho/R_0$.
The blue solid curve is proportional to $x^{-1}$.}
\label{fig_4_power_law}
\end{center}      
\end{figure}

To characterize better how the stress tensor correlation decreases, we have plotted the absolute value of the extrema of the oscillations of the red curve as a function of $\rho/R_0$ in a log-log scale (see Fig.~\ref{fig_4_power_law}).
This curve seems to indicate that the amplitude of the oscillations decay with a power law, which is a characteristic sign of long-range correlations.

\begin{figure}[H]
\begin{center}
\subfigure[Tube with $R_0 = 30 \, \mathrm{nm}$.]{
\includegraphics[width=0.45\columnwidth]{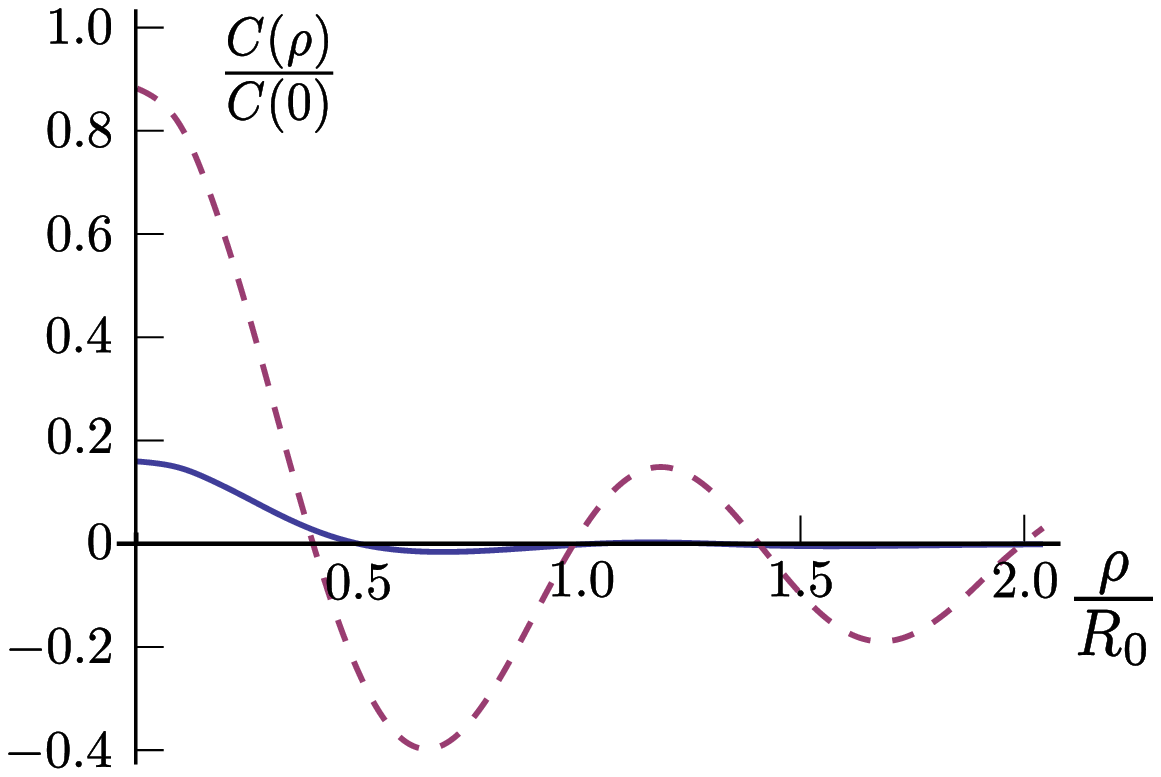}
\label{fig_4_comp_a}
}  
\subfigure[Tube with $R_0 = 100 \, \mathrm{nm}$.]{ 
  \includegraphics[width=0.45\columnwidth]{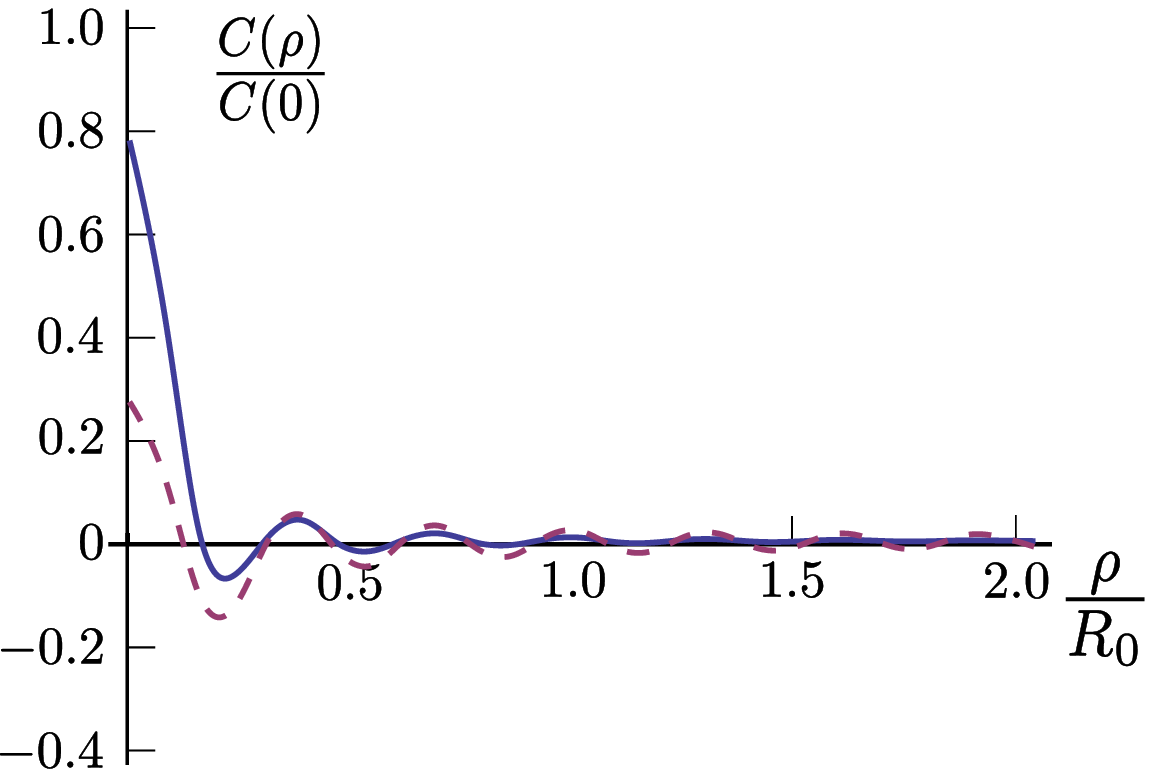}
  \label{fig_4_comp_b}
}
\caption{Contribution of the first term of eq.(\ref{eq_4_C}) (term with the double sum over the wavenumbers) to the stress tensor correlation (red dashed line) compared to the contribution of the second term of this equation, with an unique sum over the wavenumbers (blue solid curve).
In both cases, we have considered tubes with length $500 \, \mathrm{nm}$, $\kappa = 50 \, k_\mathrm{B}T$ and $\Lambda^{-1} = 5 \, \mathrm{nm}$.}
\end{center}      
\end{figure}

Finally, we have compared the contribution of the first term of eq.(\ref{eq_4_C}), involving two sums over the wavenumbers, and the contribution of the second term of eq.(\ref{eq_4_C}), with an unique sum over the wavenumbers, to the total stress tensor correlation.
As one can see in Fig.~\ref{fig_4_comp_a} and in Fig.~\ref{fig_4_comp_b}, in general, both contributions are oscillating and important.
In the following, however, we will see that the second term of eq.(\ref{eq_4_C}), with an unique sum and represented by the solid lines in these figures, gives a vanishing contribution to $\Delta f_z$.

\subsection[{Force fluctuation ${\Delta f_z}$}]{Force fluctuation $\bm{\Delta f_z}$}
\label{section_4_Df}

As we have seen in eq.(\ref{eq_4_Deltafz}), the force fluctuation is given by

\begin{eqnarray}
  (\Delta f_z)^2 &=& \int_0^{2\pi R_0} \int_0^{2\pi R_0} C(\rho, \rho') \, d\rho d\rho' \, ,\nonumber\\
  &=& \int_0^{2\pi R_0} \int_{0}^{2\pi R_0} C(\rho - \rho', 0) \, d\rho d\rho' \, .
  \label{eq_4_deltafz}
\end{eqnarray}

\noindent The last passage follows from the fact that the correlation function is invariant under translation.
The function $C(\rho, 0)$ is periodic with period $2 \pi
R_0$. Consequently, eq.(\ref{eq_4_deltafz}) is equivalent to

\begin{equation}
  (\Delta f_z)^2 = 2\pi R_0  \int_0^{2\pi R_0} C(\rho'', 0) \, d\rho'' \, .
\end{equation}

\noindent Using the fact that

\begin{equation}
  \int_0^{2\pi R_0} e^{i R_0^{-1} (m + n)\rho''} \, d\rho'' = 2\pi R_0 \, \delta_{n, -m} \, ,
\end{equation}

\noindent we obtain

\begin{eqnarray}
  (\Delta f_z)^2 &=& (2 \pi R_0)^2 \left(\frac{k_B T}{4\pi R_0 L} \right)^2 \sum_{m,\bar{q}} \sum_{\bar{k}} f_{-m,m,\bar{q},\bar{k}}  \, ,\nonumber \\
  &=& \left(\frac{k_\mathrm{B} T}{2 L} \right)^2 \sum_{m,\bar{q}} \sum_{\bar{k}} \frac{2(m^2-1)^4+4\bar{k}^2 \bar{q}^2 \left[\bar{k}^4-1+2m^2\left(1+2\bar{k}^2\right)+m^4\right]+6\bar{k}^4\bar{q}^4}{\left[(m^2-1)^2+\bar{q}^2(\bar{q}^2+2m^2)\right]\,\left[(m^2-1)^2+\bar{k}^2(\bar{k}^2+2m^2)\right]} \, . \nonumber\\
\end{eqnarray}

\noindent Note that the contribution of the last term of eq.(\ref{eq_4_C})
vanishes after integration.

\subsubsection{Long tubes}

For the case of $L > R_0$,
we can substitute the sums over $\bar{q}$ and over $\bar{k}$ by integrals, yielding

\begin{equation}
  \begin{split}
&(\Delta f_z)^2= \\
&\left(\frac{k_\mathrm{B}T}{4\pi R_0}\right)^2\!
\sum_{m=-M}^M \int_{-\Lambda R_0}^{\Lambda R_0} \int_{-\Lambda R_0}^{\Lambda R_0}
\frac{
2(m^2-1)^4
+4\bar k^2\bar q^2
\left[
\bar k^4-1+2m^2\left(1+2\bar k^2\right)+m^4
\right]+6\bar k^4\bar q^4
}{
\left[
(m^2-1)^2+\bar q^2(\bar q^2+2m^2)
\right]\,\left[
(m^2-1)^2+\bar k^2(\bar k^2+2m^2)
\right]
} \, 
d\bar q\,
d\bar k \, .
\end{split}
  \label{eq_4_Deltafzcalcul}
\end{equation}

\noindent Taking into account the fact that
eq.(\ref{eq_4_Deltafzcalcul}) depends only on $|m|$, it can be rewritten as

\begin{equation}
(\Delta f_z)^2= 
\left(\frac{k_\mathrm{B}T}{4\pi R_0}\right)^2\! \left(T_0 + 2 \, T_1 + 2 \,
  \sum_{m=2}^M T_m\right) \,,
\label{eq_4_D}
\end{equation} 

\noindent with

\begin{equation}
T_0 = \int_{-\Lambda R_0}^{\Lambda R_0} \int_{-\Lambda R_0}^{\Lambda R_0}
\frac{2 \bar{q}^4 \bar{k}^4 \left(\bar{q}^2 + \bar{k}^2\right) + \left(1 -
    \bar{q}^2 \bar{k}^2 \right)^2}{\left(1 + \bar{q}^4 \right)\left(1 +
    \bar{k}^4 \right)
} \, 
d\bar q\,
d\bar k \, ,
\label{eq_4_T0}
\end{equation}

\begin{equation}
T_1 = \int_{-\Lambda R_0}^{\Lambda R_0} \int_{-\Lambda R_0}^{\Lambda R_0}
\frac{8 + 4 \bar{k}^4 + 2 \bar{k}^2 \left(8 + 3 \bar{q}^2\right)}{\left( 2 +
    \bar{q}^2 \right)\left(2 + \bar{k}^2 \right)} \, 
d\bar q\,
d\bar k \, ,
\label{eq_4_T1}
\end{equation}

\noindent and

\begin{equation}
T_m = \int_{-\Lambda R_0}^{\Lambda R_0} \int_{-\Lambda R_0}^{\Lambda R_0}
\frac{
2(m^2-1)^4
+4\bar k^2\bar q^2
\left[
\bar k^4-1+2m^2\left(1+2\bar k^2\right)+m^4
\right]+6\bar k^4\bar q^4
}{
\left[
(m^2-1)^2+\bar q^2(\bar q^2+2m^2)
\right]\,\left[
(m^2-1)^2+\bar k^2(\bar k^2+2m^2)
\right]
} \, 
d\bar q\,
d\bar k \, .
\label{eq_4_Tm}
\end{equation}
 
\begin{figure}[H]
\begin{center}
\includegraphics[width = 0.7\columnwidth]{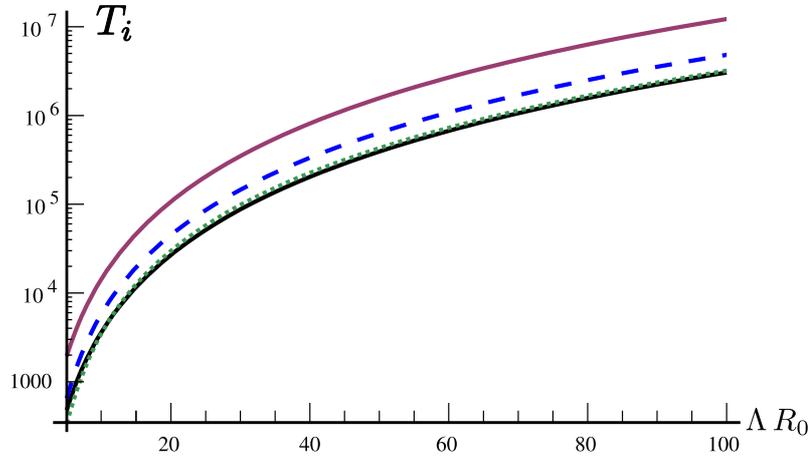}
\caption{For very long tubes, each radial mode $i$ gives a contribution $[k_\mathrm{B}T/(4\pi R_0)]^2 \times T_i$ to $(\Delta f_z)^2$, with $T_i$ given by eqs.(\ref{eq_4_T0})--(\ref{eq_4_Tm}).
  We compare here the contributions of the modes with $m=0$ (lower solid line, in black), $m = 1$ (upper solid line, in red), $m = 2$ (dashed blue line) and $m=3$ (dotted green line) as a function of the tube's radius.
  The vertical scale is shown in units of
  $[k_\mathrm{B} T/(4 \pi R_0)]^{2}$, while the radius of the tube
  is shown in units of the microscopical cutoff $a = \Lambda^{-1}$.
We see that the Goldstone modes, with $m=1$, give indeed a greater contribution to the force fluctuation.}
\label{fig_4_modes}
\end{center}      
\end{figure}

\noindent Both integrals over $\bar{q}$ and over $\bar{k}$ in
eq.(\ref{eq_4_T0}), eq.(\ref{eq_4_T1}) and eq.(\ref{eq_4_Tm}) can be performed
analytically.
One can compare the contributions of some modes to the force fluctuation in Fig.~\ref{fig_4_modes}.
Not surprisingly, the modes $|m|=1 $, which are extremely
soft~\cite{Fournier_07a}, give a greater contribution.

\begin{figure}[H]
\begin{center}
\includegraphics[width = 0.7\columnwidth]{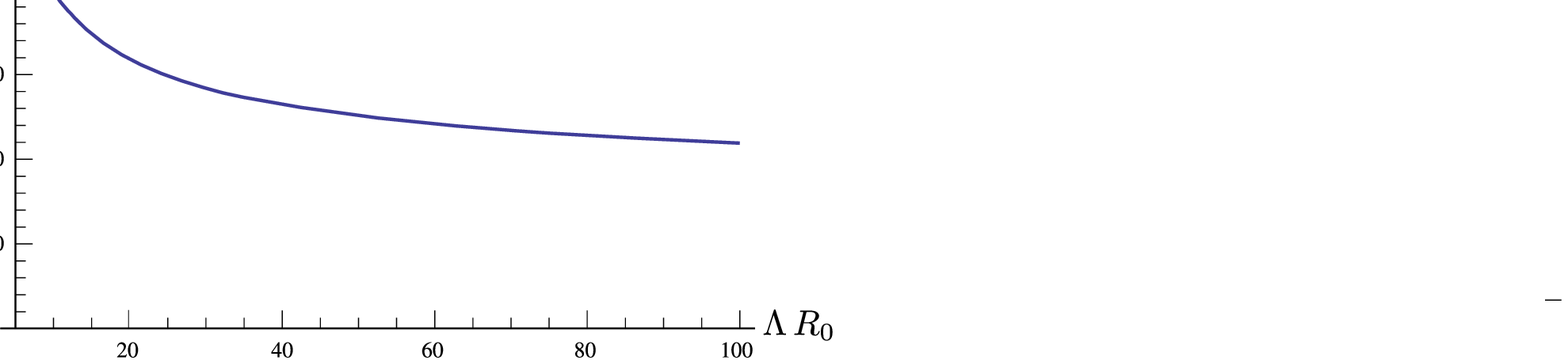}
\caption{Percent contribution of the soft modes ($|m| = 1$) to the total
  force fluctuation.}
\label{fig_4_porcent}
\end{center}      
\end{figure}

\noindent In Fig.~\ref{fig_4_porcent}, we show the percent contribution of
these soft modes to the total fluctuation.
In agreement with the curves of Fig.~\ref{fig_4_modes}, the modes with $|m| =
1$ are responsible for more than one third of the force fluctuation.

\subsubsection{Approximations and an analytical formula for $\bm{\Delta f_z}$}

In order to obtain a simple analytical expression to the force fluctuation, we
consider the limit of relatively thick tubes, with $\Lambda \, R_0 > 6$.
Considering $a = \Lambda^{-1} = 5 \, \mathrm{nm}$, this corresponds to tubes
with a radius $R_0 > 30 \, \mathrm{nm}$, which is currently observed in experiments.  
In this limit, we have

\begin{equation}
T_0 \approx \frac{2 \sqrt{2} \pi}{3} \, (\Lambda \, R_0)^3 \, ,
\end{equation}

\begin{equation}
T_1 \approx \frac{4\sqrt{2} \pi}{3} \, (\Lambda \, R_0)^3 \, ,
\end{equation}

\noindent and 

\begin{eqnarray}
T_m \approx 24 (\Lambda \, R_0)^2 + \frac{8}{3m}\left[(\Lambda \,
    R_0)^3 - 21 m^2 (\Lambda \, R_0) + 12 m^3 \arctan\left(\frac{\Lambda \,
        R_0}{m}\right) \right] \arctan\left(\frac{\Lambda \, R_0}{m} \right)
  \, . \nonumber\\
\end{eqnarray}

\noindent The sum over $m$ in
eq.(\ref{eq_4_D}) can be approximated by an integral, yielding

\begin{equation}
\sum_{m=2}^{M} T_m \approx \int_2^{\Lambda \, R_0} T_m \, dm \simeq \frac{2 \sqrt{2} \pi}{3} \, (\Lambda \, R_0)^3 \, + \frac{4
  \pi}{3} \, (\Lambda \, R_0)^3 \, \ln(\Lambda \, R_0) \, .
\end{equation}

\noindent At last, we obtain

\begin{equation}
\left(\Delta f_z\right)^2 \simeq \frac{(k_\mathrm{B}T)^2 \, R_0\, \Lambda^3}{6\pi} \left[\frac{5}{2} + \ln\left(\Lambda \, R_0\right)\right] \, .
\label{eq_4_Dfapprox}
\end{equation}

\noindent We discuss the quality of this approximation and its meaning in section~\ref{section_4_discussion}. 

\subsection{Discussion on the validity of this result}
\label{section_4_validity}

Here we remind the conditions of validity of eq.(\ref{eq_4_D}).
First, denoting $u$ the deformation of the tube relative to the mean-field cylinder, we considered here only terms up to $\mathcal{O}(u^2)$ in the Hamiltonian and in the stress tensor.
Accordingly, our result corresponds actually to the first term in a series expansion of the form

\begin{equation}
  (\Delta f_z)^2 = f_0^2 \left[\left(\frac{k_\mathrm{B} T}{\kappa}\right)^2 \, g_1(\Lambda \, R_0) + \left(\frac{k_\mathrm{B} T}{\kappa}\right)^3 \, g_2(\Lambda \, R_0) + \cdots \right] \, .
\end{equation}

\noindent The term $\propto (k_\mathrm{B} T)^2$ corresponds to the contributions of terms up to order two in $u$, coming from the diagrams of the form $\diagH{./figures_Chap4/diag.eps}$.
Further terms of higher order on $k_\mathrm{B} T$ come from the terms beyond $\mathcal{O}(u^2)$ in the Hamiltonian and in the stress tensor. 
Secondly, eq.(\ref{eq_4_D}) is valid for tubes relatively long, i. e., whose length is bigger than the radius, but still small compared to the radius of the vesicle from which it is extracted.
The simplified eq.(\ref{eq_4_Dfapprox}) is a good approximation under the supplemental condition $\Lambda \, R_0 > 6$.

Finally, let us comment on the influence of the boundary conditions.
Differently from the case of the average force, there is no conservation principle for $\Delta f_z$.
Here, we have calculated $\Delta f_z$ assuming periodic boundary conditions, or equivalently,
through an arbitrary section in the middle of a long enough tube.
The actual value of $\Delta f_z$ at the extremity of a tubule with specific
boundary conditions might be somewhat different.
Note also that we have
only calculated the fluctuation of the component of the force which is
parallel to the tube axis.

\section{Discussion and consequences for experiments}
\label{section_4_discussion}

First of all, let us discuss on the dependence of $\Delta f_z$ on $R_0$.
From eq.(\ref{eq_4_Dfapprox}), we have

\begin{equation}
  \Delta f_z \propto \sqrt{\frac{R_0}{a}} \left(\frac{k_\mathrm{B} T}{a} \right) \times \sqrt{\ln\left(\frac{R_0}{a}\right)} \, ,
\end{equation}

\noindent with $a = \Lambda^{-1}$ of the order of the membrane thickness.
The first term reminds the result obtained in chapter~\ref{Fluct_plan}.
There, we have seen that for planar membranes, the correlation of the stress tensor decreases over a very short length, whatever the membrane tension or rigidity.
One could thus consider that a piece of membrane was a composition of uncorrelated patches of size $\approx a$ and use the Central Limit Theorem to obtain the force fluctuation.
%Differently, here we see that the size of the patch depends on the membrane rigidity and tension through $R_0$.
In this case, however, the force fluctuation of tubes has a supplemental logarithmic
correction relative to force fluctuation in planar membranes.
This correction can be explained by the fact that the correlation of the stress tensor in membrane nanotubes decreases in a power law, as we have shown in section~\ref{subsection_4_corr}.

\begin{figure}[H]
\begin{center}
\includegraphics[width = 0.65\columnwidth]{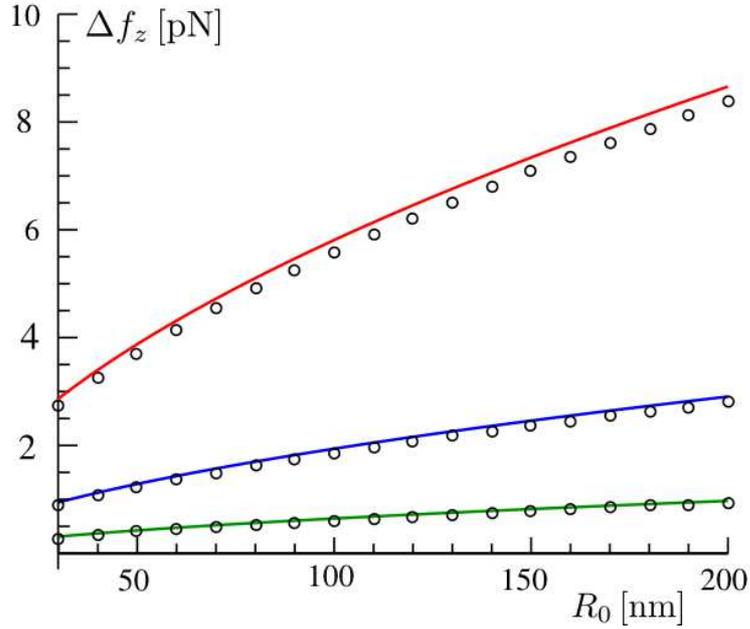}
\caption{Force fluctuation as a function of the tube's radius.
The solid lines correspond to the approximation given in eq.(\ref{eq_4_Dfapprox}).
The curves for $\Lambda \, a = 2, \, 1 \, ,0.5$ are shown, respectively, in blue (curve in the middle), red (upper curve) and green (lower curve).
The circles stand for the exact numerical sum (eq.(\ref{eq_4_D})).
Remark the good quality of the approximation.
In this plot, we have considered $k_\mathrm{B} T = 4 \times 10^{-21} \, \mathrm{J}$ and $a = 5 \, \mathrm{nm}$.}
\label{fig_4_fluctR}
\end{center}      
\end{figure}

In Fig.~\ref{fig_4_fluctR}, we show $\Delta f_z$ as a function on $R_0$ for
different values of the cutoff $\Lambda$.
The exact curve for long tubes given in eq.(\ref{eq_4_D}) is indicated by circles, while the approximation given in eq.(\ref{eq_4_Dfapprox}) corresponds to the solid lines.
We can see the good quality of the approximation.
For a given value of $\Lambda$, $\Delta f_z$ does not vary much over the experimental range of $R_0$.
On the contrary, it depends strongly on the value of the cutoff $\Lambda$. 

Numerically, we have found a force fluctuation of some $\mathrm{pN}$, which is of the
same order of magnitude of the value obtained experimentally experimentally
(see Fig.~\ref{fig_4_inaba}, \cite{Koster_05} and \cite{Cuvelier_05}).
For an accurate comparison, however,
time-resolved measurements should be performed and the Brownian force on
the pulling bead should be taken into account.

Finally, we compare the average of the force
needed to extract a tube, given by eq.(\ref{eq_3_ffin}), with it's fluctuation.
We trace both curves as a function of the effective mechanical tension $\tau$,
since this tension can be experimentally controlled (by changing the
difference of pressure in micropipettes experiments, for instance).
In agreement with the results of the previous chapters, we assume $\tau =
\sigma - \sigma_0$, where $\sigma_0 = (k_\mathrm{B} T \, \Lambda^2)/(8\pi)$.
Applying this relation to eq.(\ref{eq_4_Dfapprox}), we obtain

\begin{equation}
\left(\Delta f_z\right)^2 \simeq \frac{(k_\mathrm{B}T)^2 \, \Lambda^3}{6\pi}
\sqrt{\frac{\kappa}{2(\tau + \sigma_0)}}\left[\frac{5}{2} + \ln\left(\Lambda
    \, \sqrt{\frac{\kappa}{2(\tau + \sigma_0)}}\right)\right] \, .
\end{equation}

\begin{figure}[H]
\centerline{\includegraphics[width=.65\columnwidth]{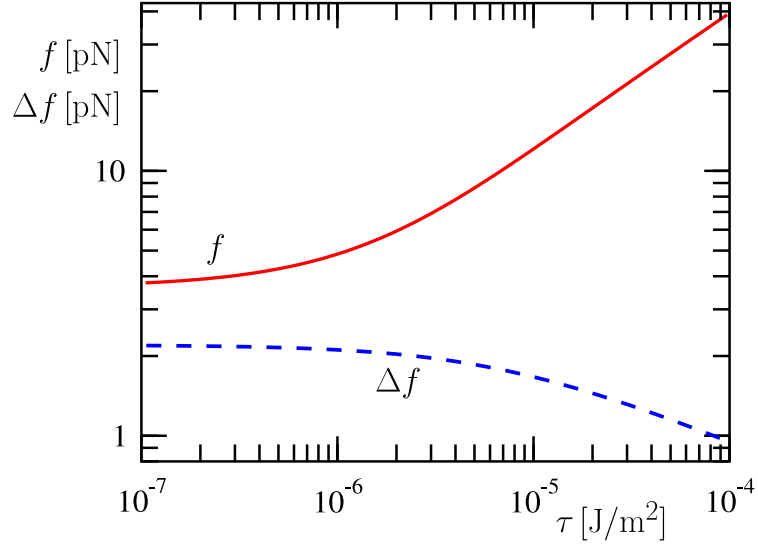}}
\caption{
Average force needed to pull a tube (solid red line) and it's fluctuation $\Delta f_z$ (dashed blue line) as a function of
the effective mechanical tension $\tau$.
The
parameters used are $\Lambda \, a=1$ with $a\simeq5\, \mathrm{nm}$,
$k_BT\simeq4\times10^{-21}\, \mathrm{J}$ and $\kappa\simeq50\,k_BT$.  
}
\label{fig_4_flucttau}
\end{figure}

\noindent The curve for $\langle f_z \rangle$, already presented in
Fig.~\ref{fig_3_fentau} and the curve of $\Delta f_z$ given above is shown in
Fig.~\ref{fig_4_flucttau}.
We can remark that $\Delta f_z$ is in general small compared to the force $f_z$ needed to extract a tube, despite the presence of soft Goldstone modes:
it is comparable to $f_z$ for
$\tau<10^{-6}\mathrm{J/m}^2$, but quite negligible for
$\tau>10^{-5}\mathrm{J/m}^2$.

Sadly, from Fig.~\ref{fig_4_fluctR} and~\ref{fig_4_flucttau}, we conclude that
$\Delta f_z$ does almost not depend on the tension of the membrane nor
on it's rigidity.
The force fluctuation seems thus of little interest in the mechanical characterization of membranes. 
On the other hand, the fact that $\Delta f_z$ does not depend neither in $\kappa$, neither on the membrane tension, could be of great interest to experiments involving active membranes.
Differently from the membranes studied in this work, which are passive, active membranes have proteins embedded in it that add non-equilibrium noise to the system. 
Experimentally, the activity of these proteins depends on a external source of energy. 
It has been observed that the protein activity causes an enhancement of the membrane fluctuations and of the excess area relative to the passive case, as if the membrane were in contact with a thermal bath of higher temperature~\cite{Manneville_01}.
Let's imagine now an experiment in which tubes were extracted from an active membrane.
If the membrane fluctuations were intensified, $\Delta f_z$ should be also affected.
Since it does almost not depend on the tension nor on the bending rigidity, it could thus be a used as a direct indicator of the proteins activity.

\section{In a nutshell}

In this chapter, we have examined the possibility of using the fluctuation of
the force along a membrane tube's axis $\Delta f_z$ as a tool to characterize
membranes.
We have only considered the contribution of the membrane's fluctuation, that can be very important
due to the presence of very soft modes.
For a weakly fluctuating tube of length $L$, with $R_\mathrm{ves} \gg L >
R_0$, where $R_\mathrm{ves}$ is the radius of the vesicle from which the tube is pulled and $R_0$ is the mean-field radius of the tube, we obtained

\begin{equation}
\left(\Delta f_z\right)^2 \simeq \frac{(k_\mathrm{B}T)^2 \, R_0\, \Lambda^3}{6\pi} \left[\frac{5}{2} + \ln\left(\Lambda \, R_0\right)\right] \, ,
\end{equation}

\noindent where $\Lambda^{-1} = a$, with $a$ of the order of the membrane thickness.
Interestingly, $\Delta f_z$ can generally be written as 

\begin{equation}
  \Delta f_z \propto \sqrt{\frac{R_0}{a}} \left(\frac{k_\mathrm{B} T}{a} \right) \times \sqrt{\ln\left(\frac{R_0}{a}\right)} \, ,
\end{equation}

\noindent which reminds the result found for the force fluctuation in planar membranes.
The logarithmic correction is a signature of the long-range correlations present in the tubes.  
Numerically, for $a \approx 5 \, \mathrm{nm}$, these equations yield $\Delta f_z
\approx 1 \mathrm{pN}$, which is compatible with experimental data.
Studying the behavior of the force fluctuation, we have found that it is extremely sensitive to the value of $\Lambda$, whereas it
does almost not depend on the bending rigidity nor on the tension.
Thus, $\Delta f_z$ seems of little usefulness to the mechanical characterization of  membranes.
It could however be used in experiments involving active membranes, i. e., membranes containing proteins whose activity can be modified, as an indicator of their activity.
Indeed, when proteins are active, the membrane fluctuations are increased, which would affect $\Delta f_z$ regardless of variations on the bending rigidity or on the tension.

%% file: chap5.tex
\chapter{Preliminary results on a $\bm{2}$-d membrane simulation}
\label{2D_Simulation}

In section~\ref{subsection_1_simu_1D}, 
we proposed a simple numerical system to verify our predictions concerning the mechanical tension $\tau$, the internal tension $\sigma$ and the tension $r$ obtained from the fluctuation spectrum of a membrane.
Our model was composed of a set of variable-sized rods, each one representing a coarse-graining of several lipids, free to move in a two-dimensional space.

In this chapter, we present a more complex numerical experiment consisting of a $2$-dimensional membrane that evolves in a three-dimensional space, which corresponds more accurately to the experimental situation.
We are motivated by the fact that a more elaborated numerical system would not only allow us to verify precisely our predictions concerning $\tau$, $\sigma$ and $r$, but it would also give access to other quantities, such as the fluctuation of the force that a frame exerts over a membrane, studied in chapter~\ref{Fluct_plan}.
Moreover, in chapters~\ref{TUBE} and~\ref{Fluct_TUBE} we predict the dependence of the force needed to extract a tube and its fluctuation on $\tau$, which could also be verified by pulling tubes from a numerical membrane.
Sadly, due to time constraints, the results presented here are far from complete and many questions are left unanswered.

\begin{figure}[H]
\begin{center}
  \vspace{2cm}
\includegraphics[width=0.4\columnwidth]{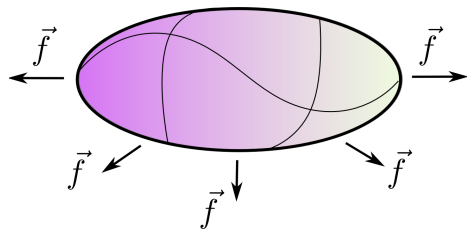}
\caption{Numerical experiment: a fluctuating membrane weakly departing from a plane held by a circular frame.
A force could be applied to the frame in order to widen its radius, increasing
thus the membrane's effective tension.
}
\label{fig_5_on_veut}
\end{center}      
\end{figure}

In our numerical experiment, we would like to study a piece of membrane held by a circular frame and weakly departing from a plane (see Fig.~\ref{fig_5_on_veut}).
%The microscopical area of the membrane should be connected to a lipid reservoir, so that its microscopical area $A$ could vary and its energy were well-described by the Helfrich Hamiltonian (eq.(\ref{Helfrich})).
We find many popular methods used to numerically simulate membranes in the scientific literature, which we sum-up briefly in section~\ref{section_5_panorama}.
We have chosen to use a phenomenological model consisting in a triangular network of extensible bonds connecting effective particles.
The connectivity of the network could be modified in order to mimic the
membrane's liquidity (see details in section~\ref{section_5_descrip}) and
a harmonic potential acted over the particles at the network's edge, forcing a circular frame.
Thus, we could measure directly the force applied to the frame and derive the effective tension $\tau$ as well as its fluctuation $\Delta \tau$ (see section~\ref{subsection_5_alpha}).
The minima of this potential could be modified to widen the frame's radius, decreasing the excess area $\alpha$ and increasing the membrane's tension.

To obtain representative averages of $\tau$, $\alpha$ and other variables, we needed to generate large sets of configurations of the numerical membrane, which was done through a Monte Carlo dynamics, described in section~\ref{section_5_dyn}.
In this section, we discuss also which were the criteria used by us to determine whether a sampling was large enough.

As usually done in laboratory experiments (see section~\ref{fluct}),  the bending rigidity $\kappa$ and tension $r$ were deduced from the average of the fluctuation spectrum of the membrane.
Since we simulate the membrane using a network, obtaining the fluctuation spectrum is somewhat complicated, as we discuss in section~\ref{subsection_5_spect}.
Finally, we explain in section~\ref{subsection_5_sigma} how we could estimate the internal tension $\sigma$.
%have not measured $\sigma$ directly in our experiment.
%We know, however, that $\sigma$ is proportional to the area of the membrane $A$ (section~\ref{mechanical}) and thus it is related to the tension of the bonds composing the network (see section~\ref{subsection_5_sigma}).
In section~\ref{section_5_prim} we discuss some preliminary results.
At last, in section~\ref{section_5_tube}, we comment briefly on extracting
tubes from our numerical membrane and we end this chapter with a brief discussion on issues that should be investigated in the future (section~\ref{section_5_persp}).

\section{Short panorama of numerical models on membranes}
\label{section_5_panorama}

Processes in membranes happen in a wide range of time, size and energy scales.
For instance, interactions between lipids and proteins inside the membrane
occur in distances of the order of the nanometer with a characteristic time of
some $\mathrm{ps}$, while the evolution of the shape of a vesicle involves scales of micrometers and may take many seconds.
Consequently, depending on the process one is interested in, several different models are used to numerically simulate biological membranes (see~\cite{Muller_06}, \cite{Orsi_07} and \cite{Noguchi_08} for some reviews).
Schematically, they can be grouped in three classes:

\begin{figure}[H]
\begin{center}
  \includegraphics[width=0.8\columnwidth]{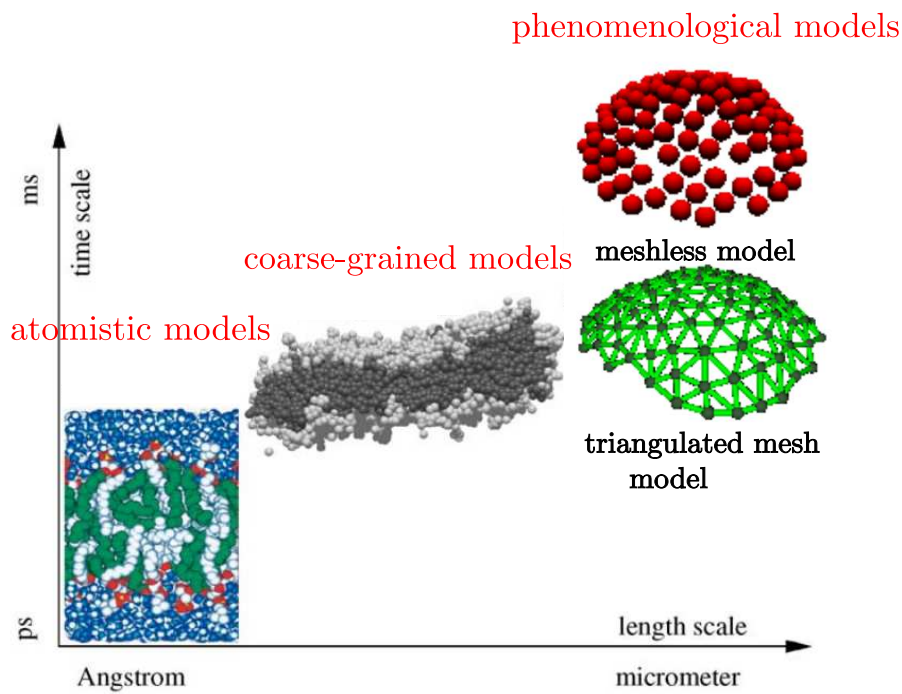}
\caption{Illustration of the three main classes of numerical models used to simulate biological membranes, ordered in terms of the characteristic time and length scale (figure based on~\cite{Muller_06}).}
\label{fig_5_time_scale}
\end{center}      
\end{figure}

\begin{enumerate}
\item {\bf atomistic models}: these models try to take into account all the chemical details of the molecules by considering the interactions between atoms.
  They are used to study how lipids interact among themselves and with proteins.
  As these simulations involve many degrees of freedom, they are very computer consuming.
  Consequently, one can at most simulate small patches of a dozen of nanometers for dozens of nanoseconds.

\item {\bf coarse-grained models}: in these models, small groups of atoms are lumped together into effective particles that interact via simplified potentials.
  The solvent can be effectively or implicitly present.
  As the number of degrees of freedom is reduced, one can observe collective movements of the membrane, such as its self-assembly, stretching~\cite{Neder_10}, pore formation~\cite{Neder_10} and thermal fluctuations~\cite{Imparato_06}.
  The main difficulty of these models is deciding which interactions are truly essential to reproduce the membrane's behavior.
  A popular model of this category is the spring-and-bead model presented in section~\ref{section_1_evidences}.
  Sadly, with these models one is still restrained to length scales of hundreds of nanometers, which is a limitation if one wants to study large-scale processes.

\item {\bf phenomenological models}: these models take coarse-graining one step further, representing several molecules as a single effective particle, which we will call a {\it bead} in the following.
  The solvent is always implicit.
  They are suitable to study the universal properties of amphiphilic systems.
  The effective particles can be attached between themselves through a triangular or square mesh or instead, the mesh can be absent~\cite{Noguchi_06}, ~\cite{Maibaum_10} (meshless models).
  In the first case, to mimic liquidity, the topology of the mesh is changed during the simulation.
  The meshwork is then called dynamic.
  Our previous simple model, presented in section~\ref{subsection_1_simu_1D}, belong to this category.
\end{enumerate}

Throughout this work, we were interested in the general properties of membranes, regardless of the molecular details, at length scales far bigger than the membrane's thickness.
Accordingly, phenomenological models are the most adapted to our case.
We give some further details on them in the following. 

\vspace{1cm}

The meshless models were first proposed by Drouffe et al. in 1991~\cite{Drouffe_91}:
the beads interact via a hard-core repulsion, an anisotropic attraction that depends on their orientation and an effective multi-body interaction favoring a closed packed environment to simulate the hydrophobic interactions between lipids and the aqueous solvent.
These models are very elegant, since one can easily observe the membrane self-assembly, topological changes, pore formation and the gel-liquid transition~\cite{Drouffe_91},~\cite{Noguchi_06},~\cite{Maibaum_10}.
As in real experiments, the bending rigidity is usually measured through the fluctuation spectrum.
Recently, however, an alternative method in which one imposes $\kappa$ directly was proposed by Noguchi et al.~\cite{Noguchi_06}.
At each point of the membrane, a quadratic curve is fitted to the beads
contained in a small region in order to obtain the local curvature.
Subsequently, the standard Helfrich Hamiltonian is used to evaluate the configuration's energy.

\vspace{1cm}

The meshwork phenomenological models are a bit older~\cite{Kantor_87} (see~\cite{Gompper_97} for a comprehensive review).
Actually, very similar models were already studied at that time in other contexts, such as lattice field theories and lattice approximations to relativistic string theories~\cite{Billoire_84},~\cite{Espriu_87}.
The beads were connected by a triangular meshwork that could have fixed topology, i. e., each bead had always six neighbors or they could be connected by a meshwork whose connectivity evolved over time, forming dynamically triangulated surfaces~\cite{David_85}, ~\cite{Ambjorn_85}.   
At this point, membranes were usually {\it phantom}, i. e., beads could superpose and the self-penetration of the network was allowed.

\vspace{1cm}

\begin{figure}[H]
\begin{center}
\includegraphics[width=0.4\columnwidth]{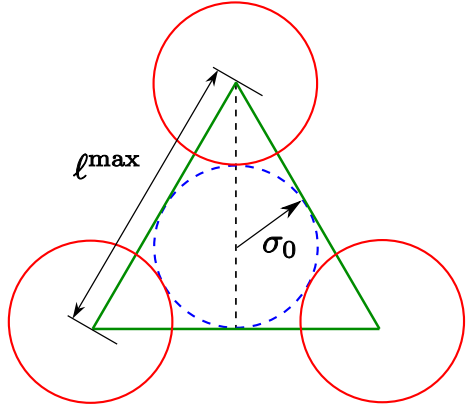}
\caption{In order to assure that the membrane cannot self-penetrate, one imposes a maximal length $\ell^{\mathrm{max}}$ to the network's bonds plus a hard-core potential between any two beads of the network.
  Here we show how $\ell^{\mathrm{max}} = 2 \sqrt{3} \, \sigma_0$ is obtained, with $\sigma_0$ the beads' radius.}
\label{fig_5_lomax}
\end{center}      
\end{figure}

In the context of biological membranes, models with fixed connectivity, representing a polymerized membrane, were first used in 1987~\cite{Kantor_87a}.
For the first time, the curvature energy was taken into account by introducing
an interaction between adjacent triangles of the network.
Many contemporary works were interested in the dependence of the gyration radius of the membrane on it's linear length~\cite{Kantor_87},~\cite{Kantor_87b} and in the crumpling transition~\cite{Kantor_87c}.
As biological membranes are self-avoiding, the effects of the self-avoidance were also studied by introducing a hard-core potential between any two beads and limiting the length of the network's bonds to $\ell^{\mathrm{max}} = 2\sqrt{3}\, \sigma_0$, where $\sigma_0$ denotes the beads radius, in order to ensure the impenetrability of the surface~\cite{Kantor_87b} (see Fig.~\ref{fig_5_lomax} for a geometrical explanation of this value).

From 1990 on, the fluidity was taken into account by dynamically modifying the triangulation, while keeping the self-avoiding restrictions~\cite{Ho_90},~\cite{Kroll_92}.
Since then, this model has been used in a wide variety of complex numerical experiments, such as studying the dynamics of vesicles and red blood cells in flows~\cite{Noguchi_05},~\cite{Noguchi_10} and the budding of vesicles mediated by proteins~\cite{Atilgan_07}.
As we explain in the next section, this well-established dynamical triangular network model was
the basis for our numerical model of membrane.

\section{Our numerical membrane}
\label{section_5_descrip}

As shown in Fig.~\ref{fig_5_on_veut}, we wanted to simulate a relatively large piece of weakly fluctuating membrane attached to a circular frame. 
Under these conditions, the probability of overhangs is very small and thus the probability that large fluctuations bring distant segments of the membrane into close spatial proximity is negligible.
Consequently, in an approximation, we decided to ignore the hard-core potential between any two beads and consider only the interactions between neighboring beads, which is much less computer consuming.
In this case, the meshwork phenomenological model presents a great advantage: with a mesh, we know at every  instant which beads are neighbors, since they are attached by bonds, whereas in meshless models determining neighbors is not straightforward.
So, we have decided to use a dynamically triangulated meshwork whose beads are phantom if they are not first neighbors.

\begin{figure}[H]
\begin{center}
\includegraphics[width=0.4\columnwidth]{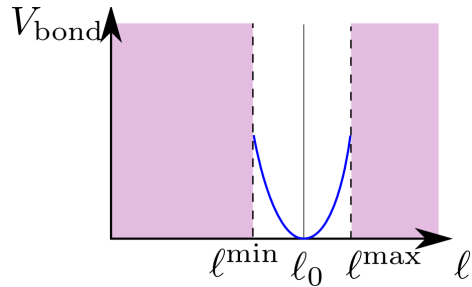}
\caption{Potential between neighboring beads as a function of their distance $\ell$.
The regions where the potential is $\infty$ are shaded.}
\label{fig_5_potpart}
\end{center}      
\end{figure}

In agreement with section~\ref{section_5_panorama}, we denote the beads' radius $\sigma_0$.
Each pair of neighboring beads interact through the potential

\begin{equation}
  V_\mathrm{bond}(\ell) = \left\{ \begin{array}{ccccc} \infty \,\,\,& $for$ &\, \ell < \ell^{\mathrm{min}}\, ,\\
    \\
    s \, \frac{(\ell - \ell_0)^2}{2} \,\,\,& $for$ &\, \ell^{\mathrm{min}} < \ell < \ell^{\mathrm{max}} ,\\
    \\
    \infty \, \, \, &$for$ &\, \ell > \ell^{\mathrm{max}} \, ,\end{array} \right.
  \label{eq_5_bond}
\end{equation}

\noindent where $\ell$ is the distance between the center of adjacent beads; $\ell^{\mathrm{min}} = 2 \, \sigma_0$ and $\ell^{\mathrm{max}}= 2 \sqrt{3} \, \sigma_0$ are, respectively, the minimal and maximal distance between the center of adjacent beads.
The length $\ell_0$ corresponds to a preferred distance that we have chosen as the average of the minimal and maximal allowed length: $\ell_0 = (\ell^{\mathrm{min}} + \ell^{\mathrm{max}})/2 = (1 + \sqrt{3}) \, \sigma_0$ (see Fig.~\ref{fig_5_potpart}).

%JUSTIFICAR PQ POTENTIAL HARMONIQUE.

In section~\ref{grand_can}, we have seen that the bending rigidity of a weakly fluctuating membrane gives a contribution

\begin{equation}
  E_\kappa = \frac{\kappa}{2} \int_{A_p} (\nabla^2 h)^2 \, dA_p \, 
\end{equation}

\noindent to the membrane's energy.
We will not consider topological changes in our simulation, so the Gaussian
contribution to the curvature energy need not to be taken into account.
In our network, we considered the commonly used bending energy discretization~\cite{Kantor_87}

\begin{equation}
  E_\kappa^{\mathrm{discret}} = \frac{k}{2} \sum_{\langle\alpha,\beta \rangle} |\bm{n}_\alpha - \bm{n}_\beta|^2 = k \sum_{\langle\alpha,\beta \rangle} (1 - \bm{n}_\alpha \cdot \bm{n}_\beta) \, ,
  \label{eq_5_Ek_discret}
\end{equation}

\noindent where the sum runs over all pairs of adjacent triangles $\alpha$ and $\beta$, with normal vectors $\bm{n}_\alpha$ and $\bm{n}_\beta$, respectively. 
This discretization, however, presents a major problem: the relationship between $\kappa$ and $k$ depends on the membrane's geometry.
Alternative more complex discretizations were proposed~(see~\cite{Gompper_97} for further details), but here we have chosen to keep this simplified discretization, since $\kappa$ will be measured through the spectrum fluctuation.
Note that eq.(\ref{eq_5_Ek_discret}) is a good approximation only for $\bm{n}_\alpha \approx \bm{n}_\beta$.
Indeed, for two triangles with $\bm{n}_\alpha = -\bm{n}_\beta$, we have a contribution $2 \, k$, while this configuration should be prohibitively costly.

\begin{figure}[H]
\begin{center}
\includegraphics[width=0.4\columnwidth]{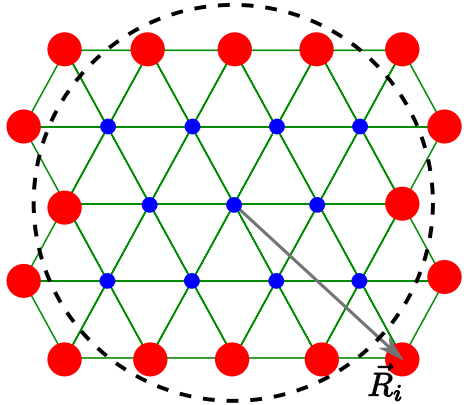}
\caption{Initial configuration of the triangular network with $N_x = 5$.
  At this point, the network is planar and each bond measures $\ell_0$.
The dashed circle represents the frame with $R_f = 6.96 \, \sigma_0$ and $\vec{R}_i$ is the distance from a boundary bead to the center of the frame.
The red large beads are subjected to the potential (\ref{eq_5_potrad}) in order to impose the circular frame.
The ratio $\ell_0 = (1 + \sqrt{3})\sigma_0$ has not been taken into account in this graphical representation.}
\label{fig_5_rede_ini}
\end{center}      
\end{figure}

At last, to impose a circular frame, each bead $i$ of the network's boundary, shown in red in Fig.~\ref{fig_5_rede_ini}, is subject to a harmonic potential

\begin{equation}
  V_f^i = k_\mathrm{f}\, \frac{(R_i - R_f)^2}{2} \, ,
  \label{eq_5_potrad}
\end{equation}

\noindent where $k_f$ is a constant that determines the rigidity of the potential, $R_i$ is the distance of the bead with respect to the center of the network and $R_f$ is the desired frame radius, imposed at the beginning of the simulation.
Note that the projected area of the membrane $A_p$ is not necessarily
equivalent to $\pi R_f^2$: it can vary more or less, depending on the
choice of $k_f$.
%As experimentally, the membrane's tension depends on the choice of $R_f$.

To initialize the network, we construct a planar triangular network alternating lines with $N_x$ and $N_x + 1$ beads, up to $N_x$ lines.
The beads are distanced by $\ell_0$ and arranged as in Fig.~\ref{fig_5_rede_ini}.

During the simulation, two kinds of moves were possible:

\begin{enumerate}
  \item Move P: the position of one bead is modified.
    The beads shown in blue in Fig.~\ref{fig_5_rede_ini} are free to move in three dimensions, while the ones belonging to the boundary can only move in the frame's plane.
    %, that we will denote the plan $\bm{e}_x \times \bm{e}_y$ in the following.
    
\item Move Flip: the network's connectivity is changed in order to represent the membrane fluidity.
  This is done by eliminating an existing bond and proposing a new one, as shown in Fig.~\ref{fig_5_flip}.

\begin{figure}[H]
\begin{center}
  \includegraphics[width=0.5\columnwidth]{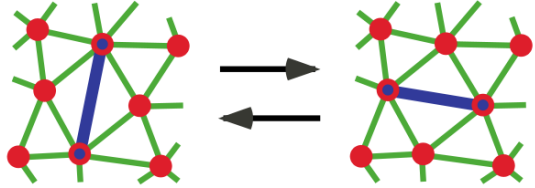}
\caption{Move called flip: a bond is deleted and a new one is proposed, changing the connectivity of the network.}
\label{fig_5_flip}
\end{center}      
\end{figure}

\end{enumerate}

In the following section, we will see how these moves were numerically implemented.

\section{Simulation dynamics}
\label{section_5_dyn}

As in section~\ref{subsection_1_simu_1D}, we used a Monte Carlo method to generate a large sample of configurations.
Again, the configurations were generated through a Markov chain algorithm: from a certain configuration $\Omega_i$, a new configuration $\Omega_{i+1}$ was accepted with a probability

\begin{equation}
  P(\Omega_i \rightarrow \Omega_{i+1}) = \min\left[1, e^{-\beta \, \Delta \mathcal{H}}\right]\, ,
\end{equation}

\noindent where $\Delta \mathcal{H} = \mathcal{H}_{i+1} - \mathcal{H}_i$ is the energy variation.
In practice, we have

\begin{enumerate}
  \item Move P: one particle $i$ is taken at random.
    If the particle belongs to the bulk of the network (blue beads in Fig.~\ref{fig_5_rede_ini}), we propose a new position $\bm{r}_i' = \bm{r}_i + \Delta \bm{r}$, where $\Delta \bm{r} = \delta r \times [\mathrm{rand}(-1,1) \, \bm{e}_x + \mathrm{rand}(-1,1) \, \bm{e}_y + \mathrm{rand}(-1,1) \, \bm{e}_z]$
    , with $\mathrm{rand}(a,b)$ a random number between $a$ and $b$, $\bm{e}_z$ the direction perpendicular to the frame's plane and $\bm{e}_x$, $\bm{e}_y$ two perpendicular directions contained in the frame's plane.
    Each bond attached to the particle $i$ has its length modified.
    The normal, area and projected area of all triangles that have the particle $i$ as a vertex must also be re-evaluated.
    The energy variation has thus two contributions: one coming from the changing on the bond's length and other coming from the curvature.
    We can see a representation of them in Fig.~\ref{fig_5_movep}.

    In the case of a boundary bead (red large beads in Fig.~\ref{fig_5_rede_ini}), one has simply $\Delta \bm{r} = \delta r \times [\mathrm{rand}(-1,1) \, \bm{e}_x + \mathrm{rand}(-1,1) \, \bm{e}_y]$.
In addition to the former contributions, one needs also in this case to consider the energy variation coming from the frame's potential.
The value of $\delta r$ was adjusted to have an acceptance rate of $\sim 50\%$.

\begin{figure}[H]
\begin{center}
\subfigure[Bonds' contribution.]{
\includegraphics[width=0.45\columnwidth]{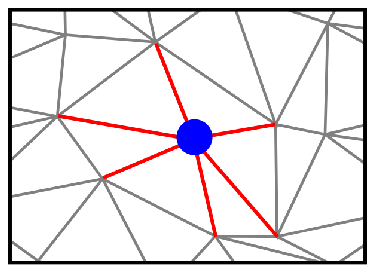}
}  
\subfigure[Curvature contribution.]{ 
  \includegraphics[width=0.45\columnwidth]{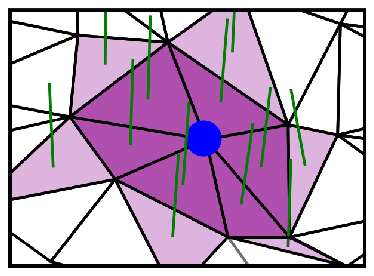}
  
}
\caption{Contributions to the energy variation when the position of a bead, here shown in blue, is modified.
  At left, we represent the energy variation due to the modification of the length of each tether attached to the bead.
  At right, we represent the curvature contribution: the normal (segment at the center of the triangles shown in green), area and projected area of the triangles that have the blue bead as a vertex (dark violet triangles) must be re-evaluated.
  One must subsequently consider the variation of the curvature energy between all adjacent violet triangles.
}
\label{fig_5_movep}
\end{center}      
\end{figure}

\begin{figure}[H]
\begin{center}
\subfigure[Before the flip.]{
\includegraphics[width=0.45\columnwidth]{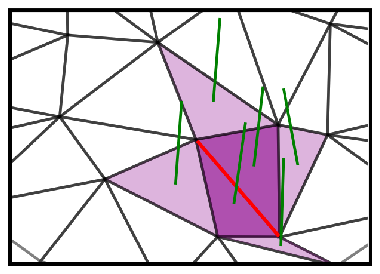}
}  
\subfigure[After the flip.]{ 
  \includegraphics[width=0.45\columnwidth]{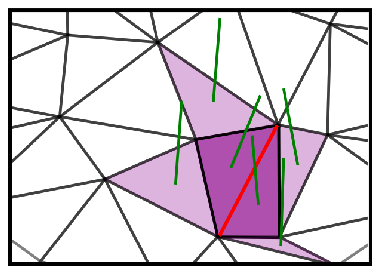}
  
}
\caption{Energy variation for a typical flip move.
  The red bond represents the bond that is suppressed (at left) / created (at right).
  The energy variation is due to the construction of a new bond of different length and to the modification of the normals (segment at the center of the triangles shown in green) of the dark violet triangles.
  Consequently, one has to consider the variation of the curvature energy between the dark violet triangles and their neighbors, shown in light violet.   
}
\label{fig_5_moveflip}
\end{center}      
\end{figure}

\item Move Flip: we randomly choose a bond belonging to the network's bulk.
   We propose a substitution to this bond, as shown in Fig.~\ref{fig_5_flip}.
   The normal, area and projected area of the two new triangles is calculated and the total energy variation involves the terms illustrated in Fig.~\ref{fig_5_moveflip}.
   Remark that the frame potential never contributes to this kind of move, since the position of the beads remain constant.
   
Note also that the acceptance rate of flip moves is completely determined by the tension applied to the network through the choice of $R_f$ and by the choice of the constants $s$ and $k$.
Typically, we have an acceptance rate between $1\%$ and $10\%$, depending on the chosen values.
For very large tensions, this can be a serious issue, since the energy variation $\Delta \mathcal{H}$ is in general very large.
Consequently, almost no flip is accepted and the network does not mimic the membrane's fluidity.
\end{enumerate}

In both kinds of move, one has to re-evaluate the normal, area and projected area of some triangles.
For each triangle, we evaluate the cross product of two of its edges to obtain the direction of its normal and its new area.
One must however pay attention to the order in which the cross product is evaluated to assure that the orientation of the normal is correct.
Similarly, to obtain the new projected area, we considered the cross product of the projection of two of the triangle's edges onto the frame's plane.

\begin{figure}[H]
\begin{center}
\subfigure[Top view.]{
\includegraphics[width=0.4\columnwidth]{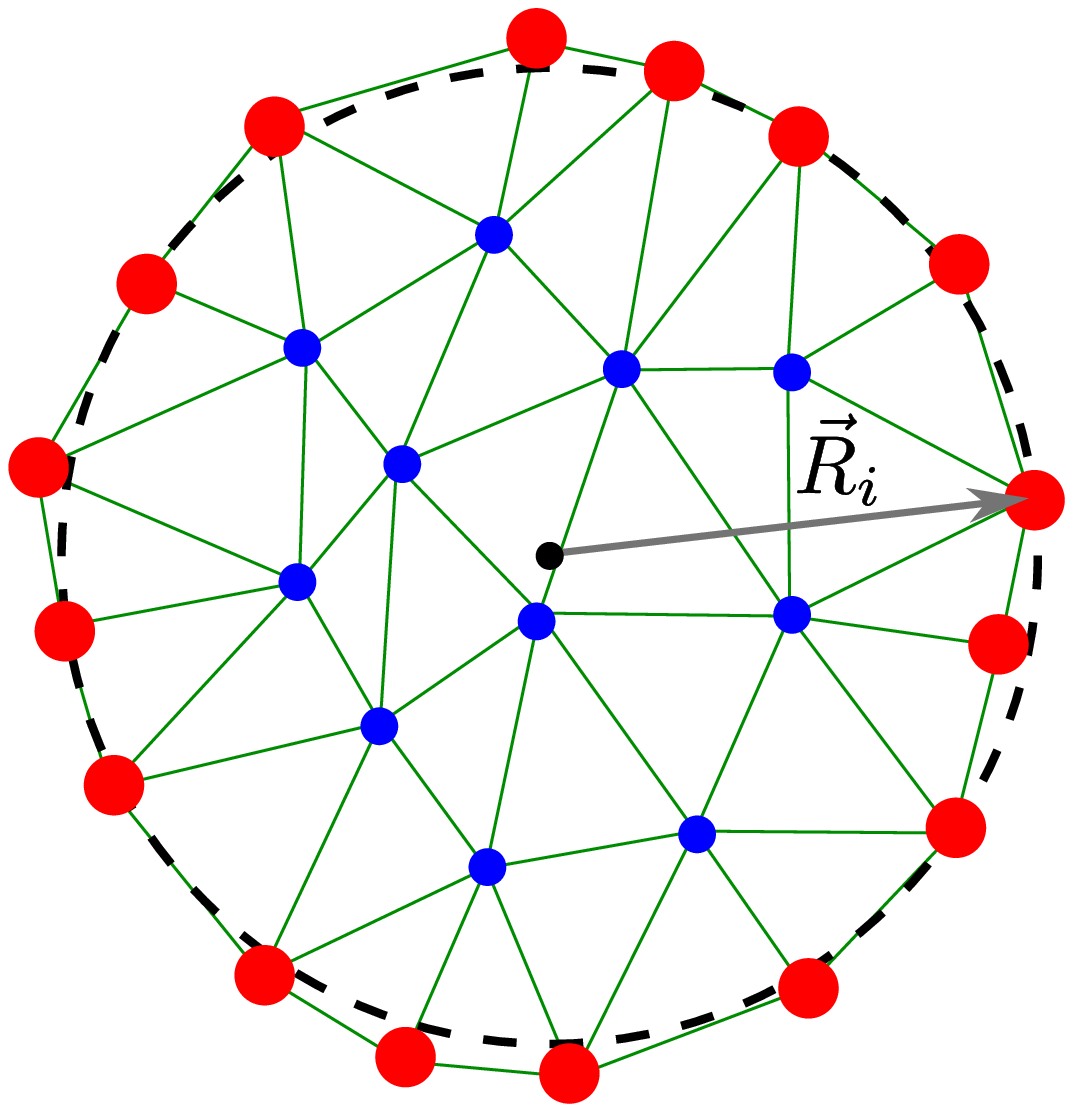}
}  
\subfigure[Side view.]{ 
  \includegraphics[width=0.55\columnwidth]{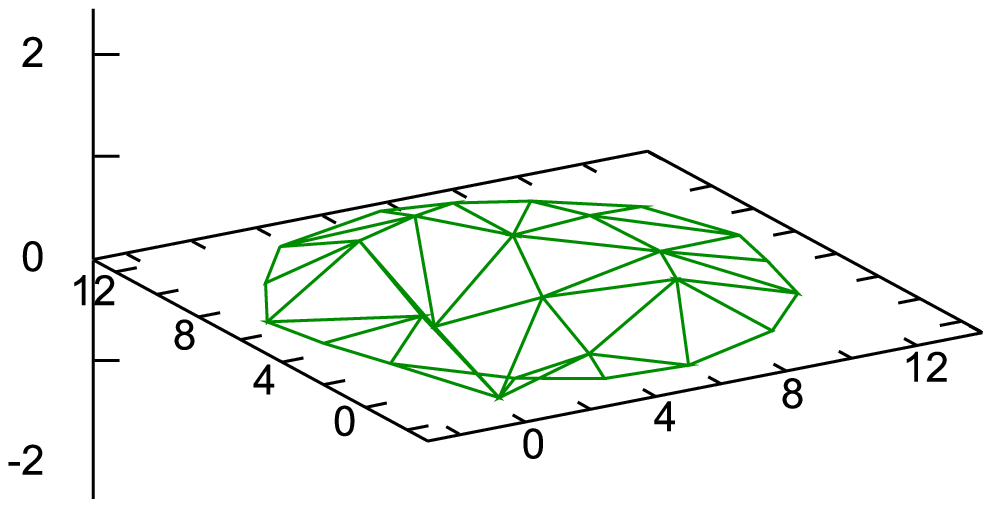}
}
\caption{Configuration for a small network with $N_x = 5$ and a total of $N=27$ beads, $\beta \, \sigma_0^2 \, s=1$, $\beta \, k=10$, $\beta \, \sigma_0^2\, k_f=30$ and $R_f = 6.96 \, \sigma_0$ after the first $2 \times 10^4$ Monte Carlo steps.
In the top view, we see that the boundaries roughly coincide with the imposed circular frame after $N_\mathrm{neg} = 2 \times 10^4$ steps (the center of the frame is indicated by the black dot).
Observe also that the topology of the network has changed: one finds beads with five and seven neighbors.
In the side view, we can see that the membrane fluctuates around the plane (note that the vertical and the horizontal scales are different).
}
\label{fig_5_cercle_ini}
\end{center}      
\end{figure}

For each attempt of move $P$, we try a flip.
We call a Monte Carlo step a set of $N$ sequences of a move P followed by a move flip, with $N$ the number of beads.
The first $N_\mathrm{neg}$ steps are not taken into account in the evaluation of averages to assure that the membrane has reached equilibrium.
In Fig.~\ref{fig_5_cercle_ini}, we show the configuration of a small network after $N_\mathrm{neg} = 2 \times 10^4$ Monte Carlo steps.
The frame is already roughly circular (the fit with the frame depends on the choice of $k_f$).
We will call a complete sequence of $N_\mathrm{neg}$ Monte Carlo steps followed by a number of equilibrium Monte Carlo steps $N_\mathrm{iter}$ a run.

\begin{figure}[H]
\begin{center}
\includegraphics[width=0.95\columnwidth]{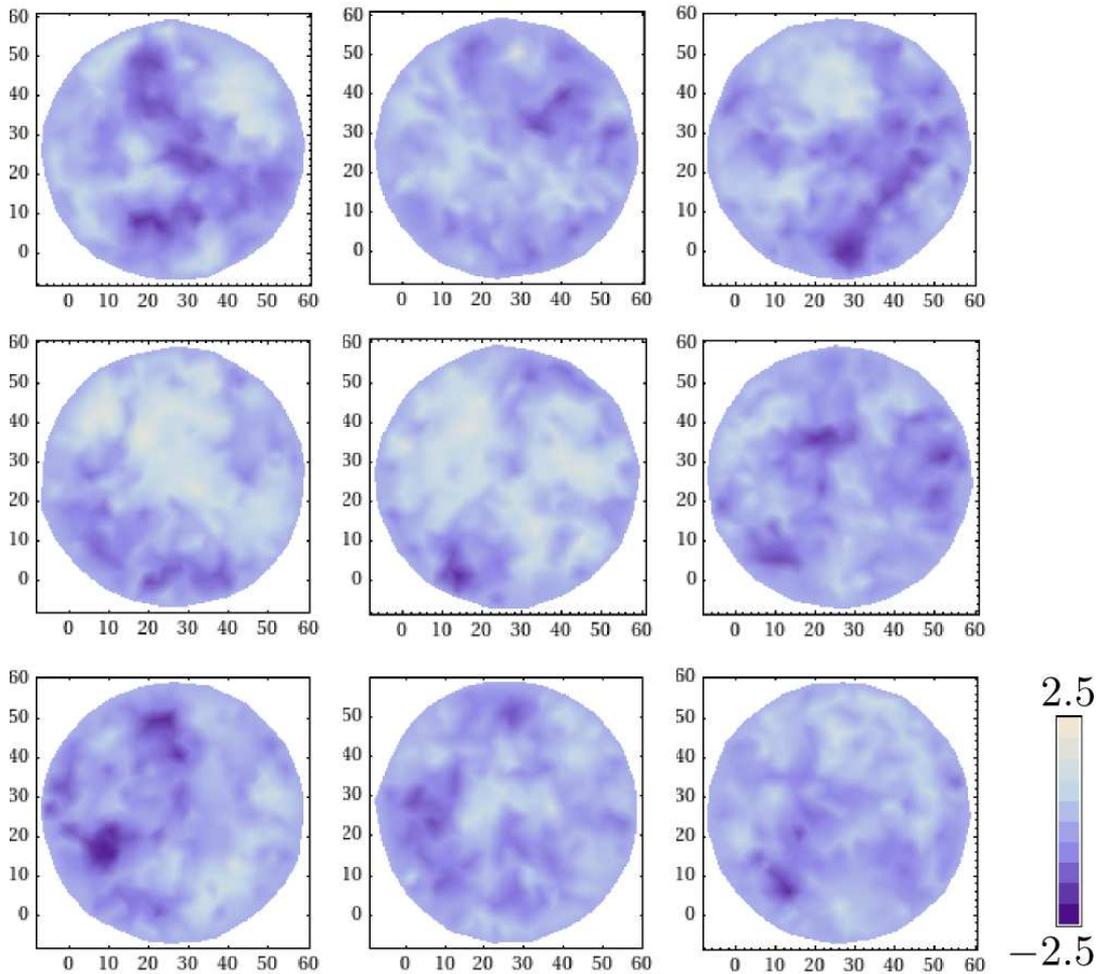}
\caption{Snapshots of the network at every $2 \times 10^{4}$ Monte Carlo steps.
  The height of the membrane is represented by the shading scale at right and the three spatial coordinates are measured in unities of $\sigma_0$.
  This image was obtained using the interpolation explained in the following (section~\ref{subsection_5_spect}) for $N_\mathrm{grid} = 128$.
Remark that the membrane weakly departs from the plane and that the configurations look uncorrelated after $2 \times 10^4$ iterations.}
\label{fig_5_evol}
\end{center}      
\end{figure}

\subsection{Verifications and equilibration criteria}
\label{subsection_5_equilibration}

In order to obtain meaningful averages, we have to assure that our configuration sampling is reasonably uniform over the space of the possible configurations, i. e, we have to assure that $N_\mathrm{iter}$ is large enough.
We have not done a systematic study on how the equilibration time depends on the network's size and constants at this preliminary stage.
We have rather evaluated the equilibration at each run.
In the following, we exemplify how we have carried this out using a typical network with $410$ beads ($N_x = 10$), $\beta \, k = 5$, $\beta \, \sigma_0^2 \,  k_f = 10$, $\beta \, \sigma_0^2 \, s = 1$ and $R_f = 33.04 \, \sigma_0$.
We assume that the system has already relaxed to its equilibrium state after $N_\mathrm{neg}$ Monte Carlo steps.
First of all, we have evaluated visually the system's evolution, as shown in Fig.~\ref{fig_5_evol}.
We can see that in this case, the configurations are already very different after $2 \times 10^4$ Monte Carlo steps.

\begin{figure}[H]
\begin{center}
\subfigure[Flip frequency.]{
\includegraphics[width=0.53\columnwidth]{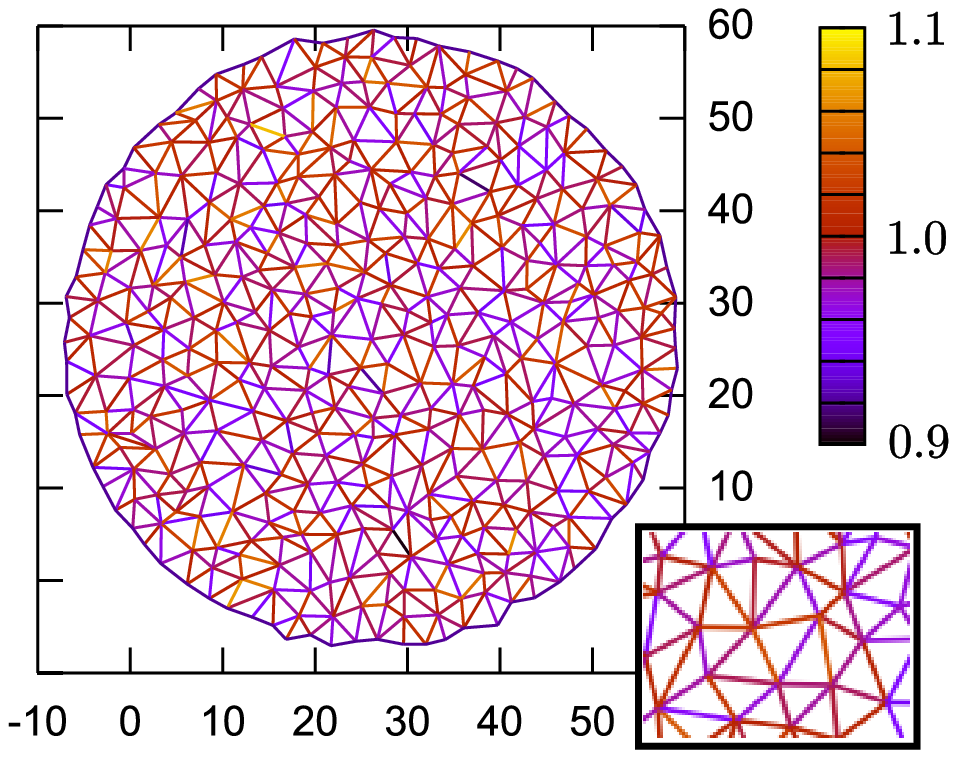}
\label{fig_5_flip_stat}
}  
\subfigure[Diffusion of one bead.]{ 
  \includegraphics[width=0.40\columnwidth]{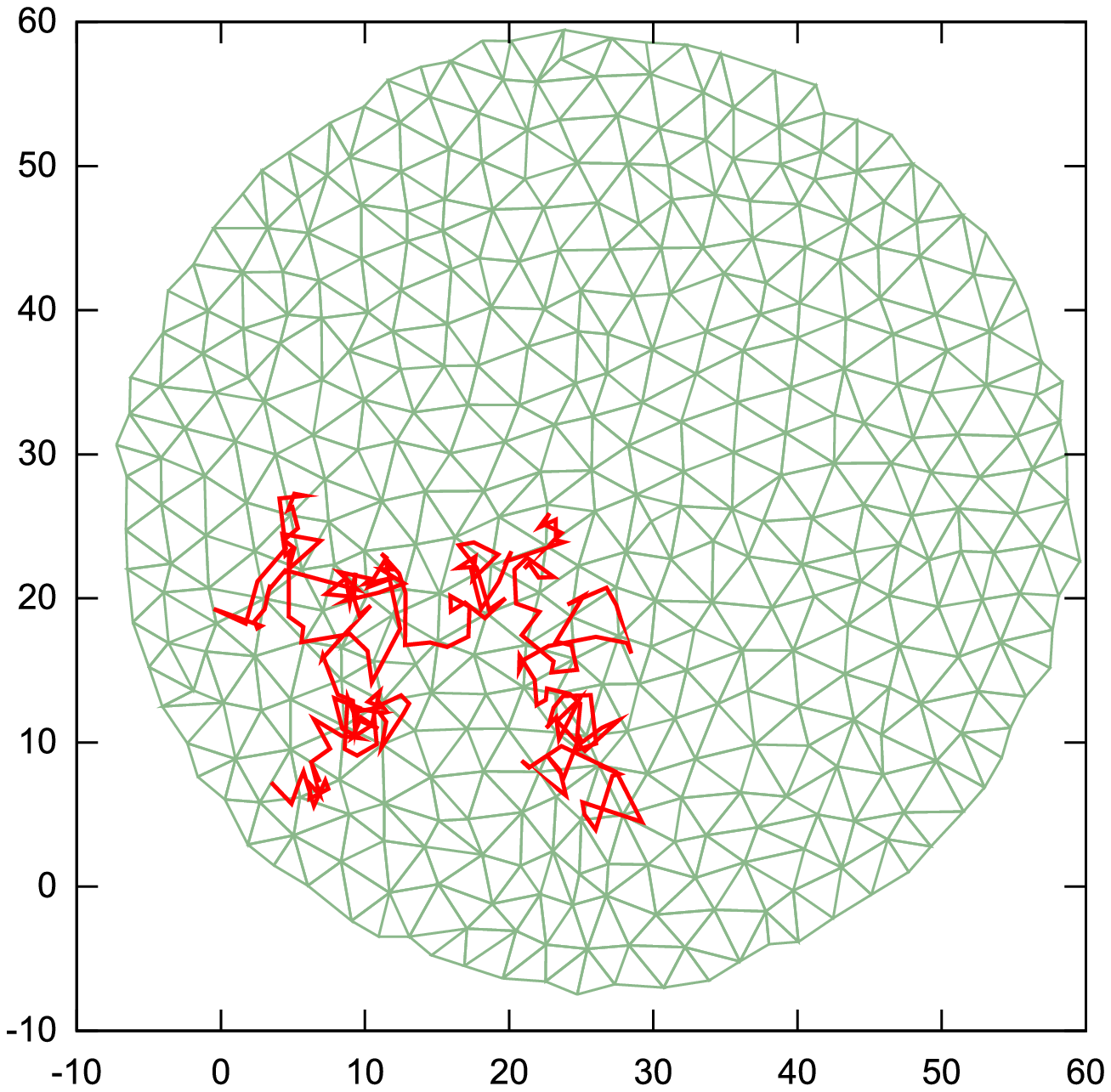}
  \label{fig_5_diff}
}
\caption{At left, we see the top view of the network after $2 \times 10^5$ iterations.
  The inset shows a detail of the network.
  Each bond was colored with the relative frequency with which flips were accepted (the average was normalized to one).
  We see that flips happened uniformly in space.
  At right, the diffusion of a bead after the same number of iterations testify of the membrane's liquidity (red curve).
  We have superposed a network snapshot (in green) for comparison.
}
\end{center}      
\end{figure}

Visually, we have also checked if all the bonds were being flipped with a similar frequency.
For the same network as before, we show in Fig.~\ref{fig_5_flip_stat} a map of the bonds colored as a function of the relative frequency with which they were flipped.
We can see that the coloring is very uniform, indicating that the network does not present regions with different liquidity.
As a supplementary check, we have also studied the diffusion of one bead over time (see Fig.~\ref{fig_5_diff}).

\begin{figure}[H]
\begin{center}
\includegraphics[width=0.45\columnwidth]{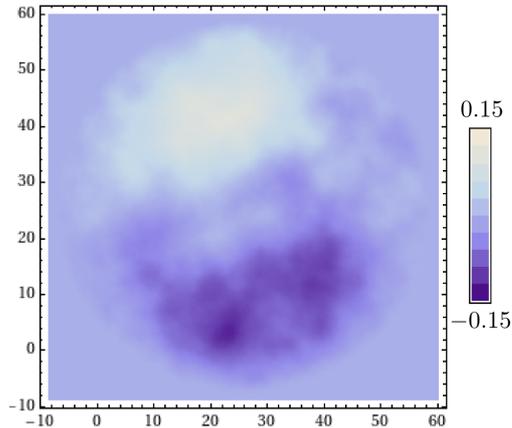}
\caption{Average shape of the membrane after $2 \times 10^{6}$ Monte Carlo steps.
  The average height of the membrane is represented by the shading scale at right and the three spatial coordinates are measured in unities of $\sigma_0$.
  Once again, we have used $N_\mathrm{grid} = 128$ to construct an interpolation grid.
Note that the vertical scale is far smaller than the horizontal.
The parameters are the same as in the last figures.}
\label{fig_5_shape_moy}
\end{center}      
\end{figure}

On a second moment, we have studied the spatial average of the membrane's height: since there is no asymmetry, after a sufficiently large number of steps, one should expect this quantity to vanish.
This condition is however not sufficient, since the membrane can have a vanishing spatial average and be non-planar.
We have thus monitored the local average of the membrane's height, i. e., we have studied the average shape of the membrane (see Fig.~\ref{fig_5_shape_moy}).
In practice, this was done by constructing an interpolation that will be explained in section~\ref{subsection_5_spect} and averaging the height over each cell of the interpolation grid.

\begin{figure}[H]
\begin{center}
\subfigure[Mode $h_{1,0}$.]{
\includegraphics[width=0.3\columnwidth]{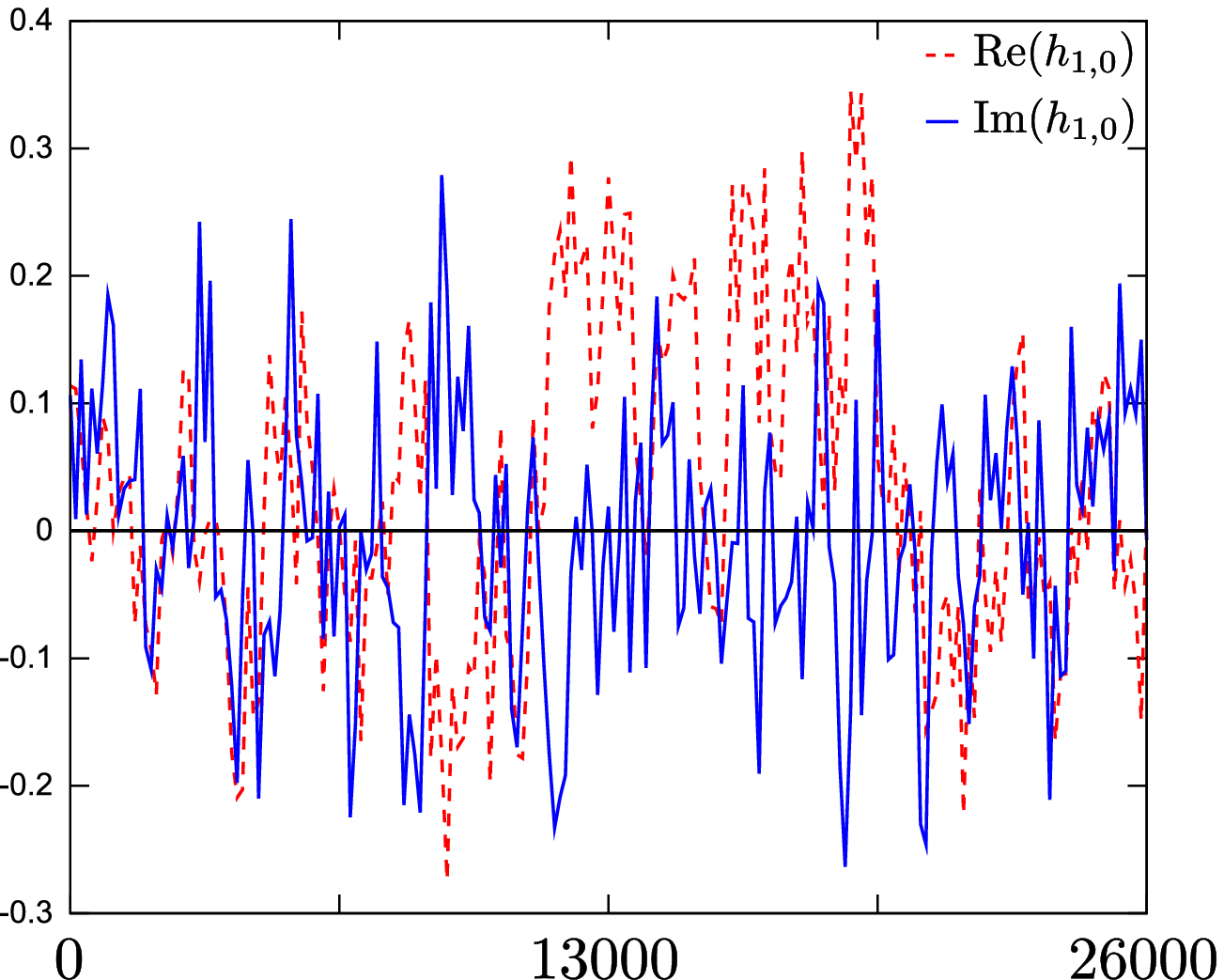}
}  
\subfigure[Mode $h_{0,1}$.]{ 
  \includegraphics[width=0.3\columnwidth]{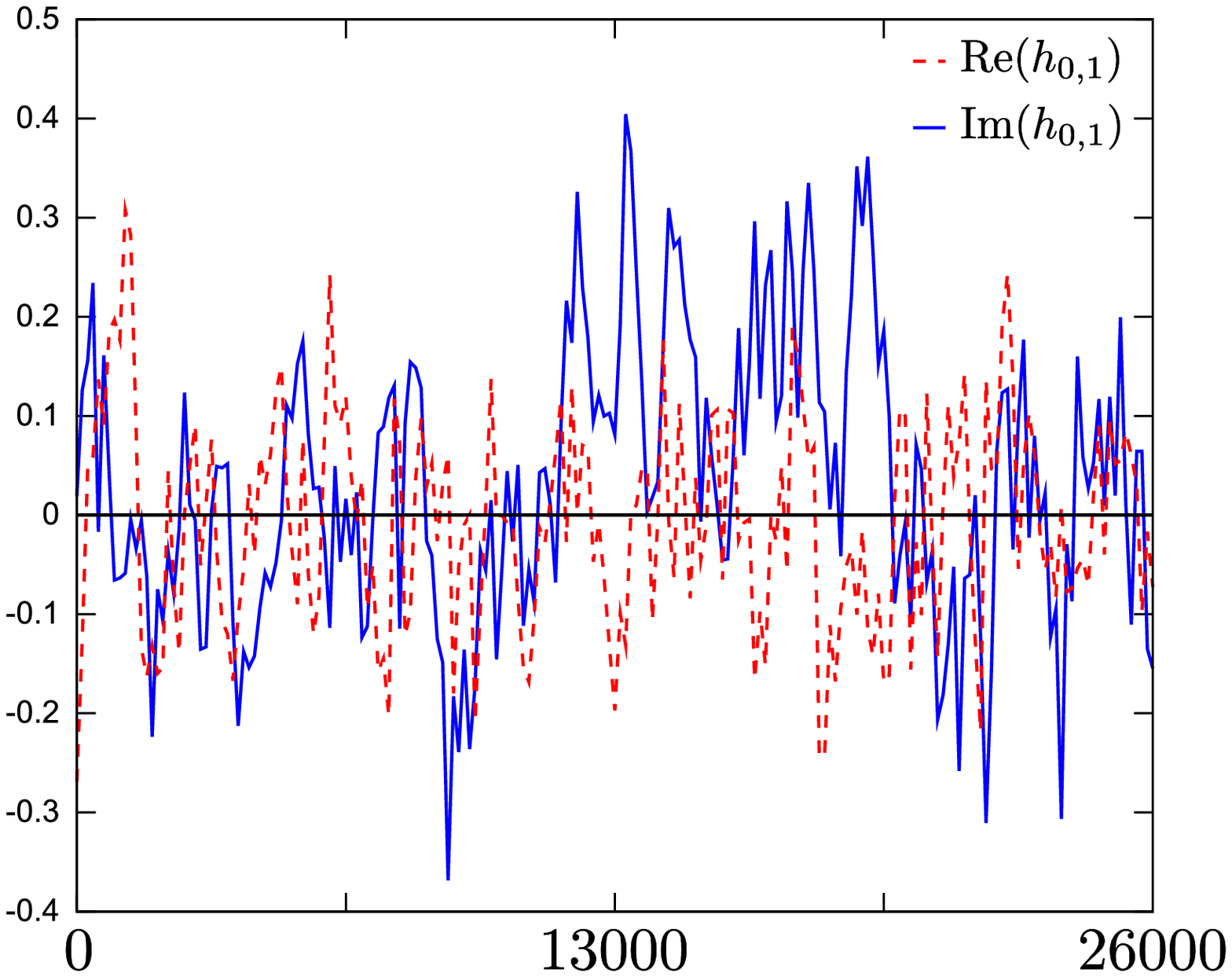}
}
\subfigure[Mode $h_{1,1}$.]{ 
  \includegraphics[width=0.3\columnwidth]{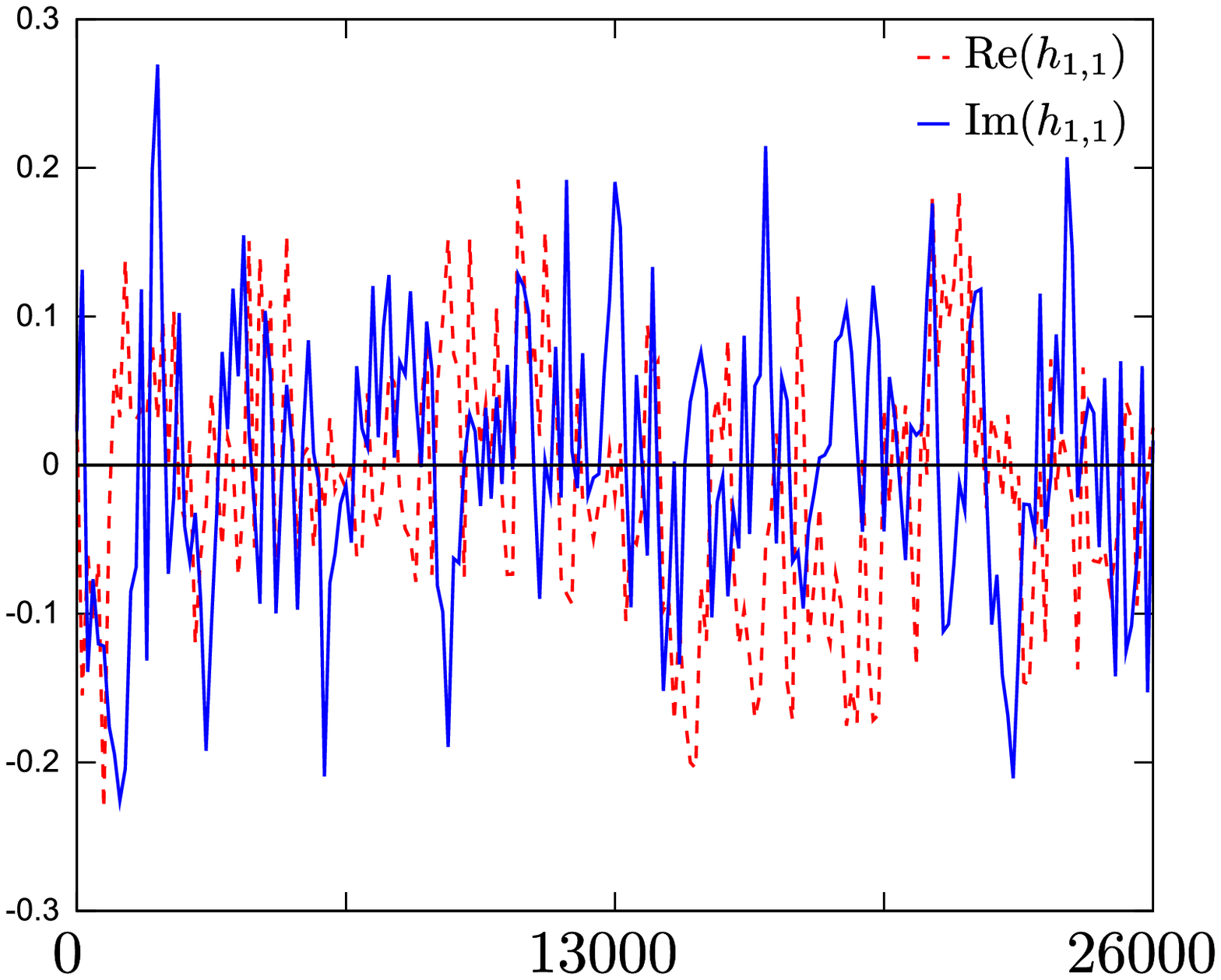}
}
\caption{Fourier coefficients as a function of the number of Monte Carlo steps.
  The parameters of the network are the same as before.
}
\label{fig_5_relax}
\end{center}      
\end{figure}

Finally, we have monitored the  evolution of longest Fourier modes $h_{1,0}$, $h_{0,1}$ and $h_{1,1}$ (in the next section we explain how we have obtained the Fourier decomposition).
Typical curves can be seen in Fig.~\ref{fig_5_relax}.
We can see that after $\approx 10^4$ steps, the coefficients are uncorrelated, which means that the longest modes have relaxed.
Accordingly, we have considered that in this case, $2 \times 10^{6}$ steps generated a sufficiently large sampling of the configuration space.

\section{Measuring tensions and the excess area}

In this section we describe how we measured the effective tension $\tau$ and the excess area $\alpha$.
We explain also the algorithm used to derive the fluctuation spectrum, from whose average we could derive $\kappa$ and the tension $r$.
Finally, we discuss the internal tension $\sigma$ in section~\ref{subsection_5_sigma}. 

\subsection{Excess area and mechanical tension measures}
\label{subsection_5_alpha}

In order to obtain $\tau$, we study the total force that the harmonic
potential given in eq.(\ref{eq_5_potrad}) exerts over the beads at the
network's boundary, which represents the force applied by the frame onto the membrane:

\begin{equation}
  f = - k_f \sum_i (R_i - R_f) \, ,
\end{equation}

\noindent where the sum runs over the beads at the network's edge and $R_i$ is
their distance to the center of the frame.
%As the system is in equilibrium, the membrane reacts with a force of the same intensity.
For $k_f$ large enough, the edge of the network fits well with the frame with
radius $R_f$ and thus the effective tension of a configuration is given by

\begin{equation}
  \tau_i = \frac{f}{2 \pi R_f} \, .
\end{equation}

\noindent During a run, $\tau_i$ was evaluated at the end of each Monte Carlo step.
At the end of it, we obtained $\tau = \langle \tau_i \rangle$ and its standard deviation $(\Delta \tau)^2 = \langle \tau_i^2 \rangle - \langle \tau_i \rangle^2$.

Concerning the excess area, we carefully updated the membrane's projected area
$A_p$ and actual area $A$ after each attempt of move.
At the end of each Monte Carlo step, the excess area of the configuration

\begin{equation}
  \alpha_i = \frac{A - A_p}{A_p} \,
\end{equation}

\noindent was added to a variable in order to obtain $\alpha = \langle \alpha_i \rangle$ in the end of the run.

\subsection{Fluctuation spectrum}
\label{subsection_5_spect}

Let's consider a square piece of membrane with lateral size $L$ weakly departing from a plane, whose shape is described in the Monge's gauge by $h(\bm{r})$.
In terms of Fourier modes, $h(\bm{r})$ can be written as

\begin{equation}
  h(\bm{r}) = \sum_{\bm{q}} h_{n,m} \, e^{i \, \bm{q} \cdot \bm{r}} \, ,
\end{equation}

\noindent with $\bm{r} = x \, \bm{e}_x + y \, \bm{e}_y$, $\bm{q} = 2 \pi/L \, (n,m)$, $n,m \in \mathbb{N}$ and

\begin{equation}
  \sum_{\bm{q}} \equiv \sum_{|n| \leq N_\mathrm{max}}\sum_{|m| \leq N_\mathrm{max}} \, ,
\end{equation}

\noindent where $N_\mathrm{max} = L/(2a)$ corresponds to the smallest possible wave length.
Note that here we have used a slightly different normalization from the rest of this work.
The coefficients $h_{n,m}$ are in general complex and $h_{-n,-m} = h_{n,m}^*$, where the superscript $*$ stands for the complex conjugate.
It is given by

\begin{equation}
  h_{n,m} = \frac{1}{L^2} \int_0^L \int_0^L h(\bm{r}) \, e^{- i \, \bm{q} \cdot \bm{r}} \, d\bm{r} \, .
  \label{eq_5_hnm}
\end{equation}

In section~\ref{grand_can}, we have seen that membranes connected to a lipid
reservoir could have their energy described by
the Helfrich Hamiltonian (eq.(\ref{Helfrich})). 
Accordingly, the average of the Fourier coefficients respects 

\begin{equation}
  \langle |h_{n,m}|^2 \rangle = \frac{1}{L^2} \, \frac{k_\mathrm{B} T}{r q^2 + \kappa q^4} \, ,
  \label{eq_5_prev}
\end{equation}

\noindent where $r$ is the macroscopic counterpart of the internal tension
$\sigma$ and $\kappa$ is the bending rigidity (in fact, as discussed before in
section~\ref{grand_can}, it corresponds more precisely to an effective
bending rigidity due to renormalization effects).
As in laboratory experiments, we would like to measure the fluctuation
spectrum of our numerical membrane in order to derive $r$ and $\kappa$.
In the following, we will explain how it was done. 

\subsubsection{Obtaining the fluctuation spectrum}

\begin{figure}[H]
\begin{center}
\includegraphics[width=0.4\columnwidth]{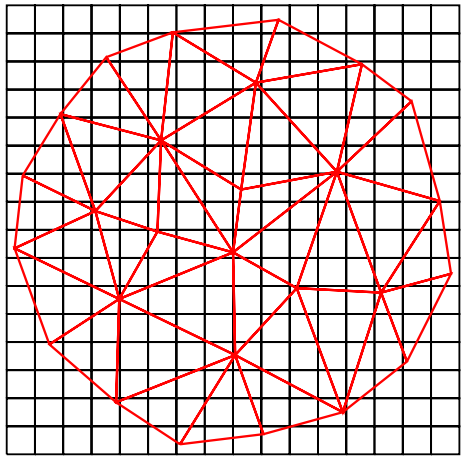}
\caption{Top view of a network with $N_x = 5$ and a total of $N = 27$ beads (red).
To evaluate the coefficients of the Fourier transform, we discretized the membrane over the grid shown in black ($N_\mathrm{grid} = 16$ here).}
\label{fig_5_grille}
\end{center}      
\end{figure}

For a general wave-vector $(n,m)$, we have to evaluate eq.(\ref{eq_5_hnm}) in order to obtain $h_{n,m}$.
The first numerical difficulty comes from the fact that instead of a continuous surface $h(\bm{r})$, we have access only to the position and height of the beads.
Consequently, the first step is to built an approximation to the network's surface by discretizing it over a square grid of $N_\mathrm{grid} \times N_\mathrm{grid}$ cells with lateral side $L$, as exemplified in Fig.~\ref{fig_5_grille}.
Each cell of has a lateral size $\Delta = L/N_\mathrm{grid}$.
We choose $L$ slightly bigger than $2 \, R_f$ to avoid problems with the discontinuities at the edges of the grid.

The discretized version of eq.(\ref{eq_5_hnm}), known as DFT (Discrete Fourier Transform) is given by

\begin{equation}
  h_{n,m} = \frac{1}{N_\mathrm{grid}^2} \sum_{\alpha = 0}^{N_\mathrm{grid} -1}\sum_{\beta = 0}^{N_\mathrm{grid}-1} h_{\alpha, \beta} \, e^{2 \pi i \, \frac{(n \alpha + m \beta)}{N_\mathrm{grid}}} \, ,
  \label{eq_5_DFT}
\end{equation}

\noindent where $h_{\alpha, \beta}$ is the height of the cell whose bottom left corner position is $\bm{r} = \Delta \times (\alpha, \beta)$.
At this point, we need to attribute a height to each cell of the grid, which is initially set to zero.
We do so in two steps:

\begin{itemize}
  \item First, we obtain the plane's equation for each triangle from the position of its three vertex.
    Using this equation, we evaluate the height of some points inside the triangle, as shown in Fig.~\ref{fig_5_dots}.

    \begin{figure}[H]
\begin{center}
\subfigure[Top view.]{
\includegraphics[width=0.4\columnwidth]{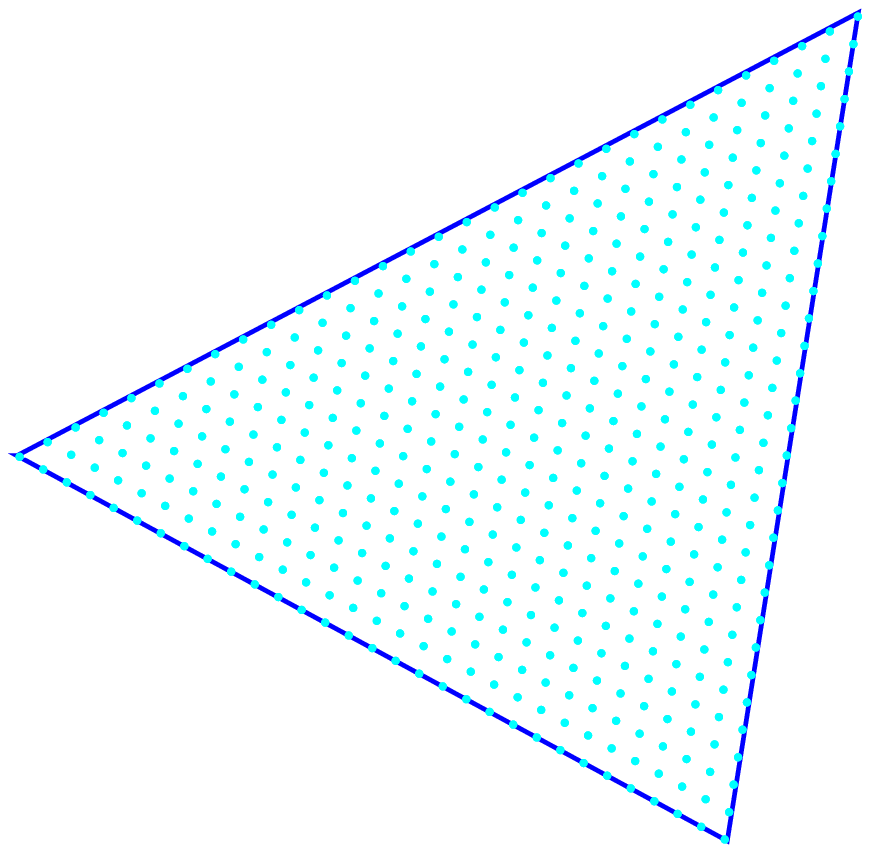}
}  
\subfigure[Side view.]{ 
  \includegraphics[width=0.4\columnwidth]{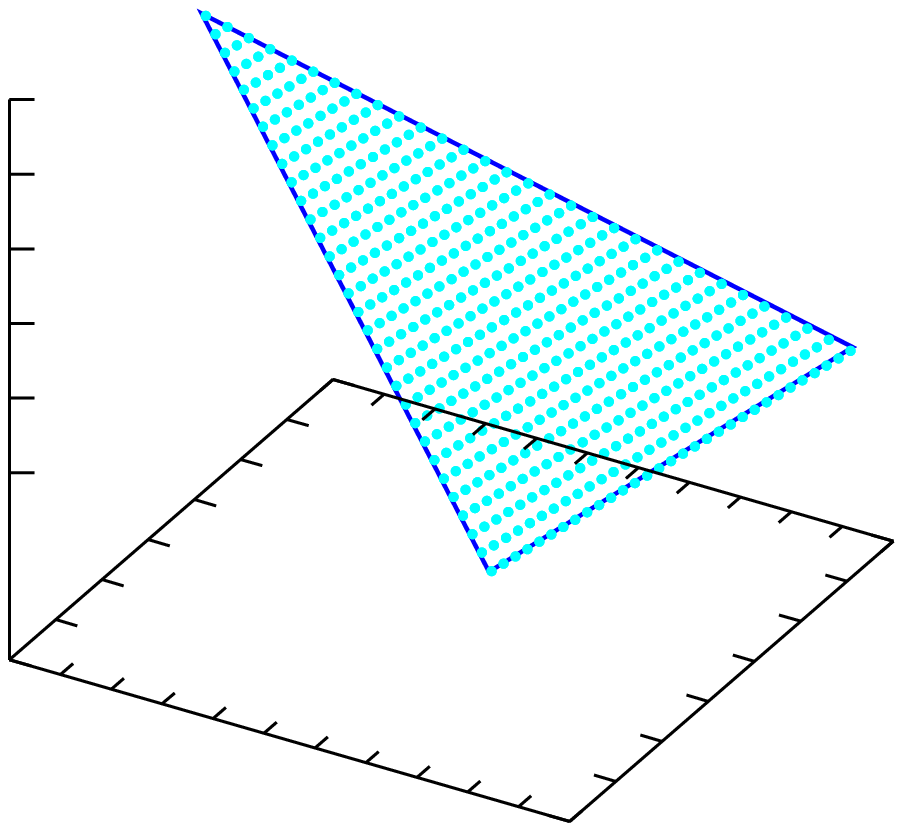}
}
\caption{Initially, we know only the position of the triangle's vertex.
  From them, the plane's equation is obtained and the height is evaluated over each dot on the triangle.
}
\label{fig_5_dots}
\end{center}      
\end{figure}
    
  \item Secondly, the cell that contains the projection of a dot receives its height.
    If the projection of more than one dot falls inside the same cell, we attribute the average of the their height to the cell (see Fig.~\ref{fig_5_dots_grid}).

\begin{figure}[H]
\begin{center}
\includegraphics[width=0.4\columnwidth]{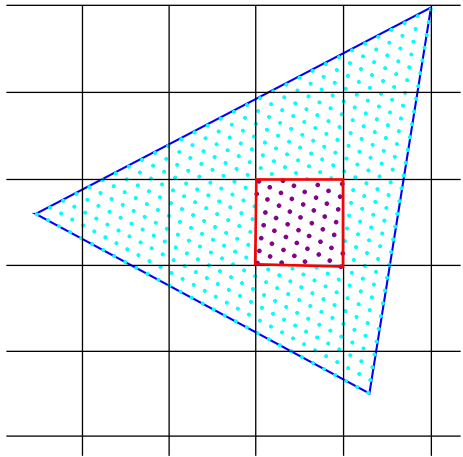}
\caption{To evaluate the height of the cell shown in red, we averaged the height of the dots whose projection are shown in violet.}
\label{fig_5_dots_grid}
\end{center}      
\end{figure}
    
\end{itemize}

\begin{figure}[H]
\begin{center}
\includegraphics[width=0.7\columnwidth]{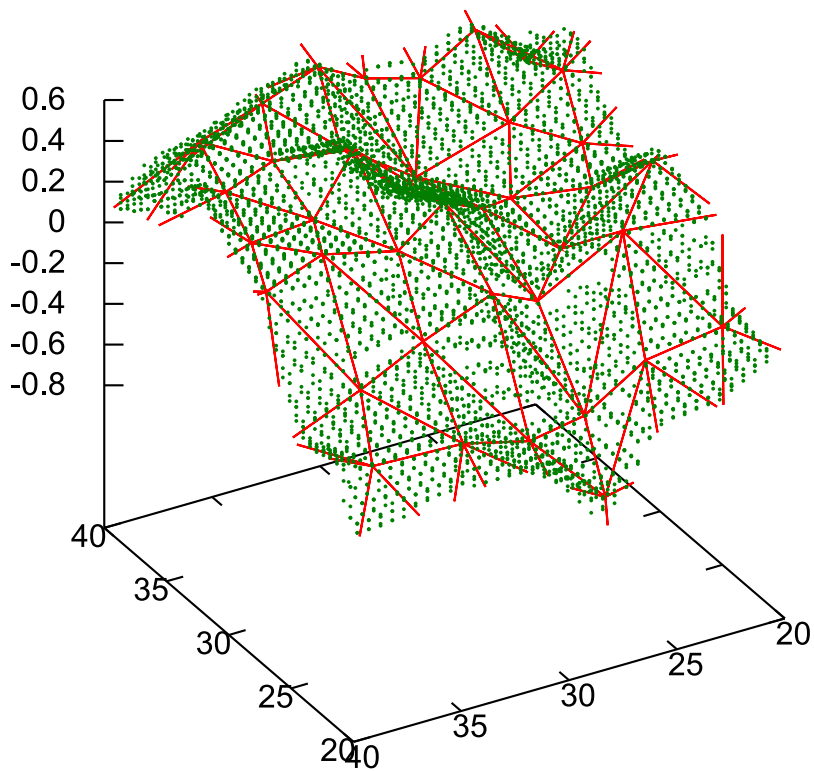}
\caption{Piece of a network with $N_x=10$ ($N = 410$ beads) and $N_\mathrm{grid} = 128$, shown in red.
The green dots correspond to the height of each cell of the grid, obtained through the method explained above.
Remark the good quality of the approximation procedure.
Note that the vertical scale is different from the horizontal.}
\label{fig_5_dots_grid1}
\end{center}      
\end{figure}

Fig.~\ref{fig_5_dots_grid1} shows an example to testify of the quality of our approximation. 

Once the approximative grid is built, we can evaluate eq.(\ref{eq_5_DFT})
which has a great advantage: it can be evaluated using the FFT (Fast Fourier Transform) algorithm with a $N_\mathrm{grid}^2 \log(N_\mathrm{grid})$ complexity, instead of a $N_\mathrm{grid}^4$ complexity for naive algorithms.
We used thus the cdft (complex discrete Fourier transform) routine of the FFT library implemented by Takuya Ooura~\cite{Ooura}, which is a general library to evaluate FFT under the condition that $N_\mathrm{grid}$ is a power of $2$.

\subsubsection{A subtlety}

The prediction given in eq.(\ref{eq_5_prev}) is valid for a squared piece of membrane with lateral size $L$.
Since our membrane is round, our situation corresponds to a squared membrane seen through a circular mask, given by 

\begin{equation}
  \mathrm{circ}(\bm{r}) = \left\{ \begin{array}{ccccc} 1 \,\,\,& $for$ &\, r < R_f\, ,\\
    \\
    0 \,\,\,& $for$ &\, r > R_f \ .\end{array} \right.
  \label{eq_5_circ}
\end{equation}

\noindent So, we are actually performing numerically the Fourier transform of the function $h(\bm{r})$ that denotes the height of the membrane multiplied by $\mathrm{circ}(\bm{r})$, instead of just performing the Fourier transform of $h(\bm{r})$.
Indicating the Fourier transform by a superscript $\, \hat{} \, $, we recall the convolution theorem:  

\begin{equation}
  \widehat{(h \, \, \mathrm{circ})} = \hat{h} * \widehat{\mathrm{circ}} \, ,
  \label{eq_5_conv}
\end{equation}

\noindent where $*$ indicates the convolution between the two functions.

\begin{figure}[H]
\begin{center}
\includegraphics[width=0.6\columnwidth]{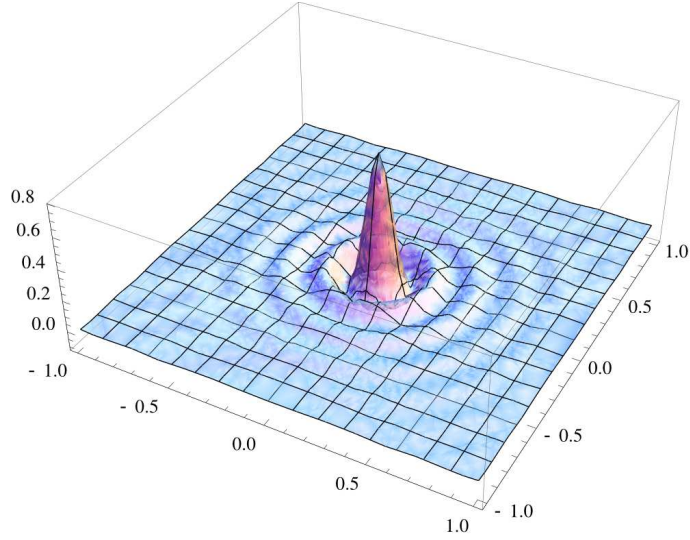}
\caption{Plot of $\widehat{\mathrm{circ}}$ as a function of the wave-vector $q$ for $R_f = 32 \, \sigma_0$.
Note that the function presents a very marked pike.}
\label{fig_5_circ}
\end{center}      
\end{figure}

In order to obtain $\hat{h}$, we will evaluate the Fourier transform of $\mathrm{circ}(\bm{r})$.
Using the above presented definition, we have~\cite{Osgood}

\begin{eqnarray}
\widehat{\mathrm{circ}} &=& \frac{1}{L^2} \int_0^L \int_0^L \mathrm{circ}(\bm{r}) \, e^{- i \, \bm{q} \cdot \bm{r}} \, d\bm{r} \, ,\nonumber\\
&=& \frac{1}{L^2} \int_0^{R_f} r \, \mathrm{circ}(\bm{r}) \left( \int_0^{2\pi} e^{-i \, \bm{q} \cdot \bm{r}} d\theta \right) \, dr \, , \nonumber\\
&=& \frac{2\pi}{L^2} \int_0^{R_f} r \, J_0(q\,r)\, dr \, , \nonumber\\
&\simeq& \frac{\pi}{2 \, R_f} \, \frac{J_1(q \, R_f)}{q}\, ,
\end{eqnarray}

\noindent where $J_i$ is the Bessel function of order $i$.
In the last passage, we have used the fact that $L \approx 2 \, R_f$.
This function has a very marked pike, as shown in Fig.~\ref{fig_5_circ}.

From eq.(\ref{eq_5_conv}), since $\widehat{\mathrm{circ}}$ is so piked, we have 

\begin{eqnarray}
  \widehat{(h \, \mathrm{circ})} &\simeq& \hat{h} \times \widehat{\mathrm{circ}}(0) \, , \nonumber\\
  &\simeq& \hat{h} \times \frac{\pi}{4} \, .
\end{eqnarray}

\noindent Finally, to obtain the Fourier transform of $h(\bm{r})$, we have to multiply the Fourier coefficients $\tilde{h}_{n,m}$ obtained numerically (considering the mask) by $4/\pi$:

\begin{equation}
  h_{n,m} = \frac{4}{\pi} \, \tilde{h}_{n,m} \, .
\end{equation}

\noindent In the following, we will keep the notation $\tilde{h}_{n,m}$ for the coefficients obtained with the mask.

\subsubsection{During a run}

As the process of grid construction is relatively computer consuming, the
fluctuation spectrum was measured over $N_{\mathrm{spec}}$ configurations
uniformly spaced during a run.
At each time, we obtained the $\mathrm{Re}(\tilde{h}_{n,m})$ and $\mathrm{Im}(\tilde{h}_{n,m})$, i. e., the real and complex parts of each Fourier coefficient, with $|n| \leq N_\mathrm{grid}$ and $|m| \leq N_\mathrm{grid}$.
In the end of the run, we evaluated $\langle |\tilde{h}_{n,m}|^2 \rangle = \langle \mathrm{Re}(\tilde{h}_{n,m})^2 \rangle + \langle \mathrm{Im}(\tilde{h}_{n,m})^2 \rangle$.

We plotted then $1/(q^2 L^2 \langle |h_{n,m}|^2 \rangle)$ as a function of $q^2$, with $q^2 = (4 \pi^2)/(L^2) \, (n^2 + m^2)$ and $h_{nm} = 4/\pi \, \tilde{h}_{n,m}$.
From eq.(\ref{eq_5_hnm}), we expect, at least for large wave-lengths, a linear
relation between these quantities:
from the $y$-intercept of the curve, we derive $r$, while from it's slope we obtain $\kappa$.
In Fig.~\ref{fig_5_spectro}, we show an example such a plot with a linear fit to the region of large wave-lengths.

\subsection[{Internal tension ${\sigma}$}]{Internal tension $\bm{\sigma}$}
\label{subsection_5_sigma}

As discussed in section~\ref{sig_intro}, the internal tension $\sigma$ is the energetic cost associated to an unitary increase in the microscopical area $A$ of the membrane.
From the energy $E$ of a system, it can be obtained through:

\begin{equation}
  \sigma = \left(\frac{\partial E}{\partial A}\right) \, .
  \label{eq_5_def}
\end{equation}

Let's consider a general triangle in the bulk of our meshwork.
In an approximation, let's consider that the triangle is equilateral and has $\ell$ as lateral size.
From eq.(\ref{eq_5_bond}), the local energy $E_\mathrm{tri}$ associated to this triangle is given by

\begin{equation}
  E_\mathrm{tri} = 3 \times \frac{1}{2} \times s \, \frac{(\ell - \ell_0)^2}{2} + E_\mathrm{curv}\, ,
\end{equation}

\noindent where $E_\mathrm{curv}$ is a contribution coming from the bending rigidity.
The factor $3$ comes from the three sides of the triangle, while the term $1/2$ comes from the fact that each side is shared by two adjacent triangles. 
Note that the first term is the only contribution involving the bead-to-bead distance $\ell$.
Under the assumption that the triangle is equilateral, its area is given by

\begin{equation}
  A_\mathrm{tri} = \frac{\sqrt{3}}{4} \, \ell^2 \, .
\end{equation}

\noindent From eq.(\ref{eq_5_def}) and under the assumption of an equilateral triangle, we can define a local internal tension:

\begin{eqnarray}
  \sigma_\mathrm{loc} &=& \left(\frac{\partial E_\mathrm{tri}}{\partial A_\mathrm{tri}}\right) \, ,\nonumber\\
  &=& \left(\frac{\partial E_\mathrm{tri}}{\partial \ell}\right) \, \left(\frac{\partial A_\mathrm{tri}}{\partial \ell}\right)^{-1} \, ,\nonumber\\
  &=& \sqrt{3} \, s \, \frac{\ell - \ell_0}{\ell} \, .
  \label{eq_5_sigma_loc}
\end{eqnarray}

Now, in our simulation, the sides of all triangles are submitted to the same harmonic potential given in eq.(\ref{eq_5_bond}).
Accordingly, the hypothesis that each triangle is in average equilateral is very reasonable.
Moreover, as the system is spatially uniform, we propose thus a generalization of eq.(\ref{eq_5_sigma_loc}) as a estimate of the internal tension:

\begin{equation}
  \sigma = \sqrt{3} \, s \, \frac{\langle\bar{\ell}\rangle - \ell_0}{\langle \bar{\ell} \rangle} \, ,
\end{equation}

\noindent where the bar over $\ell$ indicates the spatial average of the bead-to-bead distance, while $\langle \rangle$ indicates as usually the average over an ensemble of configurations.

In practice, we have kept track of the average length of the bonds over the network $\bar{\ell}$ at each Monte Carlo step.
At the end of the run, we could thus evaluate the average of $\bar{\ell}$ over a the ensemble of configurations to obtain $\sigma$.

%We could not measure $\sigma$, but we know that it is related to the
%tension of the bonds: indeed, the bigger the frame's radius, the longer the
%average length of a bond and the higher the internal tension.

\section{Some first results}
\label{section_5_prim}

The results presented here consist of a preliminary set of runs for a network with $N_x = 10$, with a total of $N = 410$ beads.
We have kept the parameters $\beta \, k = 5$, $\beta \, \sigma_0^2 \,  k_f =
10$ and $\beta \, \sigma_0^2 \, s = 1$.
As discussed in section~\ref{subsection_5_equilibration}, we let the system evolve during $N_\mathrm{neg} = 10^4$ steps in order to assure that the final frame shape had been attained.
The averages were made in a second time, over $N_\mathrm{iter} = 2 \times 10^6$ steps during which $1500$ spectra were evaluated.

We performed fifteen runs with these parameters, increasing at each run the
membrane's tension by widening the frame's radius: the initial radius $R_f = 32.34 \, \sigma_0$ was successively increased of $0.35 \, \sigma_0$ up to $R_f = 37.39 \, \sigma_0$.
As the radius increased, the excess area decreased from $\approx 3.3 \%$ to $\approx 2.4 \%$.

\begin{figure}[H]
\begin{center}
\includegraphics[width=0.55\columnwidth]{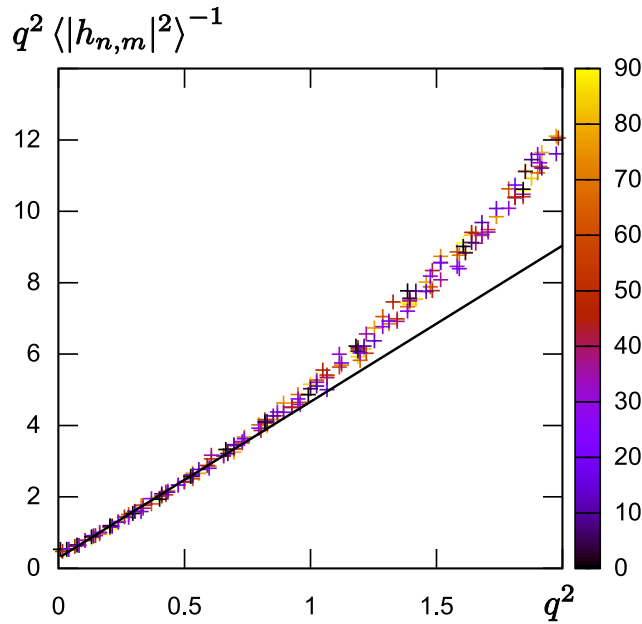}
\caption{Deducing $r$ and $\kappa$:
  the colored dots represent the inverse of the average intensity of a mode plotted as a function of the squared wave vector.
  The colored scale represents the angle that the wave-vector does with the
  horizontal direction of the grid.
  Remark that there is no color pattern, which is an evidence of the system's isotropy.
  The line represents the linear fit for large wave lengths ($\lambda > 9 \,
  \sigma_0$, corresponding to $q^2 < 0.5$) from which $r$ and $\kappa$ are deduced.
}
\label{fig_5_spectro}
\end{center}      
\end{figure}

For each run, we have plotted the fluctuation spectrum as detailed in section~\ref{subsection_5_spect} to obtain $r$ and $\kappa$.
A typical example is shown in Fig.~\ref{fig_5_spectro}, where we have colored
the points in function of the angle that the wave-vector associated to the
mode did with
the horizontal direction of the grid.
First, we can remark that there is no clear color pattern, which indicates that the membrane is indeed isotropic.
Second, we can see that in the region of large wave-lengths, the dots are well-fitted by a linear curve, from which we deduce $r$ and $\kappa$.

\begin{figure}[H]
\begin{center}
\subfigure[Bending rigidity.]{ 
  \includegraphics[width=0.45\columnwidth]{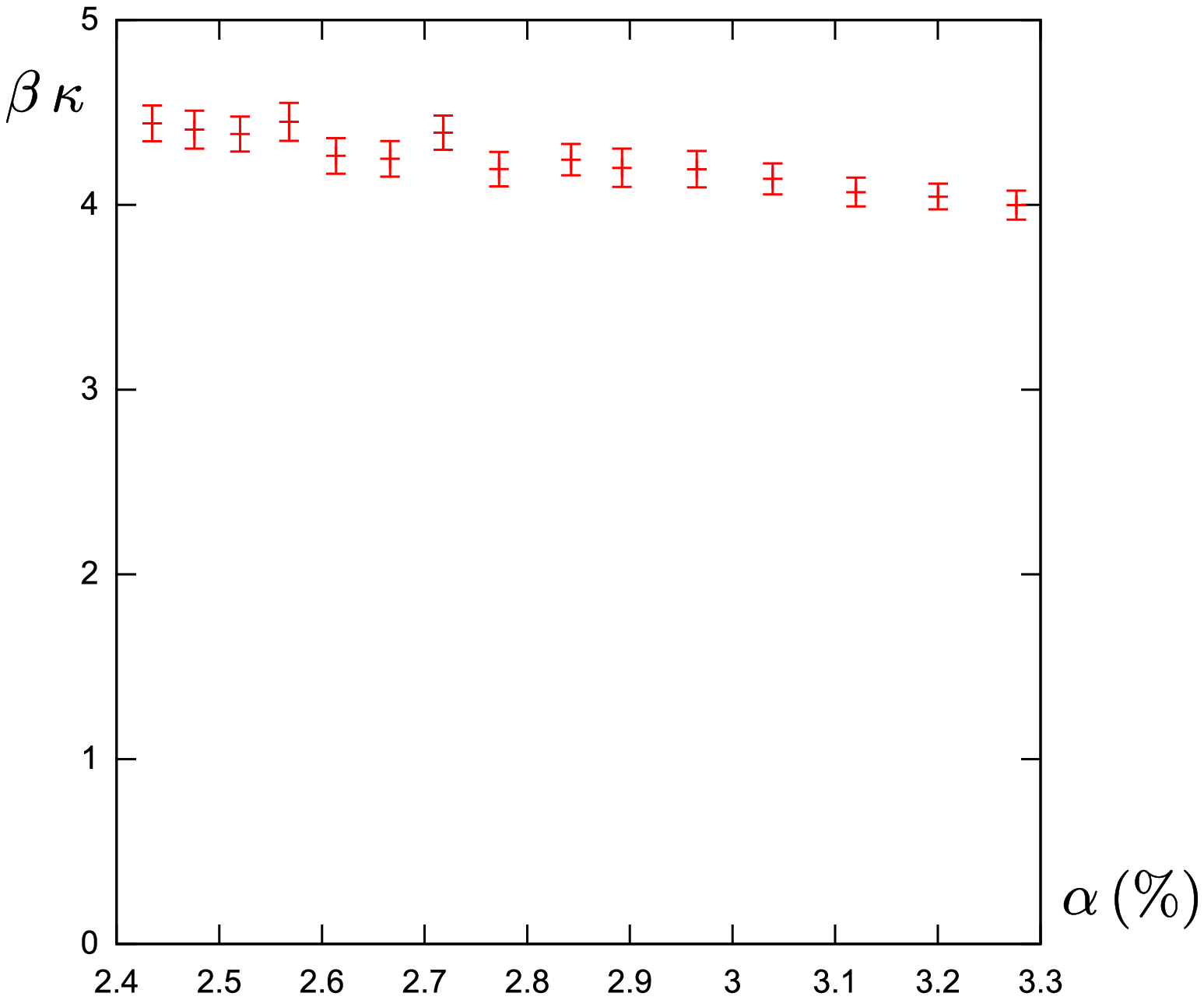}
  \label{fig_5_kappa}
}
  \subfigure[Tensions.]{
\includegraphics[width=0.45\columnwidth]{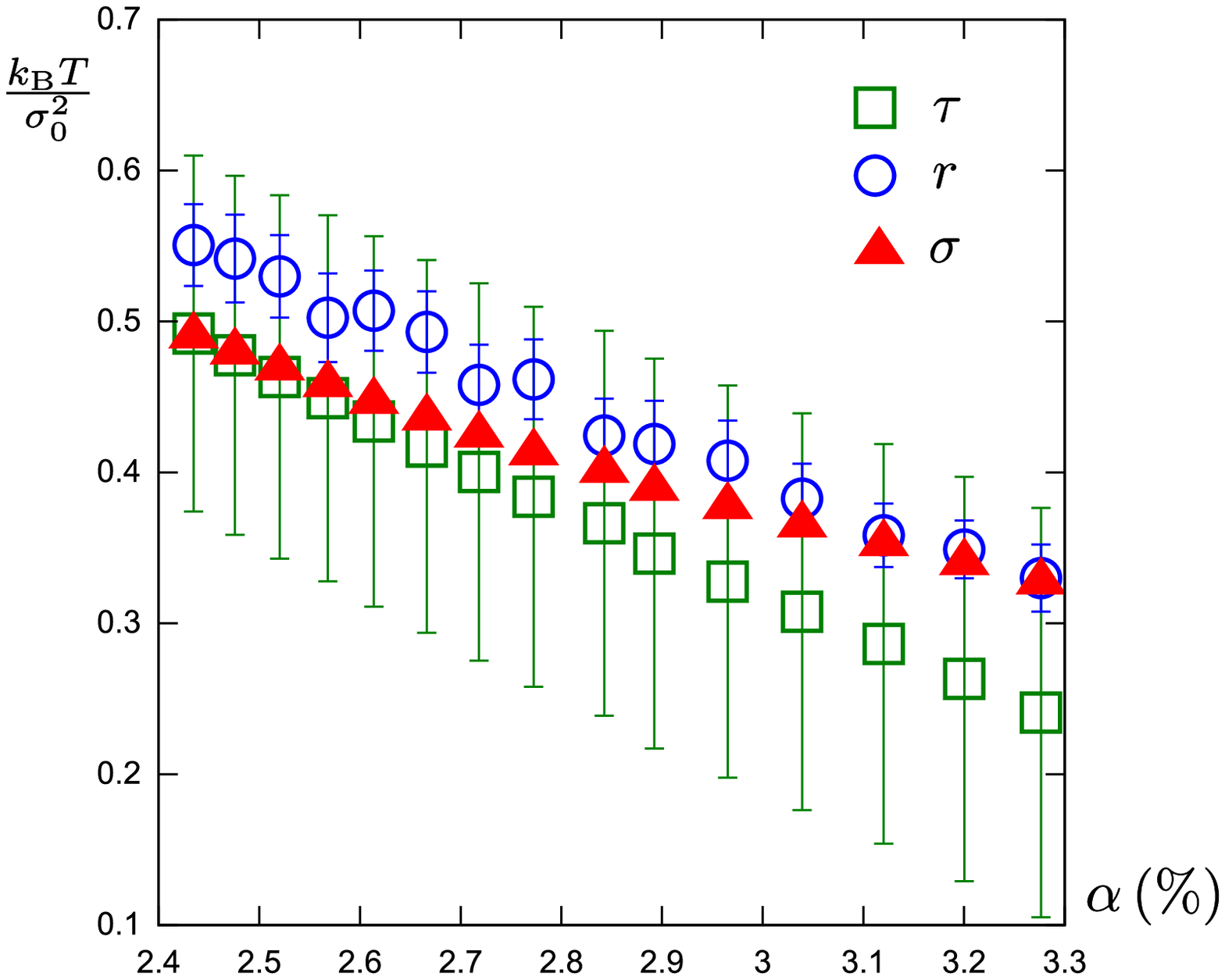}
\label{fig_5_tensions}
}  
\caption{At left: values of $\kappa$ in units of $k_\mathrm{B}T$ obtained by
  the linear fit shown in Fig.~\ref{fig_5_spectro} for different values of
  excess area.
  At right: measured tensions as a function of the excess area.
}
\label{fig_5_init}
\end{center}      
\end{figure}

We plot the results obtained for each run as a function of the corresponding
excess area in Fig.~\ref{fig_5_init} (we remind that the smaller the excess
area, the bigger the frame's radius).
In Fig.~\ref{fig_5_kappa}, we note a small dependence of $\kappa$ as a function of the excess area.
In Fig.~\ref{fig_5_tensions}, we compare the values of $\tau$, $\sigma$ and $r$: the three decrease as the excess area increases, as expected.
As we predict, we have always $\sigma$ bigger than $\tau$ and their difference is bigger for small tensions.
Concerning the renormalized tension $r$, we find values not very different from $\tau$ and $\sigma$, which is reassuring. 
The tension fluctuation $\Delta \tau$, represented by red bars, seems almost
constant, which agrees at least qualitatively with the predictions of chapter~\ref{Fluct_plan}.

At this point, we have good indications that the our network mimics well a liquid membrane under the Helfrich Hamiltonian.
In the next section, we will test quantitatively the compatibility of these results with our theoretical predictions.

\subsection[{Difference between ${\tau}$, ${\sigma}$ and ${r}$ and our predictions}]{Difference between $\bm{\tau}$, $\bm{\sigma}$ and $\bm{r}$ and our predictions}
\label{subsection_5_diff}

In chapter~\ref{chapitre/planar_membrane}, we predicted that

\begin{equation}
  \tau = \sigma - \frac{k_\mathrm{B}T \, \Lambda^2}{8 \pi}\left[1 - \frac{\sigma}{\sigma_r}\ln\left( 1 + \frac{\sigma_r}{\sigma}\right)\right]\, ,
  \label{eq_5_diff}
\end{equation}

\noindent where $\Lambda$ is the bigger wave-vector possible and $\sigma_r = \kappa \Lambda^2$.

\begin{figure}[H]
\begin{center}
\subfigure[Tension $\tau$.]{
\includegraphics[width=0.44\columnwidth]{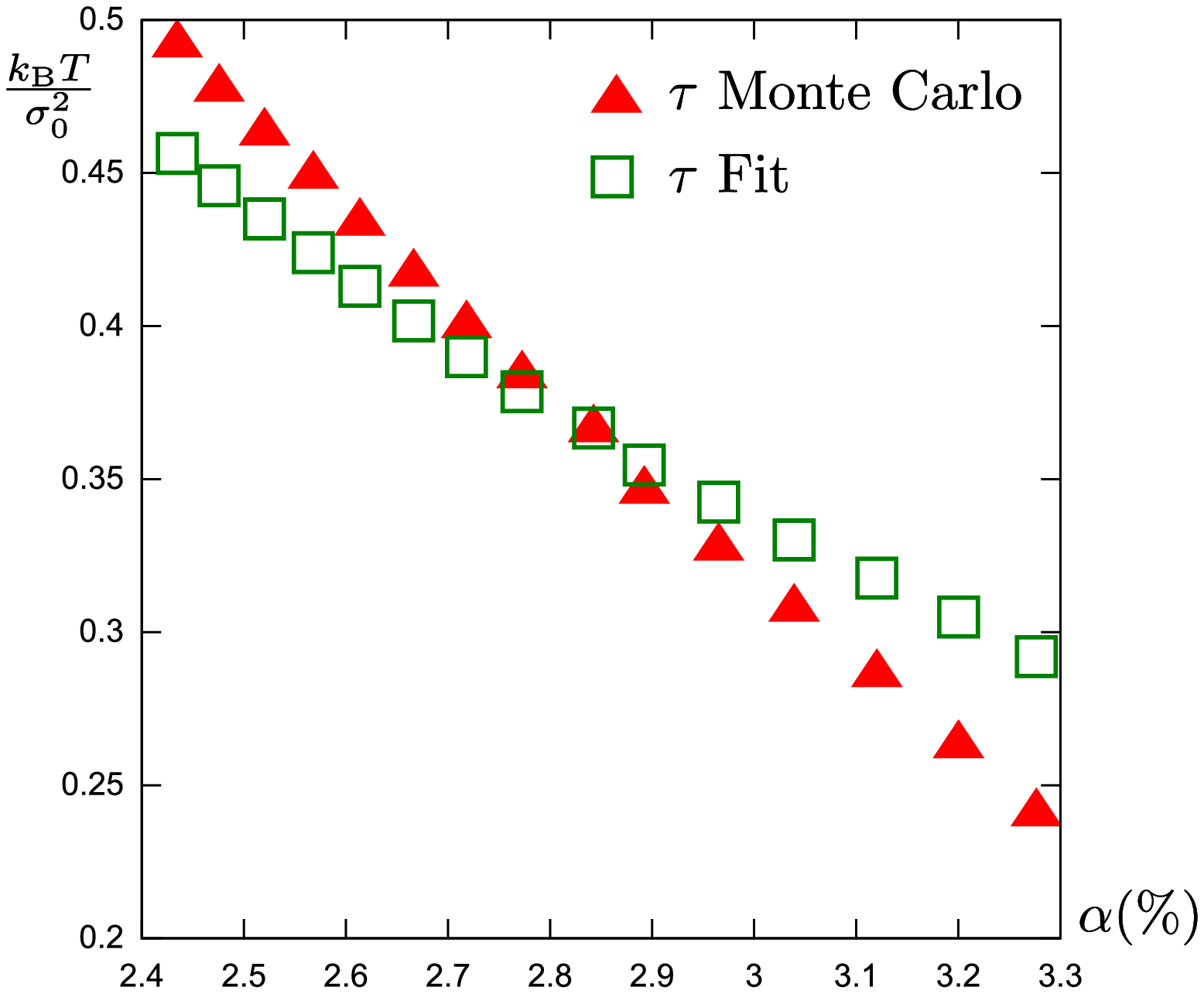}
\label{fig_5_Lambda_fit}
}  
\subfigure[Excess area.]{ 
  \includegraphics[width=0.44\columnwidth]{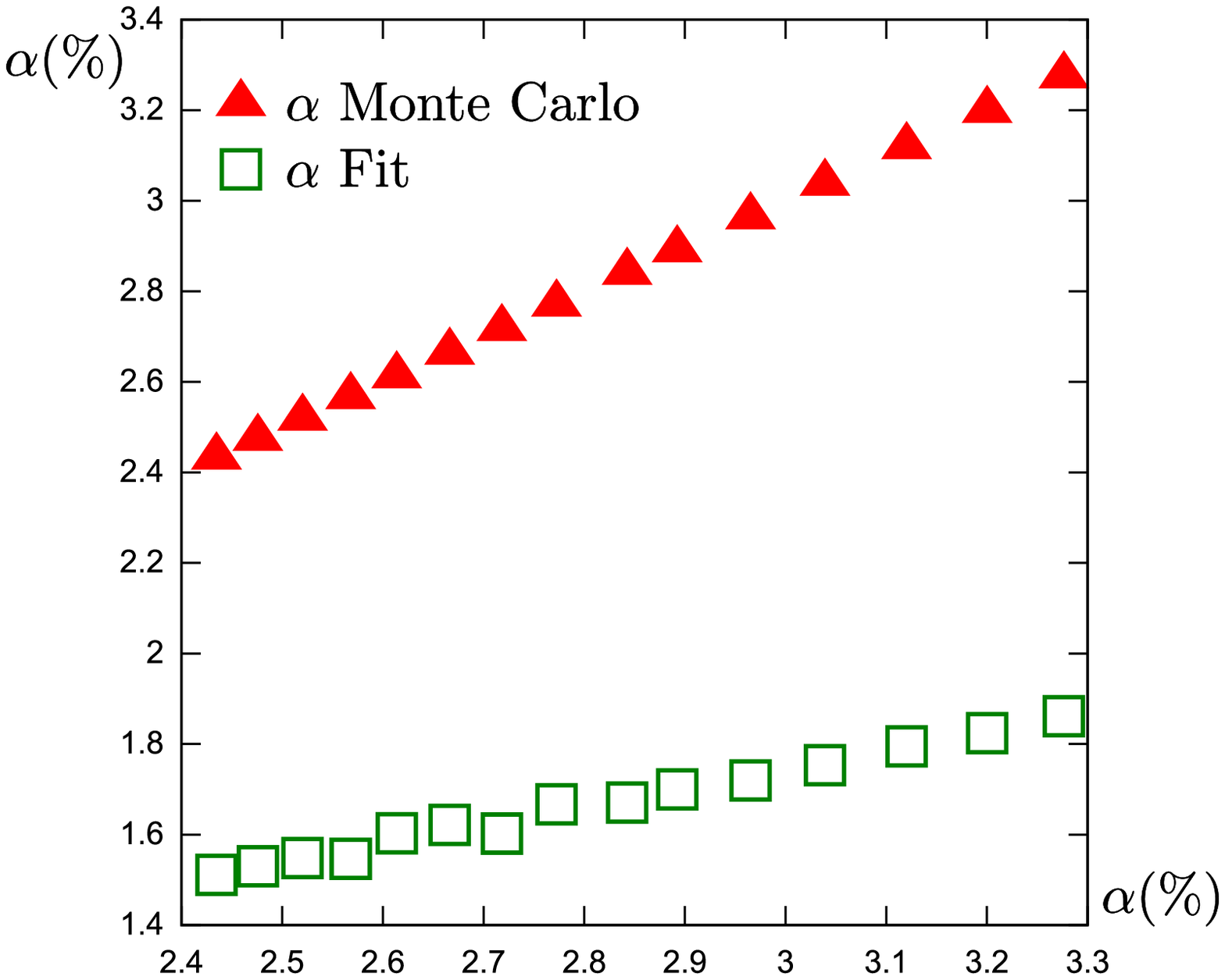}
  \label{fig_5_alpha}
}
\caption{
  At left, we show with the red triangles the measured values of $\tau$ in units of $k_\mathrm{B} T/\sigma_0^2$.
  We have fitted eq.(\ref{eq_5_diff}) to the red triangles by adjusting $\Lambda$.
  The best result, obtained for $\Lambda = 1.03 \, \sigma_0^{-1}$, is shown with green squares.
  At right, we applied this value of $\Lambda$ to eq.(\ref{eq_5_alpha_theo})
  and obtained the predicted excess area (green squares), which are clearly incompatible with the measured excess area (red triangles).
}
\label{fig_5_Fit}
\end{center}      
\end{figure}

In this section, we would like to verify quantitatively the compatibility of eq.(\ref{eq_5_diff}) with the data of the last section.
We made a one-variable fit of eq.(\ref{eq_5_diff}) by adjusting the value of $\Lambda$.
The best fit, for $\Lambda = 1.03 \, \sigma_0^{-1}$, is shown in Fig.~\ref{fig_5_Lambda_fit}. 
The compatibility between the data obtained from the simulation and the predicted values is relatively poor.

In chapter~\ref{chapitre/planar_membrane}, we have also predicted the dependence of the excess area on the tension $\sigma$:

\begin{equation}
\alpha = \frac{k_\mathrm{B} T}{8 \pi \kappa} \ln\left(\frac{\sigma_r}{\sigma}\right) \, .
\label{eq_5_alpha_theo}
\end{equation}

\noindent Both eqs.(\ref{eq_5_diff}) and (\ref{eq_5_alpha_theo}) should be valid under the same conditions.
Accordingly, we decided to make a self-consistency test by plotting the predicted values of the excess area obtained through eq.(\ref{eq_5_alpha_theo}), with the $\Lambda$ obtained above.
We observe that the predicted values for the excess area are consistently smaller than the values measured during the simulation.

Up to now, we have not a clear explanation for these results.
Two hypothesis deserve further attention:
\begin{itemize}
\item our membrane is not exactly very tense, since it is very easy to stretch the bonds.
  Indeed, stretching a bond to its maximum costs only $\sim 0.25 k_\mathrm{B} T$ for $\beta \, \sigma_0^2 \, s = 1 $.
  It is thus possible that the simulated membrane is not under the hypothesis of our theory.
  We could thus imagine further tests with a higher $s$, but in this case, as discussed earlier, the membrane would loose its liquidity.
  %Moreover, in Fig.~\ref{fig_5_tensions}, we see that the average bond tension has a slightly different dependence on the excess area as $r$.
\item the projected area of the membrane could be bigger, which would explain
  why the measured excess area is consistently bigger than the predicted.
  When we proposed the equilibration criteria in
  section~\ref{subsection_5_equilibration}, we studied the average shape of
  the membrane, as shown in Fig.~\ref{fig_5_shape_moy}.
  We have not however excluded the possibility of a rotating deformed shape:
  in this case, we would still have an average shape nearly flat, but the
  membrane would actually fluctuate around this deformed shape, with a
  projected area bigger than if it fluctuated effectively around a plane.
  To verify that, we should perform a rotation to align the configurations
  before evaluating the average shape by aligning the direction of the maximum
  height at each step, for instance.
  This would however not explain the poor fit shown in Fig.~\ref{fig_5_Lambda_fit}.
\end{itemize}

\section{Extraction of tubes}
\label{section_5_tube}

\begin{figure}[H]
\begin{center}
\subfigure[$h_0 = 10 \, \sigma_0$.]{
\includegraphics[width=0.45\columnwidth]{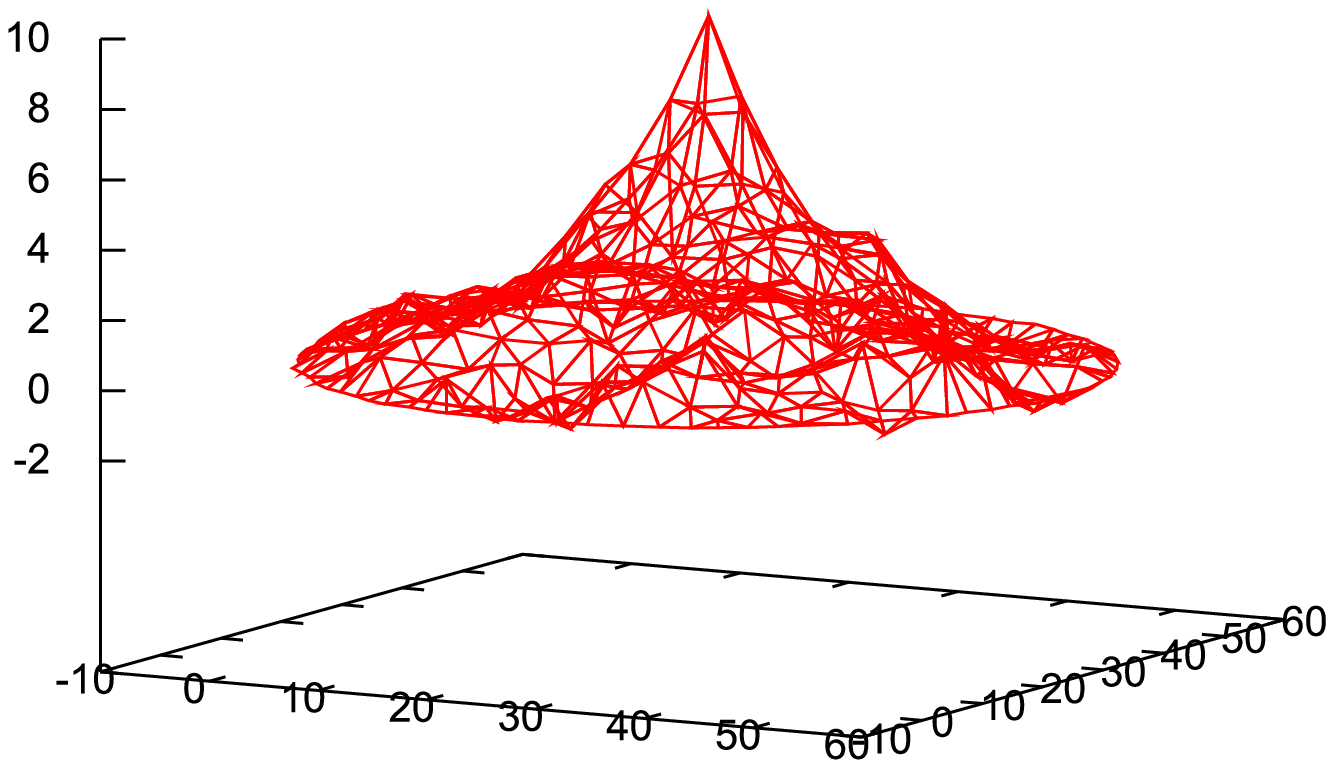}
}  
\subfigure[$h_0=15 \, \sigma_0$.]{ 
  \includegraphics[width=0.45\columnwidth]{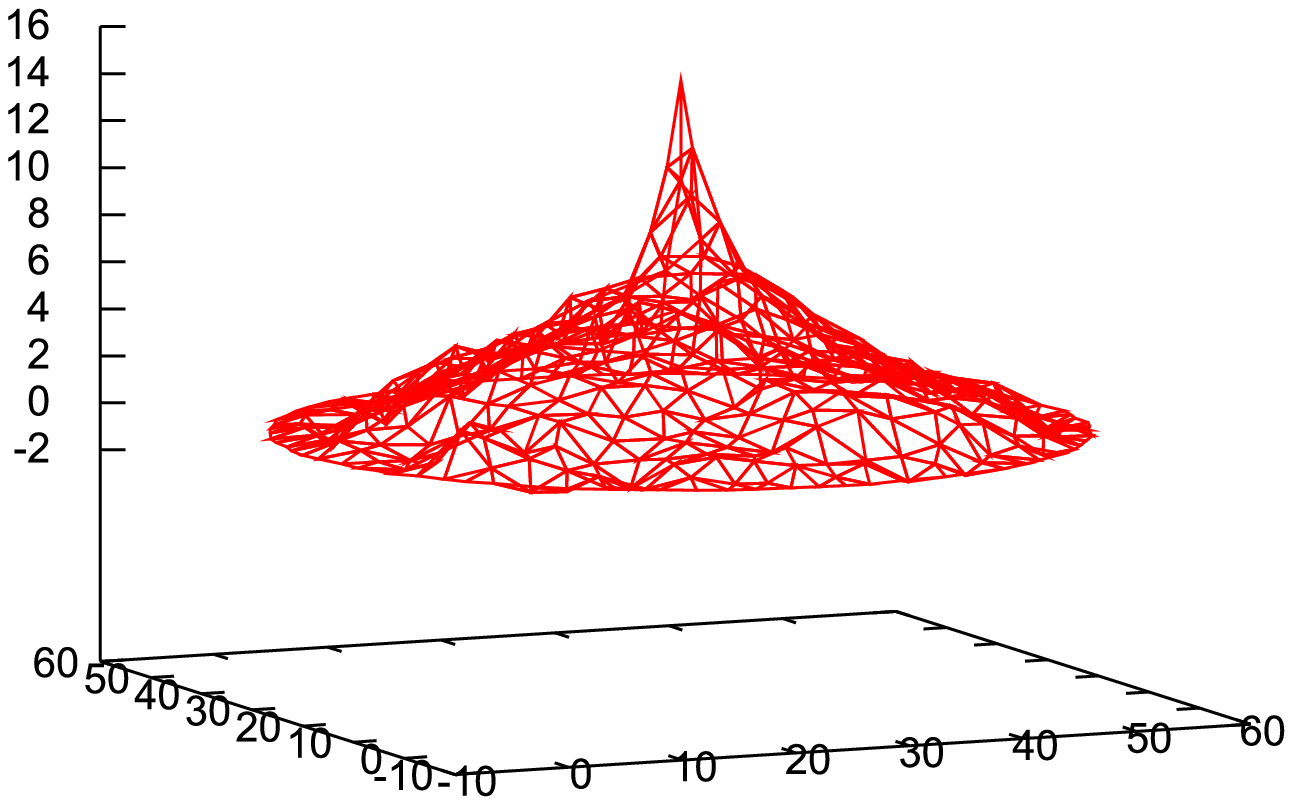}
}
\caption{Tube pulled from a membrane for two different values of $h_0$.
  At right, remark that the angle between the triangles is very important in the protruded region.
  The vertical and horizontal scales are different.
}
\label{fig_5_tube_ant}
\end{center}      
\end{figure}

In parallel with the studies on the membrane tension, we have explored the possibility of extracting tubes from our simplified membrane.
To pull a tube, we have applied a harmonic potential

\begin{equation}
  V_\mathrm{tube} = k_\mathrm{tube} \, \frac{(h - h_0)^2}{2} \, ,
\end{equation}

\noindent to a central bead whose height is denoted by $h$.
The preferred height of the tube is defined by the choice of $h_0$.
As in the case of the frame's force, we could obtain the force applied to pull the tube, as well as its fluctuation.
The first results for the same parameters of the last section, with $R_f =
33.04 \, \sigma_0$, are shown in Fig.~\ref{fig_5_tube_ant}.

\begin{figure}[H]
\begin{center}
\subfigure[$h_0 = 20 \, \sigma_0$ with eq.(\ref{eq_5_discret_ant}).]{
\includegraphics[width=0.45\columnwidth]{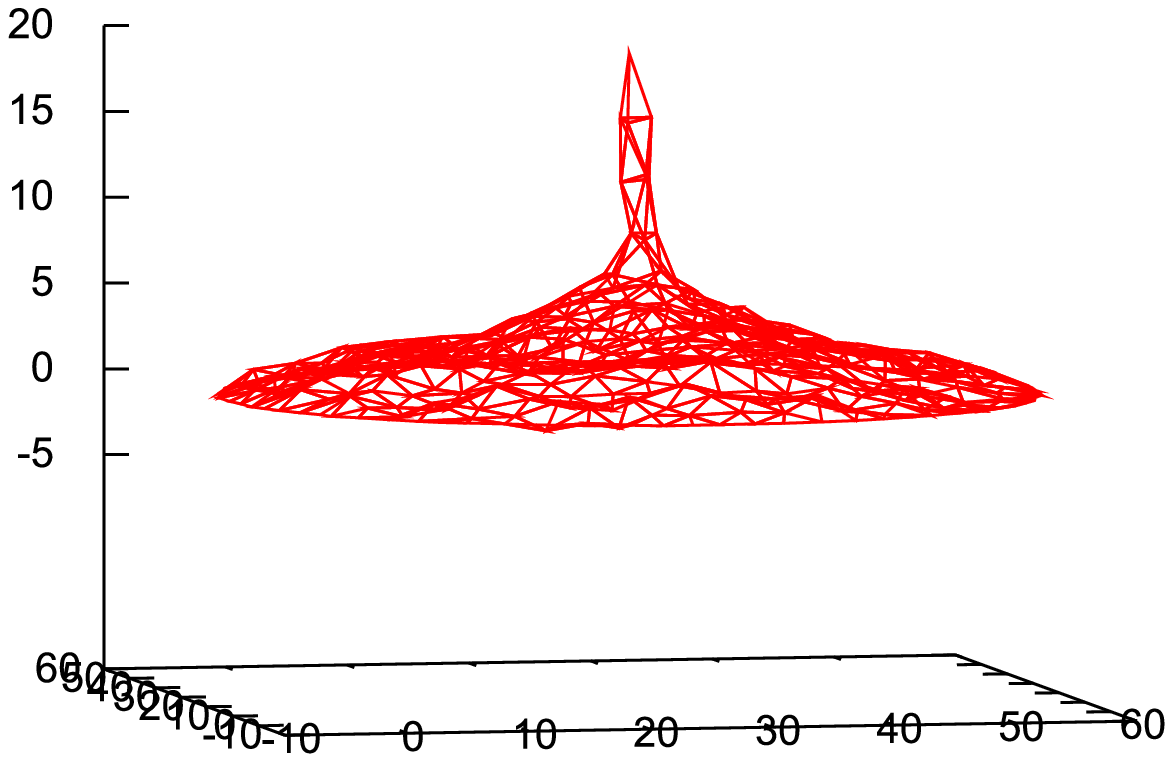}
\label{fig_5_20_feio}
}  
\subfigure[$h_0 = 20 \, \sigma_0$ with eq.(\ref{eq_5_discret_new}).]{ 
  \includegraphics[width=0.45\columnwidth]{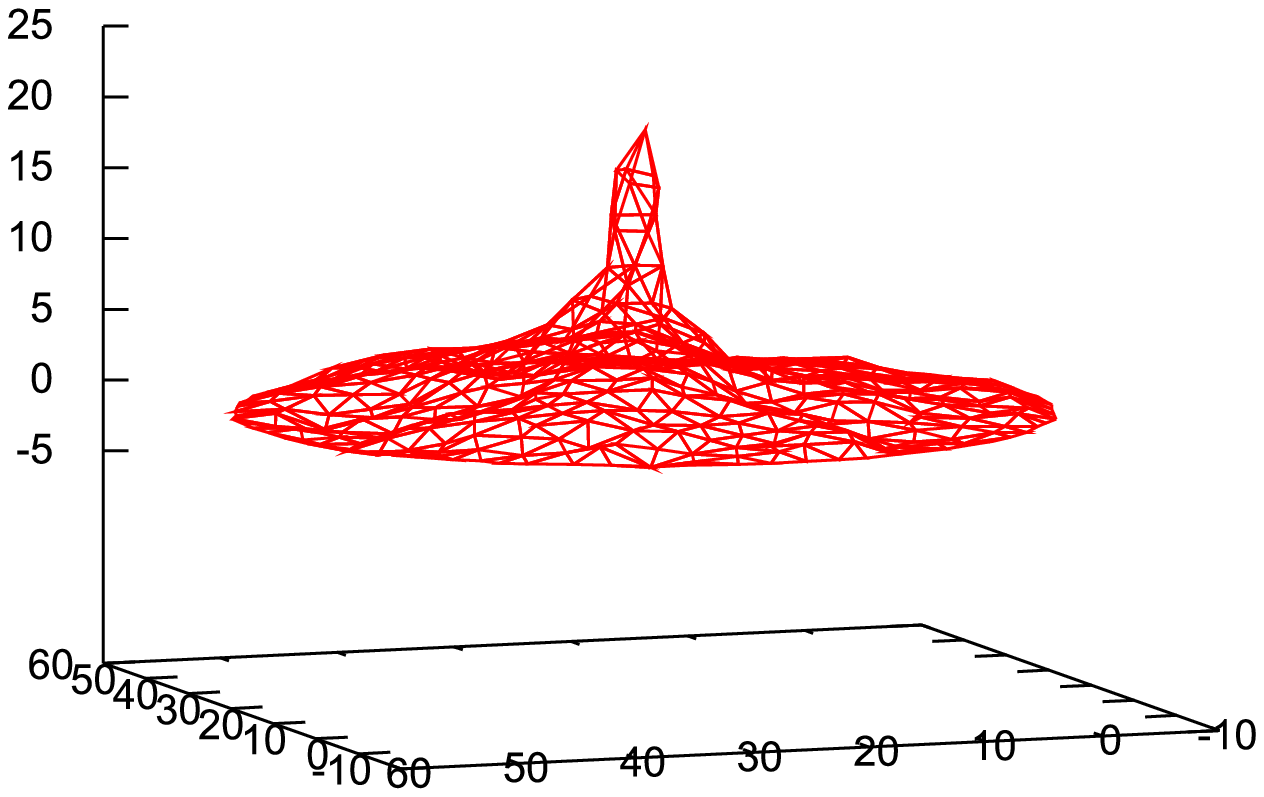}
  \label{fig_5_tube_bon}
}
\caption{Tubes pulled from membranes with the same parameters as before, but with different discretizations for the curvature energy.
  The vertical and horizontal scales are different.
At left, we see an almost flat protuberance, while at right we see a normal tube.
}
\end{center}      
\end{figure}

Looking at these images, we remark a first problem: the angle between the triangles is very large.
This phenomenon is still more marked for a bigger tubes (see
Fig.~\ref{fig_5_20_feio}), where the protuberance becomes almost flat.
The problem comes from the discretization of the curvature energy:

\begin{equation}
  E_{\kappa}^\mathrm{discret} = k \sum_{\langle \alpha,\beta \rangle} (1 - \bm{n}_\alpha \cdot \bm{n}_\beta) \, ,
  \label{eq_5_discret_ant}
\end{equation}

\noindent where $\alpha$ and $\beta$ are adjacent triangles.
As discussed before, this discretization is valid only for $\bm{n}_\alpha
\approx \bm{n}_\beta$, since large deformations bear an unphysical finite cost.
We proposed thus an alternative discretization

\begin{equation}
  {E'}_{\kappa}^\mathrm{discret} = k \sum_{\langle \alpha,\beta
    \rangle} (1 - \bm{n}_\alpha \cdot \bm{n}_\beta) \, e^{\frac{0.8}{(1 + \bm{n}_\alpha \cdot \bm{n}_\beta)}} \, .
  \label{eq_5_discret_new}
\end{equation}

\noindent With this discretization, the energy cost is roughly the same as before for $\bm{n}_\alpha \approx \bm{n}_\beta$, but it increases exponentially as $\bm{n}_\alpha$ approaches $-\bm{n}_\beta$.
The resulting tube, with the same parameters as in Fig.~\ref{fig_5_20_feio},
has a more normal appearance (see Fig.~\ref{fig_5_tube_bon}).

Due to time constraints, we have not examined the dependence of the force
needed to extract a tube as a function of its radius.
It would also be interesting to measure the tube's radius: together with the
measure of the force, one could thus deduce 
the tension $\sigma$ and the bending rigidity $\kappa$.

\section{Perspectives and discussion}
\label{section_5_persp}

As mentioned at the beginning of this chapter, we presented here only preliminary results and many issues need further attention, such as:

\begin{enumerate}
\item verify more carefully if the average shape of the membrane is indeed planar.
  \item study the dependence of $\Delta \tau$ on $\sigma$.
  \item in section~\ref{subsection_1_nat_excess_area} of chapter~\ref{chapitre/planar_membrane}, we predicted that for a membrane under no external force, i. e., with $\tau = 0$, the natural excess area was given by

    \begin{equation}
\alpha_\mathrm{eq} \simeq \frac{\ln(8\pi\beta\kappa)}{8\pi\beta\kappa} \, ,
\end{equation}

    \noindent which depends only on the membrane bending rigidity and temperature.
    If however one forgets the difference between $\tau$ and $\sigma$,
    $\alpha_\mathrm{eq}$ should also depend logarithmically on the size of the membrane (see eq.(\ref{eq_1_alpha_eq_trad})).
    Numerically, we could thus adjust the frame's radius $R_f$ for different membrane sizes in order to have $\langle \tau \rangle = 0$ and measure the excess area in each case. 
    
  \item study better the extraction of tubes and the effects of the alternative discretization of the curvature energy proposed by us.
\item perform a systematic study on the time needed for a system to equilibrate as a function of its size and parameters.
\end{enumerate}

At last, even if using a network to simulate a membrane presents many advantages, it presents a severe drawback: in order to assure liquidity, the bonds must be very easily stretched.
To make the bonds stiffer without affecting the membrane liquidity, one possible solution would be to pass to a macrocanonical ensemble of effective particles: the network would have thus a non-fixed number of beads.
A new particle could be introduced in the middle of a very stretched triangle, which would restore the liquidity for the case of high tension.
Conversely, beads should also be deleted from the network. 
In practice, this is very difficult to implement already from a data structure point of view and the results are not sure. 

\section{In a nutshell}

In this chapter we presented some preliminary results of a numerical
experiment consisting in a piece of weakly fluctuating membrane attached to a
circular frame.
Numerically, it was represented by a triangular network whose connectivity evolved to
simulate liquidity. 
At each vertex of the network, we placed effective particles
that could interact with their first neighbors. 
The bending rigidity was mimicked by an interaction between
adjacent triangles and the particles in the network's edge were submitted to a harmonic potential in
order to force the circular frame.
We used a Monte Carlo method to obtain a large sample of equilibrium
configurations, from which we could evaluate averages of the mechanical
tension, excess area and fluctuation spectrum.
Our first results seems to show that the network behaves similarly to a membrane, but we could not quantitatively verify our predictions concerning the membrane tension.
Many questions in this chapter were left untackled due to time constraints.

%% file: conclusion.tex
\chapter*{Conclusion}

Lipid membranes are very particular materials: despite being almost unstretchable microscopically, in mesoscopic scale they can be easily stretched through the flattening of thermal fluctuations.
Indeed, lipid membranes are highly fluctuating and present thus an excess area relative to its optically resolvable area.

In the beginning of this work, we have seen that the term {\it surface tension} designates several quantities in the context of lipid membranes.
First, there is the tension $\tau$ needed to increase the projected area, or equivalently, to reduce the excess area.
Secondly, there is $\sigma$, the Lagrange-multiplier introduced theoretically to impose a fixed microscopical area to the membrane.
Finally, there is the macroscopic counterpart of $\sigma$, $r$, related to the spectrum of fluctuation.

Experimentally, $r$ can be obtained directly from the fluctuation spectrum and $\tau$ can be measured through the Laplace pressure, for instance.
On the other hand, the theoretical predictions usually involve $\sigma$, which is not directly measurable.
To interpret experimental data, the equality between these quantities is often taken for granted.
Our main goal throughout this work was to determine under which conditions these suppositions are justifiable, specially the equality between $\tau$ and $\sigma$.

Firstly, we have treated the simplest case of a planar membrane.
In the literature, we find some former calculations relating $\tau$ to $\sigma$ and $r$.
There was, however, no consensus: different results were found, depending on how the calculation was made and on the precise definition of $\tau$.
Indeed, the method involved deriving the free-energy with respect to the projected area of the membrane, which we have shown here to be very tricky.
To work around this problem, we have chosen to use a more recent tool: the projected stress tensor, a tensor that relates the force exchanged through an infinitesimal cut on the membrane to the projection of this cut on the projected plane.
The definition of the mechanical tension $\tau$ is thus straightforward: it is simply given by the average of the projected stress tensor.
As supplemental advantage, the projected stress tensor can be relatively easily derived for other geometries, such as spherical and cylindrical, which we have treated in this dissertation.

After evaluating the average of the projected stress tensor, we have obtained an exact relation between $\tau$ and $\sigma$ for weakly fluctuating planar membranes.
In a general way, we have $\tau \simeq \sigma - \sigma_0$, which is the most important result of this dissertation.
The constant $\sigma_0$ depends on the temperature and on the frequency cutoff $\Lambda$, i. e., the highest wave-vector allowed.
At room temperature and considering $\Lambda = 1/(5 \, \mathrm{nm})$, we find $\sigma_0 \approx 5 \times 10^{-6} \, \mathrm{N/m}$.
Accordingly, the assumption $\sigma \approx \tau$ is justifiable only for high tensions.
Otherwise, one must consider the corrected relation to interpret correctly experimental data.
Indeed, some experiments on the adhesion of vesicles seems to agree with our predictions.

In laboratory, planar membranes are very difficult to manipulate.
Vesicles are more commonly used, specially the giant vesicles that can be easily manipulated with a micropipette.
These vesicles can be poked, i. e., free to exchange inner material with the suspension medium, or closed, i. e., with a fixed volume.
We examined thus how the volume constraint and the geometry affected the above mentioned relation for quasi-spherical vesicles.
For both poked and closed vesicles, we conclude that the relation obtained in the planar case is a very good approximation.
%The volume constraint seems to play a role on the excess area of the vesicle.
Interestingly, we predict that the internal pressure of a spherical vesicle can be smaller than the outer, which is impossible in liquid drops.

Another popularly geometry found in membrane experiments is the cylinder.
Indeed, nanotubes are extracted from a piece of membrane, typically a vesicle, by applying a point force with an optical tweezer or with a magnetic field.
Using a simplified mean-field calculations and supposing $\sigma \approx \tau$, the bending rigidity is usually obtained from the curve force versus tension. 
Recently, however, theoretical calculations have predicted that the shape fluctuations for this geometry are very strong.
We expect hence that these fluctuations may affect the interpretation of force measurements.
In this work, we have found that these fluctuation do affect indeed the value of the mean-field force.
Curiously, the effect has never been observed, since the assumption $\sigma \approx \tau$ seems to coincidently make up for the thermal fluctuations.

Aside from the evaluation of tensions and forces, we have also evaluated for the first time the standard deviation of these quantities due to thermal fluctuations.
As in tubular geometry the shape fluctuations are important, we would like to verify if the fluctuation of the force needed to extract a membrane tube could be used to characterize a membrane.
Our results show that the force fluctuation depends on the temperature and that it is very sensitive to the values of $\Lambda$, whereas it does almost not depend on the bending rigidity nor on the tension.
It should thus be of little usefulness to characterize mechanically a membrane.
On the other hand, it is possibly interesting to study the activity of active proteins embedded in the membrane, which has an effect similar to changing temperature.
%Before studying directly the case of a tube, we made a preliminary calculation for the planar case.

Finally, while we have characterized rather well the relation between $\tau$ and $\sigma$, we leave almost untackled the question of how $r$ relates to the other tensions: we have just questioned a former prediction stating that $r = \tau$ and observed a non-trivial behavior of $r$ in two numerical experiments proposed by us.
The question is however very important and needs further attention, since $r$ is a popular non-invasive method used to accede to the tension of a membrane.

%% file: appendix1.tex
\chapter{Alternative derivation of the projected stress tensor}
\label{annexe1}

We re-derive here eqs.(\ref{sigma_xx})--(\ref{sigma_zx}) directly from
eq.(\ref{Sigma_local_frame}). 
We consider a local tangent frame $(X,Y,Z)$
attached to the membrane, with the first two axes parallel to the
principal curvature directions and the third one parallel to the
membrane normal, as shown in Fig.~\ref{stress_local}. 
In this principal tangent frame, as introduced in section~\ref{stress_local_chap}, the force exchanged through a cut of length $d\ell'$ is
given by

\begin{equation}
d\bm{\phi} = \tilde{\bm{\Sigma}}\cdot \bm{\nu}\, d\ell'\,,
\label{tilde}
\end{equation}

\noindent where $\bm{\nu}$ is the normal to the cut within the tangent plane
and$\tilde{\bm{\Sigma}}$ is given by eq.(\ref{Sigma_local_frame}). 

Now, consider the fixed reference plane $\Pi$ introduced on section~\ref{section_projected_stress}. The projected stress tensor $\bm{\Sigma}$ is
defined by 

\begin{equation} 
d\bm{\phi} = \bm{\Sigma}\cdot \bm{m}\, d\ell \,,
\label{project}
\end{equation}

\noindent where $d\ell$ is the length of the cut's projection onto $\Pi$ and
$\bm{m}$ the normal to the cut's projection within $\Pi$ (see
Fig.~\ref{S1}).
 We aim to obtain
$\bm{\Sigma}$ by comparing eq.(\ref{tilde}) and eq.(\ref{project})
when the membrane normal exhibits only small deviations with respect to
the normal to $\Pi$.  

We consider an orthonormal basis $(x,y,z)$ with $\bm{e}_x$ and
$\bm{e}_y$ belonging to $\Pi$. We choose $\bm{m}=\bm{e}_x$,
the cut's projection being parallel to $\bm{e}_y$. Locally, the
membrane shape may be
approximated by a quadratic form:

\begin{equation}
z = h(x,y) \simeq h_0 + a \, x + b\, y + \frac{1}{2} \, \alpha \, x^2 + \beta
\, x y
+ \frac{1}{2}\,  \gamma \, y^2 \,, 
\end{equation} 

\noindent with $(a,b,\alpha,\beta,\gamma) \equiv
(h_x,h_y,h_{xx},h_{xy},h_{yy})=\mathcal{O}(\epsilon)$ and $\epsilon \ll
1$. 
We shall apply three successive frame rotations in order to bring
the fixed frame $(x,y,z)$ to the tangent principal frame $(X,Y,Z)$.

First, we make a rotation about $\bm{e}_x$ of angle $\theta_1=b
$, plus a vertical translation. The new coordinates, indicated by a
prime, are related to the old ones by 

\begin{eqnarray}
x &=& x'\,,\\
y &=& \left( 1 - \frac{b^2}{2} \right) y'- b \, z'+ \mathcal{O}(\epsilon^3)\, ,  \\
z &=& h_0 + b \, y' + \left( 1 - \frac{b^2}{2} \right) z'+
\mathcal{O}(\epsilon^3)\, .
\end{eqnarray}

\noindent Note that $\bm{e}_y'$ is thus parallel to the actual cut within the
membrane.  We perform a second rotation, about the new axis $\bm{e}_y'$
of angle $\theta_2=a$ in order to make the new plane
$(x'',y'')$ coincide with the tangent plane:

\begin{eqnarray}
x' &=& \left(1 - \frac{a^2}{2}\right)x'' - a \, z'' + \mathcal{O}(\epsilon^3) \, ,\\
y' &=& y''\, ,\\
z' &=& a \, x'' + \left( 1 - \frac{a^2}{2} \right) z''+ \mathcal{O}(\epsilon^3)
\, .
\end{eqnarray}

\noindent Note that $\bm{e}_y''\equiv\bm{e}_y'$ is still parallel to the actual
cut, while $\bm{e}_x''$ is now the normal to the cut within the tangent
plane.
In other words $\bm{e}_x''=\bm{\nu}$.  

In the $(x'', y'', z'')$
coordinate system, the membrane shape is given simply by 

\begin{equation} 
z'' = \frac{1}{2} \, \alpha \, x''^2 + \beta \, x''y'' 
+ \frac{1}{2} \, \gamma \, y''^2 + \mathcal{O}(\epsilon^3) \, .
\end{equation}

Finally, we arrive to the principal tangent frame $(X,Y,Z)$ by making a
rotation about $\bm{e}_{z''}$ in order to diagonalize the quadratic
form. 
Indeed, if we set

\begin{eqnarray}
x'' &=& X\cos\theta - Y\sin\theta\,,\\
y'' &=& X\sin\theta + Y\cos\theta\,,\\
z'' &=& Z\,, 
\end{eqnarray}

\noindent where $\theta$ is the solution of

\begin{equation}
2 \beta\, \cos 2\theta = (\alpha - \gamma)\sin 2\theta\,,
\label{sol}
\end{equation}

\noindent the cross term vanishes, leaving $Z = \frac{1}{2}\, C_X X^2 +
\frac{1}{2}\, C_Y Y^2$, where $C_X = \alpha \cos^2 \theta + \gamma
\sin^2\theta + \beta \sin2\theta + \mathcal{O}(\epsilon^3)$ and $C_Y =
\gamma \cos^2 \theta + \alpha \sin^2 \theta - \beta \sin2\theta +
\mathcal{O}(\epsilon^3)$ are the principal curvatures.

Comparing eqs.(\ref{tilde}) and (\ref{project}) while using
$\bm{m}=\bm{e}_x$, $\bm{\nu} = \bm{e}_{x''}$ and $d\ell' =
d\ell(1+ b^2/2)+ \mathcal{O}(\epsilon^3)$, yields

\begin{eqnarray}
\label{Sxx}
\Sigma_{xx} = \left(1 + \frac{b^2}{2}\right) \bm{e}_x \cdot
\tilde{\bm{\Sigma}} \cdot \bm{e}_{x''} + \mathcal{O}(\epsilon^3)\,, \\
\label{Syx}
\Sigma_{yx} = \left(1 + \frac{b^2}{2}\right) \bm{e}_y \cdot
\tilde{\bm{\Sigma}} \cdot \bm{e}_{x''} + \mathcal{O}(\epsilon^3)\,, \\
\label{Szx}
\Sigma_{zx} = \left(1 + \frac{b^2}{2}\right) \bm{e}_z \cdot
\tilde{\bm{\Sigma}} \cdot \bm{e}_{x''} + \mathcal{O}(\epsilon^3)\,.
\end{eqnarray}

Calculating $\tilde{\bm{\Sigma}}\cdot \bm{e}_{x''}$ from
eq.(\ref{Sigma_local_frame}), we obtain

\begin{eqnarray}
\tilde{\bm{\Sigma}}\cdot \bm{e}_{x''} &=& 
\cos\theta\left(\sigma + \frac {\kappa}{2}\, 
  C_Y^2 - \frac{\kappa}{2} \, C_X^2 \right) \bm{e}_X \nonumber \\
&-& \sin\theta\left(\sigma +\frac{\kappa}{2}\, C_X^2 - \frac{\kappa}{2}\, 
  C_Y^2\right) \bm{e}_Y \nonumber \\ 
&-&
\kappa \, \frac{\partial C}{\partial x''} \bm{e}_Z\, ,
\label{step}
\end{eqnarray}

\noindent with $C = \nabla^2 h + \mathcal{O}(\epsilon^2)$. 
We have taken advantage 
that the terms $\propto\bm{e}_Z$ in eq.(\ref{Sigma_local_frame}) are equal
$-\kappa \, \bm{e}_Z \otimes \bm{\nabla} C$. Going to the $(x'',y'',z'')$
basis and using eq.(\ref{sol}), we have simply

\begin{eqnarray}
\tilde{\bm{\Sigma}}\cdot \bm{e}_{x''} &=& \left(\sigma + \frac {\kappa}{2}\, 
  \gamma^2 - \frac{\kappa}{2} \, \alpha^2 \right) \bm{e}_{x''} \nonumber \\
&-& \frac{\kappa}{2} \, \beta \left(\alpha + \gamma \right) \bm{e}_{y''} 
-\kappa \, \frac{\partial C}{\partial x''} \, \bm{e}_{z''} 
+ \mathcal{O}(\epsilon^3)\, .\qquad
\end{eqnarray}

Since $C = \mathcal{O}(\epsilon)$, we can replace $\partial/\partial
x''$ by $\partial/\partial x$, since the difference between the latter
is of order $\epsilon^2$. 
Using now directly eqs.(\ref{Sxx})--(\ref{Szx}), with 
$\bm{e}_{x''} \cdot \bm{e}_x = (1 - a^2/2) + \mathcal{O}(\epsilon^3)$,
$\bm{e}_{y''} \cdot \bm{e}_x = 0$,
$\bm{e}_{z''} \cdot \bm{e}_x = -a + \mathcal{O}(\epsilon^3)$, etc.,
yields

\begin{eqnarray}
\Sigma_{xx}&=&\sigma+\frac{\sigma}{2}\left(b^2-a^2\right)
+\frac{\kappa}{2}\left(\gamma^2- \alpha^2\right)\nonumber\\
&&+\kappa \, a\,\partial_xC+\mathcal{O}(\epsilon^3)\,,\\
\Sigma_{yx}&=&-\sigma \, a b-\kappa\, \beta\left(\alpha+\gamma\right)
+\kappa \, b\,\partial_xC+\mathcal{O}(\epsilon^3)\,,\qquad\\
\Sigma_{zx}&=&\sigma a-\kappa\,\partial_xC+\mathcal{O}(\epsilon^3)\,,
\end{eqnarray}

\noindent which coincide with eqs.(\ref{sigma_xx})--(\ref{sigma_zx}).

%% file: appendix2.tex
\chapter{Projected stress tensor for a \bm{$1$}-d filament}
\label{annexe2}

The derivation of the projected stress tensor for a $1$-d filament follows the
same reasoning presented in section~\ref{section_projected_stress}.
The force exchanged through a cut is given by

\begin{equation}
\bm{f} = \bm{\Sigma}^\mathrm{1D} \cdot \bm{m} = \left(\Sigma^\mathrm{1D}_x \, \bm{e}_x + \Sigma^\mathrm{1D}_z \,
  \bm{e}_z \right) m_x\, ,
\label{eq_1_force}
\end{equation}

\noindent where $\bm{m} = m_x \, \bm{e_x} = \pm \, \bm{e}_x$, depending on the
orientation of the projected cut. 
Note that the stress tensor now has just two components. 
For a general filament whose energy is given by

\begin{equation}
\mathcal{H} = \int_{L_p} f(h_x, h_{xx}) \, dx \, ,
\end{equation}

\noindent an arbitrary small displacement $\delta \bm{a} = \delta a_x \, \bm{e}_x
+ \delta a_z \, \bm{e}_z$ of the membrane, at equilibrium, leads to the energy
variation

\begin{equation}
\delta\mathcal{H} = \left.\left[f \, \delta a_x + \left(\frac{\partial f}{\partial
      h_x} - \partial_x\frac{\partial f}{\partial h_{xx}}\right)\delta u -
  \frac{\partial f}{\partial h_{xx}} \, \delta u_x\right]m_x \right|_{\delta L_p}\, ,
\label{eq_1_deltaH}
\end{equation} 

\noindent where $\delta L_p$ indicates a sum over the two edges of the filament.
Keeping the tangent to the filament at the edges unchanged so that torques
perform no work, we have

\begin{equation}
\delta u = \delta a_z - \delta a_z \, h_x \, ,
\label{eq_1_deltau}
\end{equation}

\noindent and 

\begin{equation}
\delta u_x = - \, \delta a_x \, h_{xx} \, .
\label{eq_1_deltaux}
\end{equation}

Finally, comparing the work of the force $\bm{f}$ given in eq.(\ref{eq_1_force})
and the work given in eq.(\ref{eq_1_deltaH}), using eq.(\ref{eq_1_deltau})
and eq.(\ref{eq_1_deltaux}), we obtain

\begin{eqnarray}
\Sigma^\mathrm{1D}_x &=& f - \frac{\partial f}{\partial h_x} h_x - \frac{\partial
  f}{\partial h_{xx}} h_{xx} + \partial_x\left(\frac{\partial f}{\partial
    h_{xx}}\right) h_x\, ,\nonumber \\
\Sigma^\mathrm{1D}_z &=& \frac{\partial f}{\partial h_x} - \partial_x\left(\frac{\partial
    f}{\partial h_{xx}}\right)\, .
\end{eqnarray}

For $\mathcal{H}_\mathrm{1D}$ (eq.(\ref{eq_1_H1D}), we have

\begin{eqnarray}
\Sigma^\mathrm{1D}_x &=& \sigma - \frac{\sigma}{2} h_x^2 - \frac{\kappa}{2} h_{xx}^2 +
\kappa \, h_{xxx}h_x \, ,\nonumber\\
\Sigma^\mathrm{1D}_z &=& \sigma \, h_x - \kappa \, h_{xxx} \, .
\end{eqnarray}

%% file: appendix21.tex
\chapter{Estimative of $\bm{W_A^{\mathrm{theo}}}$}
\label{annexe21}

In this section we explain in details the theoretical estimative of the adhesion energy per unit area $W_A^{\mathrm{theo}}$ proposed by R\"adler et al.~\cite{Raedler_95}.
They took into account two attractive interactions, coming from the van der Waals interactions and gravity, and a repulsive interaction with entropic origins, due to the restrictions imposed on the membrane fluctuation.
They considered the screened van der Waals potential given by

\begin{equation}
  V_\mathrm{vdW} = - \frac{A_H}{12\pi}\left[\frac{1}{s^2} - \frac{1}{(s+a)^2}\right] \, \left(\frac{2s}{\lambda_D}\, e^{-2\frac{s}{\lambda_D}}\right)\, ,
  \label{eq_1_VdW}
\end{equation}

\noindent where $A_H$ is the Hamacker constant, $s$ is the distance between the membrane and the substrate, $a$ is the membrane thickness and $\lambda_D$ is the Debye screening length, given by

\begin{equation}
  \lambda_D = \sqrt{\frac{\varepsilon_0 \, \varepsilon_r k_\mathrm{B}T}{e^2 N_A \sum_i c_i z_i^2}}\, ,
\end{equation}

\noindent where $\varepsilon_0$ is the vacuum electrical permittivity,
$\varepsilon_r$ is the dielectric constant of the solvent, $e$ is the
elementary charge, $N_A$ is the Avogadro number, $z_i$ is the charge number of
a dissolved ion and $c_i$ is the respective molar concentration.
The last term in eq.(\ref{eq_1_VdW}) is a correction coming from the screening of the substrate due to the presence of ions in the solution.
Indeed, it is expected that some part of the $MgF^2$ coating of the glass
cover slip is present in small concentration in the buffer solution .  

As vesicles are prepared in a sucrose solution, there is possibly a difference of density between the internal fluid of GUVs and the buffer solution.
The potential due to gravity per unit area is given by

\begin{equation}
  V_\mathrm{grav} = g \, \Delta\rho \, h_\mathrm{CM} \, \frac{V_v}{A_C} \, ,
  \label{eq_1_grav}
\end{equation}

\noindent where $g$ is the gravity acceleration, $\Delta\rho$ is the density difference, $V_v = 4/3 \pi R_\mathrm{ves}^3$ is the vesicle volume, $A_C = \pi R_a^2$ is the contact area and $h_\mathrm{CM}$ is the height of the center of mass.
Assuming that the shape of the vesicle does not change with the distance from the substrate, we have $h_{M} \simeq R_\perp + \langle s\rangle$, where $R_\perp$ is the height of the center of mass relative to the contact region of the vesicle and so the free-energy is simply given by

\begin{equation}
  V_\mathrm{grav} = g \, \Delta\rho \, \langle s\rangle \, \frac{V_v}{A_C} \, .
\end{equation}

Finally, to evaluate the steric potential that arises when fluctuations are limited, they used the equipartition of energy to estimate the energy per uncorrelated patch of membrane of size $\xi_\parallel$: 

\begin{equation}
  V_\mathrm{steric} = b \, \frac{k_\mathrm{B}T}{\xi_\parallel^2} \, ,
  \label{eq_1_Vsteric}
\end{equation}

\noindent where $b$ is a numerical factor.
To obtain $\xi_\parallel$, the group assumed that the contact area was
equivalent to a flat membrane under a quadratic potential (Hamiltonian given
in eq.(\ref{hamilt_adhes})) which yields two limiting cases: the case where
adhesion is dominated by rigidity (with the corresponding equations shown in
the first line of table~\ref{table_1_xi}) and the case where adhesion is dominated by tension (second line of the same table).
Further details on the derivation of these equations are given in appendix~\ref{annexe3}.
%At the experiment's time, the problem of adhesion had been theoretically studied only for planar membranes\cite{Lipowsky_94},\cite{Netz_95}.
%R\"adler et al. tried so to interpret their results under this framework, that we will present in the following.

\vspace{1cm}

\begin{table}[H]
  \begin{center}
\begin{tabular}{|c|c|c|c|}
\hline
\bf{{\red Case}} & \bm{{\red $\xi_\perp^2$}} & \bm{{\red $\xi_\parallel$}} & \bf{{\red Relation}} \\ 
\hline
& & &\\
$\sigma < \sigma^*$ & $\frac{k_\mathrm{B}T}{8\sqrt{\kappa V''}}$ & $\left(\frac{4\kappa}{V''}\right)^{\frac{1}{4}}$ & $\xi_\perp^2 \approx \frac{k_\mathrm{B}T}{16\kappa} \xi_\parallel^2$ \\
& & & \\
\hline
& & & \\
$\sigma > \sigma^*$ & $\frac{k_\mathrm{B}T}{2\pi\sigma} \ln\left(\frac{2\sigma}{\sigma^*}\right)$ & $\left(\frac{\sigma}{V''}\right)^{\frac{1}{2}}$&$\xi_\perp^2\approx\frac{k_\mathrm{B}T}{4\pi\sigma}\ln\left(\frac{\sigma \, \xi_\parallel^2}{\kappa}\right)$ \\
& & & \\
\hline
\end{tabular}
\label{table_1_xi}
\caption{Theoretical previsions for $\xi_\perp$ and $\xi_\parallel$ as a function of $\sigma$, $\kappa$ and $V''$, with $\sigma^* = \sqrt{4 \kappa V''}$. The first line corresponds to the case where adhesion is dominated by rigidity, while the second line corresponds to the case where it is dominated by tension. The last column is obtained by substituting the third column on the second.}
\end{center}
\end{table}

\begin{figure}[H]
\begin{center}
\subfigure[Experimental vertical roughness $\xi_{\perp}$ as a function of the expected curve for the case dominated by tension (lower line of table~\ref{table_1_xi}).]{ 
\includegraphics[scale=.35,angle=0]{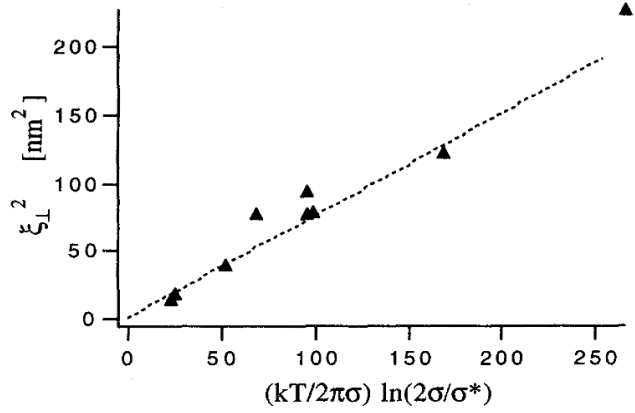}
}
\subfigure[Lateral correlation length as a function of the curve expected in the same limit (lower line of table~\ref{table_1_xi}).] {
\includegraphics[scale=.35,angle=0]{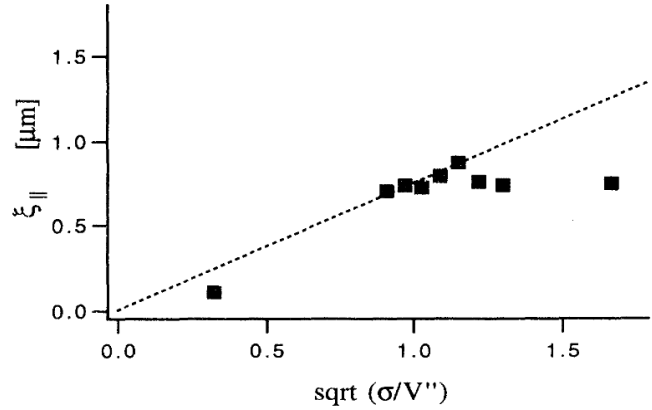}
}  
\caption{Experimental correlation lengths as a function of the expected curves for the regime  dominated by tension~\cite{Raedler_95}. Note that the curves were traced under the assumption $\sigma \approx r$.}
\label{fig_1_xi_exp}
\end{center}      
\end{figure}

To determine whether experimentally the adhesion was dominated by rigidity or tension, R\"adler et al. plotted the measured
values of $\xi_\perp^2$ as a function of $V''$, $k_\mathrm{B} T$, $\kappa$ and
$\sigma$ (assuming $r \sim \sigma$) using the equations given in the second column of table~\ref{table_1_xi}.
For the equation corresponding to the case dominated by the tension, they obtained a nice linear relation (see Fig.~\ref{fig_1_xi_exp}).
The same analysis was performed on $\xi_\parallel$, this time using the equations of the third column of table~\ref{table_1_xi}.
Again, a linear relation was obtained for the lower equation (see Fig.~\ref{fig_1_xi_exp}, lower figure).
They concluded that it was reasonable to assume $\sigma \approx r$ in this experiment and that the behavior of the membrane was dominated by tension.
%Accordingly, $\xi_\parallel$ relates to $\xi_\perp$ through the last equation of table \ref{table_1_xi1}, that also reads

%\begin{equation}
 %2 \left(\frac{\xi_\perp}{\ell_\sigma}\right)^2= \ln\left(\frac{\sigma \xi_\parallel^2}{\kappa}\right) \, ,
  %\label{eq_1_xi_1}
%\end{equation}

Besides, in this case, it is theoretically expected that $\xi_\perp$ relates to the mean separation from the substrate $\langle s \rangle$ through~\cite{Netz_95}

\begin{equation}
  2\left(\frac{\xi_\perp}{\ell_\sigma}\right)^2 = \frac{\langle s\rangle}{\ell_\sigma} + \frac{1}{4} \ln\left(\frac{\langle s\rangle}{\ell_\sigma}\right)\, .
  \label{eq_1_xi_2}
\end{equation}

\noindent where $\ell_\sigma = \sqrt{k_\mathrm{B}T/(2\pi\sigma)}$.

%\noindent Indeed, the experimental data seem to confirm this relation, even though a systematic deviation is observed, probably due to the limited spatial resolution limitations (see fig.\ref{fig_1_lsigma}). 

%\begin{figure}[H]
%\begin{center}
%\includegraphics[scale=.35,angle=0]{./figures_Chap1/lsigma.eps}
%\caption{Measured roughness $\xi_\perp$ as a function of eq.(\ref{eq_1_xi_2}). The systematic deviation is probably due to spatial resolution limitations.}
%\label{fig_1_lsigma}
%\end{center}      
%\end{figure}

\noindent From eq.(\ref{eq_1_Vsteric}), the last equation of table~\ref{table_1_xi} and eq.(\ref{eq_1_xi_2}), one obtains finally

\begin{equation}
  V_\mathrm{steric} = b \, \frac{k_\mathrm{B} T \, \sigma}{\kappa} \left(\frac{\ell_\sigma}{\langle s \rangle}\right)^{\frac{1}{4}}e^{-\frac{\langle s\rangle}{\ell_\sigma}}\, ,
\end{equation}

\noindent where $b$ is a dimensionless prefactor.
One can see the plot of $V_\mathrm{total} = V_\mathrm{vdW} + V_\mathrm{steric} + V_\mathrm{grav}$ for typical experimental values $\sigma = 1.7 \times 10^{-5} \, \mathrm{N/m}$, $\kappa = 35 \, k_\mathrm{B} T$, $b = 0.085$, $A_H = 2.6 \times 10^{-21} \, \mathrm{J}$, $D_M = 20 \, \mathrm{\mu m}$, $D_A = 10 \, \mathrm{\mu m}$, $\Delta \rho = 7 \, \mathrm{kg/m^3}$ can be seen in Fig.~\ref{radler}.

%% file: appendix3.tex
\chapter{Determination of \bm{$\xi_\perp$} and \bm{$\xi_\parallel$} for planar membranes under a quadratic potential}
\label{annexe3}

In this section we derive the correlation length $\xi_\parallel$ and the roughness $\xi_\perp$ for a planar membrane under a quadratic potential~\cite{Lipowsky_94},~\cite{Netz_95}.
Assuming that the Hamiltonian is given by eq.(\ref{hamilt_adhes}), the correlation function is given by

\begin{equation}
  G(\bm{r} - \bm{r}') = \frac{k_\mathrm{B} T}{A_p} \sum_{\bm{q}} \frac{e^{\icomp \, \bm{q}\cdot
      (\bm{r} - \bm{r'})}}{V''+ \sigma\,  q^2 + \kappa \, q^4} \, .
\end{equation}

By definition, one has

\begin{eqnarray}
  \xi_\perp^2 &=& \langle h(\bm{r})h(\bm{r})\rangle \, \nonumber\\
  &=& \frac{k_\mathrm{B} T}{A_p} \sum_{\bm{q}} \frac{1}{V''+ \sigma\,  q^2 + \kappa \, q^4} \, , \nonumber \\
  &=& \frac{k_\mathrm{B} T}{A_p} \int_{q_\mathrm{min}}^{q_\mathrm{max}} \frac{d\bm{q}}{(2\pi)^2} \frac{1}{V''+ \sigma\,  q^2 + \kappa \, q^4} \, , \nonumber \\
  &=&\frac{k_\mathrm{B}T}{2\pi \sigma} \, \Omega\left(\frac{\sigma}{\sigma^*}\right)\, ,
  \label{eq_an3_xi_perp}
\end{eqnarray}

\noindent where the last step is justified for very large $A_p$ and for $\sigma a^2/\kappa \ll 1$ ($a$ is a microscopical cut-off of order of the membrane thickness).
The crossover tension $\sigma^* = \sqrt{4 \, \kappa \, V''}$ defines two limits: one dominated by tension ($\sigma > \sigma^*$) and one dominated by the rigidity ($\sigma < \sigma^*$). The function $\Omega$ is given explicitly by

\begin{equation}
  \Omega(y) = \left\{ \begin{array}{ccc} \frac{\tan^{-1}\left(\sqrt{y^{-2}-1}\right)}{\sqrt{y^{-2}-1}} \,\,\,& $for$ &\, y < 1\, ,\\
    \\
    \frac{\tanh^{-1}\left(\sqrt{1-y^{-2}}\right)}{\sqrt{1-y^{-2}}} \,\,\, &$for$ &\, y>1\, .\end{array} \right.
\end{equation}

To estimate $\xi_\parallel$, one must evaluate the general correlation function

\begin{eqnarray}
  \langle h(\bm{r})h(0)\rangle 
  &=& \frac{k_\mathrm{B} T}{A_p} \sum_{\bm{q}} \frac{e^{\icomp \bm{q}\cdot \bm{r}}}{V''+ \sigma\,  q^2 + \kappa \, q^4} \, , \nonumber \\
  &=& \frac{k_\mathrm{B} T}{A_p} \int_{q_{min}}^{q_{max}}
  \frac{d\bm{q}}{(2\pi)^2} \frac{e^{\icomp \bm{q} \cdot \bm{r}}}{V''+ \sigma\,  q^2 + \kappa \, q^4} \, , \nonumber \\
  &=& \frac{k_\mathrm{B} T}{A_p} \int_{q_{min}}^{q_{max}} \frac{dq}{(2\pi)} \frac{q \, J_0(qr)}{V''+ \sigma\,  q^2 + \kappa \, q^4} \, ,
  \label{eq_an3_xi_parallel}
\end{eqnarray}

\noindent where $J_0(x)$ is the Bessel function of the first kind and of order $0$ and compare its asymptotic behavior with $\xi_\perp^2 e^{-r/\xi_\parallel}$.

The results coming from eq.(\ref{eq_an3_xi_perp}) and eq.(\ref{eq_an3_xi_parallel}) for the two limiting situations are summed up in table~\ref{table_an3_xi}.

\vspace{1cm}

\begin{table}[H]
  \begin{center}
\begin{tabular}{|c|c|c|c|}
\hline
\bf{{\red Case}} & \bm{{\red $\xi_\perp^2$}} & \bm{{\red $\xi_\parallel$}} & \bf{{\red Relation}} \\ 
\hline
& & &\\
$\sigma < \sigma^*$ & $\frac{k_\mathrm{B}T}{8\sqrt{\kappa V''}}$ & $\left(\frac{4\kappa}{V''}\right)^{\frac{1}{4}}$ & $\xi_\perp^2 \approx \frac{k_\mathrm{B}T}{16\kappa} \xi_\parallel^2$ \\
& & & \\
\hline
& & & \\
$\sigma > \sigma^*$ & $\frac{k_\mathrm{B}T}{2\pi\sigma} \ln\left(\frac{2\sigma}{\sigma^*}\right)$ & $\left(\frac{\sigma}{V''}\right)^{\frac{1}{2}}$&$\xi_\perp^2\approx\frac{k_\mathrm{B}T}{4\pi\sigma}\ln\left(\frac{\sigma \, \xi_\parallel^2}{\kappa}\right)$ \\
& & & \\
\hline
\end{tabular}
\label{table_an3_xi}
\caption{Theoretical previsions for $\xi_\perp$ and $\xi_\parallel$ as a function of $\sigma$, $\kappa$ and $V''$. The last column is obtained by substituting the third column on the second.}
\end{center}
\end{table}

%% file: appendix4.tex
\chapter{Verification of the projected stress tensor in spherical geometry: the planar limit}
\label{annexe4}

In order to check the expressions given in eqs.(\ref{eq_2_stt})-(\ref{eq_2_srp}), we consider the limit $R\rightarrow\infty$, which should yield the
projected stress tensor for a flat membrane given in section~\ref{section_projected_stress}.
We consider a general point $(\theta, \phi)$ on the sphere and we define a local 
system of Cartesian coordinates $(x,y,z)$, such that $d x = R d \theta$, $d y =R \sint d \phi$ and $dz = dr$.
We want to determine the projected stress tensor $\bm{\Sigma}'$ in these new coordinates.
The component of the elementary force $d f_\alpha$ along a general direction $\alpha\in \{x,y,z\}$ exerted through a cut perpendicular to $x$ reads
$d f_\alpha=\Sigma_{\alpha x}' \, d \ell=
\Sigma_{\alpha \theta} \, d \phi$, with the correspondence $d \ell=R \sint \, d \phi$. 
Thus we have $\Sigma_{\alpha x}'= \Sigma_{\alpha \theta}/(R \sint )$.
Likewise,  $\Sigma_{\alpha y}'= \Sigma_{\alpha \phi}/R$, since in this case $d\ell=R \, d \theta$.
In the limit of large $R$, the plane tangent to the sphere at the point
$(\theta,\phi)$ becomes the reference plane of the membrane, and the height of
the membrane over this plane is $h(x,y)=R \, u(\theta,\phi)$.
Hence $\ut=R \, u_x=h_x$,
$\up=R \, \sint \, u_y=\sint\,  h_y$, $u_{\theta \theta}=R^2 \, u_{xx}=R \, h_{xx}$,
$u_{\phi \phi}=R^2 \sin^2\theta \, u_{yy}=R \, \sin^2\theta \, h_{yy}$, etc.
Keeping the terms that are dominant in the limit $R \rightarrow \infty$, we obtain

\begin{eqnarray}
\Sigma_{xx}'&=& \sigma + \frac{\sigma}{2} \p{h^2_{y}-h^2_{x}}+ \frac \kappa 2
\p{h^2_{yy}-h^2_{xx}} + \kappa h_x \partial_x \nabla^2 h \, , \\
\Sigma_{yx}'&=& - \sigma h_x h_y-\kappa h_{xy} \nabla^2 h+\kappa h_y  \partial_x \nabla^2 h,\\
\Sigma_{rx}'&=&  \sigma h_x-\kappa  \partial_x \nabla^2 h,\\
\Sigma_{yy}'&=& \sigma + \frac{\sigma}{2} \p{h^2_{x}-h^2_{y}}+ \frac{\kappa}{2} \p{h^2_{xx}-h^2_{yy}} + \kappa h_y \partial_y \nabla^2 h \, ,\\
\Sigma_{xy}'&=& - \sigma h_x h_y-\kappa h_{xy} \nabla^2 h+\kappa h_x  \partial_y \nabla^2 h,\\
\Sigma_{ry}'&=&  \sigma h_y-\kappa  \partial_y \nabla^2 h,
\end{eqnarray} 

\noindent which agree with the results of section~\ref{section_projected_stress}.

%% file: appendix5.tex
\chapter{Correlation functions for vesicles and their relationship}
\label{annexe5}

There are five fundamental correlation functions, which we have calculated explicitly:

\begin{eqnarray}
\average{\up^2}&=&\sum_{\omega,\omega'} \partial_\phi \Ylm(\theta,\phi)\partial_{\phi}
Y_{l'}^{m'}(\theta,\phi) \average{\ulm u_{l'm'}} \, ,\nonumber\\
&=&\sum_\omega\frac{k_B T}{\Hl}
\partial_\phi \Ylm \partial_\phi \p{\Ylm}^*=\sum_\omega  \frac{k_B T}{\Hl} m^2
|\Ylm|^2 \, ,\label{eq_annexe_5_upup} \nonumber\\
&=&\sin^2\theta \, \frac{k_B T}{4 \pi} \sum_{l=2} \frac{l (l+1)(2l+1)}{2 \Hl}\\
\average{u_{\phi\phi}^2}&=&\sum_\omega  \frac{k_B T}{\Hl} \, m^4
|\Ylm|^2 \, ,\nonumber\\
&=&\sin^2\theta\frac{k_B T}{4 \pi} \sum_{l=2} \frac{l (l+1)(2l+1)}{8
  \Hl}\Big[4+3(-2+l+l^2)\sin^2\theta\Big]\, ,\\
\average{\ut^2}&=& \sum_\omega\frac{k_B T}{\Hl}\,  \partial_\theta \Ylm
\partial_\theta \p{\Ylm}^*=\frac{k_B T}{4 \pi} \sum_{l=2} \frac{l
  (l+1)(2l+1)}{2 \Hl} \, ,\\
\average{u_{\theta \theta}^2}&=&\sum_\omega\frac{k_B T}{\Hl} \, \partial_\theta^2
\Ylm \partial_\theta^2 \p{\Ylm}^* \, , \nonumber \\
&=&\frac{k_B T}{4 \pi} \sum_{l=2} \frac{l
  (l+1)(2l+1)(-2+3 l +3 l^2)}{8 \Hl} \, ,\\
\average{\ut \, u_{\theta \phi\phi}}&=&-\sum_\omega\frac{k_B T}{\Hl} \, m^2
\partial_\theta \Ylm \partial_\theta \p{\Ylm}^* \, ,\nonumber\\
&=&-\frac{k_B T}{4 \pi}
\sum_{l=2}\frac{l (l+1)(2l+1)}{2 \Hl} \pq{1+\frac{(l+3)(l-2)}{4} \sin^2\theta}
\, .
\label{eq_annexe_5_ututpp}
\end{eqnarray} 

The other correlation functions either vanish or may be deduced from them.
Remembering that $u_{l,-m}=(-1)^m
\ulm^*$, $\p{\Ylm}^*=(-1)^m Y^{-m}_l$ and 

\begin{equation}
\frac{\partial^n \Ylm(\theta, \phi)}{\partial \phi^n} = (i\, m)^n \, \Ylm (\theta, \phi) \, ,
\end{equation}

\noindent one can demonstrate that

\begin{equation}
\frac{\partial^{n+1}
  {\Ylm}(\theta,\phi)}{\partial \phi^{n+1}}\frac{\partial^{p-1} {\Ylm}^*
  (\theta,\phi)}{\partial \phi^{p-1}} = 
- \frac{\partial^{n}
  {\Ylm}(\theta,\phi)}{\partial \phi^n} \frac{\partial^{p} {\Ylm}^*
  (\theta,\phi)}{\partial \phi^p} \, ,
\end{equation}

\noindent which holds also in the presence of derivatives with respect to $\theta$.
From this relation, one deduces the following rule: when averaging the product of two terms, one may pass
a derivative with respect to $\phi$ from one term to the other while multiplying by
$-1$.
Hence,

\begin{eqnarray}
\langle u_{\phi}^2 \rangle &=& - \langle u
\, u_{\phi\phi}\rangle \, ,
\label{eq_annexe_5_upup1}\\
\langle u_{\phi}\, u_{\theta} \rangle &=& - \langle u\, 
u_{\phi\theta}\rangle \, ,\\
\langle u_{\phi}\, u_{\theta\theta} \rangle &=& - \langle u_\theta\, 
u_{\phi\theta}\rangle \, .
\end{eqnarray}

\noindent Consequently, averages implying an odd number of derivatives
with respect to $\phi$ vanish:

\begin{eqnarray}
\langle u\, u_\phi \rangle &=& 0\, ,\\
\langle u_\theta \, u_\phi \rangle &=& 0\, ,\\
\langle u_\phi \, u_{\phi\phi} \rangle &=& 0\, ,\\
\langle u_\theta \, u_{\theta\phi} \rangle &=& 0\, ,\\
\langle u_{\theta\theta} \, u_{\theta\phi} \rangle &=& 0\, ,\\
\langle u_{\phi\phi} \, u_{\theta\phi} \rangle &=& 0\, .
\label{eq_annexe_5_upputp}
\end{eqnarray}

Another very helpful relation is the addition theorem for spherical
harmonics:

\begin{equation}
\sum_{m = -l}^{l} \Ylm(\theta_1,\phi) {\Ylm}^* (\theta_2,\phi) = \frac{2l + 1}{4
  \pi} P_l(\cos(\gamma))\, ,
\label{eq_annexe_5_add_theo}
\end{equation}

\noindent where $P_l$ is the Legendre polynomial of order $l$ and $\gamma = \theta_2 - \theta_1$.
By differentiating this relation $k$ times with respect to $\theta_1$ and $p$ times with respect to $\theta_2$, one obtains

\begin{equation}
\sum_{m = -l}^l \frac{\partial^k \Ylm(\theta_1,\phi)}{\partial \theta_1^k}
\frac{\partial^p {\Ylm} ^* (\theta_2,\phi)}{\partial \theta_2^p} = 
 (-1)^k \frac{(2l
  + 1)}{4 \pi} \frac{d^{k+p} P_l(\cos(\gamma))}{d \gamma^{k+p}} \, .
\end{equation}

\noindent In particular, for $\theta_1 = \theta_2 = \theta$, we have  

\begin{equation}
\sum_{m = -l}^l \frac{\partial^k \Ylm(\theta,\phi)}{\partial \theta^k}
\frac{\partial^p {\Ylm}^{*}(\theta,\phi)}{\partial \theta^p} = (-1)^k \frac{(2l
  + 1)}{4 \pi} \frac{d^{k+p} P_l(1)}{d \gamma^{k+p}} \, .
\label{eq_annexe_5_deriv_1}
\end{equation}

\noindent Inversely, one can differentiate eq.(\ref{eq_annexe_5_add_theo}) $p$ times with respect to $\theta_1$ and
$k$ times with respect to $\theta_2$, and then make $\theta_1 = \theta_2 =
\theta$, which yields

\begin{equation}
\sum_{m = -l}^l \frac{\partial^p \Ylm(\theta,\phi)}{\partial \theta^p}
\frac{\partial^k {\Ylm}^{*}(\theta,\phi)}{\partial \theta^k} = (-1)^p \frac{(2l
  + 1)}{4 \pi} \frac{d^{k+p} P_l(1)}{d \gamma^{k+p}} \, .
\label{eq_annexe_5_deriv_2}
\end{equation}

\noindent For $\theta_1 = \theta_2$, we can exchange $k$ and $p$ and thus eq.(\ref{eq_annexe_5_deriv_1}) must be equal to eq.(\ref{eq_annexe_5_deriv_2}).
Accordingly, the sum

\begin{equation}
  \sum_{m=-l}^l \frac{\partial^k \Ylm (\theta,\phi)}{\partial \theta^k} \frac{\partial^p \Ylm(\theta,\phi)^*}{\partial \theta^p}
\end{equation}

\noindent vanishes for $k+p$ odd, implying:

\begin{eqnarray}
\langle u \, u_\theta \rangle &=& 0 \, ,
\label{eq_annexe_5_uut}\\
\langle u \, u_{\theta \theta \theta} \rangle &=& 0 \, .
\label{eq_annexe_5_uuttt}
\end{eqnarray}

\noindent Note that this does not hold when derivations with respect to $\phi$
are also involved.

Starting again from the addition theorem, one may show that

\begin{equation}
\sum_{m = -l}^l \frac{\partial^{k+1} \Ylm(\theta,\phi)}{\partial \theta^{k+1}}
\frac{\partial^{p-1} {\Ylm}^{*}(\theta,\phi)}{\partial \theta^{p-1}}  = -\sum_{m = -l}^l \frac{\partial^{k} \Ylm(\theta,\phi)}{\partial \theta^{k}}
\frac{\partial^{p} {\Ylm}^{*}(\theta,\phi)}{\partial \theta^{p}}\, .
\end{equation}

\noindent It follows that, when averaging the product of two terms, one may pass
the derivative on $\theta$ from one term to the other while multiplying by
$-1$.
This holds only, however, in the absence of derivatives with respect to $\phi$.
As a consequence

\begin{eqnarray}
\langle u_\theta \, u_\theta \rangle &=& - \langle u \, u_{\theta \theta}
\rangle \, ,\\
\langle u_\theta\, 
u_{\theta \theta \theta} \rangle  &=& - \langle u_{\theta \theta}
\, u_{\theta \theta } \rangle \, .
\end{eqnarray}

Finally, one may also use the fact that $\Delta \Ylm(\theta,\phi) = 0$, where $\Delta$
is the Laplacian in spherical coordinates, to obtain

\begin{equation}
\langle u_\theta \, u_{\theta \theta } \rangle = \cot\theta \langle u
\, u_{\theta \theta } \rangle - \csc^2\theta \langle u_{\theta} \, 
u_{\phi \phi} \rangle \, .
\end{equation}

\noindent Since $\langle u_\theta u_{\theta\theta} \rangle$ vanishes, one obtains

\begin{equation}
\langle u_{\theta} \, u_{\phi \phi} \rangle = - \sin\theta \cos\theta \, \langle
u_\theta^2\rangle \, .
\label{eq_annexe_5_utupp}
\end{equation}

In conclusion, eqs.(\ref{eq_annexe_5_upup})-(\ref{eq_annexe_5_ututpp}), together with eqs.(\ref{eq_annexe_5_upup1})--(\ref{eq_annexe_5_upputp}), eqs.(\ref{eq_annexe_5_uut})--(\ref{eq_annexe_5_uuttt}) and
eq.(\ref{eq_annexe_5_utupp}) give all the correlations needed to evaluate $\langle\bm{\Sigma}\rangle$.

%% file: these.bbl
\begin{thebibliography}{100}

\bibitem{Fournier_08_eu}
J.~B. FOURNIER and C.~BARBETTA.
\newblock {Direct calculation from the stress tensor of the lateral surface
  tension of fluctuating fluid membranes}.
\newblock {\em Physical review letters}, 100(7):78103, 2008.

\bibitem{Cai_94}
W.~CAI et~al.
\newblock Measure factors, tension and correlations of fluid membranes.
\newblock {\em J. Phys. II France}, 4:931--949, 1994.

\bibitem{Imparato_06}
A.~IMPARATO.
\newblock Surface tension in bilayer membranes with fixed projected area.
\newblock {\em The Journal of Chemical Physics}, 124:154714, 2006.

\bibitem{Fournier_07}
J.-B. FOURNIER.
\newblock On the stress and torque tensor in fluid membranes.
\newblock {\em Soft Matter}, 3:883--888, 2007.

\bibitem{Barbetta_10}
C.~BARBETTA et~al.
\newblock On the surface tension of fluctuating quasi-spherical vesicles.
\newblock {\em Eur. Phys. J. E}, 31(3):333--342, 2010.

\bibitem{Barbetta_09}
C.~BARBETTA and J.B. FOURNIER.
\newblock {On the fluctuations of the force exerted by a lipid nanotubule}.
\newblock {\em The European Physical Journal E: Soft Matter and Biological
  Physics}, 29(2):183--189, 2009.

\bibitem{Fournier_07a}
J.-B. FOURNIER and P.~GALATOLA.
\newblock {Critical Fluctuations of Tense Fluid Membrane Tubules}.
\newblock {\em Physical review letters}, 98(1):18103, 2007.

\bibitem{Baker_52}
J.~R. BAKER.
\newblock The cell-theory: a restatement, history, and critique.
\newblock {\em Quarterly Journal of Microscopical Science}, 93:157--190, 1952.

\bibitem{Robertson_81}
J.~D. ROBERTSON.
\newblock Membrane structure.
\newblock {\em The Journal of Cell Biology}, 91:189--204, 1981.

\bibitem{Edidin_03}
M.~EDIDIN.
\newblock Lipids on the frontier: a century of cell-membrane bilayer.
\newblock {\em Nature}, 4:414--418, 2003.

\bibitem{Heimburg_07}
T.~HEIMBURG.
\newblock {\em Thermal biophysics of membranes}.
\newblock Wiley, 2007.

\bibitem{Alberts}
B.~ALBERTS et~al.
\newblock {\em Molecular Biology of the Cell}.
\newblock Garland Science, 5th edition, 2008.

\bibitem{Bowman_bio}
G.~EKNOYAN.
\newblock Sir {W}illiam {B}owman: his contributions to the physiology and
  nephrology.
\newblock {\em Kidney International}, 50:2120--2128, 1996.

\bibitem{Bowman_1840}
W.~BOWMAN.
\newblock On the minute structure and movements of voluntary muscle.
\newblock {\em Philos. Trans. R. Soc. Lond.}, 130:457--501, 1840.

\bibitem{Overton}
E.~OVERTON.
\newblock The probable origin and physiological significance of cellular
  osmotic properties.
\newblock {\em Vierteljahrschrift der Naturfoschende Gesselschaft (Zurich)},
  44:88--135, 1899.

\bibitem{Mouritsen}
O.~MOURITSEN.
\newblock {\em Life as a matter of fat}.
\newblock Springer, 2005.

\bibitem{Fricke}
H.~FRICKE.
\newblock The electric capacity of suspensions with special reference to blood.
\newblock {\em Journal of General Physiology}, pages 137--152, 1925.

\bibitem{Robertson_59}
J.~D. ROBERTSON.
\newblock The ultrastructure of cell membranes and their derivatives.
\newblock {\em Biochem. Soc. Symp.}, 16:3--43, 1959.

\bibitem{Gorter_25}
E.~GORTER and F.~GRENDEL.
\newblock On bimolecular layers of lipoids on the chromocytes of the blood.
\newblock {\em J. Exp. Med.}, 41(4):439--443, 1925.

\bibitem{Balland_06}
M.~BALLAND et~al.
\newblock Power laws in microrheology experiments on living cells: Comparative
  analysis and modeling.
\newblock {\em Phys. Rev. E}, 74:21911, 2006.

\bibitem{Cole}
K.~S. COLE.
\newblock Surface forces of the arbacia egg.
\newblock {\em J. Cell. Comp. Physiol.}, 1:1--9, 1932.

\bibitem{Danielli_34}
F.~D. DANIELLI and E.~N. HARVEY.
\newblock The tension at the surface of mackerel egg oil, with remarks on the
  nature of the cell surface.
\newblock {\em J. Cell. Comp. Physiol.}, 5:483--494, 1934.

\bibitem{Danielli_35}
F.~D. DANIELLI and H.~DAVSON.
\newblock A contribution to the theory of permeability of thin films.
\newblock {\em J. Cell. Comp. Physiol.}, 5:495--508, 1935.

\bibitem{Fawcett}
W.~BLOOM and D.~W. FAWCETT.
\newblock {\em A Textbook of histology}.
\newblock Hodder Arnold, 12th edition, 1994.

\bibitem{Branton_66}
D.~BRANTON.
\newblock Fracture faces of frozen membranes.
\newblock {\em PNAS}, 55:1048--1056, 1966.

\bibitem{Silva_70}
P.~da~SILVA and D.~BRANTON.
\newblock Membrane splitting in freeze-etching: covalently bound ferritin as a
  membrane marker.
\newblock {\em J. Cell Biol.}, 45:598--605, 1970.

\bibitem{Frye_70}
L.~D. FRYE and M.~EDIDIN.
\newblock The rapid intermixing of cell surface antigens after formation of
  mouse-human heterokaryons.
\newblock {\em J. Cell Sci.}, 7:319--335, 1970.

\bibitem{Rothmann_77}
J.~E. ROTHMANN and J.~LENARD.
\newblock Membrane asymmetry.
\newblock {\em Science}, 195:743--753, 1997.

\bibitem{Simons_88}
K.~SIMONS and G.~VAN MEER.
\newblock Lipid sorting in epithelial cells.
\newblock {\em Biochemistry}, 12:6197--6202, 1988.

\bibitem{Edidin_97}
M.~EDIDIN.
\newblock Lipid microdomains in cell surface membranes.
\newblock {\em Current Opinion in Structural Biology}, 7:528--532, 1997.

\bibitem{Engelman_05}
D.~M. ENGELMAN.
\newblock Membranes are more mosaic than fluid.
\newblock {\em Nature}, 438:578--580, 2005.

\bibitem{Donofrio_03}
T.~G.~D'ONOFRIO et~al.
\newblock Controlling and measuring the interdependence of local properties of
  biomembranes.
\newblock {\em Langmuir}, 19:1618--1623, 2003.

\bibitem{Almeida_92}
P.~F. F.~ALMEIDA et~al.
\newblock Lateral diffusion in the liquid phases of
  dimyristoylphosphatidylcholine/cholesterol lipid bilayers: a free volume
  analysis.
\newblock {\em Biochemistry}, 31:6739--6747, 1992.

\bibitem{Veatch_03}
S.~L. VEATCH and S.~L. KELLER.
\newblock Separation of liquid phases in giant vesicles of ternary mixtures of
  lipids and cholesterol.
\newblock {\em Biophysical Journal}, 85:3074--3083, 2003.

\bibitem{Kusumi_82}
A.~KUSUMI and J.~S. HYDE.
\newblock Spin-label saturation-transfer electron spin resonance detection of
  transient association of rhodopsin in reconstituted membranes.
\newblock {\em Biochemistry}, 21:5978--5983, 1982.

\bibitem{Saxton_97}
M.~J. SAXTON and K.~JACOBSON.
\newblock Single particle tracking: application to membrane dynamics.
\newblock {\em Annu. Rev. Biophys. Biomol. Struct.}, 26:373--399, 1997.

\bibitem{Kusumi_93}
A.~KUSUMI et~al.
\newblock Confined lateral diffusion of membrane receptors as studied by single
  particle tracking (nanovid microscopy). {E}ffects of calcium-induced
  differentiation in cultured epithelial cells.
\newblock {\em Biophysical Journal}, 65:2021--2040, 1993.

\bibitem{Forstner_03}
M.~B.~FORSTNER et~al.
\newblock Simultaneous single-particle tracking and visualization of domain
  structure on lipid monolayers.
\newblock {\em Langmuir}, 19:4876--4879, 2003.

\bibitem{Gaus_03}
K.~GAUS et~al.
\newblock Visualizing lipid structures and raft domains in living cells with
  two-photon microscopy.
\newblock {\em PNAS}, 100:15554--15559, 2003.

\bibitem{Edidin_03b}
M.~EDIDIN.
\newblock The state of lipid rafts: from model membranes to cells.
\newblock {\em Annu. Rev. Biophys. Biomol. Struct.}, 32:257--283, 2003.

\bibitem{Pike_09}
L.~J. PIKE.
\newblock The challenge of lipid rafts.
\newblock {\em Journal of Lipid Research}, 50:S323, 2009.

\bibitem{Mouritsen_84}
O.~G. MOURITSEN and M.~BLOOM.
\newblock Mattress model of lipid-protein interactions in membranes.
\newblock {\em Biophys. J.}, 46:141--153, 1984.

\bibitem{Orwar_03}
O.~ORWAR et~al.
\newblock Nanofluidic networks based on surfactant membrane technology.
\newblock {\em Anal. Chem.}, 75:2529--2537, 2003.

\bibitem{Lobovkina_04}
T.~LOBOVKINA et~al.
\newblock Mechanical tweezer action of self-tightening knots in surfactant
  nanotubes.
\newblock {\em PNAS}, 101:7949--7953, 2004.

\bibitem{Misra_09}
N.~MISRA et~al.
\newblock Bioelectronic silicon nanowire devices using functional membrane
  proteins.
\newblock {\em PNAS}, 106:13780--13784, 2009.

\bibitem{Kornberg_71}
R.~D. KORNBERG and H.~M.~MC CONNELL.
\newblock Inside-outside translocation of phospholipids in vesicle membranes.
\newblock {\em Biochemistry}, 10:1111--1120, 1971.

\bibitem{Castellana_06}
E.T. CASTELLANA and P.~S. CREMER.
\newblock Solid supported lipid bilayer: from biophysical studies to sensor
  design.
\newblock {\em Surface Science Reports}, 61:429--444, 2006.

\bibitem{Sonnleitner_99}
A.~SONNLEITNER et~al.
\newblock Free brownian motion of individual lipid molecules in biomembranes.
\newblock {\em Biophysical Journal}, 77:2638--2642, 1999.

\bibitem{Winterhalter_00}
M.~WINTERHALTER.
\newblock Black lipid membranes.
\newblock {\em Current Opinion in Colloid and Interfaces Science}, 5:250--255,
  2000.

\bibitem{Aimon_per}
S.~AIMON.
\newblock private communication.

\bibitem{Hole_www}
M.~MONTAL.
\newblock http://www.whatislife.com/education/fact/making\_membrane.html, 2003.

\bibitem{Dimova_06}
R.~DIMOVA et~al.
\newblock A practical guide to giant vesicles. {P}robing the membrane
  nanoregime via optical microscopy.
\newblock {\em J. Phys.: Condens. Matter}, 18:S1151--S1176, 2006.

\bibitem{Dobereiner_97}
H.~G.~D{\"O}BEREINER et~al.
\newblock Mapping vesicle shapes into the phase diagram: a comparison of
  experiment and theory.
\newblock {\em Physical Review E}, 55(4):4458--4474, 1997.

\bibitem{Tresset_09}
G.~TRESSET.
\newblock The multiple faces of self-assembled lipidic systems.
\newblock {\em PMC Biophysics}, 2(3), 2009.

\bibitem{Rawicz_00}
W.~RAWICZ.
\newblock Effect of chain length and unsaturation on elasticity of lipid
  bilayers.
\newblock {\em Biophysical Journal}, 79:328--339, 2000.

\bibitem{Salditt_03}
T.~SALDITT et~al.
\newblock Thermal fluctuations and positional correlations in oriented lipid
  membranes.
\newblock {\em PRL}, 90:178101, 2003.

\bibitem{Nasa_03}
D.~PETTIT.
\newblock http://spaceflight.nasa.gov/station/crew/exp6/spacechronicles17.html,
  2003.

\bibitem{Chatain_04}
D.~CHATAIN and W.~C. CARTER.
\newblock {Wetting dynamics: Spreading of metallic drops}.
\newblock {\em Nature Materials}, 3(12):843--845, 2004.

\bibitem{Gruhn_07}
T.~GR{\"U}HN et~al.
\newblock Novel method for measuring the adhesion energy of vesicles.
\newblock {\em Langmuir}, 23:5423--5429, 2007.

\bibitem{Seifert_95}
U.~SEIFERT.
\newblock The concept of effective tension for fluctuating vesicles.
\newblock {\em Zeitschrift für Physik B}, 97:299--309, 1995.

\bibitem{Helfrich_73}
W.~HELFRICH.
\newblock Elastic properties of bilayers: Theory and possible experiments.
\newblock {\em Z. Naturforch.}, 28c:693--703, 1973.

\bibitem{Svetina_89}
S.~SVETINA and B.~ZEKS.
\newblock Membrane bending energy and shape determination of phospholipid
  vesicles and red blood cells.
\newblock {\em European Biophysics Journal}, 17:101--111, 1989.

\bibitem{Sheetz_74}
M.~P. SHEETZ and S.~J. SINGER.
\newblock Biological membranes as bilayer couples. a molecular mechanism of
  drug erythrocyte interactions.
\newblock {\em PNAS}, 71:4457--4461, 1974.

\bibitem{Evans_74}
E.~A. EVANS.
\newblock Bending resistance and chemically induced moments in membrane
  bilayers.
\newblock {\em Biophysical Journal}, 14:923--931, 1974.

\bibitem{Miao_94}
L.~MIAO et~al.
\newblock Budding transitions of fluid-bilayer vesicles: The effect of
  area-difference elasticity.
\newblock {\em PRE}, 49:5389--5407, 1994.

\bibitem{Dobereiner_00}
H.-G. D{\"O}BEREINER.
\newblock Properties of giant vesicles.
\newblock {\em Current Opinion in Colloid and Interface Science}, 5:256--263,
  2000.

\bibitem{Seifert_91}
U.~SEIFERT et~al.
\newblock Shape transformation of vesicles: phase diagram for
  spontaneous-curvature and bilayer-coupling models.
\newblock {\em Physical Review A}, 44:1182--1202, 1991.

\bibitem{David_91}
F.~DAVID and S.~LEIBLER.
\newblock {Vanishing tension of fluctuating membranes}.
\newblock {\em J. Phys. II France}, 1:959--976, 1991.

\bibitem{Peliti_85}
L.~PELITI and S.~LEIBLER.
\newblock Effects of thermal fluctuations on systems with small surface
  tension.
\newblock {\em PRL}, 54:1690--1693, 1985.

\bibitem{Gompper_96}
G.~GOMPPER and D.~M. KROLL.
\newblock {Random surface discretizations and the renormalization of the
  bending rigidity}.
\newblock {\em Journal de Physique I}, 6(10):1305--1320, 1996.

\bibitem{Monteiro_04}
A.~M. F.~MONTEIRO et~al.
\newblock Gliadin effect on fluctuation properties of phospholipid giant
  vesicles.
\newblock {\em Colloids and Surfaces B: Biointerfaces}, 34:53--57, 2004.

\bibitem{Henriksen_04}
J.~R. HENRIKSEN and J.~H. IPSEN.
\newblock {Measurement of membrane elasticity by micro-pipette aspiration}.
\newblock {\em The European Physical Journal E: Soft Matter and Biological
  Physics}, 14(2):149--167, 2004.

\bibitem{Evans_90}
E.~EVANS and W.~RAWICZ.
\newblock Entropy-driven tension and bending elasticity in condensed-fluid
  membranes.
\newblock {\em PRL}, 64:2094--2097, 1990.

\bibitem{Fournier_08}
J.-B. FOURNIER and P.~GALATOLA.
\newblock Corrections to the laplace law for vesicle aspiration in
  micropipettes and other confined geometries.
\newblock {\em Soft Matter}, 4:2463--2470, 2008.

\bibitem{Olbrich_00}
K.~OLBRICH et~al.
\newblock Water permeability and mechanical strength of polyunsaturated lipid
  bilayers.
\newblock {\em Biophysical Journal}, 79:321--327, 2000.

\bibitem{Dimova_02}
R.~DIMOVA et~al.
\newblock Hyperviscous diblock copolymer vesicles.
\newblock {\em Eur. Phys. J. E}, 7:241--250, 2002.

\bibitem{Hirn_99}
R.~HIRN et~al.
\newblock Collective membrane motions in the mesoscopic range and their
  modulation by the binding of a monomolecular protein layer of streptavidin
  studied by dynamic light scattering.
\newblock {\em Physical Review E}, 59:5987--5994, 1999.

\bibitem{Pecreaux_04}
J.~PECREAUX et~al.
\newblock Refined contour analysis of giant unilamellar vesicles.
\newblock {\em Eur. Phys. J. E}, 13:277--290, 2004.

\bibitem{Sheetz_01}
M.~P. SHEETZ.
\newblock Cell control by membrane-cytoskeleton adhesion.
\newblock {\em Nature Reviews Molecular Cell Biology}, 2:392--396, 2001.

\bibitem{Seifert_90}
U.~SEIFERT and R.~LIPOWSKY.
\newblock {Adhesion of vesicles}.
\newblock {\em Physical Review A}, 42(8):4768--4771, 1990.

\bibitem{Evans_85}
E.~A. EVANS.
\newblock Detailed mechanics of membrane-membrane adhesion and separation.
\newblock {\em Biophys. J.}, 48:175--183, 1985.

\bibitem{Bailey_90}
S.~M.~BAILEY et~al.
\newblock {Measurements of forces involved in vesicle adhesion using
  freeze-fracture electron microscopy}.
\newblock {\em Langmuir}, 6(7):1326--1329, 1990.

\bibitem{Evans_80}
E.~A. EVANS.
\newblock Analysis of adhesion of large vesicles to surfaces.
\newblock {\em Biophys. J.}, 31:425--432, 1980.

\bibitem{Raedler_95}
J.~O.~R{\"A}DLER et~al.
\newblock Fluctuation analysis of tension-controlled undulation forces between
  giant vesicles and solid substrates.
\newblock {\em PRE}, 51:4526--4636, 1995.

\bibitem{Bruinsma_00}
R.~BRUINSMA et~al.
\newblock {Adhesive switching of membranes: experiment and theory}.
\newblock {\em Physical Review E}, 61(4):4253--4267, 2000.

\bibitem{Puech_04}
P.~H. PUECH and F.~BROCHARD-WYART.
\newblock {Membrane tensiometer for heavy giant vesicles}.
\newblock {\em The European Physical Journal E: Soft Matter and Biological
  Physics}, 15(2):127--132, 2004.

\bibitem{Sengupta_10}
K.~SENGUPTA and L.~LIMOZIN.
\newblock Adhesion of soft membranes controlled by tension and interfacial
  polymers.
\newblock {\em PRL}, 29:345--350, 2010.

\bibitem{Albersdorfer_97}
A.~ALBERSD{\"O}RFER et~al.
\newblock {Adhesion-induced domain formation by interplay of long-range
  repulsion and short-range attraction force: a model membrane study}.
\newblock {\em Biophysical journal}, 73(1):245--257, 1997.

\bibitem{Bruinsma_95}
R.~BRUINSMA.
\newblock {Adhesion and rolling of leukocytes: a physical model}.
\newblock In {\em Proc. NATO Adv. Inst. Phys. Biomater. NATO ASI Ser}, volume
  332, pages 61--75, 1995.

\bibitem{Simson_98}
R.~SIMSON et~al.
\newblock {Membrane bending modulus and adhesion energy of wild-type and mutant
  cells of Dictyostelium lacking talin or cortexillins}.
\newblock {\em Biophysical journal}, 74(1):514--522, 1998.

\bibitem{Iglic_03}
A.~IGLIC et~al.
\newblock Possible role of phospholipid nanotubes in directed transport of
  membrane vesicles.
\newblock {\em Physics Letters A}, 310:493--497, 2003.

\bibitem{Onfelt_04}
B.~ONFELT and D.~M. DAVIS.
\newblock Can membrane nanotubes facilitate communication between immune cells?
\newblock {\em Biochemical Society Transactions}, 32:676--678, 2004.

\bibitem{Eugenin_09}
E.~A.~EUGENIN et~al.
\newblock Tunneling nanotubes ({TNT}) are induced by {HIV}-infection of
  macrophages: a potential mechanism for intercellular {HIV} trafficking.
\newblock {\em Cellular Immunology}, 254:142--148, 2009.

\bibitem{Gerdes_07}
H.-H.~GERDES et~al.
\newblock Tunneling nanotubes: a new route for the exchange of components
  between animal cells.
\newblock {\em FEBS Letters}, 581:2194--2201, 2007.

\bibitem{Waugh_95}
R.~E. WAUGH and R.~G. BAUSERMAN.
\newblock Physical measurements of bilayer-skeletal separation forces.
\newblock {\em Annals of biomedical engineering}, 23:308--321, 1995.

\bibitem{Heinrich_96}
V.~HEINRICH and R.~E. WAUGH.
\newblock {A piconewton force transducer and its application to measurement of
  the bending stiffness of phospholipid membranes.}
\newblock {\em Annals of biomedical engineering}, 24(5):595, 1996.

\bibitem{Bashirov_07}
P.~V. BASHIROV.
\newblock Membrane nanotubes in the electric field as a model for measurement
  of mechanical parameters of the lipid bilayer.
\newblock {\em Biochemistry (Moscow) Supplement Series A: Membrane and Cell
  Biology}, 1:176--184, 2007.

\bibitem{Koster_05}
G.~KOSTER et~al.
\newblock Force barrier for membrane tube formation.
\newblock {\em PRL}, 94:068101, 2005.

\bibitem{Rossier_03}
O.~ROSSIER et~al.
\newblock Giant vesicles under flows: extrusion and retraction of tubes.
\newblock {\em Langmuir}, 19:575--584, 2003.

\bibitem{Bo_89}
L.~BO and R.~E. WAUGH.
\newblock Determination of bilayer membrane bending stiffness by tether
  formation from giant, thin-walled vesicles.
\newblock {\em Biophy. J.}, 55:509--517, 1989.

\bibitem{Leduc_04}
C.~LEDUC et~al.
\newblock Cooperative extraction of membrane nanotubes by molecular motors.
\newblock {\em PNAS}, 101:17096--17101, 2004.

\bibitem{Derenyi_02}
I.~DERENYI et~al.
\newblock Formation and interaction of membrane tubes.
\newblock {\em PRL}, 88:238101--2, 2002.

\bibitem{Capovilla_02}
R.~CAPOVILLA and J.~GUVEN.
\newblock Stresses in lipid membranes.
\newblock {\em J. Phys. A: Math. Gen.}, 35:6233--6247, 2002.

\bibitem{Farago_04}
O.~FARAGO and P.~PINCUS.
\newblock {Statistical mechanics of bilayer membrane with a fixed projected
  area}.
\newblock {\em The Journal of chemical physics}, 120:2934, 2004.

\bibitem{Nelson_93}
P.~NELSON and T.~POWERS.
\newblock {Renormalization of chiral couplings in titled bilayer membranes}.
\newblock {\em J. Phys. II France}, 3:1535--1569, 1993.

\bibitem{Krauth}
W.~KRAUTH.
\newblock {\em {Statistical mechanics: algorithms and computations}}.
\newblock Oxford University Press, USA, 2006.

\bibitem{Numerical_Recipes}
W.H. PRESS, S.A. TEUKOLSKY, W.T. VETTERLING, and B.P. FLANNERY.
\newblock {\em {Numerical recipes in C}}.
\newblock Cambridge Univ. Press Cambridge MA, USA:, 1992.

\bibitem{Seifert_95b}
U.~SEIFERT.
\newblock Self-consistent theory of bound vesicles.
\newblock {\em PRL}, 74:5060--5063, 1995.

\bibitem{Neder_10}
J.~NEDER et~al.
\newblock {Coarse-Grained Simulations of Membranes under Tension}.
\newblock {\em The Journal of chemical physics}, 132:115101, 2010.

\bibitem{Helfrich_86}
W.~HELFRICH.
\newblock {Size distributions of vesicles: the role of the effective rigidity
  of membranes}.
\newblock {\em Journal de Physique}, 47(2):321--329, 1986.

\bibitem{Milner_87}
S.T. MILNER and S.A. SAFRAN.
\newblock {Dynamical fluctuations of droplet microemulsions and vesicles}.
\newblock {\em Physical Review A}, 36(9):4371--4379, 1987.

\bibitem{Powers_02}
T.R.~POWERS et~al.
\newblock {Fluid-membrane tethers: minimal surfaces and elastic boundary
  layers}.
\newblock {\em Physical Review E}, 65(4):41901, 2002.

\bibitem{Bozic_01}
B.~BO{\v{Z}}I{\v{C}} et~al.
\newblock {Shapes of nearly cylindrical, axisymmetric bilayer membranes}.
\newblock {\em The European Physical Journal E: Soft Matter and Biological
  Physics}, 6(1):91--98, 2001.

\bibitem{Ou-Zhong_89}
O.~Y. ZHONG-CAN and W.~HELFRICH.
\newblock {Bending energy of vesicle membranes: General expressions for the
  first, second, and third variation of the shape energy and applications to
  spheres and cylinders}.
\newblock {\em Physical Review A}, 39(10):5280--5288, 1989.

\bibitem{Nelson_95}
P.~NELSON et~al.
\newblock {Dynamical theory of the pearling instability in cylindrical
  vesicles}.
\newblock {\em Physical review letters}, 74(17):3384--3387, 1995.

\bibitem{Waugh_92}
R.~E.~WAUGH et~al.
\newblock {Local and nonlocal curvature elasticity in bilayer membranes by
  tether formation from lecithin vesicles}.
\newblock {\em Biophysical journal}, 61(4):974--982, 1992.

\bibitem{Heinrich_99}
et~al. V.~HEINRICH.
\newblock {Vesicle deformation by an axial load: from elongated shapes to
  tethered vesicles}.
\newblock {\em Biophysical journal}, 76(4):2056--2071, 1999.

\bibitem{Santangelo_02}
C.~D. SANTANGELO and P.~PINCUS.
\newblock {Coiling instabilities of multilamellar tubes}.
\newblock {\em Physical Review E}, 66(6):61501, 2002.

\bibitem{Cuvelier_05}
D.~CUVELIER et~al.
\newblock {Coalescence of membrane tethers: experiments, theory, and
  applications}.
\newblock {\em Biophysical journal}, 88(4):2714--2726, 2005.

\bibitem{Inaba_05}
T.~INABA et~al.
\newblock Formation and maintenance of tubular membrane projections require
  mechanical force, but their elongation and shortening do not require
  assitional force.
\newblock {\em J. Mol. Biol.}, 348:325--333, 2005.

\bibitem{Manneville_01}
J.~B.~MANNEVILLE et~al.
\newblock {Active membrane fluctuations studied by micropipet aspiration}.
\newblock {\em Phys. Rev. Lett}, 82:4356, 1999.

\bibitem{Muller_06}
M.~M{\"U}LLER et~al.
\newblock {Biological and synthetic membranes: What can be learned from a
  coarse-grained description?}
\newblock {\em Physics Reports}, 434(5-6):113--176, 2006.

\bibitem{Orsi_07}
M.~ORSI et~al.
\newblock {Coarse-grain modelling of lipid bilayers: a literature review}.
\newblock {\em Molecular Interactions: Bringing Chemistry to Life}, pages
  185--205, 2007.

\bibitem{Noguchi_08}
H.~NOGUCHI.
\newblock {Membrane simulation models from nm to $\mu$ m scale}.
\newblock {\em Arxiv preprint arXiv:0812.0055}, 2008.

\bibitem{Noguchi_06}
H.~NOGUCHI and G.~GOMPPER.
\newblock {Meshless membrane model based on the moving least-squares method}.
\newblock {\em Physical Review E}, 73(2):21903, 2006.

\bibitem{Maibaum_10}
L.~MAIBAUM et~al.
\newblock {Large-Scale Simulations of Fluctuating Biological Membranes}.
\newblock {\em Biophysical Journal}, 98:10, 2010.

\bibitem{Drouffe_91}
J.~M.~DROUFFE et~al.
\newblock {Computer simulations of self-assembled membranes}.
\newblock {\em Science}, 254(5036):1353, 1991.

\bibitem{Kantor_87}
Y.~KANTOR and D.~R. NELSON.
\newblock {Phase transitions in flexible polymeric surfaces}.
\newblock {\em Physical Review A}, 36(8):4020--4032, 1987.

\bibitem{Gompper_97}
G.~GOMPPER and D.~M. KROLL.
\newblock {Network models of fluid, hexatic and polymerized membranes}.
\newblock {\em Journal of Physics: Condensed Matter}, 9:8795--8834, 1997.

\bibitem{Billoire_84}
A.~BILLOIRE et~al.
\newblock {Simulating random surfaces}.
\newblock {\em Physics Letters B}, 139(1-2):75--80, 1984.

\bibitem{Espriu_87}
D.~ESPRIU.
\newblock {Triangulated random surfaces}.
\newblock {\em Physics Letters B}, 194(2):271--276, 1987.

\bibitem{David_85}
F.~DAVID.
\newblock {A model of random surfaces with non-trivial critical behaviour}.
\newblock {\em Nuclear Physics B}, 257:543--576, 1985.

\bibitem{Ambjorn_85}
B.~AMBJORN et~al.
\newblock {Diseases of triangulated random surface models, and possible cures}.
\newblock {\em Nuclear Physics B}, 257:433--449, 1985.

\bibitem{Kantor_87a}
Y.~KANTOR and D.~R. NELSON.
\newblock {Crumpling transition in polymerized membranes}.
\newblock {\em Physical review letters}, 58(26):2774--2777, 1987.

\bibitem{Kantor_87b}
Y.~KANTOR et~al.
\newblock {Tethered surfaces: Statics and dynamics}.
\newblock {\em Physical Review A}, 35(7):3056--3071, 1987.

\bibitem{Kantor_87c}
Y.~KANTOR and D.~R. NELSON.
\newblock {Phase transitions in flexible polymeric surfaces}.
\newblock {\em Physical Review A}, 36(8):4020--4032, 1987.

\bibitem{Ho_90}
J.~S. HO and A.~BAUMGARTNER.
\newblock {Crumpling of fluid vesicles}.
\newblock {\em Physical Review A}, 41(10):5747--5750, 1990.

\bibitem{Kroll_92}
D.~M. KROLL and G.~GOMPPER.
\newblock {The conformation of fluid membranes: Monte Carlo simulations}.
\newblock {\em Science}, 255(5047):968, 1992.

\bibitem{Noguchi_05}
H.~NOGUCHI and G.~GOMPPER.
\newblock {Shape transitions of fluid vesicles and red blood cells in capillary
  flows}.
\newblock {\em Proceedings of the National Academy of Sciences of the United
  States of America}, 102(40):14159, 2005.

\bibitem{Noguchi_10}
H.~NOGUCHI et~al.
\newblock {Dynamics of fluid vesicles in flow through structured
  microchannels}.
\newblock {\em EPL (Europhysics Letters)}, 89:28002, 2010.

\bibitem{Atilgan_07}
E.~ATILGAN and S.~X. SUN.
\newblock {Shape transitions in lipid membranes and protein mediated vesicle
  fusion and fission}.
\newblock {\em The Journal of chemical physics}, 126:095102, 2007.

\bibitem{Ooura}
T.~OOURA.
\newblock http://momonga.t.u-tokyo.ac.jp/~ooura/fft.html.

\bibitem{Osgood}
G.~OSGOOD.
\newblock {\em The Fourier Transform and its Applications}.
\newblock 2009.

\bibitem{Netz_95}
R.~R. NETZ and R.~LIPOWSKY.
\newblock Stacks of fluid membranes under pressure and tension.
\newblock {\em Europhys. Lett.}, 29:345--350, 1995.

\bibitem{Lipowsky_94}
R.~LIPOWSKY.
\newblock Generic interations of flexible membranes.
\newblock {\em in Structure and Dynamics of Membranes, edited by R. Lipowsky
  and E. Sackmann}, 1B, 1994.

\end{thebibliography}
